\documentclass{article}
\usepackage[margin=.8in,left=.8in]{geometry}

\usepackage{amsthm}
\usepackage{amssymb}
\usepackage{amsmath}
\usepackage{mathtools}
\usepackage{adjustbox}
\usepackage{xcolor}
\usepackage{stmaryrd}    
\usepackage{bbding} 
\usepackage{rotating}
\usepackage{xfrac}
\usepackage{enumitem}\setlist{nosep}
\usepackage{ wasysym } 
\usepackage{multirow}
\usepackage{multicol}
\usepackage{hhline}
\usepackage{bbm}

\usepackage{mathpartir}

\usepackage{tikz}
\usetikzlibrary{
  backgrounds,
  arrows,
  decorations.pathmorphing,
  decorations.markings,
}
\usetikzlibrary{cd}

\usepackage{mathrsfs} 
\DeclareMathAlphabet{\mathpzc}{OT1}{pzc}{m}{it} 

\usepackage[safe]{tipa} 

\newcommand{\shape}{
  \raisebox{1pt}{\rm\normalfont\textesh}
}

\usepackage{hyperref}
\hypersetup{
  colorlinks = true,
  allcolors=.
}

\usepackage{cleveref}
\crefformat{section}{\S#2#1#3} 
\crefformat{subsection}{\S#2#1#3}
\crefformat{subsubsection}{\S#2#1#3}

\definecolor{darkblue}{rgb}{0.05,0.25,0.65}
\definecolor{darkgreen}{RGB}{20,140,10}
\definecolor{lightgray}{rgb}{0.9,0.9,0.9}
\definecolor{darkorange}{RGB}{200,100,5}
\definecolor{darkyellow}{rgb}{.91,.91,0}
\definecolor{lightolive}{RGB}{189,183,107}

\newtheorem{theorem}{Theorem}[section]
\newtheorem{lemma}[theorem]{Lemma}

\newtheorem{proposition}[theorem]{Proposition}
\newtheorem{corollary}[theorem]{Corollary}
\theoremstyle{definition}
\newtheorem{definition}[theorem]{Definition}

\newtheorem{example}[theorem]{Example}

\newtheorem{remark}[theorem]{Remark}
\newtheorem{literature}[theorem]{Literature}

\newcommand*\circled[1]{\tikz[baseline=(char.base)]{\node[shape=circle,draw,inner sep=1pt] (char) {#1};}}

\newcommand{\proofstep}[1]{
  \mbox{\small #1}
}

\def\termsize{.9}
\def\termscale{.87}

\newcommand{\Sets}{
  \mathrm{Set}
}
\newcommand{\Set}{\Sets}

\newcommand{\FiniteSets}{
  \mathrm{Fin}\Sets
}

\newcommand{\Groupoids}{\mathrm{Grpd}}

\newcommand{\RealSets}
{\Sets^{\mathbf{B}\ZTwo}}

\tikzset{
    vertlabl/.style={anchor=south, rotate=90, inner sep=.5mm}
}

\newcommand{\CyclicGroup}[1]{\mathbb{Z}_{#1}}
\newcommand{\ZTwo}{
  \CyclicGroup{2}
}

\newcommand{\NaturalNumbers}{\mathbb{N}}

\newcommand{\RealNumbers}{\mathbb{R}}

\newcommand{\GroundField}{\mathbb{K}}

\newcommand{\ImaginaryUnit}{
  \mathrm{i}
}
\newcommand{\ComplexNumbers}{\mathbb{C}}

\newcommand{\Modules}[1]{\mathrm{Mod}_{\scalebox{.7}{$#1$}}}

\newcommand{\DependentModules}[2]{\Modules{#1}^{\scalebox{.7}{$#2$}}}

\newcommand{\defneq}{\equiv}

\newcommand{\bind}[2]{\mbox{\tt bind}^{\scalebox{.7}{$#1$}}_{\scalebox{.7}{$#2$}}}

\newcommand{\extend}[2]{\mbox{\tt extend}^{\scalebox{.7}{$#1$}}_{\scalebox{.7}{$#2$}}}

\newcommand{\transform}[2]{\mbox{\tt trans}^{\scalebox{.7}{$#1$}}_{\scalebox{.7}{$#2$}}}

\newcommand{\distribute}[2]{\mbox{\tt distr}^{\scalebox{.7}{$#1$}}_{\scalebox{.7}{$#2$}}}

\newcommand{\join}[2]{\mathrm{join}^{\scalebox{.7}{$#1$}}_{\scalebox{.7}{$#2$}}}

\newcommand{\duplicate}[2]{\mathrm{dupl}^{\scalebox{.7}{$#1$}}_{\scalebox{.7}{$#2$}}}

\newcommand{\pair}[2]{\mathrm{pair}^{\scalebox{.7}{$#1$}}_{\scalebox{.7}{$#2$}}}

\newcommand{\copair}[2]{\mathrm{copair}^{\scalebox{.7}{$#1$}}_{\scalebox{.7}{$#2$}}}

\newcommand{\braiding}[2]{\mathrm{braid}^{\scalebox{.7}{$#1$}}_{\scalebox{.7}{$#2$}}}
\newcommand{\braid}[2]{\braiding{#1}{#2}}

\newcommand{\handle}[3]{\mbox{\tt handle}^{\scalebox{.7}{$#1$}}_{\scalebox{.7}{$#2$}}{#3}}
\newcommand{\provide}[3]{\mbox{\tt provide}^{\scalebox{.7}{$#1$}}_{\scalebox{.7}{$#2$}}{#3}}

\newcommand{\AnyTypes}{\mathrm{Type}}

\newcommand{\BundleTypes}{\AnyTypes}

\newcommand{\ABundleType}[2]{
\left[
\def\arraystretch{1}\def\arraycolsep{1pt}
  \begin{array}{c}
    {#1}
    \\
    \downarrow
    \\
    {#2}
  \end{array}
\right]
}

\newcommand{\Types}{\AnyTypes}
\newcommand{\Type}{\Types}

\newcommand{\Propositions}{\mathrm{Prop}}
\newcommand{\Proposition}{\Propositions}
\newcommand{\Prop}{\Propositions}

\newcommand{\ClassicalTypes}{\mathrm{Cla}\Types}
\newcommand{\ClassicalType}{\ClassicalTypes}

\newcommand{\FinTypes}{\ClassicalTypes^{\mathrm{fin}}}

\newcommand{\FiniteType}{\FinTypes}
\newcommand{\FiniteTypes}{\FinTypes}

\newcommand{\FiniteClassicalType}{\FinTypes}

\newcommand{\LinTypes}{\mathrm{Qu}\AnyTypes}

\newcommand{\LinearType}{\LinTypes}
\newcommand{\LinearTypes}{\LinTypes}
\newcommand{\QuantumType}{\LinTypes}
\newcommand{\QuantumTypes}{\QuantumType}

\newcommand{\QuantumEffects}{\mathrm{QuEffect}}

\newcommand{\QuantumStateEffects}{\mathrm{QuStateEffect}}

\newcommand{\DualizableQuantumTypes}{\QuantumTypes^{\mathrm{fdm}}}
\newcommand{\DualizableQuantumType}{\DualizableQuantumTypes}

\newcommand{\HilbertSpace}[1]{\mathcal{#1}}

\newcommand{\Bit}{\mathrm{Bit}}
\newcommand{\Bits}{\Bit}

\newcommand{\QBit}{\mathrm{QBit}}
\newcommand{\QBits}{\QBit}

\newcommand{\LogicalQBit}{\mathrm{Lgcl}\QBit}

\newcommand{\Modales}[2]{{#2}^{\scalebox{.7}{${#1}$}}}
\newcommand{\FreeModales}[2]{{#2}_{\scalebox{.7}{${#1}$}}}

\newcommand{\isa}{:}
\newcommand{\isan}{\isa}
\newcommand{\isalin}{\raisebox{.75pt}{\scalebox{.6}{$\def\arraystretch{.66}\begin{array}{c}\circ\\\circ\end{array}$}}}

\newcommand{\yields}{\;\;\mathbin{\vdash}\;\;}
\newcommand{\name}{\vdash}

\newcommand{\declare}[3]{
    \def\arraystretch{1.7}
  \arraycolsep=3pt
  \begin{array}{rcc}
    {#1} &\isa& {#2}
    \\
    {#1} &\defneq& {#3}
  \end{array}
}

\newcommand{\declareLeftAligned}[3]{
    \def\arraystretch{1.7}
  \arraycolsep=3pt
  \begin{array}{rcl}
    {#1} &\isa& {#2}
    \\
    {#1} &\defneq& {#3}
  \end{array}
}

\newcommand{\maplin}{\multimap}
\newcommand{\linmap}{\multimap}

\newcommand{\declarelin}[3]{
    \def\arraystretch{1.7}
  \arraycolsep=3pt
  \begin{array}{rcc}
    {#1} &\isalin& {#2}
    \\
    {#1} &\defneq& {#3}
  \end{array}
}

\newcommand{\declarelinLeftAligned}[3]{
    \def\arraystretch{1.7}
  \arraycolsep=3pt
  \begin{array}{rcl}
    {#1} &\isalin& {#2}
    \\
    {#1} &\defneq& {#3}
  \end{array}
}

\newcommand{\linearly}{\hspace{-2pt}\raisebox{-2pt}{\scalebox{.95}{\rotatebox{30}{$\triangle$}}}\hspace{-1pt}}
\newcommand{\quantized}{\mathrm{Q}}
\newcommand{\classicized}{\mathrm{C}}

\newcommand{\ExponentialModality}{!}

\newcommand{\classically}{\natural}
\newcommand{\classical}{\classically}

\newcommand{\quantumly}{\linearly}

\newcommand{\necessarily}{\raisebox{-.9pt}{\scalebox{1.07}{$\Box$}}}
\newcommand{\necessity}{\necessarily}

\newcommand{\possibly}{\scalebox{1.06}{$\lozenge$}}

\newcommand{\randomly}{\raisebox{-1pt}{\scalebox{1}{\FiveStarOpen}}}
\newcommand{\randomness}{\randomly}
\newcommand{\indefinitely}{\raisebox{.7pt}{\scalebox{.9}{$\bigcirc$}}}
\newcommand{\indefiniteness}{\indefinitely}

\newcommand{\State}{\mathrm{State}}
\newcommand{\Store}{\mathrm{Store}}

\newcommand{\Environment}{\mathrm{Envm}}

\newcommand{\TensorUnit}{\adjustbox{scale=1.1,raise=-.7pt}{$\mathbbm{1}$}}

\newcommand{\osum}{\scalebox{1.2}{$\oplus$}}

\newcommand{\heart}{\heartsuit}

\newcommand{\unit}[2]{\mathrm{ret}^{\scalebox{.7}{$#1$}}_{\scalebox{.7}{$#2$}}}
\newcommand{\counit}[2]{\mathrm{obt}^{\scalebox{.7}{$#1$}}_{\scalebox{.7}{$#2$}}}

\newcommand{\multiplication}[2]{\mathrm{join}^{\scalebox{.7}{$#1$}}_{\scalebox{.7}{$#2$}}}
\newcommand{\comultiplication}[2]{\mathrm{dplc}^{\scalebox{.7}{$#1$}}_{\scalebox{.7}{$#2$}}}

\newcommand{\pipe}{\;\;\mbox{\tt >}\;\;}

\newcommand{\among}{\mbox{\hspace{6pt}{\tt in}\hspace{6pt}}}

\newcommand{\fordo}[2]{
{
  \color{gray}
  \left[
  \color{black}
  \hspace{-1pt}
  \def\arraystretch{1.5}
  \arraycolsep=3pt
  \begin{array}{rl}
    \scalebox{.9}{\tt for}
    &
    {#1}
    \\
    \scalebox{.9}{\tt do}
    &
    {#2}
  \end{array}
  \right.
}
}

\newcommand{\ifmeasuredthen}[2]{
{
  \color{gray}
  \left[
  \color{black}
  \def\arraystretch{1.5}
  \begin{array}{l}
  \scalebox{.9}{\tt if measured}\;\;\;{#1}
  \\
  \scalebox{.9}{\tt then}\;\;{#2}
  \end{array}
  \right.
}
}

\newcommand{\displaycode}[1]{
  \begin{equation}
    \adjustbox{}{
      \begin{minipage}{15.6cm}
        $
        {#1}
        $
      \end{minipage}
    }
  \end{equation}
}

\newcommand{\return}[2]{\mbox{\tt return}^{\scalebox{.7}{$#1$}}_{\scalebox{.7}{$#2$}}}
\newcommand{\obtain}[2]{\mbox{\tt obtain}^{\scalebox{.75}{$#1$}}_{\scalebox{.7}{$#2$}}}

\newcommand{\definitely}{\scalebox{.9}{\tt definitely}}

\newcommand{\collapse}{\scalebox{.9}{\tt collapse}}
\newcommand{\measure}{\scalebox{.9}{\tt measure}}

\newcommand{\Hadamard}{\mathrm{H}}
\newcommand{\CNOT}{\mathrm{CNOT}}
\newcommand{\PauliX}{\mathrm{X}}

\newcommand{\BellState}{\mathrm{BellState}}

\def\circuitvspace{.8}

\newcommand{\drawGate}[3]{
  \begin{scope}[shift={({#1}, {#2*\circuitvspace})}]
  \draw[
    line width=1.2,
    fill=white
  ]
    (-.45,-.35) rectangle (.45, .35);
  \draw (-0,0) node {
    \scalebox{1.1}{
      {#3}
    }
  };
  \end{scope}
}

\newcommand{\drawControlledGate}[4]{
  \draw[line width=1.1]
  ({#1}, {#4*\circuitvspace})
    to ({#1}, {#2*\circuitvspace});
  \draw[fill=black]
  ({#1},{#4*\circuitvspace}) circle (.1);
  \drawGate{#1}{#2}{#3}
}

\newcommand{\drawQBitAdder}[2]{
  \draw[line width=.8, fill=white]
    ({#1},{#2})
    circle (.2);
  \draw
    ({#1  }, {#2 +.2})
    to
    ({#1 }, {#2 - .2});
  \draw
    ({#1 +.2}, {#2 })
    to
    ({#1 - .2}, {#2 });
}

\newcommand{\drawCNOT}[3]{
  \draw[line width=1.1]
  ({#1}, {#2}) to ({#1}, {#2 - #3*\circuitvspace});
  \draw[fill=black]
  ({#1},{#2}) circle (.1);
  \drawQBitAdder{#1}{#2 - #3*\circuitvspace};
}

\newcommand{\drawStatePreparation}[2]{
  \begin{scope}[shift={({#1},{#2})}]
  \draw[
    draw=white,
    fill=white
  ]
  (-.5,-.36) rectangle (.5,.36);
  \node
    at (0,0)
    {
      \scalebox{1.5}{
        $\vert \;{\color{gray}\bullet}\; \rangle$
      }
    };
  \end{scope}
}

\newcommand{\drawMeasurement}[2]{
  \begin{scope}[shift={({#1},{#2})}]
  \draw[
    fill=white
  ]
  (-.5,-.36) rectangle (.5,.36);
  \draw[
    densely dotted,
    line width=1.2,
    gray
  ]
    (0,-.36) to (0, .36);
  \draw (-.412,-.18) node {
    \scalebox{.6}{\color{gray}0}
  };
  \draw (+.41,-.18) node {
    \scalebox{.6}{\color{gray}1}
  };
  \begin{scope}[shift={(-.01,-.2)}]
  \draw[
    line width=1.5,
    gray, densely dashed
  ]
    (156:.42) arc
    (156:24:.42);
  \begin{scope}[rotate=-38]
  \draw[
    fill=black
  ]
       (-.025,0)
    -- (+.025,0)
    -- (0,.52);
  \draw[fill=black] (0,0) circle (.06);
  \end{scope}
  \begin{scope}[rotate=+38]
  \draw[
    fill=black,
    lightgray
  ]
       (-.025,0)
    -- (+.025,0)
    -- (0,.52);
  \draw[fill=black] (0,0) circle (.06);
  \end{scope}
  \end{scope}
  \end{scope}
}

\def\LanguageNameScale{1.05}

\newcommand{\HoTT}{\scalebox{\LanguageNameScale}{\tt HoTT}}
\newcommand{\LHoTT}{\scalebox{\LanguageNameScale}{\scalebox{\LanguageNameScale}{\tt L}\HoTT}}

\newcommand{\Quipper}{\scalebox{\LanguageNameScale}{\tt Quipper}}
\newcommand{\QS}{\scalebox{\LanguageNameScale}{\tt QS}}
\newcommand{\zxCalculus}{\scalebox{\LanguageNameScale}{\tt zxCalculus}}
\newcommand{\Haskell}{\scalebox{\LanguageNameScale}{\tt Haskell}}
\newcommand{\Agda}{\scalebox{\LanguageNameScale}{\tt Agda}}
\newcommand{\Coq}{\scalebox{\LanguageNameScale}{\tt Coq}}

\newcommand{\monadology}{}

\begin{document}

\title{The Quantum Monadology}

\author{
  Hisham Sati${}^{\ast \dagger}$
  \;\;
  \;\;
  Urs Schreiber${}^{\ast}$
}

\maketitle

\thispagestyle{empty}

\begin{abstract}
  The modern theory of functional programming languages uses monads for encoding computational side-effects and side-contexts, beyond bare-bone program logic. Even though quantum computing is intrinsically side-effectful
  (as in quantum measurement) and context-dependent (as on mixed ancillary states),
  little of this monadic paradigm has previously been brought
  to bear on quantum programming languages.

  Here we systematically analyze the (co)monads on categories of parameterized module spectra which are induced by  Grothendieck's ``motivic yoga of operations'' -- for the present purpose specialized to
  $H\ComplexNumbers$-modules and further to set-indexed complex vector spaces, as discussed in a companion article \cite{EoS}.
  Interpreting an indexed vector space as a collection of alternative possible quantum state spaces parameterized by quantum measurement results, as familiar from Proto-{\Quipper}-semantics, we find that these (co)monads provide a comprehensive natural language for functional quantum programming with classical control and with ``dynamic lifting'' of quantum measurement results back into classical contexts.

  We close by indicating a domain-specific quantum programming language ({\QS})
  expressing these monadic quantum effects
  in transparent {\tt do}-notation,
  embeddable into the recently constructed {\it Linear Homotopy Type Theory} ({\LHoTT}) which interprets into parameterized module spectra.
  Once embedded into {\LHoTT}, this should make for formally verifiable universal quantum programming with
  linear quantum types, classical control, dynamic lifting, and notably also with topological effects (as discussed in the companion article \cite{TQP}).

  \medskip

  \begin{center}
    \bf Extended Abstract
  \end{center}
  Concretely, for {\it finite} classical and {\it finite-dimensional} quantum types (as of concern in quantum information theory),
  linear base change and linear internal hom constitute two ambidextrous adjunctions inducing a system of Frobenius monads which
  are linear/quantum versions of the classical Environment-, State-, and Epistemic-monads. We find that:
  \begin{itemize}[leftmargin=.65cm]
    \item[(i)]
    The QuantumEpistemic modality neatly encodes the logic of controlled quantum gates.
    \begin{itemize}[leftmargin=.5cm]
    \item[--] Its Kleisli equivalence formally proves the deferred measurement principle.
    \end{itemize}
    \item[(ii)]
    The QuantumEnvironment monad coincides with Coecke's ``classical structures'' monad used in {\zxCalculus}.
    \begin{itemize}[leftmargin=.5cm]
    \item[--] Its effect-handling computationally encodes collapsing quantum measurement

    ``dynamically lifted'' into the classical context
    akin to D. Lee's ``lifting monad''.

    \item[--] Its monoidal structure encodes enhancement of parameterized quantum circuits to mixed states.
    \end{itemize}
    \item[(iii)]
    The QuantumState monad produces spaces of density matrices.
    \begin{itemize}
    \item[--] Its monad transformations encode quantum channels acting on mixed quantum states.
    \end{itemize}
  \end{itemize}

  \noindent
  Moreover, the QuantumEnvironment and QuantumState (co)monads pairwise distribute over each other as to provide a pair of 2-sided Kleisli categories, where:
  \begin{itemize}[leftmargin=.5cm]
  \item[--] QuantumEnvironment-contextful
  and QuantumState-effectful maps
   encode mixed state preparation,

  \item[--] on which QuantumState-transformations act as quantum channels, followed by

  \item[--] QuantumState-contextful
  and QuantumEnvironment-effectful maps,
  encoding measurement and observables.
  \end{itemize}

\noindent
Notably, the action of QuantumState-transformations on QuantumState-contextful scalars (observables) is precisely Heisenberg-picture quantum evolution.

Finally, the QuantumEnvironment lifts from a monad on linear types to a (relative) monad on, in turn, QuantumState-monads, whereby the quantum effect logic for parameterized quantum circuits in the generality of mixed states becomes verbatim that for pure states, while mixed state effects such as the Born rule are brought out by the rich monadic semantics.

\end{abstract}

\vfill

\hrule
\vspace{5pt}

{
\footnotesize
\noindent
\def\arraystretch{1}
\tabcolsep=0pt
\begin{tabular}{ll}
${}^*$\,
&
Mathematics, Division of Science; and
\\
&
Center for Quantum and Topological Systems,
\\
&
NYUAD Research Institute,
\\
&
New York University Abu Dhabi, UAE.
\end{tabular}
\hfill
\adjustbox{raise=-15pt}{
\includegraphics[width=3cm]{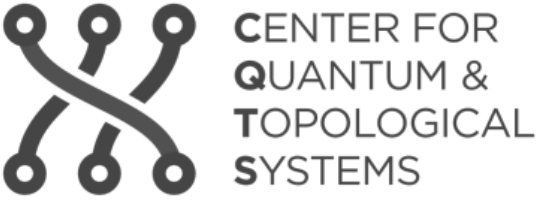}
}

\vspace{1mm}
\noindent ${}^\dagger$The Courant Institute for Mathematical Sciences, NYU, NY

\vspace{.2cm}

\noindent
The authors acknowledge the support by {\it Tamkeen} under the
{\it NYU Abu Dhabi Research Institute grant} {\tt CG008}.

}

\newpage

$\,$

\vspace{2cm}

\begin{center}
\begin{minipage}{10cm}
\setcounter{tocdepth}{2}
\tableofcontents
\end{minipage}
\end{center}

\newpage


\subsection{Motivation}
\label{Motivation}

We lay out an approach to a joint solution of the following open problems:

\medskip

\noindent
{\bf (I) The open problem of reliable quantum computing.}
While the hopes associated with quantum computing (Lit. \ref{LiteratureQuantumComputation}) are hard to overstate, experts are well-aware\footnote{
 \cite{Sau17}: ``small machines are unlikely to uncover truly macroscopic quantum phenomena, which have no classical analogs. This will likely require a scalable
 approach to quantum computation. [...] based on [...] topological quantum computation (TQC) as envisioned by Alexei Kitaev and Michael Freedman [...] The central
 idea of TQC is to encode qubits into states of topological phases of matter. Qubits encoded in such states are expected to be topologically protected, or robust,
 against the `prying eyes' of the environment, which are believed to be the bane of conventional quantum computation.''

 \cite{DasSarma22}: ``The qubit systems we have today are a tremendous scientific achievement, but they take us no closer to having a quantum computer that can
 solve a problem that anybody cares about. [...] What is missing is the breakthrough [...] bypassing quantum error correction by using far-more-stable qubits,
 in an approach called topological quantum computing.''
} that currently existing hard- and soft-ware paradigms are unlikely to support the desired heavy-duty quantum computations beyond toy examples.
The two fundamental open problems that the field still faces are both rooted in the single most enigmatic and proverbial phenomenon of quantum physics:
the {\it state collapse} or {\it decoherence} phenomenon (Lit. \ref{EpistemologyOfQuantumPhysics}), whereby the peculiar non-classical properties of
quantum systems on which rest the hopes of quantum computing are jeopardized by any measurement-like interaction of the system's environment. This means that
scalably robust quantum computing requires:

\vspace{1mm}
\begin{itemize}[leftmargin=.9cm]
\item[{\bf (i)}]  {\bf Topological hardware} (Lit. \ref{LiteratureTopologicalQuantumComputation}) given by topological quantum materials (Lit. \ref{TopologicalQuantumMaterials}) whose registry-states are
protected by an ``energy gap'' from having {\it any} interaction with the environment below that range.

\item[{\bf (ii)}] {\bf Verified software} (Lit. \ref{VerificationLiterature})
with compile-time certificates of correctness --- since the traditional run-time debugging of complex programs is
impossible for quantum programs (causing collapse),
while all the more needed due to the complexity and intransparency of gate-level quantum circuits.
\end{itemize}

\smallskip

Both of these issues have been discussed separately, but the necessary combination has remained essentially untouched until \cite{TQP}; one will need
a quantum programing language (Lit. \ref{LiteratureQuantumProgrammingLanguages}) which is
\begin{itemize}[leftmargin=.9cm]
\item[{\bf (iii)}]
{\bf certifiable and topological-hardware-aware}, allowing the programmer to formally verify
at compile-time the correctness not (just) of high-level quantum programs, but of quantum circuits consisting of the peculiar topological
quantum gates that the topological quantum hardware actually provides.
\end{itemize}

\smallskip

\noindent
For example, to state  just the most immediate problem:

\smallskip

\begin{minipage}{16cm}
{\bf Topological quantum circuit compilation problem} (Lit. \ref{TopologicalQuantumCompilation}).

\it
Suppose a topologically ordered quantum material is finally developed which features $\mathrm{su}_2$-anyon states at level $\ell$, and given any quantum circuit
written in the usual QBit-basis, then the {\it quantum compilation} of this circuit onto the given hardware is the specification of a {\it braid}
(an element of a braid group) such that the holonomy of the $\mathrm{su}_2^\ell$ Knizhnik-Zamolodchikov connection along the corresponding path in the
configuration space of defect points in the given quantum material may be conjugated onto the unitary operator
to which the quantum circuit evaluates,
within a specified accuracy.
\end{minipage}

\medskip

\noindent
Here the relevant braids are humongous while having no recognizable resemblance to the quantum algorithm which they are executing;
for instance, a single CNOT gate
\eqref{TraditionalCNOTGate}
may compile to the following braid \cite[Fig. 15]{HormoziZikosBonesteelSimon07}:

\vspace{-.2cm}
\begin{center}
  \adjustbox{
    raise=6pt
}{
  \hspace{-.3cm}
\begin{tikzpicture}[scale=.36]
  \draw[
    line width=1.5pt
  ]
  (-1,0)
  to
  (+1,0);
  \draw[
    line width=1.5pt
  ]
  (-1,-1.3)
  to
  (+1,-1.3);

  \node
   at (2.3,-.7) {$\mapsto$};

  \node
   at (0,-2.2) {
     \scalebox{.7}{
       \color{darkblue}
       CNOT gate
     }
   };

  \draw[line width=1.1] (0,0) to (0,-1.3);

  \draw[fill=white] (0,-1.3) circle (.3);
  \draw (0,-1.3-.3) to (0,-1.3+.3);
  \draw (0-.3,-1.3) to (0+.3,-1.3);

  \draw[fill=black] (0,0) circle (.15);

\end{tikzpicture}
  }
  \hspace{-2pt}
\includegraphics[width=15cm]{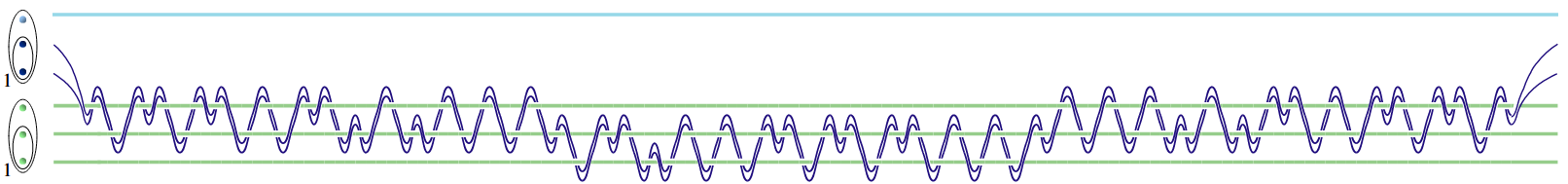}
\vspace{-.3cm}
\end{center}

\noindent
Hence future quantum programmers will need (classical) computer assistance to compile their quantum programs onto topological hardware.
To make that intricate process fail-safe to reliably run on precious scarce quantum resources, we need this computer algebra to be ``aware''
of the system specification and to certify its own correctness relative to this
specification.
And this is just for the simplest case of no classical control. The general problem is harder still:

\medskip

\label{ProblemOfCertifyingClassicalControl}{\bf The problem of certifying classical control.}
Even the most elementary quantum information protocols involve
mid-circuit measurement and classical control, such as in the
quantum teleportation protocol (cf. \cref{QuantumTeleportationProtocol}):

\vspace{.1cm}
\hspace{-1cm}
{\small
\begin{tikzpicture}[xscale={1.1}]

  \draw (-2.8, 0) node { \scalebox{1.1}{$\vert 0 \rangle$} };
  \draw (-2.8, -1) node { \scalebox{1.1}{$\vert 0 \rangle$} };

  \draw[line width=2.2]
    (11,.23) to (11,-.23);
 \draw[line width=2.2]
    (11,1+.23) to (11,1-.23);

  \draw[
    line width=1.5pt
  ]
  (-2.8,1)
  to
  (+6,1);

  \draw[
    line width=1.5pt
  ]
  (-2.5,-1)
  to
  (+10.5,-1);

  \draw[
    line width=1.5pt
  ]
  (-2.4,0)
  to
  (+5.5,0);

  \draw[
    line width=.6pt
  ]
  (5.5,+.04)
  to
  (11,+.04);
  \draw[
    line width=.6pt
  ]
  (5.5,-.04)
  to
  (11,-.04);

  \draw[
    line width=.6pt
  ]
  (5.5,1+.04)
  to
  (11,1+.04);
  \draw[
    line width=.6pt
  ]
  (5.5,1-.04)
  to
  (11,1-.04);

  \def\circuitvspace{1}

  \drawGate{-.9}{0}{H}

  \drawCNOT{0.6}{0}{1}

  \drawCNOT{2}{1}{1}

  \drawGate{3.5}{1}{H}

  \drawMeasurement{5.6}{1}

  \drawMeasurement{5.6}{0}

  \draw[line width=1.1]
    (7.5, -.37) to (7.5, -1);

  \draw[line width=5.5, color=white]
    (9.5, +.6) to (9.5, -1);
  \draw[line width=1.1]
    (9.5, +.65) to (9.5, -1);

  \drawControlledGate{7.5}{-1}{X}{0}

  \drawControlledGate{9.5}{-1}{Z}{1}

  \draw (-3.56, -.3) node[rotate=-30] {
    \scalebox{.7}{
      \color{purple}
      \bf
      \def\arraystretch{.8}
      \begin{tabular}{c}
        quantum state
        \\
        preparation
     \end{tabular}
    }
  };

  \draw (6.9, +.5) node {
    \scalebox{.7}{
      \color{purple}
      \bf
      \def\arraystretch{.8}
      \begin{tabular}{c}
        quantum
        \\
        measurement
     \end{tabular}
    }
  };

  \draw (8.45, -.32) node[rotate=-0] {
    \scalebox{.7}{
      \color{purple}
      \bf
      \def\arraystretch{.8}
      \begin{tabular}{c}
        classical
        control
     \end{tabular}
    }
  };

  \draw (1-.4, +.5) node[rotate=-6] {
    \scalebox{.7}{
      \color{darkgreen}
      \bf
      \def\arraystretch{.8}
      \begin{tabular}{c}
        plain
        quantum gates
     \end{tabular}
    }
  };

  \draw (-3.5, +.85) node[rotate=-0] {
    \scalebox{.7}{
      \color{darkblue}
      \bf
      \def\arraystretch{.8}
      \begin{tabular}{c}
        input
        \\
        quantum state
     \end{tabular}
    }
  };

  \draw (+11, -1.1) node[rotate=-0] {
    \scalebox{.7}{
      \color{darkblue}
      \bf
      \def\arraystretch{.8}
      \begin{tabular}{c}
        \colorbox{white}{
          output
        }
        \\
        state
     \end{tabular}
    }
  };

\end{tikzpicture}
}

\hspace{-.9cm}
\begin{tabular}{ll}
\label{ClassicalControlSchematics}
\begin{minipage}{6.8cm}
More importantly, beyond the currently available NISQ paradigm (Lit. \ref{LiteratureOnNISQMachines}),
serious quantum computation is expected
(Lit. \ref{ClassicalControlAndDynamicLifting})
to involve a perpetual loop of classical control operations on the quantum computer ({\it hybrid} classical/quantum computation). These are predominantly
for quantum error correction (\cref{QuantumBitFlipCode}) but also for purposes such as repeat-until-success gates
--
where subsequent quantum circuit execution is classically conditioned on run-time quantum measurement results -- also called ``dynamic lifting''
(Lit. \ref{ClassicalControlAndDynamicLifting}, namely of quantum measurement results into the classical data register).
This is schematically indicated on the right.

Last not least, for probabilistic analysis of such hybrid processes the machine state is to be modeled as a {\it mixed} classical/quantum probabilistic state (Lit. \ref{LiteratureQuantumProbability}).
\end{minipage}
&
\hspace{5mm}
\adjustbox{
  scale=.75,
  raise=-4.25cm
}{
\begin{tikzpicture}
\draw (0,0) node {
\includegraphics{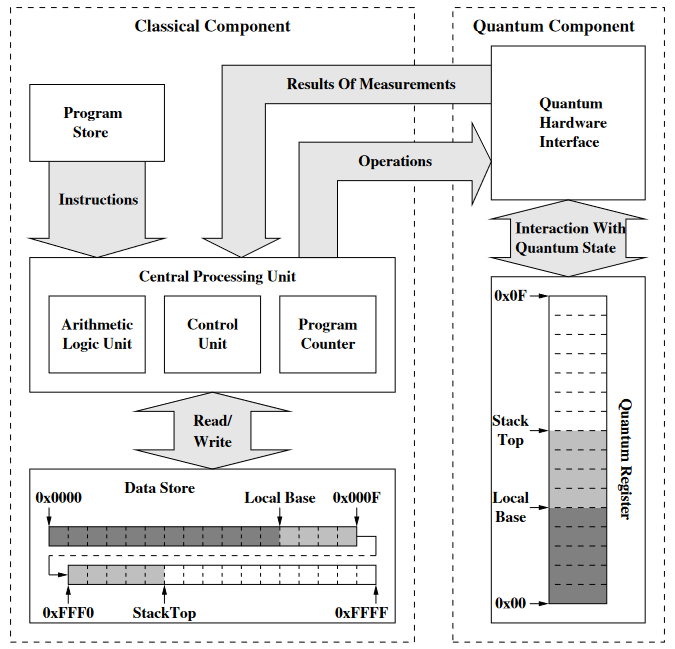}
};

\node[
  rectangle,
  line width=1pt,
  rounded corners,
  fill=lightgray,
  draw=gray
] at (-1.67,2.5)  {
  \scalebox{.8}{
    \color{purple}
    \bf
    \def\arraystretch{1}
    \hspace{-12pt}
    \scalebox{.95}{
    \begin{tabular}{c}
      dynamic
      \\
      lifting
    \end{tabular}
    }
    \hspace{-10pt}
  }
};
\node at (-6.1,-3.1) {
  \rotatebox{90}{
  \scalebox{.7}{
    \color{gray}
    diagram adapted from
    \cite[Fig. 1]{NPW07}
  }}
};
\draw[
  line width=2pt,
  lightgray
]
  (-1.7-.32, 2.5+.45) to
  (-1.7+.415, 2.5+.45);
\draw[
  line width=2pt,
  lightgray
]
  (-1.7-.32, 2.5-.45) to
  (-1.7+.415, 2.5-.45);
\end{tikzpicture}
}
\end{tabular}

\medskip
Hence for reliable heavy-duty quantum computation we need a certification language that knows about classical data
types {\it and} about linear/quantum data types {\it and} their {\it dependency} on classical data. This had been lacking:

\bigskip

\hspace{-.9cm}
\begin{tabular}{ll}
\begin{minipage}{9cm}

{\bf The problem of embedded quantum languages.}
Namely, for previous lack of a {\it universal} quantum programming language, existing quantum circuit languages are embedded into
{\it classical} host languages
(Lit. \ref{LiteratureQuantumProgrammingLanguages})
which do not have native support for linear types
(cf. Lit. \ref{VerificationLiterature}) nor for classical control of quantum circuits. For instance, basic protocol schemes such as quantum
teleportation (\cref{QuantumTeleportationProtocol}), quantum error correction (\cref{QuantumBitFlipCode}) or repeat-until-success gates
remain unverifiable with previous technology.
\end{minipage}
&
\hspace{5mm}
\adjustbox{
  scale=.9,
  raise=-2.0cm
}{
\begin{tikzpicture}
  \draw[
    line width=3.5,
    draw=gray,
    rounded corners,
    fill=darkblue,
    fill opacity=.3
  ]
    (-4, -.7)
    rectangle
    (4,4);
  \begin{scope}[yshift=-.3cm]
  \draw[yshift=3.4cm]
    (0,0) node
    {
      \def\arraystretch{1}
      \tabcolsep=3pt
      \begin{tabular}{c}
       \Haskell, \Coq, ...
       \\
       \bf Classical Type Theory
       \\
       \clap{for universal classical computation}
     \end{tabular}
    };
  \begin{scope}[xshift=0cm, yshift=+.5cm]
    \draw[
    rounded corners,
    fill=red,
    line width=1.3,
    dashed,
    draw=gray,
    opacity=.6,
  ]
      (-3-.2,-.6+.3-.2)
      rectangle
      (3+.2,2+.2);
  \draw (0,1.8) node {
    unverified linear type universe
  };
  \begin{scope}[yshift=-.55cm]
  \draw[
    rounded corners,
    fill=olive,
    draw=gray,
    opacity=1,
    line width=2.5,
  ]
      (-3,.3)
      rectangle
      (3,2);
    \draw (0,1.2)
      node
      {
        \def\arraystretch{1}
        \tabcolsep=3pt
        \begin{tabular}{cl}
          {\tt QML}, {\tt QIO} {\tt Quipper}, {\tt QWIRE}, ...
          \\
          {\bf Quantum Circuit Language}
          \\
          for quantum logic cicuits
        \end{tabular}
      };
  \end{scope}
  \end{scope}
  \end{scope}
\end{tikzpicture}
}
\end{tabular}

\bigskip

\noindent
{\bf Solution by Linear Homotopy Type Theory.}
We argue here, as announced in \cite{Schreiber22}, that the novel type theory {\LHoTT} (Lit. \ref{LiteratureLHoTT})
 recently developed in \cite{Riley22} (anticipated in \cite{QuantizationViaLHoTT})
in extension of the classical language scheme {\HoTT} (Lit. \ref{LiteratureHomotopyTypeTheory})
serves as the missing universal quantum programming/certification language.

\noindent
Our claim
\ifdefined\monadology
in \cite{QS}
\fi
is that {\LHoTT}:

\vspace{.2cm}

\hspace{-.9cm}
\begin{tabular}{ll}
\begin{minipage}{6.5cm}
  \begin{itemize}[leftmargin=.5cm]
  \item
   Solves the old problem of constructing combined classical/linear type theories (cf. Lit. \ref{VerificationLiterature}).
  \item Provides existing quantum programming languages like {\Quipper} with a certification mechanism \cite{Riley23}.
  \item Natively supports quantum effects such as dynamic lifting of run-time quantum measurement (\cref{QuantumEffects}).
  \item
  Natively supports verification of realistic topological quantum gates \cite{TQP}.
  \end{itemize}
\end{minipage}
&
\adjustbox{
  raise=-2.4cm
}{
\small
\begin{tikzpicture}
  \draw[
    line width=3.5,
    draw=gray,
    rounded corners,
    fill=olive,
    fill opacity=.5
  ]
    (-5.1, -.4)
    rectangle
    (5.5,4.2);
  \draw[yshift=3.4cm]
    (0,0) node
    {
      \def\arraystretch{1}
      \tabcolsep=3pt
      \begin{tabular}{c}
       {\tt LHoTT}
       \\
       \bf Linear Homotopy Type Theory
       \\
       for universal quantum computation
     \end{tabular}
    };
  \begin{scope}[yshift=-.1cm]
  \begin{scope}[xshift=-2.6cm, yshift=+.5cm]
    \draw[
    rounded corners,
    fill=olive,
    draw=gray,
    opacity=.5,
    line width=2.5,
  ]
      (-2.1,-.5)
      rectangle
      (2.4,2.2);
    \draw (0,1.4)
      node
      {
        \def\arraystretch{1}
        \tabcolsep=3pt
        \begin{tabular}{c}
          {\tt HoTT}
          \\
          {\bf Homotopy Type Theory}
          \\
          for topological logic gates
        \end{tabular}
      };
  \end{scope}
  \begin{scope}[
    xshift=+2.6cm,
    yshift=2cm
  ]
  \draw[
    rounded corners,
    draw=gray,
    line width=2.5,
    fill=olive,
    opacity=.5
  ]
    (-2.4, -2)
    rectangle
    (2.5,.7);
  \draw (0,-.1) node {
    \scalebox{1}{
      \def\arraystretch{1}
      \tabcolsep=3pt
      \begin{tabular}{cl}
        {\tt QS}
        \\
        {\bf Quantum Systems Language}
        \\
        for quantum logic circuits
      \end{tabular}
    }
  };
  \end{scope}
  \begin{scope}[
    xshift=+0cm,
    yshift=.7cm
  ]
  \draw[
    line width=2.7,
    draw=gray,
    rounded corners,
    fill=olive,
    fill opacity=1
  ]
    (-3, -.47)
    rectangle
    (3.1,.43);
  \node at (0,0) {
    \scalebox{1.1}{
      \def\arraystretch{1}
      \tabcolsep=3pt
      \begin{tabular}{cl}
        {\bf Topological Quantum Language}
      \end{tabular}
    }
  };
  \end{scope}
  \end{scope}
\end{tikzpicture}
}
\end{tabular}

\vspace{.3cm}

\noindent
We argue that this makes {\LHoTT} the first comprehensive paradigm for serious quantum programming beyond the
NISQ area.
\ifdefined\monadology
Here
\else
Concretely,
\fi
we describe a domain-specific language embeddable into {\LHoTT} to bring this out: {\it Quantum Systems Language} ({\QS}, \cref{Pseudocode}), based on a system of monadic effects which are definable (by admissible inference rules) in {\LHoTT} (\cref{QuantumEffects}, surveyed below in \cref{QuantumMonadology}).

\newpage

\hspace{-.9cm}
\begin{tabular}{ll}
\label{LHoTTInIntroduction}
\begin{minipage}{11.7cm}
Concretely, {\LHoTT} enhances the syntactic rules of classical {\HoTT} by further type formations which serve to exhibit every (homotopy) type $E$
of the language as secretly consisting of an underlying classical (intuitionistic) base type $B \defneq \classically E$ equipped, in a precise sense, with a microscopic
(infinitesimal) halo of linear/quantum data. As such, {\LHoTT} may neatly be thought of as the formal logical expression of a microscope
that resolves quantum aspects on structures that macroscopically appear classical. This way {\LHoTT} embeds quantum logic into
classical logic in a way reminiscent of Bohr's famous dictum\footnotemark
 that all quantum phenomena must be expressible in classical language.
\end{minipage}
&
\hspace{.2cm}
\adjustbox{
  frame,
  raise=-1.75cm
}{\includegraphics[width=5cm]{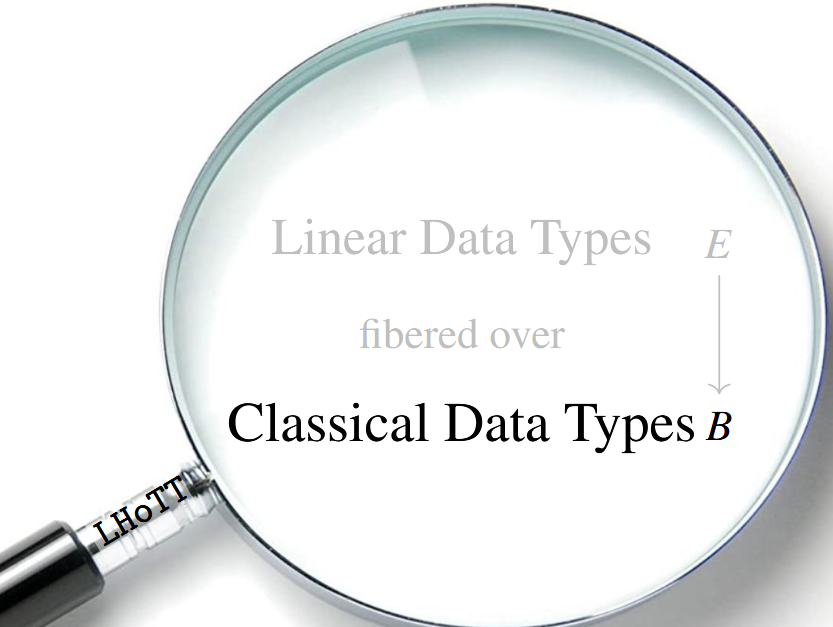}}
\end{tabular}

\footnotetext{ \cite[pp. 209]{Bohr1949}: ``however far the phenomena transcend the scope of classical physical explanation, the account of
all evidence must be expressed in classical terms''. For background and commentary see also \cite[p. 24]{Scheibe73}.}

\medskip
\noindent
{\bf Quantum halos.}
Formally this is achieved by adjoining to classical {\HoTT} an {\it ambidextrous} modal operator $\natural$
\cite{RileyFinsterLicata21} (an {\it infinitesimal cohesive modality}  \cite[Def. 3.4.12, Prop. 4.1.9]{dcct}),
whose modal types (Lit. \ref{LiteratureModalTypeTheory}) are the {\it purely classical} (ordinary) homotopy types, embedded {\it bi-reflectively}
\eqref{CoreflectionOfClassicalTypesAmongLinearBundleTypes} among all data types (see \ifdefined\monadology \cref{QuantumTypeSemantics}
\else \cref{QuantumViaLHoTTTypes}\fi):

\vspace{-.5cm}
\begin{equation}
  \adjustbox{}{
  \hspace{-.68cm}
\begin{tabular}{cc}
  \begin{minipage}{8.2cm}
The presence of the $\classically$-modality exhibits general types $E \isa \Type$ as microscopic/infinitesimal {\it halos} around their underlying purely classical
type $\classical E \isa \ClassicalType$. It is a profound fact
\eqref{TangentInfinityTopos}
of $\infty$-topos theory that models for such {\it infinitesimal cohesion}
(see Lit. \ref{TangentInfinityToposes})
are provided by parameterized module spectra, in particular by flat $\infty$-vector bundles (``$\infty$-local systems'', see  \cite{EoS})
which, in their 0-sector (Rem. \ref{Zerosector}), accommodate quantum circuit semantics (cf. \cref{ControlledQuantumGates})
in indexed sets of vector spaces
(cf. \cref{QuantumTypeSemantics})
such as known
from the Proto-{\Quipper} quantum language
(Lit. \ref{LiteratureQuantumProgrammingLanguages}).
\end{minipage}
  &
  \hspace{1.7cm}
  \begin{tikzcd}[
    column sep=-5pt,
    row sep=40pt
  ]
    \mathllap{
      \scalebox{.7}{
        \color{darkblue}
        \bf
        \def\arraystretch{.9}
        \begin{tabular}{c}
          bundles of linear
          \\
          homotopy types
        \end{tabular}
      }
    }
    \Types
     \ar[out=180-55, in=55, looseness=5,
     shift right=2pt,
     "\scalebox{1.2}{$\mathclap{
        \natural
      }$}"{description},
      "{
        \scalebox{.7}{
          \color{darkorange}
          \bf
          classical modality
        }
      }"{yshift=4pt}
    ]
    \ar[
      from=d,
      shorten >= -3pt,
      hook,
      "{ 0 }"{description}
    ]
    \ar[
      d,
      shift left=14pt,
      "{ p }"{description},
    ]
    \ar[
      d,
      shorten <= -3pt,
      phantom,
      shift left=7pt,
      "{ \scalebox{.7}{$\dashv$} }"
    ]
    \ar[
      d,
      shorten <= -3pt,
      phantom,
      shift right=8pt,
      "{ \scalebox{.7}{$\dashv$} }",
    ]
    \ar[
      d,
      shorten <= -3pt,
      shift right=14pt,
      "{ p }" description,
      "{
        \scalebox{.7}{
          \color{darkgreen}
          \bf
          \def\arraystretch{.9}
          \begin{tabular}{c}
            bireflection
          \end{tabular}
        }
      }"{yshift=-3pt, sloped, rotate=180}
    ]
    &
    \overset{
     \mathclap{
       \raisebox{2pt}{
       \scalebox{.7}{
         {
           \color{gray}
           e.g.
         }
       }
       }
     }
    }{
      =
    }
    &
    \overset{
    \mathclap{
      \raisebox{+5pt}{
      \scalebox{.7}{
        \color{darkblue}
        \bf
        \def\arraystretch{.9}
        \begin{tabular}{c}
          flat
          $\infty$-vector
          bundles
          \\
          ($\infty$-local systems)
        \end{tabular}
      }
      }
    }
    }{
    \int_{\mathbf{X}}
      \mathbf{sCh}
      ^{\mathbf{X}}
      _{\mathbb{K}}
    }
    \ar[
      d,
      shift right=5pt,
      "{
        \scalebox{.7}{
          \color{darkgreen}
          \bf
          base space
        }
      }"{yshift=-1pt, sloped, rotate=180}
    ]
    \ar[
      d,
      phantom,
      "{
        \scalebox{.7}{$\dashv$}
      }"
    ]
    \ar[
      from=d,
      shift right=5pt,
      hook,
      "{
        \scalebox{.7}{
          \color{darkgreen}
          \bf
          zero-section
        }
      }"{yshift=-1pt,swap, sloped}
    ]
    \\
    \mathllap{
      \scalebox{.7}{
        \color{darkblue}
        \bf
        \def\arraystretch{.85}
        \begin{tabular}{c}
          purely classical
          \\
          homotopy types
        \end{tabular}
      }
    }
    \ClassicalTypes
    &
    \overset{
      \raisebox{2pt}{
       \scalebox{.7}{
         {
           \color{gray}
           e.g.
         }
       }
       }
    }{
      =
    }
    &
    {
      \big\{
        \mathbf{X}
        \in
        \mathrm{sSet}\mbox{-}\mathrm{Grpd}
      \big\}
    }
  \end{tikzcd}
  \end{tabular}
  }
\end{equation}

\vspace{-.1cm}

\noindent
{\bf Motivic Yoga.} {\LHoTT} witnesses these quantum halos as {\it linear types} \eqref{LinearTypesAsSubstructuralTypes} equipped with a closed tensor product $\otimes$ and compatible base change operations which satisfy the rules of Grothendieck's ``motivic yoga of six operations'' in Wirthm{\"u}ller style (Def. \ref{MotivicYoga}, cf. \cite[\S 2.4]{Riley22}\cite[\S 3.3]{EoS}). It is this ``motivic'' structure from which the structure of quantum physics derives, as originally observed in \cite{QuantizationViaLHoTT} and here brought out in
\ifdefined\monadology
\cref{QuantumTypeSemantics}.
\else
 \cref{QuantumViaLHoTT}
 and \cref{QuantumEffects}.
\fi

\vspace{+.1cm}
\hspace{-.8cm}
\adjustbox{scale=.9}
{
\def\arraystretch{2.1}
\begin{tabular}{|c||c|c|c|}
  \hline
  \multicolumn{4}{|c|}{
    \hspace{-11.5cm}
    \scalebox{1.2}{\bf
      Linear/Quantum Data Types
    }
  }
  \\
  \hline
  \hline
  {\bf
    \def\arraystretch{1}
    \begin{tabular}{c}
      Characteristic
      \\
      Property
    \end{tabular}
  }
  &
  \def\arraystretch{1}
  \hspace{-10pt}
  \begin{tabular}{l}
     {\bf 1.} Their cartesian product
     \\
     blends into the co-product:
    $\mathclap{\phantom{\vert_{\vert_{\vert_{\vert}}}}}$
  \end{tabular}
  \hspace{-10pt}
  &
  \def\arraystretch{1}
  \hspace{-10pt}
  \begin{tabular}{l}
    {\bf 2.} A tensor product appears
    \\
    \&
    distributes over direct sum
    $\mathclap{\phantom{\vert_{\vert_{\vert}}}}$
  \end{tabular}
  \hspace{-10pt}
  &
  \def\arraystretch{1}
  \begin{tabular}{l}
    \\
    {\bf 3.}
    A linear function type
    \\
    appears adjoint to tensor
    \\
    $\mathclap{\phantom{\vert_{\vert_{\vert}}}}$
  \end{tabular}
  \\
  \hline
  {\bf Symbol}
  &
  $\bigoplus$
  \;\;
  direct sum
  &
  \scalebox{1.3}{$\otimes$}
  \;\;
  tensor product
  &
  \scalebox{1.3}{$\multimap$}
  \;\;
  linear function type
  \\
  \hline
  \def\arraystretch{.9}
  \begin{tabular}{c}
    {\bf Formula}
    \\
    \scalebox{.9}{(for $W \isa \FiniteClassicalType$)}
  \end{tabular}
  &
  $
  \mathclap{\phantom{\vert^{\vert^{\vert^{\vert^{\vert^{\vert^{\vert^{\vert^{\vert}}}}}}}}}}
    \overset{
      \mathclap{
        \;\;\;\;\;\;\;
        \scalebox{.7}{
          \raisebox{3pt}{
            \color{darkblue}
            \bf
            cart. product
          }
        }
      }
    }{
      \underset{W}{\prod}
      \HilbertSpace{H}_w
    }
    \;\simeq\;
    \underset{
      \mathclap{
        \scalebox{.7}{
          \raisebox{-2pt}{
            \color{darkorange}
            \bf
            direct sum
          }
        }
      }
    }{
      \scalebox{1.3}{$\oplus$}
      \HilbertSpace{H}_w
    }
    \;\simeq\;
    \overset{
      \mathclap{
        \hspace{-15pt}
        \scalebox{.7}{
          \raisebox{3pt}{
            \color{darkblue}
            \bf
            co-product
          }
        }
      }
    }{
      \underset{W}{\coprod}
      \HilbertSpace{H}_w
    }
  $
    $\mathclap{\phantom{\vert_{\vert_{\vert_{\vert_{\vert_{\vert_{\vert}}}}}}}}$
  &
 $
  \mathscr{V}
  \otimes
  \scalebox{1.4}{$($}
    \;
  \raisebox{1pt}{$
    \underset{w \isa W}{\scalebox{1.4}{$\oplus$}}
    \HilbertSpace{H}_{w}
  $}
    \;
  \scalebox{1.4}{$)$}
  \;\simeq\;
  \underset{w \isa W}{\scalebox{1.4}{$\oplus$}}
  \big(
    \mathscr{V}
    \otimes
    \HilbertSpace{H}_{w}
  \big)
 $
 &
 $
 \def\arraystretch{1.2}
 \begin{array}{rl}
 &
 (\mathscr{V}
 \otimes
 \HilbertSpace{H}
 )
 \multimap
 \mathscr{K}
 \\
 \simeq &
 \mathscr{V}
 \multimap
 \big(
   \HilbertSpace{H}
   \multimap
   \mathscr{K}
 \big)
 \end{array}
 $
 \\
 \hline
 \multirow{2}{*}{
 {\bf AlgTop Jargon}
 }
 &
 \multirow{2}{*}{
  \def\arraystretch{1}
  \begin{tabular}{c}
   biproduct,
   \\
   stability,
   ambidexterity
    $\mathclap{\phantom{\vert_{\vert_{\vert_{\vert}}}}}$
  \end{tabular}
  }
  &
  Frobenius reciprocity
  &
  mapping spectrum
  \\
  \cline{3-4}
  &
  &
  \multicolumn{2}{c|}{
    Grothendieck's
    \;Motivic Yoga
    of 6 oper.
    \;
    (Wirthm{\"u}ller form)
  }
  \\
  \hline
  {\bf Linear Logic}
  &
  additive disjunction
  $\mathclap{\phantom{\vert_{\vert_{\vert_{\vert}}}}}$
  &
  multiplicative conjunction
  $\mathclap{\phantom{\vert_{\vert_{\vert_{\vert}}}}}$
  &
  linear implication
  \\
  \hline
  {\bf Physics Meaning}
  &
  \def\arraystretch{1}
    \begin{tabular}{c}
    $\mathclap{\phantom{\vert^{\vert^{\vert^{\vert}}}}}$
    parallel
    \\
    quantum systems
  $\mathclap{\phantom{\vert_{\vert_{\vert_{\vert}}}}}$
    \end{tabular}
  &
  \def\arraystretch{1}
  \begin{tabular}{c}
    $\mathclap{\phantom{\vert^{\vert^{\vert^{\vert}}}}}$
  compound
  \\
  quantum systems
  $\mathclap{\phantom{\vert_{\vert_{\vert_{\vert}}}}}$
  \end{tabular}
  &
  qRAM systems
  $\mathclap{\phantom{\vert_{\vert_{\vert_{\vert}}}}}$
  \\
  \hline
\end{tabular}
}

\medskip

\noindent {\bf $H\mathbb{C}$-Linear quantum theory.}
In this scheme, conventional
quantum information theory happens in the $\mathbb{C}$-linear form of linear homotopy theory (details in \cite{EoS}) where parameterized $H\mathbb{C}$-module spectra
are equivalent to {\it flat $\infty$-bundles of chain complexes}, also known as  {\it $\infty$-local systems}. Here the higher structure of chain complexes
serves to capture topological quantum effects \cite{TQP}, but in the 0-sector (Rem. \ref{Zerosector}) these are just set-indexed complex vector spaces of the form familiar
from the categorical semantics of the quantum language {\Quipper}, this is what we discuss in detail
\ifdefined\monadology
\cref{QuantumTypeSemantics}
\else
 \cref{QuantumViaLHoTT}\fi.
But since all our quantum effects are constructed monadically (\cref{QuantumEffects}) relying just on the abstract Motivic Yoga, they apply at once to unrestricted (stable) homotopy types, providing a homotopy-theoretic form of quantum mechanics suitable for the discussion of ``topological quantum effects'' as in \cite{TQP}.

\subsection{Quantum Monadology}
\label{QuantumMonadology}

\noindent
{\bf The open problem of formalizing quantum epistemic logic.}
With the need for a universal and verifiable quantum programming language established, the next open problem is that of language design, which here
we mean in a fundamental paradigmatic way:
\begin{center}
{\it Given that dependent type theory is the fundamental paradigm for certified programming in general}
(Lit. \ref{VerificationLiterature}),

{\it what makes it applicable to certification of quantum effects such as quantum measurement} (Lit. \ref{EpistemologyOfQuantumPhysics})?
\end{center}

A universal quantum programming language has to accurately reflect the logical content of quantum physics,
where the act of formulating a quantum program is also that of recounting, in formalized language, the physical process of its execution. The execution of quantum programs {\it includes} processes of quantum measurement and therefore any formulation must handle the curious nature of quantum epistemology. In this sense, we may claim that:
\begin{center}
{\it Finding a universal quantum programming language means finding a formal language for quantum epistemology.}
\end{center}

\medskip

\noindent {\bf The role of modal logic.}
Stated this way, we need not look much further for guidance on the matter, since the formal language paradigm for
dealing with questions of epistemology has long been understood to be {\it modal logic} (Lit. \ref{ModalLogicAndManyWorlds}), where the usual logical connectives are accompanied by formal
expressions for qualified {\it modes} in which propositions may hold, such as {\it necessarily} ($\necessarily$)
or {\it possibly} ($\possibly$) namely
(which is the perspective of relevance here:) for all or any  {\it measurement outcome} that may be obtained, or {\it possible world} $w$ (as the modal
logician says) that one may find oneself in, one of the {\it many worlds} (as the quantum philosopher says):
\begin{equation}
  \label{BasicFormationInModalLogic}
  \overset{
    \mathclap{
      \raisebox{8pt}{
      \scalebox{.7}{
        \color{darkblue}
        \bf
        \def\arraystretch{.9}
\begin{tabular}{c}
          Set of
          many possible worlds
          \\
          (of measurement outcomes)
        \end{tabular}
      }
      }
    }
  }{
    W \isa \Sets
  }
  \,,
  \hspace{1.2cm}
  \overset{
    \mathclap{
      \raisebox{9pt}{
      \scalebox{.7}{
        \color{darkblue}
        \bf
        \def\arraystretch{.9}
        \begin{tabular}{c}
          $W$-dependent
          \\
          {\color{purple} proposition}
        \end{tabular}
      }
      }
    }
  }{
  P
  \,\isa\,
  \mathrm{Prop}_W
  }
  \hspace{.6cm}
     \overset{
       \raisebox{4pt}{
         \scalebox{.7}{
           yields that
         }
       }
     }{
    \yields
  }
  \hspace{1cm}
  \def\arraystretch{1.4}
  \left.
  \begin{array}{l}
    \overset{
      \mathclap{
        \raisebox{8pt}{
          \scalebox{.7}{
            \def\arraystretch{.9}
            \begin{tabular}{c}
              \color{darkorange}
              \bf
              ``$P$ holds necessarily''
              \\
              (no matter the outcome/world)
            \end{tabular}
          }
        }
      }
    }{
      \necessarily P
    \;\;\defneq\;\;
    }
    \forall_w
    \,
    P(w)
    \\
    \underset{
      \mathclap{
        \raisebox{-8pt}{
          \scalebox{.7}{
            \def\arraystretch{.9}
            \begin{tabular}{c}
              \color{darkorange}
              \bf
              ``$P$ holds possibly''
              \\
              (for some outcome/world)
            \end{tabular}
          }
        }
      }
    }{
    \possibly P
    \;\;\defneq\;\;
    }
    \exists_w
    \,
    P(w)
  \end{array}
  \right\}
  \overset{
    \raisebox{5pt}{
      \scalebox{.7}{
        is a
      }
    }
  }{
   \;\isa\;
   }
   \overset{
     \mathclap{
       \raisebox{8pt}{
         \scalebox{.7}{
           \color{darkblue}
           \bf
           \def\arraystretch{.9}
           \begin{tabular}{c}
             $W$-independent
             \\
             proposition
           \end{tabular}
         }
       }
     }
   }{
  \mathrm{Prop}
  \hookrightarrow
  \mathrm{Prop}_W
  }
\end{equation}

If here we think of classical propositions as certain data types (namely of data that certifies their assertion), then it is natural to generalize
this from modal logic to {\it modal type theory} (Lit. \ref{LiteratureModalTypeTheory}) where we consider any $W$-dependent data types:\footnote{
  We write ``$\coprod_w$'' for the (non-linear) type formation traditionally
  referred to as ``dependent sum'' and
  traditionally denoted ``$\sum_w$'', since the latter symbol is borrowed from linear algebra, an (unnecessary) abuse of notation that becomes untenable after
  our passage from classical intuitionistic to actual linear dependent type theory.
}
\begin{equation}
  \label{BasicFormationInModalTypeTheory}
  \overset{
    \mathclap{
      \raisebox{8pt}{
      \scalebox{.7}{
        \color{darkblue}
        \bf
        \def\arraystretch{.9}
        \begin{tabular}{c}
          Type of
          many possible worlds
          \\
          (of measurement outcomes)
        \end{tabular}
      }
      }
    }
  }{
    W \isa \Type
  }
  \,,
  \hspace{1.2cm}
  \overset{
    \mathclap{
      \raisebox{9pt}{
      \scalebox{.7}{
        \color{darkblue}
        \bf
        \def\arraystretch{.9}
        \begin{tabular}{c}
          $W$-dependent
          \\
          {\color{purple} data type}
        \end{tabular}
      }
      }
    }
  }{
  D
  \,\isa\,
  \mathrm{Type}_W
  }
  \hspace{.6cm}
     \overset{
       \raisebox{4pt}{
         \scalebox{.7}{
           yields that
         }
       }
     }{
    \yields
  }
  \hspace{1cm}
  \def\arraystretch{1.4}
  \left.
  \begin{array}{l}
    \overset{
      \mathclap{
        \raisebox{8pt}{
          \scalebox{.7}{
            \color{darkorange}
            \bf
            \def\arraystretch{.9}
            \begin{tabular}{c}
              type of $D$-data for
              \\
              every world/outcome
            \end{tabular}
          }
        }
      }
    }{
      \necessarily D
    \;\;\defneq\;\;
    }
    \prod_w
    \,
    D(w)
    \\
    \underset{
      \mathclap{
        \raisebox{-8pt}{
          \scalebox{.7}{
            \color{darkorange}
            \bf
            \def\arraystretch{.9}
            \begin{tabular}{c}
              type of $D$-data
              \\
              for any world/outcome
            \end{tabular}
          }
        }
      }
    }{
    \possibly D
    \;\;\defneq\;\;
    }
    \coprod_w
    \,
    D(w)
  \end{array}
  \right\}
  \overset{
    \raisebox{5pt}{
      \scalebox{.7}{
        is a
      }
    }
  }{
   \;\isa\;
   }
   \overset{
     \mathclap{
       \raisebox{8pt}{
         \scalebox{.7}{
           \color{darkblue}
           \bf
           \def\arraystretch{.9}
           \begin{tabular}{c}
             $W$-independent
             \\
             data type
           \end{tabular}
         }
       }
     }
   }{
  \mathrm{Type}
  \hookrightarrow
  \mathrm{Type}_W
  }
\end{equation}

\noindent {\bf Epistemic modal logic as Dependent type theory.}
Remarkably, in this more general form \eqref{BasicFormationInModalTypeTheory} the system {\it simplifies} since this
{\it epistemic modal type theory}
is just plain
dependent type theory with the $\mathrm{W}$-dependent type formation rules viewed not as adjoints but equivalently as (co){\it monadic} modalities (Lit. \ref{LiteratureComputationalEffectsAndModalities}, \ref{LiteratureModalTypeTheory}):
\begin{equation}
\label{MonadicModalityInIntroduction}
\adjustbox{}{
\hspace{-.6cm}
\begin{tabular}{ll}
\begin{minipage}{8.5cm}
\small
We observe in \cref{ClassicalEpistemicLogicViaDependentTypes} that possible-world semantics for modal logic (in its ``S5'' flavor with which we are concerned here)
is equivalently that induced by dependent type formation along any context extension.
Conversely, this means to observe (Rem. \ref{EpistemicInterpretationOfDependentTypes}) that one may think of standard dependent type theory as epistemic modal
type theory with a universal system of epistemic modal operators indexed by types of ``many possible worlds'' $W \isa \Type$.
From this perspective, the tradition in formal logic to refer to the large type $\mathrm{Type}$ of small types as the ``universe'' gains some vindication.
\end{minipage}
&
\begin{tikzcd}[column sep=30pt]
  \mathllap{
    \scalebox{.7}{
      \color{darkorange}
      \bf
      possibility modality
      \hspace{-15pt}
    }
  }
  &&
  \mathrlap{
    \scalebox{.7}{
      \color{darkorange}
      \bf
      \hspace{-15pt}
      randomness modality
      \hspace{-15pt}
    }
  }
  \\
  \mathrm{Type}_W
    \ar[out=160, in=60,
      looseness=4,
      shorten=-3pt,
      "\scalebox{1.6}{${
          \hspace{1pt}
          \mathclap{
            \possibly_{\!{}_{\scalebox{.45}{$W$}}}
          }
          \hspace{1pt}
        }$}"{pos=.25, description}
    ]
    \ar[out=185, in=175,
      looseness=10,
      phantom,
      "{\scalebox{.7}{$\bot$}}"{yshift=1pt, xshift=3pt}
    ]
    \ar[out=-160, in=-60,
      looseness=4,
      shorten=-3pt,
      "\scalebox{1.4}{${
          \hspace{1pt}
          \mathclap{
            \necessarily_{\mathrlap{\!{}_{\scalebox{.6}{\colorbox{white}{$\!\!\!\!W\!$}}}}}
          }
          \hspace{3pt}
        }$}"{pos=.25, description}
    ]
  \ar[
    rr,
    shift left=14pt,
    "{
      \overset{
        \mathclap{
          \raisebox{4pt}{
            \scalebox{.7}{
              \color{darkgreen}
              \bf
              dependent ``sum''
            }
          }
        }
      }{
      \coprod_W
      }
    }"
  ]
  \ar[
    from=rr,
    "{
      \times W
    }"{description}
  ]
  \ar[
    rr,
    shift right=14pt,
    "{
      \underset{
        \mathclap{
          \raisebox{-4pt}{
            \scalebox{.7}{
              \color{darkgreen}
              \bf
              dependent product
            }
          }
        }
      }{
        \prod_W
      }
    }"{swap}
  ]
  \ar[
    rr,
    phantom,
    shift left=8pt,
    "\scalebox{.7}{$\bot$}"
  ]
  \ar[
    rr,
    phantom,
    shift right=8pt,
    "\scalebox{.7}{$\bot$}"
  ]
  &&
  \mathrm{Type}
    \ar[out=20, in=130,
      looseness=5,
      shorten=-3pt,
      "\scalebox{1.2}{${
          \hspace{6pt}
          \mathclap{
            \randomly_{\!{}_{\scalebox{.45}{$W$}}}
          }
        }$}"{pos=.25, description}
    ]
    \ar[
      r,
      phantom,
      "{
        \scalebox{.7}{$\bot$}
       }"
    ]
    \ar[out=-20, in=-130,
      looseness=5,
      shorten=-3pt,
      "\scalebox{1.2}{${
          \hspace{6pt}
          \mathclap{
            \indefinitely_{\!{}_{\scalebox{.45}{$W$}}}
          }
        }$}"{pos=.25, description}
    ]
  &
  {}
  \\
  \mathllap{
    \scalebox{.7}{
     \color{darkorange}
     \bf
     necessity modality
     \hspace{-15pt}
    }
  }
  &&
  \mathrlap{
    \scalebox{.7}{
      \color{darkorange}
      \bf
      \hspace{-15pt}
      indefiniteness modality
      \hspace{-15pt}
    }
  }
\end{tikzcd}
\end{tabular}
}
\end{equation}

While for classical intuitionistic type theory, this perspective may be of interest to the analytic philosopher (see \cite[Ch. 4]{Corfield20}),
we next claim that applied to {\it linear} dependent type theory the same perspective solves the practical problem of formalized quantum epistemology
relevant for universal quantum programming/certification:

\newpage

\noindent {\bf Quantum epistemic logic as Linear dependent type theory.}
The point is that in linear dependent type theory like {\LHoTT} the situation \eqref{MonadicModalityInIntroduction} has an immediate analog (\cite[\S 2.4]{Riley22}) as $W$-dependent classical intuitionistic types are replaced by $W$-dependent {\it linear} types (quantum data types, interpreted for instance a indexed sets of vector spaces, see \cref{QuantumTypeSemantics}):
In this case and assuming $W$ is {\it finite} (as it is for any realistic quantum measurement) their linear/quantum
nature makes the dependent (co)product adjoints coincide (``ambidexterity'', Lit. \ref{QuantumMeasurementAndZXCalculus})
on the {\it direct sum} of linear types, this reflecting the superposition principle of quantum physics:

\vspace{-7mm}
\begin{equation}
\label{FrobeniusMonadicModalityInIntroduction}
\hspace{0mm}
  W \isa \FiniteType
  \,
  \yields
  \hspace{-.14cm}
\scalebox{.7}{
  \color{gray}
  \def\arraystretch{.9}
  \begin{tabular}{c}
     Frobenius monad of
     \\
     quantum epistemic logic
     \\
     (\cref{QuantumEpistemicLogicViaDependentLinearTypes})  proves
     principles
     \\
     as
     {\it deferred measurement}
     \\
     (Prop. \ref{DeferredMeasurementPrinciple})
  \end{tabular}
}
\hspace{-.8cm}
\begin{tikzcd}[column sep=23pt]
  \mathllap{
     \scalebox{.7}{
     \color{darkorange}
     \bf
      linear possibility
    }
  }
  &&
  \mathrlap{
     \scalebox{.7}{
     \color{darkorange}
     \bf
      linear randomness
    }
  }
  \\[5pt]
  \LinearTypes_W
    \ar[out=160, in=60,
      looseness=4,
      shorten=-3pt,
      "\scalebox{1.6}{${
          \hspace{1pt}
          \mathclap{
            \possibly_{\!{}_{\scalebox{.45}{$W$}}}
          }
          \hspace{1pt}
        }$}"{pos=.25, description}
    ]
    \ar[out=185, in=175,
      looseness=10,
      phantom,
      "\rotatebox{90}{{\scalebox{.8}{$\simeq$}}}"{yshift=1pt, xshift=12pt}
    ]
    \ar[out=-160, in=-60,
      looseness=4,
      shorten=-3pt,
      "\scalebox{1.4}{${
          \hspace{1pt}
          \mathclap{
            \necessarily_{\mathrlap{{}_{\scalebox{.6}{\colorbox{white}{$\!\!\!\!W\!$}}}}}
          }
          \hspace{3pt}
        }$}"{pos=.25, description}
    ]
    \;\;
  \ar[
    rr,
    shift left=14pt,
    "{
      \overset{
        \mathclap{
          \raisebox{4pt}{
            \scalebox{.7}{
              \color{darkgreen}
              \bf
              direct sum
            }
          }
        }
      }{
      \scalebox{1.2}{$\oplus$}_W
      }
    }"
  ]
  \ar[
    from=rr
  ]
  \ar[
    rr,
    shift right=14pt,
    "{
      \underset{
        \mathclap{
          \raisebox{-4pt}{
            \scalebox{.7}{
              \color{darkgreen}
              \bf
              direct sum
            }
          }
        }
      }{
        \scalebox{1.2}{$\oplus$}_W
      }
    }"{swap}
  ]
  \ar[
    rr,
    phantom,
    shift left=8pt,
    "\scalebox{.7}{$\bot$}"
  ]
  \ar[
    rr,
    phantom,
    shift right=8pt,
    "\scalebox{.7}{$\bot$}"
  ]
  &&
  \;\;
  \LinearTypes
    \ar[out=120, in=20,
      looseness=4,
      shorten <=-2pt,
      shorten >=-3pt,
      "\raisebox{-2pt}{${
          \hspace{7pt}
          \mathclap{
            \randomly_{\!\!{}_W}
          }
          \hspace{2pt}
        }$}"{pos=.76, description}
    ]
    \ar[out=5, in=-5,
      looseness=7,
      phantom,
      "{\scalebox{.9}{$\simeq$}}"{yshift=1pt, rotate=-90}
    ]
    \ar[out=-120, in=-20,
      looseness=4,
      shorten <=-2pt,
      shorten >=-3pt,
      "\scalebox{1.1}{${
          \hspace{4pt}
          \mathclap{
            \indefinitely_{\!{}_{\scalebox{.7}{$W$}}}
          }
          \hspace{1pt}
        }$}"{pos=.77, description}
    ]
    \\[5pt]
  \mathllap{
     \scalebox{.7}{
     \color{darkorange}
     \bf
      linear necessity
    }
  }
  &&
  \mathrlap{
     \scalebox{.7}{
     \color{darkorange}
     \bf
      linear indefiniteness
    }
  }
\end{tikzcd}
\hspace{-.6cm}
\rotatebox[origin=c]{0}{
  \scalebox{.7}{
    \color{gray}
    \begin{tabular}{c}
      {\it Classical context}
      (Prop. \ref{QuantumCoEffectsViaFrobeniusAlgebra})
      \\
      Frobenius monad
      \\
      as in {\zxCalculus}
      \\
      gives
      {\it effect-logic} for
      \\
      quantum gates \cref{ControlledQuantumGates}
    \end{tabular}
  }
}
\end{equation}

\vspace{-4mm}
\noindent This means equivalently that
 in the linear case the (co)monadic modal operators coincide, $\possibly_{{}_W} \simeq \necessarily_{{}_W}$, $\randomly_{{}_W} \simeq \indefinitely_{{}_W}$,
 to form a pair of {\it Frobenius monads} (cf. Prop. \ref{QuantumCoEffectsViaFrobeniusAlgebra}),
 reflecting
 the monadic nature of quantum measurement as known from the {\zxCalculus} (Lit. \ref{QuantumMeasurementAndZXCalculus}).
It may be satisfactory to observe that the modal-logical expression of this situation reflects Gell-Mann's {\it principle of quantum compulsion}
(cf: \cite[p. 31]{Bunge76}: ``{In quantum physics anything that is not forbidden [i.e., possible] is compulsory [i.e., necessary].}''):

\vspace{-.6cm}
\begin{equation}
  \label{BasicFormationInLinearModalTypeTheory}
  \overset{
    \mathclap{
      \raisebox{12pt}{
      \scalebox{.7}{
        \color{darkblue}
        \bf
        \def\arraystretch{.9}
        \begin{tabular}{c}
          Finite classical type
          \\
          of many possible worlds
          \\
          (measurement outcomes)
        \end{tabular}
      }
      }
    }
  }{
    W \isa \FiniteType
  }
  \,,
  \hspace{1.2cm}
  \overset{
    \mathclap{
      \raisebox{9pt}{
      \scalebox{.7}{
        \color{darkblue}
        \bf
        \def\arraystretch{.9}
        \begin{tabular}{c}
          $W$-dependent
          \\
          {\color{purple}
          quantum data type}
        \end{tabular}
      }
      }
    }
  }{
  D
  \,\isa\,
  \LinearType_W
  }
  \hspace{.6cm}
     \overset{
       \raisebox{4pt}{
         \scalebox{.7}{
           yields that
         }
       }
     }{
    \yields
  }
  \hspace{-.4cm}
  \adjustbox{
    raise=13pt
  }{
  \begin{tikzcd}[
    row sep=13pt,
    column sep=15pt
  ]
    &
    \mathclap{
    \scalebox{.7}{
      \color{darkorange}
      linear sum
    }
    }
    \\[-15pt]
    &
    \underset{w}{\oplus} D_w
    \ar[
      dr,
      shorten <= -4pt,
      "{\sim}"{sloped, pos=.4}
    ]
    \\
    \possibly D
    \ar[
      rr,
      "{\sim}"{swap},
      "{
        \scalebox{.7}{
          \def\arraystretch{.9}
          \begin{tabular}{c}
            \color{darkorange}
            \bf
            The possible is necessary
            \\
            Principle of quantum compulsion
          \end{tabular}
        }
      }"{swap, yshift=-5pt}
    ]
    \ar[ur, "{\sim}"{pos=.4, sloped}]
    &&
    \necessarily D
  \end{tikzcd}
  }
  \hspace{-.7cm}
  \overset{
    \raisebox{5pt}{
      \scalebox{.7}{
        is a
      }
    }
  }{
   \;\isa\;
   }
   \overset{
     \mathclap{
       \raisebox{8pt}{
         \scalebox{.7}{
           \color{darkblue}
           \bf
           \def\arraystretch{.9}
           \begin{tabular}{c}
             $W$-independent
             \\
             quantum data type
           \end{tabular}
         }
       }
     }
   }{
  \QuantumType
  \hookrightarrow
  \QuantumType_W
  }
\end{equation}

We suggest thinking of this as a Yoneda-Lemma-type statement: The derivation of \eqref{FrobeniusMonadicModalityInIntroduction} is so elementary that it borders
on being tautological, and yet as an organizing principle for quantum effects we will find it to be ubiquitous, for instance in implying the {\it deferred measurement principle} (Prop. \ref{DeferredMeasurementPrinciple})
or the commuting diagram \eqref{BranchingAndCollapseInIntroduction} below, which arguably makes precise many words \cite{Tegmark98} written in the informal literature on the matter.
This leads one to wonder (cf. \cite{AbramskyCoecky07}): Had history proceeded differently, could systematic development of combined modal and linear logic have led pure logicians to
discover the rules of quantum information theory independently of experimental input?

\smallskip

\noindent {\bf Formal logic of quantum measurement effects.}
Remarkably, unwinding the logical rules of this epistemic quantum logic \eqref{BasicFormationInLinearModalTypeTheory}
reveals that it knows all about the state collapse after quantum measurement
including formal proof of its equivalence to {\it branching} into ``many worlds'' (Lit. \ref{EpistemologyOfQuantumPhysics}):

\vspace{-.6cm}
\begin{equation}
  \label{BranchingAndCollapseInIntroduction}
  \adjustbox{}{
  \hspace{-.7cm}

      }
      }
    }
    &
    \HilbertSpace{H}
    \ar[
      rr,
      "{
        G
      }"{description}
    ]
    &&
    \HilbertSpace{K}
    \ar[
      dr,
      end anchor={[yshift=3pt]},
      "{
        \delta_0^{{\color{purple}b}}
      }"{description}
    ]
    \\[-34pt]
    {\color{purple}b}\isa\Bit
    \;\;
    \vdash
    &
    \HilbertSpace{H}
    \ar[
      ur,
      start anchor={[yshift=1pt]},
      "{
        P_{0}
      }"{description}
    ]
    \ar[
      dr,
      start anchor={[yshift=-1pt]},
      "{
        P_{1}
      }"{description}
    ]
    &
    \oplus
    \ar[
      rr,
      phantom,
      "{ \oplus }"
    ]
    &&
    \oplus
    &
    \HilbertSpace{K}
    \\[-32pt]
    &&
    \HilbertSpace{H}
    \ar[
      rr,
      "{
        G
      }"{description}
    ]
    &&
    \HilbertSpace{K}
    \ar[
      ur,
      "{
        \delta_1^{{\color{purple}b}}
      }"{description}
    ]
    \\[-20pt]
    &
    &
    \necessarily_{\Bit}
    \possibly_{\Bit}
    \HilbertSpace{H}_\bullet
    \ar[
      rr,
      "{
        \necessarily_{\Bit}
        \possibly_{\Bit}
        G_\bullet
      }",
      "{
        \scalebox{.7}{
          \color{darkorange}
          \bf
          quantum effects
          Everett-style
        }
      }"{swap, yshift=-5pt}
    ]
    &&
    \necessarily_{\Bit}
    \possibly_{\Bit}
    \HilbertSpace{K}_\bullet
    \ar[
      dr,
      "
        \counit
         {\necessarily_\Bit}
         {\possibly_\Bit \HilbertSpace{K}_\bullet}
      "{sloped}
    ]
    &&
    {}
    \\[-14pt]
    &
    \necessarily_{\Bit}
    \HilbertSpace{H}_\bullet
    \ar[
      ur,
      "{
        \necessarily_{\Bit}
        \big(
          \unit
            {\possibly_{\Bit}}
            {\HilbertSpace{H}_\bullet}
        \big)
      }"{sloped, pos=.4}
    ]
    \ar[
      dr,
      "{
        \counit
          {\necessarily_{\Bit}}
          {\HilbertSpace{H}_\bullet}
      }"{swap, sloped}
    ]
    \ar[
      rrrr,
      "{
        \adjustbox{
          rndfbox={5},
          scale=.7
        }{
          \hspace{-8pt}
          \color{gray}
          \bf
          \def\arraystretch{.9}
          \begin{tabular}{c}
            classically controlled
            \\
            quantum computing cycle
          \end{tabular}
          \hspace{-6pt}
        }
      }"{description}
    ]
    & && &
    \possibly_\Bit
    \HilbertSpace{K}_\bullet
    \\[-14pt]
    &&
    \HilbertSpace{H}_\bullet
    \ar[
      rr,
      "{
        G_\bullet
      }"{description},
      "{
        \scalebox{.7}{
          \color{darkorange}
          \bf
          quantum effects
          Copenhagen-style
        }
      }"{yshift=4pt}
    ]
    &&
    \HilbertSpace{K}_\bullet
    \ar[
      ur,
      start anchor={[yshift=-3pt]},
      "{
        \unit
          {\possibly_\Bit}
          {\HilbertSpace{K}_\bullet}
      }"{swap, sloped}
    ]
    &&
    {}
    \\[-20pt]
    {\color{purple}b}\isa\Bit
    \;\;\vdash
    &
    \HilbertSpace{H}
    \ar[
      r
    ]
    &
    \HilbertSpace{H}_{\color{purple}b}
    \ar[
      rr,
      "{
        G_{{\color{purple}b}}
      }"{description}
    ]
    &&
    \HilbertSpace{K}_{{\color{purple}b}}
    \ar[
      r,
      hook
    ]
    &
    \HilbertSpace{K}
    \\[-29pt]
    &
        \underset{
          {b' \in \Bit}
        }{\sum}
        \vert\psi_{{}_{b'}}\rangle
    \ar[
      r,
      phantom,
      "{ \mapsto }"
    ]
    &
    \vert \psi_{{}_{\color{purple}b}}\rangle
    &\mapsto&
    G_{{\color{purple}b}}
    \vert \psi_{{}_{\color{purple}b}}\rangle
    \\[-23pt]
    {}
    &
    {}
    \ar[
      r,
      phantom,
      "{
        \scalebox{.7}{
          \def\arraystretch{.9}
          \begin{tabular}{c}
          \color{darkgreen}
          \bf
          \scalebox{1.1}{$\QBit$}-measurement
          \\
          {
          \color{darkgreen}
          \bf
          collapse
          }
          \hspace{-3pt}
          \scalebox{.9}{
            (pp. \pageref{QuantumMeasurementCopenhagenStyle}
          })
          \end{tabular}
        }
      }"
    ]
    &
    {}
    \ar[
      rr,
      phantom,
      "{
        \scalebox{.7}{
          \color{darkgreen}
          \bf
          \def\arraystretch{.9}
          \begin{tabular}{c}
          quantum gate conditioned
          \\
          on classical control logic
          \end{tabular}
        }
      }"
    ]
    &
    {}
    &
    {}
    \ar[
      r,
      phantom,
      "{
        \scalebox{.7}{
          \color{darkgreen}
          \bf
          \def\arraystretch{.9}
          \begin{tabular}{c}
          dynamic
          \\
          \scalebox{1.1}{$\QBit$}-state
          \\
          preparation
          \end{tabular}
        }
      }"{yshift=6pt}
    ]
    &
    {}
    &
  \end{tikzcd}
  \hspace{-2cm}
  \end{tabular}
  }
\end{equation}

\newpage

\noindent
{\bf Monads as computational effects.} In a curious generalization of modal logic to functional programming (Lit. \ref{LiteratureFunctionalLanguages}),  monads on a category of data types serve to encode {\it computational effects} (Lit. \ref{LiteratureComputationalEffectsAndModalities}). For instance, a classical program whose output data type is
{\it nominally} $D$ but {\it de facto} the value $\indefiniteness_W D$
of the classical $W$-{\it indefiniteness monad}
\eqref{MonadicModalityInIntroduction} --- often known as the {\it Reader}- or {\it Environment}-monad \eqref{ClassicalReaderMonad} --- actually produces its $D$-valued output only conditioned on the observation (``reading'') of an indefinite variable (``environment'' state) $w \,\isa\, W$, hence on a classical $W$-{\it measurement}, so to speak.
In this sense, a program of the type $D \to \indefinitely_W D'$
has a classical {\it measurement effect} -- quite literally: in its generalized incarnation as the $\mathrm{IO}$-monad \eqref{StateMonadEndofunctor} in {\tt Haskell}, running such a procedure causes the computer to perform a read-out of its RAM-state \eqref{StateMonadAsRAMModel}:

\vspace{-9pt}
\begin{equation}
\label{KleisliCompositioInIntroduction}
  \begin{tikzcd}[
    column sep=4pt,
    row sep=-3pt
  ]
    \mathclap{
    \scalebox{.7}{
      \color{darkblue}
      \bf
      \def\arraystretch{.9}
\begin{tabular}{c}
        plain
        \\
        input data
      \end{tabular}
    }
    }
    &&
    \mathclap{
    \scalebox{.7}{
      \def\arraystretch{.9}
      \begin{tabular}{c}
        \color{darkblue}
        \bf
        indefiniteness-effectful
        \\
        {
          \color{darkblue}
          \bf
          output data
        }
        \rlap{
          \scalebox{.8}{
             \color{gray}
            \eqref{BindingAndReturning}
          }
        }
      \end{tabular}
    }
    }
    \\
    \mathllap{
    f_\bullet
    \;\isa\;\;
    }
    D
    \ar[
      rr,
      "{
        \scalebox{.7}{
          \color{darkgreen}\
          \bf
          effectful
        }
      }",
      "{
        \scalebox{.7}{
          \color{darkgreen}\
          \bf
          program
        }
      }"{swap}
    ]
    &&
    \underset{W}{\indefiniteness}
    \,
    D'
    \\
  \scalebox{0.8}{$  d $}
      &\mapsto&
  \scalebox{0.8}{$  \big(
      w \mapsto f_w(d)
    \big)
    $}
  \end{tikzcd}
  \hspace{1cm}
  \begin{tikzcd}[
    column sep=4pt,
    row sep=-3pt
  ]
    \mathclap{
    \phantom{
    \scalebox{.7}{
      \color{darkblue}
      \bf
      \def\arraystretch{.9}
      \begin{tabular}{c}
        plain
        \\
        input data
      \end{tabular}
    }}
    }
    &&
    \mathclap{
    \phantom{
    \scalebox{.7}{
      \color{darkblue}
      \bf
      \def\arraystretch{.9}
      \begin{tabular}{c}
        indefiniteness-effectful
        \\
        output data
      \end{tabular}
    }}
    }
    \\
    \mathllap{
    g_\bullet
    \;\isa\;\;
    }
    D_2
    \ar[
      rr,
      "{
        \scalebox{.7}{
          \color{darkgreen}\
          \bf
          effectful
        }
      }",
      "{
        \scalebox{.7}{
          \color{darkgreen}\
          \bf
          program
        }
      }"{swap}
    ]
    &&
    \underset{W}{\indefiniteness}
    \,
    D_3
    \\
 \scalebox{0.8}{$   d  $}
      &\mapsto&
\scalebox{0.8}{$    \big(
      w \mapsto g_w(d)
    \big)$}
  \end{tikzcd}
  \hspace{2.5cm}
  \begin{tikzcd}[
    column sep=4pt,
    row sep=-3pt
  ]
    \mathllap{
    \scalebox{.7}{
      \def\arraystretch{.9}
      %

\bigskip

\noindent
{\bf Mixed quantum measurement as monoidal-monadic effect.}
The quantum indefiniteness-monad $\indefiniteness_W$ is in fact a {\it strong monad} (Prop. \ref{IndefinitenessModalityIsStrong}). Besides guaranteeing \eqref{TypingOfStrongMonads} that it really does exist as a programming language construct, this means that it carries a symmetric monoidal monad structure \eqref{SymmetricMonoidalMonadStructure}
$\pair{ \indefiniteness_W }{}$ \eqref{SymmetricMonoidalMonadStructureOnIndefiniteness}. We observe \eqref{DecoherenceFromMonoidalMonadStructure}
that this monoidal monad structure serves to enhance the above computational typing of measurement effects from pure to mixed quantum states \eqref{DensityMatrixInIntroduction}, where it embodies the {\it Born rule} \eqref{BornRule} of quantum measurement in its form originally identified by L{\"u}ders \eqref{DynamicallyLiftedQuantumMeasurementOnDensityMatrix}:
\begin{equation}
\label{LuedersMeasurementInIntroduction}
\hspace{-4cm}
\begin{tikzcd}[
  column sep=25pt,
  row sep={between origins, 10pt}
]
  &[+5pt]
  &[-30pt]
  &[-12pt]
  &[-12pt]
  &[-15pt]
  &[-15pt]
  &[+5pt]
  \\[0pt]
  \mathclap{
  \scalebox{.7}{
    \color{darkblue}
    \bf
    \def\arraystretch{.9}
    \begin{tabular}{c}
      mixed
      \\
      states
    \end{tabular}
  }
  }
  &[-20pt]&[-30pt]
  \mathclap{
  \scalebox{.7}{
    \color{darkblue}
    \bf
    \def\arraystretch{.9}
    \begin{tabular}{c}
      density
      \\
      matrices
    \end{tabular}
  }
  }
  \ar[
    rr,
    phantom,
    "{
      \scalebox{.7}{
        \color{darkgreen}
        \bf
        \def\arraystretch{.9}
        \begin{tabular}{c}
          measure separately
          \\
          states and co-states
        \end{tabular}
      }
    }"
  ]
  &[-12pt]&[-12pt]
  {}
  \ar[
    rr,
    phantom,
    "{
      \scalebox{.7}{
        \color{darkgreen}
        \bf
        \def\arraystretch{.9}
        \begin{tabular}{c}
          decohere: discard
          \\
          off-diagonal entries
        \end{tabular}
      }
    }"{pos=.45}
  ]
  &[-23pt]&[-23pt]
  {}
  &
  \mathclap{
  \scalebox{.7}{
    \color{darkblue}
    \bf
    \def\arraystretch{.9}
    \begin{tabular}{c}
      probability
      \\
      distributions
    \end{tabular}
  }
  }
  \\[10pt]
  \quantized W
  &&
  \osum_W \ComplexNumbers
  &&
  \indefinitely_W \ComplexNumbers
  \\
  \otimes
  \ar[rr, "{ \simeq }"
  {description}]
  &&
  \otimes
  \ar[
    rr,
    "{
      \scalebox{.7}{$
        \collapse_W
      $}
      \,\defneq\;
      \join
        { \indefinitely_W }
        { \ComplexNumbers }
      \,\circ\;
      \unit
        { \indefinitely_W }
        { \osum_W \ComplexNumbers }
    }"{yshift=2pt},
    "{ \otimes }"{description},
    "{
      \scalebox{.7}{$
        \collapse_W
      $}
      \,\defneq\;
      \join
        { \indefinitely_W }
        { \ComplexNumbers^{\mathrlap{\ast}} }
      \,\circ\;
      \unit
        { \indefinitely_W }
        { \osum_W \ComplexNumbers^{\mathrlap{\ast}} }
    }"{swap, yshift=-2pt}
  ]
  &&
  \otimes
  \ar[
    rr,
    "{
      \pair
        { \indefinitely_W }
        {
          \ComplexNumbers
          ,
          \ComplexNumbers^{\mathrlap{\ast}}
        }
    }"{description}
  ]
  &&
  \indefinitely_W
  \!\!
  \def\arraystretch{.8}
  \begin{array}{c}
    \ComplexNumbers
    \\
    \otimes
    \\
    \ComplexNumbers^{\mathrlap{\ast}}
  \end{array}
  \ar[
    r,
    "{
      \indefinitely_W
      \,
      \mathrm{ev}
    }"{description}
  ]
  &[-10pt]
  \indefinitely_W
  \ComplexNumbers
  \\
  (\quantized W)^{\mathrlap{\ast}}
  \ar[
    urrrrrrr,
    rounded corners,
    to path={
         ([yshift=-00pt]\tikztostart.south)
      -- ([yshift=-19pt]\tikztostart.south)
      -- node {
         \scalebox{.7}{
           \colorbox{white}{$
             \collapse_W
           $}
         }
      }
         ([yshift=-29pt]\tikztotarget.south)
      -- ([yshift=-00pt]\tikztotarget.south)
    }
  ]
  &&
  \osum_W \ComplexNumbers^{\mathrlap{\ast}}
  &&
  \indefinitely_W \ComplexNumbers^{\mathrlap{\ast}}
  \\[5pt]
  &&
  \mathclap{
  \scalebox{.7}{
    \color{gray}
    \bf
    \def\arraystretch{.9}
\begin{tabular}{c}
      tensor product of
      \\
      free $\indefinitely_W$-modales
    \end{tabular}
  }
  }
  \ar[
    rr,
    phantom,
    "{
      \;\;\;\;
      \scalebox{.7}{
        \color{gray}
        \begin{tabular}{c}
          separately handle
          \\
          pure $\indefiniteness_W$-
          effects
        \end{tabular}
      }
    }"
  ]
  &&
  {}
  \ar[
    rr,
    phantom,
    shift left=8pt,
    "{
      \scalebox{.7}{
        \color{purple}
        \bf
        \def\arraystretch{.9}
        \begin{tabular}{c}
          monoidal monad
          \\
          structure
          on
          $\indefinitely_W$
        \end{tabular}
      }
    }"
  ]
  &&
  {}
  \\[25pt]
  \scalebox{\termscale}{$
  \vert \psi \rangle
  \langle \psi \vert
  $}
  \ar[
    rr,
    phantom,
    "{
      \mapsto
    }"{pos=.6}
  ]
  &&
  \def\arraystretch{.9}
  \begin{array}{c}
  \scalebox{\termscale}{$
  \big(
  \,
  \sum_w
  \vert w \rangle \langle w \vert \psi \rangle
  \,
  \big)
  $}
  \\
  \scalebox{\termscale}{$
    \otimes
  $}
  \\
  \scalebox{\termscale}{$
  \big(
  \,
  \sum_{w'}
  \langle \psi \vert w' \rangle
  \langle w' \vert
  \,
  \big)
  $}
  \end{array}
  &\mapsto&
  \scalebox{\termscale}{$
  \Big(
    (w,w')
    \mapsto
    \vert w \rangle
    \underset{
      \scalebox{.7}{
        \color{gray}
        coherences
      }
    }{
    \underbrace{
    \color{black}
    \langle w \vert \psi \rangle
    \langle \psi \vert w' \rangle
    }
    }
    \langle w' \vert
  \Big)
  $}
  \ar[
    rrr,
    phantom,
    "{ \mapsto }"
  ]
  &&
  &
  \scalebox{\termscale}{$
  \big(
    w
    \,\mapsto\,
    \big\vert\langle w\vert \psi \rangle\big\vert^2
  \big)
  $}
  \\[5pt]
  \mathclap{
    \scalebox{.7}{
      \color{darkblue}
      \bf
      \def\arraystretch{.9}
      \begin{tabular}{c}
        a pure state
        \\
        among mixed
      \end{tabular}
    }
  }
  &&&&&&&
  \mathclap{
    \scalebox{.7}{
      \color{purple}
      \bf
      \def\arraystretch{.9}
      \begin{tabular}{c}
        Born rule
      \end{tabular}
    }
  }
\end{tikzcd}
\hspace{-4cm}
\end{equation}

Moreover, postcomposition with the monoidal monad structure $\pair{\indefinitely_W}{}$ makes the enhancement of parameterized quantum circuits from pure to mixed states a functor of $\indefinitely_W$-effectful maps \eqref{PairedKleisliComposition},
\begin{equation}
  \label{EnhancementFromPureToMixedStatesInIntroduction}
  \hspace{-4mm}
  \begin{tikzcd}[column sep=45pt]
    {}
    \ar[
      rr,
      phantom,
      "{
        \QuantumTypes_{\indefinitely_W}
      }"
    ]
    &&
    {}
    \ar[
      rr,
      shorten >=-20pt,
      "{
        \scalebox{.7}{
          \color{darkgreen}
          \bf
          enhancement to mixed states
        }
      }"{pos=.6, yshift=1pt}
    ]
    &&
    {}
    \ar[
      rr,
      phantom,
      "{
        \QuantumTypes_{\indefinitely_W}
      }"
    ]
    &&
    {}
    \\[-11pt]
    \HilbertSpace{H}_1
    \ar[
      rr,
      "{
        G_\bullet
      }"
    ]
    &&
    \underset{W}{\indefinitely}
    \,
    \HilbertSpace{H}_2
    &\longmapsto&
    \def\arraystretch{.9}
    \begin{array}{c}
      \HilbertSpace{H}_1
      \\
      \otimes
      \\
      \HilbertSpace{H}^\ast_1
    \end{array}
    \ar[
      rr,
      rounded corners,
      to path={
           ([yshift=-00pt]\tikztostart.south)
        -- ([yshift=-10pt]\tikztostart.south)
        -- node{
          \scalebox{.8}{\colorbox{white}{$
            (G \otimes {G^\dagger}^\ast)_\bullet
          $}}
        }
           ([yshift=-11pt]\tikztotarget.south)
        -- ([yshift=-00pt]\tikztotarget.south)
      }
    ]
    \ar[
      r,
      shorten=-4pt,
      "{
        \hspace{-7pt}
        \def\arraystretch{.9}
        \begin{array}{c}
          G_\bullet
          \\
          \otimes
          \\
          {G^\dagger}^\ast_\bullet
        \end{array}
        \hspace{-7pt}
      }"{description}
    ]
    &[+10pt]
    \def\arraystretch{.9}
    \begin{array}{c}
      \underset{W}{\indefinitely}
      \HilbertSpace{H}_2
      \\
      \otimes
      \\
      \underset{W}{\indefinitely}
      \HilbertSpace{H}^\ast_2
    \end{array}
    \ar[
      r,
      "{
        \pair
          { \indefinitely_W }
          {
            \HilbertSpace{H}_1
            ,
            \HilbertSpace{H}^\ast_1
          }
      }"
    ]
    &[+15pt]
    \underset{W}{\indefinitely}
    \def\arraystretch{.9}
    \begin{array}{c}
      \HilbertSpace{H}_2
      \\
      \otimes
      \\
      \HilbertSpace{H}^\ast_2
    \end{array}
  \end{tikzcd}
\end{equation}

\vspace{-2mm}
\noindent in that it respects (Lem. \ref{EnhancingIndefinitenessEffectsToMixedStates})
their effect-bound (Kleisli) composition \eqref{KleisliCompositioInIntroduction}:
\vspace{-2mm}
\begin{equation}
  \big(
    \pair
      { \indefinitely_W }
      {
        \HilbertSpace{H}_2
        ,\,
        \HilbertSpace{H}_2^\ast
      }
    \circ
    (
    G_\bullet
    \otimes
    {G_\bullet^\dagger}^\ast
    )
  \big)
    \;\mbox{\tt >=>}\;
  \big(
    \pair
      { \indefinitely_W }
      {
        \HilbertSpace{H}_3
        ,\,
        \HilbertSpace{H}_3^\ast
      }
    \circ
    (
    H_\bullet
    \otimes
    {H_\bullet^\dagger}^\ast
    )
  \big)
  \;\;\;
  =
  \;\;\;
  \pair
    { \indefinitely_W }
    {
      \HilbertSpace{H}_3
      ,\,
      \HilbertSpace{H}_3^\ast
    }
  \circ
  \Big(
  \big(
    G_\bullet
    \;
    \mbox{\tt >=>}
    \;
    H_\bullet
  \big)
  \otimes
  \big(
    {G_\bullet^\dagger}^\ast
    \;
    \mbox{\tt >=>}
    \;
    {H_\bullet^\dagger}^\ast
  \big)
  \Big)
  \,.
\end{equation}

\vspace{-1mm}
\noindent This means that the above computational effective typing of parameterized quantum circuits
with quantum measurement enhances {\it verbatim} from pure to mixed states!

\smallskip

\noindent
{\bf The modal quantum logic QuantumState.}
We go one step further and observe (\cref{MixedQuantumTypes}) a {\it modal-logical origin} even of the notion of
mixed quantum states \eqref{DensityMatrixInIntroduction} and the quantum channel operations between them.
Namely, observing
$$
  \underset{W}{\randomly}
  \,
  \HilbertSpace{K}
  \;\;
  \defneq
  \;\;
  \underset{W}{\osum}
  \,
  \HilbertSpace{K}
  \;\;
  \simeq
  \;\;
  \underset{W}{\osum}
  \,
  \big(
  \HilbertSpace{K}
  \otimes
  \TensorUnit
  \big)
  \;\;
  \simeq
  \;\;
  \HilbertSpace{K}
  \otimes
  \big(
  \underset{W}{\osum}
  \,
  \TensorUnit
  \big)
  \;\;
  \defneq
  \;\;
  \HilbertSpace{K}
  \otimes
  \quantized W
$$
density matrices are identified among the ``indefinitely random scalars'':
$$
  \scalebox{.7}{
    \color{darkblue}
    \bf
    $\quantized W$-(density-)matrices
  }
  \;\;\;
  \quantized W
  \otimes
  \quantized W^\ast
  \;\;\;\;
   \simeq
  \;\;\;\;
  \underset{W}{\indefinitely}
  \,
  \underset{W}{\randomly}
  \,
  \TensorUnit
  \;\;
  \scalebox{.7}{
    \color{darkblue}
    \bf
    \begin{tabular}{c}
      $W$-indefinitely
      $W$-random
      \\
      scalars
    \end{tabular}
  }
$$

\vspace{-2mm}
\noindent
This equivalence ranges deeper -- it is actually an equivalence of the corresponding monads, and as such eventually is the modal-logical reason for {\it unitarity} of quantum gates -- as follows:

\smallskip

Generally, for {\it dualizable} \eqref{CoEvaluationForDualObjects}  -- namely finite-dimensional -- quantum types $\HilbertSpace{H} \,\isa\, \DualizableQuantumTypes$ their tensoring-functors again are in ambidextrous adjunction \eqref{AmbidextrousAdjunctionOfTensoringWithDualizableObjects}, yielding another Frobenius monad (cf. Rem. \ref{QuantumStateEffectAsQuantumWriter}) --- the linear/quantum version of the classical State-monad \eqref{StateMonadEndofunctor}:
\vspace{-2mm}
\begin{equation}
\label{QuantumStateMonadInIntroduction}
\hspace{-2.3cm}
\adjustbox{raise=20pt}{
\begin{tikzcd}[column sep=23pt]
  &
  \mathllap{
     \scalebox{.7}{
     \color{darkorange}
     \bf
      QuantumState
    }
  }
  &&
  \mathrlap{
     \scalebox{.7}{
     \color{darkorange}
     \bf
      QuantumStore
    }
  }
  \\[5pt]
  {}
  \ar[
    r,
    phantom,
    "{
      \scalebox{.7}{$\bot$}
    }"
  ]
  &
  \QuantumTypes
    \ar[out=160, in=60,
      looseness=4,
      shorten=-3pt,
      "{
          \mathclap{
            \scalebox{1}{\colorbox{white}{$
              \quantized W \State
            $}}
          }
        }"{pos=.25, description}
    ]
    \ar[out=-160, in=-60,
      looseness=4,
      shorten=-3pt,
      "{
          \mathclap{
            \scalebox{1}{\colorbox{white}{$
              \quantized W^\ast \Store
            $}}
          }
        }"{pos=.25, description}
    ]
    \;\;
  \ar[
    rr,
    shift left=16pt,
    "{
      \randomly_W
    }"{description}
  ]
  \ar[
    from=rr,
    "{
      \indefiniteness_W
    }"{description}
  ]
  \ar[
    rr,
    shift right=16pt,
    "{
      \scalebox{.9}{$\randomness$}_W
    }"{description}
  ]
  \ar[
    rr,
    phantom,
    shift left=8pt,
    "\scalebox{.7}{$\bot$}"
  ]
  \ar[
    rr,
    phantom,
    shift right=8pt,
    "\scalebox{.7}{$\bot$}"
  ]
  &&
  \;\;
  \QuantumTypes
    \ar[out=120, in=20,
      looseness=4,
      shorten <=-2pt,
      shorten >=-3pt,
      "{
          \mathclap{
            \scalebox{1}{\colorbox{white}{$
              \quantized W^\ast \Store
            $}}
          }
        }"{pos=.75, description}
    ]
    \ar[out=-120, in=-20,
      looseness=4,
      shorten <=-2pt,
      shorten >=-3pt,
      "{
          \mathclap{
            \scalebox{1}{\colorbox{white}{$
              \quantized W \State
            $}}
          }
        }"{pos=.75, description}
    ]
  \ar[
    r,
    phantom,
    "{
      \scalebox{.7}{$\bot$}
    }"
  ]
    &
    {}
\end{tikzcd}
\hspace{-.2cm}
\raisebox{-15pt}{generally:}
\hspace{-.2cm}
\begin{tikzcd}[
  column sep=30pt,
  row sep=20pt
]
    \mathllap{
      \scalebox{.7}{
        \color{darkorange}
        \bf
        QuantumState
      }
    }
    &&
    \mathrlap{
      \scalebox{.7}{
        \color{darkorange}
        \bf
        QuantumStore
      }
    }
    \\
    \QuantumTypes
    \ar[out=160, in=60,
      looseness=4,
      "\raisebox{-2pt}{${
          \hspace{-7pt}
          \mathclap{
            \adjustbox{
              margin=2pt,
              bgcolor=white
            }{$
              \HilbertSpace{H}\mathrm{State}
            $}
          }
        }$}"{pos=.26, description},
    ]
    \ar[out=185, in=175,
      looseness=10,
      phantom,
      "{\scalebox{.7}{$\bot$}}"{yshift=1pt, xshift=3pt}
    ]
    \ar[out=-160, in=-60,
      looseness=4,
      shorten=-0pt,
      "\raisebox{-2pt}{${
          \hspace{-7pt}
          \mathclap{
            \adjustbox{
              margin=2.5pt,
              bgcolor=white
            }{$
              \HilbertSpace{H}^\ast\mathrm{Store}
            $}
          }
        }$}"{pos=.26, description},
    ]
  \ar[
    rr,
    shift left=14pt,
    "{
        (\mbox{-})
        \otimes
        \HilbertSpace{H}
    }"
  ]
  \ar[
    from=rr,
    "{
      (\mbox{-})
      \otimes
      \HilbertSpace{H}^\ast
    }"{description}
  ]
  \ar[
    rr,
    shift right=14pt,
    "{
      (\mbox{-})
      \otimes
      \HilbertSpace{H}
    }"{swap}
  ]
  \ar[
    rr,
    phantom,
    shift left=8pt,
    "\scalebox{.7}{$\bot$}"
  ]
  \ar[
    rr,
    phantom,
    shift right=8pt,
    "\scalebox{.7}{$\bot$}"
  ]
  &&
  \QuantumTypes
    \ar[out=120, in=20,
      looseness=4,
      shorten <=-2pt,
      shorten >=-3pt,
      "\raisebox{-2pt}{${
          \hspace{7pt}
          \mathclap{
            \adjustbox{
              margin=2pt,
              bgcolor=white
            }{$
              \HilbertSpace{H}\mathrm{Store}
            $}
          }
        }$}"{pos=.7, description},
    ]
    \ar[out=5, in=-5,
      looseness=8.5,
      phantom,
      "{\scalebox{.7}{$\bot$}}"{yshift=1pt}
    ]
    \ar[out=-120, in=-20,
      looseness=4,
      shorten <=-2pt,
      shorten >=-3pt,
      "\raisebox{-2pt}{${
          \hspace{7pt}
          \mathclap{
            \adjustbox{
              margin=2pt,
              bgcolor=white
            }{$
              \HilbertSpace{H}^\ast\mathrm{State}
            $}
          }
        }$}"{pos=.7, description},
    ]
\end{tikzcd}
}
\hspace{-2cm}
\end{equation}

\vspace{-5mm}
\noindent
This identifies the $\quantized W$State-monad with the monad that is induced, in turn, by the epistemic indefiniteness/randomness
adjunction $\indefiniteness_W \dashv \randomness_W$ \eqref{FrobeniusMonadicModalityInIntroduction}:
\vspace{-2mm}
$$
  \scalebox{.7}{
    \color{darkblue}
    \bf
    \def\arraystretch{.9}
    \begin{tabular}{c}
      QuantumState
    \end{tabular}
  }
  \;\;
  \quantized W \State
  \;\;\;
    \defneq
  \;\;\;
  \quantized W
  \linmap
  \big(
    (-) \otimes \quantized W
  \big)
  \;\;
    \simeq
  \;\;
  (-)
  \otimes
  \quantized W
  \otimes
  \quantized W^\ast
  \;\;\;
    \simeq
  \;\;\;
  \underset{W}{\indefinitely}
  \,
  \underset{W}{\randomly}
  \;\;
  \scalebox{.7}{
    \color{darkblue}
    \bf
    \def\arraystretch{.9}
    \begin{tabular}{c}
      Quantum
      \\
      indefinite
      \\
      randomness
    \end{tabular}
    \color{gray}
    \def\arraystretch{1.1}
  }
$$
By itself, the QuantumState monad encodes qRAM-effects \eqref{QRAMAdjointness}, in quantization of the RAM-effect \eqref{StateMonadAsRAMModel}
of classical State-monads. But with its {\it monad transformations} \eqref{ComponentOfMonadTransformation}
taken into account it models quantum channels \eqref{QuantumChannel}:

$$
\begin{minipage}{11cm}
  \label{DistributingFrobeniusMonads}
  {\bf
  Distributing Frobenius monads at the heart of quantum information theory.}
  The QuantumState (co)monads pairwise distribute over the QuantumEnvironment (co)monads (Prop. \ref{QuantumStoreDistributesOverQuantumRreader}), which implies

  \begin{itemize}
    \item[{\bf (i)}] 2-sided Kleisli categories \eqref{CoMonadicKleisliComposition} of (Prop. \ref{QuantumStoreIndefinitenesKleisliCategory}):
    \begin{itemize}[leftmargin=.6cm]
      \item[{\bf (a)}]
      QuantumEnvironment-contextful \& QuantumState-effectful \\
      maps modelling mixed state preparation,
      eg.
      $
        \randomly_W
        \TensorUnit
        \to
        \HilbertSpace{H}
        \otimes
        \HilbertSpace{H}^\ast
      $
      \item[{\bf (b)}]
      QuantumState-effectful
      \&
      QuantumEnvironment-contextful \\
      maps modelling mixed state observables,
      eg.
      $
        \HilbertSpace{H}
        \otimes
        \HilbertSpace{H}^\ast
        \to
        \indefinitely_W
        \TensorUnit
      $
    \end{itemize}
    acted on
    by QuantumState-
    and QuantumStore-transformations,
    \\respectively.
  \item[{\bf (ii)}]
  the composite monads
  $
    \indefinitely_W
    \circ
    \HilbertSpace{H}\State
    \;\;\dashv\;\;
    \HilbertSpace{H}\Store
    \circ
    \randomly_W
  $
  exist
  \eqref{StructureMapsOfCompositeMonad}.
  \end{itemize}

\end{minipage}
\hspace{.5cm}
\adjustbox{
  fbox
}{\small
  \begin{tikzcd}[
    row sep=-3pt,
    column sep=2pt
  ]
    \mathclap{
      \scalebox{.7}{
        \color{darkblue}
        \bf
        \def\arraystretch{.9}
\begin{tabular}{c}
          Quantum
          \\
          indefiniteness
        \end{tabular}
      }
    }
    &
    \mathclap{
      \scalebox{.7}{
        \color{darkorange}
        \bf
        \def\arraystretch{.9}
        \begin{tabular}{c}
          Quantum
          \\
          environment
        \end{tabular}
      }
    }
    &
    \mathclap{
      \scalebox{.7}{
        \color{darkblue}
        \bf
        \def\arraystretch{.9}
        \begin{tabular}{c}
          quantum
          \\
          randomness
        \end{tabular}
      }
    }
    \\
    \underset{W}{\indefinitely}
    &
    (-)
    \otimes
    \quantized W
    \ar[r, phantom, "{ \simeq }"]
    \ar[l, phantom, "{ \simeq }"]
    &
    \underset{W}{\randomly}
    \\[10pt]
    \underset{W}{\indefinitely}
    \,
    \underset{W}{\randomly}
    &
    (-)
    \otimes
    \,
    \quantized W
    \otimes
    \quantized W^\ast
    \ar[l, phantom, "{ \simeq }"]
    \ar[r, phantom, "{ \simeq }"]
    &
    \underset{W}{\randomly}
    \,
    \underset{W}{\indefinitely}
    \\
    \mathclap{
      \scalebox{.7}{
        \color{darkblue}
        \bf
        \def\arraystretch{.9}
        \begin{tabular}{c}
          Quantum
          \\
          indefinite
          \\
          randomness
        \end{tabular}
      }
    }
    &
    \mathclap{
      \scalebox{.7}{
        \color{darkorange}
        \bf
        \def\arraystretch{.9}
        \begin{tabular}{c}
          quantum
          \\
          state
        \end{tabular}
      }
    }
    &
    \mathclap{
      \scalebox{.7}{
        \color{darkblue}
        \bf
        \def\arraystretch{.9}
        \begin{tabular}{c}
          Quantum
          \\
          random
          \\
          indefiniteness
        \end{tabular}
      }
    }
    \\[5pt]
    \mathrm{Monads}
    &
    \mathrm{FrobMonads}
    \ar[r]
    \ar[l]
    &
    \mathrm{CoMonads}
  \end{tikzcd}
}
$$

\newpage

\noindent
{\bf Unitary quantum channels are QuantumState-transformations.} In fact, the composition of QuantumState monads with the indefiniteness-modality
is {\it itself} a relative monad {\it on the category of QuantumState monads} (Prop. \ref{IndefinitenessEffectOnQuantumStateEffect}):
\begin{equation}
\hspace{-7mm}
  \begin{tikzcd}[row sep=-3pt, column sep=15pt]
    \scalebox{.7}{
      \color{darkblue}
      \bf
      \def\arraystretch{.9}
      \begin{tabular}{c}
        Indefiniteness
        \\
        of pure states
      \end{tabular}
    }
    &
    \underset{W}{\indefinitely}
    \;\isa\;
    &
    \QuantumTypes
    \ar[rr]
    &&
    \QuantumTypes
    &
    \scalebox{.7}{
      \color{darkorange}
      \bf
      \def\arraystretch{.9}
      \begin{tabular}{c}
        Monad on
        \\
        quantum types
      \end{tabular}
    }
    \\
    &
    &
  \scalebox{0.8}{$    \HilbertSpace{H}
  $}
    &\longmapsto&
    \scalebox{0.8}{$  \underset{W}{\indefinitely}
    \,
    \HilbertSpace{H}
    $}
    \\[+5pt]
    \scalebox{.7}{
      \color{darkblue}
      \bf
      \def\arraystretch{.9}
      \begin{tabular}{c}
        Indefiniteness
        \\
        of mixed states
      \end{tabular}
    }
    &
    \underset{W}{\indefiniteness}
    \,
    \circ
    \;\colon\;
    &
    \mathrm{StateMnd}(\QuantumTypes)
    \ar[rr]
    &&
    \mathrm{Mnd}(\QuantumTypes)
    &
    \scalebox{.7}{
      \color{darkorange}
      \bf
      \def\arraystretch{.9}
      \begin{tabular}{c}
        Relative monad on
        \\
        QuantumState monads
      \end{tabular}
    }
    \\
    &
    &
 \scalebox{0.8}{$     \HilbertSpace{H}\State
 $}
    &\longmapsto&
  \scalebox{0.8}{$  \underset{W}{\indefinitely}
    \circ
    \HilbertSpace{H}\State
    $}
  \end{tikzcd}
\end{equation}
This is such that the enhancement \eqref{EnhancementFromPureToMixedStatesInIntroduction} of indefiniteness-effectful maps from pure to mixed states is a QuantumState transformation iff the maps are unitary, $W$-wise (Prop. \ref{LiftingParamaterizedQuantumCircuitsToParameterizedQuantumChannels}):
\begin{equation}
\hspace{-2.6cm}
  \adjustbox{}{
    \begin{minipage}{6cm}
      \footnotesize
      Where pure quantum states are terms of linear (quantum) type $\HilbertSpace{H}$ \eqref{LinearTypesAsSubstructuralTypes},
      the (ambient, linear) type of {\it mixed states} in the form of (density) matrices may be identified with the QuantumState-{\it monad} $\HilbertSpace{H}\State$ \eqref{QuantumStateMonadInIntroduction}
      acting on these linear types: Where a quantum circuit of pure states is a map of linear (quantum) types, a quantum circuit of mixed states is a {\it transformation of monads} \eqref{ComponentOfMonadTransformation} of QantumState monads -- a {\it QuantumState transformation}.

      \smallskip

      It is with this natural typing of quantum circuits literally as QuantumState transformations that the unitarity axiom of quantum physics is reflected in modal quantum logic.

      \smallskip

      Moreover, the indefiniteness-modality $\indefinitely_W$ on quantum types enhances to a (relative) {\it monad on QuantumState monads} (Prop. \ref{IndefinitenessEffectOnQuantumStateEffect}), such that the $\indefiniteness$-modal typing of parameterized quantum circuits (\cref{ControlledQuantumGates}) is formally the same for pure and mixed states, under the enhancement $\HilbertSpace{H} \,\mapsto\, \HilbertSpace{H}\State$ of underlying categories of types from $\QuantumTypes$ to $\mathrm{StateMnd}(\QuantumTypes)$.
    \end{minipage}
  }
 \def\arraystretch{1}
\begin{array}{c}
 \hspace{-.9cm}
 \adjustbox{}{
    \scalebox{.7}{
      \color{darkblue}
      \bf
      \def\arraystretch{.9}
      \begin{tabular}{c}
        Parameterized
        \\
        quantum circuit
      \end{tabular}
    }
\hspace{.4cm}
\adjustbox{
  raise=-.9cm
}{
\begin{tikzpicture}
  \node at (-3, 1) {
    $W$
  };
  \node at (+3, 1) {
    $W$
  };
  \node at (-3, 0) {
    $\HilbertSpace{H}_1$
  };
  \node at (+3, 0) {
    $\HilbertSpace{H}_2$
  };
  \draw[line width=3]
    (-2.6, 1) to (2.6,1);
  \draw[white,line width=2.2]
    (-2.6, 1) to (2.6,1);
  \draw[black, line width=2]
   (-2.6, 0) to (+2.6,0);
  \draw[fill=black]
    (0,1) circle (.1);
  \draw[line width=1.3pt]
    (0,1) to (0,0);
  \draw[fill=white, line width=1.2]
    (-.5, -.5)
    rectangle
    (+.5, +.5);
  \node at (0,0)
    { $G_\bullet$ };
\end{tikzpicture}
}
}
  \\
  {}
  \\
\end{array}
  \hspace{-2cm}
\end{equation}
These unitary quantum channels are also QuantumStore-comonad transformations, and as such their action \eqref{ExtensionOfModalesAlongMonadTransformations}
on the quantum observables typed as QuantumStore-contextful scalars (Ex. \ref{AlgebraOfQuantumObservablesAsQuantumStoreContextfulMaps})
gives {\it Heisenberg evolution} (Prop. \ref{QuantumStateEvolutionIsHeisenbergEvolution}):
\begin{equation}
  \label{HeisenbergEvolutionInIntroduction}
  \hspace{-2.3cm}
  \begin{tikzcd}[
    row sep=4pt
  ]
    {}
    \ar[
      rr,
      phantom,
      "{
        \scalebox{.7}{
          \color{darkblue}
          \bf
          Observable =
          QuantumState-contextful scalar
        }
      }"
    ]
    &&
    {}
    \ar[
      rr,
      phantom,
      "{
        \scalebox{.7}{
          \color{darkgreen}
          \bf
          \begin{tabular}{c}
            acted on by unitary
            \\
            QuantumStore transformation
          \end{tabular}
        }
      }"{yshift=-4pt}
    ]
    &[+5pt]&[+5pt]
    {}
    \\
    \HilbertSpace{H}_1
    \otimes
    \HilbertSpace{H}_1^\ast
    \ar[
      rr,
      "{
        \mathcal{O}_A
      }"
    ]
    &[-13pt]&[-13pt]
    \TensorUnit
    &\mapsto&
    \HilbertSpace{H}_1
    \otimes
    \HilbertSpace{H}_1^\ast
    \ar[
      rr,
      "{
        U
          \otimes
        {U^\dagger}^\ast
      }"
    ]
    \ar[
      rrrr,
      rounded corners,
      to path={
           ([yshift=+00pt]\tikztostart.north)
        -- ([yshift=+10pt]\tikztostart.north)
        -- node {
          \scalebox{.8}{
            \colorbox{white}{$
              \mathcal{O}_{
                U^\dagger \cdot A \cdot U
              }
            $}
          }
        }
           ([yshift=+11pt]\tikztotarget.north)
        -- ([yshift=+00pt]\tikztotarget.north)
      }
    ]
    &[-13pt]&[-13pt]
    \HilbertSpace{H}_2
    \otimes
    \HilbertSpace{H}_2^\ast
    \ar[
      rr,
      "{
        \mathcal{O}_A
      }"
    ]
    &[-13pt]&[-13pt]
    \TensorUnit
    \\
   \scalebox{0.8}{$   \rho  $}
      &\longmapsto&
   \scalebox{0.8}{$   \mathrm{Tr}(\rho \cdot A)
   $}
    &&
  \scalebox{0.8}{$    \rho $}
    &\longmapsto&
 \scalebox{0.8}{$     U\cdot \rho \cdot U^\dagger $}
    &\longmapsto&
 \scalebox{0.8}{$     \mathrm{Tr} (
      \rho
        \cdot
      U^\dagger
        \cdot
      A
        \cdot
      U
    )
    $}
  \end{tikzcd}
  \hspace{-2cm}
\end{equation}

\medskip

\noindent
{\bf General quantum channels.}
The other canonical example of
a QuantumState-monad transformation is the
(quantum channel given by) coupling (tensoring) to a uniform bath state \eqref{UniformCouplingQuantumChannel},
whose formal dual is the  QuantumStore-comonad transformation given by partial trace
$$
  \mathrm{couple}^{\HilbertSpace{H}}
  \;\colon\;
  \HilbertSpace{H}\State
  \longrightarrow
  (\HilbertSpace{H} \otimes \HilbertSpace{B})\State
  \hspace{1cm}
  \mathrm{average}^{\HilbertSpace{B}}
  \;\colon\;
  (\HilbertSpace{H} \otimes \HilbertSpace{B})\Store
  \longrightarrow
  \HilbertSpace{H}\Store
  \,.
$$
This way, every {\it unistochastic quantum channel} \eqref{UnistochasticEnvironmentalRepresentation}
appears as a composite of a QuantumState transformation followed by a QuantumStore-transformation, and as such acts \eqref{CovariantFunctorOnModalesInducedByMonadTransformation}
on the 2-sided Kleisli categories
(Lem. \ref{DistributiveMonadTransformationsActOnContextEffecfulMaps})
of quantum observables and quantum state preparations.

As a simple but relevant example, the
DQC1-model of quantum computation \eqref{ProbabilityFormulaInDQC1} on a single (``clean'') qbit coupled to a uniformly distributed bath
is naturally typed in this monadic language as follows:\footnote{Notice that the environmental mixed state produced by this construction is un-normalized. This is no restriction of generality, it just means that for extracting actual probabilities one needs to normalize by the trace of the density matrix.}
\begin{equation}
  \label{MonadicDQC1}
  \hspace{-1cm}
  \begin{tikzcd}[
    column sep=32pt,
    row sep=12pt
  ]
     \randomly_\Bit \TensorUnit
     \ar[
       d,
       "{
         \overset{
           \mathclap{
           \;\;\;\;
           \raisebox{4pt}{
           \scalebox{.7}{
             \color{gray}
             (Ex. \ref{StatePreparationWithProbabilityWeights})
           }
           }
           }
         }{
         \mathrm{weigh}_{\delta_\bullet^0}
         }
         \,\mbox{\tt >=>}\,
         \mathrm{prep}
       }"{swap, pos=.0}
     ]
    \ar[
      dr,
      gray,
      rounded corners,
      to path={
           ([xshift=+00pt]\tikztostart.east)
        -- node {
          \scalebox{.7}{
            \colorbox{white}{
              \color{gray}
              QuantumState effect
            }
          }
        }
           ([xshift=+90pt]\tikztostart.east)
        -- ([xshift=-14pt]\tikztotarget.north)
      }
    ]
    \ar[
      dr,
      gray,
      "{
        \scalebox{.7}{
          \color{darkgreen}
          \bf
          preparation
        }
      }"{description, sloped}
    ]
    &[+10pt]
    &&[10pt]
    \\
     \QBit\State(\TensorUnit)
     \ar[
       r,
       "{
         \mathrm{couple}
           ^{ \HilbertSpace{B} }
       }"{swap, yshift=-1pt}
     ]
     \ar[
       rr,
       gray,
       rounded corners,
       to path={
           ([yshift=-00pt]\tikztostart.south)
        -- ([yshift=-20pt]\tikztostart.south)
        -- node {
          \scalebox{.7}{
            \colorbox{white}{
              QuantumState transformation
            }
          }
        }
           ([yshift=-20pt]\tikztotarget.south)
        -- ([yshift=-00pt]\tikztotarget.south)
      }
     ]
     &
     (\QBit \otimes \HilbertSpace{B})\State(\TensorUnit)
     \ar[
       r,
       "{
         \mathrm{chan}^{U_{\mathrm{tot}}}
       }",
       "{
         \scalebox{.7}{
           \color{darkgreen}
           \bf
           evolution
         }
       }"{swap, yshift=-2pt}
     ]
     \ar[
       rr,
       gray,
       rounded corners,
       to path={
           ([yshift=+00pt]\tikztostart.north)
        -- ([yshift=+20pt]\tikztostart.north)
        -- node {
          \scalebox{.7}{
            \colorbox{white}{
              QuantumStore transformation
            }
          }
        }
           ([yshift=+20pt]\tikztotarget.north)
        -- ([yshift=-00pt]\tikztotarget.north)
      }
     ]
     &
     (\QBit \otimes \HilbertSpace{B})\State(\TensorUnit)
     \ar[
       r,
       "{
         \mathrm{average}
           ^{ \HilbertSpace{B} }
       }"
     ]
    \ar[
      dr,
      gray,
      rounded corners,
      to path={
           ([xshift=+14pt]\tikztostart.south)
        -- ([xshift=-90pt]\tikztotarget.west)
        -- node {
          \scalebox{.7}{
            \colorbox{white}{
              \color{gray}
              QuantumStore context
            }
          }
        }
           ([xshift=0pt]\tikztotarget.west)
      }
    ]
    \ar[
      dr,
      gray,
      end anchor={[yshift=2pt]},
      "{
        \scalebox{.7}{
          \color{darkgreen}
          \bf
          measurement
        }
      }"{description, sloped}
    ]
     &
     \QBit\State(\TensorUnit)
     \ar[
       d,
       "{
         \mathcal{O}_{P_\bullet}
       }"
     ]
     \\
     &&&
     \indefiniteness_\Bit \TensorUnit
     \\
   \scalebox{0.8}{$    \vert 0 \rangle
     \langle 0 \vert
     $}
     \ar[
       r,
       phantom,
       "{ \longmapsto }"
     ]
     &
    \scalebox{0.8}{$   \vert 0 \rangle
     \langle 0 \vert
     \otimes
     I_{\HilbertSpace{B}}
     $}
     \ar[
       r,
       phantom,
       "{ \longmapsto }"{pos=.3}
     ]
     &
    \scalebox{0.8}{$   \mathclap{
     U_{\mathrm{tot}}
     \big(
     \vert 0 \rangle
     \langle 0 \vert
     \otimes
     I_{\HilbertSpace{B}}
     \big)
     U_{\mathrm{tot}}
     }
     $}
     \ar[
       dr,
       phantom,
       "{ \longmapsto }"{sloped}
     ]
     &[-200pt]
     \\
     &&&
     \hspace{-70pt}
  \scalebox{0.8}{$     \mathclap{
     b
     \,\mapsto\,
    \mathrm{Tr}^{\QBit}
    \bigg(
    \vert b \rangle \langle b \vert
    \,
     \mathrm{Tr}^{\HilbertSpace{B}}
     \Big(
       U_{\mathrm{tot}}
       \big(
       \vert 0 \rangle
       \langle 0 \vert
       \otimes
       I_{\HilbertSpace{B}}
       \big)
       U_{\mathrm{tot}}
     \Big)
     \!\! \bigg)
     }
     $}
     \ar[
       d,
       equals
     ]
     \\
     &&&
     \hspace{-70pt}
  \scalebox{0.8}{$     \mathclap{
     b
     \,\mapsto\,
    \mathrm{Tr}
    \Big(
       \big(
       \vert b \rangle \langle b \vert
       \otimes
       I_{\HilbertSpace{B}}
       \big)
       \,
       U_{\mathrm{tot}}
       (
       \vert 0 \rangle
       \langle 0 \vert
       \otimes
       I_{\HilbertSpace{B}}
       )
       U_{\mathrm{tot}}
          \Big)
     }
     $}
  \end{tikzcd}
\end{equation}

\vspace{-2.5cm}
\adjustbox{fbox}{
\begin{minipage}{8.8cm}
\footnotesize
{\bf Monadic typing of DQC1 quantum channel} \eqref{ProbabilityFormulaInDQC1}.
The quantum channel as such is the horizontal composite, consisting of an environmental coupling, followed by joint unitary evolution, followed by
bath-averaging
(and then by quantum measurement in the form \eqref{LuedersMeasurementInIntroduction}).
Here the first two steps constitute a QuantumState transformation acting compatibly \eqref{MonadTransformationRespectsKleisliComposition}
on the QuantumState-effectful maps which prepare the state (on the left).
Dually, the last two steps constitute a QuantumStore-transformation which acts compatibly on the QuantumStore-contextful observables (by Heisenberg evolution, cf. Prop. \ref{QuantumStateEvolutionIsHeisenbergEvolution}) to produce the measurement result, as in \eqref{HeisenbergEvolutionInIntroduction}.

\end{minipage}
}

\vspace{.4cm}

\noindent
{\bf Effective quantum language from Quantum modal logic.} With this thoroughly modal/monadic formulation of quantum systems in hand, standard language constructs in functional programming for handling effect monads (Lit. \ref{LiteratureProgrammingSyntaxForMonadicEffects}) become available for quantum programming. We indicate the resulting {\it Quantum Systems Language} ({\tt QS}) in \cref{Pseudocode}.

\bigskip

\ifdefined\monadology
\noindent
{\bf Outlook.} While one motivation for all these monadic constructions is the remarkable fact that they can be embedded just by suitable sugaring (Lit. \ref{LiteratureDomainSpecificLanguages}) into any dependent linear type theory which verifies the Motivic Yoga (such as {\LHoTT} does, Lit. \ref{LiteratureLHoTT}), here we speak purely in categorical semantics and relegate all discussion of type theoretic syntax to elsewhere \cite{QS} (but for a preview of the translation see \cite{Riley23}). At the same time, {\LHoTT} exists for the moment only on paper, as it is not supported yet by the {\HoTT} proof assistants such as {\Agda} or {\Coq}. There should be no fundamental obstacle to implementing a linear version of, say, {\Agda}, but this will require dedicated work. Therefore we understand our contribution here also as demonstrating that the new type system {\LHoTT} (which might superficially seem to be of only specialized interest) fundamentally deserves the attention of the computer-proof-assistant community.

Similarly, here we do not dwell on the higher homotopy theoretic aspect of  {\LHoTT}/{\QS}; but the companion article \cite{TQP} discusses in detail how anyonic topological quantum gates are naturally realized in classical {\LHoTT}, namely as twisted higher cohomology groups realized as dependent function types into higher delooping types (Eilenberg-MacLane-spaces) of the type of complex numbers. Since {\LHoTT} is conservative over {\HoTT}, this same construction from \cite{TQP} may immediately be understood as taking place in {\LHoTT}, where the type of complex numbers and hence that of anyonic quantum ground states may now be promoted to genuine linear types (Eilenberg-MacLane spectra equivalent to chain complexes, via the categorical semantics in \cite{EoS}), exhibiting the actual Hilbert space type of anyons to which quantum circuit logic may then be applied in the way we are discussing here.

Efforts are underway at CQTS\footnote{
landing page:
\href{https://nyuad.nyu.edu/en/research/faculty-labs-and-projects/cqts.html}{\tt nyuad.nyu.edu/en/research/faculty-labs-and-projects/cqts.html}
} to implement this classical {\HoTT}-realization of topological quantum gates in cubical-{\Agda} in order to demonstrate the feasibility of a formally verified topological hardware-aware quantum programming/simulation environment via dependent type theory. Our aim here is to demonstrate that the linearly-typed enhancement of such a quantum language system is theoretically viable, and naturally so, hoping to thereby spur its eventual implementation.

\vspace{1cm}

\noindent
{\bf Acknowledgements.} We thank
Thorsten Altenkirch,
David Corfield,
David Jaz Myers,
Mitchell Riley,
Sachin Valera,
and
\href{https://varkor.github.io/blog/}{Varkor}
for useful discussion concerning various aspects of this paper.
\fi

\ifdefined\monadology
\section{Background}
\label{Background}

This section provides background information and pointers to the literature on the various subjects referred to in the main text.
All items here are separately well-known to their respective experts but not always easy to comprehensively glean from the literature.
We pause at times to point out any remaining gaps that we address in the main text.

\medskip
\medskip

\cref{BackgroundQuantumComputing}: Quantum Computing

\cref{BackgroundQuantumProbability}: Quantum Probability

\cref{BackgroundMonadicEffects}: Monadic Effects

\cref{MonoidalCategories}: Monoidal Categories

\cref{BackgroundParamaterizedSpectra}: Parameterized spectra
\fi

\subsection{Quantum computing}
\label{BackgroundQuantumComputing}

\begin{literature}[\bf Quantum computation and Quantum information processing]
\label{LiteratureQuantumComputation} $\,$

\noindent
The basic idea of {\it quantum computation} and {\it quantum information processing} is to exploit,
for the purpose of machine computation and
information processing, the peculiar laws of quantum physics (Lit. \ref{EpistemologyOfQuantumPhysics}) -- which are obeyed by {\it undisturbed} (Lit. \ref{LiteratureTopologicalQuantumComputation}) microscopic systems.

The general idea of quantum computation was originally articulated by Yuri Manin \cite{Manin80}\cite{Manin00},
Paul Benioff \cite{Benioff80},
and Richard Feynman \cite{Feynman82}\cite{Feynman86}, brought into shape by David Deutsch \cite{Deutsch89}, shown to be potentially of dramatic practical
relevance by Peter Shor and others \cite{Shor94}\cite{Simon97}...
{\it if} sufficient quantum coherence can be technologically retained (cf. Lit. \ref{LiteratureTopologicalQuantumComputation}),
which has so far been achieved only marginally (Lit. \ref{LiteratureOnNISQMachines}).

Textbook accounts of the general principles of quantum computation and quantum information theory include:
\cite{NielsenChuang10}\cite{RieffelPolak11}\cite{BenentiCasatiRossini18}\cite{BEZ20}, lecture notes include \cite{Preskill04}. Impressions of the state of the field may be found in \cite{Preskill22}.
An exposition leading up to our discussion here may be found in \cite{Schreiber22}.

As usual, we are primarily concerned here with ``digital'' (or ``discrete variable'') quantum information/computa{-}tion, where all quantum state spaces are {\it finite-dimensional}, cf. \eqref{CoEvaluationForDualObjects}.
While there are notions of quantum computation on (separably) infinite-dimensional Hilbert spaces (``continuous variable'' systems, e.g. \cite{Choe22}) these represent ``analog quantum computation'' \cite{KendonNemotoMunro10} which, just as its classical analog, is typically more specialized, less reliable and less amenable to theory than ``digital'' computation on finite (dimensional) state spaces.

\medskip

\hspace{-.8cm}\label{TheIdeaOfQuantumGates}
\begin{tabular}{ll}
\begin{minipage}{8cm}
\noindent {\bf The idea of quantum gates.}
It is a standard concept in computer science to speak of {\it logic gates}
(e.g. \cite[\S 1]{GMSW21})
for operations on classical memory/registers (typically but not necessarily on a set of ``bits'', hence of Boolean ``truth values'', whence the name) -- where the terminology suggests but need not imply that this is an {\it elementary} operation performed by some computing machine under consideration. The evident analog in quantum computation (Lit. \ref{LiteratureQuantumComputation}) is that of {\it quantum logic gates} (\cite{Feynman86}\cite{Deutsch89}\cite{BBCDMSSSW95}, often called just ``quantum gates'', for short) which are {\it linear} maps acting on some quantum memory/registers -- typically imagined to be constituted by ``qbits'' \eqref{TypeOfQBits} -- cf. \eqref{CategoricalSemanticsSchematic}.

In classically controlled quantum computation (Lit. \ref{ClassicalControlAndDynamicLifting}) one is dealing with {\it classically controlled quantum gates} (e.g. \cite[\S 4.3]{NielsenChuang10}) that read/write a combination of classical and quantum data.
\end{minipage}
&
\adjustbox{  scale=.9,
  raise=4pt,
  fbox
}{
  \begin{tikzcd}[column sep=3pt]
    &&
    \mathclap{
      \begin{tikzpicture}
        \draw[line width=1.5]
          (-2,+.4) to (0,+.4);
        \draw[line width=1.5]
          (-2,-.4) to (0,-.4);
        \draw[line width=1.5]
          (0,+.3) to (2,+.3);
        \draw[line width=1.5]
          (0,-.3) to (2,-.3);
        \draw[
          fill=white,
          line width=1.3
        ]
          (-.8,-.7) rectangle (.8,.7);
        \draw (0,-.15) node {\scalebox{1}{
          \color{darkgreen}
          \bf
          GATE
         }};
        \draw (-1.4,-.2) node {\scalebox{1.2}{\bf \vdots}};
        \draw (+1.4,-.16) node {\scalebox{1}{\bf \vdots}};
        \draw (-2.8, -.1) node {
          \scalebox{.85}{
            \color{darkblue}
            $n_{{}_{\mathrm{in}}}$
            \bf
            \hspace{-13pt}
            \def\arraystretch{1}
            \begin{tabular}{c}
              Input
              \\
            registers
            \end{tabular}
          }
        };
        \draw (+2.9, -.1) node {
          \scalebox{.85}{
            \color{darkblue}
            $n_{{}_{\mathrm{out}}}$
            \bf
            \hspace{-14pt}
            \def\arraystretch{1}
            \begin{tabular}{c}
              Output
              \\
              registers
            \end{tabular}
          }
        };
      \end{tikzpicture}
    }
    \\[-7pt]
    &
    \Bit^{\times^{n_{\mathrm{in}}}}
    \ar[
      rr,
      "{
        \scalebox{.7}{
          \color{darkgreen}
          \bf
          Boolean function
        }
      }"{yshift=2pt}
    ]
    \ar[d, Rightarrow,-]
    &&
    \Bit^{\times^{n_{\mathrm{out}}}}
    \ar[d, Rightarrow,-]
    \\[-15pt]
    \colorbox{lightgray}{Logic gate}
    &
    \underset{
      \scalebox{.7}{
        \color{darkblue}
        \bf
        \def\arraystretch{.9}
        \begin{tabular}{c}
        set of
        \\
        tuples of
        \\
        $n_{\mathrm{in}}$
        bits
        \end{tabular}
      }
    }{
    \underbrace{
    \{0,1\}
      \times \cdots \times
    \{0,1\}
    }
    }
    &&
    \underset{
      \def\arraystretch{.9}
      \scalebox{.7}{
        \color{darkblue}
        \bf
        \begin{tabular}{c}
        set of
        \\
        tuples of
        \\
        $n_{\mathrm{out}}$
        bits
        \end{tabular}
      }
    }{
    \underbrace{
    \{0,1\}
      \times \cdots \times
    \{0,1\}
    }
    }
    \\[-17pt]
    \hline
    \\[-4pt]
    &
    \QBit^{\otimes^{n_{\mathrm{in}}}}
    \ar[
      rr,
      "{
        \scalebox{.7}{
          \begin{tabular}{c}
          \color{black}
          (unitary)
          \\
          \color{darkgreen}
          \bf
          linear map
          \end{tabular}
        }
      }"
    ]
    \ar[d, Rightarrow,-]
    &&
    \QBit^{\otimes^{n_{\mathrm{out}}}}
    \ar[d, Rightarrow,-]
    \\[-13pt]
    \colorbox{lightgray}{
      \hspace{-10pt}
      \def\arraystretch{.9}
      \begin{tabular}{c}
        Quantum
        \\
        logic gate
      \end{tabular}
      \hspace{-10pt}
    }
    &
    \underset{
      \mathclap{
      \scalebox{.7}{
        \color{darkblue}
        \bf
        \begin{tabular}{c}
        Hilbert space of
        \\
        tensor products of
        \\
        $n_{\mathrm{in}}$
        qbits
        \end{tabular}
      }
      }
    }{
    \underbrace{
    \mathbb{C}^{2}
      \otimes \cdots \otimes
    \mathbb{C}^2
    }
    }
    &&
    \underset{
      \mathclap{
      \scalebox{.7}{
        \color{darkblue}
        \bf
        \begin{tabular}{c}
        Hilbert space of
        \\
        tensor products of
        \\
        $n_{\mathrm{out}}$
        qbits
        \end{tabular}
      }
      }
    }{
    \underbrace{
    \mathbb{C}^{2}
      \otimes \cdots \otimes
    \mathbb{C}^2
    }
    }
  \end{tikzcd}
  \hspace{.1cm}
}
\end{tabular}

\vspace{.2cm}

\noindent
An example of a (controlled, quantum) logic gate is the {\it controlled NOT gate} \cite[Fig. 2]{Deutsch89} ({\tt CNOT} for short, cf. \cite[\S 1.3.2]{NielsenChuang10})
which operates on a pair of (q)bits by inverting the second conditioned on the first; see \eqref{TraditionalCNOTGate} and \eqref{QSCodeForCNOTGate}.

\medskip

\hspace{-.8cm}
\begin{tabular}{ll}
\begin{minipage}{11.5cm}
\noindent {\bf Quantum measurement gates.} One also wants to regard the operation of {\it quantum measurement} itself
(Lit. \ref{EpistemologyOfQuantumPhysics}) as a quantum gate (e.g. \cite[p. xxv]{NielsenChuang10}),
whose input is quantum data but whose output is the classical measurement result.
\end{minipage}
&
\;
\adjustbox{
  raise=-7pt,
  fbox
}{\!\!\!
    \begin{tikzpicture}
  \begin{scope}
      \draw (+.7, +0pt) node {
        $\quantized W$
      };
      \draw (+5.3, +0pt) node {
        $W$
      };
  \draw[
    line width=1.5pt,
  ]
  (1.1,0)
  to
  (3,0);
  \draw[
    line width=.6pt
  ]
  (3,.04)
  to
  (+5,.04);
  \draw[
    line width=.6pt
  ]
  (3,-.04)
  to
  (+5,-.04);
  \drawMeasurement{3}{0}
  \end{scope}
\end{tikzpicture}
\!\!\!}
\end{tabular}

\vspace{.1cm}

Notice that the proper data-typing (Lit. \ref{VerificationLiterature}) of a quantum measurement gate is more subtle than that of an ordinary logic gate, since the actual measurement outcome is {\it not} determined by the gate's input data (and hence {\it not} knowable at ``compile time'' of a quantum program) but is a fundamentally indefinite result, more akin to operations otherwise considered in the field of (classical but) {\it nondeterministic} computation (e.g. \cite[\S 1.2]{Sipser12}).

Beware that this is not a side issue but part of the crux of quantum computation:
On the one hand, the stochastic nature of quantum measurement is a {\it fundamental} principle of physics (certainly of presently accessible physics, see Lit. \ref{EpistemologyOfQuantumPhysics}) and not just a reflection of incomplete knowledge about a quantum system (in contrast to, for instance,  the case of classical thermodynamics). Moreover, state collapse under quantum measurement is not just a subjective update of expected probabilities, in that it objectively serves as an operational logic gate in quantum computations (such as in quantum teleportation \cref{QuantumTeleportationProtocol} and quantum error correction \cref{QuantumBitFlipCode}), to the extent that any quantum computation may be realized by {\it exclusively} using (quantum state preparation and) quantum measurement gates (known as ``measurement-based quantum computation''; cf. \cite{Nielsen03}\cite{BBDRV09}\cite{Wei21}).

We discover a natural way for dealing with formal typing of quantum measurement below in \cref{ControlledQuantumGates}.

\begin{equation}
\label{TraditionalCNOTGate}
\adjustbox{}{
\def\arraystretch{2}

  }"{description}
]
&&
  \{0,1\}
  \times
  \mathbb{C}^2
\end{tikzcd}
$
}
\\
\hline
\end{tabular}
}
\end{equation}

\medskip

\noindent {\bf Deferred measurement principle.}
Since quantum measurement turns quantum data into classical data, it intertwines quantum control with classical control. Concretely, a statement known as the {\it deferred measurement principle} asserts that any quantum circuit containing intermediate (mid-circuit) quantum measurement gates followed by gates conditioned on the measurement outcome is equivalent to a circuit where all measurements are ``deferred'' to the last step of the computation
\begin{equation}
\label{InformalDeferredMeasurementPrinciple}
\adjustbox{}{
\begin{tikzcd}
    \adjustbox{raise=20pt}{

    }
\end{tikzcd}
}
\end{equation}
(In the practice of quantum computation this principle can be used to optimize quantum circuit design. More philosophically, it is interesting to
notice that the issue of epistemological puzzlement in quantum interpretations, Lit. \ref{EpistemologyOfQuantumPhysics}, can always be thought of as postponed indefinitely.)

\smallskip
The theoretical status of the deferred measurement principle had remained somewhat inconclusive. Available textbooks (e.g. \cite[\S 4.4]{NielsenChuang10})
and numerous authors following them are content with inspecting a couple of examples while leaving it open what precisely the principle should state in generality,
a situation recently criticized in \cite[\S 1]{GurevichBlass22a}. A more precise form of the deferred measurement principle is briefly indicated
in \cite[p. 6]{Staton15} and proposed there as an ``axiom'' of quantum computation. We prove below (Prop. \ref{DeferredMeasurementPrinciple}) that the
deferred measurement principle \eqref{InformalDeferredMeasurementPrinciple} is verified in the data-typing of quantum processes provided in {\LHoTT} (Lit. \ref{LiteratureLHoTT}).

\smallskip
Notice that the content of this {\it equivalence between intermediate and deferred measurement collapse}
\eqref{InformalDeferredMeasurementPrinciple}
is not trivial without a good formalization; in fact it has historically been perceived as a {\it paradox}, namely this is essentially the paradox
of ``{\it Schr{\"o}dinger's cat}'' and of ``Wigner's friend'' (where the cat/friend plays the role of the intermediate controlled quantum gate).
Moreover, the same paradox, in different words, was influentially offered in \cite[p. 4]{Everett75a} as the main argument against the
``Copenhagen interpretation'' and for the ``many-worlds interpretation'' of quantum physics (cf. Lit. \ref{EpistemologyOfQuantumPhysics}).
Note that our same formalism which proves \eqref{InformalDeferredMeasurementPrinciple} also proves the equivalence
\eqref{BranchingAndCollapseInIntroduction} of these two ``interpretations''.

\medskip

\noindent {\bf qRAM Models.} Classical computing in its familiar {\it universal} form is based, in one way or another, on the model
of a {\it Random Access Memory} (``RAM'', also known as a {\it Mealy machine}, see \eqref{StateMonadAsRAMModel} below):
\vspace{-2mm}
\begin{equation}
  \label{ProgramInteractingWithRAM}
  \begin{tikzcd}[
    column sep=60pt
  ]
    \scalebox{.7}{
      \color{darkblue}
      \bf
      \def\arraystretch{.9}
      \begin{tabular}{c}
        read-in RAM
        \\
        \& input data
      \end{tabular}
    }
    \mathrm{RAM}
      \times
    D
    \ar[
      rr,
      "{
        \scalebox{.7}{
          \color{darkgreen}
          \bf
            \def\arraystretch{.9}
          \begin{tabular}{c}
            program interacting with
            \\
            Random Access Memory
          \end{tabular}
        }
      }"
    ]
    &&
    \mathrm{RAM}
      \times
    D'
    \hspace{-7pt}
    \scalebox{.7}{
      \color{darkblue}
      \bf
      \def\arraystretch{.9}
      \begin{tabular}{c}
        write RAM
        \\
        \& output data
      \end{tabular}
    }
  \end{tikzcd}
\end{equation}
Starting with \cite{GLM08a}\cite{GLM08b}, authors envisioned that quantum computing should similarly support a ``qRAM model''
(see \cite[p. 18]{LiuEtAl23} for implementations) the basic idea being that data in qRAM may form quantum superpositions  and may coherently be read/written in this form. As with the deferred measurement principle above, existing literature discusses this concept not in general abstraction but by way of concrete examples (see for instance \cite[Fig. 9]{ArunchalamEtAl15}\cite[Fig. 1]{PPR19}\cite[Fig. 4]{PCG23}\footnote{A transparent example is discussed at \url{https://quantumcomputinguk.org/tutorials/implementing-qram-in-qiskit-with-code}}). From these one gathers that a quantum circuit of {\it nominal} type $\HilbertSpace{H} \to \HilbertSpace{K}$ but with access to a qRAM Hilbert space $\quantized \mathrm{RAM}$ is {\it de facto} a quantum circuit of this form (a ``circuit-based qRAM'' \cite{PPR19}):
\vspace{-2mm}
\begin{equation}
  \label{QuantumProgramInteractingWithQRAm}
  \begin{tikzcd}[
    column sep=55pt
  ]
    \scalebox{.7}{
      \color{darkblue}
      \bf
      \def\arraystretch{.9}
      \begin{tabular}{c}
        read-in qRAM
        \\
        entangled with
        \\
        input quantum data
      \end{tabular}
    }
    \quantized \mathrm{RAM}
      \otimes
    \HilbertSpace{H}
    \ar[
      rr,
      "{
        \scalebox{.7}{
          \color{darkgreen}
          \bf
            \def\arraystretch{.9}
          \begin{tabular}{c}
            quantum program
            \\
            interacting with qRAM
          \end{tabular}
        }
      }"
    ]
    &&
    \quantized\mathrm{RAM}
      \otimes
    \HilbertSpace{H}'
    \hspace{-7pt}
    \scalebox{.7}{
      \color{darkblue}
      \bf
      \def\arraystretch{.9}
      \begin{tabular}{c}
        write qRAM
        \\
        entangled with
        \\
        output quantum data
      \end{tabular}
    }
  \end{tikzcd}
\end{equation}
In \cref{ControlledQuantumGates} we obtain
\eqref{QRAMAdjointness}
a formalized account/typing of qRAM and its equivalence to controlled quantum circuits.
\end{literature}

\begin{literature}[\bf Epistemology of quantum physics and its formalization]
\label{EpistemologyOfQuantumPhysics}
The curious epistemology\footnote{
  Here ``epistemology'' -- the {\it theory of knowledge} -- refers to what can {\it in principle}
  (cf. \cite[p. 121]{Fitting07})
  be known about the (quantum) universe or any model or part of it, say about a given (quantum) computing machine,
  which in practice concerns the question of what can {\it in principle} be computed with a given quantum protocol, all imperfections of
  experiments and of experimenters disregarded.
}
of quantum physics (\cite{Dirac30}\cite{vonNeumann32}, see e.g. \cite{SakuraiNapolitano94}\cite{Isham95}\cite{Landsman17}) occupied already the founding fathers
of quantum theory \cite{EPR35}\cite{Bohr1949}
and the philosophical attitudes towards them were eventually canonized as {\it interpretations of quantum physics} \cite{Mehra73}\cite{Scheibe73}.
Later experimental advances in quantum physics only verified the nature of the theory and thus reinforced the epistemological puzzlement \cite{GreenbergerReiterZeilinger99}.

\medskip

\noindent {\bf Quantum measurement.}
Concretely, the core issue is that
what otherwise appears to be the epistemologically complete {\it state} of a quantum system -- traditionally denoted ``$\vert \psi \rangle$'', being an element of some Hilbert space $\HilbertSpace{H}$ --
determines in general only the {\it probability} (see Lit. \ref{LiteratureQuantumProbability})
of which measurement outcome $w \isa W$ (which ``world'') will be observed upon measuring a given
property of the system, while only {\it right after} the observation of a given $w$ the quantum state appears to have ``collapsed'' along its linear projection onto a subspace of states with definite property $w$ (\cite[\S III.3, \S VI]{vonNeumann32}\cite{Lueders51},
cf. \cite[\S IV]{Scheibe73}\cite[p. 82]{Omnes94}\cite[(A.2)]{Renes22}):
\begin{equation}
  \label{LinearStateCollapse}
  \hspace{1.2cm}
  \begin{tikzcd}[
    row sep=-12pt,
    column sep=75pt
  ]
  &
  \HilbertSpace{H}_{w_1}
  \mathrlap{
  \scalebox{.7}{
    \color{darkblue}
    \bf
    \def\arraystretch{.9}
    \begin{tabular}{c}
      space of quantum states
      \\
      with definite property $w_1$
    \end{tabular}
  }
  }
  \ar[
    dd,
    phantom,
    "{ \vdots }"{pos=-.1}
  ]
  \\
  \mathllap{
  \scalebox{.7}{
    \color{darkblue}
    \bf
    \def\arraystretch{.7}
    \begin{tabular}{c}
      Hilbert space of all
      \\
      quantum states
      \\
      of the given system
    \end{tabular}
  }
  \HilbertSpace{H}
  \;\;
  \simeq
  \;\;
  \underset{
    \mathclap{
      \hspace{27pt}
      \raisebox{-12pt}{
      \scalebox{.7}{
        \color{darkorange}
        \bf
        \def\arraystretch{.7}
        \begin{tabular}{c}
          direct sum decomposition
          \\
          in measurement basis $W$
        \end{tabular}
      }
      }
    }
  }{
  \necessarily_{{}_W} \HilbertSpace{H}_\bullet
  \;
  \defneq
  \;\;\;
  }
  \underset{
    \mathclap{
    w' \isa W
    }
  }{\scalebox{1.3}{$\oplus$}}
  \;
  \HilbertSpace{H}_{{}_{w'}}
  }
  \;\;
  \ar[
    ur,
    start anchor={[yshift=2pt]},
    "{
      \scalebox{.7}{
        \color{darkgreen}
        \bf
        linear projection
      }
    }"{sloped}
  ]
  \ar[
    dr,
    start anchor={[yshift=-2pt]},
    "{
      \scalebox{.7}{
        \color{darkgreen}
        \bf
        linear projection
      }
    }"{sloped, swap}
  ]
  \\[-16pt]
  &
  \HilbertSpace{H}_{w_n}
  \mathrlap{
  \scalebox{.7}{
    \color{darkblue}
    \bf
    \def\arraystretch{.9}
    \begin{tabular}{c}
      space of quantum states
      \\
      with definite property $w_n$
    \end{tabular}
  }
  }
  \end{tikzcd}
\end{equation}

To some extent, this ``state collapse'' is formally just as expected (cf. \cite[\S 1.2]{Kuperberg05}\cite{Yuan12}) in a classical but probabilistic theory,
where measurement of a random variable leads one to adjust the subjectively expected probability distribution according to Bayes' Law for updating conditional
probabilities --- except that {\it Kochen-Specker-Bell theorems} (e.g. \cite{ClauserShimony78}\cite[\S 1.6.2]{Kuperberg05}\cite[\S 5.1.2]{Moretti19})
show that (under very mild assumptions) generally no actual classical probability distribution can underlie a pure quantum state, hence that quantum
states are {\it not} just a stochastic approximation to a more fundamental classical reality (cf. \cite[p. 140]{Scheibe73}).

\smallskip
Moreover, it seems untenable to regard the ``state collapse'' as just a subjective adjustment of expectation, since it is an operational component of
experimentally realizable quantum communication protocols (cf. Lit. \ref{LiteratureQuantumComputation} and
\cref{ControlledQuantumGates},
such as in the {\it quantum teleportation} protocol recalled in \cref{QuantumTeleportationProtocol}); so much so that there is a paradigm of
{\it measurement-only} quantum computation (cf. \cite{Nielsen03}\cite{BBDRV09}\cite{Wei21}) where the computational process consists entirely
of a sequence of such measurement-induced state collapses --- in this practical sense the state collapse \eqref{LinearStateCollapse}
{\it is an objective reality}.

\ifdefined\monadology
\medskip
\else
\newpage
\fi

\noindent {\bf Quantum epistemologies.}
The debates on what to make of the situation continue to this day (from the vast literature, see for instance \cite{Omnes94}\cite{Boge08}),
whence practicing physicists tend to just disregard the epistemological issue, an attitude that became proverbial under the catch-phrase
``shut up and calculate'' \cite{Mermin89}.

\noindent
Among the main attitudes of quantum philosophers towards the issues are:

\begin{itemize}[leftmargin=.5cm]

\item
{\bf Copenhagen epistemology:  Quantum/classical divide.}
The original ``Copenhagen interpretation'' (e.g. \cite[p. 99]{Primas83}\cite[p. 85]{Omnes94}) pronounces a conceptual {\it frontier} or {\it divide} between
quantum objects and their classical observers  according to which recognizable result of any quantum measurement are, and must be reasoned about as, classical states.

\item
{\bf Everett's epistemology: Branching into Many worlds.}
An increasingly popular ``many-worlds interpretation'' (following H. Everett \cite{Everett75a}\cite{Everett57}\cite{DeWittGRaham73}) rejects a separate classical
component of quantum theory and instead asserts (informally and hence ambiguously, cf. \cite{Tegmark98}) both that the quantum state does never ``really'' collapse
and at the same time that the universe successively ``branches'' into ``many-worlds'' inside which it nonetheless ``appears'' to observers to have collapsed in all possible ways.

\end{itemize}

\smallskip
\noindent
The reader uneasy with making sense of any of this we invite to \cref{QuantumEffects}, where we present a {\it modal quantum logic} (cf. Lit. \ref{ModalLogicAndManyWorlds})
which arguably makes precise these two epistemological attitudes and as such allows to prove their equivalence, cf. \eqref{BranchingAndCollapseInIntroduction}.
In particular, the perceived paradox which Everett offers \cite[p. 4]{Everett75a} to dismiss the Copenhagen interpretation and to motivate the ``many-worlds''
interpretation is arguably resolved by the {\it deferred measurement principle} \eqref{InformalDeferredMeasurementPrinciple}, which becomes {\it provable}
in quantum modal logic (Prop. \ref{DeferredMeasurementPrinciple}).

\smallskip

\noindent {\bf Many possible worlds.}
Previously, several authors
(e.g. \cite{Bunge76}\cite[\S III]{Skyrms76}\cite[p. 101]{Tappenden00}\cite[p. 22]{Nolan02}\cite[\S 8]{Girle03}\cite{Terekovich19}\cite{Wilson20}\cite{ArroyoArenhart22})
have vaguely wondered about or suggested a relation between these ``many worlds'' of quantum epistemology
and the ``possible worlds'' in the sense classical modal logic (Lit. \ref{ModalLogicAndManyWorlds})
but no formalized such discussion has previously been proposed.
  In particular, no previous author has considered this
  question with respect to a {\it linear} modal logic (cf. Lit. \ref{VerificationLiterature}).
    (Beware that philosophers also speak of a {\it modal interpretation of quantum mechanics}\footnote{Cf. \href{ https://plato.stanford.edu/entries/qm-modal}{\tt plato.stanford.edu/entries/qm-modal}} which shares some similarity in vocabulary but does not refer either to modal logic nor to many-worlds.)

\smallskip

\noindent {\bf The need for formalization.}
Indeed, in the time-honored spirit of Galileo, Kant, Hilbert, Wigner (``The book of nature is written in the language of mathematics.'') one may have suspected that the fault causing epistemological troubles is not with quantum theory itself, but with speaking about it in ordinary informal language (Bohr 1920: ``When it comes to atoms, language can only be used as in poetry.''), whence their resolution lies instead in adopting a {\it mathematical} language of {\it non-classical formal logic} more appropriate for expressing microscopic quantum reality.
In fact, a universal quantum programming language should essentially be just such a formal language, and in formulating it we do need to find a way to formally reflect the phenomenon of quantum measurement:
\vspace{-2mm}
\begin{center}
{\it
The verified programming of a quantum algorithm

is the act of accurately recounting in formalized language

the physical quantum process that executes it,
and conversely.
}
\end{center}

\vspace{-2mm}
It is towards this practical goal that here we care about quantum epistemology; and this may explain why we have more to say here about the foundations of quantum physics generally, beyond the field of quantum computation.

\smallskip

\noindent {\bf Bohr toposes.}
Another proposal in the direction of formalized quantum epistemology may be recognized in \cite{AdelmanCorbett95}
(in parallel and independently to the development of quantum/linear logic, Lit. \ref{VerificationLiterature}).
A variant of this proposal that gained some popularity is to use the internal logic of canonically ringed (co)presheaf toposes over the site of commutative subalgebras of a given $C^\ast$-algebra of quantum observables (``Bohr toposes'', following ideas of \cite{BHI98}, for review see \cite{Nuiten12}\cite[\S 12]{Landsman17}). The achievement of this approach is to show that the step from classical/commutative to quantum/noncommutative probability theory (of which a good account is in \cite{Gleason09}\cite{Gleason11}) may be understood as the logical {\it internalization} of the classical axioms into a Bohr topos \cite{HeunenLandsmanSpitters09}.
While conceptually quite satisfactory, the practical relevance of this perspective has arguably remained elusive.
In particular, it does not readily translate to a formal quantum (programming) language.

\smallskip

The approach that we take below is also ultimately (higher) topos-theoretic but otherwise rather complementary to Bohr toposes. In fact, one may understand Bohr toposes as formalizing the {\it Heisenberg picture} of quantum physics -- where conceptual primacy is given to the algebras of {\it quantum observables} -- while here we are concerned with the equivalent but ``dual'' {\it Schr{\"o}dinger picture} where the primary concept is the spaces of {\it quantum states}: These being exactly the {\it linear types} that give {\it Linear Homotopy Type Theory} its name. We relate this to algebras of observables in \cref{MixedQuantumTypes} (see Ex. \ref{AlgebraOfQuantumObservablesAsQuantumStoreContextfulMaps}).
\end{literature}

\begin{literature}[\bf Topological quantum computation]
  \label{LiteratureTopologicalQuantumComputation} $\,$

  \noindent
  (For extensive motivation, explanation and referencing of topological quantum computation see the companion article \cite{TQP}.)
    The practical promise of quantum computation (Lit. \ref{LiteratureQuantumComputation}) hinges on the achievability of fairly
    {\it undisturbed} quantum processors which are sufficiently {\it robust} against the inevitable interaction with their environment.
    There are essentially two approaches toward robust quantum computation:
    \smallskip
  \begin{itemize}
    \item[{\bf (i)}] {\bf Quantum error correction}: Operate on error-prone quantum hardware, but with software that implements enough redundancy
    to allow reading intended signals out of noisy background (cf. \cref{QuantumBitFlipCode}).
    \item[{\bf (ii)}] {\bf Topological error protection}: Operate on intrinsically stable quantum hardware (Lit. \ref{TopologicalQuantumMaterials}) which prevents errors from occurring
    in the first place.
  \end{itemize}
  \noindent
  In all likelihood, the eventual practice will be a combination of both approaches, since topological hardware error-protection achievable
  in the laboratory will itself have imperfections. Conversely, some quantum-error correction algorithms essentially consist of
  {\it simulating} topological quantum hardware on non-topological hardware, e.g. \cite{IqbalEtAl23}.
    However, the peculiarities of topological quantum gates had previously no genuine representation in quantum programming languages and were
  principally un-verifiable (cf. Lit. \ref{VerificationLiterature})
  until we argued, in the companion article \cite{TQP}, that realistic topological quantum gates are naturally modeled by {\it homotopy typed languages} (Lit. \ref{LiteratureHomotopyTypeTheory}), such as classical {\HoTT} and, more accurately, by {\tt LHoTT} (Lit. \ref{LiteratureLHoTT}).
\end{literature}

\begin{literature}[\bf Formal (quantum) software verification and dependent (linear) data typing]
 \label{VerificationLiterature}
 $\,$

\noindent
(For extensive exposition and referencing of the {\it classical} case see the companion article \cite{TQP}.)

\noindent
The benefit or even necessity of {\it formal software verification methods}
\cite{CousotCousot09}\cite{Meadows11} (often abbreviated to just ``formal methods'', cf. \cite{WLBF09}) --- hence of computer-checked proof
at compile-time of correct behavior of critical software --- is evident \cite{HuckelNeckel19} and as such increasingly of interest for
instance to the crypto-reliant industry (e.g. \cite{Hedera18}\cite{VYC22}\cite{Quantstamp23}) and the military (e.g. \href{https://app.dimensions.ai/details/grant/grant.7081074}{MURI:FA95501510053}).
Nevertheless, in less critical applications of classical computation the overhead associated with formal verification is still
widely traded for the possibility of incrementally de-bugging faulty software during application.

\smallskip

\noindent {\bf Need for verification of quantum programs.}
However, such run-time debugging is no longer a sustainable option when it comes to serious {\it quantum} computation, due
(\cite[p. 6]{VRSAS15}\cite{FHTZ15}\cite{Rand18}\footnote{
  \cite[p. iv]{Rand18}: ``We argue that quantum programs demand machine-checkable proofs of correctness. We justify this on the basis
  of the complexity of programs manipulating quantum states, the expense of running quantum programs, and the inapplicability of traditional debugging techniques to programs whose states cannot be examined. [...] Quantum programs are tremendously difficult to understand and implement, almost guaranteeing that they will have bugs. And traditional approaches to debugging will not help us: We cannot set breakpoints and look at our qubits without collapsing the quantum state. Even techniques like unit tests and random testing will be impossible to run on classical machines and too expensive to run on quantum computers -- and failed tests are unlikely to be informative. [...] Thesis Statement: {\it Quantum programming is not only amenable to formal verification: it demands it.}''
}\cite{YingFeng18}\cite{MZD20}\cite{YingFeng21})
to its:
\begin{itemize}[leftmargin=.8cm]
  \item drastically higher complexity,
  \item drastically higher run-time cost,
  \item impossibility of run-time inspection.
\end{itemize}
The last point is the fundamental one, enforced by the quantum laws of nature (state collapse under measurement, Lit. \ref{EpistemologyOfQuantumPhysics}), but the other two points will in practice be no less forbidding.

\smallskip
Accepting the need for (quantum) software verification, its implementation of choice is by {\it data typing} (which for quantum data means ``dependent linear typing'':

\medskip

\noindent {\bf Formal verification by data typing.}
 A profound confluence of computer science and pure mathematics occurs with the observation
 \cite{MartinLof82} that formal software verification is not only amenable to constructive mathematical proof but fundamentally equivalent to
 it -- every constructive mathematical proof may be understood as pseudocode for a program whose output is data of the type of certificates
 of the truth of the given statement, a profound tautology known as the {\it BHK
 (Brouwer–Heyting–Kolmogorov) correspondence}, or similar (find references around \cite[(92)]{TQP} ).

 \smallskip
Accordingly, formal verification/proof languages
 are (dependently) {\it typed} in that every piece of data they handle has assigned a precise {\it data type} which provides the strict specification
 that data has to meet in order to qualify as input or output of that type
 (\cite{MartinLof82}\cite{Thompson91}\cite{Streicher93}\cite{Luo94}\cite{Gunter95}\cite{Constable11}\cite{Harper16}).
 The abstract theory of such data typing is known as (dependent-){\it type theory} and the modern flavor relevant here is often called
 {\it Martin-L{\"o}f type theory} in honor of \cite{MartinLof71}\cite{MartinLof75}\cite{MartinLof84};
 for more elaboration and introduction
 see also \cite{Hofmann97}\cite{UFP13}.

 \smallskip
 Once this typing principle is adhered to, the distinction vanishes between writing a program and verifying its correctness.
 Moreover, such a properly typed
 functional program may equivalently be understood as a {\it mathematical} object, namely as a mathematical function \eqref{FunctionDeclaration}
 from the ``space'' of data of
 its input type to that of its output type --- called its {\it denotational semantics} (a seminal idea due to \cite{Scott70}\cite{ScottStrachey71};
 for exposition see \cite[\S 9]{SlonnegerKurtz95}):
\begin{equation}
\label{FunctionDeclaration}
\adjustbox{}{
\begin{minipage}{12cm}
\def\arraystretch{1.3}
\begin{tabular}{|l|l|}
  \hline
  \begin{minipage}{6cm}
    {\bf Syntax}
  \end{minipage}
  &
  \begin{minipage}{6cm}
    {\bf Semantics}
  \end{minipage}
  \\
  \hline
\end{tabular}

\def\arraystretch{1.3}
\begin{tabular}{|l|l|}
  \hline
 \begin{minipage}{6cm}
   $$
   \adjustbox{raise=0pt}{
   \begin{tikzcd}[
     sep=0pt
  ]
    \\[12pt]
    \overset{
      \mathclap{
        \rotatebox{30}{
          \hspace{12pt}
          \hspace{-28pt}
          \rlap{
          \scalebox{.7}{
            \color{darkblue}
            \bf
            flags
          }
          }
        }
      }
    }{
     \gamma{\isa}\Gamma
     }
     ,\;\,
    \overset{
      \mathclap{
        \rotatebox{30}{
          \hspace{12pt}
          \hspace{-28pt}
          \rlap{
          \scalebox{.7}{
            \color{darkblue}
            \bf
            input
          }
          }
        }
      }
    }{
     i{\isan}I
    }
     &
       \yields
     &
    \overset{
      \mathclap{
        \rotatebox{30}{
          \hspace{12pt}
          \hspace{-28pt}
          \rlap{
          \scalebox{.7}{
            \color{darkblue}
            \bf
            output
          }
          }
        }
      }
    }{
     p_\gamma(i) \isan O
    }
   \end{tikzcd}
   }
   $$
  \end{minipage}
 &
 \begin{minipage}{6cm}
   $$
   \begin{tikzcd}[
     column sep=30pt,
     row sep=0pt
   ]
    \\[12pt]
   &[+10pt]
    \overset{
      \mathclap{
        \rotatebox{30}{
          \hspace{12pt}
          \hspace{-28pt}
          \rlap{
          \scalebox{.7}{
            \color{darkblue}
            \bf
            domain
          }
          }
        }
      }
    }{
   \Gamma
     \times
   I
   }
   \ar[
     r,
     "{
       \name p
     }"{description}
   ]
   &
    \overset{
      \mathclap{
        \rotatebox{30}{
          \hspace{12pt}
          \hspace{-28pt}
          \rlap{
          \scalebox{.7}{
            \color{darkblue}
            \bf
            codomain
          }
          }
        }
      }
    }{
   O
   }
  \end{tikzcd}
  $$
\end{minipage}
\\[-15pt]
\scalebox{.7}{
  \color{gray}
  program
}
&
\scalebox{.7}{
  \color{gray}
  function
}
 \\
 \hline
\end{tabular}
\end{minipage}
}
\end{equation}

For classical\footnote{
  Here by {\it classical types} we mean the types of {\it intuitionistic} Martin-L{\"o}f type theory in contrast to {\it linear} (quantum) types \eqref{LinearTypesAsSubstructuralTypes}, but {\it not} in the sense of ``classical logic'': Classical types in our sense are
  ``not quantum'' in that they are
  subject to the structural inference rules \eqref{ContractionAndWeakeningRule}
  but they are still {\it constructive} in that they are  not (necessarily) subjected to the law of excluded middle and/or the axiom of choice (which distinguish ``classical logic'' from ``intuitionistic logic'').
}
data types the {\it inference rules} by which such program/function declaration may proceed equip the type universe with the structure of a Cartesian closed category \cite[\S I]{LambekScott86}, whence one also speaks of {\it categorical semantics} (see \cite{Jacobs98}\cite{Jacobs93}).
Here the inference rules for the classical logical conjunction ``$\times$'', hence for the Cartesian product, subsume the basic ``structural inference rules''
called the
{\it contraction rule} and the {\it weakening rule} (\cite[\S 1.2.1]{Gentzen35}, see \cite{Jacobs94}\cite[p. 122]{Jacobs98}\cite[\S A.2.2]{UFP13}\cite[\S 1.4]{Rijke18}), which semantically express the possibility of duplicating and of discarding classical data:
\vspace{-2mm}
\begin{equation}
  \label{ContractionAndWeakeningRule}
  \adjustbox{}{
  \hspace{-2cm}
  \begin{minipage}[c]{12cm}
\adjustbox{
  fbox,
  rotate=90,
  raise=-2.5cm
}{
  \bf
  \def\arraystretch{.9}
  \hspace{-11pt}

\end{tabular}
\end{minipage}
  }
\end{equation}

 \smallskip

 \noindent {\bf The quest for quantum data typing} was historically convoluted (starting with the much debated
 quantum logic of \cite{BvN36} and continuing with the influential ideosyncracies of \cite{Girard87}) but is, in hindsight, fairly straightforward:
  Since the hallmark of coherent quantum evolution is (see  \cite{Abramsky09} for a structural account) the pair of:
  \begin{itemize}[leftmargin=.7cm]
  \item the {\it no-cloning theorem} (\cite{WootersZurek82}, saying that quantum data cannot be {\it systematically} duplicated),
  \item the {\it no-deletion theorem} (\cite{PatiBraunstein00}, saying that quantum data cannot be {\it systematically} discarded),
  \end{itemize}
  it follows that a program handling purely quantum data types must {\it not} use the structural rules \eqref{ContractionAndWeakeningRule} for the logical conjunction of quantum data, which is then called the (non-Cartesian) {\it tensor product} $\otimes$ (Lit. \ref{LiteratureMonoidalCategories}). It is this {\it removal} of structural inference rules (``sub-structural logic'') which frees the tensor product of quantum data types from only consisting of pairs of data and hence allows for the hallmark phenomenon of {\it quantum entanglement} (see e.g. \cite{BengtssonZyczkowski06}).

  \smallskip
  Such {\it sub-structural} languages were essentially introduced in (the ``multiplicative sector'' of) the {\it linear logic} (see \cite{Seely89}\cite{Troelstra92}\cite{MihalyiNovitzka13}) originated by \cite{Girard87} (who was apparently vaguely aware of potential application to quantum logic, cf. \cite[p. 7]{Girard87}). These languages were then suggested as expressing quantum processes in \cite{Yetter90}\cite{Pratt92} and were more fully understood as quantum (programming) languages (Lit. \ref{LiteratureQuantumProgrammingLanguages})
  with {\it linear types} in
  \cite{Valiron04}\cite{SelingerValiron05}  \cite{AbramskyDuncan06}\cite{Duncan06}\cite{SelingerValiron09}. Notice that the adjective ``linear'' here refers to the preservation of the number of type factors in the absence of the structural rules \eqref{ContractionAndWeakeningRule}, which implies that functions $f \isa X \to Y$ between linear types must indeed use their argument $x \isan X$ linearly, in the algebraic sense.

  \smallskip

  \noindent
  {\bf Vector- and Hilbert-spaces as linear types.}
  Notably the usual categories $\Modules{\GroundField}$ of vector spaces over any ground field $\GroundField$, with $\GroundField$-linear maps between them, constitute categorical semantics for (the multiplicative sector) of linear logic -- and the fact that this was made explicit no earlier than in \cite{Murfet14}\cite{ValironZdancewic14} must be understood as solely reflecting the convoluted history of the subject: Constituting the heart (Lit. \ref{Zerosector}) of stable $\infty$-categories of module spectra ($H\GroundField$-modules, in this case, Lit. \ref{TangentInfinityToposes}) these categories $\Modules{\GroundField}$ appear as rather canonical models for linear types and as such we use them in \cref{QuantumTypeSemantics}.

\smallskip

\noindent
{\bf Quantum data typing.} In summary, the match between quantum phenomena, linear type theories and their semantics in categories of linear spaces
is tight (which should not be surprising in hindsight but was less than obvious for much of the history of linear logic):
\begin{equation}
\label{LinearTypesAsSubstructuralTypes}
\adjustbox{}{
{\small
  \def\arraystretch{1.6}
  \begin{tabular}{|c|c|c|}
    \hline
    \bf Quantum Phenomena
    &
    \bf Linear Type Inference
    &
    \bf Linear maps in Linear algebra...
    \\
    \hline
    \hline
    No-cloning theorem
    &
    Absence of contraction rule
    &
    ...use their argument at most once.
    \\
    \hline
    No-deleting theorem
    &
    Absence of weakening rule
    &
    ...use their argument at least once.
    \\
    \hline
  \end{tabular}
  }
  }
\end{equation}

  \noindent
  The resulting principle that

  \vspace{-.3cm}
  \begin{center}
    {\it Quantum data has linear type.}
  \end{center}
  has meanwhile come to be more commonly appreciated
  (e.g. \cite[p. 1]{DalLagoFaggian12})
  in particular in quantum language design
  (Lit. \ref{LiteratureQuantumProgrammingLanguages}, cf. in particular \cite{FuKishidaSelinger20}), where for instance the insightful \cite{Staton15} states up front that:
  \begin{center}
    {\it A quantum programming language captures the ideas of quantum computation in a linear type theory.}
  \end{center}

  \medskip

 \noindent
 {\bf Bunched classical/quantum type theory and EPR phenomena.}
  And yet, a comprehensive programming language implementing such {\it linear type theories} of {\it combined} classical and quantum data had remained elusive all along:
The type-theoretic subtlety here is that with the classical conjunction ($\times$) being accompanied by a linear multiplicative conjunction ($\otimes$), then contexts on which terms and their types should depend are no
longer just linear lists of (dependent) classical products
\vspace{-2mm}
$$
  \Gamma_1
  \,\times\,
  \Gamma_2
  \,\times\,
  \cdots
  \,\times\,
  \,
  \Gamma_n
  \hspace{1cm}
  \scalebox{.7}{
    \color{darkblue}
    \bf
    \begin{tabular}{c}
      a classical type-context
      \\
      (tuples of classical data)
    \end{tabular}
  }
$$

\vspace{-2mm}
\noindent but may be nested (``bunched'') such products, alternating with linear multiplicative conjunctions
to form tree-structured expressions like this example:
\vspace{-2mm}
$$
  \Gamma_1
  \times
  \big(
    \Gamma_2 \otimes
    (\Gamma_3 \times \Gamma_4)
  \big)
  \times
  (\Gamma_5 \otimes \Gamma_6)
  \times
  (\Gamma_7 \otimes \Gamma_8 \otimes \Gamma_9)
  \hspace{.5cm}
  \scalebox{.7}{
    \color{darkblue}
    \bf
    \begin{tabular}{c}
      a mixed classical/quantum type-context
      \\
      (tuples of classical data
      mixed with
      {\it entangled}
      quantum data).
    \end{tabular}
  }
  $$
While the idea of formulating such ``bunched'' type theories is not new \cite{OHearnPym99}\cite{Pym2002}\cite{OHearn03}, its
implementation has turned out to be tricky and the results unsatisfactory; see \cite[\S 13.6]{Pym2008}\cite[p. 19]{Riley22}.
The claim of the type theory introduced in \cite{Riley22} is to have finally resolved this long-standing issue of formulating
``bunched linear dependent type theory''. Here we understand this as saying that a verifiable universal quantum programming
language now exists -- {\LHoTT}\footnote{
  In fact, in {\LHoTT} the substructural nature of the linear types is more refined than shown in \eqref{LinearTypesAsSubstructuralTypes}: It {\it is} possible in {\LHoTT} to duplicate the {\it reference} to terms of linear type, for instance such as to assert their self-identification
  $$
    \HilbertSpace{H}
      \,\isa\,
    \LinearType
    ,\,
    \psi \,\isa\, \HilbertSpace{H}
    \hspace{.6cm}
      \yields
    \hspace{.6cm}
    \psi = \psi
    \,,
  $$
  but an accompanying ``color palette'' ensures that no such duplicate references may be used on the two sides of the tensor product.
} (Lit. \ref{LiteratureLHoTT}).

\medskip

To put this into perspective it may be noteworthy that the root of this subtlety resolved by {\LHoTT} corresponds to the hallmark
phenomenon of quantum physics which famously puzzled the subject's founding fathers (Lit. \ref{EpistemologyOfQuantumPhysics}),
namely the {\it conditioning of physics on entangled quantum states} (known as the {\it EPR phenomenon}, e.g. \cite{Selleri88}):

\vspace{.2cm}
\hspace{-.8cm}
\label{EPRFigure}
\begin{tabular}{ll}
\begin{minipage}{7.35cm}
\footnotesize
Under the correspondence between dependent linear type theory and quantum information theory, the existence of bunched typing contexts involving linearly multiplicative conjunctions $\otimes$ corresponds to the conditioning of protocols on entangled quantum states and hence to what in quantum physics are known as EPR phenomena.
\end{minipage}

&

\begin{tabular}{c}

\def\arraystretch{1.2}
\begin{tabular}{|c|c|}
  \hline
  \bf Bunched logic
  &
  \bf EPR phenomena
  \\
  \hline
  \def\arraystretch{1}
  \begin{tabular}{c}
    Typing contexts built via
    \\
    multiplicative conjunction ($\otimes$)
  \end{tabular}
 \!\!\!\!\!
  &
  \!\!\!\!\!\!
  \def\arraystretch{1}
  \begin{tabular}{c}
    Physics conditioned on
    \\
    entangled quantum states
  \end{tabular}
  \\
  \hline
\end{tabular}

\end{tabular}
\end{tabular}
\vspace{.2cm}

\smallskip

\noindent
{\bf Exponential modality.}
In the previous lack of a classically-dependent linear type theory, the strategy for recovering classical logic among a linear (quantum) type system
was to postulate a modal operator (Lit. \ref{ModalLogicAndManyWorlds}) on the linear type system -- traditionally denoted ``!'' \cite{Girard87}
and (sometimes) called the {\it exponential modality} -- where a linear type of the form $! \HilbertSpace{H}$ may be thought of (cf. Rem. \ref{ExponentialModalityGivesLinearSpanOfUnderlyingSet} below) as behaving like the linear span of the {\it underlying set} of a linear
space $\HilbertSpace{H}$, thus giving the linear type system a kind of access to this underlying classical type.
Eventually it came to be appreciated (cf. \cite[p. 36]{Mellies09}) that the exponential modality should
(this is due to \cite[\S 2]{Seely89} and \cite{dePaiva89}\cite[\S 8]{BentonBiermanDePaiva92}\cite{BentonBiermanDePaivaHyland92})
be axiomatized as a comonad (cf. Lit. \ref{LiteratureComputationalEffectsAndModalities})
 and specifically as a comonad induced by a suitably monoidal adjunction
 \eqref{MonadFromAdjunction} between linear and classical (intuitionistic) types
  (due to \cite[p. 157]{Bierman94}\cite{Benton95}):
  \vspace{-2mm}
  \begin{equation}
    \label{ExponentialModality}
    \hspace{-1cm}
    \begin{tikzcd}[
      row sep=0pt, column sep=large
    ]
      \ClassicalTypes
      \quad
      \ar[
        rr,
        shift left=7pt,
        "{
          \quantized
        }"{description},
        "{
          \scalebox{.7}{
            \color{darkgreen}
            \bf
            quantization
          }
        }"{yshift=2pt}
      ]
      \ar[
        from=rr,
        shift left=7pt,
        "{
          \classicized
        }"{description},
        "{
          \scalebox{.7}{
            \color{darkgreen}
            \bf
            classicization
          }
        }"{yshift=-2pt}
      ]
      \ar[
        rr,
        phantom,
        "{
          \scalebox{.7}{$\bot$}
        }"
      ]
      &&
      \quad
      \LinearTypes
    \ar[out=50, in=-50,
      looseness=4,
      shorten=-2pt,
      shift left=10pt,
      "{
        \scalebox{1.6}{
          ${
            \hspace{3pt}
            \mathclap{
              \ExponentialModality
            }
            \hspace{1pt}
          }$
          }
          \rlap{
          \hspace{-10pt}
    \scalebox{.7}{
      \color{darkorange}
      \bf
      exponential modality
    }
        }
      }"{description}
    ]
    \\
    \scalebox{.7}{
      \color{darkblue}
      \bf
      \def\arraystretch{.9}
      \begin{tabular}{c}
        purely
        \\
        classical
        \\
        (intuitionistic)
        \\
        types
      \end{tabular}
    }
    &&
    \scalebox{.7}{
      \color{darkblue}
      \bf
      \def\arraystretch{.9}
      \begin{tabular}{c}
        purely
        \\
        quantum
        \\
        (linear)
        \\
        types
      \end{tabular}
    }
    \end{tikzcd}
  \end{equation}

  \vspace{0mm}
\noindent   Traditionally,
  inference rules for such an exponential modality need to be adjoined to plain (non-dependent) linear type
  theories, which is laborious and not without subtleties
  (\cite{Girard93}\cite{Wadler93}\cite{Benton95}\cite{Barber96}).
  In contrast, in Prop. \ref{QuantizationAndExponentialAdjunction} we obtain (cf. \cite[Prop. 2.1.31]{Riley22}) an exponential
  modality from the basic type inference provided by a {\it dependent} linear type theory like {\LHoTT} (Lit. \ref{LiteratureLHoTT}), a possibility first highlighted in
  \cite[Ex. 4.2]{PontoShulman12}\cite[\S 4.2]{QuantizationViaLHoTT}.

\smallskip

\noindent
{\bf Full verification: Towards identity types.}
Either way, (linear) data-typing in general serves to impose and verify consistency constraints on (quantum) data. But for a fine-grained certification of
program behavior by {\it equational} constraints --- e.g. for certifying the correctness of quantum teleportation protocols
or of quantum error corrections (cf. Rem. \ref{VerificationOfBitFlipCode}) --
one specifically needs certificates of {\it identification types} (colloquially: ``identity types''), certifying the (operational) equality of pairs of
data of a given type (cf. Lit. \ref{LiteratureHomotopyTypeTheory}).

However, the correct formal treatment of data types of identifications turns out to be surprisingly subtle, which may be one reason why none of
the previously existing quantum programming languages provide such identity types --- and this includes (Proto-){\Quipper}, cf. Lit. \ref{LiteratureQuantumProgrammingLanguages}. Namely, once identifications of any data pairs $d, d' \isa D$ are promoted to data of
identification type $p \isa \mathrm{Id}_D(d,\,d')$ (``propositional equality''), the same principle applies to pairs $p, p' \isa \mathrm{Id}_D(d\, d')$
of these certificates themselves, whose
verifiable identification now requires data of {\it iterated identification type} $\mathrm{Id}_{{}_{\mathrm{Id}_{{}_D}(d,d')}}(d,\,d')$ -- {\it and so on}.
The proper handling of this phenomenon requires and leads {\it homotopy types} of data provided by classical {\HoTT} and its linear form {\LHoTT};
see the discussion in Lit. \ref{LiteratureHomotopyTypeTheory}.
\end{literature}

\begin{literature}[\bf Quantum programming languages]
  \label{LiteratureQuantumProgrammingLanguages}
  The idea of quantum programming languages
  formally expressing quantum computational processes (Lit. \ref{LiteratureQuantumComputation})
  was first systematically expressed in \cite{Knill96}, early proposals for formalization
  are due to \cite{Selinger04}\cite{Valiron04}\cite{SelingerValiron05}\cite{SelingerValiron09} (``quantum $\lambda$-calculus''), \cite{AltenkrichGrattage05} ({\tt QML}),  and \cite{AltenkirchGreen10}\cite{Green10} (via ``quantum IO'', a kind of monadic quantum effects, Lit. \ref{LiteratureComputationalEffectsAndModalities}).
  Exposition of the need and relevance of quantum programming languages (which was not originally obvious to the community, cf. the historical lead-in to \cite{Selinger16})
  specifically for quantum/classical hybrid computation, may be found in \cite{VRSAS15}.

   Based on these early developments
   (and besides a multitude of quantum circuit languages that now exist for programming available NISQ machines, Lit. \ref{LiteratureOnNISQMachines}), currently there exists essentially one quantum programming language with universal ambition: {\Quipper}\footnote{Landing page: \href{https://www.mathstat.dal.ca/~selinger/quipper}{\tt www.mathstat.dal.ca/$\sim$selinger/quipper}} \cite{GreenLumsdaineRossSelingerValiron13}\cite{GLRSV13} (for exposition see \cite{Selinger16}).
  In its formalized sector called
  ``Proto-{\Quipper}'' \cite[\S 8]{Ross15}\cite[\S 4.3]{RiosSelinger18}
  this language
  may be understood
  as involving a kind of dependent (Lit. \ref{LiteratureLHoTT}) linear types, Lit. \ref{VerificationLiterature}) with
  semantics in categories of indexed sets of linear objects
  (\cite{RiosSelinger18}\cite{FuKishidaSelinger20}\cite{Lee22}\cite{Rios21}),
  notably in indexed sets of (complex) vector spaces, of the same kind as that in \cref{QuantumTypeSemantics} we discuss as semantics for the 0-sector  (Rem. \ref{Zerosector}) of {\LHoTT} (Lit. \ref{LiteratureLHoTT}).

(Notice that  {\Quipper} (and {\tt qIO}) are embedded (Lit. \ref{LiteratureDomainSpecificLanguages}) inside the classical language {\Haskell} which means that they lack support for verification of linear (quantum) data types, cf. Lit. \ref{VerificationLiterature}.)

Another quantum programming language scheme with the ambition of certifying (Lit. \ref{VerificationLiterature}) quantum (circuit) programs is {\tt QWIRE}, see
\cite{PaykinRandZdancewic17}\cite{RandPaykinZdancewic18}\cite{PaykinZdancewic19}\cite{RennelaStaton20}\cite{HRHWH21}\cite{HRHLH21}\cite{ZBSLY23}.
\end{literature}

\begin{literature}[\bf Domain-specific embedded programming languages]
\label{LiteratureDomainSpecificLanguages}
Besides universal programming languages, more specific tasks -- such as quantum circuit programming (cf. Lit. \ref{LiteratureQuantumProgrammingLanguages}) -- often profit from non-universal languages tailor-made towards the problem at hand -- one speaks of {\it domain-specific languages} (DLS) \cite{Hudak98b}\cite{Hudak98b}. Typically these are {\it embedded} into ambient universal languages (\cite{Hudak96}), by specification of ``syntactic sugar'' (e.g. \cite[\S 1.6, \S 1.7, \S 9]{Ranta94}) for blocks of similar code in the ambient language that serve as the building blocks of the domain-specific embedded language.

\vspace{1.5cm}

An example is {\it do}-notation (Lit. \ref{LiteratureProgrammingSyntaxForMonadicEffects}) for monadic language constructs (Lit. \ref{LiteratureComputationalEffectsAndModalities}), and \cite[\S 5.3]{BentonHughesMoggi02} suggest that formulating domain-specific embedded languages is close to synonymous with identifying do-notation for suitable monads, citing the example of domain-specific parser languages identified as monadic do-notation by \cite[\S 7.1]{Wadler90}. These authors conclude:

\vspace{-.2cm}
\begin{center}
\begin{minipage}{15cm}
{\it
``Every time a functional programmer designs a combinator library, then, we might as well say that he or she designs a domain specific programming language} [...]{\it . This is a useful perspective, since it encourages programmers to produce a modular design, with a clean separation between the semantics of the DSL and the program that uses it, rather than mixing combinators and `raw' semantics willy-nilly. And since monads appear so often in programming language semantics, it is hardly surprising that they appear often in combinator libraries also!}
\end{minipage}
\end{center}

Existing functional (Lit. \ref{LiteratureFunctionalLanguages}) quantum programming languages such as {\tt qIO} and {\Quipper} (Lit. \ref{LiteratureQuantumProgrammingLanguages}) are domain-specific languages embedded in {\Haskell}, and among these Altenkrich \& Green's {\tt qIO} (the {\it quantumIO-monad}) stands out in its ambition of sticking to the monadic paradigm. However, since the ambient {\Haskell} does not verify linear (quantum) data typing (Lit. \ref{VerificationLiterature}, and no other available embedding language did), neither do these embedded languages.

In \cref{Pseudocode} we aim to show that a nice monadically-embedded quantum programming language with linear tying does exist inside {\LHoTT} (Lit. \ref{LiteratureLHoTT}).

\end{literature}

\begin{literature}[\bf Homotopically typed languages]
  \label{LiteratureHomotopyTypeTheory}
  (For extensive review cf. the companion article \cite{TQP}.)
 An operation on data so fundamental and commonplace that it is easily taken for granted is the {\it identification} of a pair of data with each other.
 But taking the idea of program verification by data typing (Lit. \ref{VerificationLiterature})
 seriously leads to consideration also of   {\it certificates of identification} of pairs of data of any given type which thus must  themselves be data
 of ``identification type'' \cite[\S 1.7]{MartinLof75}.
  Trivial as this may superficially seem,
 something profound emerges with such
 ``thoroughly typed'' programming languages (the technical term is: {\it intensional type theories} (see  \cite[p. 4, 13]{Streicher93}\cite[p. 16]{Hofmann95})
 in that now given a pair of such identification certificates the same logic applies to these and leads to the consideration of
 identifications-of-identifications (first amplified in \cite{HofmannStreicher98}), and so on to higher identifications, {\it ad infinitum}.

 Remarkably, the ``denotational semantics'' (Lit. \ref{VerificationLiterature}) of data types equipped with such towers of identification types, hence the
 corresponding pure mathematics, is  (\cite{AwodeyWarren09}\cite{Awodey12}, exposition in \cite{Shulman12}\cite{Riehl22}) just that of abstract homotopy theory (Lit. \ref{TangentInfinityToposes}) where identification types are interpreted
 as path spaces and higher-order identifications correspond to
 higher-order homotopies.
 One also expresses this state of affairs, somewhat vaguely, by saying that HoTT has {\it semantics} in homotopy theory, and
 conversely that HoTT is a {\it syntax} for homotopy theory -- we have reviewed this dictionary in \cite[\S 5.1]{TQP}.

 Ever since this has been understood, the traditional (``intuitionistic Martin-L{\"o}f''-)type theory of \cite{MartinLof75}\cite{NPS90} has essentially come to be known
 as {\it homotopy type theory} (HoTT) -- specifically so if accompanied by one further ``univalence'' axiom\footnote{
   \label{UnivalenceAttribution}
   The univalence axiom is widely attributed to \cite{Voevodsky10}, but the idea (under a different name) is actually due to
   \cite[\S 5.4]{HofmannStreicher98}, there however formulated with respect to a subtly incorrect type of equivalences (as later shown
   in \cite[Thm. 4.1.3]{UFP13}). The new contribution of \cite[p. 8, 10]{Voevodsky10} was a good definition of the types
   of (``weak'') equivalences between types.
 } (for more on this see the companion article around \cite[(105)]{TQP})
 which enforces that identification of data types themselves coincides with their operational equivalence (exposition in \cite{Aczel11}).

 The standard textbook account for ``informal'' (human-readable) {\HoTT} is \cite{UFP13}, exposition may be found in \cite{BrunerieLicataLumsdaine13},
 gentle introduction in \cite{Rijke18}\cite{Rijke23} (the former more extensive); and see the companion article \cite[\S 5]{TQP}. Available software that {\it runs} homotopically typed programs
 includes {\tt Agda}\footnote{\;{\tt Agda} landing page: \href{https://wiki.portal.chalmers.se/agda/pmwiki.php}
 {\tt wiki.portal.chalmers.se/agda/pmwiki.php}} and {\tt Coq}\footnote{\;{\tt Coq} landing page: \href{https://coq.inria.fr}{\tt coq.inria.fr}}.
\end{literature}

\begin{literature}[\bf Linear homotopically typed languge]
\label{LiteratureLHoTT}
Based on the developments of {\HoTT} (Lit. \ref{LiteratureHomotopyTypeTheory})
and in view of the idea of linear data typing for quantum languages (Lit. \ref{VerificationLiterature})
we had previously argued \cite{QuantizationViaLHoTT}\cite{Schreiber14} that there should exist a {\it linear} enhancement of {\HoTT} providing, in addition, a natural formal language for motivic (stable) homotopy (tangent $\infty$-toposes, Lit. \ref{TangentInfinityToposes}) and quantum systems.
After some partial proposals for such dependent linear type systems (\cite{KPB15}\cite[\S 3]{Vakar15}\cite{McBride16}\cite{Vakar17}\cite{Lundfall18}\cite{Atkey18}\cite{FuKishidaSelinger20}\cite{MoonEadesOrchard21}, see also earlier discussion in \cite{SchoeppStark04})\footnote{See \cite[\S 1.7]{Riley22}\cite[p. 22]{Riley22b} for critical discussion of these and other previous approaches to dependent linear types.}, a satisfactory {\it Linear Homotopy Type Theory} ({\LHoTT}) has recently been presented by M. Riley \cite{Riley22}, see also \cite{Riley22b}\cite{Riley23}.
\ifdefined\monadology

We will eventually give a detailed exposition of {\LHoTT} in \cite{QS}.
\fi
For embedding (Lit. \ref{LiteratureDomainSpecificLanguages}) the monadic quantum effects of
\ifdefined\monadology
\cref{QuantumEffects}
\else
\cref{QuantumTypeSemantics} and \cref{QuantumEffects}
\fi
into {\LHoTT}
all we need is that {\LHoTT} verifies the Motivic Yoga (Def. \ref{MotivicYoga}), which is the case by  the discussion in \cite[\S 2.4]{Riley22}.
\end{literature}

\begin{literature}[\bf Topological quantum compilation.]
\label{TopologicalQuantumCompilation}
Once serious quantum computation hardware (Lit. \ref{LiteratureTopologicalQuantumComputation})
becomes available, a central effort in quantum computation (Lit. \ref{LiteratureQuantumComputation}) concerns {\it quantum compilation}
\cite{MMRP21}, namely the translation of high-level quantum algorithms into sequences (circuits) of those logic gates that the hardware actually implements.
The seminal {\it Solovay-Kitaev theorem} (\cite[App. 3]{NielsenChuang10}\cite{DawsonNielsen06}) guarantees, under rather mild assumptions on the available gate set, that such a compilation is always possible, but optimization for scarce runtime resources requires considerable effort.

The problem of quantum computation is particularly demanding for topological quantum computation (Lit. \ref{LiteratureTopologicalQuantumComputation}), hence in the case of {\it topological quantum compilation} (e.g. \cite{HormoziZikosBonesteelSimon07}\cite{Brunekreef14}\cite{KBS14}),
since here the available gate logic is far remote from then $\QBit$-based operations \eqref{TraditionalCNOTGate} in which high-level quantum algorithms are conceived.
No attempt seems to previously have been made toward formally verifying a topological quantum compilation, and indeed the problem is not captured by classical verification strategies. Notice that:
\begin{itemize}[leftmargin=.7cm]
\item[{\bf (i)}] formal verification of quantum compilation, in general, is not a discrete but an {\it analytical} problem, whose computer verification requires {\it exact real (complex) computer arithmetic} (cf. \cite[Lit, 2.29]{TQP}),
\item[{\bf (ii)}] the generic topological quantum gate is given by a complicated analytical expression (cf. \cite[Lit. 2.24]{TQP}).
\end{itemize}

\vspace{1mm}
\noindent While here we will not further dwell on the issue explicitly, the claim of \cite{TQP} is that these two problems are addressed by homotopically-typed certification languages ({\HoTT}, Lit. \ref{LiteratureHomotopyTypeTheory}) of which the language {\LHoTT} of concern here (Lit. \ref{LiteratureLHoTT}) is an extension.
\end{literature}

\begin{literature}[\bf NISQ computers]
  \label{LiteratureOnNISQMachines}
  Currently existing quantum computers
  (such as those based on ``superconducting qbits'', see e.g. \cite{ClarkeWilhelm08}\cite{HWFZ20})
  serve as proof-of-principle of the idea of quantum computation
  (Lit.  \ref{LiteratureQuantumComputation})
  but
  offer puny computational resources, as
  they are (very) {\bf n}oisy and (at best) of {\bf i}ntermediate {\bf s}cale:  ``NISQ machines'' \cite{Preskill18}\cite{LeymannBarzen20}.
  What is currently missing are noise-protection mechanisms that would allow to scale up the size and coherence time of quantum memory.
  The foremost such protection mechanism
  arguably is {\it topological} protection (Lit. \ref{LiteratureTopologicalQuantumComputation}).
\end{literature}

\begin{literature}[\bf Classically controlled quantum computation and dynamic lifting]
\label{ClassicalControlAndDynamicLifting} $\,$
The idea of classically controlled quantum computation goes back to \cite{Knill96} and was amplified in \cite[\S 4]{NPW07} (from which we adapted the schematics graphics on p. \pageref{ClassicalControlSchematics}),
see also \cite{Devitt14}.
The term ``dynamic lifting'' for the converse control flow (where mid-circuit quantum measurement results are fed back into the classical control logic)
is due to \cite[p. 5]{GreenLumsdaineRossSelingerValiron13}, early discussion is in \cite[p. 40]{Rand18}; proposals for its categorical semantics are discussed in \cite{RennelaStaton20}\cite{LPVX21}\cite{FKRS22a}\cite{FKRS22b}\cite{ColledanDalLago22}\cite{Lee22}.

Of these, the definition in \cite[\S 4.4]{Lee22} of a monad (Lit. \ref{LiteratureComputationalEffectsAndModalities}) meant to express dynamic lifting is vaguely in the spirit of the quantum indefiniteness monad $\indefinitely_W$ from \cref{QuantumEpistemicLogicViaDependentLinearTypes} which in \cref{ControlledQuantumGates} we find to express just that: Lee's ``lifting monad'' applied to a bundle type $\footnotesize \ABundleType{\HilbertSpace{H}_\bullet}{W}$ (in the language of \cref{QuantumTypeSemantics}) produces the bundle type over the set of multisets $[w_i]_{i \in I}$ of elements of $W$ whose fibers are the direct sums $\underset{i \in I}{\osum} \HilbertSpace{H}_{w_i}$; the idea being to interpret these as the branched Hilbert spaces inside which to locate quantum states obtained after (repeated?) measurement results $w_i$.

Compare this to the indefiniteness monad, which for a (finite) set of outcomes $W$ sends a pure quantum type $\HilbertSpace{H}$ to $\indefinitely_W \HilbertSpace{H} \,\defneq\, \osum_W \HilbertSpace{H}$ -- see the typing of dynamically lifted quantum measurement results on p. \pageref{ComputationalMeasurementTyping}, and see \eqref{ComputationalTypingOfSuccessiveMeasurement} for the successive lifting of quantum measurements, accumulating the measurements results in the classical context.
\end{literature}

\subsection{Quantum probability}
\label{BackgroundQuantumProbability}

\begin{literature}[\bf Quantum probability and Quantum channels]
\label{LiteratureQuantumProbability}
Remarkably, in its relation to physical reality, quantum physics (Lit. \ref{EpistemologyOfQuantumPhysics}) is a {\it probabilistic} theory (\cite[\S III]{vonNeumann32}\cite{MehraRechenberg01}), and yet more remarkably its probabilistic aspect is tied in some deep way to the complex numbers equipped with their involution by complex conjugation:

\medskip
\noindent
{\bf Hilbert spaces of quantum states.}
The definition of {\it Hilbert spaces} $\big(\HilbertSpace{H} , \langle-\vert-\rangle\big)$ in quantum physics (\cite[\S 1]{vonNeumann30}\cite[\S II.1]{vonNeumann32})
concerns extra structure and properties on the underlying complex vector space of quantum states: (1.) A Hermitian inner product
$\langle-\vert-\rangle$ and (2.) a topological completeness condition. The latter condition is (just) to make sense of
infinite-dimensional state spaces and is of no concern for the
finite-dimensional Hilbert spaces of interest in quantum information theory (which are automatically complete).
The key structure that remains  is the Hermitian inner product structure $\langle-\vert-\rangle$ on a finite-dimensional space
$\HilbertSpace{H}$ of quantum states (e.g. \cite[\S A.1]{Landsman17}), which is (not a complex bilinear
on $\HilbertSpace{H} \otimes \HilbertSpace{H}$, but) a {\it sesquilinear} map, complex-anti linear in the first argument:
$$
  \begin{tikzcd}
    \mathllap{
  \scalebox{.7}{
    \color{darkgreen}
    \bf
    \def\arraystretch{.9}
\begin{tabular}{c}
      Hermtian
      \\
      inner product
    \end{tabular}
  }
    \;\;
    \langle - \vert - \rangle
    \;\isan\;\;
    }
    \underset{
      \mathclap{
        \hspace{-10pt}
        \rotatebox[origin=c]{-45}{
          \hspace{-18pt}
          \rlap{
            \scalebox{.7}{
              \color{darkblue}
              \bf
              \def\arraystretch{.9}
              \begin{tabular}{c}
                complex conjugate
                \\
                space
              \end{tabular}
            }
          }
        }
      }
    }{
    \overline{
      \HilbertSpace{H}
    }
    }
    \otimes
    \underset{
      \mathclap{
        \hspace{5pt}
        \rotatebox[origin=c]{-45}{
          \hspace{-31pt}
          \rlap{
            \scalebox{.6}{
              \color{darkblue}
              \bf
              \def\arraystretch{.9}
              \begin{tabular}{c}
                complex vector space
                \\
                underlying
                              Hilbert space
              \end{tabular}
            }
          }
        }
      }
    }{
    \HilbertSpace{H}
    }
    \ar[
      rr
    ]
    &&
    \ComplexNumbers
  \end{tikzcd}
$$
\vspace{1.4cm}

\noindent
namely such that
\begin{equation}
  \label{ConditionsOnHermitianInnerProduct}
  \psi, \psi' \,\isa\, \HilbertSpace{H}
  ,\;\;
  c \,\isa\, \ComplexNumbers
  \;\;\;\;\;
  \yields
  \;\;\;\;
  \def\arraystretch{1.5}
  \overset{
    \raisebox{2pt}{
      \scalebox{.7}{
        \color{darkorange}
        \bf
        Hermitian
        sesqui-linearity
      }
    }
  }{
  \begin{array}{l}
 \langle \psi' \vert c\cdot \psi \rangle
 \,=\,
 c
 \,
 \langle \psi' \vert \psi \rangle
 \\
 \langle \psi \vert \psi' \rangle
 \,=\,
 \overline{
   \langle \psi' \vert \psi \rangle
 }
 \,,
 \end{array}
 }
 \;\;
  \overset{
    \hspace{-1.9cm}
    \raisebox{2pt}{
      \scalebox{.7}{
        \color{darkorange}
        \bf
        positivity
      }
    }
  }{
  \underset{
    \raisebox{2pt}{
      \scalebox{.7}{
        \color{darkorange}
        \bf
        non-degeneracy
      }
    }
  }{
 \begin{array}{l}
   \langle \psi \vert \psi \rangle
   \,\geq 0\,
   \,,
   \\
   \langle \psi \vert \psi \rangle = 0
   \;\;
   \Rightarrow
   \;\;
   \psi = 0
   \,.
 \end{array}
 }
 }
 \end{equation}

\noindent
{\bf Bra-Ket notation.}
The non-degeneracy condition
\eqref{ConditionsOnHermitianInnerProduct} on $\langle-\vert-\rangle$ means that every element
of the linear dual space $\HilbertSpace{H}^\ast \,\defneq\, (\HilbertSpace{H} \maplin \ComplexNumbers)$ is uniquely of
the form $\langle \psi \vert - \rangle$ for some $\psi \,\in\, \HilbertSpace{H}$, which leads to the suggestive {\it bra-ket}
notation traditional in quantum physics (since \cite{Dirac39}, see e.g. \cite[\S 1.2]{SakuraiNapolitano94}\cite[\S 3]{Griffiths02}):
\vspace{-1mm}
\begin{equation}
  \label{BraKetNotation}
  \def\arraystretch{2}
  \begin{array}{c}
   \overset{
     \mathclap{
       \raisebox{4pt}{
         \scalebox{.7}{
           \color{darkblue}
           \bf
           ``ket''
           in
           Hilbert space
         }
       }
     }
   }{
   \vert \psi \rangle
   \;\,\defneq\;\,
   \psi
   \,\isa\,
   \HilbertSpace{H}
   }
   ,
   \qquad
   \overset{
     \mathclap{
       \raisebox{4pt}{
         \scalebox{.7}{
           \color{darkblue}
           \bf
           ``bra''
           in
           dual space
         }
       }
     }
   }{
   \langle \psi \vert
   \,\defneq\,
   \langle \psi \vert - \rangle
   \;\,\isa\;\,
   \HilbertSpace{H}^\ast
   }
   \,.
   \end{array}
\end{equation}

\vspace{-2mm}
\noindent

If nothing else, this notation \eqref{BraKetNotation} allows one to neatly distinguish between the element  $w \isa W$ in a
(finite) set $W$ and the corresponding vector in the linear span
$\vert w \rangle \,\in\, \quantized W \,\defneq\, \underset{W}{\oplus} \TensorUnit$ (and as such we understand
$\vert-\rangle$ as the return-operation \eqref{BindingAndReturning} of the ``quantization modality'' $\quantized$, see Def. \ref{QuantizationModality} and p. \pageref{KetAsQReturn}). Equipped with the canonical inner product this is an {\it orthonormal linear basis}:
\vspace{-3mm}
\begin{equation}
  \label{AnOrthonormalBasis}
  \def\arraystretch{2}
 \begin{array}{rrcl}
  \scalebox{.7}{
    \color{darkblue}
    \bf
    linear basis
  }
  &
  w \,\isa\, W
  &
  \yields
  &
  \vert w \rangle
  \,\isa\,
  \;
  \underset{
    \mathclap{ w \isa W }
  }{\oplus}
  \;
  \ComplexNumbers
  \,\defneq\,
  \HilbertSpace{H}
  \,,
  \\
  \scalebox{.7}{
    \color{darkblue}
    \bf
    ortho-normality
  }
  &
  w,w' \,\isa\, W
  &
    \yields
  &
  \langle w' \vert w \rangle
  \,=\,
  \delta_w^{w'}
  \,\defneq\,
  \left\{
  \def\arraystretch{.9}
  \begin{array}{cl}
    1 & \mbox{if}\; w = w'
    \\
    0 &\mbox{otherwise}
  \end{array}
  \right.
  \end{array}
\end{equation}

More profoundly, the bra-ket notation \eqref{BraKetNotation} is a lightweight precursor to the string diagram calculus in dagger-compact
closed categories \eqref{OperatorAdjoints} (as amplified by \cite[\S 7.2]{AbramskyCoecke04}\cite[p. 6]{AbramskyCoecky07}\cite[\S 3.3]{Coecke10}):
For $\HilbertSpace{H}$ a finite-dimensional Hilbert space with orthonormal basis $W$ \eqref{AnOrthonormalBasis}, the vector space of
linear maps into some $\HilbertSpace{H}'$ is canonically identified with a space of matrices as follows \eqref{LinearHomViaTensoringWithDual}:
\vspace{-2mm}
\begin{equation}
  \label{CompactClosureViaBraKets}
  \begin{tikzcd}[row sep=-3pt]
    \mathclap{
      \scalebox{.7}{
        \color{darkblue}
        \bf
        \def\arraystretch{.9}
        \begin{tabular}{c}
          linear space
          \\
          of linear maps
        \end{tabular}
      }
    }
    &&
    \mathclap{
      \scalebox{.7}{
        \color{darkblue}
        \bf
        \def\arraystretch{.9}
        \begin{tabular}{c}
          linear space
          \\
          of matrices
        \end{tabular}
      }
    }
    \\
    \big(
      \HilbertSpace{H}
      \maplin
      \HilbertSpace{H}'
    \big)
    \ar[
      rr,
      "{ \sim }"
    ]
    &&
    \HilbertSpace{H}'
    \otimes
    \HilbertSpace{H}^\ast
    \\
    \scalebox{0.9}{$
      \big(
      \underset{
        \raisebox{-4pt}{
          \scalebox{.7}{
            \color{olive}
            in
          }
        }
      }{
        \vert w \rangle
      }
      \,\mapsto\,
      \underset{
        \raisebox{-4pt}{
          \scalebox{.7}{
            \color{olive}
            out
          }
        }
      }{
      \sum_{w'}
      \vert w' \rangle
      A_{w',w}
      }
    \big)
    $}
    &
      \scalebox{\termsize}{$
        \mapsto
      $}
    &
     \scalebox{0.9}{$
       \underset{w,w'}{\sum}
       \;
      \underset{
        \raisebox{-4pt}{
          \scalebox{.7}{
            \color{olive}
            out
          }
        }
      }{
       \vert w'\rangle
      }
       A_{w',w}
      \underset{
        \raisebox{-4pt}{
          \scalebox{.7}{
            \color{olive}
            in
          }
        }
      }{
        \langle w \vert
      }
    $}
  \end{tikzcd}
\end{equation}

\vspace{-2mm}
\noindent
{\bf The Born rule.}
The Hermitian inner product $\langle-\vert-\rangle$
on spaces of quantum states serves to refine the description \eqref{LinearStateCollapse}
of the quantum measurement process by
assigning a {\it probability distribution} $\mathrm{Prob}_\psi$ to the possible measurement outcomes
on a system in state $\vert \psi \rangle \,\in\, \HilbertSpace{H}$
in a state space $\HilbertSpace{H} \,\simeq\, \osum_W\, \ComplexNumbers$ spanned by an orthonormal measurement basis $W$ \eqref{AnOrthonormalBasis}.

The  {\it Born rule} of quantum physics  postulates
(\cite[p. 805]{Born26}\cite[p. 811]{Jordan27}\cite[\S III]{vonNeumann32}, review in \cite{Landsman09})
that the probability $\mathrm{Prob}_\psi(w)$ for a quantum measurement \eqref{LinearStateCollapse}
of a system in a normalized state
\vspace{-2mm}
\begin{equation}
  \label{NormalizedStates}
  \vert \psi \rangle
  \,\isa\,
  S(\HilbertSpace{H})
  \,\defneq\,
  \overset{
    \mathclap{
      \raisebox{4pt}{
        \scalebox{.7}{
          \color{darkblue}
          \bf
          normalized states
        }
      }
    }
  }{
 \big(
   \vert \psi \rangle \,\isa\,
   \HilbertSpace{H}
 \big)
 \times
 \big(
   \langle \psi \vert \psi \rangle
   \,=\,
   1
 \big)
 }
\end{equation}
to yield the result $w \isa W$ from an orthonormal basis \eqref{AnOrthonormalBasis} is:
\vspace{-1mm}
\begin{equation}
  \label{BornRule}
  \begin{tikzcd}[
    row sep=-13pt,
    column sep=0pt
  ]
  &[-30pt]
  \scalebox{.7}{
    \color{darkblue}
    \bf
\begin{tabular}{c}
      probability to measure $w$
      \\
      on system in state $\vert \psi \rangle$
    \end{tabular}
  }
  &[-10pt]
  \scalebox{.7}{
    \color{darkorange}
    \bf
    \begin{tabular}{c}
      equals according
      \\
      to Born's rule
    \end{tabular}
  }
  &[-10pt]
  \mathrlap{
  \scalebox{.7}{
    \color{darkblue}
    \bf
    \begin{tabular}{c}
      square modulus of
      \\
      transition amplitude
    \end{tabular}
  }
  }
  \\
  \left.
  \def\arraystretch{1.2}
  \def\arraycolsep{0pt}
  \begin{array}{l}
    \;W \,\isa\, \mathrm{FinSet}
    \\
    \vert \psi \rangle
    \,\isa\,
    S\big(
    \underset{
      \mathrlap{
        \!\!\!\!
        w \isa W
      }
    }{\oplus}
    \,
    \ComplexNumbers
    \big)
    \\
    \;\,w \,\isa\, W
  \end{array}
  \;\;
  \right\}
  \;\;\yields
  &
  \mathrm{Prob}_{\psi}(w)
  &
  \defneq
  &
  \langle \psi \vert w \rangle
  \langle w \vert \psi \rangle
  &
  =
  &
  \overline{\langle w \vert \psi \rangle}
  \langle w \vert \psi \rangle
  &
  =
  &
  \big\vert
   \underset{
     \mathclap{
       \raisebox{-3pt}{
         \scalebox{.7}{
           \color{darkblue}
           \bf
           \begin{tabular}{c}
             ``transition amplitude''
             \\
             from
             $\vert \psi \rangle$
             to
             $\vert w \rangle$
           \end{tabular}
         }
       }
     }
   }{
    \underbrace{
      \langle w \vert \psi \rangle
    }
    }
  \big\vert^2
  \end{tikzcd}
\end{equation}

\vspace{-2mm}
\noindent
That the Born rule \eqref{BornRule} indeed gives a probability distribution on $W$ is intimately connected to the notion \eqref{ConditionsOnHermitianInnerProduct} of Hermitian inner products, notably via the corresponding Cauchy-Schwarz inequality:
\vspace{-2mm}
$$
  \hspace{-4.4cm}
    \scalebox{.7}{
      \color{darkblue}
      \bf
      \begin{tabular}{c}
        measurement probabilities
        \\
        indeed take values in  $[0,1]$
      \end{tabular}
    }
    \;\;\;
  \begin{tikzcd}[
    row sep=-2pt,
    column sep=3pt
  ]
    \mathrm{Prob}_\psi(w)
    &\defneq&
    \big\vert
    \langle w \vert \psi \rangle
    \big\vert^{2}
    &
      \leq
    &
    \langle w \vert w \rangle
    \langle \psi \vert \psi \rangle
    &=&
    1
    \\
    &
    \scalebox{.7}{
      \color{gray}
      \eqref{BornRule}
    }
    &&
    \mathclap{
    \scalebox{.7}{
      \color{gray}
      Cauchy-Schwarz
    }
    }
    &&
    \scalebox{.7}{
      \color{gray}
      \eqref{AnOrthonormalBasis}
      \eqref{NormalizedStates}
    }
  \end{tikzcd}
$$
$$
  \hspace{-.2cm}
    \scalebox{.7}{
      \color{darkblue}
      \bf
      \begin{tabular}{c}
        measurement probs
        \\
        indeed sum to unity
      \end{tabular}
    }
  \sum_w
  \mathrm{Prob}_\psi(w)
  \;\;
  \defneq
  \;\;
 \sum_w
 \big\vert
   \langle w \vert \psi \rangle
 \big\vert^2
 \;\;=\;\;
 \sum_w
   \langle \psi \vert w \rangle
   \langle w \vert \psi \rangle
  \;\;
  =
  \;\;
  \langle \psi \vert
  \Big(
    \underbrace{
    \sum_w
    \vert w \rangle \langle w \vert
    }_{= \mathrm{id}_{\HilbertSpace{H}} }
  \Big)
  \vert\psi \rangle
  \;\;
  =
  \;\;
  \langle \psi \vert \psi \rangle
  \;\;
  =
  \;\;
  1
  \,.
$$

\noindent
{\bf Category theory for Hermitian inner products?}
The structure of a Hermitian inner product on complex vector spaces (e.g. \cite[\S 2.1]{KadisonRingrose97}), classical as it may be, is somewhat odd
(in a precise sense, as we shall see) from the perspective of category theory: On a {\it real} vector space $\HilbertSpace{V} \,\isa\, \Modules{\RealNumbers}$
a (non-degenerate) inner product $\langle\mbox{-} \vert\mbox{-}\rangle$ is a self-duality structure in the category-theoretic sense (cf. \cite{Selinger12}):
\vspace{-2mm}
\begin{equation}
  \label{SelfDualityOnARealVectorSpace}
  \begin{tikzcd}[
    row sep=0pt
  ]
    \overset{
      \mathclap{
        \raisebox{5pt}{
          \scalebox{.7}{
            \color{darkblue}
            \bf
            \def\arraystretch{.9}
            \begin{tabular}{c}
              finite-dimensional
              \\
              vector space
            \end{tabular}
          }
        }
      }
    }{
      H
    }
    \qquad
    \ar[
      rr,
      "{ \sim }"
    ]
    &&
    \qquad
    \overset{
      \mathclap{
        \raisebox{5pt}{
          \scalebox{.7}{
            \color{darkblue}
            \bf
            \def\arraystretch{.9}
            \begin{tabular}{c}
              its dual
              \\
              vector space
            \end{tabular}
          }
        }
      }
    }{
      H^\ast
    }
    \\
    \psi \qquad
      &\longmapsto&
      \qquad
    \underset{
      \mathclap{
        \raisebox{-7pt}{
          \scalebox{.7}{
            \color{purple}
            \bf
            \def\arraystretch{.7}
            \begin{tabular}{c}
              Hermitian
              \\
              inner product
            \end{tabular}
          }
        }
      }
    }{
      \langle \psi \vert -\rangle
    }
    \,,
  \end{tikzcd}
\end{equation}

\vspace{-2mm}
\noindent
but for {\it complex} Hermitian inner product spaces the comparison map \eqref{SelfDualityOnARealVectorSpace} is {\it not complex-linear} --- it is complex
anti-linear: $c \cdot \vert \psi \rangle \,\leftrightarrow \overline{c} \cdot \langle \psi \vert$.
For this reason, finite-dimensional complex Hilbert spaces are {\it not} the self-dual objects of $\Modules{\ComplexNumbers}$, in contrast to the
situation for their real cousins.
\medskip

\noindent
{\bf Dagger categories.}
It is ultimately due to this complication \eqref{SelfDualityOnARealVectorSpace} that the category-theoretic foundations of quantum information theory have commonly come to be cast in
terms of ``dagger-categories'' (referring, since \cite{Selinger07} following \cite[Prop. 7.3]{AbramskyCoecke04}, to the notation ``$(\mbox{-})^\dagger$''
for linear operator adjoints; for review see \cite{AbramskyCoecke08}\cite{Coecke10}\cite[\S 2.3, \S 3.3]{HeunenVicary12}\cite{Karvonen18}\cite[\S 2.3]{HeunenVicary19}), namely by direct
axiomatization of the ``dagger''-involution on Hom-spaces that is (or would be, in the abstract case) induced by Hermitian inner product structure on the objects:
\vspace{-2mm}
\begin{equation}
  \label{OperatorAdjoints}
  \begin{tikzcd}
    H_1
    \ar[r, "{ g }"]
    &
    H_2
  \end{tikzcd}
  \hspace{.8cm}
  \yields
  \hspace{.8cm}
  \begin{tikzcd}
    H_1
    \ar[from=r, "{ g^\dagger }"{swap}]
    &
    H_2
  \end{tikzcd}
  \;\;\;\;\;\;\;\;\;\;
  \mbox{s.t.}
  \;\;\;\;\;\;\;\;\;\;
  \big\langle g^\dagger(\mbox{-})\big\vert -)\big\rangle_{\!{}_{H_1}}
  \;=\;
  \big\langle-\big\vert g(\mbox{-})\big\rangle_{\!{}_{H_2}}
  \mathrlap{\,.}
\end{equation}

\vspace{-2mm}
\noindent
In \ifdefined\monadology\cite[\S 3]{QS}\else\cref{QuantumProbability}\fi we discuss two ways of encoding such dagger-structure in {\LHoTT}.

\smallskip

\noindent
{\bf Mixed states and density operators.} While even a pure quantum state $\vert \psi \rangle$ (completely characterizing the state of a quantum system, cf. Lit. \ref{EpistemologyOfQuantumPhysics}) provides only a probabilistic prediction of measurement results given by the Born rule \eqref{BornRule}, in practice this {\it objective stochasticity} of nature is accompanied by {\it subjective stochasticity} due to the fact that the exact quantum state $\vert \psi\rangle$ of a system may (and typically will) not be known with certainty to the experimenter. Therefore the general state of a quantum system --- in the combined sense both of quantum physics and classical statistical physics --- is a classical probabilistic {\it mixture} of quantum states \cite[\S IV.1]{vonNeumann32}, or {\it mixed state} for short (see e.g. \cite[\S 3.4]{SakuraiNapolitano94}\cite[\S 6.1]{Isham95} and particularly \cite[\S 2.4]{NielsenChuang10}\cite[\S 1.4]{Kuperberg05}).

The exact definition notion of what this means was postulated in \cite[p. 158]{vonNeumann32} and (successfully) used ever since, but is not without conceptual subtlety worthy of consideration:
A priori, by a classical mixture of quantum states in a Hilbert space $\HilbertSpace{H}$ one might mean any probability distribution on all of (the underlying set of) the unit sphere $S\HilbertSpace{H}$ of normalized states, or just the projective space $P\HilbertSpace{H}$ of normalized states up to global phase -- this would certainly capture some idea of an ensemble of quantum states, but this is {\it not} what one considers.

Instead, \cite[p. 157]{vonNeumann32} takes the random measurement collapse \eqref{LinearStateCollapse} as the motivating source of classical uncertainty and thus takes a mixed state to be a probability distribution $p \,\isa\, W \to [0,1]$ on (only) the underlying set $W$ of an orthonormal basis $\big(\vert w \rangle \,\isa\, \HilbertSpace{H}\big)_{w \isa W}$, reflecting the pure states in which one may find the quantum system after $W$-measurement.

Finally, \cite[p. 158]{vonNeumann32} observes that it is {\it technically convenient} (our aim in \cref{MixedQuantumTypes} is to motivate this more fundamentally)
to encode this probability distribution
of basis states
as a matrix
\vspace{-2mm}
\begin{equation}
  \label{DensityMatrixInIntroduction}
  \begin{tikzcd}[sep=0pt]
    {}
    \ar[
      rr,
      phantom,
      "{
        \scalebox{.7}{
          \color{darkblue}
          \bf
          \def\arraystretch{.9}
          \begin{tabular}{c}
            probability
            distribution
            of basis states
          \end{tabular}
        }
      }"
    ]
    &&
    {}
    \\
    \mathllap{
      p_{(\mbox{-})}
      \,\isa\,
      \;
    }
    W
    \ar[
     rr
    ]
    &&
    {[0,1]}
    \,,
    \\
    w &\mapsto& p_w
  \end{tikzcd}
  \hspace{-1.1cm}
  \sum_w p_w = 1
  \hspace{1cm}
  \yields
  \hspace{1cm}
  \overset{
    \mathclap{
      \raisebox{6pt}{
        \scalebox{.7}{
          \color{darkblue}
          \bf
          ``mixed state''
          as
          ``density matrix''
        }
      }
    }
  }{
  \rho
  \;\defneq\;
  \underset{w}{\sum}
  \;
  p_w
  \cdot
  \vert w \rangle \langle w \vert
  }
  \;\;\isa\; \HilbertSpace{H} \otimes \HilbertSpace{H}^\ast
\end{equation}

\vspace{-2mm}
\noindent
because then the total probability (of combined quantum and classical origin) to find the system upon quantum
measurement of an(other) property $W'$ in the state $\vert w' \rangle$ is expressed as the {\it trace} of the
operator product of $\rho$ with the projection operator $P_{ w' } \,\defneq\, \vert w' \rangle \langle w' \vert$:
\vspace{-1mm}
\begin{equation}
  \label{TotalProbabilityViaDensityMatrix}
  \underset{
    \mathclap{
      \rotatebox{-40}{
        \hspace{-25pt}
        \rlap{
        \scalebox{.7}{
          \color{darkblue}
          \bf
          \def\arraystretch{.9}
\begin{tabular}{c}
            { \color{purple} total }
            prob.
            \\
            to measure $w'$
          \end{tabular}
        }
        }
      }
    }
  }{
    \mathrm{Prob}_{\rho}(w')
  }
  \;\;
    =
  \;\;
  \sum_w
  \underset{
    \mathclap{
      \rotatebox{-40}{
        \hspace{-25pt}
        \rlap{
        \scalebox{.7}{
          \color{darkblue}
          \bf
          \def\arraystretch{.9}
          \begin{tabular}{r}
            {  \color{purple}classical  }
            prob. that
            \\
            system is in state $\vert w \rangle$
          \end{tabular}
        }
        }
      }
    }
  }{
    p_w
  }
  \cdot
  \underset{
    \mathclap{
      \rotatebox{-40}{
        \hspace{-26pt}
        \rlap{
        \scalebox{.7}{
          \color{darkblue}
          \bf
          \def\arraystretch{.9}
          \begin{tabular}{c}
            {  \color{purple} quantum  }
            prob. to
            \\
            measure $w'$ in state $\vert w \rangle$
          \end{tabular}
        }
        }
      }
    }
  }{
    \big|\langle w' \vert w \rangle\big|^2
  }
  \;\;
  =
  \;\;
  \underset{w}{\sum}
  \,
  p_w
  \langle w'
  \vert w \rangle
  \langle w \vert w' \rangle
  \;\;
  =
  \;\;
  \langle w' \vert
  \Big(
    \underset{w}{\sum}
    \,
    p_w
    \vert w \rangle \langle w \vert
  \Big)
  \vert w' \rangle
  \;\;
  =
  \;\;
  \underset{
    \mathclap{
      \raisebox{-6pt}{
      \scalebox{.7}{
        \color{darkblue}
        \bf
        \def\arraystretch{.9}
        \begin{tabular}{c}
          { \color{olive}  trace }
          of density operator
          \\
          times observable operator
        \end{tabular}
      }
      }
    }
  }{
  \mathrm{Tr}^{\HilbertSpace{H}}
  \big(
    \rho
    \cdot
    P_{w'}
  \big)
  }
\end{equation}
\vspace{1.1cm}

In modern reformulation this means that mixed states are (represented by) {\it positive} linear operators $\HilbertSpace{H} \to \HilbertSpace{H}$ of unit trace,
often called {\it density operators} or {\it density matrices} if equivalently understood as elements of $\HilbertSpace{H} \otimes \HilbertSpace{H}^\ast$ \eqref{CompactClosureViaBraKets}:
\begin{equation}
  \label{SpaceOfMixedStates}
  \scalebox{.7}{
    \color{darkblue}
    \bf
    \def\arraystretch{.9}
    \begin{tabular}{c}
      mixed
      quantum states
    \end{tabular}
  }
  \;\;\;
  \mathrm{MxdState}(\HilbertSpace{H})
  \;\;\;\;\defneq\;\;\;\;
  \big(
    \rho
    \,\isa\,
    \HilbertSpace{H}
    \otimes
    \HilbertSpace{H}^\ast
  \big)_{{}_{\mathrm{undrl}}}
  \times
  \left(
    \def\arraystretch{1.2}
    \begin{array}{l}
      \exists_A
      \big(
        \rho \,=\, A A^\dagger
      \big)
      \\
      \mathrm{Tr}^{\HilbertSpace{H}}(\rho)
      \,=\,
      1
    \end{array}
  \right)
  \;\;
  \scalebox{.7}{
    \color{darkblue}
    \bf
    \def\arraystretch{.9}
    \begin{tabular}{c}
      density matrices
    \end{tabular}
  }
\end{equation}
This is because the {\it spectral theorem} for Hermtian operators implies that the positive unit-trace matrices $\rho$ \eqref{SpaceOfMixedStates} are precisely those which have an eigenbasis $W$ in which their diagonal form is that of \eqref{DensityMatrixInIntroduction}, with their eigenvalues forming a probability distribution

In particular, the pure states  are subsumed among the mixed states as the rank-1 projection operators
\vspace{-2mm}
\begin{equation}
  \label{PureStatesAmongMixedStates}
  \overset{
    \mathclap{
      \raisebox{12pt}{
        \scalebox{.7}{
          \bf
          \color{darkblue}
          pure state
        }
      }
    }
  }{
   \vert \psi \rangle \,\isa\, \HilbertSpace{H}
   }
   \hspace{.7cm}
   \vdash
   \hspace{.7cm}
  \overset{
    \mathclap{
      \raisebox{4pt}{
        \scalebox{.7}{
          \bf
          \color{darkblue}
          \def\arraystretch{.9}
          \begin{tabular}{c}
            regarded among
            mixed states
          \end{tabular}
        }
      }
    }
  }{
  \rho^{{}^{\vert\psi\rangle}}
  \;\;\;\;
    \defneq
  \;\;\;\;
  \frac{
    \vert \psi \rangle
    \langle \psi \vert
  }{
    \scalebox{.7}{$
    \big\vert
    \langle \psi \vert \psi \rangle
    \big\vert^{\mathrlap{2}}
    $}
  }
  }
  \;\;\;\isa\;
  \mathrm{MxdState}(\HilbertSpace{H})
  \,.
\end{equation}

\vspace{-2mm}
\noindent While further examination of this concept shows that it works beautifully and eventually provides a transparent notion of {\it non-commutative} or {\it quantum probability} in the algebraic formulation of quantum mechanics (nice review in \cite{Gleason09}\cite{Gleason11}), the curious tensor-doubling involved in passing from the pure state space $\HilbertSpace{H}$ to the density matrices inside $\HilbertSpace{H} \otimes \HilbertSpace{H}^\ast$ may seem less than obvious from first principles, especially when developing quantum physics from a formal perspective of linear logic (Lit. \ref{VerificationLiterature}). But in \cref{MixedQuantumTypes} we observe that $\HilbertSpace{H} \otimes \HilbertSpace{H}^\ast \,=\, \HilbertSpace{H} \otimes (\HilbertSpace{H} \maplin \TensorUnit)$ is naturally understood as the linear version of the costate comonad \eqref{CostateComonadEndofunctor} applied to the tensor unit, and thus in a precise logical sense as the storage of elements of the tensor unit (probability amplitudes) indexable by (pure) quantum states.

\smallskip
\noindent
{\bf Quantum channels.}
In consequence, where a coherent quantum gate or coherent quantum circuit maps directly
$$
  \begin{tikzcd}[
    column sep=40pt
  ]
  \scalebox{.7}{
    \color{darkblue}
    \bf
    pure states
  }
  \;
  \HilbertSpace{H}_1
  \ar[
    rr,
    "{
      \scalebox{.7}{
        \bf
        \color{darkgreen}
        \bf
        quantum gate
      }
    }",
    "{
      \scalebox{.7}{
        unitary map
      }
    }"{swap}
  ]
  &&
  \HilbertSpace{H}_2
  \;
  \scalebox{.7}{
    \color{darkblue}
    \bf
    pure states
  }
  \end{tikzcd}
$$
between the spaces of pure quantum states (possibly but deterministically parameterized by classical data), a combined quantum and
{\it classically probabilistic} operation on a quantum system --- such as incorporating stochastic noise due to a thermal
environment --- should instead transform the larger space of mixed states \eqref{DensityMatrixInIntroduction} or even its
ambient linear space of unconstrained matrices:
\begin{equation}
  \label{QuantumChannel}
  \begin{tikzcd}[
    column sep=50pt
  ]
  \scalebox{.7}{
    \color{darkblue}
    \bf
    mixed states
  }
  \;
    \HilbertSpace{H}_1
    \otimes
    \HilbertSpace{H}_1^\ast
    \ar[
      rr,
      "{
        \scalebox{.7}{
          \bf
          \color{darkgreen}
          quantum channel
        }
      }",
      "{
        \scalebox{.7}{
          \def\arraystretch{.9}
          \begin{tabular}{c}
            completely positive \&
            \\
            trace-preserving map
          \end{tabular}
        }
      }"{swap}
    ]
    &&
    \HilbertSpace{H}_2
    \otimes
    \HilbertSpace{H}_2^\ast
    \;
    \scalebox{.7}{
    \color{darkblue}
    \bf
    mixed states
  }
  \,.
  \end{tikzcd}
\end{equation}
but suitably preserving the subspace of density matrices, in that the linear mapping \eqref{QuantumChannel}:
\begin{itemize}
 \item[{\bf (i)}]
  preserves positivity of operators,
    in fact it should preserve positivity after coupling to any environment, hence after tensoring with any identity operator  (``complete positivity''),
 \item[{\bf (ii)}]
  preserves the trace of operators.
\end{itemize}
Under these conditions the linear maps \eqref{QuantumChannel}  are known as {\it quantum operations} \cite[\S 10]{BengtssonZyczkowski06}\cite[\S 8.2]{NielsenChuang10} or {\it quantum channels}\footnote{
Since under compact closure \eqref{CompactClosureViaBraKets} the quantum channels \eqref{QuantumChannel} are equivalently understood as linear
operations on spaces of linear operators $(\HilbertSpace{H} \maplin \HilbertSpace{H}) \to (\HilbertSpace{K} \maplin \HilbertSpace{K})$ some
authors refer to them as ``superoperators'' (in the sense of ``second order operators''), e.g. \cite[\S 6.3]{Selinger04}. But besides being
ambiguous in itself this term is used with differing conventions by differing authors and might hence better be avoided. } \cite[\S 4]{HeinosaariZiman11}, expressing the
intuition that they reflect the most general physically viable operation on a quantum system, such as when sending its states through
a physical communication channel \cite{Wilde13}\cite[\S 3.2]{KathriWilde20}.

Since the above two properties may be understood as characterizing the preservation of ``quantum probability distributions''; quantum channels
may be thought of as the {\it stochastic maps} in the context of quantum probability theory. If the mapping \eqref{QuantumChannel} in addition
\begin{itemize}
  \item [{\bf (iii)}] preserves the identity operator
\end{itemize}
then one speaks of a {\it unital quantum channel}, these being the {\it doubly stochastic maps} in quantum probability.

\smallskip

\noindent
{\bf The fundamental examples of quantum channels} are:

\vspace{.1cm}
\begin{itemize}
\item {\bf Unitary quantum channels} (e.g. \cite[Ex. 4.6]{HeinosaariZiman11}) corresponding to unitary quantum gates $U \,\isa\ \HilbertSpace{H}_1 \to \HilbertSpace{H}_2$ on pure states and given by conjugation of density matrices with that unitary operator:
\vspace{-2mm}
\begin{equation}
  \label{UnitaryQuantumChannel}
  \mathllap{
    \scalebox{.7}{
      \color{darkblue}
      \bf
      \begin{tabular}{c}
        unitary quantum gate
        \\
        as a quantum channel
      \end{tabular}
    }
    \hspace{1.4cm}
  }
  \begin{tikzcd}[sep=0pt]
    \mathllap{
      \mathrm{chan}^U
      \;\isa\;
    }
    \HilbertSpace{H}_1
    \otimes
    \HilbertSpace{H}_1^\ast
    \ar[
      rr
    ]
    &&
    \HilbertSpace{H}_2
    \otimes
    \HilbertSpace{H}_2^\ast
    \\[-2pt]
    \scalebox{\termsize}{$
      \rho
    $}
    &\scalebox{\termsize}{$\longmapsto$}&
    \scalebox{\termsize}{$
      U
        \cdot
      \rho
        \cdot
      U^\dagger
    $}
    \mathrlap{\,.}
  \end{tikzcd}
\end{equation}
\vspace{-2mm}

\noindent This is such that on pure states $\rho^{\vert\psi\rangle}$ among mixed states \eqref{PureStatesAmongMixedStates}
the unitary quantum channel acts just as the corresponding quantum gate, in that:
\vspace{-.2cm}
$$
  \mathrm{chan}^U
  \;\;:\;\;
  \rho^{ \vert \psi \rangle }
  \;\;\;
    \mapsto
  \;\;\;
  U
  \cdot
  \rho^{{\vert\psi\rangle}}
  \cdot U^\dagger
  \;\;
  =
  \;\;
  U
  \cdot
  \frac{
    \vert \psi \rangle
    \langle \psi \vert
  }{
    \big\vert
      \langle \psi \vert \psi \rangle
    \big\vert^{\mathrlap{2}}
  }
  \cdot
  U^\dagger
  \;\;
    =
  \;\;
  \frac{
    U
    \vert \psi \rangle
    \langle \psi \vert
    U^\dagger
  }{
    \big\vert
      \langle \psi \vert
      U^\dagger U
      \vert \psi \rangle
    \big\vert^{\mathrlap{2}}
  }
  \;\;
  =
  \;\;
  \rho^{{U \vert\psi\rangle}}
  \,.
$$

\item
{\bf Mixed unitary quantum channels} are probabilistic ensembles of unitary channels \eqref{UnitaryQuantumChannel} in that they are given by $S$-tuples $(U_s \isa \HilbertSpace{H}_1 \to \HilbertSpace{H}_2)_{i \isa S}$ of unitary operators indexed over an inhabited finite index-set $S$, and by a probability distribution $p_{(\mbox{-})} \isa S \to [0,1]$, as
\vspace{-2mm}
\begin{equation}
  \label{MixedUnitaryQuantumChannel}
  \hspace{.5cm}
  \mathllap{
    \scalebox{.7}{
      \color{darkblue}
      \bf
      \begin{tabular}{c}
        classical mixture of
        \\
        unitary quantum gates
        \\
        as a quantum channel
      \end{tabular}
    }
    \hspace{2.2cm}
  }
  \begin{tikzcd}[sep=0pt]
    \mathllap{
      \mathrm{chan}^{(U_\bullet, p)}
      \;\isa\;
    }
    \HilbertSpace{H}_1
    \otimes
    \HilbertSpace{H}_1^\ast
    \ar[
      rr
    ]
    &&
    \HilbertSpace{H}_2
    \otimes
    \HilbertSpace{H}_2^\ast
    \\[-2pt]
    \scalebox{\termsize}{$
      \rho
    $}
    &\scalebox{\termsize}{$\longmapsto$}&
    \scalebox{\termsize}{$
      \!\!
      \underset{s \isan S}{\sum}
      \;
      p_s
      \,
      U_s \cdot \rho \cdot U_s^\dagger
    $}
    \mathrlap{\,.}
  \end{tikzcd}
\end{equation}
For example, the
{\bf bit-flip quantum channel} is the mixed unitary channel \eqref{MixedUnitaryQuantumChannel} on single qbit states

\noindent $\QBit \,\defneq\, \underset{\{0,1\}}{\oplus} \ComplexNumbers$ \eqref{TypeOfQBits} given for $p \in [0,1]$ by (e.g. \cite[\S 8.1 \& 8.3.3]{NielsenChuang10}):
\begin{equation}
  \label{BitFlipQuantumChannel}
  \hspace{.5cm}
  \mathllap{
    \scalebox{.7}{
      \color{darkblue}
      \bf
      \begin{tabular}{c}
        qbit-flip
        \\
        quantum channel
      \end{tabular}
    }
    \hspace{2.2cm}
  }
  \begin{tikzcd}[sep=0pt]
    \mathllap{
      \mathrm{flip}_p
      \;\isa\;
    }
    \QBit
    \otimes
    \QBit^\ast
    \ar[
      rr
    ]
    &&
    \QBit
    \otimes
    \QBit^\ast
    \\[-2pt]
    \scalebox{\termsize}{$
      \rho
    $}
    &\scalebox{\termsize}{$\longmapsto$}&
    \scalebox{\termsize}{$
      (1-p)
      \,
      \rho
      \,+\,
      p
      \,
      X
        \cdot
        \rho
        \cdot
      X
    $}
    \mathrlap{\,,}
  \end{tikzcd}
\end{equation}
where $
  X
  \,\defneq\,
  \vert 0 \rangle \langle 1 \vert
  +
  \vert 1 \rangle \langle 0 \vert
$
is the ``Pauli X'' quantum gate (or {\it quantum NOT gate}) which swaps (flips) the two canonical qbit-basis elements.

Hence the bit-flip quantum channel \eqref{BitFlipQuantumChannel} models a process where a qbit the flipped with probability $p$ and retained as is with probability $(1-p)$. This is a simple model for the effect of {\it quantum noise}.

\item
{\bf Measurement quantum channels} with respect to an orthonormal linear basis $\HilbertSpace{H} \,\simeq\, \osum_W \ComplexNumbers$ \eqref{AnOrthonormalBasis}, given by
\vspace{-2mm}
\begin{equation}
  \label{MeasurementQuantumChannel}
  \mathllap{
  \scalebox{.7}{
    \color{darkblue}
    \bf
    \def\arraystretch{.9}
    \begin{tabular}{c}
      measurement statistics
      \\
      as a quantum channel
    \end{tabular}
  }
  \hspace{1.4cm}
  }
  \begin{tikzcd}[sep=0pt]
    \mathllap{
       \mathrm{chan}^W
       \;\isa\;\;
    }
    \HilbertSpace{H}
    \otimes
    \HilbertSpace{H}^\ast
    \ar[
      rr
    ]
    &&
    \HilbertSpace{H}
    \otimes
    \HilbertSpace{H}^\ast
    \\
    \scalebox{\termsize}{$
      \rho
    $}
      &\longmapsto&
      \scalebox{\termsize}{$
      \sum_w
      P_w \,\rho\, P_w
      $}
  \end{tikzcd}
\end{equation}

\vspace{-2mm}
\noindent
(where $P_w \defneq \vert w\rangle \langle w \vert$).
This description \eqref{MeasurementQuantumChannel} of quantum measurement is originally due to \cite[(8)]{Lueders51} and has become standard quantum physics lore (a nice discussion is in \cite{Wheeler12}): Notice that the density matrix on the right of \eqref{MeasurementQuantumChannel} expresses a {\it classical uncertainty regarding which measurement result was obtained} and instead provides the probabilistic mixture of collapsed quantum states for all possible measurement outcomes, weighted according to the Born rule \eqref{BornRule}:
\vspace{-.3cm}
$$
  \def\arraystretch{1.7}
  \begin{array}{rl}
  \vert \psi \rangle
  \,\isa\,
  S\big(
    \underset{
      \mathrlap{
        \!\!\!\!
        w \isa W
      }
    }{\oplus}
    \ComplexNumbers
  \big)
  \hspace{.6cm}
  \yields
  \hspace{.6cm}
  \mathrm{chan}^W
  \;\;\colon\;\;
  \rho^{ \vert \psi \rangle }
  &
  \;\mapsto\;
  \overset{
    \mathclap{
      \raisebox{9pt}{
        \scalebox{.7}{
          \color{darkblue}
          \bf
          \def\arraystretch{.9}
          \begin{tabular}{c}
            quantum measurement channel
            \\
            on a pure quantum state...
          \end{tabular}
        }
      }
    }
  }{
  \underset{w}{\sum}
  \,
  P_w
    \cdot
  \rho^{\vert \psi \rangle}
    \cdot
  P_w
  }
  \\
  &
  \;=\;
  \underset{w}{\sum}
  \,
  \vert w \rangle
  \langle w \vert
  \psi \rangle
  \langle \psi
  \vert w \rangle \langle w \vert
  \\
  &
  \;=\;
  \underset{
    \mathclap{
      \raisebox{-6pt}{
        \scalebox{.7}{
          \color{darkblue}
          \bf
          \def\arraystretch{.9}
          \begin{tabular}{c}
            ...produces the mixture of
            \\
            all possible measurement outcomes
            \\
            weighted by their Born probability
          \end{tabular}
        }
      }
    }
  }{
  \underset{w}{\sum}
  \,
  \vert w \rangle
  \,
  \mathrm{Prob}_\psi(w)
  \,
  \langle w \vert
  \mathrlap{\,.}
  }
  \end{array}
$$

\noindent
Incidentally, \eqref{MeasurementQuantumChannel} is not the only sensible modeling of quantum measurement \eqref{LinearStateCollapse}
on mixed states: If we do know and  record which specific $w \isa W$ has been measured, then the typing should rather be:
\vspace{-2mm}
\begin{equation}
  \label{DynamicallyLiftedQuantumMeasurementOnDensityMatrix}
  \mathllap{
  \scalebox{.7}{
    \color{darkblue}
    \bf
    \def\arraystretch{.9}
    \begin{tabular}{c}
      measurement of mixed states
      \\
      with dynamic lifting of results
    \end{tabular}
  }
  }
  \begin{tikzcd}[
    column sep=0pt,
    row sep=-3pt
  ]
    \HilbertSpace{H}
    \otimes
    \HilbertSpace{H}^\ast
    \ar[rr]
    &&
    \big(
      W \to \ComplexNumbers
    \big)
    \\
    \scalebox{\termsize}{$
      \rho
    $}
    &\longmapsto&
    \scalebox{\termsize}{$
    \big(
      w
      \,\mapsto\,
      P_w \cdot \rho \cdot P_w
    \big)
    $}
  \end{tikzcd}
\end{equation}

\vspace{-2mm}
\noindent
This was in fact L{\"u}ders' first proposal: \cite[(7)]{Lueders51}! In a quantum protocol, this description \eqref{DynamicallyLiftedQuantumMeasurementOnDensityMatrix}  of the measurement process retains the probabilities of the measurement outcomes but ``dynamically lifts'' \eqref{ClassicalControlAndDynamicLifting} the actual outcome to a new classical parameter (Lit. \ref{ClassicalControlAndDynamicLifting}).
We naturally recover this description \eqref{DynamicallyLiftedQuantumMeasurementOnDensityMatrix} as a monoidal-monad operation, below in \eqref{DecoherenceFromMonoidalMonadStructure}.

Later it was noticed \cite{JoosZeh85} that \eqref{MeasurementQuantumChannel} may be understood as arising from the {\bf decoherence} of the quantum state upon its coupling to an environment (here: the measurement apparatus), by which {\it the off-diagonal elements of the density matrix vanish} in the measurement basis (\cite[(3.57)]{JoosZeh85}, cf. \cite[p.277]{Omnes94}\cite[p. 95]{Schlosshauer07}\cite[(7)]{Schlosshauer19}):
\begin{equation}
  \label{MeasurementAsDecoherence}
  \def\arrayxcolsep{-1pt}
  \def\arraystretch{1.3}
\begin{array}{rcl}
  \rho^{\vert\psi\rangle}
  \;
    \defneq
  \;
  \overset{
    \mathclap{
      \raisebox{8pt}{
        \scalebox{.7}{
          \color{darkblue}
          \bf
          pure state
        }
      }
    }
  }{
    \vert \psi \rangle
    \langle \psi \vert
  }
  \;
  =
  \;
  \Big(
    \underset{w}{\sum}
    \vert w \rangle
    \langle w \vert \psi \rangle
  \Big)
  \Big(
    \underset{w'}{\sum}
    \langle \psi \vert w' \rangle
    \langle w' \vert
  \Big)
  &
  =
  &
  \underset{w,w'}{\sum}
  \vert w\rangle\!\overbrace{
  \langle w \vert \psi \rangle
  \langle \psi \vert w' \rangle
  }^{
    \mathclap{
      \scalebox{.7}{
        \color{darkblue}
        \bf
        coherent phases
      }
    }
  }
  \langle w' \vert
  \\
  &
  \overset{
    \adjustbox{
      raise=4pt,
      scale=.7
    }{$
    \mathrm{chan}^W
    $}
  }{
    \underset{
      \mathclap{
        \raisebox{-9pt}{
          \scalebox{.7}{
            \color{darkgreen}
            \bf
            \def\arraystretch{.9}
            \begin{tabular}{c}
              measurement
              \\
              channel
            \end{tabular}
          }
        }
      }
    }{
    \longmapsto
    }
  }
  &
  \;
  \,
  \underset{
   \mathclap{
     \raisebox{.4pt}{
       \scalebox{.7}{
         \color{darkblue}
         \bf
         decohered mixed state
       }
     }
   }
  }{
  \underset{w}{\sum}
  \;\,
  \vert w\rangle
  \langle w \vert \psi \rangle
  \langle \psi \vert w \rangle
  \langle w \vert
  }
  \;=\;
  \underset{w}{\sum}
  \,
  p_w \cdot
  \vert w \rangle
  \langle w \vert
  \mathrlap{\,.}
  \end{array}
\end{equation}

\item {\bf Averaging quantum channels} operate on a compound quantum systems $\HilbertSpace{H} \otimes \HilbertSpace{B}$  --- the system $\HilbertSpace{H}$ of primary interest coupled to an {\it environment} or {\it thermal bath} $\HilbertSpace{B}$ --- by retaining of the environment only the {\it expectation value} of its effects on the system, which means to form the partial trace of density matrices over $\HilbertSpace{B}$:
\vspace{-2mm}
\begin{equation}
  \label{AveraginQuantumChannel}
  {
    \scalebox{.7}{
      \color{darkblue}
      \bf
      \begin{tabular}{c}
        averaging over a subsystem
        \\
        as a quantum channel
      \end{tabular}
    }
  }
  \begin{tikzcd}[
    row sep=0pt,
    column sep=8pt
  ]
    \mathrm{chan}^{\HilbertSpace{B}}
    \,\isa\,
    (\HilbertSpace{H}
    \otimes
    \HilbertSpace{B})
    \otimes
    (\HilbertSpace{H}
    \otimes
    \HilbertSpace{B})^\ast
    \ar[
      r,
      "\sim"
    ]
    &
    \HilbertSpace{H}
    \otimes
    \HilbertSpace{B}
    \otimes
    \HilbertSpace{B}^\ast
    \otimes
    \HilbertSpace{H}^\ast
    \ar[
      rr
    ]
    &[-10pt]&[-10pt]
    \HilbertSpace{H}
    \otimes
    \HilbertSpace{H}^\ast
    \\
    &
  \scalebox{\termsize}{$  \vert \psi, \beta \rangle
    \langle \beta', \psi' \vert
    $}
    &\longmapsto&
 \scalebox{\termsize}{$   \vert \psi \rangle
    \langle \beta' \vert \beta \rangle
    \langle \psi' \vert
    $}
    \,.
  \end{tikzcd}
\end{equation}

\vspace{-2mm}
\noindent An elementary but profound insight into the structure of quantum physics ---
often referred to under the term {\bf decoherence} -- is the observation that quantum measurement channels \eqref{MeasurementQuantumChannel} may be understood as nothing but the composite of a unitary evolution \eqref{UnitaryQuantumChannel} of the system $\HilbertSpace{H}$ {\it coupled}  to its environment $\mathscr{B}$ by way of a deterministic measuring process, but then followed by an averaging \eqref{AveraginQuantumChannel} over the exact state of the measurement device:

Concretely, if $\vert b_{\mathrm{ini}} \rangle \,\isa\, \HilbertSpace{B}$ denotes the initial state of a ``device'' then any notion of this device measuring the system $\HilbertSpace{H}$ (in its measurement basis $W$) under their joint unitary quantum evolution  should be reflected in a unitary operator under which the system $\HilbertSpace{H}$ remains invariant {\it if} it is purely in any eigenstate $\vert w \rangle$ of the measurement basis, while in this case the measuring system evolves to a corresponding ``pointer state'' $\vert b_w \rangle$ \cite[(1.1)]{Zurek81}\cite[(1.1)]{JoosZeh85} (following \cite[\S VI.3]{vonNeumann32}, review includes \cite[(2.52)]{Schlosshauer07}):
\vspace{-2mm}
\begin{equation}
  \label{UnitaryMeasurementProcess}
  \begin{tikzcd}[row sep = -2pt, column sep=large]
    \mathllap{
      U_W
      \;\isa\;\;
    }
    \HilbertSpace{H}
    \otimes
    \HilbertSpace{B}
    \ar[
      rr,
      "{
        \scalebox{.7}{
          \color{darkgreen}
          \bf
          \def\arraystretch{.9}
          \begin{tabular}{c}
            unitary
            \\
            measurement process
          \end{tabular}
        }
      }"{yshift=0pt}
    ]
    &&
    \HilbertSpace{H}
    \otimes
    \HilbertSpace{B}
    \\
 \scalebox{\termsize}{$   \vert w, b_{\mathrm{ini}} \rangle
 $}
    &\longmapsto&
  \scalebox{\termsize}{$  \vert w, b_{w} \rangle
  $}
  \end{tikzcd}
\end{equation}

\vspace{-2mm}
\noindent
for $b_{\mathrm{ini}}$ and $b_w$ distinct elements of an (in practice: approximately-)orthonormal basis for $\HilbertSpace{B}$.
(There is always a unitary operator with this mapping property \eqref{UnitaryMeasurementProcess}, for instance the one which
moreover maps
$\vert w, b_{w}\rangle \mapsto \vert w, b_{\mathrm{ini}}\rangle$
and is the identity on all remaining basis elements.)
But then the composition of the corresponding unitary quantum channel with the averaging channel over $\HilbertSpace{B}$ is
indeed equal to the $W$-measurement channel (cf. e.g. \cite[(2.117)]{Schlosshauer07}, going back to \cite[(7)]{Zeh70}):
\vspace{-3mm}
\begin{equation}
  \label{EnvironmentRepresentationOfMeasurementChannel}
  \hspace{-4cm}
  \adjustbox{raise=-2.5cm}{
  \begin{tikzcd}[row sep=-2pt, column sep=8pt]
    \HilbertSpace{H}
    \otimes
    \HilbertSpace{H}^\ast
    \ar[
      rrrrrr,
      rounded corners,
      to path={
            ([yshift=+00pt]\tikztostart.north)
         -- ([yshift=+30pt]\tikztostart.north)
         -- node{
           \colorbox{white}{
             \scalebox{.7}{$
               \mathrm{chan}^W
             $}
           }
           }
           node[yshift=12pt]{
             \scalebox{.7}{
               \color{darkgreen}
               \bf
               \begin{tabular}{c}
                 quantum measurement channel
               \end{tabular}
             }
           }
            ([yshift=+30pt]\tikztotarget.north)
         -- ([yshift=+00pt]\tikztotarget.north)
      }
    ]
    \ar[
      rr,
      "{
        \mathrm{chan}^{b_{\mathrm{ini}}}
      }",
      "{
        \scalebox{.7}{
          \color{darkgreen}
          \bf
          \def\arraystretch{.9}
          \begin{tabular}{c}
            couple system to
            \\
            {\color{purple}initialized}
            meas. device
          \end{tabular}
        }
      }"{yshift=10pt}
    ]
    &&
    \HilbertSpace{H}
    \otimes
    \HilbertSpace{B}
    \otimes
    \HilbertSpace{B}^\ast
    \otimes
    \HilbertSpace{H}^\ast
    \ar[
      rr,
      "{
        \mathrm{chan}^{U_W}
      }",
      "{
        \scalebox{.7}{
          \color{darkgreen}
          \bf
          \def\arraystretch{.9}
          \begin{tabular}{c}
            evolve system \& device
            \\
            under meas. interaction
          \end{tabular}
        }
      }"{yshift=10pt}
    ]
    &&
    \HilbertSpace{H}
    \otimes
    \HilbertSpace{B}
    \otimes
    \HilbertSpace{B}^\ast
    \otimes
    \HilbertSpace{H}^\ast
    \ar[
      rr,
      "{
        \mathrm{chan}^{\HilbertSpace{B}}
      }",
      "{
        \scalebox{.7}{
          \color{darkgreen}
          \bf
          \def\arraystretch{.9}
          \begin{tabular}{c}
            average over states
            \\
            of measuring device
          \end{tabular}
        }
      }"{yshift=10pt}
    ]
    &&
    \HilbertSpace{H}
    \otimes
    \HilbertSpace{H}^\ast
    \\
    \phantom{AAA}
    \scalebox{\termsize}{$
    \underset{
      \rho
    }{
    \underbrace{
    \underset{w,w'}{\sum}
      \rho_{w,w'}
      \vert w \rangle
      \langle w' \vert
    }}
    $}
    &\longmapsto&
  \scalebox{\termsize}{$    \underset{w,w'}{\sum}
      \rho_{w,w'}
      \vert w , b_{\mathrm{ini}} \rangle
      \langle b_{\mathrm{ini}}, w' \vert
      $}
    &\longmapsto&
  \scalebox{\termsize}{$    \underset{w,w'}{\sum}
      \rho_{w,w'}
      \vert w , b_w \rangle
      \langle b_{w'} , w' \vert\
      $}
    & \hspace{-5mm} \longmapsto&[-5pt]
    \hspace{-5mm}
   \scalebox{\termsize}{$   \underset{w,w'}{\sum}
      \rho_{w,w'}
      \vert w \rangle
      \langle b_{w'} \vert b_w \rangle
      \langle w' \vert
      $}
    \\[-18pt]
    && && &=&
    \scalebox{\termsize}{$  \underset{w,w'}{\sum}
    \,
      \rho_{w,w'}
      \vert w \rangle
      \delta_w^{w'}
      \langle w' \vert
      $}
    \\
    && && &=&
  \scalebox{\termsize}{$    \underset{
      \underset{w}{\sum}
      \,
        P_w \cdot \rho \cdot P_w
    }{
    \underbrace{
    \underset{w}{\sum}
    \,
      \rho_{w, w}
      \vert w \rangle
      \langle w \vert
    }
    }
    $}
  \end{tikzcd}
  }
  \hspace{-2cm}
\end{equation}

\vspace{-2mm}
\item {\bf Coupling channels} (rarely made explicit as such, but conceptually important to notice) which for any mixed
state $\rho_{\mathrm{env}}$ of a given system $\HilbertSpace{B}$ form the tensor product state:
\vspace{-2mm}
\begin{equation}
  \label{CouplingQuantumChannel}
  \hspace{-2cm}
  \scalebox{.7}{
    \color{darkblue}
    \bf
    \begin{tabular}{c}
      coupling to ancillary system
      \\
      as a quantum channel
    \end{tabular}
  }
  \hspace{1.7cm}
  \begin{tikzcd}[sep=0pt]
    \mathllap{
      \mathrm{chan}^{
        \rho_{\mathrm{env}}
      }
      \;\isa\;\;
    }
    \HilbertSpace{H}
    \otimes
    \HilbertSpace{H}^\ast
    \ar[
      rr
    ]
    &&
    \big(
      \HilbertSpace{H}
      \otimes
      \HilbertSpace{B}
    \big)
    \otimes
    \big(
      \HilbertSpace{B}
      \otimes
      \HilbertSpace{H}
    \big)^\ast
    \\
    \scalebox{\termsize}{$   \rho $}
    &\longmapsto&
   \scalebox{\termsize}{$    \rho \otimes \rho_{\mathrm{env}}
   $}
  \end{tikzcd}
\end{equation}
\end{itemize}

\medskip

\noindent
{\bf Operator-sum decomposition of quantum channels.}
The fundamental theorem of quantum channel theory characterizes them
(\cite{Choi75}, review in \cite[Thm. 8.1]{NielsenChuang10}\cite[Thm. 4.4.1]{Wilde13})
as exactly those linear maps of the form
\vspace{-2mm}
\begin{equation}
  \label{OperatorSumDecompositionOfQuantumChannel}
  \begin{tikzcd}[sep=0pt]
    \HilbertSpace{H}_1
    \otimes
    \HilbertSpace{H}_1^\ast
    \ar[
      rr
    ]
    &&
    \HilbertSpace{H}_2
    \otimes
    \HilbertSpace{H}_2^\ast
    \\[-2pt]
    \scalebox{\termsize}{$
      \rho
    $}
    &\longmapsto&
    \scalebox{\termsize}{$
    \underset{r}{\sum}
    \,
    E_r
    \cdot
    \rho
    \cdot
    E_r^\dagger
    $}
  \end{tikzcd}
\end{equation}

\vspace{-2mm}
\noindent for non-empty tuples of linear operators
\vspace{-2mm}
$$
  R \,\isa\, \FiniteSets
  ,\;\;
  r \,\isan\, R
  \;\;\;\;
  \vdash
  \;\;\;\;
  E_r \,\isa\, \HilbertSpace{H}_1 \to \HilbertSpace{H}_2
  \hspace{1cm}
  \mbox{s.t.}
  \;\;
  \left\{
  \def\arraystretch{1.5}
  \begin{array}{ll}
     \sum_r
     \,
     E_r^\dagger \cdot E_r
     \,=\,
     \mathrm{id}
     &
     \mbox{
       (preservation of trace)
     }
     \\
     \sum_r
     \,
     E_r
     \cdot
     E_r^\dagger
     \,=\,
     \mathrm{id}
     &
     \mbox{
       (for unital channels)
     }
     \mathrlap{\,.}
  \end{array}
  \right.
$$
This looks like a purely technical lemma, but it has profound conceptual consequences, such as the following:

\smallskip

\noindent
{\bf Environmental representation of quantum channels.}
Remarkably, quantum endo-channels

\noindent $\mathrm{chan} \,\colon\, \mathscr{H} \otimes \mathscr{H}^\ast \to \mathscr{H} \otimes \mathscr{H}^\ast$
may alternatively be characterized as those linear maps which arise -- in generalization of the situation for measurement channels \eqref{EnvironmentRepresentationOfMeasurementChannel} -- under the 3-step procedure of:

\begin{itemize}
\item[{\bf (i)}] {\it coupling} the system $\rho$ to an environment system $\mathscr{B}$ in some state
$\rho_{\mathrm{env}}$ \eqref{CouplingQuantumChannel},

\item[{\bf (ii)}] {\it evolving} the compound system $\rho \otimes \rho_{\mathrm{env}}$ through a unitary quantum channel $\mathrm{chan}^U$ \eqref{UnitaryQuantumChannel}

\item[{\bf (iii)}] {\it averaging} the result over the environmental states \eqref{AveraginQuantumChannel}:
\end{itemize}
\begin{equation}
  \label{EnvironmentalRepresentation}
  \hspace{-4cm}
  \adjustbox{raise=-1cm}{
  \begin{tikzcd}[
    column sep=14pt,
    row sep=-2pt
  ]
    \HilbertSpace{H}
    \otimes
    \HilbertSpace{H}^\ast
    \ar[
      rrrrrr,
      rounded corners,
      to path={
            ([yshift=+00pt]\tikztostart.north)
         -- ([yshift=+30pt]\tikztostart.north)
         -- node{
           \colorbox{white}{
             \scalebox{.7}{$
               \mathrm{chan}
             $}
           }
           }
           node[yshift=12pt]{
             \scalebox{.7}{
               \color{darkgreen}
               \bf
               \begin{tabular}{c}
                 any quantum channel
               \end{tabular}
             }
           }
            ([yshift=+30pt]\tikztotarget.north)
         -- ([yshift=+00pt]\tikztotarget.north)
      }
    ]
    \ar[
      rr,
      "{
        \mathrm{chan}^{\rho_{\mathrm{env}}}
      }",
      "{
        \scalebox{.7}{
          \color{darkgreen}
          \bf
          \def\arraystretch{.9}
          \begin{tabular}{c}
            couple system to
            \\
            {\color{purple} initialized}
            meas. device
          \end{tabular}
        }
      }"{yshift=10pt}
    ]
    &&
    \big(
    \HilbertSpace{H}
    \otimes
    \HilbertSpace{B}
    \big)
    \otimes
    \big(
    \HilbertSpace{B}
    \otimes
    \HilbertSpace{H}
    \big)^\ast
    \ar[
      rr,
      "{
        \mathrm{chan}^{U_W}
      }",
      "{
        \scalebox{.7}{
          \color{darkgreen}
          \bf
          \def\arraystretch{.9}
          \begin{tabular}{c}
            evolve system \& environment
            \\
            under some interaction
          \end{tabular}
        }
      }"{yshift=10pt}
    ]
    &&
    \big(
    \HilbertSpace{H}
    \otimes
    \HilbertSpace{B}
    \big)
    \otimes
    \big(
    \HilbertSpace{B}
    \otimes
    \HilbertSpace{H}
    \big)^\ast
    \ar[
      rr,
      "{
        \mathrm{chan}^{\HilbertSpace{B}}
      }",
      "{
        \scalebox{.7}{
          \color{darkgreen}
          \bf
          \def\arraystretch{.9}
          \begin{tabular}{c}
            average over states
            \\
            of environment
          \end{tabular}
        }
      }"{yshift=10pt}
    ]
    &&
    \HilbertSpace{H}
    \otimes
    \HilbertSpace{H}^\ast
    \\
    \scalebox{\termsize}{$   \rho
    $}
    &\longmapsto&
    \scalebox{\termsize}{$
      \rho \otimes \rho_{\mathrm{env}}
    $}
    &\longmapsto&
    \scalebox{\termsize}{$
    U_{\mathrm{tot}}
    \big(
      \rho
        \otimes
      \rho_{\mathrm{env}}
    \big)
      U_{\mathrm{tot}}^\dagger
    $}
    &\longmapsto&
  \scalebox{\termsize}{$     \mathrm{Tr}^{\mathscr{B}}
    \big(
      U_{\mathrm{tot}}
      (
        \rho
        \otimes
        \rho_{\mathrm{env}}
      )
      U_{\mathrm{tot}}^\dagger
    \big)
    $}
  \end{tikzcd}
  }
  \hspace{-4cm}
\end{equation}

That all such averaged environment-interactions are quantum channels is immediate from the three component steps being quantum channels. That every quantum channel has an environmental representation (originally remarked by \cite[inside Lem. 5]{Lindblad75}) follows by choosing an operator-sum decomposition \eqref{OperatorSumDecompositionOfQuantumChannel}: Then taking $\mathscr{B} \,\equiv\, \oplus_r \ComplexNumbers$, singling out one of its basis vectors $\vert r_{\mathrm{ini}} \rangle$ as the pure environmental state
\vspace{-1mm}
\begin{equation}
  \label{PureEnvironmentState}
  \rho_{\mathrm{env}}
    \,\defneq\,
  \vert r_{\mathrm{init}} \rangle \langle r_{\mathrm{ini}} \vert
  \,,
\end{equation}

\vspace{-1mm}
\noindent and finally observing that any unitary operator of the form
\vspace{-2mm}
$$
  \begin{tikzcd}[sep=0pt]
    \mathllap{
      U
      \;\isa\;\;
    }
    \HilbertSpace{H}
    \otimes
    \HilbertSpace{B}
    \ar[
      rr
    ]
    &&
    \HilbertSpace{H}
    \otimes
    \HilbertSpace{B}
    \\
    \scalebox{\termsize}{$   \vert \psi \rangle
    \otimes
    \vert r_{\mathrm{ini}} \rangle
    $}
    &\longmapsto&
   \scalebox{\termsize}{$    \underset{r}{\sum}
    \,
    E_r \vert \psi \rangle
    \otimes
    \vert r \rangle
    $}
  \end{tikzcd}
$$

\vspace{-2mm}
\noindent serves the purpose (e.g. \cite[p. 365]{NielsenChuang10}\cite[Thm. 6.7]{AttalLectures}\cite[\S 10.4]{BengtssonZyczkowski06}).

(The ontological import of this theorem is profound: It is consistent to assume that the world at large fundamentally evolves according to deterministic unitary evolution of pure quantum states, while all apparent classical stochasticity in the evolution of small subsystems results entirely from ignorance about the exact microstate of their quantum environment.)

\medskip

\noindent
{\bf Noisy/unistochastic/DQC quantum channels.}
While every quantum channel is environmentally realized \eqref{EnvironmentalRepresentation} as a bath-average of a unitary evolution of the given system coupled to a {\it pure} state of the environment \eqref{PureEnvironmentState}, some quantum channels are realized even by coupling to mixed environmental states.

In the extreme but (practically highly) relevant case where the coupling is to an  environment in its maximally mixed (namely uniformly distributed) quantum state \eqref{UniformlyDistributedBathState} some authors speak of {\it noisy quantum operations} \cite{HorodeckiOppenheim03}\cite{MHP19}
others of {\it unistochastic quantum channels} \cite[p. 259]{ZyczkowskiBengtsson04}\cite{BengtssonZyczkowski06}\cite{MuszKusZyczkowski13}:
\begin{equation}
  \label{UnistochasticEnvironmentalRepresentation}
  \hspace{-4cm}
  \adjustbox{raise=-1cm}{
  \begin{tikzcd}[
    column sep=11pt,
    row sep=-2pt
  ]
    \HilbertSpace{H}
    \otimes
    \HilbertSpace{H}^\ast
    \ar[
      rrrrrr,
      rounded corners,
      to path={
            ([yshift=+00pt]\tikztostart.north)
         -- ([yshift=+30pt]\tikztostart.north)
         -- node{
           \colorbox{white}{
             \scalebox{.7}{$
               \mathrm{chan}
             $}
           }
           }
           node[yshift=12pt]{
             \scalebox{.7}{
               \color{darkgreen}
               \bf
               \begin{tabular}{c}
                 {\color{purple}unistochastic}
                 quantum channel
               \end{tabular}
             }
           }
            ([yshift=+30pt]\tikztotarget.north)
         -- ([yshift=+00pt]\tikztotarget.north)
      }
    ]
    \ar[
      rr,
      "{
        \mathrm{chan}^{\mathrm{unif}}
      }",
      "{
        \scalebox{.7}{
          \color{darkgreen}
          \bf
          \def\arraystretch{.9}
          \begin{tabular}{c}
            couple system to
            \\
            {\color{purple}
            maximally mixed}
            bath
          \end{tabular}
        }
      }"{yshift=10pt}
    ]
    &&
    \big(
    \HilbertSpace{H}
    \otimes
    \HilbertSpace{B}
    \big)
    \otimes
    \big(
    \HilbertSpace{B}
    \otimes
    \HilbertSpace{H}
    \big)^\ast
    \ar[
      rr,
      "{
        \mathrm{chan}^{U_W}
      }",
      "{
        \scalebox{.7}{
          \color{darkgreen}
          \bf
          \def\arraystretch{.9}
          \begin{tabular}{c}
            evolve system \& environment
            \\
            under some interaction
          \end{tabular}
        }
      }"{yshift=10pt}
    ]
    &&
    \big(
    \HilbertSpace{H}
    \otimes
    \HilbertSpace{B}
    \big)
    \otimes
    \big(
    \HilbertSpace{B}
    \otimes
    \HilbertSpace{H}
    \big)^\ast
    \ar[
      rr,
      "{
        \mathrm{chan}^{\HilbertSpace{B}}
      }",
      "{
        \scalebox{.7}{
          \color{darkgreen}
          \bf
          \def\arraystretch{.9}
          \begin{tabular}{c}
            average over states
            \\
            of environment
          \end{tabular}
        }
      }"{yshift=10pt}
    ]
    &&
    \HilbertSpace{H}
    \otimes
    \HilbertSpace{H}^\ast
    \\
    \scalebox{\termsize}{$
      \rho_{\mathrm{sys}}
    $}
    &\longmapsto&
    \scalebox{\termsize}{$
      \rho_{\mathrm{sys}}
        \otimes
      \rho^{\mathrm{unif}}_{\HilbertSpace{B}}
    $}
    &\longmapsto&
    \scalebox{\termsize}{$
    U_{\mathrm{tot}}
    \big(
      \rho_{\mathrm{sys}}
        \otimes
      \rho^{\mathrm{unif}}_{\HilbertSpace{B}}
    \big)
    U_{\mathrm{tot}}^\dagger
    $}
    &\longmapsto&
  \scalebox{\termsize}{$     \mathrm{Tr}^{\mathscr{B}}
    \Big(
      U_{\mathrm{tot}}
      \big(
        \rho_{\mathrm{sys}}
          \otimes
        \rho^{\mathrm{unif}}_{\HilbertSpace{B}}
      \big)
      U_{\mathrm{tot}}^\dagger
    \Big)
    $}
  \end{tikzcd}
  }
  \hspace{-4cm}
\end{equation}
But the same idea underlies already the model of quantum computation introduced under the abbreviation {\it DQC1} by \cite{KnillLaflamme98}\cite{PLMP03}\cite{ShorJordan08} (also known as the ``one clean qbit''-model), motivated by the (noisy) reality of quantum computation (specifically on NMR spin-resonance qbits). In this case $\HilbertSpace{H} \defneq \QBit$ is a single QBit, and one initializes the system in state $\vert 0\rangle$ (say) and measures the expectation value \eqref{ExpectationValueOfObservable}
of the observable $\mathcal{O}_{P_0} \,\defneq\, \vert 0 \rangle \langle 0 \vert$
\eqref{ObservablesAsQuantumChannels}
in the output of the above channel \eqref{UnistochasticEnvironmentalRepresentation}, given by the following formula (cf. \cite[(1)]{ShorJordan08}):
\begin{equation}
  \label{ProbabilityFormulaInDQC1}
  \hspace{2cm}
  \mathllap{
    \scalebox{.7}{
      \color{darkblue}
      \bf
      \begin{tabular}{c}
        probability measured by
        \\
        (repeated)
        DQC1 computations
      \end{tabular}
    }
  }
  \hspace{.5cm}
  p_0
  \;=\;
  \mathrm{Tr}^{\QBit}
  \bigg(
    P_0
    \cdot
    \mathrm{Tr}^{\HilbertSpace{B}}
    \Big(
      U_{\mathrm{tot}}
      \big(
        \vert 0 \rangle
        \langle 0 \vert
        \otimes
        \rho^{\mathrm{unif}}_{\HilbertSpace{B}}
      \big)
      U_{\mathrm{tot}}^\dagger
    \Big)
  \bigg)
  \,.
\end{equation}
The relation of this DQC1 model to unistochastic quantum channels is obvious but has been made explicit only recently \cite[\S III]{XCGX23} (and not using the ``unistochastic'' terminology). We give a natural monadic typing in Ex. \ref{MonadicDQC1}.

\smallskip

Incidentally, we may observe that among all coupling channels \eqref{CouplingQuantumChannel},  those which couple to the {\it maximally mixed state} of the environment this way, namely the one represented by a multiple of the identity matrix and representing the {\it uniform} probability distribution on (any set of) orthonormal basis states $\big(\vert b \rangle\big)_{b \isa B}$ are {\it dual} (in a precise sense) to the averaging channels \eqref{AveraginQuantumChannel}:
\vspace{-2mm}
\begin{equation}
  \label{UniformlyDistributedBathState}
  \hspace{-2cm}
  \scalebox{.7}{
    \color{darkblue}
    \bf
    \def\arraystretch{.9}
    \begin{tabular}{c}
      uniformly distributed
      \\
      mixture of bath states
    \end{tabular}
  }
  \hspace{1cm}
  \rho_{\HilbertSpace{B}}^{\mathrm{unif}}
  \;\;
  \defneq
  \;\;
  \frac{1}{
    \mathrm{dim}(\HilbertSpace{B})
  }
  \,
  \mathrm{id}_{\HilbertSpace{B}}
  \;=\;
  \underset{b}{\sum}
  \frac{
   \vert b \rangle \langle b \vert
  }{
   \mathrm{dim}(\HilbertSpace{B})
  }
  \;\isa\;
  \HilbertSpace{B}
    \otimes
  \HilbertSpace{B}^\ast
\end{equation}
\begin{equation}
  \label{UniformCouplingQuantumChannel}
  \hspace{-2cm}
  \scalebox{.7}{
    \color{darkblue}
    \bf
    \begin{tabular}{c}
      coupling to uniform bath
      \\
      as a quantum channel
    \end{tabular}
  }
  \hspace{1.7cm}
  \begin{tikzcd}[row sep=-2pt]
    \mathllap{
      \mathrm{chan}^{
        \rho_{\HilbertSpace{H}}^{\mathrm{unif}}
      }
      \;\isa\;\;
    }
    \HilbertSpace{H}
    \otimes
    \HilbertSpace{H}^\ast
    \ar[
      rr
    ]
    &&
    \big(
      \HilbertSpace{H}
      \otimes
      \HilbertSpace{B}
    \big)
    \otimes
    \big(
      \HilbertSpace{B}
      \otimes
      \HilbertSpace{H}
    \big)^\ast
    \\
    \scalebox{\termsize}{$
      \rho_{\mathrm{sys}}
    $}
    &\longmapsto&
   \scalebox{\termsize}{$
     \rho_{\mathrm{sys}}
       \otimes
    \rho_{\HilbertSpace{B}}^{\mathrm{unif}}
   $}
  \end{tikzcd}
\end{equation}

\vspace{0mm}
\noindent In \cref{MixedQuantumTypes} we understand this dual pair of quantum channels as the initial (terminal) cases among the (co)monadic QuantumState (co)monad transformations.

\smallskip

For example, every {\it uniformly} mixed unitary quantum channel \eqref{MixedUnitaryQuantumChannel} (i.e., one in which every unitary operator $U_s$ appears with the same probability $1/\mathrm{Card}(S)$) is unistochastic \eqref{UnistochasticEnvironmentalRepresentation}, with coupled-unitary given as shown below:
\begin{equation}
  \label{UniformlyMixedUnitaryQuantumChannel}
  \hspace{1.5cm}
  \mathllap{
    \scalebox{.7}{
      \color{darkblue}
      \bf
      \begin{tabular}{c}
        uniformly mixed
        \\
        unitary quantum gates
        \\
        as a quantum channel
      \end{tabular}
    }
    \hspace{2.2cm}
  }
  \begin{tikzcd}[sep=0pt]
    \mathllap{
      \mathrm{chan}^{(U_\bullet)}
      \;\isa\;
    }
    \HilbertSpace{H}_1
    \otimes
    \HilbertSpace{H}_1^\ast
    \ar[
      rr
    ]
    &&
    \HilbertSpace{H}_2
    \otimes
    \HilbertSpace{H}_2^\ast
    \\[-2pt]
    \scalebox{\termsize}{$
      \rho
    $}
    &\scalebox{\termsize}{$\longmapsto$}&
    \scalebox{\termsize}{$
      \!\!
      \underset{s \isan S}{\sum}
      \;
      \tfrac{1}{
        \mathrm{Card}(S)
      }
      \,
      U_s \cdot \rho \cdot U_s^\dagger
    $}
    \mathrlap{\,,}
  \end{tikzcd}
\end{equation}
\begin{equation}
  \label{CoupledUnitaryForUniformlyMixedUnitaryChannel}
  \begin{tikzcd}[
    row sep=-2pt
  ]
    \mathllap{
      U_{\mathrm{tot}}
      \;\isa\;
      \;
    }
    \HilbertSpace{H}
    \otimes
    \underset{S}{\oplus} \ComplexNumbers
    \ar[
      rr
    ]
    &&
    \HilbertSpace{H}
    \otimes
    \underset{S}{\oplus} \ComplexNumbers
    \\
 \scalebox{0.8}{$     \vert \psi \rangle
    \otimes
    \vert s \rangle
    $}
    &\longmapsto&
  \scalebox{0.8}{$    U_s \vert \psi \rangle
    \otimes
    \vert s \rangle
    $}
    \,.
  \end{tikzcd}
\end{equation}

In fact, on single qbits, every mixed unitary actually has such a uniformly mixed unitary presentation \cite[Thm. 1.2]{MHP19} and hence is unistochastic \eqref{UnistochasticEnvironmentalRepresentation}.

\medskip
For example, with the general argument given in \cite[Lem. 1.1]{MHP19} one finds that a {\bf unistochastic presentation of the bit-flip channel} \eqref{BitFlipQuantumChannel}
is given by the following total unitary \eqref{CoupledUnitaryForUniformlyMixedUnitaryChannel} on the single qbit-system coupled to an environment consisting of one other qbit:
\vspace{-2mm}
\begin{equation}
\hspace{-5mm}
  \scalebox{.7}{
    \color{darkblue}
    \bf
    \def\arraystretch{.9}
    \begin{tabular}{c}
      unistochastic
      \\
      environmental realization
      \\
      of bit-flip quantum channel
    \end{tabular}
  }
  \hspace{1.2cm}
  \begin{tikzcd}[row sep=-3pt, column sep=20pt]
    \mathllap{
      U^{ \mathrm{flip}_p }_{\mathrm{tot}}
      \;\colon\;\;
    }
    \QBit
    \otimes
    \QBit
    \ar[rr]
    &&
    \QBit
    \otimes
    \QBit
    \\
  \scalebox{0.8}{$    \vert 0 \rangle
    \otimes
    \vert b \rangle
    $}
    &\longmapsto&
    \def\arraystretch{1.2}
    \begin{array}{r}
   \scalebox{0.8}{$     \cos(\phi/2)
      \,
      \vert 0 \rangle
      \otimes
      \vert b \rangle
      $}
      \\
      \scalebox{0.8}{$  +(-1)^b\,
      \ImaginaryUnit
      \sin(\phi/2)
      \,
      \vert 1 \rangle
      \otimes
      \vert b \rangle
      $}
    \end{array}
    \\
    \scalebox{0.8}{$  \vert 1 \rangle
    \otimes
    \vert b \rangle
    $}
    &\longmapsto&
    \def\arraystretch{1.2}
    \begin{array}{r}
    \scalebox{0.8}{$    (-1)^b\,
      \ImaginaryUnit
      \sin(\phi/2)
      \,
      \vert 0 \rangle
      \otimes
      \vert b \rangle
      $}
      \\
     \scalebox{0.8}{$   + \cos(\phi/2)
      \,
      \vert 1 \rangle
      \otimes
      \vert b \rangle
      $}
      \mathrlap{\,,}
    \end{array}
  \end{tikzcd}
    \;\;\;
    \begin{array}{l}
    \scalebox{0.8}{$    \mbox{where} $}
      \\
    \scalebox{0.8}{$    \phi = \mathrm{arccos}(1-2p)
    $}.
    \end{array}
\end{equation}

Closely related to quantum channels:

\noindent
{\bf Quantum observables} are much like quantum channels to the trivial system,
but without the requirement that the trace be preserved:
\vspace{-2mm}
\begin{equation}
  \label{ObservablesAsQuantumChannels}
  \begin{tikzcd}[
    column sep=25pt,
    row sep=0pt
  ]
    \HilbertSpace{H}
    \otimes
    \HilbertSpace{H}^\ast
    \ar[
      rr,
      "{
        \scalebox{.7}{
          \color{darkgreen}
          \bf
          quantum observable
        }
      }",
      "{
        \scalebox{.7}{
          $\mathcal{O}_{A}$
        }
      }"{swap}
    ]
    &&
    \TensorUnit
    \\
    \scalebox{\termsize}{$
      \vert \psi \rangle
      \langle \phi \vert
    $}
    &\longmapsto&
    \scalebox{\termsize}{$
     \langle \phi \vert
      A
    \vert \psi \rangle
    $}
  \end{tikzcd}
  \hspace{1cm}
  \leftrightarrow
  \hspace{1cm}
  \begin{tikzcd}[
    column sep=30pt,
    row sep=0pt
  ]
    \HilbertSpace{H}
    \ar[
      rr,
      "{
        \scalebox{.7}{
          positive operator
        }
      }"{swap}
    ]
    &&
    \HilbertSpace{H}
    \\
    \scalebox{\termsize}{$
      \vert \psi\rangle
    $}
    &\longmapsto&
    \scalebox{\termsize}{$
    \underset{
      A
    }{
    \underbrace{
      a^\dagger a
    }
    }
    \vert \psi \rangle
    $}
  \end{tikzcd}
\end{equation}

\vspace{-2mm}
\noindent
In particular, given a mixed state represented by a density matrix $\rho \,\isa\, \HilbertSpace{H} \otimes \HilbertSpace{H}^\ast$ from \eqref{DensityMatrixInIntroduction}, then the {\it expectation value} of an observable $\mathcal{O}_A$ \eqref{ObservablesAsQuantumChannels}
in this state
is the value of the quantum operation $\mathcal{O}_{\!A}$ on $\rho$, which equals the trace \eqref{TotalProbabilityViaDensityMatrix} of the  operator product of the associated operator $A$ with the density matrix:
\vspace{-2mm}
\begin{equation}
  \label{ExpectationValueOfObservable}
  \overset{
    \mathclap{
      \raisebox{10pt}{
        \scalebox{.7}{
          \color{darkblue}
          \bf
          \def\arraystretch{.9}
          \begin{tabular}{c}
            expectation value of
            \\
            observabled $\mathcal{O}_A$
            \\
            in mixed state $\rho$
          \end{tabular}
        }
      }
    }
  }{
  \langle
    \mathcal{O}_{\!A}
  \rangle_\rho
  }
  \;\;\;\;
    \defneq
  \;\;\;\;
  \mathcal{O}_{\!A}(\rho)
  \;\;\;\;
  =
  \;\;\;\;
  \overset{
    \mathclap{
      \raisebox{10pt}{
        \scalebox{.7}{
          \color{darkblue}
          \bf
          \def\arraystretch{.9}
          \begin{tabular}{c}
            trace of product of
            \\
            associated operator $A$
            \\
            with density matrix $\rho$
          \end{tabular}
        }
      }
    }
  }{
  \mathrm{Tr}
  \big(
    A \cdot \rho
  \big)
  }
  \,.
\end{equation}

This means that after passing through a unitary quantum channel $\mathrm{chan}^U$ \eqref{UnitaryQuantumChannel} an observable $\mathcal{O}_A$ is transformed according to the {\it Heisenberg evolution formula} (e.g. \cite[p. 36]{BuschGrabowskiLahti95}\cite[(3.44)]{Preskill04})
\begin{equation}
  \label{HeisenbergEvolution}
  \mathcal{O}_A
  \;\;\longmapsto\;\;
  \mathrm{chan}^U(\mathcal{O}_A)
  \;\defneq\;
  \mathcal{O}_{U \cdot A \cdot U^\dagger}
\end{equation}
in that
$$
  \big\langle
    \mathrm{chan}^U(\mathcal{O}_A)
  \big\rangle_{\mathrm{chan}^U(\rho)}
  \;\defneq\;
  \mathrm{Tr}
  \big(
    ( U \cdot A \cdot U^\dagger )
    \cdot
    ( U \cdot \rho \cdot U^\dagger )
  \big)
  \;=\;
  \mathrm{Tr}
  \big(
    A
    \cdot
    \rho
  \big)
  \;=\;
  \big\langle
    A
  \big\rangle_{\rho}
  \,.
$$

\end{literature}

\subsection{Monadic effects}
\label{BackgroundMonadicEffects}

\begin{literature}[\bf Modal logic and Possible worlds semantics]
  \label{ModalLogicAndManyWorlds}
  The origin of modal logic of {\it necessity} ($\necessarily$) and {\it possibility} ($\possibly$) is with Aristotle, as nicely reviewed in \cite{LemmonScott77}.
  The modern formalization of modal logics originates with \cite{Becker30}\cite[pp 153 \& App II]{LewisLangford32}\cite{vonWright51}\cite{Hintikka62}.
  A good historical overview is in \cite{Goldblatt03}, a comprehensive modern account in \cite{BlackburnVanBenthemWolter07}; see also \cite{BlackburnDeRijkeVenema01}.
      Starting with \cite[App II]{LewisLangford32}, modal logicians consider a plethora of variant axiom systems, which go by a long list of alphanumerical monikers.
      We are here entirely concerned with the system known as ``S5'' modal logic \cite[p. 501]{LewisLangford32}\cite[p. 1]{Kripke63}.
  Classical S5 modal logic is widely applied as epistemic modal logic, notably in classical computer science \cite[\S 2.3]{HalpernMoses92}\cite[p. 35]{FaginHalpernMosesVardi95}\cite[\S 9]{Fitting07}\cite[\S 4]{HoekPauly07} \cite[\S 2]{DitmarschHoekKooi08}\cite{Samet10}.

  \smallskip

 \noindent  {\bf Possible worlds semantics.}
    The ``possible worlds''-semantics of modal logic is due to \cite{Kripke63} (though the basic idea is expressed already in \cite{Hintikka62}); good exposition
    is in \cite{BlackburnVanBenthem07}, modern review is in \cite[Part 5 \S 1]{BlackburnVanBenthemWolter07}. Here one speaks of {\it Kripke frames} being
    (inhabited) $W \isa \Set$ of ``possible worlds'' equipped with a binary relation $R \isa W \times W \to \Prop$, where $R(w, w')$ is interpreted as
    ``Given outcome/world $w$, the outcome/world $w'$ appears (just as) {\it possible}.'' With such a possible-worlds scenario, the modal operators
    $\necessarily_W$, $\possibly_W \,\isa\, \Prop_W \to \Prop_W$ acting on $W$-dependent propositions $P \;\isa\; \Prop_W \,\defneq\, W \to \Prop$ are
    interpreted by the following formulas (e.g. \cite[p. 10]{BlackburnVanBenthem07}):
    \vspace{-2mm}
    \begin{equation}
      \label{KripkeSemanticsOfModalOperator}
      \hspace{-15mm}
      \adjustbox{
        raise=5pt
      }{
      \begin{tikzcd}[sep=0pt]
        & &
        \mathclap{
          \scalebox{.7}{
            \color{darkblue}
            \bf
            \def\arraystretch{.9}
\begin{tabular}{c}
              A proposition $P_\bullet$
              \\
              about/dependent on
              \\
              the possible worlds $w$
            \end{tabular}
          }
        }
        \\
        P_\bullet \isa & W \ar[rr] && \Prop
        \\
        &
        w &\longmapsto&
        P_w
      \end{tikzcd}
      }
      \hspace{.5cm}
      \begin{tikzcd}[sep=0pt]
        \mathclap{
          \raisebox{10pt}{
            \scalebox{.7}{
            \color{darkorange}
            \bf
            yields
          }
          }
        }
        \\
        \yields
      \end{tikzcd}
      \hspace{.5cm}
      \begin{tikzcd}[sep=0pt]
        &&
        \mathclap{
          \scalebox{.7}{
            \color{darkblue}
            \bf
            \def\arraystretch{.9}
            \begin{tabular}{c}
              The proposition $\necessarily_{{}_W} P$
              \\
              that $P_\bullet$ holds {\it necessarily}, namely
              \\
              in/for {\it all} worlds $w'$ that appear
              \\
              as possible as the given one $w$
            \end{tabular}
          }
        }
        \\
        \necessarily_{{}_W} P
        \,\isa
        &
        W
        \ar[rr]
        && \Prop
        ,\,
        \\
        &
        w &\longmapsto&
        \underset{
          \mathclap{
            (w' \isa W) \times
            \atop
            {R(w,w')}
          }
        }{\forall}
        \;\;\;\;
        P_{w'}
      \end{tikzcd}
      \hspace{.6cm}
      \begin{tikzcd}[sep=0pt]
        \mathclap{
          \raisebox{10pt}{
            \scalebox{.7}{
            \color{darkorange}
            \bf
            and
          }
          }
        }
        \\
        \phantom{\yields}
      \end{tikzcd}
      \begin{tikzcd}[sep=0pt]
        &&
        \mathclap{
          \scalebox{.7}{
            \color{darkblue}
            \bf
            \def\arraystretch{.9}
            \begin{tabular}{c}
              The proposition $\possibly_{{}_W} P$
              \\
              that $P$ holds {\it possibly}, namely
              \\
              in/for {\it some} world $w'$ that appears
              \\
              as possible as the given one $w$
            \end{tabular}
          }
        }
        \\
        \possibly_{{}_W} P
        \,\isa
        &
        W
        \ar[rr]
        && \Prop
        \\
        &
        w &\longmapsto&
        \underset{
          \mathclap{
            (w' \isa W) \times
            \atop
            {R(w,w')}
          }
        }{\exists}
        \;\;\;\;
        P_{w'}
      \end{tikzcd}
    \end{equation}

 \noindent  {\bf Modalities as monads.} The (co)monadic nature of the necessity/possibility operators $\necessarily$/$\possibly$ in S4 (hence in S5)
  modal logic was explicitly observed in \cite{BiermanDePaiva96}\cite{BiermanDePaiva00}\cite{Kobayashi97} and the resulting relation of modalities to (computational effect-)monads in computer science (Lit. \ref{LiteratureComputationalEffectsAndModalities}) was further discussed in \cite{BentonBiermanDePaiva98}.
    The natural origin of these S5 (co)monads  $\necessarily_{{}_W} \dashv \possibly_{{}_W}$ from {\it base change} along the ``possible worlds'' was noticed in \cite[p. 279]{Awodey06} -- however the implication (which we expand on in \cref{QuantumEffects}) that, therefore, any dependent type theory may equivalently be regarded as (epistemic) {\it modal type theory} (Lit. \ref{LiteratureModalTypeTheory})
  seems not to have received attention until the note \cite{nLabNecessity} (cf. \cite[Ch. 4]{Corfield20}).
  We expand on this novel point of view in the main text around Thm. \ref{S5KripkeSemanticsViaDependentTypes}.
\end{literature}

\begin{literature}[\bf Modal type theory]
  \label{LiteratureModalTypeTheory}
  In view of the famous relation between formal logic and type theory, it is quite evident that there is an interesting generalization of modal logic (Lit. \ref{ModalLogicAndManyWorlds}) to {\it modal type theory}. After leading a niche existence for some time,
  the amplification \cite[\S 3.1]{dcct}\cite{SchreiberShulman14}
  of {\it cohesive} modalities  (see \cite{SatiSchreiber20}) in (homotopy) type theory,
  the subject of {\it modal type theory} has received much attention (e.g. \cite{RSS20}\cite{CherubiniRijke21}\cite{Jaz22}). While such modal type theory is going to be relevant for various enhancements of the computational context presented here (to be discussed elsewhere), we emphasize that the modalities we consider here are all provided already by plain (linear) dependent type theory (are definable by {\it admissible rules} inferred from just the inference rules of the dependent linear types).  This fact is what drives our observation that {\LHoTT} (Lit. \ref{LiteratureLHoTT}) already knows about quantum measurement effects -- the feature just has to be brought out by meticulous syntactic sugaring (Lit. \ref{LiteratureDomainSpecificLanguages}).

\end{literature}

\newpage

\begin{literature}[\bf Category theory]
  The point of category theory (\cite{MacLane71}\cite{Kelly82}\cite{Borceux94})
  has been said to be the notion of {\it natural transformations} between mathematical stuctures, where the concept of {\it categories} themselves just serves to allow for speaking about {\it functors} which in turn are the subjects of these natural transformations. This is implicit in the title and introduction of Eilenberg \& MacLane's original \cite{EilenbergMacLane45}, and made more explicit Freyd in \cite[p. 1]{Freyd64}.
  But this is really only half of the story.

  \smallskip
  Namely natural transformations, in turn, support the concept of {\it adjunctions} between categories, and {\it this} is where category theory becomes a theory: Adjunctions and their many equivalent incarnations such as (Kan extensions, (co)limits, (co)terminality and notably) monads (for which see Lit. \ref{LiteratureComputationalEffectsAndModalities}) are the fundamental mathematical phenomena where category theory provides its non-trivial theorems. (Curiously, adjunctions are arguably the formalization of {\it dualities}, hence it is not misleading to say that category theory is really the {\it theory of duality}. In fact, \cite{EilenbergMacLane45} motivate their introduction of category theory with the example of dualizable objects, see \eqref{CoEvaluationForDualObjects}).
  $$
  \footnotesize
    \begin{tikzcd}[
      column sep={between origins, 40pt},
      row sep=15pt
    ]
      \fbox{adjunctions}
      \ar[
        rr,
        line width=2pt,
        -
      ]
      &&
      \fbox{monads}
      \\
      &
      \fbox{natural transformations}
      \ar[
        ul,
        line width=2pt,
        -
      ]
      \ar[
        ur,
        line width=2pt,
        -
      ]
      \\
      &
      \fbox{functors}
      \ar[
        u,
        line width=2pt,
        -
      ]
      \\
      &
      \fbox{categories}
      \ar[
        u,
        line width=2pt,
        -
      ]
    \end{tikzcd}
  $$
\end{literature}

\begin{literature}[\bf Functional programming languages]
\label{LiteratureFunctionalLanguages}
In programming, it is a familiar idea (expanded on in Lit. \ref{VerificationLiterature}) that every {\it datum} $d$ be of some specified {\it data type} $D$, denoted ``$d \,\isa\, D$''.
This being so, then a {\it program} which, when run on input data of type $D_{\mathrm{in}}$ (is guaranteed to halt and then) produces
data of type $D_{\mathrm{out}}$  is thus a {\it function} of the collection of $D_{\mathrm{in}}$-data with values in the collection of $D_{\mathrm{out}}$-data ---
and we may postpone detailing what particular kind of function we might mean (for instance: {\it linear} functions for quantum programs) by speaking
of just an arrow (morphism) in the relevant {\it category of data types}:

\begin{equation}
\label{CategoricalSemanticsSchematic}
\adjustbox{}{
\def\arraystretch{1.4}
\begin{tabular}{|c|c|}
\hline
{\bf Programming syntax}
&
{\bf Categorical semantics}
\\
\hline
\hspace{3pt}
$
  \begin{tikzcd}[sep=0pt]
  \\[5pt]
  d
    \,\isa\,
  D_{\mathrm{in}}
  \ar[
    rr,
    phantom,
    "{ \vdash }"
  ]
  &&
  f(d)
  \,\isa\,
  D_{\mathrm{out}}
  \\
    \mathclap{
      \scalebox{.7}{
        \color{darkblue}
        \bf
        \def\arraystretch{.9}
        \begin{tabular}{c}
          input data
          \\
          type
        \end{tabular}
      }
    }
    &
    \scalebox{.7}{
      \color{darkgreen}
      \bf
      program
    }
    &
    \mathclap{
      \scalebox{.7}{
        \color{darkblue}
        \bf
        \def\arraystretch{.9}
        \begin{tabular}{c}
          output data
          \\
          type
        \end{tabular}
      }
    }
  \end{tikzcd}
$
\hspace{3pt}
&
\hspace{5pt}
$
  \begin{tikzcd}[
    row sep=0pt,
    column sep=15pt
  ]
    \\[5pt]
    D_{\mathrm{in}}
    \ar[
      rr,
      "{
        f
      }"
    ]
    &&
    D_{\mathrm{out}}
    \\
    \mathclap{
      \scalebox{.7}{
        \color{darkblue}
        \bf
        \def\arraystretch{.9}
        \begin{tabular}{c}
          domain
          \\
          object
        \end{tabular}
      }
    }
    &
        \scalebox{.7}{
          \color{darkgreen}
          \bf
          morphism
        }
    &
    \mathclap{
      \scalebox{.7}{
        \def\arraystretch{.9}
        \color{darkblue}
        \bf
        \begin{tabular}{c}
          codomain
          \\
          object
        \end{tabular}
      }
    }
  \end{tikzcd}
$
\hspace{5pt}
\\
\hline
\end{tabular}
}
\end{equation}

\noindent
In the simplest examples (cf. p. \pageref{TheIdeaOfQuantumGates}), the semantics of the simplest functional
\begin{itemize}[leftmargin=.5cm]
\item  {\it classical languages} may be in the {\it category of sets},

where elementary programs are interpreted as {\it logic gates}
$$
  \mathrm{Bit}^{\times^{n_{\mathrm{in}}}} \longrightarrow \mathrm{Bit}^{\times^{n_{\mathrm{out}}}}
$$
\item {\it quantum languages} may be in the category of {\it $\ComplexNumbers$-vector spaces},

where elementary programs are interpreted as {\it quantum gates}
$$
  \mathrm{QBit}^{\otimes^{n_{\mathrm{in}}}} \longrightarrow \mathrm{QBit}^{\otimes^{n_{\mathrm{out}}}}
$$
\end{itemize}

The point of {\it functional programming} (e.g. \cite{Thompson96}\cite{Thompson91}) is that
programs are such functions and {\it nothing but} such functions of data (compiled under function composition), in that they have:
\begin{itemize}
\item no side-effects -- besides producing their declared output,
\item no context-dependence --- besides on their declared input,
\end{itemize}
on the global state of the computing environment.

Therefore one also speaks of  {\it pure functions} or {\it pure programs}, for emphasis. This is in contrast to more traditionally popular ``imperative'' programming languages ---  whose programs may, while running, read unpredictable data from input devices and write to output devices in a way that is not reflected in the declaration of their input/output data types.
In contrast, the purity of functional programs is what makes them completely deterministic, hence predictable by mathematical analysis and hence formally verifiable (Lit. \ref{VerificationLiterature}).

This relation between (i) computation (ii) data typing and (iii) categorical algebra turns out to be so tight as to effectively exhibit three equivalent perspectives on the same underlying structure, a remarkable phenomenon that has been called the {\it computational trilogy} (for pointers see \cite[p. 4]{SS22TQC}):
\vspace{-.2cm}
\begin{equation}
\label{TheTrilogy}
\adjustbox{raise=-50pt, scale=.8}{
\small

}
\end{equation}

Of course, in practice one needs programs which {\it do} cause side-effects, or which {\it do} have context-dependence. Noticing the above qualifications, these may absolutely be described by functional programs, but
\begin{center}
\begin{tabular}{ccc}
    side-effects
    &
    are  to be {\it declared}
    &
    as part of the output data type,
   \\
    context-dependence
    &
    is to be {\it declared}
    &
    as part of the input data type.
\end{tabular}
\end{center}
In line with the computational trilogy \eqref{TheTrilogy}, there should be fundamental concepts in type theory and in categorical algebra which correspond to such effect/context-declaration in typed programming languages. Indeed, these correspondent ares the very concepts of
{\it modalities} (Lit. \ref{LiteratureModalTypeTheory})
and of {\it monads}, see Lit. \ref{LiteratureComputationalEffectsAndModalities}.
\begin{equation}
\label{TheEffectTrilogy}
\adjustbox{raise=-50pt, scale=.8}{
\small
%
}
\end{equation}

\end{literature}


\begin{literature}[\bf Computational Effects and Monadic modalities]
\label{LiteratureComputationalEffectsAndModalities}

We give a lightning explanation of computational effects (and computational contexts) understood as (co)monads on the type system, and of
the Eilenberg-Moore-Kleisli theory of the corresponding effect handlers (context providers) understood as (co)modules, in fact
as (co)modal types (cf. Lit. \ref{LiteratureModalTypeTheory}).

\medskip

\phantomsection\label{MonadicComputationalEffects}
\noindent {\bf Computational effects...}
The idea (\cite{Moggi89Abstract}\cite{Moggi89}\cite{Wadler90}\cite{Moggi91}\cite{PlotkinPower02}, cf. \cite[\S 6]{HylandPower07}) is that a computation which {\it nominally}
produces data of some type $D$ while however causing some computational side-effect must {\it de facto} produce data of some adjusted type
$\mathcal{E}(D)$ which is such that the effect-part of the adjusted data can be carried alongside followup programs (whence a
``notion of computation''  with ``computational side effects'', for exposition and review see \cite{BentonHughesMoggi02}\cite[\S 20]{Milewski19}\cite{Uustalu21}\cite[\S 10]{Winitzki22})
via {\tt bind}- and {\tt return}-operations, as follows:

\def\arraystretch{.9}

\vspace{0cm}
\begin{equation}
  \label{BindingAndReturning}
  \hspace{-4.4cm}
  \begin{tikzcd}[
    column sep=30pt,
    row sep=35pt
  ]
    D_1
    \ar[
      r,
      "{
        \mathrm{prog_{12}}
      }"{yshift=1pt},
      "{
        \scalebox{.7}{
          \color{darkblue}
          \bf
          first program
        }
      }"{yshift=10pt},
      "{
        \scalebox{.7}{
          \def\arraystretth{.7}
          \color{darkgreen}
          \bf
\begin{tabular}{c}
            with output data
            \\
            of nominal type
            \scalebox{1.2}{$D_2$} but
            \\
            causing effects of type \scalebox{1.15}{$\mathcal{E}(-)$}
          \end{tabular}
        }
      }"{swap, yshift=-4pt}
    ]
    &
    \mathcal{E}(D_2)
    &[-40pt]
    &[-40pt]
    D_2
    \ar[
      r,
      "\mathrm{prog_{23}}",
      "{
        \scalebox{.7}{
          \color{darkblue}
          \bf
          second program
        }
      }"{yshift=10pt},
      "{
        \scalebox{.7}{
          \def\arraystretth{.7}
          \color{darkgreen}
          \bf
          \begin{tabular}{c}
            with input data
            \\
            of type
            \scalebox{1.2}{$D_2$}
          \end{tabular}
        }
      }"{swap, yshift=-2pt}
    ]
    &[+15pt]
    \mathcal{E}(D_3)
    \ar[
      d,
      |->,
      shift left=3,
      bend left=90,
      start anchor={[yshift=-5pt]},
      end anchor={[yshift=+10pt]},
      "{
        \scalebox{.7}{
          \color{darkorange}
          \bf
          \begin{tabular}{c}
            bind previous effects
            \\
            into second program
         \end{tabular}
        }
      }"{xshift=-4pt}
    ]
    &[95pt]
    \overset{
      \mathclap{
        \raisebox{5pt}{
          \scalebox{.7}{
            \color{darkblue}
            \bf
            pure data
          }
        }
      }
    }{
      d \,\isa\,D
    }
    \ar[
      d,
      |->,
      shift left=1,
      bend left=90,
      start anchor={[yshift=-5pt, xshift=5pt]},
      end anchor={[yshift=+5pt]},
      "{
        \scalebox{.7}{
          \color{darkorange}
          \bf
          \begin{tabular}{c}
            return pure data
            \\
            with trivial effect
         \end{tabular}
        }
      }"{xshift=-4pt}
    ]
    \\
    &
    &&
    \mathcal{E}(D_2)
    \ar[
      r,
      "{
        {
        \color{purple}
        \bind
         { \color{black} \mathcal{E} }
         {}
         {
           \color{black}
           \mathrm{prog_{23}}
         }
        }
      }"{yshift=1pt},
      "{
        \scalebox{.7}{
          \def\arraystretth{.7}
          \color{darkgreen}
          \bf
          \begin{tabular}{c}
            carry any previous
            \\
            \scalebox{1.3}{$\mathcal{E}(-)$}-effects along
          \end{tabular}
        }
      }"{swap, yshift=-5pt}
    ]
    &
    \mathcal{E}(D_3)
    \ar[
      d,
      |->,
      shift left=3,
      bend left=90,
      start anchor={[yshift=-5pt]},
      end anchor={[yshift=+10pt]},
      "{
        \scalebox{.7}{
          \color{darkorange}
          \bf
          \begin{tabular}{c}
            compose
         \end{tabular}
        }
      }"{xshift=-4pt}
    ]
    &
    \;\;\;
    \mathllap{
      {
        \color{purple}
        \return
          { \color{black}\mathcal{E} }
          { \color{black} D }
      }
    (d)
    \,\isa\,
    }
    \;
    \underset{
      \mathclap{
        \raisebox{-5pt}{
          \scalebox{.7}{
            \color{darkblue}
            \bf
            \scalebox{1.2}{$\mathcal{E}$}-effectful data
          }
        }
      }
    }{
      \mathcal{E}(D)
    }
    \\[+5pt]
    D_1
    \ar[
      rrrr,
      phantom,
      "{
        \scalebox{.7}{
          \color{darkblue}
          \bf
          \scalebox{1.3}{$\mathcal{E}$}-composite program
        }
      }"{yshift=+20pt},
      "{
        \scalebox{.7}{
          \def\arraystretth{.7}
          \color{darkgreen}
          \bf
          \begin{tabular}{c}
            causing
            cumulative \scalebox{1.15}{$\mathcal{E}(-)$}-effects
          \end{tabular}
        }
      }"{swap, yshift=-15pt}
    ]
    \ar[
      rr,
      "{
        \mathrm{prog}_{12}
      }"
    ]
    &&
    \mathcal{E}(D_2)
    \ar[
      rr,
      "{
        \bind
          { \mathcal{E} }
          {  }
        \mathrm{prog}_{23}
      }"
    ]
    &&
    \mathcal{E}(D_3)
  \end{tikzcd}
  \hspace{-4cm}
\end{equation}
such that
\begin{equation}
  \label{UnitalityForBind}
  \begin{array}{c}
  \begin{tikzcd}[
    column sep=55pt
  ]
    \mathcal{E}(D)
    \ar[
      rr,
      "{
        \bind
          { \mathcal{E} }
          {  }
          \;\;\;
          {
            \return
              { \mathcal{E} }
              { D }
          }
      }",
      "{
        \hspace{-15pt}
        \scalebox{.7}{
          \color{darkorange}
          \bf
          binding effect to trivial effect
        }
      }"{ yshift=13pt }
    ]
    \ar[
      rr,
      rounded corners,
      to path={
           ([yshift=-00pt]\tikztostart.south)
        -- ([yshift=-10pt]\tikztostart.south)
        -- node[yshift=+6pt] {
             \scalebox{.7}{$
               \mathrm{id}_{\mathcal{E}}
             $}
        }
           node[yshift=-8pt]{
             \scalebox{.7}{
               \color{darkorange}
               \bf
               is to carry it along identically
             }
           }
           ([yshift=-10pt]\tikztotarget.south)
        -- ([yshift=-00pt]\tikztotarget.south)
      }
    ]
    &&
    \mathcal{E}(D)
  \end{tikzcd}
  \hspace{1.7cm}
  \begin{tikzcd}[
    column sep=35pt
  ]
    D_1
    \ar[
      rr,
      "{
        \return
          { \mathcal{E} }
          { D }
      }",
      "{
        \scalebox{.7}{
          \color{darkorange}
          \bf
          trivial effect...
        }
      }"{yshift=14pt}
    ]
    \ar[
      rrrr,
      rounded corners,
      to path={
           ([yshift=-00pt]\tikztostart.south)
        -- ([yshift=-12pt]\tikztostart.south)
        -- node[yshift=+6pt] {
            \scalebox{.7}{$
              \mathrm{prog}_{12}
            $}
        }
          node[yshift=-8pt] {
            \scalebox{.7}{
              \color{darkorange}
              \bf
              ...has no effect on program
            }
          }
           ([yshift=-12pt]\tikztotarget.south)
        -- ([yshift=-00pt]\tikztotarget.south)
      }
    ]
    &&
    \mathcal{E}(D_1)
    \ar[
      rr,
      "{
        \bind
          { \mathcal{E} }
          { }
        \mathrm{prog}_{12}
      }"{yshift=1pt},
      "{
        \scalebox{.7}{
          \color{darkorange}
          \bf
          ...bound into program
        }
      }"{yshift=14pt}
    ]
    &&
    \mathcal{E}(D_2)
  \end{tikzcd}
  \\
\end{array}
\end{equation}
One also speaks of {\it Kleisli composition}
(in honor of \cite[p. 545]{Kleisli65})
and writes (``fish notation'', e.g. \cite[p. 321]{Milewski19}):
\begin{equation}
  \label{KleisliComposition}
  \mathrm{prog}_{12}
  \;\mbox{\tt >=>}\;
  \mathrm{prog}_{23}
  \;\;\;\;
  \defneq
  \;\;\;\;
  \big(
  \bind
    { \mathcal{E} }
    {  }
  \mathrm{prog}_{23}
  \big)
  \circ
  \mathrm{prog}_{12}
\end{equation}

\smallskip

\noindent
{\bf ...as monads on the type system.}
Such $\mathcal{E}$-effect structure on the type system is equivalently \cite[p. 32]{Manes76}\cite[Prop. 1.6]{Moggi91} a
functorial operation on the category of types (given by forming ``effectless programs'')
\begin{equation}
  \label{MonadFunctorFromBindReturn}
  \begin{tikzcd}[row sep=-2pt, column sep=huge]
    \mathcal{E}
    \;\;
    :
    &[-47pt]
    \Types
    \ar[
      rr,
      "{
        \scalebox{.7}{\bf
          \color{darkblue}
          functor underlying monad
          \bf
        }
      }"{yshift=5pt}
    ]
    &&
    \Types
    \\
    &
  \scalebox{0.8}{$   \big(
      D_1
      \xrightarrow{f}
      D_2
    \big)
    $}
    &
     \underset{
      \mathclap{
        \raisebox{-5pt}{
          \scalebox{.7}{
            \color{darkorange}
            \bf
            regard $f$ as effectless program
          }
        }
      }
    }{
      \longmapsto
    }
    &
    \scalebox{0.8}{$   \bind
        {\mathcal{E}}
        {}
      {
      \Big(
      D_1
      \xrightarrow{f}
      D_2
      \xrightarrow{
        \scalebox{.7}{$
        \return{\mathcal{E}}{D}
        $}
      }
      \mathcal{E}(D_2)
      \Big)
      }$}
        \end{tikzcd}
\end{equation}
which carries the structure of a {\bf monad}\footnote{
  The terminology ``monad'' for \eqref{MonadFunctorFromBindReturn} is due to \cite[\S 5.4]{Benabou67}, together with the observation that these are equivalently lax 2-functors from the
 terminal (point) category $\ast$ to the ambient 2-category (of type universes, in our case), in which 2-category theoretic sense they are quite the ``indecomposable units'' which the ancient called monads (as in Euclid: {\it Elements}, Book VII, Defs. 1, 2, 7, 11). For the present purpose, it is useful to envision that programs running {\it in} (the Kleisli category of) an effect-monad cannot sensibly  interact with other programs until they are ``taken out'' of (the Kleisli category of) the monad by an effect handler \eqref{EffectHandler}.
}
(cf. \cite[\S VI]{MacLane71}\cite[\S 4]{Borceux94}, older terminology: ``triple''), namely natural transformations
\begin{equation}
  \label{MonadStructure}
  D \,\isa\, \Type
  \hspace{.6cm}
  \yields
  \hspace{.6cm}
  \begin{tikzcd}[column sep=large]
    D
    \ar[
      rr,
      "{
        \overset{
          \mathclap{
            \raisebox{4pt}{
              \scalebox{.7}{
                \color{darkgreen}
                \bf monad unit/return
              }
            }
          }
        }{
          \unit
            { \mathcal{E} }
            { D }
          \,\defneq\,
          \return{\mathcal{E}}{D}
        }
      }"
    ]
    &&
    \mathcal{E}(D)
    \mathrlap{\,,}
  \end{tikzcd}
  \hspace{1cm}
  \begin{tikzcd}[column sep=huge]
    \mathcal{E}
    \big(
      \mathcal{E}(D)
    \big)
    \ar[
      rr,
      "{
        \overset{
          \mathclap{
            \raisebox{4pt}{
              \scalebox{.7}{
                \color{darkgreen}
                \bf monad product/join
              }
            }
          }
        }{
        \multiplication
          { \mathcal{E} }
          { D }
        \;\defneq\;
        \bind
          {\mathcal{E}}
          {}
          {
            \mathrm{id}_{\mathcal{E}(D)}
          }
        }
      }"
    ]
    &&
    \mathcal{E}(D)
  \end{tikzcd}
\end{equation}
satisfying the axioms of a unital monoid object \eqref{MonoidObject}, in that they make the following natural diagrams commute
\begin{equation}
  \label{MonadAxioms}
  \hspace{-1cm}
  \begin{tikzcd}[column sep=large]
    \mathcal{E}(D)
    \ar[
      rr,
      "{
         \unit
           {\mathcal{E}}
           { \mathcal{E}(D) }
      }"
    ]
    \ar[
      d,
      "{
        \mathcal{E}\left(
          \unit
            { \mathcal{E} }
            { D }
        \right)
      }"{swap}
    ]
    \ar[
      drr,
      equals,
      "{
        \scalebox{.7}{
          \color{darkorange}
          \bf
          unitality
        }
      }"{description}
    ]
    &&
    \mathcal{E}\big(
     \mathcal{E}(D)
    \big)
    \ar[
      d,
      "{
        \multiplication
          { \mathcal{E} }
          { \mathcal{E}(D) }
      }"
    ]
    \\
    \mathcal{E}
    \big(
      \mathcal{E}(D)
    \big)
    \ar[
      rr,
      "{
        \multiplication
          { \mathcal{E} }
          { \mathcal{E}(D) }
      }"{swap}
    ]
    &&
    \mathcal{E}(D)
    \mathrlap{\,,}
  \end{tikzcd}
  \hspace{1cm}
  \begin{tikzcd}[column sep=large]
    \mathcal{E}
    \scalebox{1.2}{$($}
      \mathcal{E}
      \big(
        \mathcal{E}(D)
      \big)
    \scalebox{1.2}{$)$}
    \ar[
      d,
      "{
        \mathcal{E}\left(
          \join
            { \mathcal{E} }
            { \mathcal{E}(D) }
        \right)
      }"{swap}
    ]
    \ar[
      rr,
      "{
         \join
           { \mathcal{E} }
           {\mathcal{E}(D)}
       }"
    ]
    \ar[
      drr,
      phantom,
      "{
        \scalebox{.7}{
          \color{darkorange}
          \bf
          associativity
        }
      }"
    ]
    &&
    \mathcal{E}
    \big(
      \mathcal{E}(D)
    \big)
    \ar[
      d,
      "{
         \multiplication
           { \mathcal{E} }
           { D }
       }"
    ]
    \\
    \mathcal{E}
    \big(
      \mathcal{E}(D)
    \big)
    \ar[
      rr,
      "{
        \multiplication
          { \mathcal{E} }
          { D }
      }"{swap}
    ]
    &&
    \mathcal{E}(D)
    \mathrlap{\,.}
  \end{tikzcd}
\end{equation}
Namely conversely, given such a monad the bind-operation on some $\mathrm{prog} \isa D_1 \to \mathcal{E}(D_2)$
is recovered as:
\begin{equation}
  \label{RecoveringBindFromJoin}
  \begin{tikzcd}[column sep=large]
    \mathclap{
      \scalebox{.7}{
        \color{darkblue}
        \bf
        \def\arraystretch{.9}
        \begin{tabular}{c}
          already
          \\
          effectful
          \\
          data
        \end{tabular}
      }
    }
    \ar[
      rr,
      phantom,
      "{
        \scalebox{.7}{
          \color{darkgreen}
          \bf
          \def\arraystretch{.9}
          \begin{tabular}{c}
            program
            \\
            produces
            \\
            further
            \\
            effects
          \end{tabular}
        }
      }"
    ]
    &&
    \mathclap{
      \scalebox{.7}{
        \color{darkblue}
        \bf
        \def\arraystretch{.9}
        \begin{tabular}{c}
          doubly
          \\
          effectful
          \\
          data
        \end{tabular}
      }
    }
    \ar[
      rr,
      phantom,
      "{
        \scalebox{.7}{
          \color{darkgreen}
          \bf
          \def\arraystretch{.9}
          \begin{tabular}{c}
            effects
            \\
            joined
            \\
            together
          \end{tabular}
        }
      }"
    ]
    &&
    \mathclap{
      \scalebox{.7}{
        \color{darkblue}
        \bf
        \def\arraystretch{.9}
        \begin{tabular}{c}
          plain
          \\
          effectful
          \\
          data
        \end{tabular}
      }
    }
    \\[-15pt]
    \mathcal{E}(D_1)
    \ar[
      rr,
      "{
        \mathcal{E}(\mathrm{prog})
      }"
    ]
    \ar[
      rrrr,
      rounded corners,
      to path ={
           ([yshift=-00pt]\tikztostart.south)
        -- ([yshift=-9pt]\tikztostart.south)
        -- node[yshift=-7pt]{
           \scalebox{.7}{$
             \bind
               {\mathcal{E}}
               {}
              \mathrm{prog}
           $}
        }
           ([yshift=-10pt]\tikztotarget.south)
        -- ([yshift=-00pt]\tikztotarget.south)
      }
    ]
    &&
    \mathcal{E}\big(\mathcal{E}(D_2)\big)
    \ar[
      rr,
      "{
        \join
          { \mathcal{E} }
          { D_2 }
      }"
    ]
    &&
    \mathcal{E}(D_2)
    \mathrlap{\,,}
    \\[+7pt]
    \ar[
     rrrr,
     phantom,
     "{
       \scalebox{.7}{
         \color{darkgreen}
         \bf
         bind previous effects into program
       }
     }"
    ]
    &&&&
    {}
  \end{tikzcd}
\end{equation}
which shows that the {\tt join}-operation is that which {\it joins consecutive effects into a single effect}, whence then terminology.

\medskip

\noindent {\bf Monads induced by adjunctions.} Monads
arise from (cf. \cite[\S VI.1]{MacLane71}\cite{Borceux94} -- and also give rise to, see \eqref{ComparisonFunctor} below) {\it adjoint functors} (``adjunctions'' between categories, cf. \cite[\S IV]{MacLane71}),
namely pairs of back-and-forth functors (here: between categories of types)
\vspace{-3mm}
\begin{equation}
  \label{MonadFromAdjunction}
  \begin{tikzcd}[column sep=large]
    \Types'
    \ar[
      from=rr,
      shift right=6pt,
      "{
        \overset{
          \mathclap{
            \raisebox{1pt}{
            \scalebox{.7}{
              \color{darkgreen}
              \bf
              left adjoint
            }
            }
          }
        }{
          L
        }
      }"{swap}
    ]
    \ar[
      rr,
      shift right=6pt,
      "{
        \underset{
          \mathclap{
            \raisebox{-1pt}{
            \scalebox{.7}{
              \color{darkgreen}
              \bf
              right adjoint
            }
            }
          }
        }{
        R
        }
      }"{swap}
    ]
    \ar[
      rr,
      phantom,
      "{
        \scalebox{.7}{$\bot$}
      }"
    ]
    &&
    \Types
    \ar[out=+50, in=-50,
      looseness=4.5,
      shorten=-3pt,
      shift left=0pt,
      "\scalebox{1.1}{${
          \hspace{1pt}
          \mathclap{
            {
              \;
              R \circ L
              \,\defneq:\,
              \mathcal{E}
            }
          }
          \hspace{4pt}
        }$}"{
          pos=.5,
          xshift=14pt
        }
    ]
  \end{tikzcd}
  \hspace{.5cm}
  \scalebox{.7}{
    \color{darkorange}
    \bf
    induced monad
  }
\end{equation}

\vspace{-3mm}
\noindent equipped with a natural {\it hom-isomorphism} (forming ``adjuncts'')
\begin{equation}
  \label{FormingAdjuncts}
  \mathrm{Hom}_{\Types}
  \big(
    -
    ,\,
    R(-)
  \big)
  \;
  \xleftrightarrow{
    \phantom{--}
    \widetilde{(-)}
    \phantom{--}
  }
  \;
  \mathrm{Hom}_{\Types'}
  \big(
    L(-)
    ,\,
    -
  \big)
\end{equation}
and (equivalently) with natural transformations
$$
  \overset{
    \mathclap{
      \raisebox{3pt}{
        \scalebox{.7}{
          \color{darkblue}
          \bf
          \def\arraystretch{1}
          \begin{tabular}{c}
            adjunction unit /
            \\
            return operation
          \end{tabular}
        }
      }
    }
  }{
  \unit
    { R L }
    { D }
  \,\defneq\,
  \widetilde{\mathrm{id}_{L(D)}}
  }
  \;:\;
  D
  \xrightarrow{\phantom{--}}
  R \circ L (D)
  \hspace{1.5cm}
  \overset{
    \mathclap{
      \raisebox{3pt}{
        \scalebox{.7}{
          \color{darkblue}
          \bf
          \def\arraystretch{1}
          \begin{tabular}{c}
            adjunction co-unit /
            \\
            obtain operation
          \end{tabular}
        }
      }
    }
  }{
  \counit
    { L R }
    { D' }
  \,\defneq\,
  \widetilde{\mathrm{id}_{R(D')}}
  }
  \;:\;
  L \circ R (D')
  \xrightarrow{\phantom{--}}
  { D' }
$$
\vspace{-4mm}
$$
  \begin{tikzcd}[column sep=20pt]
    \overset{
      \mathclap{
        \raisebox{+1pt}{
          \scalebox{.7}{
            \color{darkorange}
            \bf
            adjunction unit
          }
        }
      }
    }{
    \scalebox{0.8}{$
    \Big(
      D
      \xrightarrow{\;
        \unit
          {R L}
          { D }
      \;}
      R \circ L(D)
    \Big)
    $}
    }
    \ar[
      rr,
      shorten=17pt,
      <-|
    ]
    &&
    \overset{
      \mathclap{
        \raisebox{+3pt}{
          \scalebox{.7}{
            \color{darkorange}
            \bf
            identity
          }
        }
      }
    }{
    \scalebox{0.8}{$
    \Big(
      L(D)
      \xrightarrow{\;
        \mathrm{id}_{L(D)}
      \;}
      L(D)
    \Big)$}
    }
    \\[-20pt]
    \underset{
      \mathclap{
        \raisebox{-1pt}{
          \scalebox{.7}{
            \color{darkorange}
            \bf
            identity
          }
        }
      }
    }{
    \scalebox{0.8}{$
    \Big(
      R(D')
      \xrightarrow{\;
        \mathrm{id}_{R(D')}
      \;}
      R(D')
    \Big)
    $}
    }
    \ar[
      rr,
      shorten=15pt,
      |->
    ]
    &&
    \underset{
      \mathclap{
      \raisebox{-1pt}{
        \scalebox{.7}{
          \color{darkorange}
          \bf
          adjunction counit
        }
      }
      }
    }{ \scalebox{0.8}{$
    \Big(
      L \circ R(D')
      \xrightarrow{\;
        \counit
          { L R }
          { D' }
      \;}
      D'
    \Big)$}
    }
  \end{tikzcd}
$$

\vspace{-1mm}
\noindent satisfying the {\it zig-zag identities}
$$
  \counit
    { L R }
    {L(D)}
    \circ
    L\big(
      \unit
        {R L}
        { D }
    \big)
  \;\;
  =
  \;\;
  \mathrm{id}_D
  \,,
  \hspace{1.5cm}
  R\big(
    \counit
      {L R}
      { D' }
  \big)
    \circ
  \unit
    { R L }
    {R(D')}
  \;\;
   =
  \;\;
  \mathrm{id}_{D'}
  \,,
$$
from which the monad structure \eqref{MonadStructure} on $\mathcal{E} := R \circ L$ is obtained as follows:
\vspace{-1mm}
\begin{equation}
  \label{MonadStructureFromAdjunction}
  \begin{tikzcd}[row sep=0pt]
    D
    \ar[
      rr,
      "{
        \unit
          { \mathcal{E} }
          { D }
      }",
      "{
        \scalebox{.7}{
          \color{darkblue}
          \bf
          monad unit/return is...
        }
      }"{yshift=14pt}
    ]
    &&
    \mathcal{E}(D)
    \\[-5pt]
    \rotatebox{-90}{$\defneq$}
    &&
    \rotatebox{-90}{$\defneq$}
    \\[-5pt]
    D
    \ar[
      rr,
      "{
        \unit
          { R L }
          { D }
      }"{swap},
      "{
        \scalebox{.7}{
          \color{darkblue}
          \bf
          \def\arraystretch{.9}
          \begin{tabular}{c}
            ...the adjunction unit
            \\
            / return
          \end{tabular}
        }
      }"{swap, yshift=-13pt}
    ]
    &&
    R \circ L (D)
  \end{tikzcd}
  \hspace{1.5cm}
  \begin{tikzcd}[
    column sep=40pt,
    row sep=0pt
  ]
    \mathcal{E}\big(
      \mathcal{E}(D)
    \big)
    \ar[
      rr,
      "{
         \multiplication
           { \mathcal{E} }
           { D }
       }",
       "{
         \scalebox{.7}{
           \color{darkblue}
           \bf
           monad product/join is...
         }
       }"{yshift=13pt}
    ]
    &&
    \mathcal{E}(D)
    \\[-5pt]
    \rotatebox{-90}{$\defneq$}
    &&
    \rotatebox{-90}{$\defneq$}
    \\[-5pt]
    R \circ
    \underbrace{
    L
      \circ
    R
    }
    \circ L(D)
    \ar[
      rr,
      "{
        R
        \big(
         \counit
           { L R }
           { L(D) }
        \big)
      }"{swap},
      "{
        \scalebox{.7}{
          \color{darkblue}
          \bf
          \def\arraystretch{.9}
          \begin{tabular}{c}
            ...value under \scalebox{1.2}{$R$} of
            \\
            adjunction counit/obtain
            \\
            on value under
            \scalebox{1.2}{$L$}
          \end{tabular}
        }
      }"{swap, yshift=-18pt}
    ]
    &&
    R \circ L
    \mathrlap{\,.}
  \end{tikzcd}
\end{equation}

\smallskip

\noindent {\bf Typing of effects via Strong monads.}
As a technical aside, beware that in describing effect monad structure this way means to view only its external action on the category
of data types. In contrast, when actually coding monadic side effects in programming language constructs (as in \cref{Pseudocode} below), the
$\mathrm{return}$- and $\mathrm{bind}$-operations \eqref{BindingAndReturning}
will be typed {\it not} externally as
\vspace{-1mm}
$$
  \return
    {\mathcal{E}}
    {D}
  \,\isa\,
  \mathrm{Hom}
  \big(
    D
    ,\,
    \mathcal{E}(D)
  \big)
  \;\;\;\;\;\;\;
  \mbox{and}
  \;\;\;\;\;\;\;
  \bind
    {\mathcal{E}}
    {D_1, D_2}
  \;\isa\;
  \mathrm{Hom}
  \big(
    D_1
    ,\,
    \mathcal{E}(D_2)
  \big)
  \longrightarrow
  \mathrm{Hom}_{\Types}
  \big(
    \mathcal{E}(D_1)
    ,\,
    \mathcal{E}(D_2)
  \big)
$$
but internally as
terms of iterated {\it function type}
(cf. \cite[Def. 5.6]{McDermottUustalu22} with \cite[\S 4.1]{BentonHughesMoggi02}\cite[\S 20.2]{Milewski19}):
\vspace{-1mm}
\begin{equation}
  \label{TypingOfStrongMonads}
  \def\arraystretch{1.4}
  \begin{array}{llrl}
  \return
    { \mathcal{E} }
    { D }
  \;\isa\;
  D \to \mathcal{E}(D)
  \,,
    &
    \hspace{.5cm}
    &
    \bind
      {\mathcal{E}}
      {D_1, D_2}
    \;:
    &
    \big(
      D_1
        \to
      \mathcal{E}(D_2)
    \big)
    \to
    \big(
      \mathcal{E}(D_1)
        \to
      \mathcal{E}(D_2)
    \big)
    \\
    &&
    =
    &
    \mathcal{E}(D_1)
    \times
    \big(
      D_1
      \to
      \mathcal{E}(D_2)
    \big)
    \to
    \mathcal{E}(D_2)
    \\
    &&
    =
    &
    \mathcal{E}(D_1)
    \to
    \color{lightgray}
    \Big(
    \color{black}
    \big(
      D_1
        \to
      \mathcal{E}(D_2)
    \big)
    \to
    \mathcal{E}(D_2)
    \color{lightgray}
    \Big)
    \color{black}
    \mathrlap{\,,}
  \end{array}
\end{equation}

\vspace{-2mm}
\noindent where
$$
  (\mbox{-}) \to (\mbox{-})
  \;\;\defneq\;\;
  [\mbox{-},\,\mbox{-}]
  \;\;\isa\;\;
  \Types^{\mathrm{op}}
  \times
  \Types
  \xrightarrow{\;\;\;}
  \Types
$$
denotes the formation of function types interpreted as the internal hom-objects in the monoidal closed category of types (e.g. \cite[\S I]{LambekScott86}\cite[\S 6.1]{Borceux94}).
(Here we stick to notation for cartesian monoidal structure just for the purpose of exposition, see \eqref{LinearTypeDeclaration} for the analogous non-classical/linear case.)

With the above monad structure phrased internally this way, it is actually richer/stronger, whence one speaks of {\it enriched} or equivalently {\it strong monads} (\cite[\S 3.2]{Moggi91}, review in \cite[\S 3.2]{Ratkovic12}\cite[Prop. 5.8]{McDermottUustalu22}), here with respect to the self-enrichment of the monoidal closed category of types.

For monads on genuinely classical types (like sets) the strength/enrichment actually exists uniquely (see \cite[Ex. 3.7]{McDermottUustalu22}),
but for cases such as linear types
\eqref{LinearTypesAsSubstructuralTypes} it needs to be established (which we do in Prop. \ref{ClassicalAndQuantumModality}).
A convenient way to obtain/verify this enriched/strong monad structure is via symmetric monoidal monad structure:

When the category of types is {\it symmetric} monoidal closed (\cite[\S III.6]{EilenbergKelly66}) --- which is the case we are concerned with throughout, cf. Prop. \ref{DoublyClosedMonoidalStructure} --- with braiding transformations
$$
  \braid{\otimes}{D, D'}
  \;\isa\;
  D \otimes D'
  \xrightarrow{\;\;\;}
  D' \otimes D
$$
then {\it symmetric monoidal} structure on a monad $\mathcal{E}$ (\cite[p. 8]{Kock70}, cf. e.g. \cite[\S 1.2]{Seal13})\footnote{
  We assume without restriction \cite{Schauenburg01} that the monoidal structure $\otimes$, $\TensorUnit$ is ``strict'', i.e. that its unitors and associators are identity morphisms, whence we do not show then in these diagrams.
}
\begin{equation}
\label{SymmetricMonoidalMonadStructure}
  \hspace{-.6cm}
  \def\arraystretch{1}
\begin{array}{cl}
  \rotatebox[origin=c]{90}{\bf
    structure
  }
  &
  \begin{tikzcd}
    \mathcal{E}(D)
    \otimes
    \mathcal{E}(D')
    \ar[
      d,
      "{
        \pair
          { \mathcal{E} }
          { D, D' }
      }"
    ]
    \\
    \mathcal{E}(D \otimes D')
  \end{tikzcd}
  \\
\end{array}
\end{equation}
bijectively induces ``commutative'' strong monad structure (\cite[Thm. 2.3]{Kock72}, detailed review in \cite[\S 7.3, \S A.4]{GLLN08} \cite[Prop. 3.3.9]{Ratkovic12}) hence in particular the required enriched monad structure \eqref{TypingOfStrongMonads}.

\medskip

\noindent {\bf Examples of effect monads.}
Fundamental examples of effect monads in classical computer science (and in their linear version of profound importance to us in \cref{QuantumEffects})
include (cf. \cite[Ex. 1.1]{Moggi91}):

\begin{itemize}[leftmargin=.4cm]

\item
 The {\bf reader-} or {\bf environment-monad}
 (e.g. \cite[\S 21.2.3]{Milewski19}\cite[p. 22]{Uustalu21}, we use ``$W$'' for the {\it worlds} being read out, cf. Lit. \ref{ModalLogicAndManyWorlds}):
 \vspace{-2mm}
 \begin{equation}
   \label{ClassicalReaderMonad}
   \begin{tikzcd}[row sep=-2pt]
     \mathllap{
       W\mathrm{Read}
       \;\isa\;\;
     }
     \Types
     \ar[
       rr
     ]
     &&
     \Types
     \\
 \scalebox{0.8}{$    D  $}
       &\longmapsto&
  \scalebox{0.8}{$   {[W,\, D]} $}
   \end{tikzcd}
 \end{equation}

 \vspace{-2mm}
\noindent  induced from the canonical {\it comonoid} structure on any cartesian type
$W$ (given by its terminal and diagonal map):
 \begin{equation}
 \hspace{-2cm}
   \begin{tikzcd}[
     column sep=28pt,
     row sep=3pt
   ]
   \scalebox{.7}{
     \def\arraystretch{.9}
     \begin{tabular}{c}
       \color{darkblue}
       \bf
       comonoid $W$
       \\
       (ambient data)
     \end{tabular}
   }
     &[35pt]
     W \times W
     \ar[
       from=rr,
       "{
         \mathrm{diag}_{{}_W}
       }"{swap}
      ]
     &&
     W
     \ar[
       rr,
       "{ \exists ! }"{yshift=2pt}
     ]
     &&
     \ast
     \\
   \scalebox{.7}{
     \def\arraystretch{.9}
     \begin{tabular}{c}
       \color{darkblue}
       \bf
       $W$-reader monad
     \end{tabular}
   }
     &
     \mathllap{
    \big[
      W
      ,\,
      [W,\,D]
    \big]
     \;\simeq
     \;\;
     }
     [W \!\times\! W ,\, D]
     \ar[
       rr,
       "{
         \multiplication
           { \scalebox{.7}{
             \scalebox{1.3}{$W$}Read
            } }
           { D }
       }",
       "{
         \defneq\,
         [
           \mathrm{diag}_{{}W}
           ,\,
           D
         ]
       }"{swap}
     ]
     &&
     {[W,\,D]}
     \ar[
       from=rr,
       "{
         \unit
           {
             \scalebox{.7}{
               \scalebox{1.3}{$W$}Read
             }
           }
           {
             D
           }
       }"{swap},
       "{
         \defneq
         \,
         \mathrm{const}
       }"
     ]
     &&
     {[\ast,\, D]}
     \mathrlap{
       \;
       \simeq
       \,
       D
     }
   \end{tikzcd}
 \end{equation}
\vspace{-1mm}

 \noindent
 Hence a $W$-Reader-effectful program is one whose nominal output is {\it indefinite} \eqref{PotentialityEpistemicEntailments} until a global parameter $w \isa W$ is read in, and the handling of $W$-Reader-effects is the handing-along of this global parameter.

 $$
   \declare
     {
       \bind
         {
           W\mathrm{Read}
         }
         { D, D' }
     }
     {
       \hspace{-.4cm}
       \big(
         D \to (W \to D')
       \big)
       \xrightarrow{\quad
         \overset{
           \mathclap{
             \raisebox{0pt}{
               \scalebox{.7}{
                 \color{darkgreen}
                 \bf
                 \def\arraystretch{.9}
\begin{tabular}{c}
                   binding of
                   \\
                   $W$Reader effects
                 \end{tabular}
               }
             }
           }
         }{
           \phantom{----}
          }
       \qquad }
       \big(
         (W \to D) \to (W \to D')
       \big)
     }
     {
       \underset{
         \mathclap{
           \scalebox{.7}{
             \color{darkblue}
             \bf
             \def\arraystretch{.9}
             \begin{tabular}{c}
               program producing output
               \\
               depending on
               \\
               a global $W$-parameter
             \end{tabular}
           }
         }
       }{
       \Big(
         d
         \,\mapsto\,
         \big(
           w \mapsto d'_w(d)
         \big)
       \Big)
       }
       \,\mapsto\,
       \Big(
         \big(
           w
           \,\mapsto\,
           d_w
         \big)
         \mapsto
         \big(
           \underset{
             \mathclap{
               \hspace{0pt}
               \rotatebox{40}{
                 \llap{
                   \scalebox{.6}{
                     \color{darkblue}
                     \bf
                     global parameter
                     \hspace{-12pt}
                   }
                 }
               }
             }
           }{
             \;w\;
           }
           \underset{
             \mathclap{
               \rotatebox{40}{
                 \llap{
                   \scalebox{.6}{
                     \color{darkblue}
                     \bf
                     gets passed to
                     \hspace{-12pt}
                   }
                 }
               }
             }
           }{
             \mapsto
            \;
          }
           \underset{
             \mathclap{
               \rotatebox{40}{
                 \llap{
                   \scalebox{.6}{
                     \color{darkblue}
                     \bf
                     \def\arraystretch{.9}
                     \begin{tabular}{r}
                       all subsequent
                       \\
                       programs \hspace{-14pt}
                     \end{tabular}
                     \hspace{-18pt}
                   }
                 }
               }
             }
           }{
             d'_w(d_w)
           }
         \big)
       \Big)
     }
 $$
 \vspace{.7cm}

 \item
 The
{\bf writer monad}  (e.g. \cite[\S 4.1 \& \S 21.2.4]{Milewski19}\cite[1, p. 23]{Uustalu21}):
\vspace{-2mm}
\begin{equation}
  \label{WriterMonad}
  \begin{tikzcd}[
    row sep=-2pt
  ]
    \mathllap{
      A\mathrm{Write}
      \;\isa\;\;
    }
    \Types
    \ar[rr]
    &&
    \Types
    \\
 \scalebox{0.8}{$   D $}
      &\longmapsto&
 \scalebox{0.8}{$     A \times D $}
    \,.
  \end{tikzcd}
\end{equation}
\vspace{-2mm}

\noindent
induced from any {\it monoid} (aka {\it unital semi-group}) structure on a type $A$,
 \vspace{-2mm}
\begin{equation}
  \label{ClassicalWriterMonad}
  \hspace{-1.5cm}
  \begin{tikzcd}[
    column sep=30pt, row sep=0pt
  ]
    \scalebox{.7}{
      \begin{tabular}{c}
      \color{darkblue}
      \bf
      monoid
      $W$
      \\
      (data output stream)
      \end{tabular}
    }
    &[-10pt]
    A \times A
    \ar[
      rr,
      "{
        \mathrm{prod}_{{}_A}
      }"
    ]
    &&
    A
    \ar[
      from=rr,
      "{ \mathrm{unit}_{{}_A} }"{swap}
    ]
    &&
    \ast
    \\
    \scalebox{.7}{
      \color{darkblue}
      \bf
    $A$-writer monad
    }
    &
    A \times A \times D
    \ar[
      rr,
      "{
        \multiplication
          {
            A\mathrm{Writ}
          }
          { D }
          \,
          \defneq
      }",
      "{
        \mathrm{prod}_{{}_A}
        \times
        \,
        \mathrm{id}_D
      }"{swap}
    ]
    &&
    A \times D
    \ar[
      from=rr,
      "{
        \unit
          {
            A\mathrm{Write}
          }
          { D }
        \,
        \defneq
      }"{swap},
      "{
        \mathrm{unit}_{{}_A}
        \times
        \,
        \mathrm{id}_D
      }"
    ]
    &&
    \ast \times D
    \mathrlap{
      \;=
      D
    }
  \end{tikzcd}
\end{equation}
(Here the unitality and associativity properties of the monoid structure on $A$ are evidently equivalent to the corresponding properties \eqref{MonadAxioms} of the associated writer monad.) In typical applications $A$ is a {\it free monoid} on an alphabet, hence is the type of {\it strings} of such characters with join product given by concatenation of strings.

Therefore a $\mathrm{Writer}$-effectful program is one which in addition to its nominal output produces a string (a log message), and the binding of cumulative such effects is by concatenating these strings (appending these messages to the log)
\vspace{-2mm}
$$
  \declare
    {
      \bind
        { A\mathrm{Write} }
        { D, D' }
    }
    {
      \big(
        D \to A \times D'
      \big)
      \xrightarrow{
        \phantom{----}
      }
      \big(
        A \times D
        \to
        A \times D'
      \big)
    }
    {
      \big(
        d \,\mapsto\, (a_d,\, d'_d)
      \big)
      \,\mapsto\,
      \big(
        (a,\,d)
          \,\mapsto\,
        (
          \underset{
            \mathclap{
              \rotatebox{40}{
                \llap{
                  \scalebox{.6}{
                    \color{darkblue}
                    \bf
                    concatenated logs
                    \hspace{-15pt}
                  }
                }
              }
            }
          }{
            a \cdot a_d
          }
          ,\,
          d'_d
        )
      \big)
    }
$$
\\

\vspace{6mm}
\item The {\bf state monad} (e.g. \cite[Ex. 42]{BentonHughesMoggi02}\cite[\S 3]{PlotkinPower02}\cite[\S 21.2.5 ]{Milewski19}\cite[1, p. 24]{Uustalu21})
 \vspace{-2mm}
\begin{equation}
  \label{StateMonadEndofunctor}
  \begin{tikzcd}[sep=0pt, row sep=0pt]
    \mathllap{
    W\mathrm{State}
    \;\isa\;
    }
    \Types
    \ar[rr]
    &&
    \Types
    \\
 \scalebox{0.8}{$     D  $}
    &\longmapsto&
 \scalebox{0.8}{$     \big[
      W
      ,\,
      W \times D
    \big]
    $}
  \end{tikzcd}
\end{equation}

 \vspace{-2mm}
\noindent
given by
\begin{equation}
  \begin{tikzcd}[
    row sep=-5pt
  ]
    \big[
    W
    ,\
    W
    \times
    [
      W
      ,\,
      W \times D
    ]
    \big]
    \ar[
      rr,
      "{
        \multiplication
          {
            \scalebox{.7}{
              \scalebox{1.3}{$W$}State
            }
          }
          {D}
      }"
    ]
    &&
    \left[
      W
      ,\,
      W \times D
    \right]
    \ar[
      from=rr,
      "{}"
    ]
    \ar[
      from=rr,
      "{
        \unit
          {\scalebox{.7}{
            \scalebox{1.3}{$W$}State
          }}
          {D}
      }"{swap}
    ]
    &&
    D
    \\
   \scalebox{0.8}{$   f $}
    &\longmapsto&
   \scalebox{0.8}{$   \mathrm{ev}\big(f(\mbox{-})\big) $}
    &&
    \\
    &&
 \scalebox{0.8}{$     \big(
      w \mapsto (w,d)
    \big)
    $}
    &\longmapsfrom&
  \scalebox{0.8}{$    d$}
  \end{tikzcd}
\end{equation}

\vspace{-2mm}
\noindent hence with bind-operation as follows:
\vspace{-2mm}
\begin{equation}
  \label{BindingOfStateEffects}
  \declare
    { \bind{W\mathrm{State}}{D_1, D_2} }
    {
      \Big(
        D_1 \to
        \big(
          W \to W \times D_2
        \big)
      \Big)
      \to
      \Big(
        \big(
          W \to W \times D_1
        \big)
        \to
        \big(
          W \to W \times D_2
        \big)
      \Big)
    }
    {
      \mathrm{prog}
      \;\mapsto\;
      \Big(
         \big(w \mapsto (w'_w, d_w)\big)
         \;\mapsto\;
         \big(
           w \mapsto \mathrm{prog}(d_w)(w'_w)
         \big)
      \Big)
      \mathrlap{\,.}
    }
\end{equation}

 \vspace{-1mm}
\noindent Such $W$State-effectful programs are adjoint \eqref{FormingAdjuncts}
to programs of the form \eqref{QuantumProgramInteractingWithQRAm}
\vspace{-1mm}
$$
  \big(\!\!
  \begin{tikzcd}
    D
    \ar[
      rr,
      "{
        \mathrm{prog}
      }"
    ]
    &&
    {[ W
      ,\,
      W \times D'
    ]}
  \end{tikzcd}
  \!\!\big)
  \;\;\;\;
  \xleftrightarrow{\;\;\;}
  \;\;\;\;
  \big(\!\!
  \begin{tikzcd}
    W \times D
    \ar[
      rr,
      "{
        \widetilde{\mathrm{prog}}
      }"
    ]
    &&
    W \times D'
  \end{tikzcd}
  \!\!\big)
$$

 \vspace{-2mm}
\noindent
(also known as {\it Mealy machines} following \cite{Mealy55}, see e.g. \cite[\S 1.1.3]{Pattison02} for the modern formulation and \cite[p. 262]{OliveiraMiraldo16}\cite[p. 3]{PeroneKarachalias23} for our state-effectful perspective)
which may be understood as producing their nominal output only after {\it reading in} data from ``memory'' type $W$ (as such like the $W$Reader monad above, but) while also re-setting (re-writing) the $W$-data that gets handed  along to a new state.

This way the state monad is the basic computational model\footnote{
 For practical purposes, the state monad is only a crude model for RAM, since it only encodes access to the entire memory at once (first read all of memory then re-write all of memory). In practice, one will want to read/write RAM only partially at a given address. This is also encoded by a (co-)monadic construction: ``lenses'' (see Rem. \ref{RelationToLenses} below), which are the modales over the dual of the state monad: The co-state co-monad \cite{OConnor11}.
} for a {\it random access memory} (``RAM'', see \cite[p. 26 \& Fig. 1.10]{Yates19}):
\vspace{-2mm}
\begin{equation}
  \label{StateMonadAsRAMModel}
  \begin{tikzcd}[
    row sep=-3pt, column sep=huge
  ]
    D
    \ar[
      rr,
      "{
        \scalebox{.7}{
          \color{darkgreen}
          \bf
          \scalebox{1.4}{$W$}-RAM effectful program
        }
      }"{yshift=1pt}
    ]
    &&
    \big[
      W
      ,\,
      W \!\times\! D'
    \big]
    &[-30pt]
    \scalebox{.7}{
      \color{darkblue}
      \bf
      \def\arraystretch{.9}
      \begin{tabular}{l}
        type of
        \\
        $W$State-effectful $D'$-data
      \end{tabular}
    }
    \\
    \underset{
      \mathclap{
        \hspace{35pt}
        \adjustbox{
         rotate=-40,
         scale=.7
        }{
          \hspace{-20pt}
          \color{darkblue}
          \bf
          \def\arraystretch{.9}
          \begin{tabular}{c}
            nominal
            \\
            input data
          \end{tabular}
        }
      }
    }{
     d
    }
      &\longmapsto&
    \scalebox{1.5}{$($}
    \underset{
      \mathclap{
        \hspace{45pt}
        \adjustbox{
         rotate=-40,
         scale=.7
        }{
          \hspace{-12pt}
          \color{darkblue}
          \bf
          \def\arraystretch{.9}
          \begin{tabular}{c}
            RAM readout
          \end{tabular}
        }
      }
    }{
      w
    }
      \mapsto
      \big(
    \underset{
      \mathclap{
        \hspace{45pt}
        \adjustbox{
         rotate=-40,
         scale=.7
        }{
          \hspace{-12pt}
          \color{darkblue}
          \bf
          \def\arraystretch{.9}
          \begin{tabular}{c}
            RAM rewrite
          \end{tabular}
        }
      }
    }{
        w'_{(w,d)}
    }
        ,\,
    \underset{
      \mathclap{
        \hspace{40pt}
        \adjustbox{
         rotate=-40,
         scale=.7
        }{
          \hspace{-23pt}
          \color{darkblue}
          \bf
          \def\arraystretch{.9}
          \begin{tabular}{c}
            nominal
            \\
            output data
          \end{tabular}
        }
      }
    }{
        d'_{(w,d)}
    }
    \big)
    \scalebox{1.5}{$)$}
  \end{tikzcd}
\end{equation}

\end{itemize}

\vspace{-4mm}
\noindent One more example (which is not central to our discussion here but is) illustrative of the general notion of computational
side effects is the {\bf throwing of exceptions} (e.g. \cite[\S 21.2.6]{Milewski19}\cite[1, p. 11]{Uustalu21}): Assuming that the
category $\Types$ has coproducts and with $\mathrm{Msg} : \Types$ some type of error messages, the exception monad is
 \vspace{-2mm}
\begin{equation}
  \label{ExceptinMonad}
  \begin{tikzcd}[row sep=-2pt]
    \mathrm{Exc}_{\mathrm{Msg}}
    &:&
    \Types
    \ar[rr]
    &&
    \Types
    \\
    &&
   \scalebox{0.8}{$ D $}
      &\longmapsto&
  \scalebox{0.8}{$  D
      \sqcup
    \mathrm{Msg}
    $}
  \end{tikzcd}
\end{equation}

 \vspace{-2mm}
\noindent
whose return-operation is the coprojection into coproduct and whose join operation is given by the co-diagonal on $\mathrm{Msg}$:
An $\mathrm{Exc}_{\mathrm{Msg}}$-effectful program with nominal output type $D_2$ is a morphism $D_1 \longrightarrow D_2 \sqcup \mathrm{Msg}$ which {\it may}
return output of type $D_2$ but might instead produce an (error-message) term of type $\mathrm{Msg}$, in which case all subsequently
$\mathrm{Exc}_{\mathrm{Msg}}$-bound programs will not execute but just hand this error message along. (Hence for $\mathrm{Msg} \defneq \ast$
the singleton type, which is also known as the {\it maybe monad}.)

In this example, it is clear that one will wish for programs that can {\it handle} the exception, and hence in general for programs that can handle
a given type of side-effect.

\medskip
\noindent {\bf Effect handling and modal types.}  Given a type of computational side effect $\mathcal{E}$ as above, a program of nominal input type $D_1$
which can {\it handle} the effect will have actual input type $\mathcal{E}(D_1)$, and handle the effect-part of $\mathcal{E}(D)$ in a way compatible
with the incremental binding of effects (\cite{PlotkinPretnar13}):
\vspace{-1mm}
\begin{equation}
  \label{EffectHandling}
  \hspace{0cm}
  \begin{tikzcd}[
    column sep=110pt,
    row sep=15pt,
  ]
    &
    D_1
    \ar[
      r,
      "{
        \mathrm{prog}_{12}
      }",
      "{
        \scalebox{.55}{
          \color{darkgreen}
          \bf
          in-effectful program
        }
      }"{swap, yshift=-4pt}
    ]
    &
    D_2
    \ar[
      d,
      |->,
      shift left=3,
      bend left=90,
      start anchor={[yshift=-5pt]},
      end anchor={[yshift=+10pt]},
      "{
        \scalebox{.75}{
          \color{darkorange}
          \bf
          \def\arraystretch{.9}
\begin{tabular}{c}
            incorporate handling
            \\
            of \scalebox{1.2}{$\mathcal{E}(-)$}-effects
         \end{tabular}
        }
      }"{xshift=-5pt}
    ]
    \\[+5pt]
    &
    \mathcal{E}(D_1)
    \ar[
      r,
      "{
        \handle
          { \mathcal{E} }
          { D_2 }
          { \mathrm{prog}_{12} }
      }"{yshift=+1pt},
      "{
        \scalebox{.55}{
          \color{darkgreen}
          \bf
          \def\arraystretch{.9}
          \begin{tabular}{c}
            in-effectful program
            \\
            handling effects of type \scalebox{1.2}{$\mathcal{E}(-)$}
          \end{tabular}
        }
      }"{swap, yshift=-4pt}
    ]
    &
    D_2
    \\[+10pt]
    D_1
    \ar[
      rr,
      rounded corners,
      to path ={
           ([yshift=-0pt]\tikztostart.south)
        -- ([yshift=-10pt]\tikztostart.south)
        -- node[yshift=-10pt] { \scalebox{.8}{$
             \underset{
               \scalebox{.7}{
                 \color{darkgreen}
                 \bf
                 no effect
               }
             }{
               \mathrm{prog}_{12}
             }
        $} }
           ([yshift=-10pt]\tikztotarget.south)
        -- ([yshift=-0pt]\tikztotarget.south)
      }
    ]
    \ar[
      r,
      "{
        \return
          { \mathcal{E} }
          { D_1 }
      }",
      "{
         \scalebox{.55}{
           \color{darkgreen}
           \bf
           \def\arraystretch{.8}
           \begin{tabular}{c}
             produce
             \\
             trivial effect
           \end{tabular}
         }
      }"{swap, yshift=-0pt}
    ]
    &
    \mathcal{E}(D_1)
    \ar[
      r,
      "{
        \handle
          { \mathcal{E} }
          { D_2 }
          { \mathrm{prog}_{12} }
      }",
      "{
         \scalebox{.55}{
           \color{darkgreen}
           \bf
           \def\arraystretch{.8}
           \begin{tabular}{c}
             handle effects
             \\
             running program
           \end{tabular}
         }
      }"{swap, yshift=-0pt}
    ]
    &
    D_2
    \\[-12pt]
    &&&[-77pt]
    \scalebox{.75}{
      \color{darkorange}
      \bf
      consistency conditions
    }
    \\[+5pt]
    \mathcal{E}(D_0)
    \ar[
      rr,
      rounded corners,
      to path ={
           ([yshift=-0pt]\tikztostart.south)
        -- ([yshift=-11pt]\tikztostart.south)
        -- node[yshift=-16pt] { \scalebox{.9}{$
              \underset{
                \raisebox{-4pt}{
                \scalebox{.65}{
                  \color{darkgreen}
                  \bf
                  \hspace{2pt}
                  handle effects...
                  \hspace{50pt}  consecutively
                }
                }
              }{
              \handle
                { \mathcal{E} }
                { D_2 }
                {
                  \scalebox{1.4}{$($}
                    \raisebox{-.5pt}{$
                    D_0
                    \xrightarrow{
                      \mathrm{prog}_{01}
                    }
                    \mathcal{E}(D_1)
                    \xrightarrow{
                      \handle
                        { \mathcal{E} }
                        { D_2 }
                        { \mathrm{prog}_{12} }
                    }
                    D_2
                    $}
                  \scalebox{1.4}{$)$}
                }
              }
           $} }
           ([yshift=-11pt]\tikztotarget.south)
        -- ([yshift=-0pt]\tikztotarget.south)
      }
    ]
    \ar[
      r,
      "{
        \bind
          { \mathcal{E} }
          {}
          { \mathrm{prog}_{01} }
      }",
      "{
        \scalebox{.55}{
          \color{darkgreen}
          \bf
          \def\arraystretch{.9}
          \begin{tabular}{c}
            carry effects
            \\
            along
          \end{tabular}
        }
      }"{swap}
    ]
    &
    \mathcal{E}(D_1)
    \ar[
      r,
      "{
        \handle
          { \mathcal{E} }
          { D_2 }
          { \mathrm{prog}_{12} }
      }",
      "{
        \scalebox{.55}{
          \color{darkgreen}
          \bf
          \def\arraystretch{.9}
          \begin{tabular}{c}
            handle
            \\
            cumulative effects
          \end{tabular}
        }
      }"{swap}
    ]
    &
    D_2
  \end{tikzcd}
\end{equation}

\vspace{-2mm}
\noindent Such $\mathcal{E}$-effect handling structure on a type $D$ is equivalent to
$\mathcal{E}$-{\bf modale}-structure on $D$ (also known as an {\it $\mathcal{E}$-module}
or {\it $\mathcal{E}$-algebra} structure), namely a morphism
\vspace{-3mm}
\begin{equation}
  \label{EffectHandler}
  \begin{tikzcd}[column sep=huge]
    \mathcal{E}(D)
    \ar[
      rr,
      "{
        \overset{
          \mathclap{
           \raisebox{3pt}{
             \scalebox{.7}{
               \color{darkblue}
               \bf
               monad action
               on
               modale
             }
           }
          }
        }{
        \rho
        \,\defneq\,
        \handle
          { \mathcal{E} }
          { D }
          { \mathrm{id}_D }
        }
      }"
    ]
    &&
    D
  \end{tikzcd}
\end{equation}
\vspace{-2mm}

\noindent satisfying the axioms of a monoid module \eqref{CategoryOfInternalModules}, in that it makes the following squares commute:
\vspace{-2mm}
\begin{equation}
  \label{ModaleAxioms}
  \begin{tikzcd}
    {}
    \ar[
      rr,
      phantom,
      "{
        \scalebox{.7}{
          \color{darkorange}
          \bf
          unitality
        }
      }"
    ]
    &&
    {}
    \\[-18pt]
    D
    \ar[
      drr,
      bend left=20,
      "{ \mathrm{id} }"
    ]
    \ar[
      d,
      "{ \eta_D }"{swap}
    ]
    \ar[
      drr,
      phantom,
      bend right=16,
      "{
        \scalebox{.7}{
          $\mathrm{utl}_{\mathcal{E}}(\rho)$
        }
      }"{pos=.35}
    ]
    &&
    \\
    \mathcal{E}(D)
    \ar[
      rr,
      "{ \rho }"{swap}
    ]
    &&
    D
  \end{tikzcd}
  \hspace{1cm}
  \begin{tikzcd}[column sep=large]
    {}
    \ar[
      rr,
      phantom,
      "{
        \scalebox{.7}{
          \color{darkorange}
          \bf
          action property
        }
      }"
    ]
    &&
    {}
    \\[-18pt]
    \mathcal{E}
    \big(
      \mathcal{E}(D)
    \big)
    \ar[
      rr,
      "{ \mathcal{E}(\rho) }"
    ]
    \ar[
      d,
      "{
        \mu_D
      }"
    ]
    \ar[
      drr,
      phantom,
      "{
        \scalebox{.7}{
          $\mathrm{act}_{\mathcal{E}}(\rho)$
        }
      }"{pos=.4}
    ]
    &&
    \mathcal{E}(D)
    \ar[
      d,
      "{ \rho }"
    ]
    \\
    \mathcal{E}(D)
    \ar[
      rr,
      "{ \rho }"{swap}
    ]
    &&
    D \mathrlap{\,.}
  \end{tikzcd}
\end{equation}

\medskip

\noindent {\bf Categories of effect-handling types.}
A {\it homomorphism} $(D_1, \rho_1) \to (D_2, \rho_2)$ of $\mathcal{E}$-effect handlers, hence of $\mathcal{E}$-modales, is a map of the underlying data types $f : D_1 \xrightarrow{\phantom{--}} D_2$ which respects the $\mathcal{E}$-action in that the following diagram commutes
\begin{equation}
  \label{ModaleHomomorphism}
  \begin{tikzcd}
    \mathcal{E}(D_1)
    \ar[
      rr,
      "{ \mathcal{E}(f) }"
    ]
    \ar[
      d,
      "{ \rho_1 }"
    ]
    &&
    \mathcal{E}(D_2)
    \ar[
      d,
      "{ \rho_2 }"
    ]
    \\
    D_1
    \ar[
      rr,
      "{ f }"
    ]
    &&
    D_2
    \mathrlap{\,,}
  \end{tikzcd}
\end{equation}
which we will indicate by the following notation (which is non-standard but nicely suggestive):
\begin{equation}
  \begin{tikzcd}[sep=-13pt]
    \scalebox{.7}{
      \color{darkblue}
      \bf
      \def\arraystretch{.7}
      \begin{tabular}{c}
        $\mathcal{E}$-modale
        \\
        stucture
      \end{tabular}
    }
    &
    D_1
     \ar[out=180-55, in=55, looseness=5, "\scalebox{1.2}{$\;\mathclap{
        \mathcal{E}
      }\;$}"{description},
     ]
    \ar[
      rr,
      "{ f }",
      "{
        \scalebox{.7}{
          \color{darkgreen}
          \bf
          modale homomorphism
        }
      }"{swap, yshift=-2pt}
    ]
    &{\phantom{-----------}}&
    D_2
     \ar[out=180-55, in=55, looseness=5, "\scalebox{1.2}{$\;\mathclap{
        \mathcal{E}
      }\;$}"{description},
     ]
     &
    \scalebox{.7}{
      \color{darkblue}
      \bf
      \def\arraystretch{.7}
      \begin{tabular}{c}
        $\mathcal{E}$-modale
        \\
        stucture
      \end{tabular}
    }
  \end{tikzcd}
\end{equation}
This makes a {\bf category of $\mathcal{E}$-modales} (traditionally known as the {\it Eilenberg-Moore category} of $\mathcal{E}$ and) traditionally denoted by super-scripting: $\Types^{\mathcal{E}}$.

For example, for any $B : \Types$, the type $\mathcal{E}(B)$ carries $\mathcal{E}$-modale structure, with $\rho \defneq \mu_B$. These are called the {\it free} $\mathcal{E}$-modales and the full sub-category they form is traditionally denoted by sub-scripting, $\Types_{\mathcal{E}}$:
\begin{equation}
  \label{CategoriesOfModales}
  \hspace{-5mm}
  \begin{tikzcd}[
    row sep=5pt,
    column sep=10pt
  ]
  \Types
  \ar[
    r,
    ->>,
    "{
      \overset{
        \mathclap{
          \raisebox{7pt}{
            \scalebox{.7}{
              \color{darkgreen}
              \bf
              free construction
            }
          }
        }
      }{
        F_{\mathcal{E}}
      }
    }"{pos=.35}
  ]
  \ar[
    rrr,
    rounded corners,
      to path={
           ([yshift=+0pt]\tikztostart.south)
        -- ([yshift=-6pt]\tikztostart.south)
        -- node[
             yshift=6pt,
             pos=.14
           ] {
               \scalebox{.7}{
                 $F^{\mathcal{E}}$
               }
             }
           ([yshift=-6pt]\tikztotarget.south)
        -- ([yshift=+0pt]\tikztotarget.south)
      }
    ]
  &
  \overset{
    \mathclap{
      \raisebox{8pt}{
      \scalebox{.7}{
        \def\arraystretch{.9}
        \begin{tabular}{c}
          \color{darkblue}
          \bf
          free $\mathcal{E}$-modales in $\Types$
          \\
          (``Kleisli category'')
        \end{tabular}
      }
      }
    }
  }{
    \Types_{\mathcal{E}}
  }
  \ar[
    rr,
    hook,
    "{
      \overset{
        \mathclap{
          \raisebox{6pt}{
            \scalebox{.7}{
              \color{darkgreen}
              \bf
              total comparison functor
            }
          }
        }
      }{
        K_{ U^{\mathcal{E}} F^{\mathcal{E}} }
      }
    }"
  ]
  &&
  \overset{
    \mathclap{
      \scalebox{.7}{
        \def\arraystretch{.9}
        \begin{tabular}{c}
          \color{darkblue}
          \bf
          $\mathcal{E}$-modales in $\Types$
          \\
          (``Eilenberg-Moore category'')
        \end{tabular}
      }
    }
  }{
    \Types^{\mathcal{E}}
  }
  \\
  \big\{
    B : \Types
  \big\}
  &
  \big\{
    \mathcal{E}(B)
    ,\,
    \rho_B
      \defneq
    \mu_B
    :
    \mathcal{E}
    \big(
      \mathcal{E}(B)
    \big)
    \to
    \mathcal{E}(B)
  \big\}
  &&
  \big\{
    D : \Types
    ,\,
    \rho :
    \mathcal{E}(D)
    \to D
    \;\big\vert\;
    \mathrm{untl}_{\mathcal{E}}(\rho),
    \mathrm{act}_{\mathcal{E}}(\rho)
  \big\}
  \end{tikzcd}
\end{equation}
Incidentally, notice that thereby every $\mathcal{E}$-effect handler $\rho$
\eqref{ModaleAxioms}
is itself a modale-homomorphism
\eqref{ModaleHomomorphism}
from a free modale \eqref{CategoriesOfModales}:
\begin{equation}
  \label{ModaleActionsAreModaleHomomorphisms}
  \begin{tikzcd}[row sep=-12pt, column sep=small]
    \scalebox{.7}{
      \color{darkblue}
      \bf
      \def\arraystretch{.9}
      \begin{tabular}{c}
        free modale
        \\
        structure
      \end{tabular}
    }
    &
    \mathcal{E}(D)
     \ar[out=180-55, in=55, looseness=5, "\scalebox{1.2}{$\;\mathclap{
        \mathcal{E}
      }\;$}"{description},
     ]
    \ar[
      rr,
      "{
        \overset{
          \mathclap{
            \raisebox{1pt}{
              \scalebox{.7}{
                \color{darkgreen}
                \bf
                given effect handler
              }
            }
          }
        }{
          \rho \,\defneq\,
          \handle{\mathcal{E}}{D}{}
        }
      }",
      "{
        \scalebox{.7}{
          \color{darkgreen}
          \bf
          modale homomorphism
        }
      }"{swap, yshift=-2pt}
    ]
    &\phantom{----------}&
    D
     \ar[out=180-55, in=55, looseness=5, "\scalebox{1.2}{$\;\mathclap{
        \mathcal{E}
      }\;$}"{description},
     ]
     &
    \scalebox{.7}{
      \color{darkblue}
      \bf
      \def\arraystretch{.9}
      \begin{tabular}{c}
        given modale
        \\
        structure
      \end{tabular}
    }
  \end{tikzcd}
\end{equation}
(and regarding it this way is crucial for the monadic typing of quantum measurement, see p. \pageref{QuantumMeasurementCopenhagenStyle} below).

Concretely, the {\it Kleisli equivalence} re-identifies the homomorphism between free $\mathcal{E}$-modales with the
$\mathcal{E}$-effectful programs that we started with \eqref{BindingAndReturning},
as follows (e.g. \cite[Prop. 1.4.6]{Borceux94}):
\vspace{-2mm}
\begin{equation}
  \label{KleisliEquivalence}
  \begin{tikzcd}[row sep=-3pt, column sep=large]
    \Types_{\mathcal{E}}
    \ar[
      rr,
      hook
    ]
    &&
    \Types^{\mathcal{E}}
    \\
  \scalebox{0.8}{$  D $}
  &\longmapsto&
   \scalebox{0.8}{$  \big(
      \mathcal{E}(D)
      ,\,
      \mu_D
    \big)
    $}
    \\[+5pt]
    \Types_{\mathcal{E}}
    (D, D')
    \ar[
      rr,
      <->,
      "{ \sim }",
      "{
        \scalebox{.7}{
          \color{darkgreen}
          \bf
          Kleisli equivalence
        }
      }"{swap, yshift=-1pt}
    ]
    &&
    \Types^{\mathcal{E}}
    \Big(
      \big(\mathcal{E}(D),\mu_D\big)
      ,\,
      \big(\mathcal{E}(D'),\mu_{D'}\big)
    \Big)
    \\
    \scalebox{0.8}{$ \Big(
      D
      \xrightarrow{
        \hspace{22pt}
        f
        \hspace{22pt}
      }
      \mathcal{E}(D')
    \Big)
    $}
    &\longmapsto&
     \scalebox{0.8}{$\Big(
      \mathcal{E}(D)
      \xrightarrow{ \mathcal{E}(f)}
      \mathcal{E}\big(\mathcal{E}(D')\big)
      \xrightarrow{ \mu_{D'} }
      \mathcal{E}(D')
    \Big)$}
    \\
  \scalebox{0.8}{$   \Big(
    D
    \xrightarrow{
      \unit{\mathcal{E}}{D}
    }
    \mathcal{E}(D)
    \xrightarrow{ \phi }
    \mathcal{E}(D')
    \Big)    $}
    &\longmapsfrom&
     \scalebox{0.8}{$\Big(
    \mathcal{E}(D)
    \xrightarrow{
      \hspace{39pt}
      \phi
      \hspace{39pt}
    }
    \mathcal{E}(D')
    \Big)$}.
  \end{tikzcd}
\end{equation}

 \vspace{-2mm}
\noindent
This free construction is readily checked to be left adjoint to evident forgetful functors
 \vspace{-2mm}
\begin{equation}
  \begin{tikzcd}[row sep=0pt, column sep=huge]
    \Types_{\mathcal{E}}
    \ar[
      r,
      hook,
      "{
        K_{U^{\mathcal{E}} F^{\mathcal{E}}}
      }"{description}
    ]
    \ar[
      rrr,
      rounded corners,
      to path={
           ([yshift=+0pt]\tikztostart.north)
        -- ([yshift=+8pt]\tikztostart.north)
        -- node[yshift=7pt, pos=.71] {
             \scalebox{.7}{$
               U_{\mathcal{E}}
             $}
           }
           ([yshift=+8pt]\tikztotarget.north)
        -- ([yshift=+0pt]\tikztotarget.north)
      }
    ]
    &
    \Types^{\mathcal{E}}
    \ar[
      rr,
      " U^{\mathcal{E}} ",
      "{
        \scalebox{.7}{
          \color{darkgreen}
          \bf
          forgetful functor
        }
      }"{swap, yshift=-1pt}
    ]
    &&
    \Types
    \\
    &
   \scalebox{0.8}{$  \big(
      D,
      \rho
        :
      \mathcal{E}(D)
        \to
      D
    \big)
    $}
    &\longmapsto&
  \scalebox{0.8}{$   D $}
  \end{tikzcd}
\end{equation}

 \vspace{-2mm}
\noindent
and both adjunctions $F_{\mathcal{E}} \dashv U_{\mathcal{E}}$ and $F^{\mathcal{E}} \dashv U^{\mathcal{E}}$ re-induce \eqref{MonadFromAdjunction}
the original monad, with the modale structure  recovered from the adjunction counit $\counit{}{}$
(e.g. \cite[\S VI.2, Thm. 1, \S IV.5, Thm. 1]{MacLane71}):
 \vspace{-1mm}
\begin{equation}
  \label{ModaleActionFromEMAdjunctionCOhunit}
  (D,\rho)
  \,\isa\,
  \Type^{\mathcal{E}}
  \hspace{1cm}
  \yields
  \hspace{1cm}
  \begin{tikzcd}[row sep=small,
    column sep=20pt
  ]
    U^{\mathcal{E}}
    F^{\mathcal{E}}
    U^{\mathcal{E}}(D,\rho)
    \ar[
      dd,
      "{
        U^{\mathcal{E}}
        (
        \counit
          {  }
          { (D,\rho) }
        )
      }"
      {swap}
    ]
    \ar[
      r,
      equals
    ]
    &
    \mathcal{E}(D)
    \ar[
      dd,
      "{ \rho }"
    ]
    \\
    \\
    U^{\mathcal{E}}(D,\rho)
    \ar[
      r,
      equals
    ]
    &
    D \mathrlap{\,.}
  \end{tikzcd}
\end{equation}

 \vspace{-1mm}
\noindent
In fact, {\it every} adjunction which induces $\mathcal{E}$ is ``in between'' these two adjunctions,
in that it fits into a commuting diagram of the following form (e.g. \cite[\S VI.3]{MacLane71}):
\begin{equation}
\label{ComparisonFunctor}
\hspace{-1cm}
\adjustbox{
  raise=-0pt
}{
$
  \begin{tikzcd}[
    row sep=40pt, column sep=85pt
  ]
    &&
 \quad    \FreeModales
      { \mathcal{E} }
      { \Types }
    \mathrlap{
      \scalebox{.7}{
        \def\arraystretch{.9}
        \begin{tabular}{c}
          \color{darkblue}
          \bf
          free
          $\mathcal{E}$-modales in $\Types$
          \\
          (``Kleisli category'')
        \end{tabular}
      }
    }
    \ar[
      d,
      hook,
      "{ K_{U F} }",
      "{
        \scalebox{.6}{
          \color{darkgreen}
          \bf
          \def\arraystretch{.9}
          \begin{tabular}{c}
            initial
            \\
            comparison
            \\
            functor
          \end{tabular}
        }
      }"{sloped, swap}
    ]
    \ar[
      from=dll,
      ->>,
      shorten <=4pt,
      shift left=9.5pt,
      "{
        F_{\mathcal{E}}
      }"{description, sloped},
      "{
        B
        \,\mapsto\,
        \left(
          \mathcal{E}(B)
          ,\,
          \rho \defneq \mu_B
        \right)
      }"{yshift=5pt, sloped}
    ]
    \ar[
      dll,
      shift left=4pt,
      shorten >=5pt,
      "{
        U_{\mathcal{E}}
      }"{description, sloped}
    ]
    \ar[
      dll,
      phantom,
      shift right=3.6pt,
      "{\scalebox{.6}{$\bot$}}"{pos=.5,sloped}
    ]
    \\
    \mathllap{
      \scalebox{.7}{
        \color{darkorange}
        \bf
        induced monad
      }
      \hspace{22pt}
    }
    \Types
    \ar[out=180-50, in=180+50,
      looseness=5,
      shorten=-3pt,
      "\scalebox{1.6}{${
          \hspace{1pt}
          \mathclap{
            \mathcal{E}
          }
          \hspace{1pt}
        }$}"{pos=.5, description},
    ]
    &&
  \quad   \Types'
    \mathrlap{
      \scalebox{.7}{
        \def\arraystretch{.9}
        \begin{tabular}{c}
          {\color{darkorange}
          \bf
          any adjunction
          }
          \hspace{-4pt}
          for $\mathcal{E}$
        \end{tabular}
      }
    }
    \ar[
      d,
      "{ K^{U F} }",
      "{
        \scalebox{.6}{
          \color{darkgreen}
          \bf
          \def\arraystretch{.9}
          \begin{tabular}{c}
            terminal
            \\
            comparison
            \\
            functor
          \end{tabular}
        }
      }"{sloped, swap}
    ]
    \ar[
      from=ll,
      shift left=7pt,
      shorten <=27pt,
      "{
        F
      }"{description, sloped}
    ]
    \ar[
      ll,
      shift left=5.5pt,
      shorten >=27pt,
      "{
        U
      }"{description, sloped}
    ]
    \ar[
      ll,
      phantom,
      "{
        \scalebox{.6}{$
          \bot
        $}
      }"{sloped, yshift=.8pt, xshift=-.5pt}
    ]
    \\
    &&
  \quad   \Modales
     { \mathcal{E} }
     { \Types }
    \mathrlap{
      \scalebox{.7}{
        \def\arraystretch{.9}
        \begin{tabular}{c}
          \color{darkblue}
          \bf
          $\mathcal{E}$-modales in $\Types$
          \\
          (``EM-category'')
        \end{tabular}
      }
    }
    \ar[
      from=ull,
      shift left=3pt,
      shorten <=5pt,
      "{
        F^{\mathcal{E}}
      }"{description, sloped}
    ]
    \ar[
      ull,
      shift left=9.5pt,
      "{
        U^{\mathcal{E}}
      }"{description, sloped},
      "{
        \scalebox{.7}{
          \color{darkorange}
          \bf
          monadic adjunction
        }
      }"{swap, , yshift=-3pt, sloped}
    ]
    \ar[
      ull,
      phantom,
      shift left=3.5pt,
      "{
        \scalebox{.6}{$
          \bot
        $}
      }"{pos=.52, sloped}
    ]
  \end{tikzcd}
$
}
\end{equation}

\medskip

\noindent {\bf The monadicity theorem} (cf. \cite[Thm. 4.4.4]{Borceux94}) characterizes the monadic adjunctions
on the bottom of diagram \eqref{ComparisonFunctor}: For a functor $U$ to be {\it monadic} in that it is of the form $U^{\mathcal{E}}$ in \eqref{ComparisonFunctor},
it is sufficient\footnote{
  The necessity clause involves the preservation of those coequalizers that are ``split'',  which we disregard here for brevity
  since we will not need it.
} that
\begin{itemize}
  \item[{\bf (i)}]
    $U$ is conservative (reflects isomorphisms),
  \item[{\bf (ii)}]
    $U$ has a left adjoint $F$,
  \item[{\bf (iii)}]
    $\mathrm{dom}(U)$ has coequalizers and
    $U$ preserves them;
\end{itemize}

\smallskip
\noindent and hence for a functor $U$ between cocomplete categories monadic it is, in particular, sufficient that:
\smallskip
\begin{itemize}
  \item[{\bf (i)}]
    $U$ is conservative,
  \item[{\bf (ii)}]
    $U$ has besides the left adjoint $F$ also a right adjoint,
\end{itemize}
in which case:
\vspace{-.4cm}
\begin{equation}
  \label{MonadicityTheorem}
  \hspace{1cm}
    \Rightarrow
  \hspace{1cm}
  \adjustbox{raise=2pt}{
  \begin{tikzcd}[
    row sep=20pt
  ]
    \mathcal{D}
    \ar[
      d,
      shift left=7pt,
      "{ U }"
    ]
    \ar[
      from=d,
      shift left=7pt,
      "{ F }"
    ]
    \ar[
      d,
      phantom,
      "{
        \scalebox{.7}{$\dashv$}
      }"
    ]
    \\
    \Types
  \end{tikzcd}
  }
  \hspace{.1cm}
  \mbox{is monadic}
  \hspace{1cm}
    \Rightarrow
  \hspace{1cm}
  \adjustbox{raise=12pt}{
  \begin{tikzcd}[
    row sep=20pt,
    column sep=15pt
  ]
    \mathcal{D}
    \ar[
      dr,
      shift left=7pt,
      shorten <=3pt,
      "{ U }"{description}
    ]
    \ar[
      from=dr,
      shift left=7pt,
      "{ F }"{description}
    ]
    \ar[
      dr,
      phantom,
      "{
        \scalebox{.7}{$\dashv$}
      }"{sloped, rotate=90}
    ]
    \ar[
      rr,
      shift left=2pt,
      "{
        \sim
      }"
    ]
    &&
    \Modales
      { \mathcal{E} }
      { \Types }
    \ar[
      dl,
      shift left=7pt,
      "{
         U^{\mathcal{E}}
      }"{description}
    ]
    \ar[
      from=dl,
      shift left=7pt,
      shorten >=3pt,
      "{
        F^{\mathcal{E}}
      }"{description}
    ]
    \ar[
      dl,
      phantom,
      "{
        \scalebox{.7}{$\dashv$}
      }"{sloped, rotate=90}
    ]
    \\
    &
    \;\;\;\;\;\;\;
    \Types
    \;\;\;\;\;\;\;
     \ar[out=-180+55, in=-55, looseness=5, "\scalebox{1.2}{$\;\mathclap{
        \mathcal{E}
      }\;$}"{description},
    ]
  \end{tikzcd}
  }
\end{equation}

\medskip

\noindent
{\bf Relative monads.}
While monads are equivalent to computational effects as in \eqref{BindingAndReturning}, the latter notion has a suggestive generalization to what are called {\it relative monads}  \cite{ACU15}(see also \cite{ArkorMcDermott23}), where the effect-attaching functor $\mathcal{E}$ \eqref{MonadFunctorFromBindReturn} is not an endofunctor but maps between two different categories of types
$$
  \mathcal{E}
  \;\isa\;
  \Types \to \Types'
  \,.
$$
An common situation is where $\Types \hookrightarrow \Types'$ is a subcategory inclusion, where it just means that $\mathcal{E}$-effects are attachable only to types in this subcategory. Generally one can consider any comparison functor
\begin{equation}
  \label{ComparisonFunctorForRelativeMonads}
 J
 \;\isa\;
 \Type \to \Type'
\end{equation}
and define a {\it $J$-relative monad} structure to be given by {\it $J$-relative} return- and bind-operations:
\vspace{-3mm}
\begin{equation}
  \label{RelativeBindOperation}
  \def\arraystretch{2}
  \def\arraycolsep{10pt}
  \begin{array}{lcl}
  D
    \,\isa\,
  \Type
  &
  \yields
  &
  \return
    { \mathcal{E} }
    { D }
  \;\isa\;
  J(D)
  \to
  \mathcal{E}(D)
  \hspace{1cm}
  \\
  D_1, D_2 \,\isa\, \Type
  &\yields&
  \bind
    { \mathcal{E} }
    { D_1, D_2 }
  \;\isa\;
  \big(
    J(D_1)
    \to
    \mathcal{E}(D_2)
  \big)
  \longrightarrow
  \big(
    \mathcal{E}(D_1)
    \to
    \mathcal{E}(D_2)
  \big)
  \end{array}
\end{equation}
otherwise satisfying the same form of consistency conditions as in the non-relative case \eqref{UnitalityForBind}.

As a simple but relevant example, for every actual monad $\mathcal{E}'$ on $\Types'$, its precomposition with any functor $J \,\isa\, \Type \to \Types'$ \eqref{ComparisonFunctorForRelativeMonads} yields a $J$-relative monad (\cite[Prop. 2.3]{ACU15}) via:
\begin{equation}
  \label{RelativeMonadByPrecomposition}
  \mathcal{E}
  \;\defneq\;
  \mathcal{E} \circ J
  \,,
  \;\;\;\;\;\;\;\;
  \return
    { \mathcal{E} }
    { D }
  \,\defneq\,
  \return
    { \mathcal{E}' }
    { J(D) }
  \,,
  \;\;\;\;\;\;\;\;
  \bind
    { \mathcal{E} }
    { D_1, D_2 }
  \,\defneq\,
  \bind
    { \mathcal{E}' }
    { J(D_1), D_2 }
  \,.
\end{equation}

\medskip

\noindent
{\bf Monad transformations.}
With monads encoding effectful programs, one is bound to consider several monadic effects $\mathcal{E}$, $\mathcal{E}'$, ... at once, and procedures that {\it transform} these into each other:
\begin{equation}
  \label{ComponentOfMonadTransformation}
  D \,\isa\, \Type
  \;\;\;\;\;\;
  \yields
  \;\;\;\;\;
  \transform
    { \mathcal{E} \to \mathcal{E}' }
    {D}
  \,\isa\,
  \begin{tikzcd}
    \mathcal{E}(D)
    \ar[rr]
    &&
    \mathcal{E}'(D)
    \,.
  \end{tikzcd}
\end{equation}
For consistency these transformations
\eqref{ComponentOfMonadTransformation}
ought to respect the return- and bind-operations \eqref{BindingAndReturning}, in that the following diagrams commute:
\begin{equation}
\label{ConditionsOnMonadTransformer}
\def\arraystretch{1}
\def\arraycolsep{1.9pt}
\hspace{-3.5mm}
\begin{array}{lcl}
    D \,\isa\, \Type
    &
      \yields
    &
    \begin{tikzcd}
      D
      \ar[
        rr,
        "{
          \return
            { \mathcal{E} }
            { D }
        }"
      ]
      \ar[
        rrrr,
        rounded corners,
        to path={
             ([yshift=-00pt]\tikztostart.south)
          -- ([yshift=-17pt]\tikztostart.south)
          -- node[yshift=7.5pt]{
              \scalebox{.9}{$
                \return
                  { \mathcal{E}' }
                  { D }
              $}
          }
             ([yshift=-15pt]\tikztotarget.south)
          -- ([yshift=-00pt]\tikztotarget.south)
        }
      ]
      &&
      \mathcal{E}(D)
      \ar[
        rr,
        "{
          \transform
            { \mathcal{E} \to \mathcal{E}' }
            {D}
        }"
      ]
      &&
      \mathcal{E}'(D)
    \end{tikzcd}
  \\
  \phantom{A}
  \\
  \def\arraystretch{1.2}
  \begin{array}{c}
    \mathrm{prog}_{12}
    \,\isa\,
    D_1 \to \mathcal{E}(D_2)
    \\
    \mathrm{prog}_{23}
    \,\isa\,
    D_2 \to \mathcal{E}(D_3)
  \end{array}
  &\yields&
  \begin{tikzcd}[column sep=30pt]
    D_1
    \ar[
      rr,
      "{
        \mathrm{prog}_{12}
      }"
    ]
    \ar[
      dd,
      equals
    ]
    &&
    \mathcal{E}(D_2)
    \ar[
      rr,
      "{
        \bind
          { \mathcal{E} }
          {  }
          { \mathrm{prog}_{23} }
      }"{yshift=1pt}
    ]
    &&
    \mathcal{E}(D_3)
    \ar[
      rr,
      "{
        \transform
          { \mathcal{E} \to \mathcal{E}' }
          { D_3 }
      }"
    ]
    &&
    \mathcal{E}'(D_3)
    \ar[
      dd,
      equals
    ]
    \\
    {}
    \\
    D_1
    \ar[
      r,
      "{
        \mathrm{prog}_{12}
      }"
    ]
    &
    \mathcal{E}(D_2)
    \ar[
      r,
      "{
        \transform
          { \mathcal{E} \to \mathcal{E}' }
          { D_2 }
      }"
    ]
    &
    \mathcal{E}'(D_2)
    \ar[
      rrrr,
      "{
      \scalebox{0.8}{$
        \bind
          { \mathcal{E}' }
          {  }
        \Big(
          D_2
            \xrightarrow{ \mathrm{prog}_{23}  }
          \mathcal{E}(D_3)
            \xrightarrow{
              \hspace{-3pt}
              \transform
                { \mathcal{E} \to \mathcal{E}' }
                { D_3 }
              \hspace{-5pt}
            }
          \mathcal{E}'(D_3)
        \Big)
        $}
      }"
    ]
    && &&
    \mathcal{E}'(D_3)
    \mathrlap{\,,}
  \end{tikzcd}
  \end{array}
\end{equation}
hence such that the Kleisli composition \eqref{KleisliComposition} is respected:
\begin{equation}
  \label{MonadTransformationRespectsKleisliComposition}
  \big(
  \transform
    { \mathcal{E} \to \mathcal{E}' }
  { D_2 }
  \circ
  \mathrm{prog}_{12}
  \big)
  \;\;
  \mbox{\tt >=>}
  \;\;
  \big(
  \transform
    { \mathcal{E} \to \mathcal{E}' }
  { D_3 }
  \circ
  \mathrm{prog}_{23}
  \big)
  \;\;\;\;\;\;
  =
  \;\;\;\;\;\;
  \transform
    { \mathcal{E} \to \mathcal{E}' }
  { D_2 }
  \circ
  \big(
  \mathrm{prog}_{12}
  \;\;
    \mbox{\tt >=>}
  \;\;
  \mathrm{prog}_{23}
  \big)
  \mathrlap{\,,}
\end{equation}
exhibiting a covariant functor on free modales \eqref{CategoriesOfModales}
\vspace{-2mm}
\begin{equation}
  \label{CovariantFunctorOnModalesInducedByMonadTransformation}
  \begin{tikzcd}[
    row sep=2pt
  ]
    &
    \Types
    \ar[
      dl,
      end anchor={[yshift=2pt]},
      "{ F_{\mathcal{E}} }"{swap}
    ]
    \ar[
      dr,
      end anchor={[yshift=4pt]},
      "{ F_{\mathcal{E}'} }"
    ]
    \\
    \Types_{\mathcal{E}}
    \ar[
      rr
    ]
    &&
    \Types_{\mathcal{E}'}
    \\
  \scalebox{0.8}{$ \mathcal{E}(D_1)
    \xrightarrow{ \phi }
    \mathcal{E}(D_2)
    $}
    &\longmapsto&
\scalebox{0.8}{$ \bind
      { \mathcal{E}' }
      {}
    \bigg(
    D_1
    \xrightarrow{
      \return
      { \mathcal{E} }
      { D_1 }
    }
    \mathcal{E}(D_1)
    \xrightarrow{ \phi }
    \mathcal{E}(D_2)
    \xrightarrow{
      \transform
        { \mathcal{E} \to \mathcal{E}' }
        { D_2 }
    }
    \mathcal{E}'(D_2)
    \bigg)
    $}
  \end{tikzcd}
\end{equation}

\vspace{-2mm}
\noindent which preserves (as indicated on top) the free maps \eqref{CategoriesOfModales},
$\phi = \mathcal{E}(f) \mapsto \mathcal{E}'(f)$, due to the commutativity of the following pasting
diagram (the left square being the unitality condition in \eqref{ConditionsOnMonadTransformer},
the right square the implied naturality property \eqref{MonadTransformation}):
\vspace{-2mm}
$$
  \begin{tikzcd}[row sep=small, column sep=large]
    D_1
    \ar[dd, equals]
    \ar[
      rr,
      "{
        \return
          { \mathcal{E} }
          { D_1 }
      }"
    ]
    &&
    \mathcal{E}(D_1)
    \ar[
      rr,
      "{
        \mathcal{E}(f)
      }"
    ]
    \ar[
      dd,
      "{
        \transform
          { \mathcal{E} \to \mathcal{E}' }
          { D_1 }
      }"
    ]
    &&
    \mathcal{E}(D_2)
    \ar[
      dd,
      "{
        \transform
          { \mathcal{E} \to \mathcal{E}' }
          { D_2 }
      }"
    ]
    \\
    \\
    D_1
    \ar[
      rr,
      "{
        \return
          { \mathcal{E} }
          { D_1 }
      }"
    ]
    &&
    \mathcal{E}'(D_1)
    \ar[
      rr,
      "{
        \mathcal{E}'(f)
      }"
    ]
    &&
    \mathcal{E}'(D_2)
    \,.
  \end{tikzcd}
$$

\vspace{-2mm}
\noindent This notion of {\it monad transformers}
originates with \cite[\S 2.6]{Espinosa95}, the explicit definition \eqref{ConditionsOnMonadTransformer} is due to
\cite[p. 339]{LiangHudakJones95}
now commonly used in {\Haskell}\footnote{\href{https://hackage.haskell.org/package/transformers-0.5.6.2/docs/Control-Monad-Trans-Class.html\#g:1}{\tt hackage.haskell.org/package/transformers-0.5.6.2/docs/Control-Monad-Trans-Class.html\#g:1}}.
But we may observe that the equivalent definition not in terms of the bind- but the join-operation (considered within {\Haskell} in \cite[\S 2.2]{SPWJ19}) is much older:

Namely in category theory, a {\it morphism of monads} is known to be a natural transformation of their underlying functors \eqref{MonadFunctorFromBindReturn}
\vspace{0mm}
\begin{equation}
  \label{MonadTransformation}
  \transform
    { \mathcal{E} \to \mathcal{E}' }
    {  }
  \,\isa\,
  \mathcal{E}
  \xrightarrow{\quad}
  \mathcal{E}'
\end{equation}
which is compatible with the return- and join-operations
\eqref{MonadStructure} as follows:
\begin{equation}
  \label{MonadTransformationConditions}
  \hspace{-4mm}
 D \,\isa\, \Type
 \;\;\;\;\;\;
 \vdash
 \;\;\;\;\;\;
  \begin{tikzcd}[column sep=large]
    D
    \ar[
      d,
      "{
        \unit
          { \mathcal{E} }
          { D }
      }"{swap}
    ]
    \ar[r, equals]
    &
    D
    \ar[
      d,
      "{
        \unit
          { \mathcal{E}' }
          { D }
      }"
    ]
    \\
    \mathcal{E}(D)
    \ar[
      r,
      "{
        \mathrm{trans}^{\mathcal{E} \to \mathcal{E}'}_D
      }"{pos=.6}
    ]
    &
    \mathcal{E}'(D)
    \mathrlap{\,,}
  \end{tikzcd}
  \hspace{.7cm}
  \begin{tikzcd}[column sep=50pt]
    \mathcal{E}
    \circ
    \mathcal{E}(D)
    \ar[
      r,
      "{
        \mathcal{E}\big(
          \mathrm{trans}^{
            \mathcal{E} \to \mathcal{E}'
          }
            _D
        \big)
      }"
    ]
    \ar[
      d,
      "{
        \join
          { \mathcal{E} }
          {D}
      }"{swap}
    ]
    &
    \mathcal{E}
    \circ
    \mathcal{E}'(D)
    \ar[
      r,
      "{
        \mathrm{trans}^{
          \mathcal{E} \to \mathcal{E}'
        }
          _{\mathcal{E}'(D)}
     }"
    ]
    &
    \mathcal{E}'
    \circ
    \mathcal{E}'(D)
    \ar[
      d,
      "{
        \join
          { \mathcal{E}' }
          {D}
      }"
    ]
    \\
    \mathcal{E}(D)
    \ar[
      rr,
      "{
        \mathrm{trans}
          ^{ \mathcal{E} \to \mathcal{E}' }
          _D
      }"
    ]
    &&
    \mathcal{E}'(D)   \mathrlap{\,.}
  \end{tikzcd}
\end{equation}

\vspace{-2mm}
\noindent Notice here that the order of the composites at the top right of \eqref{MonadTransformationConditions} is arbitrary,
since the naturality of $\transform{\mathcal{E} \to \mathcal{E}'}{}$ implies that the following diagram commutes:

\newpage
$$
  \begin{tikzcd}[row sep=small, column sep=huge]
    \mathcal{E}\big(
      \mathcal{E}(D)
    \big)
    \ar[
      rr,
      "{
        \mathcal{E}\big(
          \transform
            { \mathcal{E} \to \mathcal{E}' }
            { D }
        \big)
      }"
    ]
    \ar[
      dd,
      "{
        \transform
          { \mathcal{E} \to \mathcal{E}' }
          { \mathcal{E}(D) }
      }"{swap}
    ]
    &&
    \mathcal{E}\big(
      \mathcal{E}'(D)
    \big)
    \ar[
      dd,
      "{
        \transform
          { \mathcal{E} \to \mathcal{E}' }
          { \mathcal{E'}(D) }
      }"
    ]
    \\
    \\
    \mathcal{E}'\big(
      \mathcal{E}(D)
    \big)
    \ar[
      rr,
      "{
        \mathcal{E}'\big(
          \transform{ \mathcal{E} \to \mathcal{E}' }
          { D }
        \big)
      }"{yshift=0pt}
    ]
    &&
    \mathcal{E}'\big(
      \mathcal{E}'(D)
    \big)
  \end{tikzcd}
$$

This definition \eqref{MonadTransformationConditions}
of monad morphism is implicit in \cite[pp. 39]{Benabou67} (whose identification of monads as lax 2-functors out of the terminal category implies that their morphisms should be the corresponding lax natural transformations), first made explicit in \cite{Maranda66} and then in \cite{Frei69}\cite[p. 330]{Pumpluen70}\cite[pp. 150]{Street72}\footnote{Beware that \cite{Street72} says ``transformation'' for the 2-morphisms in the 2-category of monads,  while we use it for the 1-morphisms, matching the completely standard terminology for the 1-morphisms of their underlying
endofunctors and staying close to the established use of ``monad transformers'' \eqref{PointedEndofunctorOnCategoryOfMonads}.}, often in slight further generality. A transparent textbook account is in \cite[\S 6.1]{BarrWells85}, discussion in the context of monadic computations effects is in \cite[Def. 4.0.11]{Moggi89Abstract}.

One readily checks\footnote{
  We are not aware of an explicit reference providing this equivalence; for the record we have spelled it out at: \\ \href{https://ncatlab.org/nlab/show/monad+transformer\#EquivalenceOfDefinitions}{\tt ncatlab.org/nlab/show/monad+transformer\#EquivalenceOfDefinitions}.
} that the conditions \eqref{ConditionsOnMonadTransformer} and \eqref{MonadTransformationConditions} are equivalent under the translation \eqref{MonadStructure};  in particular the naturality of the transformation \eqref{MonadTransformation} is already implied by \eqref{ConditionsOnMonadTransformer}.

If we denote by $\mathrm{Mnd}(\Type)$ the category whose objects are the monads on the type system and whose morphisms are monad transformations in the form \eqref{MonadTransformation}, then their equivalence with \eqref{CovariantFunctorOnModalesInducedByMonadTransformation} means that we have a faithful functor from monad transformations to functors between free modales:
$$
  \begin{tikzcd}[row sep=-2pt]
    \mathrm{Mnd}(\Types)
    \ar[rr]
    &&
    \Types/\mathrm{Cat}
    \\
  \scalebox{0.8}{$  \mathcal{E}  $}
    &\longmapsto&
    \scalebox{0.8}{$   \Types_{\mathcal{E}} $}
    \mathrlap\,.
  \end{tikzcd}
$$
(This is known to experts but scarcely represented in the literature: The functor is alluded to in \cite[Lem. 10.2]{Linton69}
and only recently was discussed \cite[Cor. 6.49]{ArkorMcDermott23} in detail but much more abstractly.)

\smallskip

For example, there is a {\it unique} transformation from the identity monad (the trivial effect) to any other monad $\mathcal{E}$,
making the identity monad the initial object in the category of monads:
\vspace{-1mm}
\begin{equation}
  \label{InitialMonadTransformation}
  \exists !
  \;
 \transform
   { \mathrm{Id} \to \mathcal{E} }
  \,\isa\,
  \mathrm{Id} \to \mathcal{E}
  ,\,
  \;\;
  \mbox{since}\;\;
  \begin{tikzcd}
    D
    \ar[
      rr,
      equals
    ]
    \ar[
      d,
      equals
    ]
    &&
    D
    \ar[
      d,
      "{
        \unit
          { \mathcal{E} }
          { D }
      }"
    ]
    \\
    D
    \ar[
      rr,
      "{
        \mathrm{trans}^{\mathrm{Id} \to \mathcal{E}}_D
      }",
      "{
        :=
        \unit
          { \mathcal{E} }
          { D }
      }"{swap}
    ]
    &&
    \mathcal{E}(D)
    \mathrlap{\,,}
  \end{tikzcd}
  \hspace{.7cm}
  \begin{tikzcd}[column sep=45pt]
    D
    \ar[
      r,
      "{
        \unit
          { \mathcal{E} }
          { D }
      }"
    ]
    \ar[
      d,
      equals
    ]
    &
    \mathcal{E}(D)
    \ar[
      r,
      "{
        \unit
          { \mathcal{E} }
          { \mathcal{E}(D) }
     }"
    ]
    \ar[
      dr,
      gray,
      equals
    ]
    &
    \mathcal{E}
    \circ
    \mathcal{E}(D)
    \ar[
      d,
      "{
        \join
          { \mathcal{E} }
          {D}
      }"
    ]
    \\
    D
    \ar[
      rr,
      "{
        \unit
          { \mathcal{E} }
          { D }
      }"
    ]
    &&
    \mathcal{E}(D)
    \mathrlap{\,.}
  \end{tikzcd}
\end{equation}

But in fact, \cite[p. 339]{LiangHudakJones95} and the functional programming/{\Haskell}-community following them impose a further condition
on monad transformers $\transform{\mathcal{E} \to \mathcal{E}'}{D}$, namely that they themselves arrange into the component maps of a pointed endofunctor
\begin{equation}
  \label{PointedEndofunctorOnCategoryOfMonads}
  \mathrm{Id}
    \xrightarrow{ \mathrm{trans} }
  (\mbox{-})' \,\isa\,
 \mathrm{Mnd} \to \mathrm{Mnd}
\end{equation}
on the category of monads (made explicit in this form in \cite[p. 474]{Winitzki22}).
This is tailored towards the application of {\it combining} monadic effects and hence regarding $\mathcal{E}'$ as behaving like the composition of $\mathcal{E}$ with another effect.

In addition to the covariant functor on free modales \eqref{CovariantFunctorOnModalesInducedByMonadTransformation},
a transformation between monads \eqref{MonadTransformation}
{\it contra}variantly induces (\cite[Thm. 2]{Frei69}, cf. \cite[Thm. 6.3]{BarrWells85}) a functor between
their general modales  \eqref{ModaleHomomorphism} by what we may recognize as the usual ``extension of scalars''-formula from algebra:
\begin{equation}
  \label{ExtensionOfModalesAlongMonadTransformations}
  \begin{tikzcd}[
    column sep=60pt,
    row sep=1pt
  ]
    \mathcal{E}'
    \ar[
      from=rr,
      "{
        \mathrm{trans}^{ \mathcal{E} \to \mathcal{E}' }
      }"{swap}
    ]
    &&
    \mathcal{E}
    \\
    \Types^{\mathcal{E}'}
    \ar[
      rr,
      "{
      }"
    ]
    &&
    \Types^{\mathcal{E}}
    \\
  \end{tikzcd}
\end{equation}
\vspace{-5pt}
$$
{\small
  \begin{tikzcd}
    \mathcal{E}'(D_1)
    \ar[
      d,
      "{
        \rho'_1
      }"
    ]
    \ar[
      r,
      "{
        \mathcal{E}'(\phi)
      }"
    ]
    &
    \mathcal{E}'(D_2)
    \ar[
      d,
      "{
        \rho'_2
      }"
    ]
    \\
    D_1
    \ar[
      r,
      "{
        \phi
      }"
    ]
    &
    D_2
  \end{tikzcd}
  }
  \hspace{10pt}
  \longmapsto
  \hspace{17pt}
  {\footnotesize
  \begin{tikzcd}[
    row sep=20pt
  ]
    \mathcal{E}(D_1)
    \ar[
      r,
      "{
        \mathcal{E}(\phi)
      }"
    ]
    \ar[
      d,
      "{
        \mathrm{trans}
          ^{ \mathcal{E} \to \mathcal{E}' }
          _{D_1}
      }"{description, pos=.46}
    ]
    \ar[
      dd,
      rounded corners,
      to path ={
           ([xshift=-00pt]\tikztostart.west)
        -- ([xshift=-8pt]\tikztostart.west)
        -- node[xshift=-5pt] {
               \scalebox{.7}{$
                 \rho_1
               $}
            }
            ([xshift=-14pt]\tikztotarget.west)
        -- ([xshift=-00pt]\tikztotarget.west)
      }
    ]
    &
    \mathcal{E}(D_2)
    \ar[
      d,
      "{
        \mathrm{trans}
          ^{ \mathcal{E} \to \mathcal{E}' }
          _{D_2}
      }"{description, pos=.46}
    ]
    \ar[
      dd,
      rounded corners,
      to path ={
           ([xshift=+00pt]\tikztostart.east)
        -- ([xshift=+8pt]\tikztostart.east)
        -- node[xshift=+5pt] {
               \scalebox{.7}{$
                 \rho_2
               $}
            }
            ([xshift=+14pt]\tikztotarget.east)
        -- ([xshift=-00pt]\tikztotarget.east)
      }
    ]
    \\[+3pt]
    \mathcal{E}'(D_1)
    \ar[
      d,
      "{
        \rho'_1
      }"
    ]
    \ar[
      r,
      "{
        \mathcal{E}'(\phi)
      }"
    ]
    &
    \mathcal{E}'(D_2)
    \ar[
      d,
      "{
        \rho'_2
      }"
    ]
    \\
    D_1
    \ar[
      r,
      "{
        \phi
      }"
    ]
    &
    D_2
  \end{tikzcd}
  }
$$

\smallskip

\noindent
{\bf Composite effect monads.} With computational side-effects encoded by monads $\mathcal{E}$, $\mathcal{E}'$, ...,
one is bound to consider {\it combined effects} encoded by {\it composite monads}
\vspace{-1mm}
\begin{equation}
  \label{CompositeMonad}
  \mathcal{E}' \circ \mathcal{E}
  \;\colon\;
  \Type \to \Type
  \,.
\end{equation}

\vspace{-1mm}
\noindent In order for the combined join-operation on the composite underlying functors to exist in
an evident way, one needs a natural transformation between the two possible orders of composition
\vspace{-1mm}
\begin{equation}
  \label{DistributivityOfMonadOverMonad}
  \distribute
    {
      \mathcal{E}
      ,\,
      \mathcal{E}'
    }
  \,\colon\,
  \mathcal{E} \circ \mathcal{E}'
  \longrightarrow
  \mathcal{E}' \circ \mathcal{E}
  \,,
\end{equation}

\vspace{-1mm}
\noindent because then the candidate composite join-operation is this:
\vspace{-2mm}
\begin{equation}
  \label{StructureMapsOfCompositeMonad}
  \hspace{-3mm}
  \begin{tikzcd}[
    row sep=30pt,
    column sep=32pt
  ]
    &&[-5pt] &[-20pt]
    \mathcal{E}' \circ
    \mathcal{E} \circ \mathcal{E}
    \ar[
      dr,
      "{
        \mathcal{E}'\big(
          \join
            { \mathcal{E} }
            { (-) }
        \big)
      }"{sloped, pos=.6}
    ]
    &[-10pt]
    \\
    \mathcal{E}' \circ \mathcal{E}
    \circ
    \mathcal{E}' \circ \mathcal{E}
    \ar[
      rr,
      "{
        \mathcal{E}'\big(
          \distribute
            {
              \mathcal{E},
              \mathcal{E}'
            }
            { \mathcal{E}(-) }
        \big)
      }"{swap}
    ]
    &\phantom{-}&
    \mathcal{E}' \circ
    \mathcal{E}'
    \circ
    \mathcal{E}'
    \circ
    \mathcal{E}
    \ar[
      ur,
      shorten <=-4,
      "{
        \join
          { \mathcal{E}' }
          {
            \mathcal{E}
            \circ
            \mathcal{E}(-)
          }
      }"{sloped}
    ]
    \ar[
      dr,
      "{
        \mathcal{E}'
        \circ
        \mathcal{E}'
        \big(
        \join
          { \mathcal{E} }
          {
            (-)
          }
        \big)
      }"{swap, sloped, pos=.35}
    ]
    &
    &
    \mathcal{E}'
    \circ
    \mathcal{E}
    \ar[
      from=llll,
      crossing over,
      rounded corners,
      to path={
           ([yshift=+00pt]\tikztostart.north)
        -- ([yshift=+16pt]\tikztostart.north)
        --  node[yshift=7pt, sloped, pos=.3] {
           \scalebox{.7}{$
              \join
                {
                  \mathcal{E}'
                  \circ
                  \mathcal{E}
                }
                { (-) }
           $}
        }
            ([yshift=+04pt]\tikztotarget.west)
      }
    ]
    \mathrlap{\,,}
    \\
    &&
    &
    \mathcal{E}'
      \circ
    \mathcal{E}'
      \circ
    \mathcal{E}
    \ar[
      ur,
      "{
        \join
          {
            \mathcal{E}'
          }
          {
            \mathcal{E}(-)
          }
      }"{swap, sloped}
    ]
  \end{tikzcd}
  \hspace{.7cm}
  \begin{tikzcd}
    &
    \mathcal{E}'
    \ar[
      dr,
      "{
        \mathcal{E}'\big(
          \unit
            { \mathcal{E} }
            { (-) }
        \big)
      }"{sloped}
    ]
    \\
    \mathrm{id}
    \ar[
      rr,
      "{
        \unit
          {
            \mathcal{E}' \circ \mathcal{E}
          }{
            (-)
          }
      }"{description}
    ]
    \ar[
      dr,
      "{
        \unit
          { \mathcal{E} }
          { (-) }
      }"{swap, sloped}
    ]
    \ar[
      ur,
      "{
        \unit
          { \mathcal{E}' }
          { \mathcal{E}(-) }
      }"{sloped}
    ]
    &&
    \mathcal{E}'
    \circ
    \mathcal{E}
    \mathrlap{\,.}
    \\
    &
    \mathcal{E}
    \ar[
      ur,
      "{
        \unit
          { \mathcal{E}' }
          { (-) }
      }"{swap, sloped}
    ]
  \end{tikzcd}
\end{equation}

\vspace{-1mm}
\noindent For this construction to satisfy the monad axioms \eqref{MonadAxioms}, the distributivity
transformation \eqref{DistributivityOfMonadOverMonad} needs to make the following diagrams commute (\cite[\S 1]{Beck69},
review in \cite[\S 9 2.1]{BarrWells85}):

\vspace{-15mm}
\begin{equation}
\label{MonadOverMonadDistributivityAxioms}
  \def\arraystretch{6}
  \begin{array}{c}
  \begin{array}{cc}
  \begin{tikzcd}[row sep=small, column sep=large]
    \mathcal{E}
    \ar[
      rr,
      equals
    ]
    \ar[
      dd,
      "{
        \mathcal{E}\big(
          \unit
            { \mathcal{E}' }
            { (-) }
        \big)
      }"{swap}
    ]
    &&
    \mathcal{E}
    \ar[
      dd,
      "{
        \unit
          { \mathcal{E}' }
          { \mathcal{E}(-) }
      }"
    ]
    \\
    \\
    \mathcal{E}
    \circ
    \mathcal{E}'
    \ar[
      rr,
      "{
        \distribute
          { \mathcal{E}, \mathcal{E}' }
          { (-) }
      }"
    ]
    &&
    \mathcal{E}'
    \circ
    \mathcal{E}
  \end{tikzcd}
  &
  \begin{tikzcd}[row sep=small, column sep=large]
    \mathcal{E}'
    \ar[
      rr,
      equals
    ]
    \ar[
      dd,
      "{
        \unit
          { \mathcal{E} }
          { \mathcal{E}'(-) }
      }"{swap}
    ]
    &&
    \mathcal{E}'
    \ar[
      dd,
      "{
        \mathcal{E}'
        \big(
        \unit
          { \mathcal{E} }
          { (-) }
        \big)
      }"
    ]
    \\
    \\
    \mathcal{E}
    \circ
    \mathcal{E}'
    \ar[
      rr,
      "{
        \distribute
          { \mathcal{E}, \mathcal{E}' }
          { (-) }
      }"
    ]
    &&
    \mathcal{E}'
    \circ
    \mathcal{E}
  \end{tikzcd}
  \end{array}
  \\
  \begin{tikzcd}[row sep=small, column sep=large]
    \mathcal{E}
    \circ
    \mathcal{E}'
    \circ
    \mathcal{E}'
    \ar[
      rr,
      "{
        \distribute
          { \mathcal{E}, \mathcal{E}' }
          { \mathcal{E}'(-) }
      }"
    ]
    \ar[
      dd,
      "{
        \mathcal{E}
        \big(
        \join
          { \mathcal{E}' }
          { (-) }
        \big)
      }"{swap}
    ]
    &&
    \mathcal{E}'
    \circ
    \mathcal{E}
    \circ
    \mathcal{E}'
    \ar[
      rr,
      "{
        \mathcal{E}'
        \big(
        \distribute
          { \mathcal{E}, \mathcal{E}' }
          { (-) }
        \big)
      }"
    ]
    &&
    \mathcal{E}'
    \circ
    \mathcal{E}'
    \circ
    \mathcal{E}
    \ar[
      dd,
      "{
        \join
          { \mathcal{E}' }
          { \mathcal{E}(-) }
      }"
    ]
    \\
    \\
    \mathcal{E}
    \circ
    \mathcal{E}'
    \ar[
      rrrr,
      "{
        \distribute
          { \mathcal{E}, \mathcal{E}' }
          { (-) }
      }"
    ]
    &&&&
    \mathcal{E}'
    \circ
    \mathcal{E}
  \end{tikzcd}
  \\
  \begin{tikzcd}[row sep=small, column sep=large]
    \mathcal{E}
    \circ
    \mathcal{E}
    \circ
    \mathcal{E}'
    \ar[
      rr,
      "{
        \mathcal{E}\big(
        \distribute
          {
            \mathcal{E},
            \mathcal{E}'
          }
          { (-) }
        \big)
      }"
    ]
    \ar[
      dd,
      "{
        \join
          { \mathcal{E}, \mathcal{E}' }
          { \mathcal{E}'(-) }
      }"{swap}
    ]
    &&
    \mathcal{E}
    \circ
    \mathcal{E}'
    \circ
    \mathcal{E}
    \ar[
      rr,
      "{
        \distribute
          { \mathcal{E}, \mathcal{E}' }
          {
            \mathcal{E}(-)
          }
      }"
    ]
    &&
    \mathcal{E}'
    \circ
    \mathcal{E}
    \circ
    \mathcal{E}
    \ar[
      dd,
      "{
         \mathcal{E}'\big(
         \join
          { \mathcal{E}, \mathcal{E}' }
          { (-) }
        \big)
      }"
    ]
    \\
    \\
    \mathcal{E}
    \circ
    \mathcal{E}'
    \ar[
      rrrr,
      "{
        \distribute
          { \mathcal{E}, \mathcal{E}' }
          { (-) }
      }"
    ]
    &&&&
    \mathcal{E}'
    \circ
    \mathcal{E}
  \end{tikzcd}
  \end{array}
\end{equation}

\medskip

\noindent  {\bf Computational contexts and co-monads on the type system.}
All of the above discussion of effect-monads has a formally dual incarnation
(by reversal of all arrows in the above diagrams),
now given by {\it co-monads} on the type system, which some authors refer to as ``computational co-effect'' but which may
naturally be understood as expressing {\it computational contexts} \cite{UustaluVene08}\cite{POM13}.
The idea now is, dually, that a program which {\it nominally reads in} data of some type $D$ while however executing
in dependence on some further context must {\it de facto} read in data of some adjusted type $\mathcal{C}(D)$ which
is such that the context-part of the adjusted data is being transferred (extended) to followup programs:
\vspace{0mm}
\begin{equation}
  \label{CoBindingAndCoReturning}
  \hspace{-3mm}
  \begin{tikzcd}[
    column sep=50pt,
    row sep=35pt
  ]
    \mathcal{C}(D_1)
    \ar[
      r,
      "{
        \mathrm{prog_{12}}
      }",
      "{
        \scalebox{.7}{
          \color{darkblue}
          \bf
          first program
        }
      }"{yshift=10pt},
      "{
        \scalebox{.55}{
          \def\arraystretth{.7}
          \color{darkgreen}
          \bf
\begin{tabular}{c}
            output data
            \\
            of type $D_2$
            \\
            obtained in context of type $\mathcal{C}(-)$
          \end{tabular}
        }
      }"{swap, yshift=-2pt}
    ]
    &
    D_2
    &[-40pt]
    \mathcal{C}(D_2)
    \ar[
      r,
      "\mathrm{prog_{23}}",
      "{
        \scalebox{.7}{
          \color{darkblue}
          \bf
          second program
        }
      }"{yshift=10pt},
      "{
        \scalebox{.55}{
          \def\arraystretth{.7}
          \color{darkgreen}
          \bf
          \begin{tabular}{c}
            input data
            \\
            of nominal type
            \scalebox{1.15}{$D_2$}
            \\
            having context of type \scalebox{1.15}{$\mathcal{C}(-)$}
          \end{tabular}
        }
      }"{swap, yshift=-2pt}
    ]
    &
    D_3
    \ar[
      d,
      |->,
      shift left=3,
      bend left=90,
      start anchor={[yshift=-5pt]},
      end anchor={[yshift=+10pt]},
      "{
        \scalebox{.6}{
          \color{darkorange}
          \bf
          \begin{tabular}{c}
            extend previous context
            \\
            over second program
         \end{tabular}
        }
      }"{xshift=-4pt}
    ]
    &
    &[+20pt]
    \mathcal{C}(D)
    \ar[
      r,
      "{
        \obtain
          { \mathcal{C} }
          { D }
      }",
      "{
        \scalebox{.55}{
          \color{darkgreen}
          \bf
          obtain plain data
          from
          \scalebox{1.2}{$\mathcal{C}(-)$}-context
        }
      }"{swap, yshift=-16pt, pos=.4}
    ]
    &
    D
    \\
    \mathcal{C}(D_1)
    \ar[
      r,
      "{
        \extend
         { \mathcal{E} }
         {}
         {
           \mathrm{prog_{12}}
         }
      }"{yshift=2pt},
      "{
        \scalebox{.55}{
          \def\arraystretth{.7}
          \color{darkgreen}
          \bf
          \begin{tabular}{c}
            extend any previous \\
            \scalebox{1.3}{$\mathcal{C}(-)$}-context going forward
          \end{tabular}
        }
      }"{swap, yshift=-5pt}
    ]
    &
    \mathcal{C}(D_2)
    &
    \mathcal{C}(D_2)
    \ar[
      r,
      "{
        \mathrm{prog}_{23}
      }"
    ]
    &
    D_3
    \ar[
      d,
      |->,
      shift left=3,
      bend left=90,
      start anchor={[yshift=-5pt]},
      end anchor={[yshift=+10pt]},
      "{
        \scalebox{.7}{
          \color{darkorange}
          \bf
          \begin{tabular}{c}
            compose
         \end{tabular}
        }
      }"{xshift=-4pt}
    ]
    &&
    \mathcal{E}(D)
    \ar[
      r,
      "{
        \extend
          { \mathcal{C} }
          {}
          {
            \obtain
              { \mathcal{C} }
              { D }
          }
      }"{yshift=4pt},
      "{
        \;=\;
        \mathrm{id}_{\mathcal{E}(D)}
      }"{swap}
     ]
    &
    \mathcal{E}(D)
    \\
    \mathcal{C}(D_1)
    \ar[
      rrr,
      "{
        \mathrm{prog}_{23}
        \;\circ\;
        \extend
          { \mathcal{C} }
          {}
          { \mathrm{prog}_{12} }
      }",
      "{
        \scalebox{.7}{
          \color{darkblue}
          \bf
          \scalebox{1.3}{$\mathcal{C}$}-composite program
        }
      }"{yshift=14pt},
      "{
        \scalebox{.55}{
          \def\arraystretth{.7}
          \color{darkgreen}
          \bf
          \begin{tabular}{c}
            in shared
            \scalebox{1.15}{$\mathcal{C}(-)$}-context
          \end{tabular}
        }
      }"{swap, yshift=-4pt}
    ]
    &&&
    D_3
  \end{tikzcd}
\end{equation}

Further, by formal duality, all the above discussion for monadic effects and their modal types gives rise to analogous phenomena of comonadic contexts
and their (co)modal types. In particular, comonads are induced on the other sides of an adjunction
 \vspace{-5mm}
\eqref{MonadFromAdjunction}:
\begin{equation}
  \label{CoMonadFromAdjunction}
  \begin{tikzcd}[column sep=large]
    \Types'
    \ar[
      rr,
      shift right=6pt,
      "{
        \underset{
          \mathclap{
            \raisebox{-1pt}{
            \scalebox{.7}{
              \color{darkgreen}
              \bf
              right adjoint
            }
            }
          }
        }{
          R
        }
      }"{swap}
    ]
    \ar[
      from=rr,
      shift right=6pt,
      "{
        \overset{
          \mathclap{
            \raisebox{+1pt}{
            \scalebox{.7}{
              \color{darkgreen}
              \bf
              right adjoint
            }
            }
          }
        }{
        L
        }
      }"{swap}
    ]
    \ar[
      rr,
      phantom,
      "{
        \scalebox{.7}{$\bot$}
      }"
    ]
    &&
    \Types
    \ar[out=+50, in=-50,
      looseness=4.5,
      shorten=-3pt,
      shift left=0pt,
      "\scalebox{1.1}{${
          \hspace{1pt}
          \mathclap{
            {
              L \circ R
              \,=:\,
              \mathcal{C}
            }
          }
          \hspace{4pt}
        }$}"{
          pos=.5,
          xshift=14pt
        }
    ]
  \end{tikzcd}
  \hspace{.5cm}
  \scalebox{.7}{
    \color{darkorange}
    \bf
    induced co-monad
  }
\end{equation}
 \vspace{-3mm}

\noindent
{\bf Examples of context comonads.}
Dualizing the example of the state monad \eqref{StateMonadEndofunctor} yields the {\bf costate comonad}
(or {\it store comonad}, cf. \cite{Milewski19}\cite[3, p. 14]{Uustalu21}):
\vspace{-2mm}
\begin{equation}
  \label{CostateComonadEndofunctor}
  \begin{tikzcd}[sep=0pt]
    \mathllap{W \mathrm{Store}\;:\;}
    \Types
    \ar[rr]
    && \Types
    \\
 \scalebox{0.8}{$   D  $}
    &\longmapsto&
 \scalebox{0.8}{$    W \times [W,\,D] $}
  \end{tikzcd}
\end{equation}

\vspace{-3mm}
\noindent with operations
\vspace{-2mm}
\begin{equation}
  \label{ObtainAndExtendOfStoreComonad}
  \hspace{-5mm}
  \declare
    {
      \obtain
        { W\mathrm{Store} }
        { D }
    }
    {
      W \times [W,D]
      \to
      D
    }
    {
      (w,f)
      \,\mapsto\,
      f(w)
    }
  \hspace{.5cm}
  \declare
    {
      \extend
        {
          W\mathrm{Store}
        }
        {}
        {D}
    }
    {
      \big(W \times [W,D] \to D\big)
      \,\mapsto\,
      \big(
        W \times [W,D]
      \big)
    }
    {
      \big(
        (w,f)
        \,\mapsto\,
        \mathcal{O}(w,f)
      \big)
      \,\mapsto\,
      \Big(
        (w,f)
        \,\mapsto\,
        \big(
          w
          ,\,
          \mathcal{O}(\mbox{-},f)
        \big)
      \Big)
    }
\end{equation}
which means that $W\mathrm{Store}(D)$ is the type of $W$-indexed supply (``storage'') $f \isa W \to D$  of $D$-data equipped with an address $w \isa W$ of one such $D$-datum, which is the one that is {\tt obtained} from such a computational context.

\smallskip

Similarly, dualizing the previous examples
\eqref{ClassicalWriterMonad}\eqref{WriterMonad}
of read/write-effect monads this way, one obtains the following list of
{\bf examples of reader/writer (co)monads}:
\begin{equation}
\label{RederWriterCoMonads}
\adjustbox{}{
\def\arraystretch{1.4}
\begin{tabular}{|l|l|l|}
  \hline
  \bf
  (Co)monad name
  &
  \bf
  Underlying endofunctor
  &
  \bf
  (Co)monad structure induced by
  \\
  \hline
  \hline
  Reader monad
  &
  [W,\,\,\mbox{-}\,]
  on cartesian types
  &
  unique comonoid structure on $W$
  \\
  \hline
  CoReader comonad
  &
  $W \!\times\! (\mbox{-})$
  on cartesian types
  &
  unique comonoid structure on $W$
  \\
  \hline
  Writer monad
  &
  $A \!\otimes\! (\mbox{-})$
  on monoidal types
  &
  chosen monoid structure on $A$
  \\
  \hline
  \
  \multirow{2}{*}{
  \hspace{-10pt}
  CoWriter comonad
  }
  &
  $[A,\,\, \mbox{-}\,]$
  on monoidal types
  &
  chosen monoid structure on $A$
  \\
  &
  $A \!\otimes\! (\mbox{-})$
  on monoidal types
  &
  chosen comonoid structure on $A$
  \\
  \hline
  \hline
  \hspace{-10pt}
  \def\arraystretch{.9}
  \begin{tabular}{c}
    Writer/CoWriter
    \\
    Frobenius monad
  \end{tabular}
  &
  $A \!\otimes\! (\mbox{-})$
  on monoidal types
  &
  chosen Frob. monoid
  structure on $A$
  \\
  \hline
\end{tabular}
}
\end{equation}

\noindent {\bf Adjoint (co)monads.} In the case of an {\it adjoint triple} of adjoint functors the induced (co)monads
are themselves pairwise adjoint --- as in  \eqref{MonadicModalityInIntroduction}, a situation central to our discussion in \cref{QuantumEffects}.
In this case their categories of (co)modales \eqref{CategoriesOfModales} are isomorphic (e.g. \cite[\S V.8, Thm. 2]{MacLaneMoerdijk92}):
\vspace{-2mm}
\begin{equation}
  \label{AdjointCoMonadsHaveIsomorphicModales}
  \overset{
    \mathclap{
      \raisebox{5pt}{
        \scalebox{.7}{
          \color{darkblue}
          \bf
          adjoint (co)monads
        }
      }
    }
  }{
  \mathcal{E}
  \;\dashv\;
  \mathcal{C}
  }
  \hspace{1cm}
  \overset{
    \mathclap{
      \raisebox{5pt}{
        \scalebox{.7}{
          have
        }
      }
    }
  }{
    \yields
  }
  \hspace{1cm}
  \adjustbox{
    raise=-6pt
  }{
  \begin{tikzcd}[
    column sep=45pt,
    row sep=5pt
  ]
    \Types^{\mathcal{E}}
    \ar[
      rr,
      <->,
      "{ \sim }",
      "{
        \scalebox{.7}{
          \color{darkblue}
          \bf
            equivalent categories
            of modales
        }
      }"{yshift=8pt}
    ]
    \ar[
      dr,
      shorten=-2pt,
      "{
        U^{\mathcal{E}}
      }"{swap}
    ]
    &&
    \Types^{\mathcal{C}}
    \ar[
      dl,
      shorten=-2pt,
      "{
        U^{\mathcal{C}}
      }"
    ]
    \\
    &
    \Types
  \end{tikzcd}
  }
\end{equation}

\noindent {\bf Frobenius monads.}
Something special happens here when the underlying endo-functors in \eqref{AdjointCoMonadsHaveIsomorphicModales} are not just adjoints but also identified, $\mathcal{E} \,\simeq\, \mathcal{C}$.
In this case, their (co)monad structures fuse to a single {\it Frobenius monad}-structure
\cite[pp. 151]{Lawvere69EquationalDoctrines}\cite{Street04}\cite{Lauda06}
--- induced  via \eqref{ComparisonFunctor} and \eqref{CoMonadFromAdjunction} from an ``ambidextrous'' adjunction, where the left and the right adjoint of a middle functor agree
\vspace{-2mm}
\begin{equation}
\label{FrobeniusMonadFromAmbidextrousAdjunction}
\begin{tikzcd}[column sep=40pt]
    \ar[
      rr,
      phantom,
      "{
        \scalebox{.7}{
          \color{darkgreen}
          \bf
          ambidextrous adjunction
        }
      }"
    ]
    &&
    {}
    \\
    \mathllap{
      \scalebox{.7}{
        \color{darkblue}
        \bf
        Frobenius monad
      }
      \hspace{.8cm}
    }
    \Types
    \ar[out=160, in=60,
      looseness=4,
      "{
         \scalebox{1.3}{$
           \mathcal{E}
         $}
      }"{pos=.26, description},
    ]
    \ar[out=185, in=175,
      looseness=10,
      phantom,
      "{\scalebox{.7}{$\simeq$}}"{yshift=1pt, xshift=3pt, rotate=90}
    ]
    \ar[out=-160, in=-60,
      looseness=4,
      shorten=-0pt,
      "{
         \scalebox{1.3}{$
           \mathcal{C}
         $}
      }"{pos=.26, description},
    ]
  \ar[
    rr,
    shift left=14pt,
    "{
       L \,\defneq\, R
    }"
  ]
  \ar[
    from=rr,
    "{
      C
    }"{description}
  ]
  \ar[
    rr,
    shift right=14pt,
    "{
       R \,\defneq\, L
    }"{swap}
  ]
  \ar[
    rr,
    phantom,
    shift left=8pt,
    "\scalebox{.7}{$\bot$}"
  ]
  \ar[
    rr,
    phantom,
    shift right=8pt,
    "\scalebox{.7}{$\bot$}"
  ]
  &&
  \Types
  \mathrlap{\,,}
\end{tikzcd}
\end{equation}

\vspace{-4mm}
\noindent so-called because these monads are {\it Frobenius algebras} (Frobenius monoids, see e.g. \cite[\S 5]{HeunenVicary19})
internal to the category of endofunctors: Combined (co)algebras whose (co)products are compatible in the sense that all ways that map $n$
input elements to $m$ output elements by $(n-1)$ products and $(m-1)$-coproducts coincide.
For {\bf example} -- shown in the last line of \eqref{CoMonadFromAdjunction}: if type $A$ carries Frobenius algebra structure, then the
induced (Co)Reader (co)monad $A\otimes (\mbox{-})$ carries induced Frobenius monad structure.

\newpage

\noindent
{\bf Combined contextful and effectful programs.}
We have seen effectful programs typed as maps into monad types $\mathcal{E}(-)$ \eqref{BindingAndReturning}
and contextful programs typed as maps out of comonad types $\mathcal{C}(-)$ \eqref{CoBindingAndCoReturning}.
Of course, in general a program may be {\it both} effectful as well as context-dependent, in which case it should clearly be a map of the form
\vspace{-2mm}
\begin{equation}
  \label{ContextEffectfulProgram}
  \mathrm{prog}_{12}
  \;\;\;\colon\;
  \begin{tikzcd}
    \mathcal{C}(D_1)
    \ar[
      r
    ]
    &
    \mathcal{E}(D_2)
    \mathrlap{\,.}
  \end{tikzcd}
\end{equation}

\vspace{-2mm}
\noindent In order for such procedures to have a consistent composition, the context-comonad $\mathcal{C}$ needs to be compatible
with the effect-monad $\mathcal{E}$ in the following way, known as a {\it distributivity law} for comonads over monads (\cite[Def. 3]{BrookesVanStone93}\footnote{Beware that \cite[Def. 3]{BrookesVanStone93} refer to \eqref{DistributivityTransformation} as the {\it monad distributing over the comonad} instead of the other way around (therein following convention for the original discussion of monads distributing over monads in \cite[\S 1]{Beck69}); but comparison with the eponymous case in arithmetic --- $a \times \sum_i b_i \;\mapsto\; \sum_i a \times b_i $ --- as well as with our main Ex. \ref{QuantumStoreDistributesOverQuantumRreader} makes our converse terminology more natural, which also coincides with the terminology used in \cite[p. 138]{PowerWatanabe02}. In any case, the formulas will always make unambiguously clear what is meant.}).
Namely, the order of application of the (co)monads must be interchangeable via a natural transformation
\vspace{-2mm}
\begin{equation}
  \label{DistributivityTransformation}
  D \,\isa\,
  \Type
  \qquad
    \yields
  \qquad
  \begin{tikzcd}
      \mathrm{distr}
        ^{ \mathcal{C}, \mathcal{E} }
        _{ D }
      \;\isa\;\;
    \mathcal{C}
    \big(\mathcal{E}(D)\big)
    \ar[
      r
    ]
    &
    \mathcal{E}
    \big(
     \mathcal{C}(D)
    \big)
  \end{tikzcd}
\end{equation}

\vspace{-2mm}
\noindent
that make the following diagrams commute, not unlike the conditions on monad transformations \eqref{ComponentOfMonadTransformation}:
\vspace{-3mm}
\begin{equation}
  \label{CoMonadDistributivityConditions}
  \def\arraystretch{5}
\begin{array}{c}
  \begin{tikzcd}[row sep=small]
    \mathcal{C}(D)
    \ar[
      rr,
      equals
    ]
    \ar[
      dd,
      "{
        \mathcal{C}(
          \unit
            { \mathcal{E} }
            { D }
        )
      }"
      {swap}
    ]
    &&
    \mathcal{C}(D)
    \ar[
      dd,
      "{
          \unit
            { \mathcal{E} }
            { \mathcal{C}(D) }
      }"
    ]
    \\
    \\
    \mathcal{C}
    \big(
      \mathcal{E}(D)
    \big)
    \ar[
      rr,
      "{
        \mathrm{distr}
          ^{ \mathcal{C}, \mathcal{E} }
          _{ D }
      }"
    ]
    &&
    \mathcal{E}
    \big(
      \mathcal{C}(D)
    \big)
  \end{tikzcd}
  \hspace{.8cm}
  \begin{tikzcd}[row sep=small]
    \mathcal{C}\big(
      \mathcal{E}(D)
    \big)
    \ar[
      rr,
      "{
        \mathrm{distr}
          ^{\mathcal{C}, \mathcal{E}}
          _D
      }"
    ]
    \ar[
      dd,
      "{
        \counit
          { \mathcal{C} }
          { \mathcal{E}(D) }
      }"
      {swap}
    ]
    &&
    \mathcal{E}\big(
      \mathcal{C}(D)
    \big)
    \ar[
      dd,
      "{
        \mathcal{E}
        (
        \counit
          { \mathcal{C} }
          { D }
        )
      }"
    ]
    \\
    \\
    \mathcal{E}(D)
    \ar[
      rr,
      equals
    ]
    &&
    \mathcal{E}(D)
  \end{tikzcd}
  \\
\end{array}
\end{equation}

With such distributivity structure, the $\mathcal{C}$-context-dependent $\mathcal{E}$-effectful programs \eqref{ContextEffectfulProgram} have a consistent composition (\cite[Thm. 3]{BrookesVanStone93}\cite[Prop. 7.4]{PowerWatanabe02}) by combining the $\mathcal{C}$-context extension \eqref{CoBindingAndCoReturning} of the first with the $\mathcal{E}$-effect binding \eqref{BindingAndReturning} of the second, concatenated via the distributivity transformation \eqref{DistributivityTransformation}:
\begin{equation}
  \label{CoMonadicKleisliComposition}
  \hspace{-4mm}
  \begin{tikzcd}[
    column sep=20pt
  ]
    &&
    \mathcal{C}(D_1)
    \ar[
      rr,
      "{
        \mathrm{prog}_{12}
      }"
    ]
    &&
    \mathcal{E}(D_2)
    &&
    \mathcal{C}(D_2)
    \ar[
      rr,
      "{
        \mathrm{prog}_{23}
      }"
    ]
    &&
    \mathcal{E}(D_3)
    \\[20pt]
    \mathcal{C}(D_1)
    \ar[
      rr,
      "{
        \duplicate
          { \mathcal{C} }
          { D }
      }"
    ]
    \ar[
      rrrr,
      rounded corners,
      to path={
           ([yshift=+00pt]\tikztostart.north)
        -- ([yshift=+10pt]\tikztostart.north)
        -- node[yshift=7pt]{
             \scalebox{.7}{$
               \extend
                 { \mathcal{C} }
                 { }
               \mathrm{prog}_{12}
              $}
           }
           ([yshift=+09pt]\tikztotarget.north)
        -- ([yshift=+00pt]\tikztotarget.north)
      }
    ]
    \ar[
      rrrrrrrrrr,
      rounded corners,
      to path={
           ([yshift=+00pt]\tikztostart.south)
        -- ([yshift=-10pt]\tikztostart.south)
        -- node[yshift=-8pt]{$
              \mathrm{prog}_{12}
              \;\;\mbox{\tt >=>}\;\;
              \mathrm{prog}_{23}
           $}
           ([yshift=-09pt]\tikztotarget.south)
        -- ([yshift=+00pt]\tikztotarget.south)
      }
    ]
    &&
    \mathcal{C}\big(
      \mathcal{C}(D_1)
    \big)
    \ar[
      rr,
      "{
        \mathcal{C}(
          \mathrm{prog}_{12}
        )
      }"{yshift=1.5pt}
    ]
    &&
    \mathcal{C}\big(
      \mathcal{E}(D_2)
    \big)
    \ar[
      rr,
      "{
        \mathrm{distr}
          ^{ \mathcal{C}, \mathcal{E} }
          _{ D_2 }
      }"{yshift=1.5pt}
    ]
    &&
    \mathcal{E}\big(
      \mathcal{C}(D_2)
    \big)
    \ar[
      rr,
      "{
        \mathcal{E}(
          \mathrm{prog}_{23}
        )
      }"{yshift=1.5pt}
    ]
    \ar[
      rrrr,
      rounded corners,
      to path={
           ([yshift=+00pt]\tikztostart.north)
        -- ([yshift=+09pt]\tikztostart.north)
        -- node[yshift=7pt]{
             \scalebox{.7}{$
               \bind
                 { \mathcal{E} }
                 { }
               \mathrm{prog}_{23}
              $}
           }
           ([yshift=+10pt]\tikztotarget.north)
        -- ([yshift=+00pt]\tikztotarget.north)
      }
    ]
    &&
    \mathcal{E}\big(
      \mathcal{E}(D_3)
    \big)
    \ar[
      rr,
      "{
        \join
          { \mathcal{E} }
          { D_3 }
      }"
    ]
    &&
    \mathcal{E}(D_3)
  \end{tikzcd}
\end{equation}
Notice that for $\mathcal{C} = \mathrm{Id}$ or $\mathcal{E} = \mathrm{Id}$ the trivial (co)monad also the distributivity may be taken to be the identity and then this composition reduces to the Kleisli composition \eqref{KleisliComposition} of purely contextful- or purely effectful programs, whence we may use the same notation $\mbox{\tt >=>}$ also for this general case.

\end{literature}

\begin{literature}[\bf Classical structures via Frobenius monads]
  \label{QuantumMeasurementAndZXCalculus}

 The QuantumEnvironment (co)monad expressing quantum measurement effects which we derive in Prop. \ref{QuantumCoEffectsViaFrobeniusAlgebra} (cf. Rem. \ref{QuantumStateEffectAsQuantumWriter} and p. \pageref{DistributingFrobeniusMonads}) was originally considered for this purpose in \cite{CoeckePavlovic08}\cite{CoeckePaquette08}\cite{CPP09}\cite{CPV12}, partial review in \cite{HeunenVicary19}.
 Its graphical formalization as part of the \zxCalculus\footnote{{\zxCalculus} landing page: \href{https://zxcalculus.com}{\tt zxcalculus.com}} (review in: \cite{Wetering20}\cite{Coecke23})
 originates in \cite[\S 3]{CoeckeDuncan08}\cite[Def. 6.4]{CoeckeDuncan11}\cite[\S\S 2]{Kissinger2008}\cite[\S 4]{Kissinger09}.
\end{literature}

\begin{literature}[\bf Programming language for monadic effects]
\label{LiteratureProgrammingSyntaxForMonadicEffects}

With a good categorical semantics in hand for effectful functional programs via monads (Lit. \ref{LiteratureComputationalEffectsAndModalities}) one is left with finding a good syntax for neatly expressing  such constructions inside a given programming language (a ``domain-specific embedded language'', Lit. \ref{LiteratureDomainSpecificLanguages}). We review the traditional such syntax known as ``{\tt do}-notation''  but highlight that --- for conceptual clarity and for generalization to linear data types (Lit. \ref{VerificationLiterature}) --- this is better cast in {\tt for}...{\tt do}-form,
which is what we use for our quantum pseudo-code in \cref{Pseudocode}.

\medskip

\noindent
{\bf Traditional \mbox{\tt do}-notation.}
The main example of an existing programming language with support for monadic effects is  {\Haskell}.
\footnote{{\Haskell} landing page: \href{https://www.haskell.org}{\tt www.haskell.org}}
Here the (Kleisli-)composition of $\mathcal{E}$-effectful programs via effect-binding \eqref{BindingAndReturning} is encoded by
``\mbox{\tt do}-notation'' (due to \cite[\S 3.3]{Launchbury93}, see \cite[p. 25]{HudakHughesPeytonJonesWadler07}, and adopted in {\Haskell} since  v1.3, \footnote{\href{https://www.haskell.org/definition/from12to13.html\#do}{\tt www.haskell.org/definition/from12to13.html\#do}} for review see
\cite[p. 70]{BentonHughesMoggi02}\cite[\S 20.3]{Milewski19}).
First of all, \mbox{\tt do}-notation is suggestive syntax for the operation of effect-binding \eqref{BindingAndReturning}
\vspace{-3mm}
\begin{equation}
  \label{BindingInDoNotation}
   \declareLeftAligned
   { \bind{\mathcal{E}}{}{\mathrm{prog}} }
   { \mathcal{E}D \to \mathcal{E}D' }
   {
  \mbox{\tt E}
  \;\mapsto\;
  \color{gray}
  \left[
  \color{black}
  \def\arraystretch{1}
  \begin{array}{l}
    \mbox{\tt do}
    \\
    \;\;\;  d \leftarrow
    \mbox{\tt E}
    \\
    \;\;\;\mathrm{prog}(d)
  \end{array}
  \right.
  }
\end{equation}

\vspace{-.5mm}
\noindent but thereby it furthermore provides a convenient means of expressing successive Kleisli-composition simply by successive ``calling'' of separate procedures,
much in the style of ``imperative'' programming (which is thereby emulated into functional programming, Lit. \ref{LiteratureFunctionalLanguages}):
\vspace{-2mm}
\begin{equation}
  \label{DoNotationGenericExample}
   \begin{tikzcd}[
    column sep=80pt,
    row sep=18pt
  ]
    \mathclap{
    \mbox{
      \hspace{60pt}
      \bf
      \def\arraystretch{.9}
      \begin{tabular}{c}
        Composite
        \\
        Kleisli morphism
      \end{tabular}
    }
    }
    &
    &[-40pt]
    \mathclap{
    \mbox{
      \hspace{-40pt}
      \bf
      \def\arraystretch{.9}
      \begin{tabular}{c}
        Corresponding
        \\
        do-notation
      \end{tabular}
    }
    }
    \\[-10pt]
    &&[-40pt]
    \mbox{
      \hspace{-50pt}
      \tt
      d1 ->
      do
    }
    \\[-5pt]
    \phantom{\mathcal{E}(}D_1\phantom{)}
    \ar[
      d,
      rounded corners,
       to path ={
           ([xshift=1pt]\tikztostart.east)
        -- node[yshift=7pt]{
               \scalebox{.9}{$
                 \mathrm{id}
                 \big(
                   D_1
                    \xrightarrow{
                       \scalebox{.9}{\tt this}
                    }
                   \mathcal{E}(D_2)
                 \big)
               $}
            }
           ([xshift=100pt]\tikztostart.east)
        -- ([xshift=100pt, yshift=-10pt]\tikztostart.east)
        -- ([xshift=-50pt, yshift=-10pt]\tikztostart.east)
        -- ([xshift=-50pt, yshift=-35pt]\tikztostart.east)
        -- ([xshift=0pt, yshift=-00pt]\tikztotarget.west)
      },
     "{
        \bind
          { \mathcal{E} }
          {}
        \big(
          D_2
          \xrightarrow{
            \mbox{\tt that}
          }
          \mathcal{E}(D_3)
        \big)
      }"
    ]
    &&
    \mbox{
      \tt
      d2 <-
      this d1
    }
    \\
    \mathcal{E}(D_2)
    \ar[
      d,
      rounded corners,
       to path ={
           ([xshift=0pt]\tikztostart.east)
        -- node[yshift=9pt]{
               \scalebox{.9}{$
                 \bind
                   {\mathcal{E}}
                   {}
                 \big(
                   D_2
                    \xrightarrow{
                       \scalebox{.9}{\tt that}
                    }
                   \mathcal{E}(D_3)
                 \big)
               $}
            }
           ([xshift=100pt]\tikztostart.east)
        -- ([xshift=100pt, yshift=-10pt]\tikztostart.east)
        -- ([xshift=-50pt, yshift=-10pt]\tikztostart.east)
        -- ([xshift=-50pt, yshift=-35pt]\tikztostart.east)
        -- ([xshift=0pt, yshift=-00pt]\tikztotarget.west)
      }
    ]
    &
    &
    \mbox{
      \tt
      d3 <-
      that d2
    }
    \\
    \mathcal{E}(D_3)
    \ar[
      r,
      equals,
      "{
        \bind
          { \mathcal{E} }
          {  }
        \big(
          \return
            {\mathcal{E}}
            {D_3}
        \big)
      }"{yshift=0pt}
    ]
    &
    \mathcal{E}(D_3)
    &
    \mbox{
      \hspace{-30pt}
      \tt
      return d3
    }
  \end{tikzcd}
\end{equation}
(For the moment we closely stick to {\Haskell} typewriter-style typesetting on the right, just for ease of comparison, but in \cref{Pseudocode} we use more fonts to better guide the eye.)

This notation is particularly suggestive due to the further convention that the variable names may be suppressed for functions with trivial in- or out-put (i.e. of unit type $\ast$,
such for programs whose only purpose is write to a log as in \eqref{WriterMonad}) besides their $\mathcal{E}$-effect:
\vspace{-2mm}
$$
  \begin{tikzcd}[
    column sep=80pt,
    row sep=18pt
  ]
    \mathclap{
    \mbox{
      \hspace{60pt}
      \bf
      \def\arraystretch{.9}
      \begin{tabular}{c}
        Composite
        \\
        Kleisli morphism
      \end{tabular}
    }
    }
    &
    &[-20pt]
    \mathclap{
    \mbox{
      \hspace{-40pt}
      \bf
      \def\arraystretch{.9}
      \begin{tabular}{c}
        Corresponding
        \\
        do-notation
      \end{tabular}
    }
    }
    \\[-15pt]
    &&[-40pt]
    \mbox{
      \hspace{-50pt}
      \tt
      do
    }
    \\[-5pt]
    \phantom{\mathcal{E}(}\ast\phantom{)}
    \ar[
      d,
      rounded corners,
       to path ={
           ([xshift=0pt]\tikztostart.east)
        -- node[yshift=7pt]{
               \scalebox{.9}{$
                 \mathrm{id}
                 \big(
                   \ast
                    \xrightarrow{
                       \scalebox{.9}{\tt this}
                    }
                   \mathcal{E}(\ast)
                 \big)
               $}
            }
           ([xshift=100pt]\tikztostart.east)
        -- ([xshift=100pt, yshift=-10pt]\tikztostart.east)
        -- ([xshift=-50pt, yshift=-10pt]\tikztostart.east)
        -- ([xshift=-50pt, yshift=-35pt]\tikztostart.east)
        -- ([xshift=0pt, yshift=-00pt]\tikztotarget.west)
      },
     "{
        \bind
          { \mathcal{E} }
          {}
        \big(
          \ast
          \xrightarrow{
            \mbox{\tt that}
          }
          \mathcal{E}(\ast)
        \big)
      }"
    ]
    &&
    \mbox{
      \hspace{-20pt}
      \tt
      this
    }
    \\
    \mathcal{E}(\ast)
    \ar[
      d,
      rounded corners,
       to path ={
           ([xshift=0pt]\tikztostart.east)
        -- node[yshift=9pt]{
               \scalebox{.9}{$
                 \bind
                   {\mathcal{E}}
                   {}
                 \big(
                   \ast
                    \xrightarrow{
                       \scalebox{.9}{\tt that}
                    }
                   \mathcal{E}(\ast)
                 \big)
               $}
            }
           ([xshift=100pt]\tikztostart.east)
        -- ([xshift=100pt, yshift=-10pt]\tikztostart.east)
        -- ([xshift=-48pt, yshift=-10pt]\tikztostart.east)
        -- ([xshift=-48pt, yshift=-35pt]\tikztostart.east)
        -- ([xshift=0pt, yshift=-00pt]\tikztotarget.west)
      }
    ]
    &
    &
    \mbox{
      \hspace{-20pt}
      \tt
      that
    }
    \\
    \mathcal{E}(\ast)
    \ar[
      r,
      equals,
      "{
        \bind
          { \mathcal{E} }
          {  }
        \big(
          \return
            {\mathcal{E}}
            {\ast}
        \big)
      }"{yshift=0pt}
    ]
    &
    \mathcal{E}(\ast)
    &
    \mbox{
      \hspace{-30pt}
      \tt
      return
    }
  \end{tikzcd}
$$
Here it is manifest how the outer {\tt do{\ldots}return}-block syntax expresses the consecutive Kleisli-composition of any number effectful procedures.

\smallskip
On top of that, the ``{\tt <-}''-syntax is meant to be suggestive of {\it reading out} a value from an effectful datum.
This imagery is accurate in case of the $\mathrm{State}$-monad \eqref{StateMonadEndofunctor} (particularly in its incarnation as the {\tt IO}-monad \cite{PeytonJonesWalder} modelling actual machine reading from an input device such as a keyboard and machine writing to an output device such as a file).
To make this explicit, consider the following stateful programs for reading/writing the state of a global variable of type $W$:
\vspace{-2mm}
\begin{equation}
  \label{ReadWriteGlobalState}
   \declare
    { \mbox{\tt read}_{W} }
    { W\mathrm{State}(W) }
    { w \mapsto (w,w) }
  \hspace{1.5cm}
  \declare
    { \mbox{\tt write}_W }
    { W \to W\mathrm{State}(\ast)  }
    { w \mapsto \big( w' \mapsto w\big) }
\end{equation}
From these, all other stateful operations may be composed via {\tt do}-notation. For instance,
the operation which increments a global integer variable
\vspace{-2mm}
$$
  \declare
    { \mbox{\tt inc} }
    { \mathbb{Z}\mathrm{State}(\ast) }
    { n \mapsto n + 1 }
$$
may be coded as follows, cf. \eqref{BindingOfStateEffects},
and the example in \cite[p. 68 \& 71]{BentonHughesMoggi02}:
\vspace{-2mm}
$$
  \begin{tikzcd}[
    column sep=80pt,
    row sep=18pt
  ]
    \mathclap{
    \mbox{
      \hspace{60pt}
      \bf
      \def\arraystretch{.9}
      \begin{tabular}{c}
        Composite
        \\
        Kleisli morphism
      \end{tabular}
    }
    }
    &
    &[-50pt]
    \mathclap{
    \mbox{
      \hspace{-40pt}
      \bf
      \def\arraystretch{.9}
      \begin{tabular}{c}
        Corresponding
        \\
        do-notation
      \end{tabular}
    }
    }
    \\[-15pt]
    &&[-40pt]
    \mbox{
      \hspace{-80pt}
      \tt
      do
    }
    \\[-5pt]
    \phantom{\mathcal{E}(}\ast\phantom{)}
    \ar[
      d,
      rounded corners,
       to path ={
           ([xshift=0pt]\tikztostart.east)
        -- node[yshift=8pt]{
               \scalebox{.9}{$
                 \mathrm{id}
                 \big(
                   \ast
                    \xrightarrow{
                       \scalebox{.9}{\tt read}
                    }
                   \mathbb{Z}\mathrm{State}(\mathbb{Z})
                 \big)
               $}
            }
           ([xshift=100pt]\tikztostart.east)
        -- ([xshift=100pt, yshift=-10pt]\tikztostart.east)
        -- ([xshift=-53pt, yshift=-10pt]\tikztostart.east)
        -- ([xshift=-53pt, yshift=-35pt]\tikztostart.east)
        -- ([xshift=0pt, yshift=-00pt]\tikztotarget.west)
      },
    ]
    &&
    \mbox{
      \hspace{-5pt}
      \tt
      n <- read
    }
    \\
    \mathbb{Z}\mathrm{State}(\mathbb{Z})
    \ar[
      d,
      rounded corners,
       to path ={
           ([xshift=0pt]\tikztostart.east)
        -- node[yshift=9pt]{
               \scalebox{.9}{$
                 \mathbb{Z}\mathrm{State}
                 \big(
                   \mathbb{Z}
                   \xrightarrow{+1}
                   \mathbb{Z}
                 \big)
               $}
            }
           ([xshift=100pt]\tikztostart.east)
        -- ([xshift=100pt, yshift=-10pt]\tikztostart.east)
        -- ([xshift=-64pt, yshift=-10pt]\tikztostart.east)
        -- ([xshift=-64pt, yshift=-35pt]\tikztostart.east)
        -- ([xshift=0pt, yshift=-00pt]\tikztotarget.west)
      }
    ]
    &
    &
    \mbox{
      \hspace{-25pt}
    }
    \\
    \mathbb{Z}\mathrm{State}(\mathbb{Z})
    \ar[
      d,
      rounded corners,
       to path ={
           ([xshift=0pt]\tikztostart.east)
        -- node[yshift=9pt]{
               \scalebox{.9}{$
                 \bind
                   { }
                   {  }
                  \big(
                    \mathbb{Z}
                    \xrightarrow{
                      \scalebox{.7}{\tt write}
                    }
                    \mathbb{Z}\mathrm{State}(\ast)
                  \big)
               $}
            }
           ([xshift=100pt]\tikztostart.east)
        -- ([xshift=100pt, yshift=-10pt]\tikztostart.east)
        -- ([xshift=-64pt, yshift=-10pt]\tikztostart.east)
        -- ([xshift=-64pt, yshift=-35pt]\tikztostart.east)
        -- ([xshift=0pt, yshift=-00pt]\tikztotarget.west)
      }
    ]
    &
    &
    \mbox{
      \hspace{0pt}
      \tt
      write n + 1
    }
    \\
    \mathbb{Z}\mathrm{State}(\ast)
    \ar[
      r,
      equals,
      "{
        \bind
          {  }
          {  }
        \big(
          \return
            {\mathbb{Z}\mathrm{State}}
            {\ast}
        \big)
      }"{yshift=0pt}
    ]
    &
    \mathbb{Z}\mathrm{State}(\ast)
    &
    \mbox{
      \hspace{-60pt}
      \tt return}
  \end{tikzcd}
$$
In this case it is nicely suggestive that the line ``{\tt n <- read}'' instructs to read out the given state and to bind its value to the variable {\tt n}.
However,  already for similar effect monads such as the list monad (\cite[2.1]{Wadler90}\cite[pp 304]{Milewski19})
\vspace{-3mm}
\begin{equation}
  \label{ListMonad}
  \begin{tikzcd}[sep=0pt]
    \mathllap{
      \mathrm{List}
      \;\isa\;
    }
    \Type
    \ar[rr]
    &&\Type
    \\
    D &\longmapsto&
    \underset{n \isa \NaturalNumbers}{\coprod}
    D^{\times n}
  \end{tikzcd}
  \hspace{.5cm}
  \begin{tikzcd}[sep=0pt]
    D
    \ar[
      rr,
      "{
        \unit
          { \mathrm{List} }
          { D }
      }"
    ]
    &&
    \mathrm{List}(D)
    \\
    d &\mapsto& (d)
  \end{tikzcd}
  \hspace{.5cm}
  \begin{tikzcd}[sep=0pt]
    \mathrm{List}
    \big(
      \mathrm{List}(D)
    \big)
    \ar[
      rr,
      "{
        \join
          { \mathrm{List} }
          { D }
      }"
    ]
    &&
    \mathrm{List}(D)
    \\
    \scalebox{.8}{$
    \left(\!\!
      \def\arraystretch{1}
      \begin{array}{l}
      (d_{11}, \cdots, d_{1 n_1}),
      \\
      \;\;\;\;\vdots
      \\
      (d_{k 1},  \cdots, d_{1 n_k})
      \end{array}
    \!\!\right)
    $}
    &\scalebox{.8}{$\mapsto$}&
    \scalebox{.8}{$
    \big(
      d_{11}, \cdots, d_{1 n_1}
      ,\,
      \cdots
      ,\,
      d_{k 1}, \cdots, d_{k n_k}
    \big)
    $}
  \end{tikzcd}
\end{equation}
the idea of Kleisli composition as being about ``reading out'' intermediate variables is a little inaccurate. For example, the operation of incrementing
all entries in a list of integers is coded in {\tt do}-notation as follows:
\vspace{-2mm}
\begin{equation}
  \label{ElementwiseListIncrementInDoNotation}
    \begin{tikzcd}[
    column sep=80pt,
    row sep=18pt
  ]
    \mathclap{
    \mbox{
      \hspace{60pt}
      \bf
      \def\arraystretch{.9}
      \begin{tabular}{c}
        Composite
        \\
        Kleisli morphism
      \end{tabular}
    }
    }
    &
    &[-50pt]
    \mathclap{
    \mbox{
      \hspace{-40pt}
      \bf
      \def\arraystretch{.9}
      \begin{tabular}{c}
        Corresponding
        \\
        do-notation
      \end{tabular}
    }
    }
    \\[-15pt]
    &&[-40pt]
    \mbox{
      \hspace{-80pt}
      \tt
      do
    }
    \\[-5pt]
    \phantom{\mathcal{E}(}\ast\phantom{)}
    \ar[
      d,
      rounded corners,
       to path ={
           ([xshift=0pt]\tikztostart.east)
        -- node[yshift=8pt]{
               \scalebox{.9}{$
                 \mathrm{id}
                 \big(
                   \ast
                    \xrightarrow{
                       \scalebox{.9}{\tt MyList}
                    }
                   \mathrm{List}(\mathbb{Z})
                 \big)
               $}
            }
           ([xshift=100pt]\tikztostart.east)
        -- ([xshift=100pt, yshift=-10pt]\tikztostart.east)
        -- ([xshift=-47pt, yshift=-10pt]\tikztostart.east)
        -- ([xshift=-47pt, yshift=-35pt]\tikztostart.east)
        -- ([xshift=0pt, yshift=-00pt]\tikztotarget.west)
      },
    ]
    &&
    \mbox{
      \tt
      n <- MyList
    }
    \\
    \mathrm{List}(\mathbb{Z})
    \ar[
      d,
      rounded corners,
       to path ={
           ([xshift=0pt]\tikztostart.east)
        -- node[yshift=9pt]{
               \scalebox{.9}{$
                 \mathrm{List}
                 \big(
                   \mathbb{Z}
                   \xrightarrow{+1}
                   \mathbb{Z}
                 \big)
               $}
            }
           ([xshift=100pt]\tikztostart.east)
        -- ([xshift=100pt, yshift=-10pt]\tikztostart.east)
        -- ([xshift=-50pt, yshift=-10pt]\tikztostart.east)
        -- ([xshift=-50pt, yshift=-35pt]\tikztostart.east)
        -- ([xshift=0pt, yshift=-00pt]\tikztotarget.west)
      }
    ]
    &
    &
    \mbox{
      \hspace{-25pt}
      \tt
      return n + 1
    }
    \\
    \mathrm{List}(\mathbb{Z})
    \ar[
      r,
      equals,
      "{
        \bind
          { \mathrm{List} }
          {  }
        \big(
          \return
            {\mathrm{List}}
            {\mathbb{Z}}
        \big)
      }"{yshift=0pt}
    ]
    &
    \mathrm{List}(\mathbb{Z})
    &
  \end{tikzcd}
\end{equation}
\noindent Here the code on the right nicely evokes the idea that we are ``reading out'' an element from the list and {\tt return}ing
its increment --- but it leaves linguistically implicit the crucial fact that this process is to be applied {\it for all} elements of
the list, and that the results be re-compiled into an output list: Instead of just ``{\tt do this}'', the natural-language rendering
of the above list algorithm would be more like ``{\tt do this for any element}''.

\medskip

\noindent
{\bf For-Do-Notation.}
Indeed, we may observe in generality that it is misleading to think of effect-composition as being about ``reading out'' data elements: Rather,
Kleisli morphisms, in their nature as $(U^\mathcal{E} \dashv F^{\mathcal{E}})$-adjuncts \eqref{ComparisonFunctor}\eqref{FormingAdjuncts}
of modale homomorphisms out of {\it free} modales
\vspace{-2mm}
$$
  \begin{tikzcd}[column sep=large]
    \underset{
      \mathclap{
        \scalebox{.7}{
          \color{darkblue}
          \bf
          free modale
        }
      }
    }{
    \mathcal{E}(D_1)
    }
     \ar[out=180-55, in=55, looseness=5, "\scalebox{1.2}{$\;\mathclap{
        \mathcal{E}
      }\;$}"{description},
     ]
      \ar[
        rr,
        "{
          \overset{
            \mathclap{
              \scalebox{.7}{
                \color{darkgreen}
                \bf
                \def\arraystretch{.9}
                \begin{tabular}{c}
                  \scalebox{1.2}{$\mathcal{E}$}-modale
                  \\
                  homomorphism
                \end{tabular}
              }
            }
          }{
            \widetilde{\mathrm{prog}}
          }
        }"
      ]
      &&
    \mathcal{E}(D_2)
     \ar[out=180-55, in=55, looseness=5, "\scalebox{1.2}{$\;\mathclap{
        \mathcal{E}
      }\;$}"{description},
    ]
  \end{tikzcd}
  \qquad
    \xleftrightarrow{\;\;}
  \qquad
  \begin{tikzcd}
    \underset{
      \mathclap{
        \scalebox{.7}{
          \color{darkblue}
          \bf
          \def\arraystretch{.9}
          \begin{tabular}{c}
            generating
            \\
            data
          \end{tabular}
        }
      }
    }{
      D_1
    }
    \quad
      \ar[
        rr,
        "{
          \overset{
            \scalebox{.7}{
              \color{darkgreen}
              \bf
              Kleisli map
            }
          }{
            \mathrm{prog}
          }
        }"
      ]
      &&
    \mathcal{E}(D_2)
  \end{tikzcd}
$$

\vspace{-3mm}
\noindent are about acting on freely {\it generated data} types $\mathcal{E}(D)$ by declaring how to
operate {\it on generators} $d \isa D$, hence about what to do {\it for} a given generator.

\medskip

Therefore, we may argue that the program-linguistically more evocative rendering of what is going on in monadic effect-binding
operation is a slight enrichment of the traditional {\tt do}-notation to a "{\tt for}...{\tt do}"-block, as follows:
\begin{equation}
  \label{ForDoNotationForEffectBinding}
    \def\arraystretch{1}
\begin{array}{ccc}
    \mbox{\bf
      \def\arraystretch{.9}
      \begin{tabular}{c}
        Monadic
        \\
        effect binding
      \end{tabular}
    }
    &&
    \mbox{\bf
      \def\arraystretch{.9}
      \begin{tabular}{c}
        Corresponding
        \\
        for-do-notation
      \end{tabular}
    }
    \\
    \\[-10pt]
    \begin{tikzcd}
      \mathcal{E}(D)
      \ar[
        rr,
        "{
          \bind
            { \mathcal{E} }
            {  }
            (
              \mathrm{prog}
            )
        }"
      ]
      &&
      \mathcal{E}(D')
    \end{tikzcd}
    &&
    \mbox{\tt E}
    \mapsto
    \fordo
      { \mbox{\tt d} \among \mbox{\tt E}  }
      { \mbox{\tt prog d} }
  \end{array}
\end{equation}
(Notice that in imperative languages the {\tt for}...{\tt do}-syntax is traditionally used to code loops, but in the functional languages that we are
concerned with such loops are instead coded by recursion, so that the {\tt for}...{\tt do}-syntax does remain free to be used for the purpose of effect binding.)

In this notation, the generic example \eqref{DoNotationGenericExample} is rendered into code as follows:
\vspace{-1mm}
$$
  \begin{array}{ccc}
    \mathclap{
    \mbox{
      \hspace{60pt}
      \bf
      \def\arraystretch{.9}
      \begin{tabular}{c}
        Composite
        \\
        Kleisli morphism
      \end{tabular}
    }
    }
    &
    {\phantom{AA}}
    &
    \mathclap{
    \mbox{
      \hspace{-40pt}
      \bf
      \def\arraystretch{.9}
      \begin{tabular}{c}
        Corresponding
        \\
        for-do-notation
      \end{tabular}
    }
    }
  \\
\end{array}
$$
This may be notationally less concise than \eqref{DoNotationGenericExample} but in its close relation to natural language rendering of the computational process it
lends itself to the formulation of transparent pseudocode such as we consider in \cref{Pseudocode}, especially when it comes to operations on linear types,
cf. \eqref{ComparisonOfMonadicDeclarationsOfLinearMaps}.

\smallskip
For instance, in this {\tt for}...{\tt do}-notation the previous example
\eqref{ElementwiseListIncrementInDoNotation}
of entry-wise increments in a list  now reads as follows, neatly indicative of how the incremenet is applied {\it for} every element $n$
found {\it in} the given list $L$:
\vspace{-1mm}
$$
  \begin{array}{ccc}
    \mathclap{
    \mbox{
      \bf
      \def\arraystretch{.9}
      \begin{tabular}{c}
        Composite Kleisli morphism
      \end{tabular}
    }
    }
    &
    {\phantom{AA}}
    &
    \mathclap{
    \mbox{
      \hspace{-40pt}
      \bf
      \def\arraystretch{.9}
      \begin{tabular}{c}
        Corresponding
        \\
        for-do-notation
      \end{tabular}
    }
    }
  \\
    \mathrm{List}(\mathbb{Z})
    \xrightarrow{
      \bind{}{}
      \big(
        \mathbb{Z}
        \xrightarrow{+1}
        \mathbb{Z}
        \xrightarrow{
          \scalebox{.6}{
            \tt return
          }
        }
        \mathrm{List}(\mathbb{Z})
      \big)
    }
    \mathrm{List}(\mathbb{Z})
  &&
  \declareLeftAligned
    {
      \mbox{\tt inc}
    }
    {
      \mathrm{List}(\mathbb{Z})
      \to
      \mathrm{List}(\mathbb{Z})
    }
    {
      \mbox{\tt L ->}
      \fordo
        {
          \mbox{\tt n}
            \among
          \mbox{\tt L}
        }
        {
          \mbox{
            return n + 1
          }
        }
    }
  \end{array}
$$

\end{literature}

\subsection{Monoidal categories}
\label{MonoidalCategories}

\begin{literature}[\bf Monoidal categories of quantum types]
\label{LiteratureMonoidalCategories}
One of the key distinctions between classical and quantum types (Lit. \ref{VerificationLiterature}) is the nature of their logical conjunction, reflected in a {\it monoidal structure} (\cite[\S II.1]{EilenbergKelly66}\cite[\S VII]{MacLane71} \cite[\S 6.1]{Borceux94}) on the categories that they form.

Purely classical types should form a (locally) {\it cartesian} closed category, while purely quantum types should form a symmetric monoidal closed category which is non-cartesian \eqref{LinearTypesAsSubstructuralTypes} to admit a good supply of dualizable (finite-dimensional) types:

\smallskip

\noindent
{\bf Dualizable/Finite-dimensional linear types.}
Somewhat in contrast to quantum theory in general, the focus of quantum computation/information-theory is on quantum systems with {\it finite-dimensional} {(Hilbert-)}spaces $\HilbertSpace{H}$ of quantum states (Lit. \ref{LiteratureQuantumComputation}), whose characteristic property is that they are the dual spaces $\big(\HilbertSpace{H}^\ast\big)^\ast$ of their own dual spaces.

Abstractly, the characterization of finite-dimensionality of an object $\HilbertSpace{H}$ in a symmetric monoidal category is its {\it strong dualizability} \cite[\S 1]{DoldPuppe84} (indeed originally called ``finite objects'' in \cite[p. 113]{Pareigis76}), given equivalently \cite[Thm. 1.3]{DoldPuppe84} by the existence of an object $\HilbertSpace{H}^\ast$ (to be called its {\it dual object}) and of morphisms
\vspace{-2mm}
\begin{equation}
  \label{CoEvaluationForDualObjects}
  \begin{tikzcd}
    \TensorUnit
    \ar[
      rr,
      "{
        \mathrm{cev}_{\HilbertSpace{H}}
      }"
    ]
    &&
    \HilbertSpace{H}
    \otimes
    \HilbertSpace{H}^\ast
    \mathrlap{\,,}
  \end{tikzcd}
  \hspace{1cm}
  \begin{tikzcd}
    \HilbertSpace{H}^\ast
    \otimes
    \HilbertSpace{H}
    \ar[
      rr,
      "{
        \mathrm{ev}_{\HilbertSpace{H}}
      }"
    ]
    &&
    \TensorUnit
  \end{tikzcd}
\end{equation}

\vspace{-2mm}
\noindent such that the following diagrams commute:
\vspace{-2mm}
\begin{equation}
  \label{ZigZagLawsForDualObjects}
  \begin{tikzcd}
    \HilbertSpace{H}
    \ar[
      r,
      "{
        l_{\HilbertSpace{H}}
      }",
      "{ \sim }"{swap}
    ]
    \ar[
      rrrrrr,
      rounded corners,
      to path={
           ([yshift=-00pt]\tikztostart.south)
        -- ([yshift=-13pt]\tikztostart.south)
        -- node[yshift=6pt]{
          \scalebox{.7}{$\mathrm{id}$}
        }
           ([yshift=-13pt]\tikztotarget.south)
        -- ([yshift=-00pt]\tikztotarget.south)
      }
    ]
    &
    \TensorUnit
    \otimes
    \HilbertSpace{H}
    \ar[
      rr,
      "{
        \mathrm{cev}_{\HilbertSpace{H}}
        \,\otimes\,
        \mathrm{id}_{\HilbertSpace{H}}
      }"
    ]
    &&
    \HilbertSpace{H}
      \otimes
    \HilbertSpace{H}^\ast
      \otimes
    \HilbertSpace{H}
    \ar[
      rr,
      "{
        \mathrm{id}_A
        \,\otimes\,
        \mathrm{ev}_{\HilbertSpace{H}}
      }"
    ]
    &&
    \HilbertSpace{H}
      \otimes
    \TensorUnit
    \ar[
      r,
      "{ r_{\HilbertSpace{H}}^{-1} }",
      "{ \sim }"{swap}
    ]
    &
    \HilbertSpace{H}
    \\[15pt]
    \HilbertSpace{H}^\ast
    \ar[
      r,
      "{
        r_{\HilbertSpace{H}^\ast}
      }",
      "{ \sim }"{swap}
    ]
    \ar[
      rrrrrr,
      rounded corners,
      to path={
           ([yshift=-00pt]\tikztostart.south)
        -- ([yshift=-13pt]\tikztostart.south)
        -- node[yshift=6pt]{
          \scalebox{.7}{$\mathrm{id}$}
        }
           ([yshift=-13pt]\tikztotarget.south)
        -- ([yshift=-00pt]\tikztotarget.south)
      }
    ]
    &
    \HilbertSpace{H}^\ast
      \otimes
    \TensorUnit
    \ar[
      rr,
      "{
        \mathrm{id}_{\HilbertSpace{H}}
          \,\otimes\,
        \mathrm{cev}_{\HilbertSpace{H}}
      }"
    ]
    &&
    \HilbertSpace{H}^\ast
      \otimes
    \HilbertSpace{H}
      \otimes
    \HilbertSpace{H}^\ast
    \ar[
      rr,
      "{
        \mathrm{ev}_{\HilbertSpace{H}}
        \,\otimes\,
        \mathrm{id}_A
      }"
    ]
    &&
    \TensorUnit
      \otimes
    \HilbertSpace{H}^\ast
    \ar[
      r,
      "{ l_{\HilbertSpace{H}^\ast}^{-1} }",
      "{ \sim }"{swap}
    ]
    &
    \HilbertSpace{H}^\ast
  \end{tikzcd}
\end{equation}

This implies\footnote{Beware that for $\HilbertSpace{H}$ to be strong dualizable it is {\it not sufficient} that $(\mbox{-}) \otimes \HilbertSpace{H}$ be a left adjoint. But an evaluation-type map on $\HilbertSpace{H}$ does exhibit a strong duality iff it induces the counit of such an adjunction, this is \cite[Thm. 1.3 (b) \& (c)]{DoldPuppe84}.}
that the tensor product functors with these objects are adjoint to each other \eqref{MonadFromAdjunction} as
\vspace{-2mm}
\begin{equation}
\label{AmbidextrousAdjunctionOfTensoringWithDualizableObjects}
\begin{tikzcd}
  (\mbox{-})
  \otimes
  \HilbertSpace{H}
  \;\;
  \dashv
  \;\;
  (\mbox{-})
  \otimes
  \HilbertSpace{H}^\ast
\end{tikzcd}
\end{equation}

\vspace{-2mm}
\noindent with adjunction counit given by the evaluation map. By uniqueness of adjoints this means that when the ambient
category is {\it closed} monoidal (as it is in all our applications) with internal hom $(\mbox{-}) \maplin (\mbox{-})$ then
\vspace{-2mm}
\begin{equation}
  \label{LinearHomViaTensoringWithDual}
  (\mbox{-})
  \otimes
  \HilbertSpace{H}^\ast
  \;\;\;
   \simeq
  \;\;\;
  \HilbertSpace{H}
    \maplin
  (\mbox{-})
\end{equation}
and hence in particular that
\begin{equation}
  \label{StrongDualIsAClosedDual}
  \HilbertSpace{H}^\ast
  \;\;
  \simeq
  \;\;
  \TensorUnit
    \otimes
  \HilbertSpace{H}
  \;\;
  \simeq
  \;\;
  \HilbertSpace{H} \maplin \TensorUnit
  \,.
\end{equation}

But by symmetry, the conditions \eqref{ZigZagLawsForDualObjects} imply that $\HilbertSpace{H} \,\simeq\, (\HilbertSpace{H}^\ast)^\ast$
is the dual of its dual object, to that this adjunction is actually ambidextrous, in that
\vspace{-2mm}
\begin{equation}
\label{AmbidextrousAdjunctTensorFunctorsForStrongDualObjects}
\begin{tikzcd}
  (\mbox{-})
  \otimes
  \HilbertSpace{H}
  \;\;
  \dashv
  \;\;
  (\mbox{-})
  \otimes
  \HilbertSpace{H}^\ast
  \;\;
  \dashv
  \;\;
  (\mbox{-})
  \otimes
  \HilbertSpace{H}
  \,.
\end{tikzcd}
\end{equation}

\noindent
{\bf Categories of internal modules.}
Sometimes it is useful to produce new categories of linear types from given ones by {\it internal algebra} (eg. \cite{Boardman95}): If $(\mathcal{C}, \otimes, \TensorUnit)$ is a bicomplete symmetric monoidal closed category \cite[\S III]{EilenbergKelly66}, then for
\begin{equation}
\label{MonoidObject}
A \,\in\,\mathrm{Mon}(\mathcal{C}, \otimes, \TensorUnit)
\end{equation}
an internal {\it monoid object} \cite[VII.3]{MacLane71}, its category of internal module objects \cite[VII.4]{MacLane71}
\begin{equation}
    \label{CategoryOfInternalModules}
    \big(
      \Modules{A}
      ,\,
      \otimes_A
      ,\,
      A
    \big)
    \;\defneq\;
    \Modules{A}
    \big(
      \mathcal{C}
      ,
      \otimes
      ,
      \TensorUnit
    \big)
  \end{equation}
  is itself
  \begin{itemize}
  \item[{\bf (i)}]
  bicomplete, where the forgetful functor
  $U \,\isa\,\Modules{A} \to \mathcal{C}$ creates both limits and colimits \cite[Lem. 1.2.14]{Marty09}, in particular:
  \begin{equation}
    \label{CoLimitsOfInternalModules}
    U
      \circ
    \underset{\longrightarrow}{\lim}(\,\mbox{-}\,)
    \;\simeq\;
    \underset{\longrightarrow}{\lim}(
      U
        \circ
      \mbox{-}
    )
    \,,
    \hspace{20pt}
    U
      \circ
    \underset{\longleftarrow}{\lim}(\mbox{-})
    \;\simeq\;
    \underset{\longleftarrow}{\lim}(
      U
        \circ
      \mbox{-}
    )
    \,,
  \end{equation}
  \item[{\bf (ii)}]
  symmetric monoidal closed \cite[Lem. 2.2.2 \& 2.2.8]{HoveyShipleySmith00}\cite[Lem. 1.2.15-17]{Marty09}\cite[Prop. 4.1.10]{Brandenburg14},
  with tensor unit $A$ and tensor product the evident coequalizer:
  \begin{equation}
    \label{CoequalizerForInternalTensorProduct}
    N, N'
    \,\isa\,
    \Modules{A}
    \;\;\;\;
    \yields
    \;\;\;\;
    \begin{tikzcd}
    N \otimes A \otimes N'
    \ar[
      r, shift left=3pt
    ]
    \ar[
      r, shift right=3pt
    ]
    &
    N \otimes N'
    \ar[
      rr,
      "{
        \mathrm{coeq}
      }"
    ]
    &&
    N \otimes_A N'
    \mathrlap{\,.}
    \end{tikzcd}
  \end{equation}
  \end{itemize}

\end{literature}

\subsection{Parameterized spectra}
\label{BackgroundParamaterizedSpectra}

\begin{literature}[\bf Parameterized stable homotopy theory, Tangent $\infty$-toposes \& Twisted cohomology]
  \label{TangentInfinityToposes}

  The language of {\LHoTT}
  (Lit. \ref{LiteratureLHoTT})
  syntactically captures the following striking confluence of fundamental structures in algebraic topology and homotopy theory:

 \noindent  {\bf The dichotomy between spaces and motives.}
  One may observe that the following two fundamental types of 1-categories (cf. \ref{VerificationLiterature}):
  \begin{itemize}
    \item[{\bf (i)}] {\it toposes} -- which are the home of
     geometry and classical intuitionistic logic,
    \item[{\bf (ii)}] {\it abelian categories} -- which are the home of linear algebra and forms of linear logic,
  \end{itemize}

  \vspace{1mm}
  \noindent
  while antithetical (for instance in that only the terminal category is an example of both), secretly share a sizeable list of exactness
  properties \cite{Freyd99}.
  The analogous situation for $\infty$-categories may appear similar, since here the two notions of
  \begin{itemize}
    \item [{\bf (i)}] {\it $\infty$-toposes} -- which are the home of higher geometric and of classical
    (intuitionistic) homotopy type theory,
    \item [{\bf (ii)}] {\it stable $\infty$-categories} -- which are the home of higher algebra,
  \end{itemize}

  \vspace{1mm}
\noindent
do remain as antithetical,
(even though both satisfy analogous Giraud-type axioms in that both arise, when locally presentable, as accessible left-exact localizations of $\infty$-categories of
  presheaves: the former with values in $\infty$-groupoids, the latter with values in spectra).

  \smallskip

  {\bf But a miracle happens} after the passage to $\infty$-category theory, in that here a non-trivial
  unification of the two notions does exist
  for a large class of stable $\infty$-categories (``Joyal loci'') including those of module spectra. Namely,
  the collection of {\it parameterized spectra} \cite{MaySigurdsson06}\cite{Malkiewich23}
  over varying base types $\mathcal{X} \,\in\, \mathrm{Grpd}_\infty$ --- i.e.,
  the $\infty$-Grothendieck construction on the $\infty$-functor categories to
  $R\mathrm{Mod}(\mathrm{Spctr})$ --- is itself an $\infty$-topos:
  \vspace{-1mm}
  \begin{equation}
    \label{TangentInfinityTopos}
    R \,\in\, E_\infty\mathrm{Ring}\big(
      \mathrm{Spctr}
    \big)
    \hspace{1cm}
     \vdash
    \hspace{1.3cm}
    T^R \mathrm{Grpd}_\infty
    \;\;
      :\equiv
    \quad
    \underset{
      \mathclap
      {W \in \mathrm{Grpd}_\infty}
    }{\int} \;\;
    \Modules{R}^{W}
    \;\;\;
    \in
    \;
    \mathrm{Topos}_\infty \;.
  \end{equation}

  \vspace{-1mm}
\noindent This observation is originally due to \cite{Biedermann07}, was noted down in \cite[\S 35]{Joyal08} and
received a dedicated discussion in \cite{Hoyois19}. The special case for plain spectra (i.e. with $R = \mathbb{S}$
the sphere spectrum), is touched upon in \cite[Rem. 6.1.1.11]{Lurie17}, where
$\int_{\mathcal{X}} \mathrm{Spectra}^{\mathcal{X}}$ would be called the {\it tangent bundle} to
$\mathrm{Grpd}_\infty$ \cite[\S 7.3.1]{Lurie17} when thought of as equipped with the canonical projection
to the base topos \eqref{FunctorsOnTangentTopos}.
We may thus think of \eqref{TangentInfinityTopos} as something like the {\it $R$-linear tangent $\infty$-topos} to $\mathrm{Grpd}_\infty$ \cite[Prop. 4.1.8]{dcct}
(all these considerations work for base $\infty$-toposes other than $\mathrm{Grpd}_\infty$; which we
disregard just for sake of exposition).

\smallskip

{\bf Infinitesimal cohesion and classicality.}
To pinpoint the nature of this logical context, notice that there is a canonical inclusion of $\mathrm{Grpd}_\infty$ into its tangent $\infty$-topos \eqref{TangentInfinityTopos} by assigning the 0-spectrum everywhere. Since the 0-spectrum is a zero-object, it readily follows that this inclusion is bireflective in that it is both left and right adjoint to the ``tangent projection''
\begin{equation}
  \label{FunctorsOnTangentTopos}
  \hspace{-1cm}
  \begin{tikzcd}[
    row sep=30pt
  ]
    \mathllap{
      \scalebox{.7}{
        \color{darkblue}
        \bf
        \def\arraystretch{.9}
        \begin{tabular}{c}
          $R$-linear
          \\
          tangent $\infty$-topos
        \end{tabular}
      }
    }
    T^R \mathrm{Grpd}_\infty
     \ar[out=180-55, in=55, looseness=5, "\scalebox{1.2}{$\mathclap{
        \natural
      }$}"{description},
      "{
        \scalebox{.7}{
          \color{darkorange}
          \bf
          classical modality
        }
      }"{yshift=4pt}
    ]
    \ar[
      from=d,
      shorten >= -3pt,
      hook,
      "{ 0 }"{description}
    ]
    \ar[
      d,
      shift left=14pt,
      "{ p }"{description}
    ]
    \ar[
      d,
      shorten <= -3pt,
      phantom,
      shift left=7pt,
      "{ \scalebox{.7}{$\dashv$} }"
    ]
    \ar[
      d,
      shorten <= -3pt,
      phantom,
      shift right=8pt,
      "{ \scalebox{.7}{$\dashv$} }"
    ]
    \ar[
      d,
      shorten <= -3pt,
      shift right=14pt,
      "{ p }" description
    ]
    \ar[r, equals]
    &[-20pt]
    \int_{\mathcal{X}} R \mathrm{Mod}^{\mathcal{X}}
    \mathrlap{
      \scalebox{.7}{
        \color{darkblue}
        \bf
        \def\arraystretch{.9}
        \begin{tabular}{c}
          flat $\infty$-bundles of
          \\
          $R$-module spectra
        \end{tabular}
      }
    }
    \\
    \mathllap{
      \scalebox{.7}{
        \color{darkblue}
        \bf
        \def\arraystretch{.85}
        \begin{tabular}{c}
          classical
          \\
          base $\infty$-topos
        \end{tabular}
      }
    }
    \mathrm{Grpd}_\infty
  \end{tikzcd}
\end{equation}
In \cite[Prop. 4.1.9]{dcct} this situation is interpreted as exhibiting {\it infinitesimal cohesive structure} on $T^R \mathrm{Grpd}_\infty$ relative to $\mathrm{Grpd}_\infty$, meaning that, in some precise abstract sense, the objects of $\mathrm{T}^R\mathrm{Grpd}_\infty$ may be regarded as equipped with an {\it infinitesimal thickening} of sorts: In the notation there, the adjoint pair of (co)monads induced by the adjoint triple \eqref{FunctorsOnTangentTopos} is denoted $\shape \dashv \, \flat$, expressing the {\it shape} and the {\it underlying points} of an object, respectively; and the ambidexterity of the adjunction implies that the canonical {\it points-to-pieces transform} is an equivalence
$
  \begin{tikzcd}
    \flat \ar[r, "{ \sim }"] & \shape
  \end{tikzcd}
$
hence reflecting the idea that the extra geometric substance which the objects of $\mathrm{T}^R \mathrm{Grpd}_\infty$
carry on their classical underlying skeleta in $\mathrm{Grpd}_\infty$ is  ``infinitesimal'' (think: ``microscopic'')
so that  it cannot be noticed from looking just at the macroscopic shape of these objects.

As a result, these two cohesive modalities $\flat$ and $\shape$ unify into a single ambidextrous modality as shown in \eqref{FunctorsOnTangentTopos}, now to be denoted``$\natural$''  (following \cite{RileyFinsterLicata21}),  which we
may think of as retaining the underlying classical aspect of types while discarding their infinitesimal/microscopic
(quantum) aspects, see Prop. \ref{ClassicalAndQuantumModality} for more.

\smallskip
\noindent {\bf Flat vector bundles and Indexed vector spaces.}
Specifically when $R = H\GroundField$ is the Eilenberg-MacLane spectrum over a ring or even a field $\GroundField$, then there is an equivalence (\cite{Robinson87}\cite[Thm. 5.1.6]{SchwedeShipley03})
between the homotopy theory of $H\GroundField$-module spectra
and that of $\GroundField$-chain complexes, hence between that of $W$-parameterized $H\GroundField$-module spectral and that of {\it flat $\infty$-vector bundles} over $W$, also known as {\it $\infty$-local systems} over $W$ (see \cite[\S 3.1]{EoS} for more):
\vspace{-2mm}
$$
  \overset{
    \mathclap{
      \raisebox{8pt}{
        \scalebox{.7}{
          \color{darkblue}
          \bf
          \def\arraystretch{.9}
          \begin{tabular}{c}
            parameterized
            \\
            $H\GroundField$-module spectra
          \end{tabular}
        }
      }
    }
  }{
  \Modules{
    H\GroundField
  }^{W}
  }
  \;\;\;\;\;\;\;\;
  \simeq
  \;\;\;\;\;\;\;\;
  \overset{
    \mathclap{
      \raisebox{8pt}{
        \scalebox{.7}{
          \color{darkblue}
          \bf
          \begin{tabular}{c}
            $\infty$-local systems of
            \\
            chain complexes
          \end{tabular}
        }
      }
    }
  }{
   \mathrm{Ch}_\GroundField^W
  }
$$
and the {\it hearts} (Lit. \ref{Zerosector}) of these stable $\infty$-categories are the 1-categories of ordinary flat vector bundles
hence of ordinary local systems of vector spaces:
\vspace{-3mm}
$$
  \begin{tikzcd}[
    row sep=1pt,
    column sep=2pt
  ]
  \scalebox{.7}{
    \color{darkblue}
    \bf
    Vector spaces
  }
  \ar[
    r, phantom,
    "{
      \scalebox{.7}{are}
    }"{yshift=-1pt}
  ]
  &
  \scalebox{.7}{
    \color{darkblue}
    \bf
    the heart
  }
  \ar[
    r,
    phantom,
    "{
      \scalebox{.7}{of}
    }"
  ]
  &
  \scalebox{.7}{
    \color{darkblue}
    \bf
    $H\GroundField$-module spectra
  }
  \\
  \Modules{\GroundField}
  \ar[r, phantom, "{ \simeq }"]
  &
  \heart
  \big(
    \Modules{H\GroundField}
  \big)
  \ar[r, hook]
  &
  \Modules{H\GroundField}
  \end{tikzcd}
  \hspace{.6cm}
  \begin{tikzcd}[
    row sep=-1pt,
    column sep=2pt
  ]
  \scalebox{.7}{
    \color{darkblue}
    \bf
    \def\arraystretch{.9}
    \begin{tabular}{c}
      Flat
      \\
      vector bundles
    \end{tabular}
  }
  \ar[
    r, phantom,
    "{
      \scalebox{.7}{are}
    }"{yshift=-1pt}
  ]
  &
  \scalebox{.7}{
    \color{darkblue}
    \bf
    the heart
  }
  \ar[
    r,
    phantom,
    "{
      \scalebox{.7}{of}
    }"
  ]
  &
  \scalebox{.7}{
    \color{darkblue}
    \bf
    \begin{tabular}{c}
      parameterized
      \\
      $H\GroundField$-module spectra
    \end{tabular}
  }
  \\
  \Modules{\GroundField}^W
  \ar[r, phantom, "{ \simeq }"]
  &
  \heart
  \big(
    \Modules{H\GroundField}^W
  \big)
  \ar[r, hook]
  &
  \Modules{H\GroundField}^W
  \end{tikzcd}
$$

\vspace{-2mm}
\noindent
Over $W \isa \Set \subset \Groupoids$ these are plain vector bundles over the discrete spaces $W$, hence $W$-indexed vector
spaces, whence their Grothendieck construction is the free coproduct completion
$\mathrm{Fam}_{\GroundField}$  of vector spaces providing the categorical semantics of (Proto-){\Quipper} (Lit. \ref{LiteratureQuantumProgrammingLanguages})
and the 0-sector of {\LHoTT}, which we discuss in detail in \cref{QuantumTypeSemantics}:
\vspace{-2mm}
\begin{equation}
  \label{VectorSpacesAmongInfinityLocalSystems}
  \begin{tikzcd}[
    column sep=0pt,
    row sep=2pt
  ]
  \scalebox{.7}{
    \color{darkblue}
    \bf
    \def\arraystretch{1}
    \begin{tabular}{c}
      Categorical semantics of
      \\
      (Proto-){\Quipper} \&
      \\
      0-sector of {\LHoTT}
    \end{tabular}
  }
  &
  \scalebox{.7}{
    \color{darkblue}
    \bf
    \def\arraystretch{1}
    \begin{tabular}{c}
      Categorical semantics of
      \\
      heart-sector of {\LHoTT}
      \\
      including Hermitian spaces
    \end{tabular}
  }
  &
  \scalebox{.7}{
    \color{darkblue}
    \bf
    \def\arraystretch{1}
    \begin{tabular}{c}
      Categorical semantics
      \\
      of {\LHoTT} including
      \\
      topological effects
    \end{tabular}
  }
  \\
  \mathrm{Fam}_\GroundField
  \qquad
  \ar[r, hook]
  &
  \qquad \quad
  \mathrm{Loc}_{\GroundField}
  \qquad \quad
  \ar[r, hook]
  &
  \qquad
  T^{H\GroundField}
  \Groupoids_\infty
  \\
  \underset{
    \mathclap{
    W \,\isa\, \Sets
    }
  }{\int}
  \;\;\;
    \Modules{\GroundField}^W
    \qquad
  \ar[
    r,
    hook
  ]
  &
  \qquad \quad
  \underset{
    \mathclap{
    W \,\isa\, \Groupoids
    }
  }{\int}
  \;
    \Modules{\GroundField}^W
    \qquad \quad
  \ar[
    r,
    hook
  ]
  &
  \qquad
  \underset{
    \mathclap{
      W \,\isa\, \Groupoids_\infty
    }
  }{\int}
  \;\;\;
  \Modules{H\GroundField}^W
  \end{tikzcd}
\end{equation}

\vspace{-2mm}
\noindent
In the middle, we are showing an intermediate ground which turns out to be useful for typing Hermitian structure on quantum types and hence
captures the probabilistic aspect of quantum theory (Lit. \ref{LiteratureQuantumProbability}):
\smallskip

\noindent {\bf Equivariance by homotopy type-dependency.}
For $G$ a group, a spectrum parameterized over its delooping (its 1st Eilenberg-Maclane space) $\mathbf{B}G$ is equivalently a $G$-action on
the underlying spectrum (also known as a ``na{\"i}vely $G$-equivariant spectrum''). Generally, the slice over $\mathbf{B}G$, hence the types
{\it dependent on} variables in context $\mathbf{B}G$ are types equipped with a $G$-action
(see \cite[Prop. 0.2.1]{EquivariantBundles}\cite[\S 2.2]{SatiSchreiber20}):

\begin{equation}
\label{EquivarianceBySlicing}
\footnotesize
\adjustbox{}{
\begin{tabular}{|c|c|}
\hline
\multicolumn{2}{|c|}{
  \bf
  Equivariance by dependency on
  delooping
}
\\
{\bf Syntax}
&
{\bf Semantics}
\\
\hline
\hline
$
\def\arraystretch{1.2}
\begin{array}{l}
  \vdash
  \;\;\;\;\;\;
  \mathrm{pt}
  \isa
  \mathbf{B}G
  \\
  \vdash
  \;\;\;\;\;\;
  \mathrm{Id}_{\mathbf{B}G}
  (\mathrm{pt}, \mathrm{pt})
  \,\simeq\,
  G
\end{array}
$
&
$
  \begin{tikzcd}[row sep=13pt, column sep=large]
    \\[-7pt]
    \mathllap{
      \scalebox{.7}{
        \color{darkblue}
        \bf
        group
      }
    }
    G
    \ar[
      rr
    ]
    \ar[
      dd
    ]
    \ar[
      ddrr,
      phantom,
      "{
        \scalebox{.7}{(pb)}
      }"
    ]
    &&
    \ast
    \ar[
      dd,
      "{
        \vdash \mathrm{pt}
      }"{description}
    ]
    \\
    \\
    \ast
    \ar[
      rr,
      "{
        \vdash \mathrm{pt}
      }"{description}
    ]
    &&
    \mathbf{B}G
    \mathrlap{
      \scalebox{.7}{
        \color{darkblue}
        \bf
        delooping
      }
    }
    \\[-12pt]
  \end{tikzcd}
$
\\
\hline
$
  \mathrm{pt}
  \,\isa\,
  \mathbf{B}G
  \;\;\;\;
  \yields
  \;\;\;\;
  E_{\mathrm{pt}} \isa \Type
$
&
\hspace{.6cm}
$
  \begin{tikzcd}[row sep=13pt, column sep=huge]
    \\[-10pt]
    \mathllap{
      \scalebox{.7}{
        \color{darkblue}
        \bf
        $G$-action
      }
    }
    E_{\mathrm{pt}}
     \ar[out=180-55, in=55, looseness=5, "\scalebox{1.2}{$\;\mathclap{
        G
      }$\;}"{description},
    ]
    \ar[r]
    \ar[d]
    \ar[
      dr,
      phantom,
      "{
        \scalebox{.7}{(pb)}
      }"
    ]
    &
    \overset{
      \mathclap{
        \scalebox{.7}{
          \color{darkblue}
          \bf
          \color{darkblue}
          \bf
          \begin{tabular}{c}
            homotopy quotient/
            \\
            Borel construction
          \end{tabular}
        }
      }
    }{
    E_{\mathrm{pt}}
    \sslash
    G
    }
    \ar[
      d
    ]
    \\
    \ast
    \ar[
      r,
      "{
        \vdash \, \mathrm{pt}
      }"{description}
    ]
    &
    \mathbf{B}G
  \end{tikzcd}
$
\hspace{1cm}
\\[-10pt]
&
\\
\hline
\end{tabular}
}
\end{equation}

\medskip

\noindent
{\bf Twisted cohomology.}
Interestingly, the hom-spaces in the $R$-tangent $\infty$-topos
\eqref{TangentInfinityTopos}
are sections of $R$-module bundles $\tau_{\mathcal{X}}$, which means
\cite{ABGHR14}\cite[Prop. 3.5]{SSS23Character}\cite[p. 6]{SatiSchreiber20}
that their connected components form the  {\it $\tau_{\mathcal{X}}$-twisted $R$-cohomology} $R^\tau(\mathcal{X})$ of $\mathcal{X}$ \cite[\S 22.11]{MaySigurdsson06}:
\vspace{-2mm}
\begin{equation}
  \label{TwistedCohomologyFromTangentToposes}
  \left.
  \def\arraystretch{1.1}
  \begin{array}{l}
    \mathcal{X}
    \,\in\,
    \mathrm{Grpd}_\infty
    \\
    R \,\in\, E_\infty\mathrm{Rng}(\mathrm{Spctr})
  \end{array}
  \! \right\}
  \hspace{1cm}
  \vdash
  \hspace{1cm}
  \mathrm{Maps}\big(
    0_{\mathcal{X}}
    ,\,
    R \!\sslash\! \mathrm{GL}_1(R)
  \big)
  \;\;
    =
  \;\;
  \left\{\!
  \begin{tikzcd}[
    row sep=30pt,
    column sep=45pt
  ]
    &
    R \!\sslash\! \mathrm{GL}_1(R)
    \ar[
      d,
      ->>
    ]
    \\
    \mathscr{X}
    \ar[
      r,
      dashed,
      "{
         \tau_{{}_{\mathcal{X}}}
      }"{description},
      "{
        \scalebox{.7}{
          \color{darkgreen}
          twist
        }
      }"{swap, yshift=-2pt}
    ]
    \ar[
      ur,
      dashed,
      "{
        \scalebox{.7}{
          \color{darkgreen}
          cocycle in
          \scalebox{1.2}{$R^\tau(\mathcal{X})$}
        }
      }"{yshift=1pt, sloped, pos=.42}
    ]
    &
    B \mathrm{GL}_1(R)
  \end{tikzcd}
 \!\!\! \right\}
  \,.
\end{equation}
This already suggests \cite{Schreiber14}
that tangent $\infty$-toposes are a natural logical context
for describing strongly-coupled quantum systems,
since twisted $R$-cohomology theories play a key role in their holographic (stringy) formulations (Lit. \ref{TopologicalQuantumMaterials}).

\end{literature}

\smallskip

\begin{remark}[\bf 0-sector and Heart-sector of {\LHoTT}]
  \label{Zerosector}
  $\,$

 \noindent {\bf (i)} By the {\it 0-sector} of {\LHoTT} (Lit. \ref{LiteratureLHoTT}) we mean more than just its 0-truncated types (which are just the classical hSets of {\LHoTT}). Namely, in the
  {\it stable} homotopy theory which is incorporated in {\LHoTT}, the classical notion of $n$-truncation becomes almost meaningless (due to the
  existence of spectra with homotopy groups in arbitrary {\it negative} degree, cf. \cite[Warning 1.2.1.9]{Lurie17}), its proper replacement instead
  being the notion of {\it t-structure} (eg. \cite[\S 1.2.1]{Lurie17}).

   \noindent {\bf (ii)}  The {\it heart} of the t-structure (formed by the spectra whose homotopy
  groups are concentrated in degree 0) reflects the intended 0-sector of the given stable homotopy theory.
  Hence by the 0-sector of {\LHoTT} we mean those types which are in the heart and whose {\it underlying} purely classical type is 0-truncated.

   \noindent {\bf (iii)}  For the discussion of Hilbert space structure and quantum probability (Lit. \ref{LiteratureQuantumProbability}) in \ifdefined\monadology\cite[\S 3.2]{QS}\else\cref{RealQuantumTypes}\fi, we
  employ a slightly larger sector of {\LHoTT}, where the purely classical types are allowed to be homotopy 1-types while the purely quantum types
  are still required to be in the heart. We may call this the {\it heart-sector} of {\LHoTT}, for short (leaving the 1-truncation of the classical
  types understood, because beyond 1-truncated classical types it makes little sense to constrain the quantum types to the heart.
\end{remark}

\begin{literature}[\bf Topological quantum materials and Topological K-theory]
\label{TopologicalQuantumMaterials}
For extensive background and referencing see \cite{Ord}.
\end{literature}

\section{Quantum Effects}
\label{QuantumEffects}

We show that a system of basic (co)monads which is canonically {\it defineable} (via admissibke inference rule) in any dependent linear homotopy type theory which satisfies the Motivic Yoga (Def. \ref{MotivicYoga}) equips the underlying (independent) linear type theory with the computational effects which otherwise have to be postulated in (typed) quantum programming languages: besides a quantization modality ($\quantized$) (turning bits into q-bits, etc.), these effects notably include quantum measurement ($\indefinitely$) and conditional quantum state preparation ($\randomly$),  which turn out to correspond to Coecke et al.'s ``classical structures'' Frobenius monad.

\medskip

\ifdefined\monadology
\cref{QuantumTypeSemantics} -- Semantics of Dependent linear types
\else
\fi

\cref{ClassicalEpistemicLogicViaDependentTypes} -- Classical epistemic logic via Dependent classical types;

\cref{QuantumEpistemicLogicViaDependentLinearTypes} -- Quantum epistemic logic via Dependent linear types;

\cref{ControlledQuantumGates} -- Controlled quantum gates via Quantum effect logic;

\cref{MixedQuantumTypes} -- Controlled quantum channels via QuantumState effects.

\subsection{Quantum Semantics}
\label{QuantumTypeSemantics}

We lay out a concrete example (Def. \ref{CategoryOfBundleTypes} below ) of a category that interprets
\ifdefined\monadology
\else
(as we shall see in \cref{QuantumTypesSyntax})
\fi
the 0-sector (Rem. \ref{Zerosector}) of {\LHoTT} relevant for expressing quantum circuits (in \cref{ControlledQuantumGates}).
Category-theoretically this example is elementary and standard (going back to \cite[\S 3.3]{Benabou85}\cite[pp. 281]{HuTholen95}),
but it is important in applications, e.g. as the established model for Proto-{\Quipper}
(Lit. \ref{LiteratureQuantumProgrammingLanguages}, where it appears as \cite[Def. 3.3]{RiosSelinger18} for the case that their fiber
category $\overline{M}$ is the category $\Modules{\GroundField}$ of $\mathbb{K}$vector bundles). Here we highlight previously
underappreciated aspects of this model (all shared by its homotopy-theoretic generalizations in \cite{EoS}):

\smallskip
\begin{itemize}
\item[$\circ$]
its doubly closed monoidal structure (Prop. \ref{DoublyClosedMonoidalStructure}),
\item[$\circ$] its doubly strong monadic reflections (Prop. \ref{ClassicalAndQuantumModality}),
\item[$\circ$] its quantization/exponential modality (Prop. \ref{QuantizationAndExponentialAdjunction}),
\item[$\circ$] its support of 6-operations motivic yoga (Prop. \ref{LinearBundleTypesSatisfyMotivicYoga}),
\end{itemize}
which make the model interpret an expressive modal/monadic/effectful quantum language {\tt QS}, in \cref{Pseudocode}.

\begin{definition}[\bf Category of linear bundle types]
\label{CategoryOfBundleTypes}
 $\,$

 \noindent
 For the purpose of this section, we write
 ``$\Types$'' for the category equivalently described as follows

 \noindent
 (cf. \cite{EoS}, where this category is denoted ``$\mathrm{Fam}_{\GroundField}$''):

 \smallskip
 \begin{itemize}[leftmargin=.5cm]
   \item[$\circ$] $\Type$ is the free coproduct completion of $\Modules{\GroundField}$,
   \item[$\circ$] $\Type$ is the category of {\it indexed sets} of $\mathbb{K}$-vector spaces,
   \item[$\circ$]
   $\Type$ is the category of vector bundles over varying discrete base spaces,
   \item[$\circ$]
   ${\Type}$ is the 0-sector of the $\infty$-category of $\infty$-local systems over varying general base spaces,
   \item[$\circ$]
   $\Type$ is the Grothendieck construction of the $\Sets$-indexed category whose fiber over $W \isa \Set$ is the category
   $
     \DependentModules{\GroundField}{W}
     \,\defneq\,
     \mathrm{Func}(W,\,\Modules{\GroundField})
   $
   of $W$-indexed vector spaces (vector bundles over $W$):
 \end{itemize}
\end{definition}

\begin{equation}
\label{CategoryOfIndexedSetsOfComplexVectorSpaces}
\adjustbox{}{
\def\arraystretch{.8}

}
\end{equation}

\smallskip
\noindent When describing their linear fiber types concretely, we also denote linear bundle types and their hom-sets as follows (the bottom lines exhibiting the type-theoretic syntax, see Rem. \ref{NotationForInternalHoms}):
\begin{equation}
\label{HomSetsOfBundleTypes}
  \def\arraystretch{2}
  %

\medskip
\begin{remark}[{\bf Notation for internal homs}]
\label{NotationForInternalHoms}
$\,$

\noindent {\bf (i)}
The arrow-notation for the hom-sets in $\LinearTypes$ and $\LinearTypes_{W}$ is that inherited from $\Type$ under
the embeddings $\ClassicalTypes, \LinearTypes \hookrightarrow \Types$ \eqref{SubcategoriesOfBundleTypes}, in that:
\vspace{-3mm}
$$
  \def\arraystretch{3.5}
  \begin{array}{rcl}
  \classically
  \left(
    \ABundleType
      { \HilbertSpace{H} }
      { \ast }
    \to
    \ABundleType
      { \HilbertSpace{H}' }
      { \ast }
  \right)
  &
    =
  &
  \classically
  (\HilbertSpace{H} \to \HilbertSpace{H}')
  \\
  \hspace{.4cm}
  \left(
    \ABundleType
      { \HilbertSpace{H} }
      { \ast }
    \maplin
    \ABundleType
      { \HilbertSpace{H}' }
      { \ast }
  \right)
  &
    =
  &
  (\HilbertSpace{H} \maplin \HilbertSpace{H}')
  \\[-30pt]
  {}
  \end{array}
$$

\vspace{-2mm}
\noindent where on the right the embeddings
\eqref{SubcategoriesOfBundleTypes}
are understood.

\noindent {\bf (ii)}  This way, e.g. the natural hom-isomorphism
expressing the closed monoidal structure on $\QuantumTypes$ reads
\begin{equation}
  \label{TensorHomIsomorphism}
  \classically
  \big(
  \HilbertSpace{H}
  \otimes
  \HilbertSpace{H}'
  \to
  \HilbertSpace{H}''
  \big)
  \;\simeq\;
  \classically
  \big(
  \HilbertSpace{H}
  \to
  (
    \HilbertSpace{H}'
    \maplin
    \HilbertSpace{H}''
  )
  \big)
\end{equation}
\noindent {\bf (iii)} But we now also have mixed  classical/quantum expressions, notably this one, which is going to be important:
\begin{equation}
  \label{ClassicalToQuantumHom}
  \adjustbox{
    margin=3pt,
    fbox
  }{$
  \big(
    W \to \HilbertSpace{H}
  \big)
  \;\;\;\defneq\;\;\;
  \ABundleType
    {0}{W}
  \to
  \ABundleType
    { \HilbertSpace{H} }
    { \ast }
  \;\;\;=\;\;\;
  \ABundleType
    { \prod_W \HilbertSpace{H} }
    { \ast }
  \;\;\;=\;\;\;
  \ABundleType
    { \TensorUnit_\bullet }
    { W }
  \maplin
  \ABundleType
    { \HilbertSpace{H} }
    { \ast }
  \;\;\;=\;\;\;
  \big(
     \TensorUnit
     \!\!\times\!\!
     W
      \maplin
    \HilbertSpace{H}
  \big)
  $}
\end{equation}
\end{remark}

\medskip

\newpage

\begin{proof}[Proof of Prop. \ref{DoublyClosedMonoidalStructure}]
By standard arguments \cite{Schauenburg01} we may assume the unitors and associators to be identities.
The symmetric braiding is given by the evident exchange of variables
$$
  \declare
    {
      \braiding
        { \otimes }
        {
          \HilbertSpace{H}_W
          ,
          \HilbertSpace{H}'_{W'}
        }
    }
    {
    \ABundleType
      { \HilbertSpace{H}_\bullet }
      { W }
    \otimes
    \ABundleType
      { \HilbertSpace{H}'_\bullet }
      { W' }
    \to
    \ABundleType
      { \HilbertSpace{H}'_\bullet }
      { W' }
    \otimes
    \ABundleType
      { \HilbertSpace{H}_\bullet }
      { W }
    }
    {
    \vert \psi_w \rangle
    \otimes
    \vert \psi'_{w'} \rangle
    \;\mapsto\;
    \vert \psi'_{w'} \rangle
    \otimes
    \vert \psi_w \rangle
    }
$$
To see the internal-hom adjunction it is clearly sufficient
(since our classical base category is $\ClassicalTypes \,\defneq\, \Sets$)
to check
the defining hom-isomorphism for the case that $W = \ast$.
In this case, we have the following sequences of natural isomorphisms:
$$
\hspace{-3mm}
  \def\arraystretch{2}
\begin{array}{lll}
    \mathrm{Hom}
    \left(
    \ABundleType
      {
       \HilbertSpace{H}
         \oplus
        \HilbertSpace{H}'_\bullet
      }
      {W'}
    ,
    \ABundleType
      {\HilbertSpace{H}''_\bullet}
      {W''}
    \right)
  & \hspace{-4mm} \simeq
    (f \isa W' \to W'')
    \times
    \underset{w'}{\prod}
    \classically
    \big(
     \HilbertSpace{H}
     \oplus
     \HilbertSpace{H}'_{w'}
     \to
     \HilbertSpace{H}''_{f(f')}
    \big)
    &
  \hspace{-2mm}   \proofstep{
      by
      \eqref{HomSetsOfBundleTypes}
    }
    \\
 &   \hspace{-4mm} \simeq
    (f \isa W' \to W'')
    \times
    \underset{w'}{\prod}
    \Big(
    \classically
    \big(
     \HilbertSpace{H}'_{w'}
     \to
     \HilbertSpace{H}''_{f(w')}
    \big)
    \times
    \classically
    \big(
      \HilbertSpace{H}
      \to
     \HilbertSpace{H}''_{f(w')}
     \big)
    \Big)
    &
  \hspace{-2mm}   \proofstep{
      by
      coproduct property of $\oplus$
    }
    \\
  &  \hspace{-4mm} \simeq
    (f \isa W' \to W'')
    \times
    \underset{w'}{\prod}
    \classically
    \big(
     \HilbertSpace{H}'_{w'}
     \to
     \HilbertSpace{H}''_{f(w')}
    \big)
    \times
    \underset{w'}{\prod}
    \classically
    \big(
      \HilbertSpace{H}
      \to
     \HilbertSpace{H}''_{f(w')}
     \big)
    &
   \hspace{-2mm}  \proofstep{
      since $\prod_{W}(-)$ is right adjoint
    }
    \\
   & \hspace{-4mm} \simeq
    (f \isa W' \to W'')
    \times
    \underset{w'}{\prod}
    \classically
    \big(
     \HilbertSpace{H}'_{w'}
     \to
     \HilbertSpace{H}''_{f(w')}
    \big)
    \times
    \classically
    \Big(
    \HilbertSpace{H}
      \to
    \underset{w'}{\prod}
     \HilbertSpace{H}''_{f(w')}
     \Big)
    &
   \hspace{-2mm} \proofstep{
      since $\HilbertSpace{H} \to (\mbox{-})$ is right adjoint
    }
    \\
\end{array}
$$
and
$$
  \def\arraystretch{2.}
  \begin{array}{lll}
    \mathrm{Hom}
    \left(
      \ABundleType
      {
      \HilbertSpace{H}
        \otimes
      \HilbertSpace{H}'_{\bullet}
      }
      {W'}
    ,\,
    \ABundleType
      {\HilbertSpace{H}''_{\bullet}}
      {W''}
    \right)
    &
    \;\simeq\;
    \big(
     f \isa W' \to W''
    \big)
    \times
    \underset{w'}{\prod}
    \classically
    \Big(
      \HilbertSpace{H}
      \otimes
      \HilbertSpace{H}'_{w'}
      \to
      \HilbertSpace{H}''_{f(w')}
    \Big)
    &
    \proofstep{
      by
      \eqref{HomSetsOfBundleTypes}
    }
    \\
   & \;\simeq\;
    \big(
     f \isa W' \to W''
    \big)
    \times
    \underset{w'}{\prod}
    \classically
    \Big(
      \HilbertSpace{H}
      \to
      \big(
      \HilbertSpace{H}'_{w'}
      \maplin
      \HilbertSpace{H}''_{f(w')}
      \big)
    \Big)
    &
    \proofstep{
      by
      \eqref{TensorHomIsomorphism}
    }
    \\
    & \;\simeq\;
    \big(
     f \isa W' \to W''
    \big)
    \times
    \classically
    \Big(
      \HilbertSpace{H}
      \to
      \underset{w'}{\prod}
      \big(
      \HilbertSpace{H}'_{w'}
      \maplin
      \HilbertSpace{H}''_{f(w')}
      \big)
    \Big)
    &
    \proofstep{
      since
      $\HilbertSpace{H}\to (\mbox{-})$
      is right adjoint
    }
    \\[+10pt]
    & \;\simeq\;
    \mathrm{Hom}
    \left(
      \ABundleType
      { \HilbertSpace{H} }
      { \ast }
      ,\,
      \ABundleType
      {
      \prod_{w'}
      \big(
       \HilbertSpace{H}'_{w'}
       \maplin
       \HilbertSpace{H}''_{f(w')}
      \big)
      }
      {
      \big(
        f \isa W' \to W''
      \big)
      }
    \right)
    &
    \proofstep{
      by \eqref{HomSetsOfBundleTypes}
    }
  \end{array}
$$
which proves the claim.
\end{proof}

\medskip

\noindent

\noindent
{\bf Classical and Quantum Modality.}

\begin{proposition}[\bf Reflective subcategories of purely classical/quantum
modal types]
\label{ClassicalAndQuantumModality}
The category of Def. \ref{CategoryOfBundleTypes} has
monadic \eqref{ComparisonFunctor}
reflective subcategory inclusions
as follows:
\begin{equation}
  \label{SubcategoriesOfBundleTypes}
  \hspace{-1cm}
  \begin{tikzcd}[row sep=0pt, column sep=10pt]
    \scalebox{\termscale}{$W$}
    &\mapsfrom&
    \scalebox{\termscale}{$
    \ABundleType
      { \HilbertSpace{H}_\bullet }
      { W }
    $}
    \\
    \ClassicalTypes
    \ar[
      from=rr,
      shift right=6pt,
    ]
    \ar[
      rr,
      phantom,
      "{
        \scalebox{.7}{$\bot$}
      }"
    ]
    \ar[
      rr,
      shift right=6pt,
      hook
    ]
    &&
    \Types
    \ar[out=50, in=-50,
      looseness=4,
      shorten=-3pt,
      shift left=5pt,
      "{
        \scalebox{1.6}{
          ${
            \hspace{1pt}
            \mathclap{
              \classically
            }
            \hspace{1pt}
          }$
        }
      }"{description}
    ]
    \ar[out=50, in=-50,
      phantom,
      looseness=4,
      shift left=14pt,
      "{
        \color{darkorange}
        \bf
        classically
      }"{
        scale=.7,
        description, rotate=-90
        }
    ]
    \\
    \scalebox{\termscale}{$W$}
    &\mapsto&
    \scalebox{\termscale}{$
     \ABundleType
      { 0_\bullet }
      { W }
    $}
  \end{tikzcd}
  \hspace{1.5cm}
  \begin{tikzcd}[row sep=0pt, column sep=10pt]
    \underset{w}{\osum}
    \HilbertSpace{H}_w
    &\mapsfrom&
    \scalebox{\termscale}{$
      \ABundleType
        { \HilbertSpace{H}_\bullet }
        { W }
    $}
    \\
    \LinearTypes
    \ar[
      from=rr,
      shift right=6pt,
    ]
    \ar[
      rr,
      phantom,
      "{
        \scalebox{.7}{$\bot$}
      }"
    ]
    \ar[
      rr,
      shift right=6pt,
      hook
    ]
    &&
    \Types
    \ar[out=50, in=-50,
      looseness=4,
      shorten=-3pt,
      shift left=5pt,
      "{
        \scalebox{1.4}{
          ${
            \hspace{1pt}
            \mathclap{
              \quantumly
            }
            \hspace{1pt}
          }$
        }
      }"{description}
    ]
    \ar[out=50, in=-50,
      phantom,
      looseness=4,
      shift left=14pt,
      "{
        \color{darkorange}
        \bf
        quantumly
      }"{
        scale=.7,
        description, rotate=-90
        }
    ]
    \\
    \scalebox{\termscale}{$
      \HilbertSpace{H}
    $}
    &\mapsto&
    \scalebox{\termscale}{$
    \ABundleType
      { \HilbertSpace{H} }
      { \ast }
    $}
  \end{tikzcd}
\end{equation}
\end{proposition}
\noindent
Moreover, the induced classical/quantum-modalities are strong monads \eqref{TypingOfStrongMonads}
with respect to the monoidal structures of Prop. \ref{DoublyClosedMonoidalStructure}, whence we have $\return{}{}$- and $\bind{}{}$-operations \eqref{BindingAndReturning} as follows, using  \eqref{LinearTypeDeclaration}:

\vspace{-.3cm}
\begin{equation}
\label{ClassicalQuantumBindReturn}
\adjustbox{
}{
\hspace{-.6cm}
\def\arraystretch{5}
\def\tabcolsep{1pt}
\begin{tabular}{|l||ll|}
\hline
\raisebox{4pt}{
\rotatebox
  [origin=c]{90}
  {\bf Classically}
  $\classically$
}
&
$
  \declare
    {
      \return
        { \classical }
        { W }
    }
    {
      \ABundleType
        {
          \HilbertSpace{H}_\bullet
        }
        {
          W
        }
      \to
      \ABundleType
        {
          0_\bullet
        }
        {
          W
        }
    }
    {
      \vert \psi_w \rangle
      \;\;\mapsto\;\;
      0_w
    }
$
&
$
  \declare
    { \bind{\classically}{} }
    {
      \left(
        \ABundleType
          {
            \HilbertSpace{H}_\bullet
          }
          {
            W
          }
        \to
        \ABundleType
          {
            0_\bullet
          }
          {
            W'
          }
      \right)
      \to
      \left(
        \ABundleType
          {
            0_\bullet
          }
          {
            W
          }
        \to
        \ABundleType
          {
            0_\bullet
          }
          {
            W'
          }
      \right)
    }
    {
      \;\;
      \big(
        \vert \psi_w \rangle
        \,\mapsto\,
        0_{f(w)}
        ,\,
        0
      \big)
      \;\;
      \mapsto
      \;\;
      \big(
        0_w
        \;\;\mapsto\;\;
        0_{f(w)}
        ,\,
        0
      \big)
    }
$
\\
\hline
\end{tabular}
}
\end{equation}
\begin{proof}
It is evident that the inclusions are fully faithful and reflective. Formally we may check the required hom-isomorphisms
\eqref{FormingAdjuncts}
using \eqref{HomSetsOfBundleTypes}:
\begin{align*}
\hspace{-3mm}
    \mathrm{Hom}\left(
      \ABundleType
        { \HilbertSpace{H}_\bullet }
        { W }
      ,
      \ABundleType
        { 0_\bullet }
        { W' }
    \right)
&    \;\simeq\;
    \mathrm{Hom}(W,W')
    \times
    \underset{w}{\prod}
    \mathrm{Hom}
    \big(
      \HilbertSpace{H}_w
      ,
      0
    \big)
    \;\simeq\;
    \mathrm{Hom}
    \big(
      W,\, W'
    \big)
    \;\simeq\;
    \mathrm{Hom}
    \left(
      \ABundleType
        { 0_\bullet }
        { W }
      ,\,
      \ABundleType
        { 0_\bullet }
        { W' }
    \right)
\\
    \mathrm{Hom}
    \left(
      \ABundleType
        { \HilbertSpace{H}_\bullet }
        { W }
      ,
      \ABundleType
        { \HilbertSpace{H}' }
        { \ast }
    \right)
&    \;\simeq\;
    \mathrm{Hom}\big(
      \HilbertSpace{H}
      ,\,
      \ast
    \big)
    \times
    \underset{w}{\prod}
    \mathrm{Hom}
    \big(
      \HilbertSpace{H}_w
      ,\,
      \HilbertSpace{H}'
    \big)
    \;\simeq\;
    \mathrm{Hom}
    \big(
      \coprod_w
      \HilbertSpace{H}_w
      ,\,
      \HilbertSpace{H}'
    \big)
    \;\simeq\;
    \mathrm{Hom}
    \left(
      \ABundleType
        { \osum_w \HilbertSpace{H}_w }
        { \ast }
      ,\,
      \ABundleType
        { \HilbertSpace{H}' }
        { \ast }
    \right)
\end{align*}
Monadicity follows because every reflective inclusion is monadic (e.g. \cite[Cor. 4.2.4]{Borceux94}). Alternatively, we may invoke the
monadicity theorem in the form \eqref{MonadicityTheorem}: Since both inclusions are
right adjoints and evidently conservative, it is sufficient to observe that they both preserve all coequalizers.
For this we can appeal to \cite[Prop. A.9]{EoS}.

Finally, to check that the induced monads are strong, we may equivalently check that they are monoidal \eqref{SymmetricMonoidalMonadStructure}: The (strong) monoidal structure on the underlying functors is indicated vertically in the following diagrams. Since the monads are idempotent,
it is sufficient to check furthermore that their unit transformations are monoidal, hence that these squares commute, which is
immediate in components \eqref{ClassicalQuantumBindReturn}:

\vspace{2mm}

  \hspace{-.3cm}
  &
  $
  \begin{tikzcd}[
    row sep=-2pt
  ]
    \LinearTypes
    \;\defneq\;
    \Modales
      { \linearly }
      { \Types }
    \\
    \hspace{-30pt}
    \ABundleType
      { \HilbertSpace{H} }
      { \ast }
  \end{tikzcd}
  $
  &
  $
  \begin{tikzcd}[
    row sep=1pt
  ]
    \HilbertSpace{H}
    \ar[rr]
    &&
    \HilbertSpace{H}'
    \\
    \ABundleType
      { \HilbertSpace{H} }
      { \ast }
    \ar[
      rr,
      shift left=10pt,
      "{ \phi }"
    ]
    \ar[
      rr,
      shift right=10pt,
      "{  }"
    ]
    &&
    \ABundleType
      { \HilbertSpace{H}' }
      { \ast }
  \end{tikzcd}
  $
  \\
  \hline
\end{tabular}
\end{tabular}
\end{center}

In fact, the purely classical types are also coreflective, whence the classical-modality $\classically$
is in fact a {\it bireflective Frobenius modality} \ifdefined\monadology\cite[Def. 8]{FHPTST99}\else (cf. \cref{BireflectiveFrobeniusMonads})\fi:
\begin{proposition}[{\bf Coreflection of classical types among linear bundle types}] We have an ambidextrous reflection:
\vspace{-2mm}
\begin{equation}
  \label{CoreflectionOfClassicalTypesAmongLinearBundleTypes}
  \begin{tikzcd}
  \ClassicalTypes
  \ar[
    from=rr,
    shift left=8pt,
    "{
      \classically
    }"{description}
  ]
  \ar[
    rr,
    hook
  ]
  \ar[
    from=rr,
    shift right=8pt,
    "{
      \classically
    }"{description}
  ]
  &&
  \Types \;.
  \end{tikzcd}
\end{equation}
\end{proposition}


\noindent
{\bf Quantization and Exponential modality.}
Composing the Cartesian hom-adjunction for $\TensorUnit$ (from Prop. \ref{DoublyClosedMonoidalStructure}) with the classicality-coreflection \eqref{CoreflectionOfClassicalTypesAmongLinearBundleTypes} gives another adjunction between linear bundle types and purely classical types:
\vspace{-2mm}
\begin{equation}
  \label{TheOtherClassicalReflection}
  \begin{tikzcd}[column sep=large]
    \scalebox{\termscale}{$
      W
    $}
    \ar[
      rrrr,
      phantom,
      "{
        \longmapsto
      }"
    ]
    &&&&
    \scalebox{.7}{$
    \ABundleType
      { \TensorUnit_\bullet }
      { W }
    $}
    \\
    \ClassicalTypes
    \ar[
      from=rr,
      shift left=7pt,
      "{
        \classically
      }"
    ]
    \ar[
      rr,
      shift left=7pt,
      hook
    ]
    \ar[
      rr,
      phantom,
      "{
        \scalebox{.7}{$
          \bot
        $}
      }"
    ]
    \ar[
      rrrr,
      rounded corners,
      to path={
         ([yshift=+00pt]\tikztostart.north)
      -- ([yshift=+12pt]\tikztostart.north)
      -- ([yshift=+12pt]\tikztotarget.north)
      -- ([yshift=+00pt]\tikztotarget.north)
      }
    ]
    \ar[
      from=rrrr,
      rounded corners,
      to path={
         ([yshift=-00pt]\tikztostart.south)
      -- ([yshift=-12pt]\tikztostart.south)
      -- ([yshift=-12pt]\tikztotarget.south)
      -- ([yshift=-00pt]\tikztotarget.south)
      }
    ]
    &&
    \Types
    \ar[
      from=rr,
      shift left=7pt,
      "{
         \TensorUnit
         \to
         (\mbox{-})
      }"
    ]
    \ar[
      rr,
      shift left=7pt,
      "{
        \TensorUnit
        \times
        (\mbox{-})
      }"
    ]
    \ar[
      rr,
      phantom,
      "{
        \scalebox{.7}{$
          \bot
        $}
      }"
    ]
    &&
    \Types
    \\
    \scalebox{\termscale}{$
    \mathclap{
    (w \isa W)
    \times
    \classically\big(
      \TensorUnit
        \to
      \HilbertSpace{H}_w
    \big)
    }
    $}
    \ar[
      rrrr,
      phantom,
      "{\longmapsfrom}"
    ]
    &&&&
    \scalebox{.7}{$
    \ABundleType
      { \HilbertSpace{H}_\bullet }
      { W }
    $}
  \end{tikzcd}
\end{equation}
(cf. Rem. \ref{ExponentialModalityInQuipper}).
Further composing \eqref{TheOtherClassicalReflection} with the reflection of purely quantum types
\eqref{SubcategoriesOfBundleTypes} gives (cf. Rem. \ref{ExponentialModalityGivesLinearSpanOfUnderlyingSet}):

\begin{proposition}[\bf Quantization and  Classicization]
\label{QuantizationAndExponentialAdjunction}
$\,$

\noindent {\bf (i)} We have a pair of adjoint functors between purely classical and purely quantum
types \eqref{SubcategoriesOfBundleTypes} of this form
\vspace{-2mm}
\begin{equation}
  \label{QuantizationAdjunction}
  \begin{tikzcd}[row sep=22pt, column sep=huge]
    \scalebox{\termscale}{$
       W
    $}
    \ar[
      rrrr,
      phantom,
      "{ \longmapsto }"
    ]
    &&&&
    \scalebox{\termscale}{$
      \underset{W}{\osum}
      \TensorUnit
    $}
    \\
    \ClassicalTypes
    \ar[
      rr,
      shift left=7pt,
      "{
        \TensorUnit
          \times
        (\mbox{-})
      }"
    ]
    \ar[
      from=rr,
      shift left=7pt,
      "{
        \classically
        (
        \TensorUnit
        \to
        (\mbox{-})
        )
      }"
    ]
    \ar[
      rr,
      phantom,
      "{
        \scalebox{.7}{$\bot$}
      }"
    ]
    \ar[
      rrrr,
      rounded corners,
      to path ={
         ([yshift=+00pt]\tikztostart.north)
      -- ([yshift=+15pt]\tikztostart.north)
      -- node[yshift=+3pt]{
       \colorbox{white}{
       \scalebox{.7}{$
          \overset{
            \mathclap{
              \raisebox{3pt}{
                \scalebox{.9}{
                  \color{darkorange}
                  \bf
                  quantized
                }
              }
            }
          }{
            \quantized
          }
          \;\;\;\;\;\;
          \defneq
          \;\;\;\;\;\;
          \overset{
            \mathclap{
              \raisebox{3pt}{
                \scalebox{.9}{
                  \color{darkblue}
                  \bf
                  motive
                }
              }
            }
          }{
            \Sigma^\infty_+
          }
        $}
        }
      }
         ([yshift=+14pt]\tikztotarget.north)
      -- ([yshift=-00pt]\tikztotarget.north)
      }
    ]
    &&
    \ar[
      rr,
      shift left=7pt,
      "{
        \quantumly
      }",
    ]
    \ar[
      from=rr,
      shift left=7pt,
      hook'
    ]
    \ar[
      rr,
      phantom,
      "{
        \scalebox{.7}{$\bot$}
      }"
    ]
    \BundleTypes
    &&
    \QuantumTypes
    \ar[
      llll,
      rounded corners,
      to path ={
         ([yshift=-00pt]\tikztostart.south)
      -- ([yshift=-14pt]\tikztostart.south)
      -- node[yshift=-4pt]{
        \colorbox{white}{
        \scalebox{.7}{$
          \underset{
            \mathclap{
              \raisebox{-5pt}{
              \scalebox{.9}{
                \color{darkorange}
                \bf
                classicized
              }
              }
            }
          }{
            \classicized
          }
          \;\;\;
          \defneq
          \;\;\;
          \Omega^\infty_+
        $}
        }
      }
         ([yshift=-14pt]\tikztotarget.south)
      -- ([yshift=-00pt]\tikztotarget.south)
      }
    ]
    \ar[out=-50, in=+50,
      looseness=4.5,
      shorten=-3pt,
      shift right=4,
      "\scalebox{1.7}{${
          \hspace{1pt}
          \mathclap{
            {\ExponentialModality}
          }
        }$}"{pos=.5, description},
    ]
    \ar[out=-50, in=+50,
      looseness=4.5,
      phantom,
      shift right=10,
      "{
        \scalebox{.7}{
          \color{darkorange}
          \bf
          \def\arraystretch{.9}
          \begin{tabular}{c}
            exponential
            \\
            modality
          \end{tabular}
        }
      }"{
        description,
        rotate=-90
        },
    ]
    \\[+5pt]
    \scalebox{\termscale}{$
    \classically(
      \TensorUnit
      \to
      \HilbertSpace{H}
    )
    $}
    \ar[
      rrrr,
      phantom,
      "{ \longmapsfrom }"
    ]
    &&&&
    \scalebox{\termscale}{$
      \HilbertSpace{H}
    $}
  \end{tikzcd}
\end{equation}

\vspace{-2mm}
\noindent where the composite $\ExponentialModality \defneq QC$ is the ``exponential modality'' (\emph{Rem. \ref{ExponentialModalityGivesLinearSpanOfUnderlyingSet})}.

\noindent {\bf (ii)} These are monoidal with respect to the classical/quantum monoidal structures (Prop. \ref{DoublyClosedMonoidalStructure})
via natural transformations of
the following form:
\vspace{-2mm}
\begin{equation}
  \label{QuantizationClassicizationAreStrongMonoidal}
  \def\arraycolsep{0pt}
  \def\arraystretch{1.2}
  \begin{array}{rcl}
  W, W' \,\isa\, \ClassicalType
  &
  \hspace{.7cm}
  \yields
  \hspace{.7cm}
  &
  (\quantized W) \otimes (\quantized W')
  \;\simeq\;
  \quantized(W \times W')
  \\
  \HilbertSpace{H},
  \HilbertSpace{H}' \,\isa\,
  \QuantumType
  &
  \hspace{.7cm}
  \yields
  \hspace{.7cm}
  &
  (\classicized \,\HilbertSpace{H}\,)
  \times
  (\classicized \,\HilbertSpace{H}'\,)
  \,\simeq\,
  \classicized
  (\,\HilbertSpace{H}\,
    \times
  \,\HilbertSpace{H}'\,)
  \\
  \HilbertSpace{H},
  \HilbertSpace{H}' \,\isa\,
  \QuantumType
  &
  \hspace{.7cm}
  \yields
  \hspace{.7cm}
  &
  (\classicized \,\HilbertSpace{H}\,)
  \times
  (\classicized \,\HilbertSpace{H}'\,)
  \,\to\,
  \classicized
  (\,\HilbertSpace{H}\,
    \otimes
  \,\HilbertSpace{H}'\,)
  \end{array}
\end{equation}

\vspace{-2mm}
\begin{equation}
  \def\arraystretch{1.6}
  \begin{array}{l}
  \quantized \, \ast
  \;\simeq\;
  \TensorUnit
  \,,\hspace{.7cm}
  \classicized
  \,
  0
  \;\simeq\;
  \TensorUnit
  \,,\hspace{.7cm}
  \classicized \TensorUnit
  \;\to\;
  \TensorUnit \;.
  \end{array}
\end{equation}
\noindent {\bf (iii)} In particular, the induced modality
\eqref{QuantizationAdjunction}
sends (direct) sums to (tensor) products
\vspace{-2mm}
$$
  \ExponentialModality\big(
    \HilbertSpace{H}
    \oplus
    \HilbertSpace{H}'
  \big)
  \;\defneq\;
  \quantized \classicized
  \big(
    \HilbertSpace{H}
    \oplus
    \HilbertSpace{H}'
  \big)
  \;\simeq\;
  \quantized
  \big(
    (\classicized\HilbertSpace{H})
    \times
    (\classicized\HilbertSpace{H}')
  \big)
  \;\simeq\;
    (\quantized\classicized\HilbertSpace{H})
    \times
    (\quantized\classicized\HilbertSpace{H}')
  \;\defneq\;
  (\ExponentialModality \HilbertSpace{H})
  \otimes
  (\ExponentialModality \HilbertSpace{H}')
$$

\vspace{-2mm}
\noindent
and zero (objects) to unit (objects)
\vspace{-2mm}
$$
  \ExponentialModality
  \,0
  \;\defneq\;
  \quantized
  \classicized
  \,0
  \;\simeq\;
  \quantized
  \, \ast
  \;\simeq\;
  \TensorUnit
  \,,
$$
\vspace{-2mm}

\noindent as befits an exponential map.
\end{proposition}
\begin{proof}
  The adjunction itself is the composite
  of \eqref{TheOtherClassicalReflection} with \eqref{SubcategoriesOfBundleTypes}, as shown.

  That $\quantized$ is strong monoidal follows for instance from the fact that $\HilbertSpace{H} \otimes (\mbox{-})$ is a left adjoint and hence distributes over the coproduct $\osum_{W}$:
  \vspace{-2mm}
  $$
      (\quantized W)
      \otimes
      (\quantized W')
      \;\defneq\;
      (\osum_W \TensorUnit)
      \otimes
      (\osum_{W'} \TensorUnit)
      \;\defneq\;
      \underset{W \times W'}{\osum}
      (\TensorUnit
      \otimes
      \TensorUnit)
      \;=\;
      \underset{W \times W'}{\osum}
      \TensorUnit
      \;\defneq\;
      \quantized(W \times W')
      \,.
  $$

  \vspace{-2mm}
\noindent
  Similarly, $\classicized$ is strong monoidal with respect to the Cartesian product on both sides, since $\classically(\TensorUnit \to (\mbox{-}))$ is a right adjoint, whence it becomes lax monoidal with respect to the tensor product by composition with the universal bilinear map:
  \vspace{-2mm}
  $$
    \def\arraystretch{1.5}
    \begin{array}{lll}
      (\classicized \HilbertSpace{H})
      \times
      (\classicized \HilbertSpace{H})
    &      \;\defneq\;
      \classically
      \big(
        \TensorUnit \to \HilbertSpace{H}
      \big)
      \times
      \classically
      \big(
        \TensorUnit \to \HilbertSpace{H}'
      \big)
      &
      \\
     & \;\simeq\;
      \classically
      \Big(
      \big(
        \TensorUnit \to \HilbertSpace{H}
      \big)
      \times
      \big(
        \TensorUnit \to \HilbertSpace{H}'
      \big)
      \Big)
      &
      \proofstep{
        since $\classically$ is right adjoint
      }
      \\
    &  \;\simeq\;
      \classically
      \Big(
      \big(
        \TensorUnit \to
        (
        \HilbertSpace{H}
        \times
        \HilbertSpace{H}'
        )
      \big)
      \Big)
      &
      \proofstep{
        since
          $\TensorUnit \to (\mbox{-})$
        is right adjoint
      }
      \\
     & \;\defneq\;
      \classicized
      (\HilbertSpace{H}
        \times
      \HilbertSpace{H}')
      \\
    &  \;\to\;
      \classicized(
        \HilbertSpace{H}
          \otimes
        \HilbertSpace{H}'
      )
      &
      \proofstep{
        univ. bilin.
      }
    \end{array}
  $$

  \vspace{-7mm}
\end{proof}

\begin{remark}[\bf Exponential modality, traditionally]
\label{ExponentialModalityGivesLinearSpanOfUnderlyingSet}
Prop \ref{QuantizationAndExponentialAdjunction}
recovers -- via dependent linear type formations -- the {\it exponential modality}
\eqref{ExponentialModality}
usually postulated in linear logic/type theory (Lit. \ref{VerificationLiterature}).\footnote{
  Beware that \cite[p. 9]{FuKishidaSelinger20} instead use ``$!$'' to denote the comonad induced by the adjunction $\TensorUnit \times (\mbox{-}) \,\dashv\, \TensorUnit \to (\mbox{-})$ inside \eqref{TheOtherClassicalReflection}.
}
In the model
$\QuantumTypes \,\defneq\, \Modules{\GroundField}$
\eqref{CategoryOfIndexedSetsOfComplexVectorSpaces},
the operation $\HilbertSpace{H} \mapsto \classically(\TensorUnit \to \HilbertSpace{H})$ \eqref{TheOtherClassicalReflection}
produces the {\it underlying set} of vectors in the vector space $\HilbertSpace{H}$, whence the exponential
modality \eqref{QuantizationAdjunction} sends a vector space to the linear span of its underlying set of vectors
$$
  \HilbertSpace{H}
  \,\isa\,
  \Modules{\GroundField}
  \hspace{.7cm}
  \yields
  \hspace{.7cm}
  ! \HilbertSpace{H}
  \,=\,
  \underset{\HilbertSpace{H}}{\osum}
  \TensorUnit
  \,.
$$

\vspace{-2mm}
\noindent As an aside it is interesting that in the homotopy-theoretic semantics of {\HoTT} in parameterized spectra,
the exponential modality \eqref{QuantizationAdjunction} on, in that case, $\QuantumTypes \defneq \mathrm{Spectra}$ is
known to behave like an exponential function in the sense of ``Goodwillie calculus'', see  \cite[Ex. 2.6]{AroneChing19}.
\end{remark}

\begin{remark}[\bf Exponential modality, in {\Quipper}]
  \label{ExponentialModalityInQuipper}
  In contrast to Rem. \ref{ExponentialModalityGivesLinearSpanOfUnderlyingSet},
  beware that the literature on {\Quipper} (Lit. \ref{LiteratureQuantumProgrammingLanguages}) instead chooses to write ``!'' for the comonad induced in \eqref{TheOtherClassicalReflection}, see \cite[\S 3.5, \S 3.7 \& Def. 3.7]{RiosSelinger18}.
\end{remark}

\begin{remark}[\bf Role of the exponential modality]
Below in \cref{Pseudocode} we will not have much use for the exponential modality: Its purpose in traditional linear logic/type theory is to get
access to a stand-in for classical types in a theory that natively only knows about linear types. But this becomes a moot point in a
classically-dependent linear type theory like {\LHoTT}, as formally reflected by the above construction showing that the exponential
modality is derivable from dependent linear type formation. For our purpose here this construction serves to show that {\LHoTT}
is backwards-compatible with previous discussion of linear type theory via an exponential modality, cf. \cite[p. 9]{Riley23}.
\end{remark}

\noindent {\bf Quantum type declaration.}
For transparent distinction between the classical and quantum monoidal structures from Prop. \ref{DoublyClosedMonoidalStructure} it is convenient to use, besides the standard notation for

\begin{itemize}[leftmargin=.5cm]
\item
 the classical type declaration in the empty context
 \vspace{-2mm}
$$
  \phantom{\ast}
  \hspace{.4cm}
  \yields
  \hspace{.4cm}
  w \isa W
  \mathrlap{\,,}
$$

\vspace{-2mm}
\noindent which is equivalently type declaration in the context of the cartesian monoidal unit $\ast \isa \ClassicalType$
$$
  \ast
  \hspace{.4cm}
  \yields
  \hspace{.4cm}
  w \isa W
  \mathrlap{\,,}
$$
\end{itemize}

\vspace{-2mm}
\noindent
also notation for
\begin{itemize}[leftmargin=.5cm]
\item
 a linear (quantum) type declaration
\[
  \phantom{\TensorUnit}
  \hspace{.4cm}
  \yields
  \hspace{.4cm}
  \vert \psi \rangle
  \isalin
  \HilbertSpace{H}
  \mathrlap{\,,}
\]
to be understood as syntactic sugar for (ordinary) type declaration in the context of the tensor monoidal unit:
\[
  \TensorUnit
  \hspace{.3cm}
  \yields
  \hspace{.4cm}
  \vert \psi \rangle
    \isa
  \HilbertSpace{H}
  \mathrlap{\,,}
  \mathrlap{\,.}
\]
\end{itemize}
This little notational device will be particularly useful when declaring data of type $W \to \HilbertSpace{H}$ \eqref{ClassicalToQuantumHom}.

\begin{equation}
\label{LinearTypeDeclaration}
\footnotesize
\adjustbox{}{
\def\arraystretch{1.8}
\tabcolsep=4pt

  $
  &
  $
    \begin{tikzcd}[column sep=large]
      \ABundleType
        { \TensorUnit }
        { \ast }
      \ar[
        r,
        "{
          \sum_w
          \!
          \vert w \rangle
        }",
        shift left=8pt
      ]
      \ar[
        r,
        "{\ast}"{description},
        shift right=9pt
      ]
      &
      \ABundleType
        { \underset{W}{\prod}
          \HilbertSpace{H} }
        { \ast }
    \end{tikzcd}
  $
  \\
  \hline
\end{tabular}
}
\end{equation}

\medskip
\smallskip

For open terms (terms in an arbitrary context), it will will be most convenient to say that $f \isalin B$ (for arbitrary linear bundle type $B$) means $f \isa C B \defneq \classically(\TensorUnit \to B)$; equivalently, we will assume that in $f \isalin B$, the expression $f$ is ``dull'', or an element of the external hom from the tensor unit $\TensorUnit$ to $B$.

\begin{proposition}[{\bf Modality as free vector space}]
\label{prop:Q.modality.as.free.vector.space}

For any classical type $W$ and linear type $\HilbertSpace{H}$, we have an equivalence of classical types
\[
  C(W \to \HilbertSpace{H})
   \;\simeq\;
  C(QW \multimap \HilbertSpace{H})
\]
In particular, we get a bijection between judgements
\[f \isalin W \to \HilbertSpace{H} \quad\mbox{and}\quad f \isalin QW \multimap \HilbertSpace{H}.\]
\end{proposition}
\begin{proof}
 This follows by using the expressions in \eqref{ClassicalToQuantumHom}.
\end{proof}

We will have much use in \cref{Pseudocode} for the following:

\begin{definition}[\bf Quantization modality]
\label{QuantizationModality}
We will regard quantization
\eqref{QuantizationAdjunction} as the {\it relative} monad \eqref{RelativeBindOperation}
obtained by restricting
\eqref{RelativeMonadByPrecomposition}
the quantum-modality $\quantumly$ \eqref{ClassicalAndQuantumModality}
along precomposition with \eqref{TheOtherClassicalReflection}:
\vspace{-2mm}
\begin{equation}
    \label{QuantizationFunctor}
    \begin{tikzcd}[row sep=0pt, column sep=10pt]
      \quantized
      \;\isa
      &
      \ClassicalTypes
      \ar[
        rr,
        "{
          \TensorUnit
          \times
          (\mbox{-})
        }"
      ]
      &&
      \BundleTypes
      \ar[
        rr,
        "{ \quantumly }"
      ]
      &&
      \BundleTypes
      \\
      &
      \scalebox{\termscale}{$W$}
      &\mapsto&
      \scalebox{0.7}{$
      \ABundleType
        { \TensorUnit_\bullet }
        { W }
      $}
      &\mapsto&
      \scalebox{\termscale}{$
      \underset{W}{\oplus}\TensorUnit
      $}
    \end{tikzcd}
  \end{equation}

\vspace{-2mm}
\noindent This (just) means that we take the $\return{}{}$- and $\bind{}{}{}$-operations
\eqref{BindingAndReturning}
of $\quantized$ to be special instances of those of $\quantumly$, as follows, where we use
the linear type declaration from \eqref{LinearTypeDeclaration}:
\vspace{-2mm}
\begin{equation}
  \label{QuantizationBind}
  \declarelin
    {
      \return{\quantized}{W}
    }
    {
      \big(
      \TensorUnit
        \!\times\!
      W
      \maplin
      \quantized W
      \big)
    }
    {
      \return
        { \quantumly }
        { \TensorUnit \!\times\! W }
    }
  \hspace{1.2cm}
  \declarelin
    {
      \bind
        { \quantized }
        {
          W, \quantized W'
        }
    }
    {
      \big(
        \TensorUnit
        \!\times\!
        W
        \maplin
        \quantized W'
      \big)
      \maplin
      \big(
        \quantized W
        \maplin
        \quantized W'
      \big)
    }
    {
      \bind
        { \quantumly }
        {
          \TensorUnit
          \times
          W,
          \quantized W'
        }
    }
\end{equation}

\vspace{-2mm}
\noindent But in these special cases of $\quantumly$-operations we may, by
\eqref{ClassicalToQuantumHom}, equivalently write this pleasantly suggestively as follows:
\begin{equation}
\label{BindReturnForQuantizationModality}
\adjustbox{
}{
\begin{tabular}{|l||l|}
\hline
\rotatebox[origin=c]{90}{
  \bf Quantized
}
$\quantized$
 &
 $
  \declarelin
    {
      \return{\quantized}{W}
    }
    {
      \big(
      W
      \to
      \quantized W
      \big)
    }
    {
      \;\,
      w
      \,\mapsto\,
      \vert w \rangle
    }
  \hspace{1.2cm}
  \declarelin
    {
      \bind
        { \quantized }
        {  }
    }
    {
      \big(
        W
        \to
        \quantized W'
      \big)
      \maplin
      \big(
        \quantized W
        -\!\!\!-\!\!\!-\!\!\!-\!\!\!-\!\!\!-\!\!\!\maplin
        \quantized W'
      \big)
    }
    {
      \!
      \big(
        w
        \,\mapsto\,
        \vert \psi_w \rangle
      \big)
      \mapsto
      \Big(
      \underset{w}{\sum}
      q_{{}_w}
      \vert w \rangle
      \,\mapsto\,
      \underset{w}{\sum}
      q_w
      \vert \psi_w
      \rangle
      \Big)
    }
$
\\
\hline
\end{tabular}
}
\end{equation}
\end{definition}

\noindent
Hence the quantization monad, when handed a classical state $w$, ${\tt return}s$ the corresponding quantum state $\vert w \rangle$.
In quantum information theory, this is commonly used in the following:

\begin{example}[\bf Type of qbits]
The notation for the quantization-monad
(Def. \ref{QuantizationModality}) is such as to reproduce the standard notation ``$\QBits$'' for
the type of q-bits
(e.g. \cite[\S 1.2]{NielsenChuang10}, often also ``$\mathrm{qubit}$'', e.g. \cite{Rios21})
as the quantum analog of the type $\Bits \,\defneq\, \{0,1\}$ of classical bits (cf. \cite[(110)]{TQP}):

\begin{equation}
  \label{TypeOfQBits}
  \QBits
  \,\defneq\,
  \quantized(\Bits)
  \,\defneq\,
  \linearly (\TensorUnit_{\Bit})
  \,\defneq\,
  \underset{\Bit}{\osum}
  \,
  \underset{\Bit}{\TensorUnit}
  \,\defneq\,
  \underset{\{0,1\}}{\osum}
  \,
  \underset{\{0,1\}}{\TensorUnit}
  \,\defneq\,
  \TensorUnit_{{}_0}
  \oplus
  \TensorUnit_{{}_1}
  \,=\,
  \big\{
    q_{{}_0}
    \,
    \vert 0 \rangle
    \,+\,
    q_{{}_1}
    \,
    \vert 1 \rangle
  \big\}.
\end{equation}
\end{example}
Similarly we have the restriction of the quantum-modality to tensor products, hence to entangled states:

\begin{definition}[\bf Entanglement modality]
\label{TensorProductAsRelativeMonad}
Recalling the cartesian product of classical types and the tensor product
(Prop. \ref{DoublyClosedMonoidalStructure})
of quantized linear types (Def. \ref{QuantizationModality})
\vspace{-2mm}
$$
  \def\arraystretch{1.3}
\begin{array}{r}
  (\mbox{-}) \times (\mbox{-})
  \;\colon\;
  \ClassicalTypes
    \times
  \ClassicalTypes
  \longrightarrow
  \Types
  \\
  \quantized(\mbox{-})
  \otimes
  \quantized(\mbox{-})
  \;\colon\;
  \ClassicalTypes
    \times
  \ClassicalTypes
  \longrightarrow
  \Types
  \end{array}
$$
\vspace{-2mm}

\noindent
the restriction of the $\quantumly$-monad along $\quantized (-) \otimes \quantized (-)$ yields a relative monad of this form (recalling that $\quantumly$ is the identity on linear types)
\vspace{-1mm}
\begin{equation}
  \label{TensorProductBind}
  \adjustbox{}{
  \begin{tabular}{|c|c|}
  \hline
  \rotatebox[origin=c]{90}{
    \bf entangled
  }
  &
  $
  \begin{array}{l}
  \declarelin
    { \return{\otimes}{(B_1, B_2)} }
    { B_1 \times B_2 \to \quantized B_1 \otimes \quantized B_2 }
    { (b_1, b_2) \mapsto \vert b_1 \rangle \otimes \vert b_2 \rangle }
  \\
  \\[-5pt]
  \declarelin
    {
      \bind{\otimes}{(B_1,B_2), \HilbertSpace{H}}
    }
    {
      \big(
        B_1 \times B_2
          \to
        \HilbertSpace{H}
      \big)
      \maplin
      \big(
        \quantized B_1 \otimes \quantized B_2
        -\!\!\!-\!\!\!-\!\!\!-\!\!\!-\!\!\!-\!\!\!-\!\!\!-\!\!\!\maplin
        \HilbertSpace{H}
      \big)
    }
    {
      \big(
        (b_1, b_2)
        \mapsto
        \vert \psi_{{}_{b_1, b_2}}\rangle
      \big)
      \mapsto
      \Big(
      \big(
        \underset{b_1,b_2}{\sum}
        q_{{}_{b_1,b_2}}
        \cdot
        \vert b_1 \rangle
          \otimes
        \vert b_2 \rangle
      \big)
      \mapsto
      \underset{
        b_1, b_2
      }{
        \sum
      }
      q_{{}_{b_1, b_2}}
      \cdot
      \vert \psi_{{}_{b_1, b_2}}\rangle
      \Big)
    }
  \end{array}
  $
  \\
  \hline
  \end{tabular}
  }
\end{equation}
\end{definition}
In summary so far, we have the following fundamental quantum modalities:
\begin{center}
\adjustbox{scale=.9}{
\def\arraystretch{1.4}
\begin{tabular}{|c||l|l|}
 \hline
 \multicolumn{3}{|c|}{
   \bf
   The Quantum/Classical Divide
 }
 \\
 \hline
 \multicolumn{1}{|c||}{
   {\bf Modality}
 }
 &
 \multicolumn{1}{c|}
   {
     Idempotent
     \bf monad
  }
 &
 \multicolumn{1}{c|}
   {\bf Pure effect}
 \\
 \hline
 \hline
 {\bf
   Classical
 }
 &
 $
   \def\arraystretch{2.3}
   \def\arraycolsep{2pt}
   \begin{array}{ccccccc}
     \classically
     &\isa&
     \BundleTypes
     &\twoheadrightarrow&
     \ClassicalTypes
     &\hookrightarrow&
     \BundleTypes
     \\
     \classically
     &\defneq&
     \left[
     \def\arraystretch{.8}
     \begin{array}{c}
     \HilbertSpace{H}_\bullet
     \\
     \downarrow
     \\
     W
     \end{array}
     \right]
     &\mapsto&
     W
     &\mapsto&
     \left[
     \def\arraystretch{.8}
     \begin{array}{c}
     0_\bullet
     \\
     \downarrow
     \\
     W
     \end{array}
     \right]
   \end{array}
 $
 &
 $
  \begin{tikzcd}[
    column sep={between origins, 60pt},
    row sep=2pt
  ]
  \\[10pt]
      \unit
        { \classically  }
        { \HilbertSpace{H}_\bullet }
    &[-45pt]
    \isa
    &[-45pt]
    \HilbertSpace{H}_\bullet
    \ar[
      r,
      "{
      }"
    ]
    &
    \classically \HilbertSpace{H}_\bullet
    \\
    &&
    \def\arraystretch{.9}
    \def\arraycolsep{2pt}
    {\left[
    \begin{array}{c}
      \HilbertSpace{H}_\bullet
      \\
      \downarrow
      \\
      W
    \end{array}
    \right]}
    \ar[
      r,
      shift left=10pt,
      "{ 0 }"
    ]
    \ar[
      r,
      shift right=10pt,
      "{ \mathrm{id} }"
    ]
    &
    \def\arraystretch{.9}
    \def\arraycolsep{2pt}
    {\left[
    \begin{array}{l}
      0_\bullet
      \\
      \,\downarrow
      \\
      W
    \end{array}
    \right]}
  \end{tikzcd}
 $
 \\
 &
 \;\small (strong wrt $\times$)
 &
 \\
\hline
\end{tabular}
}
\end{center}

\newpage

\noindent
{\bf Base change and dependent classical/linear type formation.}
In a parameterized generalization of the reflection of quantum types inside all bundle types (Prop. \ref{ClassicalAndQuantumModality}),
also the $W$-parameterized linear types \eqref{CategoryOfIndexedSetsOfComplexVectorSpaces} are reflective in the {\it slice category}
$\Types_{/W}$ of bundle types over the given classical type $W = \scalebox{.7}{$\ABundleType{0_\bullet}{W}$}$:
\begin{equation}
  \label{WQuantumlyReflection}
  \begin{tikzcd}[row sep=0pt, column sep=large]
    \scalebox{\termscale}{$
    \ABundleType
      {
       \underset{
         p'(w') = w
       }{\osum}
       \!\!\!\!\!\!
       \HilbertSpace{H}'_{w'}
      }
      {
        (w \isa W)
      }
    $}
    &\mapsfrom&
    \scalebox{\termscale}{$
    \def\arraycolsep{1pt}
    \def\arraystretch{1}
    \begin{array}{c}
      \big[
      \HilbertSpace{H}'_\bullet
      \to
      W'
      \big]
      \\
      \downarrow
      \;\;\;\;\;\;\;\;\;
      \downarrow
      \mathrlap{
        \raisebox{1pt}{\scalebox{.6}{$p'$}}
      }
      \\
      \,
      \big[
      \,0_\bullet\,
      \to
      W
      \big]
    \end{array}
    $}
    \\[+6pt]
    \LinearTypes_W
    \ar[
      from=rr,
      shift right=6pt,
    ]
    \ar[
      rr,
      shift right=6pt,
      hook
    ]
    \ar[
      rr,
      phantom,
      "{
        \scalebox{.7}{
          $\bot$
        }
      }"
    ]
    &&
    \Types_{{}_{/W}}
    \ar[out=50, in=-50,
      looseness=4,
      shorten=-3pt,
      shift left=10pt,
      "{
        \;\;\;
        \scalebox{1.3}{
          ${
            \hspace{1pt}
            \mathclap{
              \quantumly_{{}_W}
            }
            \hspace{1pt}
          }$
        }
      }"{description}
    ]
    \ar[out=50, in=-50,
      looseness=5,
      phantom,
      shift left=10pt,
      "{
        \scalebox{.7}{
          \color{darkorange}
          \bf
          $W$-quantumly
        }
      }"{
        description,
        rotate=-90,
        yshift=14pt
      }
    ]
    \\
    \scalebox{\termscale}{$
    \ABundleType
     { \HilbertSpace{H}_\bullet }
     { W }
    $}
    &\mapsto&
    \scalebox{\termscale}{$
    \def\arraycolsep{1pt}
    \def\arraystretch{1}
    \begin{array}{c}
      \big[
      \HilbertSpace{H}_\bullet
      \to
      W
      \big]
      \\
      \downarrow
      \;\;\;\;\;\;\;\;\;
      \downarrow
      \\
      \,
      \big[
      \,0_\bullet\,
      \to
      W
      \big]
    \end{array}
    $}
  \end{tikzcd}
\end{equation}

\noindent However, the category of linear bundle types is locally cartesian closed; in particular:

\begin{proposition}[{\bf Left and right adjoints for linear bundle types}]
For $W,\, \Gamma \isa \ClassicalTypes$ and  $p \isa W \to \Gamma$, the pullback base change operation $W \times_\Gamma (\mbox{-})$
between the respective slices of the category of linear bundle types (Def. \ref{CategoryOfBundleTypes})
\vspace{-2mm}
$$
  \begin{tikzcd}[
    row sep=5pt
  ]
    W
    \ar[
      rr,
      "{
        p
      }"
    ]
      &&
    \Gamma
    \\[+5pt]
    \Types_{/W}
    \ar[
      from=rr,
      "{
        W \times_\Gamma (\mbox{-})
      }"{description},
      "{
        \scalebox{.65}{
          \color{darkgreen}
          \bf
          contex extension
        }
      }"{yshift=-3pt}
    ]
    &&
    \Types_{/\Gamma}
    \\
    \scalebox{\termscale}{$
    \begin{array}{c}
      \big[
        \HilbertSpace{H}'_{w'}
        \to
        (w' \isa W'_{p(w)})
      \big]
      \\
      \!\!\!
      \downarrow
        \;\;\;\;\;\;\;\;\;\;\;\;\;\;
      \downarrow
      \\
      \big[
        0_w
        \to
        (w \isa W)
      \big]
    \end{array}
    $}
    &
    \scalebox{\termscale}{$\mapsfrom$}
    &
    \scalebox{\termscale}{$
    \begin{array}{c}
      \big[
        \HilbertSpace{H}'_\bullet
        \to
        W'
      \big]
      \\
      \downarrow
        \;\;\;\;\;\;\;\;
      \downarrow
      \\
      \big[
        0_\bullet
        \to
        \Gamma
      \big]
    \end{array}
    $}
  \end{tikzcd}
$$
has both a left adjoint (``dependent coproduct \footnote{Of course, in type theory this dependent coproduct $\textstyle{\coprod}_W$ is traditionally called the ``dependent sum'' and denoted ``$\Sigma_W$''. But this (quite unnecessary but deeply engrained) abuse of terminology/notation from linear algebra becomes problematic in the context of dependent linear type theory with its actual (direct) {\it sums} $\osum_W$ of linear types.  }'') and a right adjoint (``dependent product''), given as follows:
\vspace{-2mm}
\begin{equation}
  \label{BaseChangeAdjunctionForLinearBundleTypes}
  \begin{tikzcd}[column sep=large]
    \scalebox{\termscale}{$
      \def\arraystretch{1}
\begin{array}{c}
      \big[
        \HilbertSpace{H}'_\bullet
        \,\to\,
        W'
      \big]
      \\
      \downarrow
         \;\;\;\;\;\;\;\;\;
      \downarrow
      \mathrlap{
        \!\!\!
        \raisebox{1pt}{
        \scalebox{.7}{$p'$}
        }
      }
      \\
      \big[
        0_\bullet
        \,\to\,
        W
      \big]
      \end{array}
    $}
    &\scalebox{\termscale}{$\mapsto$}&
    \scalebox{\termscale}{$
      \def\arraystretch{1}
      \begin{array}{c}
      \Big[
        \HilbertSpace{H}'_{w'_w}
        \,\to\,
        \big(
        (w, w'_w)
        \isa
        \underset{
          \mathclap{
          p(w) = \gamma
          }
        }{\coprod}
        \,
        W'_w
        \big)
      \Big]
      \\
      \downarrow
         \;\;\;\;\;\;\;\;\;\;\;\;\;\;\;\;\;\;\;\;\;\;\;\;\;\;\;\;\;\;
      \downarrow
      \\
      \big[
        0_\bullet
        \,\xrightarrow{\;\;\;\;\;\;\;\;\;\;\;\;\;\;\;\;\;\;\;\;\;\;\;}\,
        (\gamma \isa \Gamma)
      \big]
      \end{array}
    $}
    \\[+3pt]
    \Types_{/W}
    \ar[
      rr,
      shift left=18pt,
      "{
        \coprod_W
      }"{description},
      "{
        \scalebox{.65}{
          \color{darkgreen}
          \bf
          dependent coproduct
        }
      }"{yshift=4pt}
    ]
    \ar[
      from=rr,
      "{
        W \times_\Gamma (\mbox{-})
      }"{description}
    ]
    \ar[
      rr,
      shift right=18pt,
      "{
        \prod_W
      }"{description},
      "{
        \scalebox{.65}{
          \color{darkgreen}
          \bf
          dependent product
        }
      }"{swap, yshift=-4pt}
    ]
    \ar[
      rr,
      phantom,
      shift left=9pt,
      "{
        \scalebox{.7}{$\bot$}
      }"
    ]
    \ar[
      rr,
      phantom,
      shift right=9pt,
      "{
        \scalebox{.7}{$\bot$}
      }"
    ]
    &&
    \Types_{/\Gamma}
    \\
    \scalebox{\termscale}{$
      \def\arraystretch{1}
      \begin{array}{c}
      \big[
        \HilbertSpace{H}'_\bullet
        \,\to\,
        W'
      \big]
      \\
      \downarrow
         \;\;\;\;\;\;\;\;\;
      \downarrow
      \\
      \big[
        0_\bullet
        \,\to\,
        W
      \big]
      \end{array}
    $}
    &\scalebox{\termscale}{$\mapsto$}&
    \scalebox{\termscale}{$
      \def\arraystretch{1}
      \begin{array}{c}
      \bigg[
        \;\;\;\;
        \underset
        {
          \mathclap{
            p(w) = \gamma
          }
        }
        {
        \prod
        }
        \,
        \HilbertSpace{H}'_{w'_w}
        \,\to\,
        \Big(
        w'_\bullet
        \isa
        \underset
        {
          \mathclap{
            p(w) = \gamma
          }
        }
        {\prod}
        \,
        W'_{w}
        \Big)
      \bigg]
      \\
      \downarrow
         \;\;\;\;\;\;\;\;\;\;\;\;\;\;
         \;\;\;\;\;\;\;\;\;\;\;\;\;\;\;\;
      \downarrow
      \\
      \big[
        0_\bullet
        \,
        \xrightarrow
        {\;\;\;\;\;\;\;\;\;\;\;\;\;\;\;\;\;\;\;}
        \,
        (\gamma \isa \Gamma)
      \big]
      \end{array}
    $}
  \end{tikzcd}
\end{equation}
\end{proposition}
\begin{proof}
We may formally check the hom-isomorphisms, using \eqref{HomSetsOfBundleTypes}. It is sufficient to consider the case that $\Gamma = \ast$:
$$
\hspace{-2mm}
%
\end{array}
$$

\vspace{-.7cm}
\end{proof}

\newpage

The (co)restriction of the base change adjoint triple \eqref{BaseChangeAdjunctionForLinearBundleTypes}
along the reflective inclusion of $W$-quantum types \eqref{WQuantumlyReflection}
yields base change of dependent linear types:
\begin{equation}
  \label{BaseChangeOnDependentLinearTypes}
  \begin{tikzcd}[column sep=huge]
    \Types_{/W}
    \ar[
      from=rr,
      "{
        W \times_\Gamma (\mbox{-})
      }"{description}
    ]
    \ar[
      rr,
      shift left=16pt,
      "{
        \coprod_W
      }"{description}
    ]
    \ar[
      rr,
      shift right=16pt,
      "{
        \prod_W
      }"{description},
      shorten <=16pt
   ]
   \ar[
     rr,
     shift left=7pt,
     phantom,
     "{
       \scalebox{.6}{$\bot$}
     }"
   ]
   \ar[
     rr,
     shift right=7pt,
     phantom,
     "{
       \scalebox{.6}{$\bot$}
     }"
   ]
    &[-15pt]
    &
    \Types_{/\Gamma}
    \ar[
      dr,
      shift left=18pt,
      "{
        \underset{\Gamma}{\quantumly}
      }"{description}
    ]
    \ar[
      dr,
      phantom,
      shift left=7pt,
      "{
        \scalebox{.6}{$\bot$}
      }"{sloped}
    ]
    &[-15pt]
    \\[25pt]
    &
    \QuantumTypes_W
    \ar[ul, hook']
    \ar[
      rr,
      shift left=16pt,
      shorten >=10pt,
      "{ \osum_W }"{description}
    ]
    \ar[
      from=rr,
      "{
        \underset{W}{\TensorUnit}
        \otimes (\mbox{-})
      }"{description, yshift=-2pt}
    ]
    \ar[
      rr,
      shift right=16pt,
      "{ \prod_W }"{description}
    ]
    \ar[
      rr,
      shift left=8pt,
      phantom,
      "{
        \scalebox{.6}{$\bot$}
      }"
    ]
    \ar[
      rr,
      shift right=8pt,
      phantom,
      "{
        \scalebox{.6}{$\bot$}
      }"
    ]
    &&
    \QuantumTypes_\Gamma
    \ar[
      ul,
      hook',
      shorten <=00pt,
      shorten >=0pt
    ]
    \\
    &
    \scalebox{\termscale}{$
    \big(
      w \isa W
      \yields
      \HilbertSpace{H}_w
    \big)
    $}
    &
    \scalebox{\termscale}{$\mapsto$}
    &
    \scalebox{\termscale}{$
    \big(
      \gamma \isa \Gamma
      \yields
      \underset{
        \mathclap{
          p(w)=\gamma
        }
      }{\prod}
      \HilbertSpace{H}_w
    \big)
    $}
  \end{tikzcd}
\end{equation}

\vspace{-2mm}
\noindent Now something special happens: Since $\Modules{\GroundField}$ is an additive category,
it has {\it biproducts}, meaning that finite coproducts are finite products.
This is a key aspect of what it means for its objects to be {\it linear} types.

\begin{proposition}[{\bf Identifying structures and ambidexterity}]
If $W$ is {\it finite} (over $\Gamma$) then
the direct sum and the direct product of linear spaces coincide, $\osum_W \,\simeq\, \prod_W$, and so
the base change adjunction
\eqref{BaseChangeOnDependentLinearTypes}
on linear types becomes ambidextrous:
\vspace{-2mm}
\begin{equation}
  \label{AmbidextrousBaseChange}
  \Gamma \isa \ClassicalType
  ,\,\;\;
  W \isa \FiniteType
  \;\;\;\;\;\;\;\;
  \yields
  \;\;\;\;\;\;\;\;
  \begin{tikzcd}
    \scalebox{\termscale}{$
      \big(
        w \isa W
        \,\yields\,
        \HilbertSpace{H}_w
      \big)
    $}
    &\scalebox{\termscale}{$\longmapsto$}&
    \scalebox{\termscale}{$
      \big(
      \gamma \isa \Gamma
      \,\yields\,
      \underset{
        \mathclap{
          p(w)=\gamma
        }
      }{\osum}
      \,
      \HilbertSpace{H}_w
      \big)
    $}
    \\[-2pt]
    \QuantumTypes_W
    \ar[
      rr,
      shift left=16pt,
      "{
        \osum_W
      }"{description}
    ]
    \ar[
      from=rr,
      "{
        \underset{W}{\TensorUnit}
        \otimes (\mbox{-})
      }"{description, yshift=-2pt}
    ]
    \ar[
      rr,
      shift right=16pt,
      "{
        \osum_W
      }"{description}
    ]
    \ar[
      rr,
      shift left=8pt,
      phantom,
      "{
        \scalebox{.65}{$\bot$}
      }"
    ]
    \ar[
      rr,
      shift right=8pt,
      phantom,
      "{
        \scalebox{.65}{$\bot$}
      }"
    ]
    &&
    \QuantumTypes_\Gamma
  \end{tikzcd}
\end{equation}
\end{proposition}

\medskip

All these structures and properties are elementary to see in the concrete model of indexed sets of vector spaces, but they hold quite generally for  (higher) categories of  parameterized linear (homotopy) types. In fact, much of this structure is traditionally encoded by {\it Grothendieck's yoga of six operations} used in motivic (homotopy) theory.

\medskip

\noindent
{\bf Motivic yoga.}
For the purposes of the present discussion, we make the following definition (cf. \cite[p. 41]{EoS}):
\begin{definition}[\bf Motivic yoga]
  \label{MotivicYoga}
  Let $\BundleTypes$
  be a locally cartesian closed category with coproducts.
  We say that a
  {\it Grothendieck-Wirthm{\"u}ller motivic yoga of operations}
  on $\BundleTypes$ -- or just {\it motivic yoga}, for short -- is:

  \begin{itemize}[leftmargin=.7cm]
  \item[{\bf (i)}]
  an ambidextrously reflected subcategory $\ClassicalTypes$ (``of classical base types''), hence a functor $\classically$ onto a full
  subcategory such that it is both left and right adjoint to the inclusion functor:
   \vspace{-4mm}
  \begin{equation}
    \begin{tikzcd}
      \ClassicalTypes
      \ar[
        from=rr,
        shift right=14pt,
        "{ \classically }"{description}
      ]
      \ar[rr, hook]
      \ar[
        from=rr,
        shift left=14pt,
        "{ \classically }"{description}
      ]
      \ar[
        rr,
        phantom,
        shift left=6pt,
        "{\scalebox{.7}{$\bot$}}"
      ]
      \ar[
        rr,
        phantom,
        shift right=6pt,
        "{\scalebox{.7}{$\bot$}}"
      ]
      &&
      \BundleTypes
    \ar[out=-50, in=+50,
      looseness=4.5,
      shorten=-3pt,
      shift left=1,
      "\scalebox{1.1}{${
          \hspace{1pt}
          \mathclap{
            {\classically}
          }
          \hspace{4pt}
        }$}"{pos=.5, swap},
    ]
    \end{tikzcd}
  \end{equation}

   \vspace{-3mm}
\noindent
  This implies in particular that $\ClassicalTypes$ has all
  (fiber-)products and coproducts, and we write
  \begin{equation}
    \label{FiniteClassicalTypes}
    \FiniteTypes
    \;\xhookrightarrow{\;\;\;}\;
    \ClassicalTypes
  \end{equation}
  for the further full subcategory on the finite coproducts of the terminal object with itself.

 \item[{\bf (ii)}] For each $W \colon \ClassicalTypes$ a symmetric closed monoidal structure
 $(\LinearTypes_B, \otimes_B, \mathbbm{1}_B)$
 on the iso-comma categories (``of linear bundles over $W$''):
 \vspace{-2mm}
 \begin{equation}
   \LinearTypes_W
   \;\;\defneq\;\;
   \classically/W
   \;\;
   =
   \;\;
   \left\{ \hspace{-4pt}
    \begin{tikzcd}
    \ABundleType
      { \HilbertSpace{H}_\bullet }
      { W }
    \ar[
      r,
      Rightarrow,-,
      shift right=10pt
    ]
    \ar[
      r,
      shift left=10pt,
      "{
        \phi_\bullet
       }"
    ]
    &
    \ABundleType
      { \HilbertSpace{H}'_\bullet }
      { W }
    \end{tikzcd}
  \hspace{-4pt}
  \right\}
   \,,
 \end{equation}

\vspace{-2mm}
\item[{\bf (iii)}] For each morphism in $\ClassicalTypes$ an adjoint triple of (``base change'') functors:
 \vspace{-2mm}
\begin{equation}
  \mbox{for}
  \hspace{.6cm}
  B \xrightarrow{\; f \;} B'
  \hspace{.6cm}
  \mbox{we have}
  \hspace{.6cm}
  \begin{tikzcd}
    \LinearTypes_W
    \ar[
      rr,
      shift left=12pt,
      "{ f_! }"
    ]
    \ar[
      from=rr,
      " f^\ast "{description}
    ]
    \ar[
      rr,
      shift right=12pt,
      "{ f_\ast }"{swap}
    ]
    \ar[
      rr,
      phantom,
      "{ \scalebox{.7}{$\bot$} }",
      shift left=7pt
    ]
    \ar[
      rr,
      phantom,
      "{ \scalebox{.7}{$\bot$} }",
      shift right=7pt
    ]
    &&
    \LinearTypes_{W'}
  \end{tikzcd}
\end{equation}

\end{itemize}
such that the following conditions hold:

\newpage

\begin{itemize}[leftmargin=.6cm]
\item[{\bf (a)}] {\bf Linearity}:
the left and right base change along finite types $W \xrightarrow{p_W} \ast$ (see \eqref{FiniteClassicalTypes})
are naturally equivalent:
$$
  W
  \;\isa\;
  \FiniteTypes
  \hspace{.5cm}
  \yields
  \hspace{.5cm}
  (p_w)_!
    \;\simeq\;
  (p_w)_\ast
$$
\item[{\bf (b)}]
{\bf Functoriality}: for composable morphisms $f, g$ of base objects we have
\begin{equation}
(f^\ast \circ g^\ast)
\;\simeq\; g^\ast \circ f^\ast \mbox{\hspace{.4cm}and\hspace{.4cm}}  \mathrm{id}^\ast = \mathrm{id}
\end{equation}

\item[{\bf (c)}]  {\bf Monoidalness}:
the pullback functors are strong monoidal in that there are natural equivalences:
\begin{equation}
  \label{StrongMonoidalPullbackFunctors}
  f^\ast
  \big(
    \HilbertSpace{H}
    \underset{W'}{\otimes}
    \HilbertSpace{H}'
  \big)_\bullet
  \;\;
  \simeq
  \;\;
  \Big(
  f^\ast
  \big(
    \HilbertSpace{H}
  \big)
  \underset{W'}{\otimes}
  f^\ast
  \big(
    \HilbertSpace{H}'
  \big)
  \Big)_\bullet
\end{equation}

\item[{\bf (d)}]
{\bf Beck-Chevalley condition}:
for a pullback square in $\ClassicalTypes$
the ``pull-push operations'' across one tip are naturally equivalent to those across the other:
\begin{equation}
\label{BCCondition}
\hspace{-6mm}
  \mbox{For}
  \hspace{.3cm}
  \begin{tikzcd}[
    column sep=1pt,
    row sep=10pt
  ]
    &
    B
      \times_{B_0}
    B'
    \ar[
      dl,
      shorten <=-2,
      "{ \mathrm{pr}_{B} }"{swap}
    ]
    \ar[
      dr,
      shorten <=-2,
      "{ \mathrm{pr}_{B'} }"
    ]
    \ar[
      dd,
      phantom,
      "{ \scalebox{.7}{(pb)} }"{pos=.4}
    ]
    \\
    B
    \ar[
      dr,
      "{ p_B }"{swap}
    ]
    &&
    B'
    \ar[
      dl,
      "{ p_{B'} }"
    ]
    \\
    &
    B_0
  \end{tikzcd}
  \hspace{.2cm}
  \mbox{we have}
  \hspace{.2cm}
  \begin{tikzcd}[
    column sep=-20pt,
    row sep=10pt,
  ]
    &
    \LinearTypes_{B \times_{B_0} B'}
    \ar[
      dl,
      "{ (\mathrm{pr}_{B})_! }"{swap}
    ]
    \\
    \LinearTypes_B
    &&
    \LinearTypes_{B'}
    \ar[
      ul,
      "{ (\mathrm{pr}_{B'})^\ast }"{swap}
    ]
    \ar[
      dl,
      "{ (p_{B'})_! }"
    ]
    \\
    &
    \LinearTypes_{B_0}
    \ar[
      ul,
      "{ (p_B)^\ast }"
    ]
  \end{tikzcd}
  \hspace{.2cm}
  \mbox{and}
  \hspace{.2cm}
  \begin{tikzcd}[
    column sep=-20pt,
    row sep=10pt,
  ]
    &
    \LinearTypes_{B \times_{B_0} B'}
    \ar[
      dl,
      "{ (\mathrm{pr}_{B})_\ast }"{swap}
    ]
    \\
    \LinearTypes_B
    &&
    \LinearTypes_{B'}
    \ar[
      ul,
      "{ (\mathrm{pr}_{B'})^\ast }"{swap}
    ]
    \ar[
      dl,
      "{ (p_{B'})_\ast }"
    ]
    \\
    &
    \LinearTypes_{B_0}
    \ar[
      ul,
      "{ (p_B)^\ast }"
    ]
  \end{tikzcd}
\end{equation}

\item[{\bf (e)}]  {\bf Frobenius reciprocity} / {\bf projection formula}: the left pushforward of a tensor with a pullback is naturally equivalent to the tensor with the left pushforward (equivalent to $f^\ast$ being also strong closed):
\begin{equation}
  \label{FrobeniusReciprocity}
  f_!
  \big(
    \HilbertSpace{H}
    \underset{W}{\otimes}
    f^\ast(\HilbertSpace{H}')
  \big)_\bullet
  \;\;
  \simeq
  \;\;
  f_!(\HilbertSpace{H})
  \underset{W'}{\otimes}
  \HilbertSpace{H}'
\end{equation}
\item[{\bf (f)}] {\bf Stability}: Over finite classical types $f_!$ and $f_\ast$ agree to make an ambidextrous adjunction:
\begin{equation}
  W \,\isa\,
  \FiniteClassicalType
  \hspace{1cm}
  \yields
  \hspace{1cm}
  f_! \,\simeq\, f_\ast
  \;\;\isa\;\;
  \QuantumTypes_W \to \QuantumTypes
  \,.
\end{equation}
\end{itemize}

\end{definition}

\medskip
\begin{proposition}[\bf Linear bundle types satisfy Motivic Yoga]
  \label{LinearBundleTypesSatisfyMotivicYoga}
  The indexed category $W \mapsto \LinearTypes_{{}_W}$ of Def. \ref{CategoryOfBundleTypes}
 satisfies the motivic yoga (Def. \ref{MotivicYoga}) with respect to the fiberwise tensor product:
\vspace{-2mm}
 $$
   \begin{tikzcd}[row sep=0pt, column sep=10pt]
     \LinearTypes_{W}
     \times
     \LinearTypes_W
     \ar[
       rr,
       "{
         \underset{W}{\otimes}
       }"
     ]
     &&
     \LinearTypes_W
     \\
  \scalebox{0.75}{$   \left(
     \ABundleType
       { \HilbertSpace{H}_\bullet }
       { W }
     ,
     \ABundleType
       { \HilbertSpace{H}'_\bullet }
       { W }
     \right)
     $}
     &\longmapsto&
   \scalebox{0.75}{$   \ABundleType
       {
         \HilbertSpace{H}_w
         \otimes
         \HilbertSpace{H}'_w
       }
       { (w \isa W) }
       $}
   \end{tikzcd}
 $$
\end{proposition}
\begin{proof}
  This is straightforward to check. Details for this case and its higher generalization are spelled out in \cite[\S 3.3]{EoS}.
\end{proof}

\begin{remark}[{\bf Modalities via mortivic yoga}]
We may alternatively see the monoidality of $\quantumly$ and $\quantized$ just using the motivic yoga (Def. \ref{MotivicYoga}).
For this purpose we shall denote the projection maps involved in a cartesian product as follows:
\begin{equation}
  \label{ProductDiagramFormBC}
  \hspace{-5mm}
  \begin{tikzcd}[
    column sep=19pt,
    row sep=2pt
  ]
    &
    W \times W'
    \ar[
      dl,
      "{
        \mathrm{pr}_W
      }"{swap}
    ]
    \ar[
      dr,
      "{
        \mathrm{pr}_{W'}
      }"
    ]
    \ar[
      dd,
      "{ p_{W \times W'} }"{description}
    ]
    \\
    W
    \ar[
      dr,
      "{ p_W }"{swap}
    ]
    &&
    W'
    \ar[
      dl,
      "{ p_{W'} }"
    ]
    \\
    &
    \ast
  \end{tikzcd}
\end{equation}
$$
\hspace{-3mm}
\def\arraystretch{1.6}
\begin{array}{lll}
    &
    \linearly
    \big(
      \HilbertSpace{E}
        \otimes
      \HilbertSpace{E}'
    \big)
    \\
    &\;=\;
    (p_{B \times B'})_!
    \big(
     (\mathrm{pr}_{B})^\ast \HilbertSpace{E}
     \otimes
     (\mathrm{pr}_{B'})^\ast
     \HilbertSpace{E}'
    \big)
    &
    \proofstep{ def }
    \\
    &\;\simeq\;
    (p_{B})_!
    (\mathrm{pr}_{B})_!
    \big(
     (\mathrm{pr}_{B})^\ast
     \HilbertSpace{E}
     \otimes
     (\mathrm{pr}_{B'})^\ast
     \HilbertSpace{E}'
    \big)
    &
    \proofstep{ \eqref{ProductDiagramFormBC} }
    \\
    &\;\simeq\;
    (p_{B})_!
    \Big(
     \HilbertSpace{E}
     \otimes
     \big(
       (\mathrm{pr}_{B'})_!
       (\mathrm{pr}_{B'})^\ast
       E
     \big)
    \Big)
    &
    \proofstep{
      \eqref{FrobeniusReciprocity}
    }
    \\
    &\;\simeq\;
    (p_{B})_!
    \Big(
     \HilbertSpace{E}
     \otimes
     \big(
       (p_{B})^\ast
       (p_{B'})_!
       E
     \big)
    \Big)
    &
    \proofstep{
      \eqref{BCCondition}
    }
    \\
    &\;\simeq\;
     \big(
       (p_{B})_!
       E
     \big)
     \otimes
     \big(
       (p_{B'})_!
       \HilbertSpace{E}'
     \big)
     &
     \proofstep{
       \eqref{FrobeniusReciprocity}
     }
  \end{array}
 \hspace{1cm}
\def\arraystretch{1.6}
  \begin{array}{lll}
    \quantized(B \times B)
    &    \;=\;
    (p_{B \times B})_!
    (p_{B \times B})^\ast
    \mathbbm{1}
    &
    \proofstep{ def }
    \\
  &  \;\simeq\;
    (p_{B'})_!
    (\mathrm{pr}_{B'})_!
    (\mathrm{pr}_{B})^\ast
    (p_B)^\ast
    \mathbbm{1}
    &
    \proofstep{ \eqref{ProductDiagramFormBC} }
    \\
  &  \;\simeq\;
    (p_{B'})_!
    (p_{B'})^\ast
    (p_B)_!
    (p_B)^\ast
    \mathbbm{1}
    &
    \proofstep{
      \eqref{BCCondition}
    }
    \\
    &
    \;\simeq\;
    (p_{B'})_!
    \big(
    \mathbbm{1}_{B'}
    \otimes
    (p_{B'})^\ast
    (p_B)_!
    (p_B)^\ast
    \mathbbm{1}
    \big)
    &
    \proofstep{ unit law }
    \\
&    \;\simeq\;
    \big(
      (p_{B'})_!
      \mathbbm{1}_{B'}
    \big)
    \otimes
    \big(
      (p_B)_!
      (p_B)^\ast
      \mathbbm{1}
    \big)
    &
    \proofstep{
      \eqref{FrobeniusReciprocity}
    }
    \\
    &
    \;\simeq\;
    \big(
      (p_{B'})_!
      (p_{B'})^\ast
      \mathbbm{1}
    \big)
    \otimes
    \big(
      (p_B)_!
      (p_B)^\ast
      \mathbbm{1}
    \big)
    &
    \proofstep{
      \eqref{StrongMonoidalPullbackFunctors}
    }
    \\
    &
    \;=\;
    (\quantized B)
    \otimes
    (\quantized B')
    &
    \proofstep{ def\,. }
  \end{array}
$$
\end{remark}

\subsection{Classical Epistemic Logic}
\label{ClassicalEpistemicLogicViaDependentTypes}

We lay out our perspective (following \cite{nLabNecessity}\cite[Ch. 4]{Corfield20}) on (S5 Kripke semantics for) modal logic/type theory
(Lit. \ref{ModalLogicAndManyWorlds}). This is naturally realized (see Rem. \ref{ModalTypeTheoryAsPerspectiveOnDependentTypeTheory} below) by {\it dependent}
type theory (Lit. \ref{VerificationLiterature}), with ``possible worlds'' given by terms of base types and with modal operators given by the (co)monads
induced by dependent (co)product\footnote{
  We say {\it dependent co-product} ``$\coprod_B$'' for what is traditionally called the {\it dependent sum} ``$\sum_B$'' in intuitionistic type theory.
  Apart from being the more descriptive term, this avoids a clash of terminology after passage to {\it linear} type theory where actual linear sums of
  types (``direct sums'') do play a(nother) role.
}
type formation followed by context re-extension.
The discussion prepares the ground for our formal
quantum epistemic logic in \cref{QuantumEpistemicLogicViaDependentLinearTypes}.

\smallskip
For expository convenience, we speak in the 1-categorical semantics where the type universe ``$\ClassicalType$'' refers to a topos of types
(e.g.: $\Sets$) and for $B \isa \Type$ the universe $\ClassicalTypes_B$ of $B$-dependent types refers to the slice topos over $B$.
All of the discussion is readily adapted to homotopy type theory proper and its $\infty$-topos semantics without any relevant changes,
whence we do not dwell on it here (the homotopy theoretic aspect does become relevant further below).
The crux is that all the constructions considered now are readily available inside a dependently typed language such as {\HoTT} or {\LHoTT}.

\medskip

\noindent
{\bf Dependent type formation by base change.}
The starting point is the basic fact that any $W \isa \Type_\Gamma$, hence any {\it display map} $p_{{}_W} \,\isa\, W \to \Gamma$, induces
a {\it base change adjoint triple} between $W$-dependent types and bare types in the default context $\Gamma$:
\begin{equation}
\label{BasicBaseChangeAdjointtriple}
\begin{tikzcd}[
  column sep=70pt
]
  W
  \ar[
    rr,
    "{ p_w }"
  ]
  &&
  \Gamma
  \\[+10pt]
  \mathllap{
      \raisebox{1pt}{
      \scalebox{.7}{
        \color{darkblue}
        \bf
        \def\arraystretch{.7}
        \begin{tabular}{c}
        \scalebox{1}{$W$}-dependent
        \\
        types
        \end{tabular}
      }
      \hspace{5pt}
    }
  }
  \ClassicalTypes_W
    \quad
  \ar[
    rr,
    shift left=14pt,
    "{
      \overset{
        \mathclap{
          \raisebox{3pt}{
            \scalebox{.7}{
              \color{darkgreen}
              \bf
              dependent co-product
            }
          }
        }
      }{
        \coprod_W
      }
    }"
  ]
  \ar[
    from=rr,
    "{
      \scalebox{.7}{
        \color{darkgreen}
        \bf
        context\;
      }
      \times W
      \scalebox{.7}{
        \color{darkgreen}
        \bf
        \;extension
      }
    }"{description}
  ]
  \ar[
    rr,
    shift right=14pt,
    "{
      \underset{
        \mathclap{
          \raisebox{-3pt}{
            \scalebox{.7}{
              \color{darkgreen}
              \bf
              dependent product
            }
          }
        }
      }{
        \prod_W
      }
    }"{swap}
  ]
  \ar[
    rr,
    phantom,
    shift left=8pt,
    "\scalebox{.7}{$\bot$}"
  ]
  \ar[
    rr,
    phantom,
    shift right=8pt,
    "\scalebox{.7}{$\bot$}"
  ]
  &&
  \quad
  \ClassicalTypes_\Gamma
  \mathrlap{
      \hspace{5pt}
      \raisebox{1pt}{
      \scalebox{.7}{
        \color{darkblue}
        \bf
        \def\arraystretch{.9}
        \begin{tabular}{c}
          types in
          \\
          default context
        \end{tabular}
      }
    }
  }
\end{tikzcd}
\end{equation}
via
\begin{equation}
  \label{BaseChangeInComponents}
  \def\arraystretch{3}
  \begin{array}{rcl}
  D\,\isa\, \Type_W
  &
  \hspace{.5cm}
  \vdash
  \hspace{.5cm}
  &
  \begin{tikzcd}[row sep=0pt, column sep=20pt]
    \coprod_{{}_W} D
    \,\isa\,
    &
    \Gamma
    \ar[rr]
    &&
    \ClassicalTypes
    \\
    &
\scalebox{0.7}{$    \gamma $}
      &\longmapsto&
    \;\;\;
\scalebox{0.7}{$     \underset{
      \mathclap{
        w \,\isa\, \mathrm{fib}_\gamma(p_{{}_W})
      }
    }{\coprod}
    \;\;\;
    D_w
    $}
  \end{tikzcd}
  \\
  D\,\isa\, \Type_\Gamma
  &
  \hspace{.5cm}
  \vdash
  \hspace{.5cm}
  &
  \begin{tikzcd}[row sep=0pt, column sep=20pt]
    D \times W
    \,\isa\,
    &
    W
    \ar[rr]
    &&
    \ClassicalTypes
    \\
    &
\scalebox{0.7}{$     w $}
      &\longmapsto&
  \scalebox{0.7}{$   D_{p_{\!{}_{{}_W}}\!\!(w)} $}
  \end{tikzcd}
  \\
  D\,\isa\, \Type_W
  &
  \hspace{.5cm}
  \vdash
  \hspace{.5cm}
  &
  \begin{tikzcd}[row sep=0pt, column sep=20pt]
    \prod_{{}_W} D
    \,\isa\,
    &
    \Gamma
    \ar[rr]
    &&
    \ClassicalTypes
    \\
    &
   \scalebox{0.7}{$  \gamma $}
      &\longmapsto&
    \;\;\;
 \scalebox{0.7}{$    \underset{
      \mathclap{
        w \,\isa\, \mathrm{fib}_\gamma(p_{{}_W})
      }
    }{\prod}
    \;\;\;
    D_w
    $}
  \end{tikzcd}
  \end{array}
\end{equation}
whose (co)restriction along
\begin{equation}
\label{PropositionalTruncation}
\begin{tikzcd}[column sep=100pt]
  \scalebox{.7}{
    \color{darkblue}
    \bf
    types
  }
  \hspace{.2cm}
  \ClassicalTypes_\Gamma
  \ar[
    from=r,
    shift left=6pt,
    hook'
  ]
  \ar[
    r,
    shift left=7pt,
    "{
      [\mbox{-}]_0
    }"{description},
    "{
      \scalebox{.7}{
        \color{darkgreen}
        \bf
        propositional truncation
      }
    }"{yshift=4pt}
  ]
  \ar[
    r,
    phantom,
    shift right=1pt,
    "{
      \scalebox{.7}{$\bot$}
    }"
  ]
  &
  \Propositions_\Gamma
  \hspace{.2cm}
  \scalebox{.7}{
    \color{darkblue}
    \bf
    propositions
  }
\end{tikzcd}
\end{equation}
gives the quantifiers of first-order logic:
\vspace{-2mm}
\begin{equation}
\label{QuantifiersAdjointTriple}
\begin{tikzcd}[
  column sep=80pt
]
      \raisebox{1pt}{
      \scalebox{.7}{
        \color{darkblue}
        \bf
        \def\arraystretch{.7}
        \begin{tabular}{c}
        \scalebox{1}{$W$}-dependent
        \\
        propositions
        \end{tabular}
      }
      \hspace{5pt}
    }
  {\color{purple}\Propositions}_W
    \quad
  \ar[
    rr,
    shift left=14pt,
    "{
      \overset{
        \mathclap{
          \raisebox{3pt}{
            \scalebox{.7}{
              \color{darkgreen}
              \bf
              existential quantification
            }
          }
        }
      }{
        \exists_W
        \;=\;
        \big[\coprod_W(\mbox{-})\big]_0
      }
    }"
  ]
  \ar[
    from=rr,
    "{
      \scalebox{.7}{
        \color{darkgreen}
        \bf
        context\;
      }
      \times W
      \scalebox{.7}{
        \color{darkgreen}
        \bf
        \;extension
      }
    }"{description}
  ]
  \ar[
    rr,
    shift right=14pt,
    "{
      \underset{
        \mathclap{
          \raisebox{-3pt}{
            \scalebox{.7}{
              \color{darkgreen}
              \bf
              universal quantification
            }
          }
        }
      }{
        \forall_W \,=\, \prod_W
      }
    }"{swap}
  ]
  \ar[
    rr,
    phantom,
    shift left=8pt,
    "\scalebox{.7}{$\bot$}"
  ]
  \ar[
    rr,
    phantom,
    shift right=8pt,
    "\scalebox{.7}{$\bot$}"
  ]
  &&
  \quad
  {\color{purple}\Propositions}_\Gamma
      \hspace{5pt}
      \raisebox{1pt}{
      \scalebox{.7}{
        \color{darkblue}
        \bf
        \def\arraystretch{.9}
        \begin{tabular}{c}
          propositions in
          \\
          default context
        \end{tabular}
      }
    }
\end{tikzcd}
\end{equation}

It is immediate (and generally well-known but has previously received little attention in modal type theory) that by composing
the adjoint type constructors \eqref{BasicBaseChangeAdjointtriple} to endo-functors yields a pair of adjoint pairs of (co)monads:
\begin{equation}
\label{CoMonadsOfClassicalDependentTypeFormation}
\begin{tikzcd}[column sep=80pt]
   W
   \ar[
     rr,
     "{ p_{{}_W} }"
   ]
   &&
   \Gamma
   \\[20pt]
    \mathllap{
      \raisebox{1pt}{
      \scalebox{.7}{
        \color{darkblue}
        \bf
        actual data
      }
      }
      \hspace{25pt}
    }
    \ClassicalTypes_W
    \ar[out=160, in=60,
      looseness=4,
      shorten=-3pt,
      "\scalebox{1.6}{${
          \hspace{1pt}
          \mathclap{
            \possibly_{\!{}_{\scalebox{.45}{$W$}}}
          }
          \hspace{1pt}
        }$}"{pos=.25, description},
      "{
        \scalebox{.7}{
          \hspace{-20pt}
          \color{darkorange}
          \bf
          possibly
        }
      }"{pos=.55, yshift=2pt}
    ]
    \ar[out=185, in=175,
      looseness=10,
      phantom,
      "{\scalebox{.7}{$\bot$}}"{yshift=1pt, xshift=3pt}
    ]
    \ar[out=-160, in=-60,
      looseness=4,
      shorten=-3pt,
      "\scalebox{1.4}{${
          \hspace{1pt}
          \mathclap{
            \necessarily_{\mathrlap{\!{}_{\scalebox{.6}{\colorbox{white}{$\!\!\!W\!$}}}}}
          }
          \hspace{3pt}
        }$}"{pos=.25, description},
      "{
        \scalebox{.7}{
          \hspace{-40pt}
          \color{darkorange}
          \bf
          necessarily
        }
      }"{swap, pos=.59, yshift=-2.5pt}
    ]
    \quad
  \ar[
    rr,
    shift left=14pt,
    "{
      \overset{
        \mathclap{
          \raisebox{3pt}{
            \scalebox{.7}{
              \color{darkgreen}
              \bf
              dependent co-product
            }
          }
        }
      }{
        \coprod_W
      }
    }"
  ]
  \ar[
    from=rr,
    "{
      (\mbox{-}) \times W
    }"{description}
  ]
  \ar[
    rr,
    shift right=14pt,
    "{
      \underset{
        \mathclap{
          \raisebox{-3pt}{
            \scalebox{.7}{
              \color{darkgreen}
              \bf
              dependent product
            }
          }
        }
      }{
        \prod_W
      }
    }"{swap}
  ]
  \ar[
    rr,
    phantom,
    shift left=8pt,
    "\scalebox{.7}{$\bot$}"
  ]
  \ar[
    rr,
    phantom,
    shift right=8pt,
    "\scalebox{.7}{$\bot$}"
  ]
  &&
  \quad
  \ClassicalTypes_\Gamma
    \mathrlap{
      \hspace{20pt}
      \raisebox{1pt}{
      \scalebox{.7}{
        \color{darkblue}
        \bf
        potential data
      }
      }
    }
    \ar[out=120, in=20,
      looseness=4,
      shorten <=-2pt,
      shorten >=-3pt,
      "\raisebox{-2pt}{${
          \hspace{7pt}
          \mathclap{
            \randomly_{\!\scalebox{.75}{$W$}}
          }
          \hspace{2pt}
        }$}"{pos=.7, description},
      "{
        \scalebox{.7}{
          \hspace{23pt}
          \color{darkorange}
          \bf
          randomly
        }
      }"{pos=.4, yshift=2pt}
    ]
    \ar[out=5, in=-5,
      looseness=8.5,
      phantom,
      "{\scalebox{.7}{$\bot$}}"{yshift=1pt}
    ]
    \ar[out=-120, in=-20,
      looseness=4,
      shorten <=-2pt,
      shorten >=-3pt,
      "\scalebox{1.1}{${
          \hspace{4pt}
          \mathclap{
            \indefinitely_{\!{}_{\scalebox{.7}{$W$}}}
          }
          \hspace{1pt}
        }$}"{pos=.7, description},
      "{
        \scalebox{.65}{
          \hspace{40pt}
          \color{darkorange}
          \bf
          indefinitely
        }
      }"{swap, pos=.398, yshift=-3pt}
    ]
\end{tikzcd}
\end{equation}

\vspace{-2mm}
\noindent whose (co)restriction along propositional truncation \eqref{PropositionalTruncation} we shall denote by the same symbols:
\begin{equation}
\label{CoMonadsOfClassicalQuantification}
\begin{tikzcd}[column sep=80pt]
  W
  \ar[
    rr,
    "{ p_{{}_W} }"
  ]
  &&
  \Gamma
   \\[25pt]
    \mathllap{
      \raisebox{1pt}{
      \scalebox{.7}{
        \color{darkblue}
        \bf
        actual propositions
      }
      }
      \hspace{25pt}
    }
    {\color{purple}\Propositions}_W
    \ar[out=160, in=60,
      looseness=4,
      shorten=-3pt,
      "\scalebox{1.6}{${
          \hspace{1pt}
          \mathclap{
            \possibly_{\!{}_{\scalebox{.45}{$W$}}}
          }
          \hspace{1pt}
        }$}"{pos=.25, description},
      "{
        \scalebox{.7}{
          \hspace{-20pt}
          \color{darkorange}
          \bf
          possibly
        }
      }"{pos=.55, yshift=2pt}
    ]
    \ar[out=185, in=175,
      looseness=10,
      phantom,
      "{\scalebox{.7}{$\bot$}}"{yshift=1pt, xshift=3pt}
    ]
    \ar[out=-160, in=-60,
      looseness=4,
      shorten=-3pt,
      "\scalebox{1.4}{${
          \hspace{1pt}
          \mathclap{
            \necessarily_{\mathrlap{\!{}_{\scalebox{.6}{\colorbox{white}{$\!\!\!W\!$}}}}}
          }
          \hspace{3pt}
        }$}"{pos=.25, description},
      "{
        \scalebox{.7}{
          \hspace{-40pt}
          \color{darkorange}
          \bf
          necessarily
        }
      }"{swap, pos=.59, yshift=-2.5pt}
    ]
    \quad
  \ar[
    rr,
    shift left=14pt,
    "{
      \overset{
        \mathclap{
          \raisebox{4pt}{
            \scalebox{.7}{
              \color{darkgreen}
              \bf
              \def\arraystretch{.85}
              \begin{tabular}{c}
                0-truncated
                \\
                dependent co-product
              \end{tabular}
            }
          }
        }
      }{
        \big[\coprod_W(\mbox{-})\big]_0
      }
    }"
  ]
  \ar[
    from=rr,
    "{
      (\mbox{-}) \times W
    }"{description}
  ]
  \ar[
    rr,
    shift right=14pt,
    "{
      \underset{
        \mathclap{
          \raisebox{-3pt}{
            \scalebox{.7}{
              \color{darkgreen}
              \bf
              dependent product
            }
          }
        }
      }{
        \prod_W
      }
    }"{swap}
  ]
  \ar[
    rr,
    phantom,
    shift left=8pt,
    "\scalebox{.7}{$\bot$}"
  ]
  \ar[
    rr,
    phantom,
    shift right=8pt,
    "\scalebox{.7}{$\bot$}"
  ]
  &&
  \quad
  {\color{purple}\Propositions}_\Gamma
    \mathrlap{
      \hspace{20pt}
      \raisebox{1pt}{
      \scalebox{.7}{
        \color{darkblue}
        \bf
        potential propositions
      }
      }
    }
    \ar[out=120, in=20,
      looseness=4,
      shorten <=-2pt,
      shorten >=-3pt,
      "\raisebox{-2pt}{${
          \hspace{7pt}
          \mathclap{
            \randomly_{\!\scalebox{.75}{$W$}}
          }
          \hspace{2pt}
        }$}"{pos=.7, description},
      "{
        \scalebox{.7}{
          \hspace{23pt}
          \color{darkorange}
          \bf
          randomly
        }
      }"{pos=.4, yshift=2pt}
    ]
    \ar[out=5, in=-5,
      looseness=8.5,
      phantom,
      "{\scalebox{.7}{$\bot$}}"{yshift=1pt}
    ]
    \ar[out=-120, in=-20,
      looseness=4,
      shorten <=-2pt,
      shorten >=-3pt,
      "\scalebox{1.1}{${
          \hspace{4pt}
          \mathclap{
            \indefinitely_{\!{}_{\scalebox{.7}{$W$}}}
          }
          \hspace{1pt}
        }$}"{pos=.7, description},
      "{
        \scalebox{.65}{
          \hspace{40pt}
          \color{darkorange}
          \bf
          indefinitely
        }
      }"{swap, pos=.398, yshift=-3pt}
    ]
\end{tikzcd}
\end{equation}

\medskip

\noindent
{\bf Actuality logic.}
The terminology on the left of  diagram\eqref{CoMonadsOfClassicalDependentTypeFormation} is justified by the following Remark \ref{EpistemicInterpretationOfDependentTypes} and the observation of  Theorem \ref{S5KripkeSemanticsViaDependentTypes} below, which we articulate as a {\it theorem} not because its proof would be much more than an unwinding of definitions (nor surprising, in view of \cite{Lawvere69}), but to highlight its Yoneda-Lemma-like conceptual importance:

\begin{remark}[{\bf Epistemic interpretation of dependent types}]
\label{EpistemicInterpretationOfDependentTypes}
Concretely, we may read these modal operators
\eqref{CoMonadsOfClassicalDependentTypeFormation}
as follows, where we use the traditional language of ``possible worlds'' (Lit. \ref{ModalLogicAndManyWorlds}) but suggest to think of these ``worlds'' quite concretely as classical states of an observed universe to the extent partially revealed by a particular measurement, hence like the ``many worlds'' of quantum epistemology (Lit. \ref{EpistemologyOfQuantumPhysics}).

\noindent {\bf (i)} Given a proposition $P_\bullet$ which depends on which world $w$ is or has been measured:

\vspace{-.23cm}
\begin{center}
\small
\def\arraystretch{2}
\def\tabcolsep{5pt}
\begin{tabular}{|c|c|c|}
\hline
$\necessarily_{{}_W} P_\bullet$ means:
&
$P_w$ means:
&
$\possibly_{{}_W} P_\bullet$ as:
\\
\begin{minipage}{4.1cm}

  ``$P_w$ {\it does} or {\it is known to}

  \;\;hold {\it necessarily}''

  namely, no matter which

  world $w$ is measured.
\end{minipage}
&
\begin{minipage}{4.1cm}
  ``$P_w$ {\it does} or {\it is known} to

  \;\;hold {\it actually}''

  namely for the {\it given}

  world $w$ measured.
\end{minipage}
&
\begin{minipage}{4.1cm}
  ``$P_w$ {\it does} or {\it is known} to

  \;\;hold {\it possibly}''

  namely for {\it some} possibly

  measured world $w$.
\end{minipage}
\\[-10pt]
&&
\\
\hline
\end{tabular}
\end{center}
\noindent {\bf (ii)} Moreover, the (co)unit $\unit{\possibly}{}$ ($\counit{\necessarily}{}$) of the above (co)monads
reflect the logical entailment of these modal propositions:
\vspace{-2mm}
\begin{equation}
  \label{EpistemicEntailments}
  \hspace{-8mm}
  \begin{tikzcd}[
    column sep=60pt,
    row sep=6pt
  ]
    &[-32pt]
    \overset{
      \mathclap{
      \raisebox{3pt}{
        \scalebox{.8}{
          \color{darkblue}
          \bf
          necessarily $D_\bullet$
        }
      }
      }
    }{
      \necessarily_{{}_W} \, D_\bullet
    }
    \ar[
      rr,
      "{
        \scalebox{.8}{
          \color{darkorange}
          \bf
          entails
        }
      }"{yshift=8pt, pos=.48},
      "{
        \counit
          {\necessarily_{{}_W}}
          {D_\bullet}
      }"{description}
    ]
    &&
    \overset{
      \mathclap{
      \raisebox{4pt}{
        \scalebox{.8}{
          \color{darkblue}
          \bf
          actually $D_\bullet$
        }
      }
      }
    }{
      D_\bullet
    }
    \ar[
      rr,
      "{
        \scalebox{.8}{
          \color{darkorange}
          \bf
          entails
        }
      }"{yshift=8pt, pos=.55},
      "{
        \unit
          {\possibly_{{}_W}}
          {D_\bullet}
      }"{description}
    ]
    &&
    \overset{
      \mathclap{
      \raisebox{3pt}{
        \scalebox{.8}{
          \color{darkblue}
          \bf
          possibly $D_\bullet$
        }
      }
      }
    }{
      \possibly_{{}_W} \, D_\bullet
    }
    \\
    {\color{purple}w} \isa W \;\;\yields
    &
    \underset{w' \isa W}{\prod} D_{{}_{w'}}
    \ar[
      rr,
      "{
        (d_{w' \isa W})
        \;\,\mapsto\;\,
        d_{\color{purple}w}
      }"
    ]
    &&
    D_w
    \ar[
      rr,
      "{
        d_{\color{purple}w}
        \;\,\mapsto\;\,
        (
          {\color{purple}w}
         ,\,
         d_{\color{purple}w}
        )
      }"
    ]
    &&
    \underset{w' \isa W}{\coprod} D_{w'}
  \end{tikzcd}
\end{equation}
\end{remark}

\begin{remark}[{\bf Hexagon of epistemic entailments}]
  The {\it naturality} of the transformations \eqref{EpistemicEntailments} is reflected in commuting squares as shown in the following diagram
  \eqref{HexagonalCompositionOfEpistemicEntailments}, whose hexagonal composition gives the diagram \eqref{BranchingAndCollapseInIntroduction} announced in the Introduction (there evaluated for linear/quantum types, which we come to in \cref{QuantumEpistemicLogicViaDependentLinearTypes}, but the existence of the commuting hexagon as such depends only on the naturality of the epistemic entailments):
\begin{equation}
  \label{HexagonalCompositionOfEpistemicEntailments}
  \def\arraystretch{1.3}

\end{equation}
\end{remark}

For emphasis, the following theorem highlights that this epistemic logic of dependent types recovers what is traditionally understood in modal logic:
\begin{theorem}[\bf S5 Kripke semantics as co-monadic descent]
\label{S5KripkeSemanticsViaDependentTypes}
  The possible-worlds Kripke semantics
  \eqref{KripkeSemanticsOfModalOperator}
  for S5 modal logic are precisely given by dependent type formation \eqref{CoMonadsOfClassicalDependentTypeFormation} (for $\ClassicalTypes \defneq \Sets$)
  where a Kripke frame $\big(W \isa \Set,\, R \isa W \times W \to \Prop\big)$ corresponds to that display map \eqref{BasicBaseChangeAdjointtriple}
  which is its quotient projection
  $
    p_W
    \;\isa\;
    W \twoheadrightarrow \Gamma \defneq W_{/R}
  $.
\end{theorem}
\begin{proof}
  A classical theorem (\cite{Kripke63}\cite[Thm. 3.1.5]{FaginHalpernMosesVardi95}, cf. \cite{Samet10}) identifies the Kripke semantics for S5 modal logic
  with precisely those Kripke frames $\big(W , R\big)$ where $R$ is an equivalence relation. The equivalence classes $\Gamma$ of $R$ hence form a partition of $W$ as
  $$
    W
      \;=\;
    \coprod_{\gamma \isa \Gamma}
    \mathrm{fib}_\gamma(p_{{}_W})
    \,,
  $$
  which gives the incarnation of $W$ as a $\Gamma$-dependent type. By \eqref{BaseChangeInComponents},
  the induced comonad \eqref{CoMonadsOfClassicalDependentTypeFormation}  acts as
  \begin{equation}
    \label{InducedNecessityComonad}
    P \isa \Proposition_W
    \hspace{.7cm}
      \yields
    \hspace{.7cm}
    \begin{tikzcd}[sep=0pt]
      \necessarily_{{}_W} P
      \;\isa
      &
   W
      \ar[rr]
      &&
      \Propositions
      \\
      &
    \scalebox{0.8}{$  w $}
      &\mapsto&
   \scalebox{0.8}{$   \underset{
        \scalebox{.7}{$
        w'
          \isa
        \mathrm{fib}_{p_{{}_W}\!(w)}
        (
          p_{{}_W}
        )
        $}
      }{\forall}
      \,
      P(w')
      $}
    \end{tikzcd}
  \end{equation}

  \vspace{-3mm}
\noindent   But with $p_W$ identified as the quotient coprojection of $R$, we have
  $$
    \mathrm{fib}_{p_{{}_W}\!(w)}
    (
      p_{{}_W}
    )
    \;\;=\;\;
    (w' \isa W)
    \times
    R(w,w')
  $$
  whence \eqref{InducedNecessityComonad} equals the traditional formula \eqref{KripkeSemanticsOfModalOperator} for the Kripke semantics of the modal operator.
\end{proof}

\begin{remark}[{\bf Dependent type theory as universal Epistemic modal type theory}]
  \label{ModalTypeTheoryAsPerspectiveOnDependentTypeTheory}
  $\,$

\noindent {\bf (i)}   Thm. \ref{S5KripkeSemanticsViaDependentTypes} suggests that one may regard dependent type theory equivalently
as a universal form of epistemic type theory (Lit. \ref{LiteratureModalTypeTheory}) in generalization of how modal logic may be
viewed as an equivalent perspective on (fragments) of first-order logic   (cf. \cite[pp. xiii]{BlackburnVanBenthemWolter07}).
In both cases, one switches perspective from type formation by base change adjoint triples
  \eqref{BasicBaseChangeAdjointtriple}\eqref{QuantifiersAdjointTriple} to the associated adjoint pairs of (co)monads
  \eqref{CoMonadsOfClassicalDependentTypeFormation}\eqref{CoMonadsOfClassicalQuantification}. (An analogous change in perspective happens in
  (algebraic) geometry when expressing {\it descent theory} in terms of {\it monadic descent}.)

 \noindent {\bf (ii)}   Noticing that the development of general modal type theory is still in its infancy with its general
 {\it linear} form hardly known at all,
  this change of perspective allows us to use
  (in \cref{QuantumEpistemicLogicViaDependentLinearTypes})
  well-developed (linear) dependent type theory to realize the epistemic form of modal type theory that we need for certifying quantum protocols.
\end{remark}

\medskip

\noindent
{\bf Potentiality logic.}
The (co)monads on the right side of \eqref{CoMonadsOfClassicalDependentTypeFormation} are known
in effectful classical computer science (Lit. \ref{LiteratureComputationalEffectsAndModalities})
as the $W$-{\it (co)reader (co)monad},
\eqref{RederWriterCoMonads}
often denoted as on the right here:
\begin{equation}
  \label{ReaderMonad}
  \def\arraystretch{1.4}
  \begin{array}{ll}
  \indefinitely_{{}_W}
  D
  \;\defneq\;
  [W,\,D]
  &
  \scalebox{.7}{
    \color{darkblue}
    \bf
    $W$-reader monad
  }
  \\
  \randomly_{{}_W}
  D
  \;\defneq\;
  W \!\times\! D
  &
  \scalebox{.7}{
    \color{darkblue}
    \bf
    $W$-coreader comonad
  }
  \end{array}
\end{equation}
What has not previously found attention is the corresponding modal/epistemic perspective on these operators. It will be useful to dwell
on this point a little.
Our suggestion
in \eqref{CoMonadsOfClassicalDependentTypeFormation}
of {\it potentiality} as the antonym to {\it actuality} (the latter well-established in modal logic) follows Aristotle and Heisenberg (as recounted in \cite{Jaeger17}). In further support of this nomenclature, we offer the following fact, which gives a precise sense that:
\vspace{-1mm}
\begin{equation}
  \label{PotentialDataAsPossibilityModalDataExposition}
  \begin{tikzcd}[
    column sep=60pt,
    row sep=0pt
  ]
    \ar[
      rr,
      phantom,
      "    \mathclap{
      \mbox{
      {\it Potential data is equivalently
data whose possibility entails its actuality, consistently}
      }
    }
"
    ]
    &
    {}
    {}
    &
    {}
    \\
    \ClassicalTypes_\Gamma
    \ar[rr, <->, "{ \sim }"]
    &&
    \ClassicalTypes_{{}_W}^{\scalebox{.7}{$\possibly_{{}_W}$}}
    \\
    \underset{
      \mathclap{
      \raisebox{-4pt}{
      \scalebox{.7}{
        \color{darkblue}
        \bf
        potential data
      }
      }
      }
    }{
      D
      \isa
      \Type_\Gamma
    }
    \ar[
      rr,
      <->,
      shorten=40pt,
      "{ \sim }",
      "{
        \scalebox{.7}{
          \color{darkorange}
          \bf
          is equivalently
        }
      }"{swap, yshift=-8pt}
    ]
    &&
    \big(
      \underset{
        \mathclap{
          \raisebox{-3pt}{
            \scalebox{.7}{
              \color{darkblue}
              \bf
              data whose
            }
          }
        }
      }{
      D_\bullet
      \isa
      \Type_{{}_W}
      }
      ,\,
      \rho
      \,\isa\,
      \underset{
        \mathclap{
          \raisebox{-3pt}{
            \scalebox{.7}{
              \color{darkblue}
              \bf
              possibility entails its
              actuality,
            }
          }
        }
      }{
      \possibly_{{}_W}
      D_\bullet
      \xrightarrow{\phantom{----}}
      D_\bullet
      }
      ,\,
      \underset{
        \mathclap{
          \raisebox{-2pt}{
            \scalebox{.7}{
              \color{darkblue}
              \bf
              consistently
            }
          }
        }
      }{
      \mathrm{utl}_{\scalebox{.7}{$\possibly_{{}_W}$}}\!(\rho)
      ,\,
      \mathrm{act}_{\scalebox{.7}{$\possibly_{{}_W}$}}\!(\rho)
      }
    \big)
  \end{tikzcd}
\end{equation}

\vspace{-2mm}
\noindent (This compares favorably with the traditional informal intention of the ``potentiality'' modality, cf. \cite[\S 44]{FerroniGili16}.)
Namely, we have:
\begin{proposition}[{\bf Potential data as possibility modal data}]
  \label{PotentialDataAsPossibilityModalData}
  For $p_W \isa W \twoheadrightarrow \Gamma$ an epimorphism (as in Thm. \ref{S5KripkeSemanticsViaDependentTypes}), the context extension
  $(\mbox{-}) \times W \isa \ClassicalTypes_\Gamma \to \ClassicalTypes_{{}_W}$ is monadic \eqref{ComparisonFunctor} whence the {\it potential types} \eqref{CoMonadsOfClassicalDependentTypeFormation} are identified with the (free) {\it possibility-modal types} \eqref{CategoriesOfModales} and hence \eqref{AdjointCoMonadsHaveIsomorphicModales} also with the {\it necessity-modal types}:
  \vspace{-3mm}
\begin{equation}
\label{PotentialDataIdentifiedWithPossibilityModalDat}
\hspace{-2cm}
  \begin{tikzcd}[column sep=60pt]
    &
    \ClassicalTypes
      ^{ \scalebox{.7}{$\possibly_{{}_W}$} }
      _{{}_W}
    \mathrlap{
      \;
      \scalebox{.7}{
        \color{darkblue}
        \bf
        possibility modal data
      }
    }
    \ar[
      d,
      phantom,
      "{ \simeq }"{rotate=-90}
    ]
    \\[-14pt]
    \mathllap{
      \raisebox{2pt}{
      \scalebox{.7}{
        \color{darkblue}
        \bf
        actual data
      }
      }
      \hspace{17pt}
    }
    \ClassicalTypes_{{}_W}
    \ar[out=160, in=60,
      looseness=4,
      shorten=-3pt,
      "\scalebox{1.6}{${
          \hspace{1pt}
          \mathclap{
            \possibly_{\!{}_{\scalebox{.45}{$W$}}}
          }
          \hspace{1pt}
        }$}"{pos=.25, description},
      "{
        \scalebox{.7}{
          \color{darkorange}
          \bf
          possibly
        }
      }"{pos=.54, xshift=-5pt, yshift=2pt}
    ]
    \ar[out=185, in=175,
      looseness=9,
      phantom,
      "{\scalebox{.7}{$\bot$}}"{yshift=1pt}
    ]
    \ar[out=-160, in=-60,
      looseness=4,
      shorten=-3pt,
      "\scalebox{1.4}{${
          \hspace{1pt}
          \mathclap{
            \necessarily_{\mathrlap{\!{}_{\scalebox{.6}{\colorbox{white}{$\!\!\!W\!$}}}}}
          }
          \hspace{3pt}
        }$}"{pos=.25, description},
      "{
        \scalebox{.7}{
          \color{darkorange}
          \bf
          necessarily
        }
      }"{swap, pos=.575, yshift=-2.5pt}
    ]
    \ar[
      r,
      shift left=22pt,
      "{
        \coprod_W
      }"
    ]
    \ar[
      r,
      shift right=24pt,
      "{
        \prod_W
      }"{swap}
    ]
    \ar[
      from=r,
      "{
        \times W
      }"{description}
    ]
    \ar[
      r,
      shift left=12pt,
      phantom,
      "{ \scalebox{.7}{$\bot$} }"{}
    ]
    \ar[
      r,
      shift right=12pt,
      phantom,
      "{ \scalebox{.7}{$\bot$} }"
    ]
    &
    \ClassicalTypes_\Gamma
    \mathrlap{
      \;\;
      \scalebox{.7}{
        \color{darkblue}
        \bf
        potential data
      }
    }
    \\[-14pt]
    &
    \ClassicalTypes
      ^{ \scalebox{.7}{$\necessarily_{{}_W}$}}
      _{{}_W}
    \mathrlap{
      \;
      \scalebox{.7}{
        \color{darkblue}
        \bf
        necessity modal data
      }
    }
    \ar[
      u,
      phantom,
      "{ \simeq }"{rotate=-90}
    ]
  \end{tikzcd}
\end{equation}
\end{proposition}

\vspace{-5mm}
\begin{proof}
  By the Monadicity Theorem \eqref{ComparisonFunctor} and since the functor $(\mbox{-}) \times W$ has both a left and a right adjoint,
  it is sufficient to see that it reflects isomorphisms; but this follows immediately from the assumption that $p_W$ is surjective. Compare to \cite[Lem. 1.3.2]{Johnstone02}, namely if $(f \times W)_w \,\defneq\,  f_{p_{{}_W}(w)}$ is an isomorphism for $w \isa W$ then
  surjectivty of $p_{{}_W}$ implies that $f_{\gamma}$ is an isomorphism for $\gamma \isa \Gamma$.
\end{proof}
\begin{remark}[{\bf Relation to monadic descent}]
 The statement and proof of Prop. \ref{PotentialDataAsPossibilityModalData} correspond to what in (algebraic) geometry is known as
  {\it monadic descent} (e.g. \cite[\S 2.1]{JanelidzeTholen94}):
  In this context, the display map $p_{{}_W}$ would be called an
  {\it effective descent morphism}, and $\possibly_{{}_W}$-modale structure would be called {\it descent data} along $p_W$.
\end{remark}
\begin{remark}[{\bf Relation to lenses}]
  \label{RelationToLenses}
 In the case $\Types = \Sets$, the statement of Prop. \ref{PotentialDataAsPossibilityModalData} is known in the theory of {\it lenses} in computer science. Here one regards $\possibly_{{}_W}$-modale structure
 as a data base-type $S$ equipped with functionality to read out ($\mathrm{get}$) and to over-write ($\mathrm{put}$)  $W$-data subject to consistency conditions (``lawful lenses''):
 \begin{equation}
 \hspace{-4mm}
   \left(\!
   \def\arraystretch{1}
   \def\arraycolsep{0pt}

 \!\!  \right)
   \isa\,
   \big(
     \Types_{{}_W}
   \big)^{\lozenge_{{}_W}}
 \end{equation}
 and the upshot of the monadicity statement
 (Prop. \ref{PotentialDataAsPossibilityModalData}, \cite[Thm. 12]{JohnsonRosebrughWood10}\footnote{
   \cite{Spivak19} concludes from this situation that the theory of ``lenses'' is best regarded as an aspect of the much broader and classical theory of indexed categories (Grothendieck fibrations).
   Syntactically this means to  regard them as an aspect of the theory of dependent types which -- when also taking into account the related system of (co)monads -- is the thesis that we are developing here.
 })
 is that this describes ``addressed'' access to a data sub-base type, in that such $S$ are necessarily of product form $S \,\simeq\, W \times D$ with $\mathrm{get} = \mathrm{pr}_{{}_W}$, etc.
\end{remark}

\noindent {\bf Random and (in)definite data.}
The (co)monads
$\indefinitely$ ($\randomly$)
on the right of \eqref{CoMonadsOfClassicalDependentTypeFormation} are well-known in terms of (co)effects in computer science (Lit. \ref{LiteratureComputationalEffectsAndModalities}) as the ``(co)reader (co)monad'' \eqref{RederWriterCoMonads}, referring to the idea of a program {\it reading} ({\it providing}) a global variable $w \isa W$. However, for staying true to the spirit of modal logic, here we refer to these as the modalities of {\it indefiniteness} ({\it randomness}), in the following sense:

\medskip

\vspace{-.23cm}
\begin{center}
\adjustbox{}{
\small
\def\arraystretch{2}
\def\tabcolsep{6pt}
\begin{tabular}{|c|c|c|}
\hline
$\randomly_{\!{}_W} D$ is the type of
&
$D$ is the type of
&
$\indefinitely_{{}_W} P_\bullet$ is the type of:
\\
\begin{minipage}{4.1cm}

  $D$-data $d$ in a {\it definite}

  but {\it random} world $w$

  (as in ``random access'')

\end{minipage}
&
\begin{minipage}{4.1cm}
  plain $D$-data $d$

  only {\it potentially} in

  some possible world
\end{minipage}
&
\begin{minipage}{4.1cm}

  {\it indefinite} $D$-data $w \mapsto d_w$

  contingent on a pending

  choice of possible world $w$.

\end{minipage}
\\[-10pt]
&&
\\
\hline
\end{tabular}
}
\end{center}

\begin{equation}
  \label{PotentialityEpistemicEntailments}
  \hspace{-2cm}
  \begin{tikzcd}[
    column sep=60pt,
    row sep=6pt
  ]
    &
    \overset{
      \mathclap{
        \raisebox{3pt}{
          \scalebox{.8}{
            \color{darkblue}
            \bf
            randomly $P$
          }
        }
      }
    }{
      \randomly_{\!W} P
    }
    \ar[
      rr,
      "{
        \unit
          { \randomly_{\!W} }
          { P }
      }"{description, pos=.45},
      "{
        \scalebox{.8}{
          \color{darkorange}
          \bf
          entails
        }
      }"{yshift=9pt, pos=.45}
    ]
    &&
    \overset{
      \mathclap{
      \raisebox{5pt}{
        \scalebox{.8}{
          \color{darkblue}
          \bf
          potentially $P$
        }
      }
      }
    }{
      P
    }
    \ar[
      rr,
      "{
        \counit
          { \indefinitely_W }
          { P }
      }"{description, pos=.55},
      "{
        \scalebox{.8}{
          \color{darkorange}
          \bf
          entails
        }
      }"{yshift=9pt, pos=.55}
    ]
    &&
    \overset{
      \mathclap{
        \raisebox{4pt}{
          \scalebox{.8}{
            \color{darkblue}
            \bf
            indefinitely $P$
          }
        }
      }
    }{
      \indefinitely_{{}_W} P
    }
    \\
    &
    \underset{w' \isa W}{\coprod}
    \, P
    \ar[
      rr,
      "{
        \scalebox{1.2}{$($}w,\,p\scalebox{1.2}{$)$}
        \;\mapsto\;
        p
      }"
    ]
    &&
    P
    \ar[
      rr,
      "{
        p
        \;\mapsto\;
        \scalebox{1.2}{$($}w' \mapsto p\scalebox{1.2}{$)$}
      }"
    ]
    &&
    \underset{w' \isa W}{\prod}
    \,
    P
  \end{tikzcd}
\end{equation}

 In particular, the monadic effect model (cf. Lit. \ref{LiteratureComputationalEffectsAndModalities})
 for operating on the parameter space $W$ as on a {\it random access memory} (RAM) is the state monad
 \eqref{StateMonadEndofunctor},
 which we may realize as the composite
 \vspace{-2mm}
 \begin{equation}
   \label{RandomAccessAsIndefiniteRandomness}
   \underset{W}{\indefinitely}
   \,
   \underset{W}{\randomly}
   \,
   D
   \;\;\simeq\;\;
   \underset{W}{\prod}
   \;
   \underset{W}{\coprod}
   \,
   D
   \;\;\simeq\;\;
   \big[
     W
     ,\,
     W \times D
   \big]
   \;\;\defneq\;\;
   W\mathrm{State}(D)
   ,
   \qquad
  \begin{tikzcd}
    \Types
    \ar[
      rr,
      shift left=6pt,
      "{
        \randomly_{{}_{W}}
      }"
    ]
    \ar[
      from=rr,
      shift left=6pt,
      "{
        \indefinitely_{{}_W}
      }"
    ]
    \ar[
      rr,
      phantom,
      "{
        \scalebox{.7}{$\bot$}
      }"
    ]
    \ar[out=140, in=-140,
      looseness=4.5,
      shorten=-3pt,
      "{
        W\mathrm{State}
      }"{description},
    ]
    &&
    \Types
  \end{tikzcd}
   \,.
\end{equation}

\vspace{-2mm}
\noindent It is in this common sense of {\it random access} as about ``choice'' (instead of ``chance'')
that one should think about $\randomly_{W}$ as the modality of ``being random''.

\medskip

\noindent
{\bf In summary} so far, we have found that any classical (intuitionistic) dependently typed language may be regarded as a rich epistemic
modal type theory with, for every inhabited type $W$ (in any ambient context $\Gamma$), the following identifications:

\begin{equation}
\label{SummaryClassicalEpistemicTypeTheory}
\adjustbox{}{
\hspace{-.4cm}
\def\arraystretch{3}
\tabcolsep=1pt

\\
\hline
  $
  \hspace{0cm}
  \begin{tikzcd}[
    column sep=60pt,
    row sep=6pt
  ]
    &[-32pt]
    \overset{
      \mathclap{
      \raisebox{3pt}{
        \scalebox{.8}{
          \color{darkblue}
          \bf
          necessarily $P_\bullet$
        }
      }
      }
    }{
      \necessarily_{{}_W} \, P_\bullet
    }
    \ar[
      rr,
      "{
        \scalebox{.8}{
          \color{darkorange}
          \bf
          entails
        }
      }"{yshift=8pt, pos=.48},
      "{
        \epsilon^{\necessarily_{{}_W}}_{P_\bullet}
      }"{description}
    ]
    &&
    \overset{
      \mathclap{
      \raisebox{4pt}{
        \scalebox{.8}{
          \color{darkblue}
          \bf
          actually $P_\bullet$
        }
      }
      }
    }{
      P_\bullet
    }
    \ar[
      rr,
      "{
        \scalebox{.8}{
          \color{darkorange}
          \bf
          entails
        }
      }"{yshift=8pt, pos=.55},
      "{
        \eta^{\possibly_{{}_W}}_{P_\bullet}
      }"{description}
    ]
    &&
    \overset{
      \mathclap{
      \raisebox{3pt}{
        \scalebox{.8}{
          \color{darkblue}
          \bf
          possibly $P_\bullet$
        }
      }
      }
    }{
      \possibly_{{}_W} \, P_\bullet
    }
    \\
    {\color{purple}w} \isa W \;\;\vdash
    &
    \underset{w' \isa W}{\prod} P_{w'}
    \ar[
      rr,
      "{
        \scalebox{1.1}{$($}w' \mapsto p_{w'}\scalebox{1.1}{$)$}
        \;\,\mapsto\;\,
        p_{\color{purple}w}
      }"
    ]
    &&
    P_{\color{purple}w}
    \ar[
      rr,
      "{
        p_{\color{purple}w}
        \;\,\mapsto\;\,
        \scalebox{1.2}{$($}{\color{purple}w},\,p_{\color{purple}w}\scalebox{1.2}{$)$}
      }"
    ]
    &&
    \underset{w' \isa W}{\coprod} P_{w'}
    \\[-9pt]
    {}
    \ar[
      rrrrr,
      -,
      shorten=-30pt,
      shift right=1pt,
      dashed,
      gray
    ]
    &&&&&
    {}
    \\[-10pt]
    &
    \overset{
      \mathclap{
        \raisebox{3pt}{
          \scalebox{.8}{
            \color{darkblue}
            \bf
            randomly $P$
          }
        }
      }
    }{
      \randomly_{\!W} P
    }
    \ar[
      rr,
      "{
        \unit
          { \randomly_{\!W} }
          { P }
      }"{description, pos=.45},
      "{
        \scalebox{.8}{
          \color{darkorange}
          \bf
          entails
        }
      }"{yshift=9pt, pos=.45}
    ]
    &&
    \overset{
      \mathclap{
      \raisebox{5pt}{
        \scalebox{.8}{
          \color{darkblue}
          \bf
          potentially $P$
        }
      }
      }
    }{
      P
    }
    \ar[
      rr,
      "{
        \counit
          { \indefinitely_W }
          { P }
      }"{description, pos=.55},
      "{
        \scalebox{.8}{
          \color{darkorange}
          \bf
          entails
        }
      }"{yshift=9pt, pos=.55}
    ]
    &&
    \overset{
      \mathclap{
        \raisebox{4pt}{
          \scalebox{.8}{
            \color{darkblue}
            \bf
            indefinitely $P$
          }
        }
      }
    }{
      \indefinitely_{{}_W} P
    }
    \\
    &
    \underset{w' \isa W}{\coprod}
    \, P
    \ar[
      rr,
      "{
        \scalebox{1.2}{$($}w,\,p\scalebox{1.2}{$)$}
        \;\mapsto\;
        p
      }"
    ]
    &&
    P
    \ar[
      rr,
      "{
        p
        \;\mapsto\;
        \scalebox{1.2}{$($}w' \mapsto p\scalebox{1.2}{$)$}
      }"
    ]
    &&
    \underset{w' \isa W}{\prod}
    \,
    P
  \end{tikzcd}
$
\hspace{.3cm}
\\
\hline
\end{tabular}
}
\end{equation}

Next we proceed to find the quantum analog \eqref{SummaryQuantumEpistemicTypeTheory} of this logic.

\medskip

\subsection{Quantum Epistemic Logic}
\label{QuantumEpistemicLogicViaDependentLinearTypes}

On the backdrop (\cref{ClassicalEpistemicLogicViaDependentTypes}) of classical (intuitionistic) epistemic type theory understood as an equivalent re-interpretation of classical
(intuitionistic) dependent type theory,  and in view (\ifdefined\monadology \cref{QuantumTypeSemantics}\else\cref{QuantumViaLHoTT}\fi) of the existence of dependent {\it linear} type theory {\LHoTT}, we are led
to expect that {\it quantum epistemic type theory} ought to analogously be obtained by re-regarding the base change adjunction
\eqref{AmbidextrousBaseChange}
of dependent {\it linear} type formation
$$
\hspace{-2mm}
  \begin{tikzcd}[
    row sep=34pt
  ]
      \mathclap{
      \raisebox{+5pt}{
        \scalebox{.7}{
          \color{darkblue}
          \bf
          \def\arraystretch{.8}

    }
    &
    \big(
      \ClassicalTypes_{{}_W},
      \,
      \overset{
        \mathclap{
          \hspace{-10pt}
          \rotatebox{40}{
            \rlap{
              \hspace{-10pt}
              \scalebox{.7}{
                \color{darkorange}
                product
              }
            }
          }
        }
      }{
        \times_{{}_W}
      }
    \big)
    \ar[
      rr,
      shift left=16pt,
      "{
        \coprod_{{}_W}
      }"{description},
      "{
        \scalebox{.7}{
          \color{darkgreen}
          \def\arraystretch{.9}
          \begin{tabular}{c}
            dependent
            \\
            co-product
          \end{tabular}
        }
      }"{yshift=4pt}
    ]
    \ar[
      from=rr,
      "{
        \times W
      }"{description}
    ]
    \ar[
      rr,
      shift right=16pt,
      "{
        \prod_{{}_W}
      }"{description},
      "{
        \scalebox{.7}{
          \color{darkgreen}
          dependent product
        }
      }"{swap, yshift=-4pt}
    ]
    \ar[
      rr,
      phantom,
      shift left=8pt,
      "\scalebox{.6}{$\bot$}"
    ]
    \ar[
      rr,
      phantom,
      shift right=8pt,
      "\scalebox{.6}{$\bot$}"
    ]
    &&
    \big(
      \ClassicalTypes_{{}_\Gamma},
      \,
      \overset{
        \mathclap{
          \rotatebox{40}{
            \rlap{
              \hspace{-10pt}
              \scalebox{.7}{
                \color{darkorange}
                product
              }
            }
          }
        }
      }{
        \times_{{}_\Gamma}
      }
    \big)
    &
    \scalebox{.7}{
      \def\arraystretch{.9}
      \color{darkblue}
      \bf
      %
    }
    &
    \big(
      \LinTypes_{{}_W},
      \,
      \overset{
        \mathclap{
          \rotatebox{40}{
            \rlap{
              \hspace{-10pt}
              \scalebox{.7}{
                \color{darkorange}
                tensor
              }
            }
          }
        }
      }{
        \otimes_{{}_W}
      }
    \big)
    \;\;
    \ar[
      rr,
      shift left=16pt,
      "{
          {\scalebox{1.3}{$\oplus$}}_{{}_W}
      }"{description},
      "{
        \scalebox{.7}{
          \color{darkgreen}
          \def\arraystretch{.9}
          %
        }
      }"{yshift=5pt}
    ]
    \ar[
      from=rr,
      "{ \otimes \mathbbm{1}_{\!W} }"{description}
    ]
    \ar[
      rr,
      shift right=16pt,
      "{
        {\scalebox{1.3}{$\oplus$}}_{{{}_W}}
      }"{description}
    ]
    \ar[
      rr,
      phantom,
      shift left=8pt,
      "\scalebox{.6}{$\bot$}"
    ]
    \ar[
      rr,
      phantom,
      shift right=8pt,
      "\scalebox{.6}{$\bot$}"
    ]
    &&
    \;\;
    \big(
      \LinTypes_{{}_\Gamma},
      \,
      \overset{
        \mathclap{
          \rotatebox{40}{
            \rlap{
              \hspace{-10pt}
              \scalebox{.7}{
                \color{darkorange}
                tensor
              }
            }
          }
        }
      }{
        \otimes_{{}_\Gamma}
      }
    \big)
    &
    \scalebox{.7}{
      \def\arraystretch{.9}
      \color{darkblue}
      \bf
      %
    }
   \hspace{-15pt}
   }
  \end{tikzcd}
$$
by passing to the induced (co)monads \eqref{MonadFromAdjunctions}, which we denote by the same symbols as their classical counterparts \eqref{CoMonadsOfClassicalDependentTypeFormation}:
\begin{equation}
\label{CoMonadsOfLinearDependentTypeFormation}
\begin{tikzcd}[column sep=huge]
   W
   \ar[
     rr,
     "{ p_{{}_W} }"
   ]
   &&
   \Gamma
   \\[20pt]
    \mathllap{
      \raisebox{1pt}{
      \scalebox{.7}{
        \color{darkblue}
        \bf
        Actual quantum data
      }
      }
      \hspace{25pt}
    }
    \QuantumTypes_W
    \ar[out=160, in=60,
      looseness=4,
      shorten=-3pt,
      "\scalebox{1.6}{${
          \hspace{1pt}
          \mathclap{
            \possibly_{\!{}_{\scalebox{.45}{$W$}}}
          }
          \hspace{1pt}
        }$}"{pos=.25, description},
      "{
        \scalebox{.7}{
          \hspace{-20pt}
          \color{darkorange}
          \bf
          possibly
        }
      }"{pos=.55, yshift=2pt}
    ]
    \ar[out=185, in=175,
      looseness=5,
      phantom,
      "{\scalebox{.7}{$\bot$}}"{yshift=1pt, xshift=3pt}
    ]
    \ar[out=-160, in=-60,
      looseness=4,
      shorten=-3pt,
      "\scalebox{1.4}{${
          \hspace{1pt}
          \mathclap{
            \necessarily_{\mathrlap{\!{}_{\scalebox{.6}{\colorbox{white}{$\!\!\!W\!$}}}}}
          }
          \hspace{3pt}
        }$}"{pos=.25, description},
      "{
        \scalebox{.7}{
          \hspace{-40pt}
          \color{darkorange}
          \bf
          necessarily
        }
      }"{swap, pos=.59, yshift=-2.5pt}
    ]
    \quad
  \ar[
    rr,
    shift left=14pt,
    "{
      \overset{
        \mathclap{
          \raisebox{3pt}{
            \scalebox{.7}{
              \color{darkgreen}
              \bf
              dependent direct sum
            }
          }
        }
      }{
        \osum_W
      }
    }"
  ]
  \ar[
    from=rr,
    "{
      \otimes \mathbbm{1}_{W}
    }"{description}
  ]
  \ar[
    rr,
    shift right=14pt,
    "{
      \underset{
        \mathclap{
          \raisebox{-3pt}{
            \scalebox{.7}{
              \color{darkgreen}
              \bf
              dependent direct sum
            }
          }
        }
      }{
        \osum_W
      }
    }"{swap}
  ]
  \ar[
    rr,
    phantom,
    shift left=8pt,
    "\scalebox{.7}{$\bot$}"
  ]
  \ar[
    rr,
    phantom,
    shift right=8pt,
    "\scalebox{.7}{$\bot$}"
  ]
  &&
  \quad
  \QuantumTypes_\Gamma
    \mathrlap{
      \hspace{20pt}
      \raisebox{1pt}{
      \scalebox{.7}{
        \color{darkblue}
        \bf
        Potential quantum data
      }
      }
    }
    \ar[out=120, in=20,
      looseness=4,
      shorten <=-2pt,
      shorten >=-3pt,
      "\raisebox{-2pt}{${
          \hspace{7pt}
          \mathclap{
            \randomly_{{}_W}
          }
          \hspace{2pt}
        }$}"{pos=.8, description},
      "{
        \scalebox{.7}{
          \hspace{23pt}
          \color{darkorange}
          \bf
          \;\;\;\;\;\;randomly
        }
      }"{pos=.42, yshift=2pt}
    ]
    \ar[out=5, in=-5,
      looseness=4.5,
      phantom,
      "{\scalebox{.7}{$\bot$}}"{yshift=1pt}
    ]
    \ar[out=-120, in=-20,
      looseness=4,
      shorten <=-2pt,
      shorten >=-3pt,
      "\scalebox{1.1}{${
          \hspace{4pt}
          \mathclap{
            \indefinitely_{\!{}_{\scalebox{.7}{$W$}}}
          }
          \hspace{1pt}
        }$}"{pos=.8, description},
      "{
        \scalebox{.65}{
          \hspace{40pt}
          \color{darkorange}
          \bf
          \;\;\;\;\;\;\;indefinitely
        }
      }"{swap, pos=.398, yshift=-3pt}
    ]
\end{tikzcd}
\end{equation}

\noindent
A key point now is the {\it ambitexterity} \eqref{AmbidextrousBaseChange} of the base change for dependent linear
types along a finite classical type $W$:
\begin{equation}
  \label{AmbidexterityInQuantumEpistemology}
  W
  \,\isa\,
  \FiniteType
  \hspace{.8cm}
  \yields
  \hspace{.8cm}
  \Big(
  \;
  \underset{W}{\osum}
  \;\dashv\;
  \otimes \mathbbm{1}_{{}_W}
  \,\dashv\,
  \underset{W}{\osum}
  \;
  \Big)
\end{equation}

\medskip
\medskip

It is now as elementary to work out  the (co)units of these (co)monads (they are the universal maps of the direct sum construction)
as it is interesting -- in view of quantum epistemology (Lit. \ref{LiteratureQuantumComputation}):
\newpage

\begin{proposition}[{\bf Component expressions of the (co)monad (co)units}]
\label{ComponentExpressionsOfTheCoMonadCoUnits}
The (co)units and (co)joins of the (co)monads in \eqref{CoMonadsOfLinearDependentTypeFormation} are given,
in components, as follows:

\begin{equation}
\label{EntailmentsOfQuantumEpistemology}
\hspace{-.7cm}
\adjustbox{}{
\begin{minipage}{16cm}
\def\arraystretch{2}


\vspace{.03cm}

\hspace{-.09cm}
\adjustbox{
  fbox
}{
$
  \begin{tikzcd}[
    sep=0pt,
    column sep=16.3pt
  ]
    &[-20pt]
    \underset{W}{\indefinitely}
    \underset{W}{\indefinitely}
    \HilbertSpace{H}
    \ar[
      rr,
      "{
        \multiplication
          { \indefinitely_{{}_W} }
          { \HilbertSpace{H} }
      }",
      "{
        \scalebox{.7}{
          \color{darkgreen}
          \bf
          indefiniteness join
        }
      }"{swap}
    ]
    &&
    \underset{W}{\indefinitely}
    \HilbertSpace{H}
    &[-20pt]
    \ar[
      dddddd,
      -,
      shorten <=-24pt,
      shorten >=-20pt,
      shift left=11pt,
      dashed,
      line width=.pt,
      gray
    ]
    &
    &[-20pt]
    \underset{W}{\randomly}
    \HilbertSpace{H}
    \ar[
      rr,
      "{
        \comultiplication
          { \randomly_{{}_W} }
          { \HilbertSpace{H} }
      }",
      "{
        \scalebox{.7}{
          \color{darkgreen}
          \bf
          randomness cojoin
        }
      }"{swap}
    ]
    &&
    \underset{W}{\randomly}
    \underset{W}{\randomly}
    \HilbertSpace{H}
    &[-20pt]
    \\
    \underset{
      {\color{purple}w}
    }{\osum}
    \Big(
        \hspace{-2.4cm}
    &
    \underset{W}{\necessarily}
    \HilbertSpace{H}
    \ar[
      rr,
      "{
        \counit
          { \necessarily_{{}_W} }
          { \HilbertSpace{H} }
      }",
      "{
        \scalebox{.7}{
          \rm
          quantum
          \color{darkorange}
          \bf
          state collapse
        }
      }"{swap}
    ]
    &&
    \HilbertSpace{H}
    &
   \hspace{-1.6cm}  \Big)
    &
    \underset{
      {\color{purple}w}
    }{\osum}
    \Big(
    \hspace{-1.9cm}
    &
    \HilbertSpace{H}
    \ar[
      rr,
      "{
        \counit
          { \possibly_{{}_W} }
          { \HilbertSpace{H} }
      }",
      "{
        \scalebox{.7}{
          \rm
          quantum
          \color{darkorange}
          \bf
          state prepar.
        }
      }"{swap}
    ]
    &&
    \underset{W}{\possibly}
    \HilbertSpace{H}
    &
   \hspace{-1.2cm}  \Big)
    \\
    &
    \oplus_{w'}
    \vert \psi_{{\color{purple}w}, w'} \rangle
    &
    \mapsto
    &
    \vert \psi_{{\color{purple}w}, {\color{purple}w}}\rangle
    &&
    &
    \vert \psi_{\color{purple}w} \rangle
    &
    \mapsto
    &
    \oplus_{w'}
    \delta_{{\color{purple}w}}^{w'}
    \vert \psi_{{\color{purple}w}} \rangle
    \\[+10pt]
    \ar[
      rrrrrrrrr,
      -,
      shorten <=-35pt,
      shorten >=-17pt,
      gray,
      dashed,
      line width=1.8pt,
    ]
    &&&&&&&&&
    {}
    \\[+10pt]
    &[-15pt]
    \underset{W}{\possibly}
    \underset{W}{\possibly}
    \HilbertSpace{H}_\bullet
    \ar[
      rr,
      "{
        \multiplication
          { \possibly_{{}_W} }
          { \HilbertSpace{H}_\bullet }
      }",
      "{
        \scalebox{.7}{
          \color{darkgreen}
          \bf
          possibility join
        }
      }"{swap}
    ]
    &&
    \underset{W}{\possibly}
    \HilbertSpace{H}_\bullet
    &&
    &[-5pt]
    \underset{W}{\necessarily}
    \HilbertSpace{H}_\bullet
    \ar[
      rr,
      "{
        \comultiplication
          { \necessarily_{{}_W} }
          { \HilbertSpace{H}_\bullet }
      }",
      "{
        \scalebox{.7}{
          \color{darkgreen}
          \bf
          necessity cojoin
        }
      }"{swap}
    ]
    &&
    \underset{W}{\necessarily}
    \underset{W}{\necessarily}
    \HilbertSpace{H}_\bullet
    \\[+8pt]
    {\color{purple}w} \isa W
    \;
    \yields
    &
    \underset{W}{\randomly}
    \,
    \underset{
      W
    }{\osum}
    \HilbertSpace{H}_{\bullet}
    \ar[
      rr,
      "{
        \counit
          { \randomly_{{}_W} }
          { \oplus_{{}_W} \HilbertSpace{H}_{\bullet} }
      }",
      "{
        \scalebox{.7}{
          \rm quantum
          \color{darkorange}
          \bf
          superposition
        }
      }"{swap}
    ]
    &&
    \underset{
      W
    }{\scalebox{1.2}{$\oplus$}}
    \HilbertSpace{H}_{\bullet}
    &&
    {\color{purple}w} \isa A
    \;
    \yields
    &
    \underset{
      W
    }{
      \osum
    }
    \HilbertSpace{H}_{\bullet}
    \ar[
      rr,
      "{
        \unit
          { \indefinitely_{{}_W} }
          {
            \oplus_{{}_W}
            \HilbertSpace{H}_{\bullet}
          }
      }",
      "{
        \scalebox{.7}{
          \rm quantum
          \color{darkorange}
          \bf
          parallelism
        }
      }"{swap, yshift=-1pt}
    ]
    &&
    \underset{W}{\indefinitely}
    \,
    \underset{
      {W}
    }{
      \osum
    }
    \HilbertSpace{H}_{\bullet}
    \\
    &
    \underset{w''}{\oplus}
    \vert \psi_{{\color{purple}w}, w', w''} \rangle
    &\mapsto&
    \underset{w''}{\sum}
    \,
    \vert \psi_{{\color{purple}w}, w', w''} \rangle
    &{}&{}&{}
    \vert \psi_{{\color{purple} w}, w'}  \rangle
    &\mapsto&
    \underset{w''}{\oplus}
    \vert \psi_{{\color{purple}w}, w'} \rangle
  \end{tikzcd}
$
}
\end{minipage}
}
\end{equation}
\end{proposition}
Here the (co)joins in the lower half follow from the (co)units in the top half via \eqref{MonadStructureFromAdjunction}.


\medskip

\noindent
{\bf Monadicity of quantum data.} We observe that quantum data as in \eqref{CoMonadsOfLinearDependentTypeFormation} is characterized by two monadicity theorems:

-- Prop. \ref{QuantumPotentialDataAsPossibilityModalData}: Potential quantum data is possibility-modal actual data.

-- Prop. \ref{ActualQuantumDataIsInfinitenessModalPotentialData}: Actual quantum data is indefiniteness-modal potential data.

\medskip

\noindent
First, we have the following quantum analog of the classical situation from Prop. \ref{PotentialDataAsPossibilityModalData}:

\begin{proposition}[{\bf Potential quantum data as possibility-modal actual data}]
  \label{QuantumPotentialDataAsPossibilityModalData}
  For $p_W \isa W \twoheadrightarrow \Gamma$ an epimorphism (as in Thm. \ref{S5KripkeSemanticsViaDependentTypes})
  the context extension $(\mbox{-}) \otimes \mathbbm{1}_{{}_W} \isa \QuantumTypes_\Gamma \to \QuantumTypes_{{}_W}$ is monadic \eqref{ComparisonFunctor}
  whence the {\it potential quantum types} \eqref{CoMonadsOfLinearDependentTypeFormation} are identified with the (free) {\it possibility/necessity modal types} \eqref{CategoriesOfModales} (just as classically \eqref{PotentialDataIdentifiedWithPossibilityModalDat}):
\begin{equation}
\label{QuantumPotentialDataIdentifiedWithPossibilityModalDat}
\hspace{-3cm}
  \begin{tikzcd}[column sep=55pt]
    &[12pt]
    \;\;
    \QuantumTypes
      ^{ \scalebox{.7}{$\possibly_{{}_W}$} }
      _{{}_W}
    \mathrlap{
      \;
      \scalebox{.7}{
        \color{darkblue}
        \bf
        Possibility modal data
      }
    }
    \ar[
      d,
      phantom,
      "{ \simeq }"{rotate=-90}
    ]
    \\[-14pt]
    \mathllap{
      \raisebox{2pt}{
      \scalebox{.7}{
        \color{darkblue}
        \bf
        Actual
        quantum data
      }
      }
      \hspace{17pt}
    }
    \QuantumTypes_{{}_W}
    \ar[out=160, in=60,
      looseness=4,
      shorten=-3pt,
      "\scalebox{1.6}{${
          \hspace{1pt}
          \mathclap{
            \possibly_{\!{}_{\scalebox{.45}{$W$}}}
          }
          \hspace{1pt}
        }$}"{pos=.25, description},
      "{
        \scalebox{.7}{
          \color{darkorange}
          \bf
          possibly
        }
      }"{pos=.54, xshift=-5pt, yshift=2pt}
    ]
    \ar[out=185, in=175,
      looseness=6,
      phantom,
      "{\scalebox{.7}{$\bot$}}"{yshift=1pt}
    ]
    \ar[out=-160, in=-60,
      looseness=3.3,
      shorten=-3pt,
      "\scalebox{1.4}{${
          \hspace{1pt}
          \mathclap{
            \necessarily_{\mathrlap{\!{}_{\scalebox{.6}{\colorbox{white}{$\!\!\!W\!$}}}}}
          }
          \hspace{3pt}
        }$}"{pos=.25, description},
      "{
        \scalebox{.7}{
          \color{darkorange}
          \bf
          necessarily
        }
      }"{swap, pos=.57, yshift=-2.5pt}
    ]
    \ar[
      r,
      shift left=22pt,
      "{
        \scalebox{1.2}{$\oplus$}_{W}
      }"
    ]
    \ar[
      r,
      shift right=24pt,
      "{
        \scalebox{1.2}{$\oplus$}_{W}
      }"{swap}
    ]
    \ar[
      from=r,
      "{
        \otimes \mathbbm{1}_W
      }"{description}
    ]
    \ar[
      r,
      shift left=12pt,
      phantom,
      "{ \scalebox{.7}{$\bot$} }"{}
    ]
    \ar[
      r,
      shift right=12pt,
      phantom,
      "{ \scalebox{.7}{$\bot$} }"
    ]
    &
    \;\;
    \QuantumTypes_\Gamma
    \mathrlap{
      \;\;
      \scalebox{.7}{
        \color{darkblue}
        \bf
        Potential quantum data
      }
    }
    \\[-14pt]
    &
    \;\;
    \QuantumTypes
      ^{ \scalebox{.7}{$\necessarily_{{}_W}$}}
      _{{}_W}
    \mathrlap{
      \;
      \scalebox{.7}{
        \color{darkblue}
        \bf
        Necessity modal data
      }
    }
    \ar[
      u,
      phantom,
      "{ \simeq }"{rotate=-90}
    ]
  \end{tikzcd}
\end{equation}
\end{proposition}
\begin{proof}
This statement has
verbatim the same abstract proof -- via the monadicity theorem \eqref{MonadicityTheorem}
and the comparison statement \eqref{AdjointCoMonadsHaveIsomorphicModales} -- as its classical counterpart
in Prop. \ref{PotentialDataAsPossibilityModalData}, relying on the fact that $\otimes \mathbbm{1}_{W}$
is conservative (by the same argument as before) and both a left and a right adjoint.
\end{proof}
\begin{remark}[{\bf Homomorphisms of free $\possibly$/$\necessarily$-modales}]
\label{HomomorphismsOfFreeNecessityModales}
More explicitly,

\noindent {\bf (i)} for some $G_\bullet \isa \possibly_{{}_W} \HilbertSpace{H}_\bullet \to \possibly_{{}_W} \HilbertSpace{K}_\bullet$
to be a homomorphism of (free) $\possibly$-modales, it needs to make the following square commute:
$$
  \begin{tikzcd}[row sep=small,
    column sep=5pt
  ]
    \underset{W}{\possibly}
    \,
    \underset{W}{\possibly}
    \,
    \HilbertSpace{H}_\bullet
    \ar[
      rrrrr,
      "{
        \multiplication
          { \possibly_{{}_W} }
          { \HilbertSpace{H}_\bullet }
      }"{description}
    ]
    \ar[
      ddddd,
      "{
        \underset{W}{\possibly}
        \,
        G_\bullet
      }"{description}
    ]
    &&&&[-28pt]&
    \underset{W}{\possibly}
    \,
    \HilbertSpace{H}_\bullet
    \ar[
      ddddd,
      "{ G_\bullet }"{description}
    ]
    \\
    &
  \scalebox{.8}{$  \underset{w''}{\oplus}
    \vert \psi_{ {\color{purple}w}, w', w'' } \rangle
    $}
    \ar[rrr, |->, shorten=5pt]
    \ar[
      ddd,
      |->,
      shorten=5pt
    ]
    &&&
   \scalebox{.8}{$   \underset{w''}{\sum} \underset{w'}{\oplus}
    \,
    \vert \psi_{ {\color{purple}w}, w', w'' } \rangle
    $}
    \ar[
      dd,
      |->
    ]
    \\[+1pt]
    \\[+1pt]
    &
    &&&
   \scalebox{.8}{$   G_{{\color{purple}w}}
    \underset{w''}{\sum} \underset{w'}{\oplus}
    \,
    \vert \psi_{ {\color{purple}w}, w', w'' } \rangle
    $}
    \ar[
      dl,
      equals,
      rounded corners,
      to path ={
           ([yshift=0pt]\tikztostart.south)
        -- ([yshift=-13pt]\tikztostart.south)
        -- ([xshift=10pt]\tikztotarget.east)
        -- ([xshift=00pt]\tikztotarget.east)
        }
    ]
    \\
    &
    \scalebox{.8}{$  \underset{w''}{\oplus}
    G_{w''} \underset{w'}{\oplus}
    \vert \psi_{ {\color{purple}w}, w', w'' } \rangle
    $}
    \ar[
      rr,
      |->
    ]
    &&
  \scalebox{.8}{$    \underset{w''}{\sum}
    G_{w''} \underset{w'}{\oplus}
    \vert \psi_{ {\color{purple}w}, w', w'' } \rangle
    $}
    &
    \\
    \underset{W}{\possibly}
    \,
    \underset{W}{\possibly}
    \,
    \HilbertSpace{K}_\bullet
    \ar[
      rrrrr,
      "{
        \multiplication
          { \possibly_{{}_W} }
          { \HilbertSpace{K}_\bullet }
      }"{description}
    ]
    &&&&&
    \underset{W}{\possibly}
    \,
    \HilbertSpace{H}_\bullet
  \end{tikzcd}
$$
This is clearly possible only if $G_w$ is actually independent of $w$, i.e.,
if $G_\bullet \,=\, G \,:=\, G \otimes \mathbbm{1}_{{}_W}$.

\noindent {\bf (ii)} Analogously for homomorphisms of free $\necessarily$-modales:
$$
  \begin{tikzcd}[sep=5pt]
    \underset{W}{\necessarily}
    \HilbertSpace{H}_\bullet
    \ar[
      rrrrr,
      "{
        \comultiplication
          { \necessarily_{{}_W} }
          { \HilbertSpace{H}_\bullet }
      }"{description}
    ]
    \ar[
      ddddd,
      "{
        G_\bullet
      }"{description}
    ]
    &&&[+5pt]&[-30pt]&
    \underset{W}{\necessarily}
    \underset{W}{\necessarily}
    \HilbertSpace{H}_\bullet
    \ar[
      ddddd,
      "{
        \underset{W}{\necessarily}
        G_\bullet
      }"{description}
    ]
    \\
    &
 \scalebox{1}{$  \underset{w'}{\scalebox{1.2}{$\oplus$}}
    \vert \psi_{{\color{purple}w},w'}\rangle
    $}
    \ar[rrr, |->, shorten=5pt]
    \ar[ddd, |->, shorten=5pt]
    &&&
  \scalebox{1}{$   \underset{w''}{\scalebox{1.2}{$\oplus$}}
    \,
    \underset{w'}{\scalebox{1.2}{$\oplus$}}
    \vert \psi_{{\color{purple}w},w'}\rangle
    $}
    \ar[dd, |-> , shorten <=-2pt]
    \\
    &&&&
    \\
    & &&&
  \scalebox{1}{$   \underset{w''}{\scalebox{1.2}{$\oplus$}}
    \,
    G_{w''}
    \,
    \underset{w'}{\scalebox{1.2}{$\oplus$}}
    \,
    \vert \psi_{{\color{purple}w},w'}\rangle
    $}
    \ar[
      dl,
      equals,
      rounded corners,
      to path ={
           ([yshift=0pt]\tikztostart.south)
        -- ([yshift=-13pt]\tikztostart.south)
        -- ([xshift=10pt]\tikztotarget.east)
        -- ([xshift=00pt]\tikztotarget.east)
        }
    ]
    \\
    &
\scalebox{1}{$     G_{{\color{purple}w}}
    \,
    \underset{w'}{\scalebox{1.2}{$\oplus$}}
    \,
    \vert
      \psi_{{\color{purple}w},w'}
    \rangle
    $}
    \ar[rr, |->, shorten=1pt]
    &&
\scalebox{1}{$     \underset{w''}{\scalebox{1.2}{$\oplus$}}
    \,
    G_{{\color{purple}w}}
    \,
    \underset{w'}{\scalebox{1.2}{$\oplus$}}
    \vert
      \psi_{{\color{purple}w},w'}
    \rangle
    $}
    \\
    \underset{W}{\necessarily}
    \HilbertSpace{K}_\bullet
    \ar[
      rrrrr,
      description,
      "{
        \comultiplication
          { \necessarily_{{}_W} }
          { \HilbertSpace{K}_\bullet }
      }"{description}
    ]
    &&&&&
    \underset{W}{\necessarily}
    \underset{W}{\necessarily}
    \HilbertSpace{K}_\bullet   \\
  \end{tikzcd}
$$
\end{remark}


\noindent
{\bf In summary} so far, we have found a quantum epistemic logic with the following interpretations, analogous to \eqref{SummaryClassicalEpistemicTypeTheory}:

\begin{equation}
\label{SummaryQuantumEpistemicTypeTheory}
\hspace{-.4cm}
\adjustbox{
  fbox
}{
$
  \begin{tikzcd}[
    row sep=6pt,
    column sep=40pt
  ]
    &[-30pt]
    \overset{
      \mathclap{
      \raisebox{4pt}{
        \scalebox{\termscale}{
          \color{darkblue}
          \bf
          necessarily $\HilbertSpace{H}_\bullet$
        }
      }
      }
    }{
      \necessarily_{{}_W} \, \HilbertSpace{H}_\bullet
    }
    \ar[
      rr,
      "{
        \scalebox{\termscale}{
          \color{darkorange}
          \bf
          entails
        }
      }"{yshift=8pt, pos=.48},
      "{
        \counit
          {\necessarily_{{}_W}}
          {\HilbertSpace{H}_\bullet}
      }"{description}
    ]
    &&
    \overset{
      \mathclap{
      \raisebox{4pt}{
        \scalebox{\termscale}{
          \color{darkblue}
          \bf
          actually $\HilbertSpace{H}_\bullet$
        }
      }
      }
    }{
      \HilbertSpace{H}_\bullet
    }
    \ar[
      rr,
      "{
        \scalebox{\termscale}{
          \color{darkorange}
          \bf
          entails
        }
      }"{yshift=8.5pt, pos=.55},
      "{
        \unit
          {\possibly_{{}_W}}
          {\HilbertSpace{H}_\bullet}
      }"{description}
    ]
    &&
    \overset{
      \mathclap{
      \raisebox{4pt}{
        \scalebox{\termscale}{
          \color{darkblue}
          \bf
          possibly $\HilbertSpace{H}_\bullet$
        }
      }
      }
    }{
      \possibly_{{}_W} \, \HilbertSpace{H}_\bullet
    }
    \ar[
      r,
      phantom,
      "{
        \simeq
      }",
      "{
        \scalebox{.7}{
          \color{black}
          \def\arraystretch{.8}
          \begin{tabular}{c}
            principle
            of quantum compulsion:
          \end{tabular}
        }
      }"{xshift=0pt, yshift=+24pt},
      "{
        \scalebox{\termscale}{
          \color{darkorange}
          \bf
          is
        }
      }"{xshift=0pt, yshift=13.7pt}
    ]
    &[-35pt]
    \overset{
      \mathclap{
      \raisebox{4pt}{
        \scalebox{\termscale}{
          \color{darkblue}
          \bf
          \;\;
          necessarily $\HilbertSpace{H}_\bullet$
        }
      }
      }
    }{
      \necessarily_{{}_W} \, \HilbertSpace{H}_\bullet
    }
    \\
    \overset{
      \mathclap{
      \raisebox{3pt}{
        \scalebox{.7}{
          \color{darkorange}
          \bf
          In world
        }
      }
      }
    }{
      {\color{purple}w} \isa W
    }
    \hspace{12pt}
    \overset{
      \mathclap{
      \;\;\;\;\;\;\;\;
      \raisebox{3pt}{
        \scalebox{\termscale}{
          \color{darkorange}
          \bf
          observe...
        }
      }
      }
    }{
    \vdash
    }
    \hspace{-2pt}
    &
    \scalebox{1.2}{$\HilbertSpace{H}$}
    \mathrlap{\phantom{,}}
    \ar[
      rr,
      ->>,
      "{
        \scalebox{1.2}{$\oplus$}_{w'}
        \vert \psi_{w'} \rangle
        \;\;\mapsto\;\;
        \vert \psi_{{}_{\color{purple}w}} \rangle
      }"{yshift=1pt},
      "{
        \scalebox{.7}{
          \color{darkgreen}
          \bf
          measurement collapse
        }
      }"{swap, yshift=-1pt}
    ]
    \ar[
      rrrr,
      rounded corners,
      to path={
           ([yshift=-0pt]\tikztostart.south)
        -- ([yshift=-6pt]\tikztostart.south)
        -- node[yshift=-6pt]{
            \scalebox{.7}{
              {\color{darkgreen}
              \bf
              linear projector onto
              sub-Hilbert space} $\HilbertSpace{H}_{\color{purple}w}$
            }
          }
        ([yshift=-6pt]\tikztotarget.south)
        -- ([yshift=-0pt]\tikztotarget.south)
      }
    ]
    &&
    \HilbertSpace{H}_{\color{purple}w}
    \ar[
      rr,
      hook,
      "{
        \vert \psi_{\color{purple}w} \rangle
        \mapsto
        \,
        \scalebox{1.2}{$\oplus$}_{w'}
        \delta^{w'}_{{\color{purple}w}}
        \vert \psi_{\color{purple}w} \rangle
      }",
      "{
        \scalebox{.7}{
          \color{darkgreen}
          \bf
          state preparation
        }
      }"{swap, yshift=-1pt}
    ]
    &{\phantom{-}}&
    \scalebox{1.2}{$
      \HilbertSpace{H}
    $}
    \mathrlap{,}
    &
  \mbox{
    where
    $
      \scalebox{1.2}{$\HilbertSpace{H}$}
      \;:=\;
      \underset{w' \isa W}{\scalebox{1.3}{$\oplus$}}
      \HilbertSpace{H}_{w'}
    $
  }
  \\[+10pt]
  {}
  \ar[
    rrrrr,
    -,
    dashed,
    shorten <=-1.0cm,
    shorten >=-4cm,
    gray
  ]
  &&&&&
  {}
  \\[-8pt]
  &
  \overset{
    \mathclap{
      \raisebox{4pt}{
        \scalebox{\termscale}{
          \color{darkblue}
          \bf
          randomly $\HilbertSpace{H}$
        }
      }
    }
  }{
    \randomly_{{}_W} \HilbertSpace{H}
  }
  \ar[
    rr,
    "{
      \counit
        { \randomly_{{}_W} }
        { \HilbertSpace{H} }
    }"{description, pos=.4},
    "{
      \scalebox{\termscale}{
        \color{darkorange}
        \bf
        entails
      }
    }"{pos=.4, yshift=9pt}
  ]
  &&
  \overset{
    \mathclap{
      \raisebox{5pt}{
        \scalebox{\termscale}{
          \color{darkblue}
          \bf
          potentially $\HilbertSpace{H}$
        }
      }
    }
  }{
    \HilbertSpace{H}
  }
  \ar[
    rr,
    "{
      \unit
        { \indefinitely_{{}_W} }
        { \HilbertSpace{H} }
    }"{description},
    "{
      \scalebox{\termscale}{
        \color{darkorange}
        \bf
        entails
      }
    }"{pos=.5, yshift=10pt}
  ]
  &&
  \overset{
    \mathclap{
      \raisebox{4pt}{
        \scalebox{\termscale}{
          \color{darkblue}
          \bf
          indefinitely $\HilbertSpace{H}$
        }
      }
    }
  }{
    \indefinitely_{{}_W} \HilbertSpace{H}
  }
  \\[+10pt]
  &
  \underset{w \isa W}{\scalebox{1.2}{$\oplus$}}
  \HilbertSpace{H}
  \ar[
    rr,
    "{
      \scalebox{1.2}{$\oplus$}_{{}_W}
      \,
      \vert \psi_{{}_W} \rangle
      \;\mapsto\;
      \sum_{{}_W} \vert \psi_{{}_W} \rangle
    }",
    "{
      \scalebox{.7}{
        \color{darkgreen}
        \bf
        quantum superposition
      }
    }"{swap, yshift=-1pt}
  ]
  &&
  \HilbertSpace{H}
  \ar[
    rr,
    "{
      \vert \psi \rangle
      \;\mapsto\;
      \scalebox{1.2}{$\oplus$}_{w'\isa W} \vert \psi \rangle
    }",
    "{
      \scalebox{.7}{
        \color{darkgreen}
        \bf
        quantum parallelism
      }
    }"{swap, yshift=-1pt}
  ]
  &&
  \underset{b : B}{\scalebox{1.2}{$\oplus$}}
  \HilbertSpace{H}
  \end{tikzcd}
$
\hspace{-10pt}
}
\end{equation}

\medskip

However, for linear types, we have yet another monadicity statement:

\begin{proposition}[{\bf Actual quantum data as indefiniteness-modal potential data}]
  \label{ActualQuantumDataIsInfinitenessModalPotentialData}
  For $W \isa \FiniteType_{\Gamma}$ and $p_{{}_W} \isa W \to \Gamma$ an epimorphism,
 the dependent sum $\scalebox{1.2}{$\oplus$}_{{}_W} \isa \QuantumTypes_{{}_W} \to \QuantumTypes_{\Gamma}$ is
 also monadic,
 whence the actual quantum types are identified with the (free) randomness/infiniteness modal types:
\begin{equation}
\label{ActualQuantumTypesIndentifiedWithInfinitenessModalTypes}
 \hspace{.5cm}
\begin{tikzcd}[
  column sep=30pt,
  row sep=4pt
]
    \mathllap{
      \raisebox{1pt}{
      \scalebox{.7}{
        \color{darkblue}
        \bf
        Randomness modal data
      }
      }
      \hspace{5pt}
    }
   \QuantumTypes
     _{\Gamma}
     ^{\!\scalebox{\termscale}{$\randomly_{{}_W}$}}
   \ar[
     d,
     phantom,
     "{ \simeq }"{rotate=-90}
   ]
   \\
    \mathllap{
      \raisebox{1pt}{
      \scalebox{.7}{
        \color{darkblue}
        \bf
        Actual quantum data
      }
      }
      \hspace{5pt}
    }
    \QuantumTypes_W
    \quad
  \ar[
    rr,
    shift left=22pt,
    "{
      \oplus_W
    }"
  ]
  \ar[
    from=rr,
    "{
      \otimes \mathbbm{1}_{W}
    }"{description}
  ]
  \ar[
    rr,
    shift right=20pt,
    "{
      \oplus_W
    }"{swap}
  ]
  \ar[
    rr,
    phantom,
    shift left=11pt,
    "\scalebox{.7}{$\bot$}"
  ]
  \ar[
    rr,
    phantom,
    shift right=11pt,
    "\scalebox{.7}{$\bot$}"
  ]
  &&
  \quad
  \QuantumTypes_\Gamma
    \mathrlap{
      \hspace{20pt}
      \raisebox{1pt}{
      \scalebox{.7}{
        \color{darkblue}
        \bf
        Potential quantum data
      }
      }
    }
    \ar[out=120, in=20,
      looseness=4,
      shorten <=-2pt,
      shorten >=-3pt,
      "\raisebox{-2pt}{${
          \hspace{7pt}
          \mathclap{
            \randomly_{\!\scalebox{\termscale}{$W$}}
          }
          \hspace{2pt}
        }$}"{pos=.7, description},
      "{
        \scalebox{.7}{
          \hspace{23pt}
          \color{darkorange}
          \bf
          randomly
        }
      }"{pos=.4, yshift=2pt}
    ]
    \ar[out=5, in=-5,
      looseness=8.5,
      phantom,
      "{\scalebox{.7}{$\bot$}}"{yshift=1pt}
    ]
    \ar[out=-120, in=-20,
      looseness=4,
      shorten <=-2pt,
      shorten >=-3pt,
      "\scalebox{1.1}{${
          \hspace{4pt}
          \mathclap{
            \indefinitely_{\!{}_{\scalebox{.7}{$W$}}}
          }
          \hspace{1pt}
        }$}"{pos=.7, description},
      "{
        \scalebox{.65}{
          \hspace{40pt}
          \color{darkorange}
          \bf
          indefinitely
        }
      }"{swap, pos=.398, yshift=-3pt}
    ]
  \\[-3pt]
    \mathllap{
      \raisebox{1pt}{
      \scalebox{.7}{
        \color{darkblue}
        \bf
        Indefiniteness modal data
      }
      }
      \hspace{5pt}
    }
   \QuantumTypes
     _{\Gamma}
     ^{\scalebox{\termscale}{$\indefinitely_{{}_W}$}}
   \ar[
     u,
     phantom,
     "{ \simeq }"{rotate=-90, pos=-3}
   ]
\end{tikzcd}
\end{equation}
\end{proposition}
\begin{proof}
  Due to ambidexterity
  \eqref{AmbidexterityInQuantumEpistemology}
  for finite $W$, in the quantum case also $\scalebox{1.2}{$\oplus$}_{{}_W}$ is both a left and right adjoint, as shown. Therefore the monadicity theorem \eqref{MonadicityTheorem}
  implies the claim for $\indefinitely_{{}_W}$
  by observing that $\scalebox{1.2}{$\oplus$}_{{}_W}$ is conservative. This is indeed the case, as it sends a morphism to its world-wise
  application, which is an isomorphism of dependent types if and only if it is so world-wise, hence if and only the original morphisms was so.
  The dual claim for the adjoint comonad $\randomly_{{}}$ now follows by \eqref{AdjointCoMonadsHaveIsomorphicModales}.
\end{proof}

\begin{remark}[Effective perspective on quantum epistemology]
  Prop. \ref{ActualQuantumDataIsInfinitenessModalPotentialData} says that (over a finite inhabited type of classical worlds $W$) dependent linear types are
  $\indefinitely$-monadic! But since we have seen that dependent linear types may be thought of as quantum states in ``many worlds'', this
  gives a monadic perspective on quantum epistemology which allows for speaking about it in terms of {\it computational effects} (Lit. \ref{LiteratureComputationalEffectsAndModalities}).
    Hence we shall refer to these equivalent perspectives as the {\it epistemic} and the {\it effective} perspective, respectively:
\vspace{-2mm}
  \begin{equation}
  \label{EpistemicEffectivePerspective}
  \hspace{-4mm}
  \adjustbox{
    fbox
  }{
  \hspace{-18pt}
  $
    \begin{tikzcd}[
      column sep=10pt,
      row sep=8pt
    ]
      \scalebox{\termscale}{\bf
       \def\arraystretch{.9}
       \begin{tabular}{c}
          Epistemic
          \\
          perspective
        \end{tabular}
      }
      \ar[
        dd,
        <->,
        shorten=3pt,
        "{
          \scalebox{.7}{
            \color{darkorange}
            \bf
            monadicity
          }
        }"{
          yshift=-1pt,
          sloped,
          rotate=180
        },
        "{
          \scalebox{.7}{
            \color{darkorange}
            \bf
            of $\osum_{{}_W}$
          }
        }"{
          yshift=+1pt,
          swap,
          sloped,
          rotate=180
        }
      ]
      &
      \QuantumTypes_{{}_W}
      \ar[
        dd,
        shift left=7pt,
        "{
          \underset{W}{\osum}
        }"{}
      ]
      \ar[
        from=dd,
        shift left=7pt,
        "{
          \otimes
          \mathbbm{1}_{\!{}_W}
        }"{}
      ]
      \ar[
        dd,
        phantom,
        "{
          \scalebox{.7}{$\dashv$}
        }"
      ]
      &[+5pt]
      &[+5pt]
      \HilbertSpace{H}_\bullet
      \ar[
        rr,
        "{
          G_\bullet
        }"{description},
        "{
          \scalebox{.7}{
            \color{darkgreen}
            \bf
            map of $W$-dependent types
          }
        }"{swap, yshift=-3pt}
      ]
      &
      \phantom{--}
      &
      \HilbertSpace{K}_\bullet
      &[25pt]
      \underset{
        \mathclap{
        \scalebox{.7}{
        \color{darkgreen}
        \bf
        in-dependent type
        }}
      }{
      \HilbertSpace{H}
      }
      &[25pt]
      \HilbertSpace{H}
      \ar[
        rr,
        "{
          G_\bullet
        }",
        "{
          \scalebox{.7}{
            \color{darkgreen}
            \bf
            \def\arraystretch{.9}
            \begin{tabular}{c}
              $W$-dependent map of
              \\
              in-dependent types
            \end{tabular}
          }
        }"{swap, yshift=-3pt}
      ]
      &[+6pt]
      &[+6pt]
      \HilbertSpace{H}
      \\
      &&&&
      \rotatebox{90}{$\longleftrightarrow$}
      &&
      \rotatebox{90}{$\longleftrightarrow$}
      &&
      \rotatebox{90}{$\longleftrightarrow$}
      \\
      \scalebox{\termscale}{\bf
       \def\arraystretch{.9}
       \begin{tabular}{c}
          Effective
          \\
          perspective
        \end{tabular}
      }
      &
      \QuantumTypes_\Gamma
     \ar[
       out=55-180, in=-55, looseness=5,
       shift right=3pt,
       "\scalebox{1.2}{\colorbox{white}{$\mathclap{
        \;
        \underset{W}{\indefiniteness}
        \;
      }$}}"{description},
    ]
    &&
    \underset{W}{\osum}
    \HilbertSpace{H}_\bullet
     \ar[out=64-180, in=-64, looseness=3.2, "\scalebox{1.2}{\colorbox{white}{$\mathclap{
        \;
        \underset{W}{\indefiniteness}
        \;
      }$}}"{description},
    ]
    \ar[
      rr,
      "{
        \underset{W}{\osum}
        G_\bullet
      }"{description, yshift=-3pt},
      "{
        \scalebox{.7}{
          \color{darkgreen}
          \bf
          homomorphism of
          \scalebox{1.3}{$\indefinitely_{{}_W}$}-modales
        }
      }"{yshift=4pt}
    ]
    &{}&
    \underset{W}{\osum}
    \HilbertSpace{K}_\bullet
     \ar[out=64-180, in=-64, looseness=3.2, "\scalebox{1.2}{\colorbox{white}{$\mathclap{
        \;
        \underset{W}{\indefiniteness}
        \;
      }$}}"{description},
    ]
    &
    \overset{
      \mathclap{
        \raisebox{3pt}{
          \scalebox{.7}{
            \color{darkgreen}
            \bf
            free
            \scalebox{1}{$\indefinitely_{{}_W}$}-modale
          }
        }
      }
    }{
      \underset{W}{\indefinitely}
      \HilbertSpace{H}
    }
    &
    \mathrm{bind}
  \Big(
      \HilbertSpace{H}
      \ar[
       rr,
       "{
         \underset{W}{\osum}
         G_\bullet
         \,\circ\,
         \unit
           { \indefinitely_{{}_W} }
           { \HilbertSpace{H} }
       }"{swap},
       "{
         \scalebox{.7}{
           \color{darkgreen}
           \bf
           \scalebox{1.2}{$\indefinitely_{{}_W}$}-Kleisli
           map
         }
       }"{yshift=6pt}
      ]
      &&
      \underset{W}{\indefinitely}
      \HilbertSpace{K}
    \Big)
    \end{tikzcd}
  $
 \hspace{-8pt}}
 \end{equation}

The effective perspective on the epistemic entailments \eqref{SummaryQuantumEpistemicTypeTheory} yields an effect-language for
quantum measurement and controlled quantum gates -- this we discuss next in \cref{ControlledQuantumGates}.

\end{remark}

\begin{remark}[{\bf Relation to {\tt \zxCalculus}}]
  Something close to the identification
  $
    \Modales
    {
      \randomly_{{}_W}
    }
    {
    \big(
      \QuantumTypes_{{}_\Gamma}
    \big)
    }
    \,\simeq\;
    \QuantumTypes_{{}_W}
  $
  (in Prop. \ref{ActualQuantumDataIsInfinitenessModalPotentialData}) has previously been observed in \cite[Thm. 1.5]{CoeckePavlovic08} (cf. Lit. \ref{QuantumMeasurementAndZXCalculus}), subject to some translation which we discuss now.
\end{remark}

\medskip

\noindent
{\bf Frobenius-algebraic formulation.}
Remarkably, the above modal quantum logic gives rise to the
``classical-structures'' Frobenius monads used in the {\zxCalculus} (Lit. \ref{QuantumMeasurementAndZXCalculus}).
In particular, this shows that/how {\LHoTT}/{\QS} can be used for certifying (type-checking) {\zxCalculus}-protocols:

\medskip

\begin{proposition}[\bf Quantum (co)effects via Frobenius algebra]
\label{QuantumCoEffectsViaFrobeniusAlgebra}
$\,$

\begin{itemize}
\item[{\bf (i)}] For $W \isa \ClassicalType$, the $W$-(co)reader (co)monad on linear types
(\cref{QuantumEpistemicLogicViaDependentLinearTypes})
is equivalent to the linear version
$\quantized W \otimes (\mbox{-})$
of the
(co)writer (co)monad
\eqref{ClassicalWriterMonad}
induced by the canonical (co)algebra structure on $\quantized W \,\defneq\, \osum_{{}_W} \mathbbm{1}$;

\item[{\bf (ii)}] If $W \isa \FiniteType$ is finite then
the underlying functors of all these (co)monads agree and make a single {\it Frobenius monad} induced from the canonical Frobenius-algebra structure on
$
  \quantized W
  \,=\,
  \underset{W}{\osum}
  \mathbbm{1}
$ (cf. Lit. \ref{QuantumMeasurementAndZXCalculus}):
\end{itemize}
\end{proposition}

\vspace{-.2cm}
{\small

}
\begin{proof}
With Prop. \ref{ComponentExpressionsOfTheCoMonadCoUnits}, this is a straightforward matter of unwinding the definitions:
\begin{equation}
\hspace{-4mm}
\adjustbox{}{
\def\arraystretch{1.4}
%
}
\end{equation}
\end{proof}
In fact, this Frobenius structure is ``{\it special}'' in that
\begin{equation}
  \label{SpecialityOfFrobeniusStructure}
  \begin{tikzcd}
    \underset{W}{\randomly}
    \ar[
      rr,
      "{
        \comultiplication
          { \randomly_{\!{}_W} }
          {  }
      }"
    ]
    \ar[
      rrrrr,
      rounded corners,
      to path ={
           ([yshift=-0pt]\tikztostart.south)
        -- ([yshift=-5pt]\tikztostart.south)
        -- node[yshift=-5pt]{\scalebox{.7}{$\sim$}}
         ([yshift=-5pt]\tikztotarget.south)
        -- ([yshift=-0pt]\tikztotarget.south)
      }
    ]
    &&
    \underset{W}{\randomly}
    \,
    \underset{W}{\randomly}
    \ar[
      r,
      phantom,
      "{ \simeq }"
    ]
    &[-10pt]
    \underset{W}{\indefinitely}
    \,
    \underset{W}{\indefinitely}
    \ar[
      rr,
      "{
        \multiplication
          { \indefinitely_{{}_W} }
          {  }
      }"
    ]
    &&
    \underset{W}{\indefinitely}
  \end{tikzcd}
\end{equation}

\begin{remark}[Frobenius property and Spider theorem]
\label{FrobeniusPropertyAndSpiderTheorem}
The Frobenius property of $\indefinitely \simeq \randomly$ (Prop. \ref{QuantumCoEffectsViaFrobeniusAlgebra})
says explicitly that this diagram commutes:
\vspace{-2mm}
$$
  \begin{tikzcd}[
    row sep=5pt, column sep=50pt
  ]
    &[+5pt]
    \underset{W}{\randomly}
    \underset{W}{\randomly}
    \underset{W}{\indefinitely}
    \ar[
      r,
      phantom,
      "{ \simeq }"
    ]
    &[-16pt]
    \underset{W}{\randomly}
    \underset{W}{\indefinitely}
    \underset{W}{\indefinitely}
    \ar[
      dr,
      "{
        \underset{W}{\randomly}
        \multiplication
          { \indefinitely_{{}_W} }
          { (\mbox{-}) }
      }"{sloped}
    ]
    &[+5pt]
    \\
    \underset{W}{\randomly}
    \underset{W}{\indefinitely}
    \ar[
      d,
      phantom,
      "{ \simeq }"{rotate=90}
    ]
    \ar[
      ur,
      "{
        \comultiplication
          { \randomly_{{}_W} }
          { \indefinitely_{{}_W}(\mbox{-})}
      }"{sloped}
    ]
    \ar[
      drrr,
      phantom,
      "{
\adjustbox{
  scale=.8,
  raise=-12pt
}{
\begin{tikzpicture}[
  xscale=2.2,
  gray
]
  \begin{scope}[
    yscale=.3,
    yshift=2.5cm
  ]
  \draw[line width=1.2, -]
   (-1,1) -- (1,1);
  \draw[line width=1.2, -]
   (-1,-1) -- (1,-1);
   \draw[line width=1.2, -]
     (-.9,1)
       .. controls
         (0,1) and (0,-1)
       ..
     (.9,-1);
  \end{scope}
  \begin{scope}[
    yscale=.3,
    yshift=-2.5cm
  ]
  \draw[line width=1.2, -]
   (-1,1) -- (1,1);
  \draw[line width=1.2, -]
   (-1,-1) -- (1,-1);
   \draw[line width=1.2, -]
     (-.9,-1)
       .. controls
         (0,-1) and (0,+1)
       ..
     (.9,+1);
  \end{scope}
\end{tikzpicture}
}
      }"
    ]
    &&&
    \underset{W}{\randomly}
    \underset{W}{\indefinitely}
    \ar[
      d,
      phantom,
      "{ \simeq }"{rotate=-90}
    ]
    \\
    \underset{W}{\indefinitely}
    \underset{W}{\randomly}
    \ar[
      dr,
      "{
        \underset{W}{\indefinitely}
        \comultiplication
          { \randomly_{{}_W} }
          { (\mbox{-}) }
      }"{swap, sloped}
    ]
    &&&
    \underset{W}{\indefinitely}
    \underset{W}{\randomly}
    \\
    &
    \underset{W}{\indefinitely}
    \underset{W}{\randomly}
    \underset{W}{\randomly}
    \ar[
      r,
      phantom,
      "{ \simeq }"{rotate=180}
    ]
    &
    \underset{W}{\indefinitely}
    \underset{W}{\indefinitely}
    \underset{W}{\randomly}
    \ar[
      ur,
      "{
        \multiplication
          { \indefinitely_{{}_W} }
          { \randomly_{{}_W}(\mbox{-}) }
      }"{swap, sloped}
    ]
    &&
  \end{tikzcd}
$$

\vspace{-2mm}
\noindent But this already implies (by the theory of {\it normal forms} \cite[Prop. 12, Fig. 3]{Abrams96}\cite{Kock04}, together with specialty \eqref{SpecialityOfFrobeniusStructure}) the equality of all those transformations of the form
\vspace{-2mm}
\begin{equation}
  \label{SpiderMap}
  \begin{tikzcd}
    \indefinitely^{n}
    \ar[r]
    &
    \randomly^{n'}
  \end{tikzcd}
\end{equation}

\vspace{-2mm}
\noindent which arise as composites of $\indefinitely$-joins and of $\randomly$-duplicates
and which are {\it connected}
in that there is no non-trivial
horizontal decomposition ---
such as in this simple example:
\vspace{-2mm}
$$
\hspace{-2mm}
  \begin{tikzcd}[
    column sep=28pt,
    row sep=10pt
  ]
    \underset{W}{\indefinitely}
    \,
    \underset{W}{\indefinitely}
    \,
    \underset{W}{\indefinitely}
    \,
    \HilbertSpace{H}
    \ar[
      rr,
      "{
        \multiplication
          { \indefinitely_{{}_W} }
          {
            \indefinitely_{{}_W}
            \HilbertSpace{H}
          }
      }"
    ]
    &&
    \underset{W}{\indefinitely}
    \,
    \underset{W}{\indefinitely}
    \,
    \HilbertSpace{H}
    \ar[
      rr,
      "{
        \multiplication
          { \indefinitely_{{}_W} }
          { \HilbertSpace{H} }
          {}
      }"
    ]
    &&
    \underset{W}{\indefinitely}
    \,
    \HilbertSpace{H}
    &[-40pt]
    \simeq
    &[-40pt]
    \underset{W}{\randomly}
    \,
    \HilbertSpace{H}
    \ar[
      rr,
      "{
        \comultiplication
          { \randomly_{{}_W} }
          { \HilbertSpace{H} }
          {}
      }"
    ]
    &&
    \underset{W}{\randomly}
    \,
    \underset{W}{\randomly}
    \,
    \HilbertSpace{H}
    \\
    \quantized W
    \otimes
    \quantized W
    \otimes
    \quantized W
    \otimes
    \HilbertSpace{H}
    \ar[
      rr,
      "{
        \mathrm{prod}_{ \quantized W }
        \,\otimes\,
        \mathrm{id}_{\quantized W}
      }"{yshift=1pt}
    ]
    &&
    \quantized W
    \otimes
    \quantized W
    \otimes
    \HilbertSpace{H}
    \ar[
      rrr,
      "{
        \mathrm{prod}_{{}_{\quantized W} }
        \,\otimes\,
        \mathrm{id}_{\HilbertSpace{H}}
      }"{yshift=1pt}
    ]
    &&&
    \quantized W
    \!\otimes\!
    \HilbertSpace{H}
    \ar[
      rrr,
      "{
        \mathrm{coprod}_{{}_{\quantized W}}
        \,\otimes\,
        \mathrm{id}_{\HilbertSpace{H}}
      }"{yshift=1pt}
    ]
    &&&
    \quantized W
    \otimes
    \quantized W
    \otimes
    \HilbertSpace{H}
  \end{tikzcd}
$$

\vspace{-2mm}
\noindent
This classical fact of Frobenius algebra theory has been called the {\it spider theorem} in \cite[Thm. 1]{CoeckeDuncan08}, since it means that in string diagram notation, all the operations \eqref{SpiderMap} may uniquely by depicted by a diagram of this form:

\begin{equation}
n
\left\{
\mathclap{
  \def\arraystretch{3.5}
  \begin{array}{c}
    {} \\ {}
  \end{array}
}
\right.
\adjustbox{raise=-1.32cm}{
\begin{tikzpicture}[scale=1]
  \begin{scope}[yscale=.7]
  \draw[line width=1]
    (-2,2)
     .. controls (-1,2) and (-1,0)  ..
     (-.2,0)
     --
    (0,0);
  \draw[line width=1]
    (-2,1.5)
     .. controls (-1,1.5) and (-1,0)  ..
     (-.2,0)
     --
    (0,0);
  \draw[line width=1]
    (-2,1)
     .. controls (-1,1) and (-1,0)  ..
     (-.2,0)
     --
    (0,0);

  \draw[line width=1]
    (-2,-2)
     .. controls (-1,-2) and (-1,0)  ..
     (-.2,0)
     --
    (0,0);
  \draw[line width=1]
    (-2,-1.5)
     .. controls (-1,-1.5) and (-1,0)  ..
     (-.2,0)
     --
    (0,0);
  \draw[line width=1]
    (-2,-1)
     .. controls (-1,-1) and (-1,0)  ..
     (-.2,0)
     --
    (0,0);

  \draw[line width=1]
    (+2,1.5)
     .. controls (+1,1.5) and (+1,0)  ..
     (-.2,0)
     --
    (0,0);
  \draw[line width=1]
    (+2,1)
     .. controls (+1,1) and (+1,0)  ..
     (+.2,0)
     --
    (0,0);

  \draw[line width=1]
    (+2,-1.5)
     .. controls (+1,-1.5) and (+1,0)  ..
     (+.2,0)
     --
    (0,0);
  \draw[line width=1]
    (+2,-1)
     .. controls (+1,-1) and (+1,0)  ..
     (+.2,0)
     --
    (0,0);
\end{scope}
\draw[fill=black]
  (0,0)
  circle
  (.15);

\node
  at
  (-1.9,.2)
  {
    \scalebox{1.6}{\vdots}
  };
\node
  at
  (+1.85,.2)
  {
    \scalebox{1.6}{\vdots}
  };
\end{tikzpicture}
}
\left.
\mathclap{
  \def\arraystretch{2.75}
  \begin{array}{c}
    {} \\ {}
  \end{array}
}
\right\}
n'
\end{equation}
These are the {\it spider diagrams} used in {\zxCalculus} (Lit. \ref{QuantumMeasurementAndZXCalculus}).
\end{remark}

\medskip

\medskip

\noindent
{\bf Indefiniteness as a computational effect.} We may now cast these structures into natural programming language constructs for {\it computational effects} used in \cref{ControlledQuantumGates} to encode (quantum gates controlled by) quantum measurement.

\begin{proposition}[\bf Indefiniteness modality is strong]
\label{IndefinitenessModalityIsStrong}
 $\,$

 \noindent
  For $W \isa \ClassicalType$ the indefiniteness-modality $\indefiniteness_W \,\colon\, \QuantumTypes \to \QuantumTypes$ carries symmetric monoidal
  structure \eqref{SymmetricMonoidalMonadStructure} as shown in \eqref{SymmetricMonoidalMonadStructureOnIndefiniteness}
  exhibiting it as a computational effect
  \eqref{TypingOfStrongMonads}:
  \vspace{-2mm}
\begin{equation}
  \label{IndefinitenssMonadicEffect}
  \begin{array}{l}
  \declarelin
    {
      \return
        { \indefinitely_W }
        {
          \HilbertSpace{H}
        }
    }
    {
      \HilbertSpace{H}
      \maplin
      \underset{W}{\indefinitely}
      \HilbertSpace{H}
    }
    {
      \vert \psi \rangle
      \,\mapsto\,
      \big(
        w \mapsto
        \vert \psi \rangle
      \big)
    }
   \\
   \\
  \declarelin
    {
      \bind
        { \indefinitely_W }
        {
          \HilbertSpace{H},
          \HilbertSpace{H}'
        }
    }
    {
      \big(
        \HilbertSpace{H}
        \maplin
        \underset{W}{\indefinitely}
        \HilbertSpace{H}'
      \big)
      \maplin
      \big(
        \underset{W}{\indefinitely}
        \HilbertSpace{H}
        \maplin
        \underset{W}{\indefinitely}
        \HilbertSpace{H}'
      \big)
    }
    {
      \Big(
        \vert \psi \rangle
        \mapsto
        \big(
          w \mapsto
          G_w\vert \psi \rangle
        \big)
      \Big)
      \mapsto
      \Big(
        \big(
          w \mapsto
          \vert \psi_w \rangle
        \big)
        \mapsto
        \big(
          w \mapsto
          G_w\vert \psi_w \rangle
        \big)
      \Big)
    }
\end{array}
\end{equation}
\end{proposition}
As such, this monadic effect is the part of the {\QS} language in \cref{Pseudocode} responsible for quantum measurement and classical control.

Dually:
\begin{proposition}[\bf Randomness modality is costrong]
  \label{RandomnessModalityISCostrong}
  For $W \,\isa\, \ClassicalType$ the randomness-modality $\randomly_W \,\isa\, \QuantumTypes \to \QuantumTypes$ carries symmetric comonoidal comonad structure as shown in \eqref{SymmetricCoMonoidalCOMonadStructureOnRandomness}.
\end{proposition}

\newpage

\begin{equation}  \label{SymmetricMonoidalMonadStructureOnIndefiniteness}
  \def\arraystretch{2}

  \\[-15pt]
  \\
    \rotatebox[origin=c]{90}{\bf symmetric}
    &
  \begin{tikzcd}[
    column sep=0pt,
    row sep=0pt
  ]
    \big(
    \underset{W}{\indefinitely}
    \HilbertSpace{H}
    \big)
    \otimes
    \big(
    \underset{W}{\indefinitely}
    \HilbertSpace{H}'
    \big)
    \ar[
      rrrr,
      "{
        \braid
        {\otimes}
        {
          \indefinitely_W
          \HilbertSpace{H},
          \indefinitely_W
          \HilbertSpace{H}'
        }
      }"
    ]
    \ar[
      dddd,
      "{
        \pair
          { \indefinitely }
          {
            \HilbertSpace{H}
            ,
            \HilbertSpace{H}'
          }
      }"{swap}
    ]
    &[-35pt]&&&[-35pt]
    \big(
    \underset{W}{\indefinitely}
    \HilbertSpace{H}'
    \big)
    \otimes
    \big(
    \underset{W}{\indefinitely}
    \HilbertSpace{H}
    \big)
    \ar[
      dddd,
      "{
        \pair
          { \indefinitely }
          {
            \HilbertSpace{H}'
            ,
            \HilbertSpace{H}
          }
      }"
    ]
    \\[-2pt]
    &
    \scalebox{\termscale}{$
    \big(
      w \mapsto
      \vert \psi_w\rangle
    \big)
    \otimes
    \big(
      w \mapsto
      \vert \psi'_w\rangle
    \big)
    $}
    &\scalebox{\termscale}{$\mapsto$}&
    \scalebox{\termscale}{$
    \big(
      w \mapsto
      \vert \psi'_w\rangle
    \big)
    \otimes
    \big(
      w \mapsto
      \vert \psi_w\rangle
    \big)
    $}
    \\
    &
    \scalebox{\termscale}{\rotatebox[origin=c]{-90}{$\mapsto$}}
    &&
    \scalebox{\termscale}{\rotatebox[origin=c]{-90}{$\mapsto$}}
    \\
    &
    \scalebox{\termscale}{$
    \big(
      w \mapsto
      \vert \psi_w \rangle
      \otimes
      \vert \psi'_w \rangle
    \big)
    $}
    &\scalebox{\termscale}{$\mapsto$}&
    \scalebox{\termscale}{$
    \big(
      w \mapsto
      \vert \psi'_w \rangle
      \otimes
      \vert \psi_w \rangle
    \big)
    $}
    \\[-5pt]
    \underset{W}{\indefinitely}
    \big(
      \HilbertSpace{H}
      \otimes
      \HilbertSpace{H}'
    \big)
    \ar[
      rrrr,
      "{
        \indefinitely_W
        \big(
          \braid
          {\otimes}
          {
            \HilbertSpace{H}
            ,
            \HilbertSpace{H}'
          }
        \big)
      }"{swap}
    ]
    &&&&
    \underset{W}{\indefinitely}
    \big(
      \HilbertSpace{H}'
      \otimes
      \HilbertSpace{H}
    \big)
  \end{tikzcd}
  \end{array}
\end{equation}

\newpage

\begin{equation}  \label{SymmetricCoMonoidalCOMonadStructureOnRandomness}
  \def\arraystretch{2}

  \\[-15pt]
  \\
    \rotatebox[origin=c]{90}{\bf symmetric}
    &
  \begin{tikzcd}[
    column sep=0pt,
    row sep=0pt
  ]
    \big(
    \underset{W}{\randomly}
    \,
    \HilbertSpace{H}
    \big)
    \otimes
    \big(
    \underset{W}{\randomly}
    \HilbertSpace{H}'
    \big)
    \ar[
      from=rrrr,
      "{
        \braid
        {\otimes}
        {
          \randomly_W
          \HilbertSpace{H},
          \randomly_W
          \HilbertSpace{H}'
        }
      }"{swap}
    ]
    \ar[
      from=dddd,
      "{
        \copair
          { \randomly }
          {
            \HilbertSpace{H}
            ,
            \HilbertSpace{H}'
          }
      }"
    ]
    &[-30pt]&&&[-30pt]
    \big(
    \underset{W}{\randomly}
    \,
    \HilbertSpace{H}'
    \big)
    \otimes
    \big(
    \underset{W}{\randomly}
    \,
    \HilbertSpace{H}
    \big)
    \ar[
      from=dddd,
      "{
        \copair
          { \randomly }
          {
            \HilbertSpace{H}'
            ,
            \HilbertSpace{H}
          }
      }"{swap}
    ]
    \\[-5pt]
    &
    \scalebox{\termscale}{$
    \big(
      w,\,
      \vert \psi \rangle
    \big)
    \otimes
    \big(
      w,\,
      \vert \psi'\rangle
    \big)
    $}
    &\scalebox{\termscale}{$\mapsfrom$}&
    \scalebox{\termscale}{$
    \big(
      w,\,
      \vert \psi' \rangle
    \big)
    \otimes
    \big(
      w,\,
      \vert \psi \rangle
    \big)
    $}
    \\
    &
    \scalebox{\termscale}{\rotatebox[origin=c]{-90}{$\mapsfrom$}}
    &&
    \scalebox{\termscale}{\rotatebox[origin=c]{-90}{$\mapsfrom$}}
    \\
    &
    \scalebox{\termscale}{$
    \big(
      w,\,
      \vert \psi \rangle
      \otimes
      \vert \psi' \rangle
    \big)
    $}
    &\scalebox{\termscale}{$\mapsfrom$}&
    \scalebox{\termscale}{$
    \big(
      w,\,
      \vert \psi' \rangle
      \otimes
      \vert \psi \rangle
    \big)
    $}
    \\[-5pt]
    \underset{W}{\randomly}
    \big(
      \HilbertSpace{H}
      \otimes
      \HilbertSpace{H}'
    \big)
    \ar[
      from=rrrr,
      "{
        \randomly_W
        \big(
          \braid
          {\otimes}
          {
            \HilbertSpace{H}
            ,
            \HilbertSpace{H}'
          }
        \big)
      }"
    ]
    &&&&
    \underset{W}{\randomly}
    \big(
      \HilbertSpace{H}'
      \otimes
      \HilbertSpace{H}
    \big)
  \end{tikzcd}
  \end{array}
\end{equation}

\newpage

In outlook to the discussion of mixed quantum states in \cref{MixedQuantumTypes} we close this section on quantum epistemology by observing that indefiniteness- and randomness-effects lift from pure to mixed quantum states via the above (co)monoidal (co)monad structure, via the monoidal monad structure
$\pair{ \indefiniteness_W }{}$
\eqref{SymmetricMonoidalMonadStructureOnIndefiniteness} on the indefinite modality and the comonoidal comonad structure $\copair{\randomly_W }{}$ \eqref{SymmetricCoMonoidalCOMonadStructureOnRandomness} on the random modality.

\smallskip

\noindent
{\bf Indefinite mixed states.}
A quantum system with pure state space $\HilbertSpace{H} \,\isa\, \DualizableQuantumTypes$ a dualizable \eqref{CoEvaluationForDualObjects} quantum type
generally has {\it mixed} states \eqref{DensityMatrixInIntroduction} in $\HilbertSpace{H} \otimes \HilbertSpace{H}^\ast \,\isa\, \QuantumType$, such that a quantum gate on pure states induces a {\it quantum channel} on mixed states, of the form
\begin{equation}
  \label{TensoringLinearMapsWithTheirAdjointDuals}
 A
 \,\isa\,
 \HilbertSpace{H}_1
 \to
 \HilbertSpace{H}_2
 \hspace{.7cm}
 \yields
 \hspace{.7cm}
 \mathrm{chan}^A
    \,\isa\,
  \begin{tikzcd}[column sep=large]
    \def\arraystretch{.9}
    \begin{array}{c}
      \HilbertSpace{H}_1
      \\
      \otimes
      \\
      \HilbertSpace{H}^\ast_1
    \end{array}
    \ar[
      rr,
      "{
        \def\arraystretch{.9}
        \begin{array}{c}
          f
          \\
          \otimes
          \\
          {f^\dagger}^\ast
        \end{array}
      }"{description}
    ]
    &&
    \begin{array}{c}
      \HilbertSpace{H}_2
      \\
      \otimes
      \\
      \HilbertSpace{H}^\ast_2
      \mathrlap{\,,}
    \end{array}
  \end{tikzcd}
\end{equation}
(for the moment the  dagger-$(-)^\dagger$ operation may be treated as a black box, we discuss this in \ifdefined\monadology\cite{QS}\else\cref{QuantumProbability}\fi).

\smallskip

\begin{lemma}[\bf Enhancing indefiniteness-effects to Mixed states]
\label{EnhancingIndefinitenessEffectsToMixedStates}
The assignment which sends an $\indefinitely_W$-effectful map to its tensor product with its adjoint dual \eqref{TensoringLinearMapsWithTheirAdjointDuals} followed by the $\indefiniteness_W$-pairing \eqref{SymmetricMonoidalMonadStructureOnIndefiniteness}
\begin{equation}
  \label{TensorPairingOfIndefinitenessEffectfulMaps}
  \begin{tikzcd}
    \HilbertSpace{H}_1
    \ar[
      rr,
      "{
        G_\bullet
      }"
    ]
    &
      {}
      \ar[
        d,
        phantom,
        "{\longmapsto}"{
          sloped,
          pos=.3
        }
      ]
    &
    \underset{W}{\indefinitely}
    \,
    \HilbertSpace{H}_2
    \\
    \def\arraystretch{.9}
    \begin{array}{c}
      \HilbertSpace{H}_1
      \\
      \otimes
      \\
      \HilbertSpace{H}^\ast_1
    \end{array}
    \ar[
      rr,
      "{
        G_\bullet
      }"{yshift=2.5pt},
      "{
        \otimes
      }"{description},
      "{
        {G_\bullet^\dagger}^\ast
      }"{swap, yshift=-2.5pt}
    ]
    &{}&
    \def\arraystretch{.9}
    \begin{array}{c}
      \underset{W}{\indefinitely}
      \,
      \HilbertSpace{H}_2
      \\
      \otimes
      \\
      \underset{W}{\indefinitely}
      \,
      \HilbertSpace{H}^\ast_2
    \end{array}
    \ar[
      rr,
      "{
        \pair
          { \indefinitely_W }
          {}
      }"
    ]
    &&
    \underset{W}{\indefinitely}
    \def\arraystretch{.9}
    \begin{array}{c}
      \HilbertSpace{H}_2
      \\
      \otimes
      \\
      \HilbertSpace{H}^\ast_2
    \end{array}
  \end{tikzcd}
\end{equation}
preserves $\indefinitely_W$-Kleisli-composition \eqref{KleisliComposition}, in that:
\begin{equation}
  \label{PairedKleisliComposition}
  \big(
    \pair
      { \indefinitely_W }
      {
        \HilbertSpace{H}_2
        ,\,
        \HilbertSpace{H}_2^\ast
      }
    \circ
    (
    G_\bullet
    \otimes
    {G_\bullet^\dagger}^\ast
    )
  \big)
    \;\mbox{\tt >=>}\;
  \big(
    \pair
      { \indefinitely_W }
      {
        \HilbertSpace{H}_3
        ,\,
        \HilbertSpace{H}_3^\ast
      }
    \circ
    (
    H_\bullet
    \otimes
    {H_\bullet^\dagger}^\ast
    )
  \big)
  \;\;\;\;
  =
  \;\;\;\;
  \pair
    { \indefinitely_W }
    {
      \HilbertSpace{H}_3
      ,\,
      \HilbertSpace{H}_3^\ast
    }
  \circ
  \Big(
  \big(
    G_\bullet
    \;
    \mbox{\tt >=>}
    \;
    H_\bullet
  \big)
  \otimes
  \big(
    {G_\bullet^\dagger}^\ast
    \;
    \mbox{\tt >=>}
    \;
    {H_\bullet^\dagger}^\ast
  \big)
  \Big)
\end{equation}
and hence defines a faithful endofunctor on the free $\indefinitely_W$-modales \eqref{CategoriesOfModales}
\begin{equation}
  \label{EnhancingControlledGatesToControlledChannelsAsFaithfulFunctor}
  \pair
    { \indefinitely_W }
    {  }
  \circ
  (-)_\bullet\otimes {(-)_\bullet^{\dagger}}^\ast
  \;\isa\;
  \begin{tikzcd}
    \QuantumTypes_{\indefinitely_W}
    \ar[
      rr
    ]
    &&
    \QuantumTypes_{\indefinitely_W}
  \end{tikzcd}
\end{equation}
\end{lemma}
\begin{proof}
This is an argument analogous to that for monad transformations \eqref{MonadTransformationRespectsKleisliComposition}.
Consider the following diagram:
\begin{equation}
\hspace{-4mm}
  \begin{tikzcd}[
    column sep=20pt
  ]
    \HilbertSpace{H}_1
    \otimes
    \HilbertSpace{H}_1^\ast
    \ar[
      rr,
      "{
        G_\bullet
          \otimes
        {G_\bullet^\dagger}^\ast
      }"
    ]
    &[-4pt]&[-4pt]
    \big(
    \underset{W}{\indefinitely}
    \,
    \HilbertSpace{H}_2
    \big)
    \otimes
    \big(
    \underset{W}{\indefinitely}
    \,
    \HilbertSpace{H}_2^\ast
    \big)
    \ar[
      rr,
      "{
       \scalebox{0.8}{$   \big(
        \indefinitely_W
          H_\bullet
        \big)
        \otimes
        \big(
        \indefinitely_W
          {H_\bullet^{\dagger}}^\ast
        \big)
        $}
      }"{yshift=1pt}
    ]
    \ar[
      dd,
      "{
        \pair
          { \indefinitely_W }
          {
            \HilbertSpace{H},
            \,
            \HilbertSpace{H}^\ast
          }
      }"{description}
    ]
    &\phantom{--}&
    \big(
    \underset{W}{\indefinitely}
    \,
    \underset{W}{\indefinitely}
    \,
    \HilbertSpace{H}_3
    \big)
    \otimes
    \big(
    \underset{W}{\indefinitely}
    \,
    \underset{W}{\indefinitely}
    \,
    \HilbertSpace{H}_3^\ast
    \big)
    \ar[
      rr,
      "{
        \join
          {
            \indefinitely_W
          }
          { \HilbertSpace{H} }
        \otimes
        \join
          {
            \indefinitely_W
          }
          { \HilbertSpace{H}^\ast }
      }"
    ]
    \ar[
      dd,
      "{
        \pair
          { \indefinitely_W }
          {
            \indefinitely_W
            \HilbertSpace{H}_3,
            \,
            \indefinitely_W
            \HilbertSpace{H}_3^\ast
          }
      }"{description}
    ]
    &\phantom{-}&
    \big(
      \underset{W}{\indefinitely}
      \,
      \HilbertSpace{H}
    \big)
    \otimes
    \big(
      \underset{W}{\indefinitely}
      \,
      \HilbertSpace{H}^\ast
    \big)
    \ar[
      dddd,
      "{
        \pair
          { \indefinitely_W }
          {
            \HilbertSpace{H}_3
            ,\,
            \HilbertSpace{H}_3^\ast
          }
      }"{description}
    ]
    \\
    \\
    &&
    \HilbertSpace{H}_2
    \otimes
    \HilbertSpace{H}_2^\ast
    \ar[
      rr,
      "{
        \indefinitely_W
        \big(
        H_\bullet
        \otimes
        \,
        {H_\bullet^\dagger}^\ast
        \big)
      }"
    ]
    &&
    \underset{W}{\indefinitely}
    \Big(
      \big(
      \underset{W}{\indefinitely}
      \,
      \HilbertSpace{H}_3
      \big)
      \otimes
      \big(
      \underset{W}{\indefinitely}
      \,
      \HilbertSpace{H}_3^\ast
      \big)
    \Big)
    \ar[
      dd,
      "{
        \underset{W}{\indefinitely}
        \Big(
          \pair
            { \indefinitely_W }
            {
              \HilbertSpace{H}_3
              ,\,
              \HilbertSpace{H}_3^\ast
            }
        \Big)
      }"{description}
    ]
    \\
    \\
    &&
    &&
    \underset{W}{\indefinitely}
    \,
    \underset{W}{\indefinitely}
    \big(
      \HilbertSpace{H}_3
      \otimes
      \HilbertSpace{H}_3^\ast
    \big)
    \ar[
      rr,
      "{
        \join
          { \indefinitely_W }
          {
            \HilbertSpace{H}_3
            ,\,
            \HilbertSpace{H}^\ast_3
          }
      }"
    ]
    &&
    \underset{W}{\indefinitely}
    \big(
      \HilbertSpace{H}_3
      \otimes
      \HilbertSpace{H}_3^\ast
    \big)
  \end{tikzcd}
\end{equation}
Here the middle square commutes by the naturality of the pairing map, while the right square commutes as part of the monoidal monad structure \eqref{SymmetricMonoidalMonadStructureOnIndefiniteness}
exhibited by the pairing. Therefore the full diagram commutes. Since its total top and right composite is the right hand side of \eqref{PairedKleisliComposition} while its total left and bottom (diagonal) composite is the left hand side of \eqref{PairedKleisliComposition}, this proves the claim.
\end{proof}

\subsection{Quantum Gates \& Measurement}
\label{ControlledQuantumGates}

We explain how {\it controlled quantum gates} and {\it quantum measurement gates} (Lit. \ref{LiteratureQuantumComputation})
are naturally represented in the quantum modal logic of \cref{QuantumEpistemicLogicViaDependentLinearTypes} and give (Prop. \ref{DeferredMeasurementPrinciple}) a formal proof of the {\it deferred measurement principle} \eqref{InformalDeferredMeasurementPrinciple}.

\medskip

\noindent
{\bf Data-typing of controlled quantum gates via quantum modal types.}

\hspace{-.8cm}

\end{center}

Here the ``epistemic''-typing of controlled quantum gates shown in the middle row is manifest: For classical control the quantum gate is a $W$-dependent linear map, while for quantum
\normalfont
control it is a genuine linear map on the $W$-indexed direct sum.
The equivalent \eqref{EpistemicEffectivePerspective} ``effective'' typing in the top line of the bottom row follows by monadicity of $\osum_{{}_W}$
(see Prop. \ref{ActualQuantumDataIsInfinitenessModalPotentialData}).
The very last line shows the corresponding Kleisli-triple formulation of ``programs with side effects'' \eqref{BindingAndReturning}. On  the left this requires assuming that the dependent linear type is constant, $\HilbertSpace{H}_\bullet = \HilbertSpace{H}$ (which typically is the case in practice, see the example on p. \pageref{ExampleOfCNOTGateTyping}) since that makes it correspond to a free $\indefiniteness$-modale. On the right we see the effectless operation \eqref{MonadFunctorFromBindReturn}.

\hspace{-.8cm}
\begin{tabular}{cc}
\label{QuantumMeasurementCopenhagenStyle}
\begin{minipage}{8.3cm}
  {\bf Quantum measurement -- Copenhagen-style.}
  Last but not least, we obtain this way a natural typing of the otherwise subtle case of
  quantum measurement gates: These are now given simply by the $\necessity$-counit and, equivalently, by the $\indefiniteness$-join (cf. Prop. \ref{ComponentExpressionsOfTheCoMonadCoUnits}), as shown on the right.

  \medskip

  Via the language of effectful computation (Lit. \ref{LiteratureComputationalEffectsAndModalities}) and with the ``reader-monad'' $\indefinitely$ modally pronounced as ``indefiniteness'' \eqref{PotentialityEpistemicEntailments},
  this translates to the pleasant statement that:

  \medskip

  ``For effectively-typed quantum data, quantum measurement is nothing but the {\it handling of indefiniteness-effects}'' (regarded as modale homorphisms via \eqref{ModaleActionsAreModaleHomomorphisms}).
  \medskip

  In more detail:

  ``Before measurement, quantum data is indefinite(-effectful),
  while quantum measurement actualizes the data by handling of its indefiniteness(-effect)''

  \medskip

  This way the puzzlement of the ``state collapse'' \eqref{LinearStateCollapse} is resolved into an appropriate quantum effect language equivalent
  \eqref{EpistemicEffectivePerspective}
  to quantum modal logic.
\end{minipage}

&

\def\arraystretch{2}
\begin{tabular}{|c||r|}
  \hline
  \multicolumn{2}{|c|}{
  {\bf Quantum measurement gate}
  }
  \\
  \hline
  \hline
  \rotatebox[origin=c]{90}{
    {
      \scalebox{.9}{
        \bf
        Symbolic
      }
    }
  }
  &
    \adjustbox{raise=-20pt}{
    \begin{tikzpicture}
  \begin{scope}
      \draw (+.7, +0pt) node {
        \scalebox{.9}{$\quantized W$}
      };
      \draw (+5.3, +0pt) node {
        \scalebox{.9}{$W$}
      };
      \draw (+.75, -1) node {
        \scalebox{.9}{$\HilbertSpace{H}$}
      };
      \draw (+5.3, -1) node {
        \scalebox{.9}{$\HilbertSpace{H}$}
      };
  \draw[
    line width=1.5pt,
  ]
  (1.1,0)
  to
  (3,0);
  \draw[
    line width=.6pt
  ]
  (3,.04)
  to
  (+5,.04);
  \draw[
    line width=.6pt
  ]
  (3,-.04)
  to
  (+5,-.04);
  \drawMeasurement{3}{0}
  \end{scope}
    \draw[
      line width=1.5pt
    ]
    (1,-1)
    to
    (5,-1);
\end{tikzpicture}
}
\\
\hline
  \rotatebox[origin=c]{90}{
    {
      \scalebox{.9}{
        \bf
        Epistemic
      }
    }
  }
&
\begin{tikzcd}[
  column sep=44pt,
  row sep=2pt
]
  \\[+10pt]
  &[-45pt]
  \underset{W}{\necessarily}
  \HilbertSpace{H}_\bullet
  \ar[
    rr,
    "{
      \counit
        {\underset{W}{\necessarily}}
        { \HilbertSpace{H}_\bullet }
    }",
    "{
      \scalebox{.7}{
        \color{gray}
        the necessary becomes actual
      }
    }"{swap}
  ]
  &&
  \HilbertSpace{H}_\bullet
  \\
  {\color{purple}w}
  \isa W
  \;\yields
  &
  \underset{w'}{\osum}
  \HilbertSpace{H}_{w'}
  \ar[
    rr,
    "{
      \mathrm{pr}_{\color{purple}w}
    }",
    "{
      \scalebox{.7}{
        \rm quantum
        \color{darkorange}
        \bf
        state collapse
      }
    }"{swap}
  ]
  &&
  \HilbertSpace{H}_{{\color{purple}w}}
  \\
  &
  \underset{w'}{\oplus}
  \,
  \vert \psi_{w'} \rangle
  \ar[rr, |->, shorten=40pt]
  &&
  \vert \psi_{{\color{purple}w}} \rangle
\end{tikzcd}
\\
\hline
  \rotatebox[origin=c]{90}{
    {
      \scalebox{.9}{
        \bf
        Effective
      }
    }
  }
&
\begin{tikzcd}[
  column sep=30pt,
  row sep=10pt
]
  \\[+12pt]
  \underset{W}{\osum}
  \,
  \underset{W}{\necessarily}
  \,
  \HilbertSpace{H}_\bullet
    \ar[out=180-65, in=65,
      looseness=2,
      shorten=-1pt,
      "{
        \hspace{6pt}
        \mathclap{
        \scalebox{1.1}{$
          \indefinitely_{{}_W}
        $}
        }
        \hspace{6pt}
      }"{pos=.5, description},
      start anchor={[xshift=-8pt]},
      end anchor={[xshift=+8pt]},
    ]
  \ar[
    rr,
    "{
      \underset{W}{\osum}
      \,
      \counit
        { \underset{W}{\necessarily} }
        { \HilbertSpace{H}_\bullet }
    }"
  ]
  \ar[
    dd, equals
  ]
  &&
  \underset{W}{\osum}
  \HilbertSpace{H}_{\bullet}
    \ar[out=180-65, in=65,
      looseness=2,
      shorten=-1pt,
      "{
        \hspace{6pt}
        \mathclap{
        \scalebox{1.1}{$
          \indefinitely_{{}_W}
        $}
        }
        \hspace{6pt}
      }"{pos=.5, description},
      start anchor={[xshift=-8pt]},
      end anchor={[xshift=+8pt]},
    ]
  \ar[
    dd, equals
  ]
  \\[-10pt]
  &
  \scalebox{.7}{
    \color{gray}
    \eqref{ModaleActionFromEMAdjunctionCOhunit}
  }
  \\
    \underset{W}{\indefinitely}
    \,
    \underset{W}{\osum}
    \,
    \HilbertSpace{H}_{\bullet}
    \ar[
      rr,
      "{
        \handle
          {\indefinitely_{{}_W}}
          { \oplus_{{}_W}\! \HilbertSpace{H}_\bullet }
          {}
      }",
      "{
        \scalebox{.7}{
          \color{darkorange}
          \bf
          \scalebox{1.2}{$\indefinitely_{{}_W}$}-effect
          handling
        }
      }"{swap}
    ]
    &&
    \underset{W}{\osum}
    \,
    \HilbertSpace{H}_{\bullet}
\end{tikzcd}
\\
\hline
\end{tabular}
\end{tabular}

\medskip

\smallskip

Before looking at examples (p. \pageref{ExampleOfCNOTGateTyping}), we record a basic structural result immediately implied by this typing, which may evidently be understood as formalizing the {\it deferred measurement principle} \eqref{InformalDeferredMeasurementPrinciple}, thus making this principle verifiable in {\LHoTT} as \cite{Staton15} envisioned should be the case for any respectable quantum programming language:

\noindent
\begin{proposition}[{\bf Deferred measurement principle}]
\label{DeferredMeasurementPrinciple}
With respect to the above typing of quantum gates,
the $\necessity$-Kleisli equivalence \eqref{KleisliEquivalence} is the following transformation of quantum circuits:
$$
\hspace{-6mm}
  \begin{tikzcd}[
    column sep=8pt,
    row sep=6pt
  ]
    \underset{
      \mathclap{
        \raisebox{-2pt}{
          \scalebox{.7}{
            \color{darkblue}
            \bf
            $\underset{W}{\necessarily}$-Kleisli morphisms
          }
        }
      }
    }{
     \FreeModales
       {\necessarily_{{}_W}}
       {(\QuantumTypes_{{}_W})}
       \!\!
      \big(
        \HilbertSpace{H}_\bullet
        ,\,
        \HilbertSpace{H}''_\bullet
      \big)
    }
    \ar[
      rr,
      "\sim",
      "{
        \underset{W}{\necessarily}(\mbox{-})
          \;\circ\;
        \comultiplication
          { \necessarily_{{}_W}}
          {(\mbox{-})}
      }"{swap}
    ]
    \ar[
      rrrr,
      rounded corners,
      to path ={
           ([yshift=+0pt]\tikztostart.north)
        -- ([yshift=+12pt]\tikztostart.north)
        -- node[yshift=6pt]{$\mathrm{id}$}
           node[yshift=-7pt]{
             \scalebox{.7}{
               \color{darkgreen}
               \bf
               Kleisli equivalence
             }
           }
           ([yshift=+12pt]\tikztotarget.north)
        -- ([yshift=+0pt]\tikztotarget.north)
      }
    ]
    &&
    \underset{
      \mathclap{
        \raisebox{-2pt}{
          \scalebox{.7}{
            \color{darkblue}
            \bf
            \def\arraystretch{.9}
            \begin{tabular}{c}
            homomorphisms of
            \\
            free
            $\underset{W}{\necessarily}$-coalgebras
            \end{tabular}
          }
        }
      }
    }{
     \QuantumTypes
       _{{}_W}
       ^{ \scalebox{.7}{$\necessarily_{{}_W}$} }
     \big(
       \underset{W}{\necessarily}
       \HilbertSpace{H}_\bullet
       ,\,
       \underset{W}{\necessarily}
       \HilbertSpace{H}''_\bullet
     \big)
   }
    \ar[
      rr,
      "\sim",
      "{
        \counit
           { \necessarily_{{}_W} }
           {(\mbox{-})}
         \,\circ\,
         (-)
      }"{swap}
    ]
    &&
    \underset{
      \mathclap{
        \raisebox{-2pt}{
          \scalebox{.7}{
            \color{darkblue}
            \bf
            $\underset{W}{\necessarily}$-Kleisli morphisms
          }
        }
      }
    }{
     \FreeModales
       {\necessarily_{{}_W}}
       {(\QuantumTypes_{{}_W})}
       \!\!
      \big(
        \HilbertSpace{H}_\bullet
        ,\,
        \HilbertSpace{H}''_\bullet
      \big)
    }
    \\[-10pt]
    \underset{
      \mathclap{
      \raisebox{-4pt}{
        \scalebox{.7}{
          \color{darkorange}
          \bf
          \def\arraystretch{.9}
          \begin{tabular}{c}
            measurement-controlled
            quantum gate
          \end{tabular}
        }
      }
      }
    }{\scalebox{0.8}{$
    \Big(
    \underset{W}{\necessarily}  \HilbertSpace{H}_\bullet
    \xrightarrow{F}
    \underset{W}{\necessarily}  \HilbertSpace{H}'_\bullet
    \xrightarrow{
      \scalebox{.7}{$
      \counit
        { {\necessarily_{{}_W}} }
        { \HilbertSpace{H}_\bullet }
      $}
    }
    \HilbertSpace{H}'_\bullet
        \xrightarrow{G_\bullet}
    \HilbertSpace{H}''_\bullet
    \Big)
    $}
    }
    &\mapsto&
    \underset{
      \mathclap{
      \raisebox{-3pt}{
        \scalebox{.7}{
          \color{darkorange}
           \bf
          quantum-controlled quantum gate\rlap{...}
        }
      }
      }
    }{\scalebox{0.8}{$
      \Big(
      \underset{W}{\necessarily} \HilbertSpace{H}_\bullet
      \xrightarrow{
        F
      }
      \underset{W}{\necessarily} \HilbertSpace{H}'_\bullet
      \xrightarrow{
        {\necessarily_{{}_W}}
        G_\bullet
      }
      \underset{W}{\necessarily} \HilbertSpace{H}''_\bullet
      \Big)
      $}
    }
    &\mapsto&
    \underset{
      \mathclap{
      \raisebox{-3pt}{
        \scalebox{.7}{
          \color{darkorange}
           \bf
          \llap{...}followed by measurement
        }
      }
      }
    }{\scalebox{0.8}{$
     \Big(
      \underset{W}{\necessarily} \HilbertSpace{H}_\bullet
      \xrightarrow{
        F
      }
      \underset{W}{\necessarily}
      \HilbertSpace{H}'_\bullet
      \xrightarrow{
        {\necessarily_{{}_W}}
        G_\bullet
      }
      \underset{W}{\necessarily} \HilbertSpace{H}''_\bullet
      \xrightarrow{
        \scalebox{.7}{$
          \counit
            { \necessarily_{{}_W} }
            { \HilbertSpace{H}_\bullet }
        $}
      }
      \HilbertSpace{H}''_\bullet
      \Big)
      $}
    }
    \\
    \adjustbox{raise=20pt}{

    }
    &&
    \mathclap{
      \xleftrightarrow{
         \scalebox{.7}{
           \color{darkgreen}
           \bf
           \;
           deferred measurement principle
           \;
         }
      }
    }
    &&
 \adjustbox{raise=20pt}{
      %
    }
  \end{tikzcd}
$$
\end{proposition}
\begin{proof}
  It just remains to  see that the Kleisli equivalence $\underset{W}{\necessarily}(\mbox{-}) \circ \comultiplication{\necessarily_{{}_W}}{(\mbox{-})}$ acts in the first step as claimed,
  hence that the following diagram commutes:
  \vspace{-2mm}
  $$
    \begin{tikzcd}[row sep=30pt, column sep=50pt]
      \underset{W}{\necessarily}
      \,
      \HilbertSpace{H}_\bullet
      \ar[
        d,
        "{
          \comultiplication
            { \necessarily_{{}_W} }
            { \HilbertSpace{H}_\bullet }
        }"{swap}
      ]
      \ar[
        r,
        "{ F }"
      ]
      &
      \underset{W}{\necessarily}
      \,
      \HilbertSpace{H}'_\bullet
      \ar[
        d,
        "{
          \comultiplication
            { \necessarily_{{}_W} }
            { \HilbertSpace{H}'_\bullet }
        }"{swap}
      ]
      \ar[
        dr,
        equals
      ]
      \\
      \underset{W}{\necessarily}
      \,
      \underset{W}{\necessarily}
      \,
      \HilbertSpace{H}_\bullet
      \ar[
        r,
        "{
          \necessarily_{{}_W}
          F
        }"{swap}
      ]
      &
      \underset{W}{\necessarily}
      \,
      \underset{W}{\necessarily}
      \,
      \HilbertSpace{H}'_\bullet
      \ar[
       r,
       "{
          \underset{W}{\necessarily}
          \big(
          \counit
            { \necessarily_{{}_W} }
            { \HilbertSpace{H}'_\bullet }
          \big)
       }"{swap}
      ]
      &
      \underset{W}{\necessarily}
      \,
      \HilbertSpace{H}'_\bullet
      \ar[
        r,
        "{
          \necessarily_{{}_W}
          G_\bullet
        }"{swap}
      ]
      &
      \underset{W}{\necessarily}
      \,
      \HilbertSpace{H}''_\bullet
    \end{tikzcd}
  $$

  \vspace{-2mm}
\noindent  But the square commutes since the gate $F$ is independent of the measurement result $w \isa W$ and hence is a homomorphism of free $\necessity$-coalgebras (by Rem. \ref{HomomorphismsOfFreeNecessityModales}), while the triangle commutes by the comonad axioms \eqref{MonadAxioms}.
\end{proof}

\medskip

\noindent

\label{ExampleOfCNOTGateTyping}
\hspace{-.8cm}


\noindent
For the record, we also spell out the two possible combinations of the above CNOT- and $\QBit$-measurement gates:

\vspace{.1cm}

\hspace{-.955cm}
\adjustbox{
  scale=.94
}{
\def\tabcolsep{1pt}
\def\arraystretch{2}
%
}
\vspace{.2cm}

\noindent
Notice here how the component expressions on the left and right agree, in accord with the {\it deferred measurement principle} (Prop. \ref{DeferredMeasurementPrinciple}). In components this is an elementary triviality, but the point is that by making this triviality follow from typing rules it becomes machine-verifiable also in more complex cases.

\medskip

\noindent
{\bf qRAM.} As a byproduct of the modal typing of controlled quantum gates, we may notice a formal reflection of the idea of {\it circuit models for qRAM} \eqref{QuantumProgramInteractingWithQRAm}. Namely if, with \eqref{StateMonadAsRAMModel}, we recall that RAM-effects are typed by the state monad $\underset{W}{\indefinitely}\,\underset{W}{\randomly}$ \eqref{RandomAccessAsIndefiniteRandomness} --- which immediately makes sense linearly just as it does classically---, then quantumly controlled quantum circuits in the above sense (p. \pageref{ControlledQuantumGatesTyping}) are formally identified with QRAM-effective quantum programs as follows, where the first transformation is for effectless programs \eqref{MonadFunctorFromBindReturn} while the second is $\randomly_{{}_W} \dashv \indefinitely_{{}_W}$-adjointness \eqref{FormingAdjuncts}:
\begin{equation}
  \label{QRAMAdjointness}
  \hspace{-4cm}
  \adjustbox{fbox}{
    \begin{minipage}{6cm}
      \footnotesize
      The passage to circuit models for
      qRAM \eqref{QuantumProgramInteractingWithQRAm} may formally be understood as the modal adjointness  between
      \begin{itemize}
        \item[(i)] QRAM-effective quantum programs $\HilbertSpace{H} \longmapsto \underset{W}{\indefinitely}\,\underset{W}{\randomly}\, \HilbertSpace{K}$
        \item[(ii)]
        quantumly controlled quantum circuits

        $\underset{W}{\osum}\HilbertSpace{H} \longmapsto \underset{W}{\osum}\HilbertSpace{K}$
      \end{itemize}
    \end{minipage}
  }
  \hspace{1cm}
  \adjustbox{}{
  \begin{tikzcd}[
    column sep=30pt,
    row sep=0pt
  ]
    \underset{W}{\indefinitely}
    \,
    \underset{W}{\osum}
    \HilbertSpace{H}
    \ar[out=180-65, in=65,
      looseness=2,
      shorten=-1pt,
      "{
        \hspace{6pt}
        \mathclap{
        \scalebox{1.1}{$
          \indefinitely_{{}_W}
        $}
        }
        \hspace{6pt}
      }"{pos=.5, description},
      start anchor={[xshift=-8pt]},
      end anchor={[xshift=+8pt]},
    ]
    \ar[
      rr,
      "{
        \underset{W}{\indefinitely}
        \,
        \underset{W}{\osum}
        \,
        G_\bullet
      }"{description, yshift=-2pt}
    ]
    &&
    \underset{W}{\indefinitely}
    \,
    \underset{W}{\osum}
    \HilbertSpace{K}
    \mathrlap{
      \hspace{.3cm}
      \scalebox{.7}{
        \def\arraystretch{.9}
        \begin{tabular}{c}
        \color{darkblue}
        \bf
        $\quantized W$-controlled
        \\
        {
        \color{darkblue}
        \bf
        quantum gate}
        (p. \pageref{ControlledQuantumGatesTyping})
        \end{tabular}
      }
    }
    \ar[out=180-65, in=65,
      looseness=2,
      shorten=-1pt,
      "{
        \hspace{6pt}
        \mathclap{
        \scalebox{1.1}{$
          \indefinitely_{{}_W}
        $}
        }
        \hspace{6pt}
      }"{pos=.5, description},
      start anchor={[xshift=-8pt]},
      end anchor={[xshift=+8pt]},
    ]
    \\
    \cline{1-6}
    \underset{W}{\osum}
    \HilbertSpace{H}
    \ar[d, shorten <=-2pt, equals]
    \ar[
      rr,
      "{
        \underset{W}{\osum}
        G_\bullet
      }"{description, yshift=-2pt}
    ]
    &&
    \underset{W}{\osum}
    \,
    \HilbertSpace{K}
    \ar[d, shorten <=-2pt, equals]
    \\[+4pt]
    \underset{W}{\randomly}
    \HilbertSpace{H}
    \ar[
      rr,
      "{
        \underset{W}{\osum}
        G_\bullet
      }"{description, yshift=-2pt}
    ]
    &&
    \underset{W}{\randomly}
    \,
    \HilbertSpace{K}
    \\
    \cline{1-6}
    \HilbertSpace{H}
    \ar[
      rr,
      "{
        \widetilde
        {
          \underset{W}{\osum}
          G_\bullet
        }
      }"{description, yshift=-3pt}
    ]
    &&
    \underset{W}{\indefinitely}
    \,
    \underset{W}{\randomly}
    \,
    \HilbertSpace{K}
    \mathrlap{
      \hspace{.3cm}
      \scalebox{.7}{
        \color{darkblue}
        \bf
        \def\arraystretch{.9}
        \begin{tabular}{c}
          quantum circuit interacting
          \\
          with a QRAM space $\quantized W$
        \end{tabular}
      }
    }
  \end{tikzcd}
  }
\end{equation}
At the same time, this QuantumState-monad
$$
  \quantized W\State
  \;\;
    \simeq
  \;\;
  \underset{W}{\indefinitely}
  \,
  \underset{W}{\randomly}
$$
reflects mixed $\quantized W$-states, discussed in \cref{MixedQuantumTypes}.

\newpage

\noindent
{\bf Quantum contexts.} The formal dual of the previous discussion of quantum measurement realized as a monadic computational effect yields {\it quantum state preparation} realized as a {\it comonadic computational context} \eqref{CoBindingAndCoReturning}:
Shown on the left below is the modal typing of {\it quantum state preparation} in the generality of classical control, namely quantum state preparation conditioned on a classical parameter $w \isa W$.
In the practice of quantum circuits, this typically applies to quantum types of the form $\underset{W}{\mathbbm{1}}$ in which case the traditional notion of state preparation is manifest: In world $w$ the result of the preparation is the quantum state $\vert w \rangle$. This is shown for the example of $\QBit$-preparation on the right:

\begin{center}

}

\end{tabular}
\end{center}

\label{QuantumMeasurementEverettStyle}
\noindent
{\bf Quantum measurement -- Everett style.}
But we may observe that quantum state preparation in the above classically-controlled generality can itself be used to model quantum
measurement, namely as the {\it preparation of the collapsed state conditioned on the classical measurement outcome}!

This is seen from the last line of the co-effective typing above, which we recognize as the branching perspective on quantum measurement -- if
only we disregard the $\randomly_{{}_W}$-modale homomorphism property of this map -- which formally corresponds to pulling this map back up
by applying $(\mbox{-}) \otimes \mathbbm{1}_{{}_W}$.
This yields the following purple map and hence the {\it Everett-style} typing of quantum measurement mentioned in the introduction \eqref{BranchingAndCollapseInIntroduction} --- which is related to the above Copenhagen-style typing (from p. \pageref{QuantumMeasurementCopenhagenStyle})
by the {\it hexagon of epistemic entailments} \eqref{QuantumEpistemicLogicViaDependentLinearTypes}:
\begin{equation}
  \label{EverettStyleTypingOfMeasurement}
  \hspace{-3mm}
  \begin{tikzcd}[
    column sep=32pt,
    row sep=1pt
  ]
    &[-28pt]
    \underset{W}{\necessarily}
    \HilbertSpace{H}_\bullet
    \ar[
      rr,
      "{
        \underset{W}{\necessarily}
        G_\bullet
      }"
    ]
    &&
    \underset{W}{\necessarily}
    \HilbertSpace{H}_\bullet
    \ar[
      rr,
      "{
        \underset{W}{\necessarily}
        \unit
          { \possibly_{{}_W} }
          { \HilbertSpace{H}_\bullet }
      }",
      "{
        \scalebox{.7}{
          \color{purple}
          \bf
          \def\arraystretch{.9}

        }
      }"{swap}
    ]
    &&
    \underset{W}{\randomly}
    \underset{W}{\osum}
    \HilbertSpace{H}_\bullet
    \ar[
      d,
      phantom,
      "{\defneq}"{rotate=90}
    ]
    \ar[out=180-65, in=65,
      looseness=2,
      shorten=-1pt,
      "{
        \hspace{6pt}
        \mathclap{
        \scalebox{1.1}{$
          \randomly_{{}_W}
        $}
        }
        \hspace{6pt}
      }"{pos=.5, description},
      start anchor={[xshift=-8pt]},
      end anchor={[xshift=+8pt]},
    ]
    \\[+5pt]
    &
    \HilbertSpace{H}
    \ar[
      rr,
      "{ G }"{description}
    ]
    &&
    \HilbertSpace{H}
    \ar[
      rr,
      shorten >=-4pt,
      start anchor={[yshift=4pt]},
      end anchor={[yshift=23pt]},
      "{
        P_{ 1 }
      }"{description}
    ]
    \ar[
      rr,
      phantom,
      "{
        \vdots
        \mathrlap{
          \;\;
          \adjustbox{
            scale=.7,
            raise=2pt
          }{
            \color{purple}
            \bf
            branching
          }
        }
      }"{yshift=4pt}
    ]
    \ar[
      rr,
      shorten >=-4pt,
      start anchor={[yshift=-4pt]},
      end anchor={[yshift=-20pt]},
      "{
        P_{ |W| }
      }"{description}
    ]
    &&
    \def\arraystretch{.9}
    \begin{array}{c}
      \HilbertSpace{H}
      \\
      \oplus
      \\[-4pt]
      \vdots
      \\
      \oplus
      \\
      \HilbertSpace{H}
    \end{array}
    \ar[
      rr,
      shift left=24pt,
      "{ G }"{description}
    ]
    \ar[
      rr,
      phantom,
      shift left=16pt,
      "{ \oplus }"{scale=.9}
    ]
    \ar[
      rr,
      phantom,
      "{ \vdots }"
    ]
    \ar[
      rr,
      phantom,
      shift right=16pt,
      "{ \oplus }"{scale=.9}
    ]
    \ar[
      rr,
      shift right=24pt,
      "{ G }"{description}
    ]
    &&
    \def\arraystretch{.9}
    \begin{array}{c}
      \HilbertSpace{H}
      \\
      \oplus
      \\[-4pt]
      \vdots
      \\
      \oplus
      \\
      \HilbertSpace{H}
    \end{array}
    \\
    &
    &&
    \underset{w'}{\sum}
    \,
    \vert \psi_{w'} \rangle
    &\mapsto&
    \underset{w''}{\osum}
    \,
    \vert \psi_{w''} \rangle
    &
    \mapsto
    &
    \underset{w''}{\osum}
    \,
    G_{w''} \vert \psi_{w''} \rangle
  \end{tikzcd}
\end{equation}

\begin{remark}[{\bf No classical control appears in Everett-typing}]
$\,$

 \noindent {\bf (i)} Comparing the epistemic hexagon \eqref{BranchingAndCollapseInIntroduction}, we find that
  where the Copenhagen-style typing
  sees a classically-controlled quantum gate (cf. p. \pageref{ControlledQuantumGatesTyping}) the Everett-style typing
  \eqref{EverettStyleTypingOfMeasurement}
  sees (no classical control) but the corresponding quantumly-controlled quantum gate --- but applied in each of several ``branches''.

  \noindent {\bf (ii)} This primacy of the non-classical quantum perspective and the disregard for the need for any classical
  contexts is what Everett amplified when speaking of the ``universality'' of the quantum state (this being the very title of his
  thesis \cite{Everett75a}). The modal typing of quantum processes in \eqref{EverettStyleTypingOfMeasurement} provides a formalization
  of this intuition in a precise and machine-verifiable form.
\end{remark}

\begin{remark}[{\bf Everett-style measurement typing in the literature}]
\label{EverettStyleMeasurementTypingInTheLiterature}
$\,$

\noindent {\bf (i)} Essentially the typing-by-branching of quantum measurement in the bottom of \eqref{EverettStyleTypingOfMeasurement} may be recognized
in the early proposal for quantum programming language syntax in \cite[p. 568]{Selinger04}.

\noindent {\bf (ii)} The observation (apparently independently of \cite{Selinger04}) that this may usefully be understood as the $\provide{}{}{}$-operation of modales (coalgebras) over the comonad $\randomly{{}_W} \simeq \quantized W \otimes (\mbox{-})$
(Prop. \ref{QuantumCoEffectsViaFrobeniusAlgebra})
is due to \cite[Thm. 1.5]{CoeckePavlovic08} (cf. \cite[pp. 28]{CPP09}) --- this being the origin of the Frobenius-monadic formalization of ``classical structures'' in the {\zxCalculus}
(Rem. \ref{FrobeniusPropertyAndSpiderTheorem}).

\noindent {\bf (iii) } While  --- in formulating the quantum language {\QS} below in \cref{Pseudocode} --- we focus on language constructs for the Copenhagen-style typing (since this brings out the desired {\it dynamic lifting} of quantum-to-classical control, Lit. \ref{ClassicalControlAndDynamicLifting}), the situation \eqref{EverettStyleTypingOfMeasurement} shows that and how the ambient {\LHoTT} language may in principle also be used to verify protocols in Everett-style formalisms such as the {\zxCalculus}.
\end{remark}

\medskip

\noindent
{\bf Computational quantum measurement as entering the Indefiniteness-monad.} In summary, we have seen that coherent quantum gates are naturally typed as {\it free} indefinite-effectful linear maps, with quantum measurement given by the handling of the free indefiniteness-effect. Computationally this means equivalently that coherent quantum gates are equivalently the plain linear maps that one expects them to be, with quantum measurement being the step of ``entering the indefiniteness''-monad, in the sense of the commutativity of the following diagram:

\hspace{-.8cm}
\def\tabcolsep{4pt}
\begin{tabular}{cc}
\begin{minipage}{8.5cm}
  Computationally,
  the $\indefiniteness$-effective typing of quantum gates with quantum measurement amounts to regarding the map
  \begin{equation}
    \label{ComputationalMeasurementTyping}
    \hspace{.8cm}
    \begin{tikzcd}[
      sep = 0pt
    ]
      \mathllap{
        \collapse_W
        \;\isa\;
      }
      \quantized W
      \otimes
      \HilbertSpace{H}
      \ar[
        rr
      ]
      &&
      \underset{W}{\indefinitely}
      \,
      \HilbertSpace{H}
      \\
      \vert w \rangle
      \otimes
      \vert \psi \rangle
      &\mapsto&
      \big(
        w,\,
        \vert \psi \rangle
      \big)
    \end{tikzcd}
  \end{equation}
  (whose underlying function is the identity, up to re-typing)
  as passing into (the category of free modales over) the $\indefiniteness_W$-monad, as shown by the commuting diagram on the right.

  It is this final computational typing of quantum measurement which neatly lends itself to programming language-articulation in \cref{Pseudocode}, see p. \pageref{CodingQuantumMeasurement}.

  \smallskip

  Notice that while the epistemic, effective and computational perspectives are all equivalent, they superficially express a different ontology of the measurement collapse:

  \smallskip

  In the epistemic and effective perspective the {\it eventual} measurement in the $W$-basis is declared (possibly long) before that measurement takes places: In this perspective all possible future measurement outcomes are pre-emptively allocated in classical data.
\end{minipage}
&
\adjustbox{
  raise=+20pt
}{
$
  \begin{tikzcd}[sep=30pt]
    &[-27pt]
    &&
    \mathclap{
    \hspace{-20pt}
    \adjustbox{
      margin=4pt,
      fbox
    }{
      \bf
      Quantum Gate
      followed by
      Measurement
    }
    }
    \\[-33pt]
    \rotatebox[origin=c]{90}{\bf symbolic}
    &
    &&
    \hspace{-7pt}
    \mathclap{
      \adjustbox{}{
        \begin{tikzpicture}
          \draw[
            line width=1.8pt
          ]
            (-3.5,0)
            to
            (2,0);
          \draw[
            line width=.2pt
          ]
            (2,+.04)
            to
            (3.5,+.04);
          \draw[
            line width=.2pt
          ]
            (2,-.04)
            to
            (3.5,-.04);
          \drawMeasurement
            {1.9}{0}
        \node[
          fill=white
        ]
        at (-3.2,-.1)
        { \scalebox{1}{$
          \HilbertSpace{H}
        $} };
        \node[
          fill=white
        ]
        at (.2,-.1)
        { \scalebox{1}{$
          \quantized W
        $} };
        \node[
          fill=white
        ]
        at (3.5,-.1)
        { \scalebox{1}{$
          W
        $} };
        \node[
          draw=black,
          fill=white
        ]
        at (-1.4,-.1)
        { \adjustbox{
          margin=3pt
        }{$
          G
        $} };
        \end{tikzpicture}
      }
    }
    \hspace{+7pt}
    \\[-23pt]
    \rotatebox[origin=c]{90}{\bf epistemic}
    &
    \quantized V
    \!\otimes\!
    \underset{W}{\TensorUnit}
    \ar[
      rr,
      "{
        G
        \otimes\!
        \underset{W}{\TensorUnit}
      }"{description, yshift=-2pt}
    ]
    \ar[
      d,
      phantom,
      "{\mapsto}"{sloped},
      "{
        \underset{W}{\osum}
      }"{xshift=-8pt}
    ]
    &&
    \underset{W}{\necessarily}
    \underset{W}{\TensorUnit}
    \ar[
      rr,
      "{
        \counit
          { \necessarily_W }
          { \TensorUnit_W }
      }"{description}
    ]
    \ar[
      d,
      phantom,
      "{\mapsto}"{sloped},
      "{
        \underset{W}{\osum}
      }"{xshift=-8pt}
    ]
    &&
    \underset{W}{\TensorUnit}
    \ar[
      d,
      phantom,
      "{\mapsto}"{sloped},
      "{
        \underset{W}{\osum}
      }"{xshift=-8pt}
    ]
    \\[-10pt]
    \rotatebox[origin=c]{90}{\llap{\bf effective}\hspace{-30pt}}
    &
    \underset{W}{\indefinitely}
    \quantized V
    \ar[
      rr,
      "{
        \underset{W}{\indefinitely}
        G
      }"{description, yshift=-2pt}
    ]
    &&
    \underset{W}{\indefinitely}
    \quantized W
    \ar[
      rr,
      "{
        \join
         { \indefinitely_W }
         { \TensorUnit }
        {}
      }"{description},
      "{
        \scalebox{1.05}{
          \measure
        }
      }"{swap, yshift=-6pt, pos=.4  }
    ]
    &&
    \underset{W}{\indefinitely}
    \TensorUnit
    \\
    \\
    \rotatebox[origin=c]{90}{\clap{\bf computational\hspace{-15pt}}}
    &
    \quantized V
    \ar[
      uu,
      "{
        \unit
          { \indefinitely_W }
          {
            \quantized V
          }
      }"{swap},
      "{
        \scalebox{.7}{
          \color{darkgreen}
          \bf
          dynamic lifting
        }
      }"{sloped, yshift=2pt}
    ]
    \ar[
      rr,
      "{ G }"{description},
      "{
        \scalebox{.7}{
          \color{darkgreen}
          \bf
          coherent quantum gate
        }
      }"{yshift=+2pt, pos=.52}
    ]
    &&
    \quantized W
    \ar[
      uu,
      "{
        \unit
          { \indefinitely_W }
          { \quantized W }
      }",
      "{
        \scalebox{.9}{
          \definitely
        }
      }"{swap, yshift=-2pt, sloped}
    ]
    \ar[
      uurr,
      rounded corners,
      to path={
           ([xshift=+00pt]\tikztostart.east)
        -- node[yshift=6pt] {
              \collapse
           }           ([xshift=+76pt]\tikztostart.east)
        -- ([yshift=-00pt]\tikztotarget.south)
      }
    ]
  \end{tikzcd}
$
}
\end{tabular}
\vspace{.1cm}

  In contrast, in the computational typing the ``dynamically lifted'' classical measurement outcomes are syntactically referenced only the moment that the measurement actually takes place (computationally).
In particular, as successive quantum measurements are made, the computational typing of the quantum circuit accumulates the corresponding indefiniteness-modalities, reflecting the fact that more and more measurement outcomes $w_i \,\isa\, W_i$ become ``dynamically lifted''  into the classical register (Lit. \ref{ClassicalControlAndDynamicLifting}):
\newpage

\begin{equation}
\label{ComputationalTypingOfSuccessiveMeasurement}
\hspace{-5mm}

\end{equation}

\smallskip

\noindent
{\bf Enhancing dynamically lifted quantum measurement from pure to mixed states.}
Remarkably, the above effective and computational typing of quantum measurement and controlled quantum gates is enhanced {\it verbatim} to quantum channels on mixed states \eqref{DensityMatrixInIntroduction}, due to the faithful functor \eqref{EnhancingControlledGatesToControlledChannelsAsFaithfulFunctor}
\vspace{-1mm}
$$
  \begin{tikzcd}[sep=5pt]
  \mathllap{
  \pair
    { \indefinitely_W }
    {  }
  \circ
  (-)\otimes {(-)^{\dagger}}^\ast
  \;\isa\;\;
  }
    \QuantumTypes_{\indefinitely_W}
    \ar[
      rr
    ]
    &&
    \QuantumTypes_{\indefinitely_W}
    \\
    \HilbertSpace{H}_1
    \ar[
      dd,
      "{
        A_\bullet
      }"{swap}
    ]
    &&
    \HilbertSpace{H}_1
    \otimes
    \HilbertSpace{H}_1^\ast
    \ar[
      d,
      "{
        A_\bullet
        \otimes
        {A^\dagger_\bullet}^\ast
      }"
    ]
    \\[+20pt]
    &&
    \Big(
    \underset{W}{\indefinitely}
    \,
    \HilbertSpace{H}_2
    \Big)
    \otimes
    \Big(
    \underset{W}{\indefinitely}
    \,
    \HilbertSpace{H}_2^\ast
    \Big)
    \ar[
      d,
      "{
        \pair
          { \indefinitely_W }
          {
            \HilbertSpace{H}_2
            ,
            \HilbertSpace{H}^\ast_2
          }
      }"
    ]
    \\[+20pt]
    \underset{W}{\indefinitely}
    \,
    \HilbertSpace{H}_2
    &&
    \underset{W}{\indefinitely}
    \big(
    \HilbertSpace{H}_2
    \otimes
    \HilbertSpace{H}_2^\ast
    \big)
    \,.
  \end{tikzcd}
$$
In the same manner,
the computational typing \eqref{ComputationalMeasurementTyping} of quantum measurements enhances to mixed states, by first applying
$\collapse_W$ \eqref{ComputationalMeasurementTyping}
to states and co-states in parallel, and then $\indefiniteness-$\mbox{\tt pair}ing \eqref{SymmetricMonoidalMonadStructureOnIndefiniteness} the result, whence we may and will denote this operation by the same symbol ``$\collapse_W$'':
\begin{equation}
\label{DecoherenceFromMonoidalMonadStructure}
\hspace{-4cm}
\begin{tikzcd}[
  column sep=25pt,
  row sep={between origins, 10pt}
]
  &[+5pt]
  &[-30pt]
  &[-12pt]
  &[-12pt]
  &[-15pt]
  &[-15pt]
  &[+5pt]
  \\[0pt]
  \mathclap{
  \scalebox{.7}{
    \color{darkblue}
    \bf
    \def\arraystretch{.9}
    \begin{tabular}{c}
      mixed
      \\
      states
    \end{tabular}
  }
  }
  &[-20pt]&[-30pt]
  \mathclap{
  \scalebox{.7}{
    \color{darkblue}
    \bf
    \def\arraystretch{.9}
    \begin{tabular}{c}
      density
      \\
      matrices
    \end{tabular}
  }
  }
  \ar[
    rr,
    phantom,
    "{
      \scalebox{.7}{
        \color{darkgreen}
        \bf
        \def\arraystretch{.9}
        \begin{tabular}{c}
          measure separately
          \\
          states and co-states
        \end{tabular}
      }
    }"
  ]
  &[-12pt]&[-12pt]
  {}
  \ar[
    rr,
    phantom,
    "{
      \scalebox{.7}{
        \color{darkgreen}
        \bf
        \def\arraystretch{.9}
        \begin{tabular}{c}
          decohere: discard
          \\
          off-diagonal entries
        \end{tabular}
      }
    }"{pos=.45}
  ]
  &[-23pt]&[-23pt]
  {}
  &
  \mathclap{
  \scalebox{.7}{
    \color{darkblue}
    \bf
    \def\arraystretch{.9}
    \begin{tabular}{c}
      probability
      \\
      distributions
    \end{tabular}
  }
  }
  \\[10pt]
  \quantized W
  &&
  \osum_W \ComplexNumbers
  &&
  \indefinitely_W \ComplexNumbers
  \\
  \otimes
  \ar[rr, "{ \simeq }"{description}]
  &&
  \otimes
  \ar[
    rr,
    "{
      \scalebox{.7}{$
        \collapse_W
      $}
      \,\defneq\;
      \join
        { \indefinitely_W }
        { \ComplexNumbers }
      \,\circ\;
      \unit
        { \indefinitely_W }
        { \osum_W \ComplexNumbers }
    }"{yshift=2pt},
    "{ \otimes }"{description},
    "{
      \scalebox{.7}{$
        \collapse_W
      $}
      \,\defneq\;
      \join
        { \indefinitely_W }
        { \ComplexNumbers^{\mathrlap{\ast}} }
      \,\circ\;
      \unit
        { \indefinitely_W }
        { \osum_W \ComplexNumbers^{\mathrlap{\ast}} }
    }"{swap, yshift=-2pt}
  ]
  &&
  \otimes
  \ar[
    rr,
    "{
      \pair
        { \indefinitely_W }
        {
          \ComplexNumbers
          ,
          \ComplexNumbers^{\mathrlap{\ast}}
        }
    }"{description}
  ]
  &&
  \indefinitely_W
  \!\!
  \def\arraystretch{.8}

  \ar[
    r,
    "{
      \indefinitely_W
      \,
      \mathrm{ev}
    }"{description}
  ]
  &[-10pt]
  \indefinitely_W
  \ComplexNumbers
  \\
  (\quantized W)^\ast
  \ar[
    urrrrrrr,
    rounded corners,
    to path={
         ([yshift=-00pt]\tikztostart.south)
      -- ([yshift=-25pt]\tikztostart.south)
      -- node {
         \scalebox{.7}{
           \colorbox{white}{$
             \collapse_W
           $}
         }
      }
         ([yshift=-35pt]\tikztotarget.south)
      -- ([yshift=-00pt]\tikztotarget.south)
    }
  ]
  &&
  \osum_W \ComplexNumbers^\ast
  &&
  \indefinitely_W \ComplexNumbers^\ast
  \\[10pt]
  &&
  \mathclap{
  \scalebox{.7}{
    \color{gray}
    \bf
    \def\arraystretch{.9}
    %
    }
  }
\end{tikzcd}
\hspace{-4cm}
\end{equation}
Notice two remarkable aspects of this $\indefiniteness_W$-effectful map:
\begin{itemize}[leftmargin=.8cm]
  \item[{\bf (i)}] in this form, the coherent quantum phases drop out, as expected for a realistic quantum measurement

  (the failure of which to happen for the analogous process on pure states was highlighted in \cite[\S 1.6]{CoeckePaquette08}, where a different solution was discussed),
\item[{\bf (ii)}]
in fact, \eqref{DecoherenceFromMonoidalMonadStructure} reproduces exactly the typing of the quantum measurement process in {\it L{\"u}ders' first form} \eqref{DynamicallyLiftedQuantumMeasurementOnDensityMatrix}, neatly embodying the Born rule \eqref{BornRule}.
\end{itemize}

\smallskip

In conclusion: Due to the symmetric monoidal monad structure on the indefiniteness-modality $\bigcirc$, the monadic typing of classically controlled quantum circuits with dynamically lifted quantum measurement gates has {\it syntactically} the same form whether applied to pure or to mixed states.

The difference with interpreting quantum circuits in the generality of mixed states is that here further stochastic
quantum operations become available, the {\it quantum channels}. We discuss this in \cref{MixedQuantumTypes}.

\subsection{Mixed Quantum Types}
\label{MixedQuantumTypes}

We discuss a natural monadic formalization of mixed quantum states \eqref{DensityMatrixInIntroduction} and their quantum channels \eqref{QuantumChannel}.
The key observation is once again that the main structure happens to come for free as (co)monadic (co)effects that need not be postulated but are definable (admissible) in a suitably expressive linear type theory:

\begin{itemize}
  \item [{\bf (i)}]
  quantum channel dynamics
  \eqref{QuantumChannel}
  on mixed quantum states
  \eqref{DensityMatrixInIntroduction}
  and their quantum observables
  \eqref{ObservablesAsQuantumChannels}
  is all encoded by {\it transformations} \eqref{ComponentOfMonadTransformation} of the QuantumState (co)monads $\HilbertSpace{H}\mathrm{State}$

  \item[{\bf (ii)}] the collapsing measurement process on such mixed states is given by the monoidal monadic structure on the $\indefiniteness$-modality \eqref{DecoherenceFromMonoidalMonadStructure}.
\end{itemize}

What requires a little extra work to formalize is, finally:

\begin{itemize}
  \item [{\bf (iii)}]
  the dagger-structure $(-)^\dagger$ \eqref{OperatorAdjoints} on quantum types.
  \ifdefined\monadology
  This has a rather beautiful (homotopy-)type-theoretic solution, which however is beyond the scope of this article and instead relegated to
  \cite[\S 3]{QS}.
  \else
  whose discussion is relegated to \cref{QuantumProbability}.
  \fi

For the present discussion, we assume the existence of operator adjoints as a black box; in fact, we exclusively need {\it dual} operator adjoints.
\begin{equation}
  \label{FormingAdjointDuals}
  \HilbertSpace{H}_1
  ,\,
  \HilbertSpace{H}_2
  \;
  \isa
  \;
  \DualizableQuantumType
  \,,
  \hspace{.6cm}
  f \,\isa\, \HilbertSpace{H}_1 \to \HilbertSpace{H}_2
  \;\;\;\;\;\;
  \yields
  \;\;\;\;\;\;
  {f^\dagger}^\ast
    \,\isa\,
  \HilbertSpace{H}^\ast_1
    \to
  \HilbertSpace{H}^\ast_2 \;.
\end{equation}
\end{itemize}

\medskip

\hspace{-.7cm}
\begin{tabular}{ll}
\begin{minipage}{6.6cm}
  \footnotesize
  We find that the structure of quantum probability
  theory (Lit. \ref{LiteratureQuantumProbability}) --- where quantum gates operating on pure quantum states are generalized to
  quantum channels operating on mixed quantum states (density matrixes) --- is closely reflected in the monadic computational
  theory (Lit. \ref{LiteratureComputationalEffectsAndModalities}) of the linear analog of the classical State/Store (co)monads, namely the QuantumState Frobenius monads
  $
    \HilbertSpace{H}\mathrm{State}
    \;\defneq\;
    (\mbox{-})
    \otimes
    \HilbertSpace{H}
    \otimes
    \HilbertSpace{H}^\ast
  $
\end{minipage}
&
\footnotesize
\def\arraystretch{1.7}
\begin{tabular}{|r|l|}
  \hline
  {\bf Quantum Probability Theory}
  &
  {\bf QuantumState monadic computation}
  \\
  \hline
  \hline
  Quantum channels
  &
  QuantumState transformations
  \\
  \hline
  Mixed quantum states
  &
  QuantumState effectful scalars
  \\
  \hline
  Quantum observables
  &
  QuantumState contextful scalars
  \\
  \hline
  Evolution
  of quantum observables
  &
  QuantumState transformation
  on modales
  \\
  \hline
\end{tabular}
\end{tabular}

\bigskip


\smallskip

\noindent
{\bf QuantumState modality.}
We consider the evident linear version of the classical state monad \eqref{StateMonadEndofunctor}
and the classical store comonad \eqref{CostateComonadEndofunctor},
which over a {\it finite}-dimensional quantum state space fuse to a Frobenius monad \eqref{FrobeniusMonadFromAmbidextrousAdjunction}
that, we will see, quite deserves to be called the {\it QuantumState modality}.

\medskip

\begin{definition}[\bf QuantumState]
\label{QuantumStateCoMonads}
For $\HilbertSpace{H} \,\isa\, \DualizableQuantumType$ a strongly dualizable linear type
\eqref{CoEvaluationForDualObjects} (hence a finite-dimensional vector space in the model of Def. \ref{CategoryOfBundleTypes})
with dual $\HilbertSpace{H}^\ast \simeq \HilbertSpace{H} \maplin \TensorUnit$ \eqref{StrongDualIsAClosedDual},
we say that the corresponding
{\it QuantumState} (co)monads are
the Frobenius monads \eqref{FrobeniusMonadFromAmbidextrousAdjunction}
induced \eqref{MonadFromAdjunction}
by the corresponding ambidextrous adjunction
of tensoring functors \eqref{AmbidextrousAdjunctTensorFunctorsForStrongDualObjects}:
\vspace{-4mm}
\begin{equation}
\label{QuantumStateInducedFromLinearHomAdjunction}
\mathllap{
\scalebox{.7}{
  \color{darkblue}
  \bf
  \def\arraystretch{.9}
  \begin{tabular}{c}
    QuantumState
    \\
    Frobenius
    \\
    monad
  \end{tabular}
}
}
\begin{tikzcd}[column sep=40pt]
    \QuantumTypes
    \ar[out=160, in=60,
      looseness=4,
      "\raisebox{-2pt}{${
          \hspace{-7pt}
          \mathclap{
            \adjustbox{
              margin=2pt,
              bgcolor=white
            }{$
              \HilbertSpace{H}\mathrm{State}
            $}
          }
        }$}"{pos=.26, description},
    ]
    \ar[out=185, in=175,
      looseness=10,
      phantom,
      "{\scalebox{.7}{$\bot$}}"{yshift=1pt, xshift=3pt}
    ]
    \ar[out=-160, in=-60,
      looseness=4,
      shorten=-0pt,
      "\raisebox{-2pt}{${
          \hspace{-7pt}
          \mathclap{
            \adjustbox{
              margin=2.5pt,
              bgcolor=white
            }{$
              \HilbertSpace{H}^\ast\mathrm{Store}
            $}
          }
        }$}"{pos=.26, description},
    ]
  \ar[
    rr,
    shift left=14pt,
    "{
        (\mbox{-})
        \otimes
        \HilbertSpace{H}
    }"
  ]
  \ar[
    from=rr,
    "{
      (\mbox{-})
      \otimes
      \HilbertSpace{H}^\ast
    }"{description}
  ]
  \ar[
    rr,
    shift right=14pt,
    "{
      (\mbox{-})
      \otimes
      \HilbertSpace{H}
    }"{swap}
  ]
  \ar[
    rr,
    phantom,
    shift left=8pt,
    "\scalebox{.7}{$\bot$}"
  ]
  \ar[
    rr,
    phantom,
    shift right=8pt,
    "\scalebox{.7}{$\bot$}"
  ]
  &&
  \QuantumTypes
    \ar[out=120, in=20,
      looseness=4,
      shorten <=-2pt,
      shorten >=-3pt,
      "\raisebox{-2pt}{${
          \hspace{7pt}
          \mathclap{
            \adjustbox{
              margin=2pt,
              bgcolor=white
            }{$
              \HilbertSpace{H}\mathrm{Store}
            $}
          }
        }$}"{pos=.7, description},
    ]
    \ar[out=5, in=-5,
      looseness=8.5,
      phantom,
      "{\scalebox{.7}{$\bot$}}"{yshift=1pt}
    ]
    \ar[out=-120, in=-20,
      looseness=4,
      shorten <=-2pt,
      shorten >=-3pt,
      "\raisebox{-2pt}{${
          \hspace{7pt}
          \mathclap{
            \adjustbox{
              margin=2pt,
              bgcolor=white
            }{$
              \HilbertSpace{H}^\ast\mathrm{State}
            $}
          }
        }$}"{pos=.7, description},
    ]
\end{tikzcd}
\end{equation}

\vspace{-4mm}
\noindent Since the ambidexterity means that $\HilbertSpace{H}\mathrm{State}$ and $\HilbertSpace{H}^\ast\mathrm{Store}$ fuse to a single Frobenius monad \eqref{FrobeniusMonadFromAmbidextrousAdjunction}, we will often refer to both or either as {\it QuantumState modalities} and
speak of the {\it QuantumStore modality} when referring specifically only to the comonad structure.
\end{definition}

For the record, in bra-ket notation \eqref{BraKetNotation}
the return/obtain-operations
of QuantumState are as follows  (where $W$, $\quantized W \,\simeq\, \HilbertSpace{H}$ denotes any orthonormal basis
for $\HilbertSpace{H}$ with respect to any chose Hermitian inner product $\langle\cdot\vert\cdot\rangle$):
\begin{equation}
\label{CoUnitsOfQuantumState}
\adjustbox{}{
\def\arraystretch{3}
\def\arraycolsep{120pt}

}
\end{equation}
so that the join/duplicate-operations are as follows:
\begin{equation}
\label{JoinAndDuplicateOfQuantumState}
\hspace{-5mm}
\adjustbox{}{
\def\tabcolsep{0pt}
%
}
\end{equation}

Notice that these operations all express in one way or another the basic bra-ket manipulations known from quantum mechanics textbooks (evaluation and ``insertion of an identity'').
In particular, the zig-zag identities which witness the adjunctions in \eqref{QuantumStateInducedFromLinearHomAdjunction} are nothing but the following familiar basic identities:

$$
  \def\arraystretch{2}
  %
$$


\begin{remark}[\bf QuantumState as QuantumWriter]
  \label{QuantumStateEffectAsQuantumWriter}
  The QuantumState Frobenius monad of Def. \ref{QuantumStateCoMonads} is equivalently the linear (co)Writer monad \eqref{WriterMonad} over $\HilbertSpace{H}\otimes \HilbertSpace{H}^\ast$, the latter understood with its canonical Frobenius monoid structure
  of endomorphism objects in compact closed categories (see e.g. \cite[Lem. 3.17]{Vicary11}):
  \vspace{-2mm}
  $$
    \begin{tikzcd}[
      row sep=-2pt
    ]
      \scalebox{.7}{
        \color{darkblue}
        \bf
        \def\arraystretch{.9}
        \begin{tabular}{c}
          quantum
          \\
          state
        \end{tabular}
      }
      &
      \scalebox{.7}{
        \color{purple}
        \bf
        \def\arraystretch{.9}
        \begin{tabular}{c}
          quantum
          \\
          (co)writer
        \end{tabular}
      }
      &
      \scalebox{.7}{
        \color{darkblue}
        \bf
        \def\arraystretch{.9}
        \begin{tabular}{c}
          quantum
          \\
          store
        \end{tabular}
      }
      \\
      \HilbertSpace{H}\mathrm{State}
      &
      (\mbox{-})
      \otimes
      \big(
      \HilbertSpace{H}
      \otimes
      \HilbertSpace{H}^\ast
      \big)
      &
      \HilbertSpace{H}^\ast\mathrm{Store}
      \\
      \mathrm{Monads}
      &
      \mathrm{FrobMonads}
      \ar[r]
      \ar[l]
      &
      \mathrm{CoMonads}
    \end{tikzcd}
  $$

\vspace{-2mm}
\noindent  In particular, if $\HilbertSpace{H} \,\simeq\, \quantized W$ then QuantumState is the (co)Writer monad for
  $\quantized W \otimes \quantized W^\ast$, in which form it is interesting to compare to the quantum
  indefiniteness/randomness modality, which is the (co)writer for a single copy $\quantized W$, according to Prop. \ref{QuantumCoEffectsViaFrobeniusAlgebra}.

  \begin{center}
  \footnotesize
  \def\arraystretch{1.8}
  \begin{tabular}{|c|c|c|}
    \hline
    {\bf Frobenius algebra}
    &
    {\bf Quantum modalities}
    &
    {\bf Quantum effects}
    \\
    \hline
    \hline
    $\quantized W$
    &
    indefiniteness/randomness
    &
    \def\arraystretch{.9}
    \begin{tabular}{c}
      collapsing
      \\
      quantum measurement
    \end{tabular}
    \\
    \hline
    $\quantized W \otimes \quantized W^\ast$
    &
    quantum state/store
    &
    quantum probability
    \\
    \hline
  \end{tabular}
  \end{center}
\end{remark}

\begin{proposition}[\bf QuantumState effect-/contextful maps are Linear operators]
  \label{QuantumStoreComonad}
$\,$

\noindent {\bf (i)}   The $\HilbertSpace{H}^\ast\mathrm{State}$ modality of Def. \ref{QuantumStateCoMonads}
has (co)Kleisli morphisms of the form
\vspace{-2mm}
\begin{equation}
  \label{HStoreKleisliMap}
  \begin{tikzcd}[sep=0pt]
    \HilbertSpace{K}
      \otimes
    \HilbertSpace{H}
      \otimes
    \HilbertSpace{H}^\ast
    \ar[
      rr,
      "{
        \mathcal{O}_A
      }"
    ]
    &&
    \HilbertSpace{L}
    \\
    \scalebox{\termscale}{$  \vert \kappa \rangle
    \otimes
    \vert \psi \rangle
    \langle \phi \vert
    $}
    &\mapsto&
    \scalebox{\termscale}{$  \langle \phi, - \vert
      \,A\,
    \vert \kappa, \psi \rangle
    $}
  \end{tikzcd}
  \hspace{.6cm}
  \leftrightarrow
  \hspace{1.2cm}
  \begin{tikzcd}
    \mathllap{
      A
      \;\isa\;
    }
    \HilbertSpace{K}
    \otimes
    \HilbertSpace{H}
    \to
    \HilbertSpace{L}
    \otimes
    \HilbertSpace{H}
  \end{tikzcd}
  \hspace{.8cm}
  \leftrightarrow
  \hspace{.8cm}
  \begin{tikzcd}[
    sep=0pt
  ]
    \HilbertSpace{K}
    \ar[
      rr,
      "{
        \mathcal{S}_A
      }"
    ]
    &&
    \HilbertSpace{L}
    \otimes
    \HilbertSpace{H}
    \otimes
    \HilbertSpace{H}^\ast
    \\
    \scalebox{.85}{$
      \vert \kappa \rangle
    $}
    &\mapsto&
    \scalebox{.85}{$
    \langle -,- \vert
      \,A\,
    \vert \kappa, -  \rangle
    $}
  \end{tikzcd}
\end{equation}

\vspace{-2mm}
\noindent {\bf (ii)}  on which the bind/extend- operations are given by
\vspace{-2mm}
\begin{equation}
  \label{HDualStoreExtend}
  \declarelin
    {
      \extend
        {
          \HilbertSpace{H}^\ast\mathrm{Store}
        }
        {
          \HilbertSpace{K}
          ,
          \HilbertSpace{L}
        }
    }
    {
      \big(
        \HilbertSpace{K}
        \otimes
        \HilbertSpace{H}
        \otimes
        \HilbertSpace{H}^\ast
        \maplin
        \HilbertSpace{L}
      \big)
      \maplin
      \big(
        \HilbertSpace{K}
        \otimes
        \HilbertSpace{H}
        \otimes
        \HilbertSpace{H}^\ast
        \maplin
        \HilbertSpace{L}
          \otimes
        \HilbertSpace{H}
          \otimes
        \HilbertSpace{H}^\ast
      \big)
    }
    {
        \scalebox{\termscale}{$   \big(
      \vert \kappa \rangle
      \,
      \vert \psi \rangle
      \langle \phi \vert
      \,\mapsto\,
      \langle \phi, - \vert
        {A}
      \vert \kappa, \psi \rangle
      \big)
      $}
      \;\mapsto\;
          \scalebox{\termscale}{$
      \big(
      \vert \kappa \rangle
      \,
      \vert \psi \rangle
      \langle \phi \vert
      \,\mapsto\,
        {A}
      \vert \kappa, \psi \rangle
      \langle \phi \vert
      \big)$}
    }
\end{equation}
\begin{equation}
  \declarelin
    {
      \bind
        { \HilbertSpace{H}\mathrm{State} }
        { \HilbertSpace{K}, \HilbertSpace{L} }
    }
    {
      \big(
        \HilbertSpace{K}
        \maplin
        \HilbertSpace{L}
        \otimes
        \HilbertSpace{H}
        \otimes
        \HilbertSpace{H}^\ast
      \big)
      \maplin
      \big(
        \HilbertSpace{K}
        \otimes
        \HilbertSpace{H}
        \otimes
        \HilbertSpace{H}^\ast
        \maplin
        \HilbertSpace{L}
        \otimes
        \HilbertSpace{H}
        \otimes
        \HilbertSpace{H}^\ast
      \big)
    }
    {
      \scalebox{.85}{$
      \big(
      \vert \kappa \rangle
      \,\mapsto\,
      \langle -,-\vert \,A\, \vert \kappa, - \rangle
      \big)
      \,\mapsto\,
      \big(
      \vert \kappa \rangle
      \vert \psi \rangle \langle \phi \vert
      \,\mapsto\,
      A\, \vert \kappa, \psi \rangle
      \langle \phi \vert
      \big)
      $}
    }
\end{equation}
\noindent {\bf (iii)}  Hence we have bijections
\vspace{-2mm}
$$
  \overset{
    \mathclap{
      \raisebox{7pt}{
        \scalebox{.7}{
          \color{darkblue}
          \bf
          \def\arraystretch{.9}
          \begin{tabular}{c}
            $\HilbertSpace{H}\mathrm{State}$-contextful
            maps
          \end{tabular}
        }
      }
    }
  }{
  \mathcal{O}_A
  \;\isa\;
  \HilbertSpace{H}\mathrm{State}(\HilbertSpace{K})
    \to
  \HilbertSpace{L}
  }
  \hspace{.8cm}
  \leftrightarrow
  \hspace{.8cm}
  \overset{
    \mathclap{
      \raisebox{7pt}{
        \scalebox{.7}{
          \color{darkblue}
          \bf
          \def\arraystretch{.9}
          \begin{tabular}{c}
            linear operators
          \end{tabular}
        }
      }
    }
  }{
  A
    \,\isa\,
  \HilbertSpace{K}
    \otimes
  \HilbertSpace{H}
  \to
  \HilbertSpace{L}
    \otimes
  \HilbertSpace{H}
  }
  \hspace{.8cm}
  \leftrightarrow
  \hspace{.8cm}
  \overset{
    \mathclap{
      \raisebox{7pt}{
        \scalebox{.7}{
          \color{darkblue}
          \bf
          \def\arraystretch{.9}
          \begin{tabular}{c}
            $\HilbertSpace{H}\mathrm{State}$-effectful
            maps
          \end{tabular}
        }
      }
    }
  }{
  \mathcal{S}_A
  \;\isa\;
  \HilbertSpace{K}
    \to
  \HilbertSpace{H}\mathrm{State}(\HilbertSpace{L})
  }
$$

\vspace{-2mm}
\noindent under which Kleisli composition corresponds to ordinary composition of linear operators:
\vspace{-2mm}
$$
  \overset{
    \mathclap{
      \raisebox{4pt}{
        \scalebox{.7}{
          \color{darkblue}
          \bf
          composition of
          $\HilbertSpace{H}\mathrm{State}$-contextful maps
        }
      }
    }
  }{
  \mathcal{O}_{A}
  \;\circ\;
  \big(
  \extend
    {
      \HilbertSpace{H}^\ast\mathrm{Store}
    }
    {}
    \big(
      \mathcal{O}_B
    \big)
    \big)
  }
  \;\;\;
  =
  \;\;\;
  \underset{
    \mathclap{
      \raisebox{-4pt}{
        \scalebox{.7}{
          \color{darkblue}
          \bf
          composition of
          linear operators
        }
      }
    }
  }{
  \mathcal{O}_{ A \cdot  B  }
  \;\;\;
  \leftrightarrow
  \;\;\;
  \mathcal{S}_{A \cdot B}
  }
  \;\;\;
  =
  \;\;\;
  \overset{
    \mathclap{
      \raisebox{4pt}{
        \scalebox{.7}{
          \color{darkblue}
          \bf
          composition of
          $\HilbertSpace{H}\mathrm{State}$-effectful maps
        }
      }
    }
  }{
  \mathcal{S}_{A}
  \;\circ\;
  \big(
  \bind
    {
      \HilbertSpace{H}\mathrm{State}
    }
    {}
    \big(
      \mathcal{S}_B
    \big)
    \big)
  }
  \,.
$$
\end{proposition}
\begin{proof}
By direct unwinding of the formulas
\eqref{RecoveringBindFromJoin}
and \eqref{JoinAndDuplicateOfQuantumState}:
\vspace{-3mm}
$$
  \hspace{1cm}
  \begin{tikzcd}[
    row sep=-1pt,
    column sep=15pt
  ]
    \HilbertSpace{K}
    \otimes
    \HilbertSpace{H}
    \otimes
    \HilbertSpace{H}^\ast
    \ar[
      rr,
      "{
        \duplicate
          { \HilbertSpace{H}^\ast\mathrm{Store} }
          { \HilbertSpace{K}^\ast }
      }"
    ]
    &&
    \HilbertSpace{K}
      \otimes
    \HilbertSpace{H}
      \otimes
    \HilbertSpace{H}^\ast
      \otimes
    \HilbertSpace{H}
      \otimes
    \HilbertSpace{H}^\ast
    \ar[
      rr,
      "{
        \mathcal{O}_A
        \,\otimes\,
        \HilbertSpace{H}
        \otimes
        \HilbertSpace{H}^\ast
      }"
    ]
    &&
    \HilbertSpace{L}
      \otimes
    \HilbertSpace{H}
      \otimes
    \HilbertSpace{H}^\ast
    \\
    \mathllap{
      \extend
        { \HilbertSpace{H}^\ast\mathrm{Store} }
        { \HilbertSpace{K}, \HilbertSpace{L} }
      \mathcal{O}_A
      \;:\;
      \;\;\;
    }
    \scalebox{\termscale}{$
    \vert \kappa \rangle
    \otimes
    \vert \psi \rangle
    \langle \phi \vert
    $}
    &\mapsto&
    \scalebox{\termscale}{$
    \underset{w}{\sum}
    \,
    \vert \kappa \rangle
    \,
    \vert \psi \rangle
    \langle w \vert
      \otimes
    \vert w \rangle
    \langle \phi \vert
    $}
    &\mapsto&
    \scalebox{\termscale}{$
    \underset{w}{\sum}
    \,
    \vert w \rangle
    \langle
      w,-
    \vert
     \,A\,
    \vert
      \kappa, \psi
    \rangle
    \langle  \phi \vert
    $}
    \\[-6pt]
    &&
    &=&
    \scalebox{\termscale}{$
    A \, \vert \kappa , \psi \rangle
    \langle \phi \vert
    $}
    \\[10pt]
    \HilbertSpace{K}
    \otimes
    \HilbertSpace{H} \otimes \HilbertSpace{H}^\ast
    \ar[
      rr,
      "{
        \mathcal{S}_A
        \otimes
        \HilbertSpace{H} \otimes \HilbertSpace{H}^\ast
      }"
    ]
    &&
    \HilbertSpace{L}
    \otimes
    \HilbertSpace{H} \otimes \HilbertSpace{H}^\ast
    \otimes
    \HilbertSpace{H} \otimes \HilbertSpace{H}^\ast
    \ar[
      rr,
      "{
        \join
          { \HilbertSpace{H}\mathrm{State} }
          { \HilbertSpace{L} }
      }"
    ]
    &&
    \HilbertSpace{L}
    \otimes
    \HilbertSpace{H} \otimes \HilbertSpace{H}^\ast
    \\
    \mathllap{
      \bind
        { \HilbertSpace{H}\mathrm{State} }
        { \HilbertSpace{K}, \HilbertSpace{L} }
      \mathcal{S}_A
      \;:\;
      \;\;\;
    }
    \scalebox{.85}{$
    \vert \kappa \rangle
    \otimes
    \vert \psi \rangle
    \langle \phi \vert
    $}
    &\mapsto&
    \scalebox{.85}{$
    \langle -,- \vert
      \,A\, \vert \kappa, - \rangle
    \otimes
    \vert \psi \rangle
    \langle \phi \vert
    $}
    &\scalebox{.85}{$\mapsto$}&
    \scalebox{.85}{$
    A\, \vert \kappa, \psi \rangle
    \langle \phi \vert
    $}
  \end{tikzcd}
$$

\vspace{-7mm}
\end{proof}


\noindent
{\bf Quantum observables.} We show that the core structure of {\it quantum observables} is reflected in the QuantumState-contextful scalars (Ex. \ref{AlgebraOfQuantumObservablesAsQuantumStoreContextfulMaps}) including:
\begin{itemize}[leftmargin=.5cm]
\item their expectation values \eqref{QuantumObservablesAsCoKleisliMorphisms},  \item their algebra structure \eqref{AlgebraOfQuantumObservablesAsKleisliEndomorphsims},
\item their Heisenberg-evolution (Prop. \ref{QuantumStateEvolutionIsHeisenbergEvolution}).
\end{itemize}

\begin{example}[{\bf Quantum observables are the QuantumState contextful scalars}]
\label{AlgebraOfQuantumObservablesAsQuantumStoreContextfulMaps}
Notice that in any monoidal category like $(\QuantumTypes, \otimes, \TensorUnit)$ it makes sense to refer to the endomorphisms $c \,\isa\,\TensorUnit \to \TensorUnit$ of the tensor unit as the {\it scalars} of the theory (\cite[\S 6]{AbramskyCoecke04}\cite[2.1]{HeunenVicary12}).
Therefore, with the understanding of comonadic computational contexts (Lit. \ref{LiteratureComputationalEffectsAndModalities}) and given a comonad $\mathcal{C}$ on $\QuantumTypes$, the Kleisli-endomorphisms of the tensor unit $\mathcal{C}(\TensorUnit) \to \TensorUnit$ may be thought of \eqref{CoBindingAndCoReturning}
as the {\it $\mathcal{C}$-contextful scalars}.
Now Prop. \ref{QuantumStoreComonad} says that the {\it $\HilbertSpace{H}\mathrm{State}$-contextful scalars} are equivalently the linear operators on $\HilbertSpace{H}$, here seen to be representing quantum observables \eqref{ObservablesAsQuantumChannels}
incarnated via their system of expectation values \eqref{ExpectationValueOfObservable}:
\vspace{-2mm}
\begin{equation}
  \label{QuantumObservablesAsCoKleisliMorphisms}
  \begin{tikzcd}[row sep=-2pt, column sep=0pt]
    \mathllap{
      \mathcal{O}_A
      \;\isa\;
    }
    \HilbertSpace{H}
    \otimes
    \HilbertSpace{H}^\ast
    \ar[
      rr
    ]
    &&
    \TensorUnit
    \\
    \scalebox{\termscale}{$  \vert \psi \rangle
    \langle \phi \vert
    $}
    &\mapsto&
   \scalebox{\termscale}{$  \langle \phi \vert
      \,A\,
    \vert \psi \rangle
    $}
    \\
    \scalebox{.85}{$\rho$}
    &
    \scalebox{.85}{$\mapsto$}
    &
    \scalebox{.85}{$
    \mathrm{Tr}
    \big(
     \rho \cdot A
    \big)
    $}
  \end{tikzcd}
  \;\;\;\;\;
  \leftrightarrow
  \;\;\;\;\;
  A \,\isa\, \HilbertSpace{H} \to \HilbertSpace{H}
  \,.
\end{equation}

\vspace{-3mm}
\noindent Moreover, the (Kleisli-)composition of such QuantumState-contextful scalars reproduces the ordinary operator product of the corresponding linear operators:
\vspace{-5mm}
\begin{equation}
  \label{AlgebraOfQuantumObservablesAsKleisliEndomorphsims}
  \mathcal{O}_A
  \circ
  \extend
    { \HilbertSpace{H}\mathrm{Store} }
    { \TensorUnit }
  \mathcal{O}_B
  \;\;
  =
  \;\;
  \mathcal{O}_{A \cdot B}
  \,,
  \;\;\;\;\;\;\;\;
  \mbox{so that}
  \;\;\;\;\;\;\;\;
  \overset{
    \mathclap{
      \raisebox{10pt}{
      \scalebox{.7}{
        \color{darkblue}
        \bf
        \def\arraystretch{.9}
        \begin{tabular}{c}
          QuantumState
          \\
          Kleisli-endomorphism
          \\
          algebra of tensor unit
        \end{tabular}
      }
      }
    }
  }{
    \QuantumTypes_{
      \HilbertSpace{H}\mathrm{Store}
    }(\TensorUnit, \TensorUnit)
  }
  \;\simeq\,
  \overset{
    \mathclap{
      \raisebox{7pt}{
      \scalebox{.7}{
        \color{darkblue}
        \bf
        \def\arraystretch{.9}
        \begin{tabular}{c}
          algebra of
          \\
          linear operators
        \end{tabular}
      }
      }
    }
  }{
   \mathrm{End}(\HilbertSpace{H})
  }
  \;\;\;
  \mbox{(as algebras)}
  \,.
\end{equation}
\end{example}

\begin{remark}[\bf The operational/logical meaning of operator products of quantum observables]
$\,$

  \noindent {\bf (i)}  It is commonplace in modern quantum physics that the {\it algebra of quantum observables} is indeed that:
  an associative algebra under operator products of the corresponding linear operators. However, while mathematically suggestive,
  it is subtle to decide which aspect of quantum reality is really modeled by forming the plain operator product of a pair of
  {\it non-commuting} observables $\mathcal{O}_{A}$, $\mathcal{O}_{A'}$; because in this case a prescription for measuring
  them separately (namely via their respective eigenbases $W$, $W'$) does not readily yield a prescription for measuring
  their operator product $\mathcal{O}_{A B}$.

 \noindent {\bf (ii)}  This issue was felt to be severe enough of a conceptual problem by the founding fathers of quantum
 physics that another non-associative notion of algebras of quantum observables was proposed
 \cite{Jordan32}\cite{JordanVonNeumannWigner34}, now known as {\it Jordan algebras} (see \cite{Baez20}
 for more on the quantum foundational motivation of Jordan algebras).
However, while the concept of Jordan algebras turned out to be useful in various areas of mathematics, its relevance for
conceptualizing quantum observables has remained inconclusive.

\noindent {\bf (iii)}  Indeed, the highly successful modern algebraic formulation of quantum physics (for a good exposition see \cite{Gleason09}\cite{Gleason11}) is entirely based on the associative algebra structure on observables (further promoted
to a $C^\ast$-algebra structure for infinite-dimensional algebras) and has no use of Jordan algebras.

\noindent {\bf (iv)}  This begs the question that may originally have motivated Jordan et al.: To give a {\it logical}
justification from first principles for considering quantum observables as an associative algebra under operator products.
But if we grant (with Lit. \ref{VerificationLiterature}, \ref{ModalLogicAndManyWorlds} and \ref{LiteratureComputationalEffectsAndModalities}) a foundational logical content to natural (co)monadic structures on linear types, then Ex. \ref{AlgebraOfQuantumObservablesAsQuantumStoreContextfulMaps} provides a satisfactory answer.

\end{remark}

 For the following Proposition \ref{UnitaryQuantumChannelsAreQuantumStateTransformations}, recall (Lit. \ref{LiteratureQuantumProbability}) that for a pair of quantum systems (represented by) $\HilbertSpace{H}_1, \HilbertSpace{H}_2 \,\isa\, \QuantumType$,
 a {\it quantum channel} \eqref{QuantumChannel} between them is a (linear) map of the form
 \vspace{-3mm}
 $$
   \begin{tikzcd}
   \HilbertSpace{H}_1
   \otimes
   \HilbertSpace{H}_1^\ast
   \ar[
     rr,
     "{ \mathrm{chan} }"
   ]
   &&
   \HilbertSpace{H}_2
   \otimes
   \HilbertSpace{H}_2^\ast
   \end{tikzcd}
 $$

 \vspace{-3mm}
\noindent satisfying some properties; and that in general such a channel may act among further ``ancillary'' systems $\HilbertSpace{K}$ (such as $\HilbertSpace{K}= \HilbertSpace{B} \otimes \HilbertSpace{B}^\ast$, for $\HilbertSpace{B}$ a ``bath'' environment), being more generally a tensor map of the form
\vspace{-2mm}
 $$
   \begin{tikzcd}
   \HilbertSpace{K}
   \otimes
   \HilbertSpace{H}_1
   \otimes
   \HilbertSpace{H}_1^\ast
   \ar[
     rr,
     "{
       \mathrm{id}_{
        \HilbertSpace{K}
       }
       \otimes
       \,
       \mathrm{chan}
     }"
   ]
   &&
   \HilbertSpace{K}
   \otimes
   \HilbertSpace{H}_2
   \otimes
   \HilbertSpace{H}_2^\ast \;.
   \end{tikzcd}
 $$


\begin{proposition}[\bf Unitary quantum channels are quantum state transformations]
\label{UnitaryQuantumChannelsAreQuantumStateTransformations}
The unitary quantum channel $U \otimes U^{\dagger^\ast}$ \eqref{UnitaryQuantumChannel} corresponding to a unitary operator\footnote{
  For the statement of the proposition at this point it just matters that $U$ is an invertible linear map with inverse denoted $U^\dagger$.
}
$U \,\isa\, \HilbertSpace{H}_1 \to \HilbertSpace{H}_2$
induces a (co)monad transformation \eqref{MonadTransformation} between the corresponding Quantum State (co)monads, in that
\vspace{-2mm}
$$
  \begin{tikzcd}[row sep=3pt, column sep=large]
    \ar[
      rr,
      phantom,
      "{
        \scalebox{.7}{
          \color{darkgreen}
          \bf
          QuantumState transformation
        }
      }"
    ]
    &&
    {}
    \\
    \HilbertSpace{H}_1\mathrm{State}
    \ar[
      rr,
      "{
        \mathrm{chan}^{U}
      }"
    ]
    &&
    \HilbertSpace{H}_2\mathrm{State}
    \\
    (\mbox{-})
    \otimes
    \HilbertSpace{H}_1
    \otimes
    \HilbertSpace{H}_1^\ast
    \ar[
      rr,
      "{
        (\mbox{-})
        \otimes
        U
          \otimes
        {U^{\dagger}}^\ast
      }"
    ]
    &&
    (\mbox{-})
    \,\otimes\,
    \HilbertSpace{H}_2
    \otimes
    \HilbertSpace{H}_2^\ast
    \\
 \scalebox{0.8}{$    \vert \kappa \rangle
    \otimes
    \rho
    $}
    &\longmapsto&
  \scalebox{0.8}{$   \vert \kappa \rangle
    \otimes
    \big(
      U \cdot \rho \cdot U^\dagger
    \big)
    $}
    \\
    \ar[
      rr,
      phantom,
      "{
        \scalebox{.7}{
          \color{darkgreen}
          \bf
          unitary quantum channel
        }
      }"
    ]
    &&
    {}
  \end{tikzcd}
$$
\end{proposition}
\begin{proof}
  We need to check the compatibility conditions \eqref{MonadTransformationConditions}.
  But since the (co)unit of $\HilbertSpace{H}\mathrm{State}$ is given
  \eqref{CoUnitsOfQuantumState}
  by inserting an identity and by inner product, respectively,
  their preservation is essentially the definition of two-sided inverse operators. As a warmup for the following computations, we spell this out.

  $$
    \begin{tikzcd}[
      row sep=7pt,
      column sep=20pt
    ]
      (\mbox{-})
      \ar[rrrrr, equals]
      \ar[
        ddddd,
        "{
          \unit
            { \HilbertSpace{H}_1\mathrm{State} }
            { (\mbox{-}) }
        }"{swap}
      ]
      &[-30pt]
      &&
      &[-50pt]
      &[-40pt]
      (\mbox{-})
      \ar[
        ddddd,
        "{
          \unit
            { \HilbertSpace{H}_2\mathrm{State} }
            { (\mbox{-}) }
        }"
      ]
      \\[-10pt]
      &
      \scalebox{\termscale}{$  \vert\mbox{-}\rangle $}
      \ar[
        ddd,
        phantom,
        "{\longmapsto}"{sloped}
      ]
      \ar[
        rrr,
        phantom,
        "{ \longmapsto }"
      ]
      &&&
      \scalebox{\termscale}{$ \vert\mbox{-}\rangle $}
      \ar[
        dd,
        phantom,
        "{ \longmapsto }"{sloped}
      ]
      \\
      \\
      &&&&
      \scalebox{\termscale}{$
        \vert\mbox{-}\rangle
        \otimes
        \underset{w_2}{\sum}
        \,
        \vert w_2 \rangle
        \langle w_2 \vert
     $}
     \\
      &
    \scalebox{\termscale}{$
      \vert\mbox{-}\rangle
      \otimes
      \underset{w_1}{\sum}
      \,
      \vert w_1 \rangle
      \langle w_1 \vert
      $}
      \ar[
        rr,
        phantom,
        "{\longmapsto}"
      ]
      &&
      \scalebox{\termscale}{$
         \vert\mbox{-}\rangle
         \otimes
         \underset{w_1}{\sum}
         \,
         U \vert w_1 \rangle
         \langle w_1 \vert U^\dagger
      $}
      \ar[
        ur,
        equals,
        "{
          \scalebox{.6}{
            \color{gray}
            \eqref{ConjugationOfAnInsertedIdentity}
          }
        }"{sloped}
      ]
      \\[-12pt]
      (\mbox{-})
        \otimes
      \HilbertSpace{H}_1
        \otimes
      \HilbertSpace{H}_1^\ast
      \ar[
        rrrrr,
        "{
          (\mbox{-})
          \,\otimes
          \,
          U
          \,\otimes\,
          {U^{\dagger}}^\ast
        }"{swap}
      ]
      &&&&&
      (\mbox{-})
      \otimes
      \HilbertSpace{H}_2
        \otimes
      \HilbertSpace{H}_2^\ast
    \end{tikzcd}
  $$
$$
  \begin{tikzcd}[column sep=25pt, row sep=7pt]
    (\mbox{-})
    \otimes
    \HilbertSpace{H}_1
    \otimes
    \HilbertSpace{H}_1^\ast
    \otimes
    \HilbertSpace{H}_1
    \otimes
    \HilbertSpace{H}_1^\ast
    \ar[
      rrrrr,
      "{
        (\mbox{-})
        \,\otimes\,
        U \otimes {U^{\dagger}}^\ast
        \otimes
        U \otimes {U^{\dagger}}^\ast
      }"
    ]
    \ar[
      ddddd,
      "{
        \join
          { \HilbertSpace{H}_1\mathrm{State} }
          { (\mbox{-}) }
      }"{swap}
    ]
    &[-60pt]
    &&&[-90pt]
    &[-60pt]
    (\mbox{-})
    \otimes
    \HilbertSpace{H}_2
    \otimes
    \HilbertSpace{H}_2^\ast
    \otimes
    \HilbertSpace{H}_2
    \otimes
    \HilbertSpace{H}_2^\ast
    \ar[
      ddddd,
      "{
        \join
          { \HilbertSpace{H}_2\mathrm{State} }
          { (\mbox{-}) }
      }"
    ]
    \\
    &
    \scalebox{\termscale}{$
      \vert \mbox{-} \rangle
      \otimes
      \vert \mbox{-} \rangle
      \langle \phi \vert
      \otimes
      \vert \psi \rangle
      \langle \mbox{-} \vert
    $}
    \ar[
      rrr,
      phantom,
      "{ \longmapsto }"
    ]
    \ar[
      ddd,
      phantom,
      "{ \longmapsto }"{sloped}
    ]
    &&&
    \scalebox{\termscale}{$
      \vert \mbox{-} \rangle
      \otimes
      U \vert \mbox{-} \rangle
      \langle \psi \vert
      U^\dagger
      \otimes
      U \vert \psi \rangle
      \langle \mbox{-} \vert
      U
    $}
    \ar[
      dd,
      phantom,
      "{ \longmapsto }"{sloped}
    ]
    \\
    \\
    &&&&
    \scalebox{\termscale}{$
      \vert \mbox{-} \rangle
      \otimes
      U \vert \mbox{-} \rangle
      \langle \phi \vert
        U^\dagger
        U
      \vert \psi \rangle
      \langle \mbox{-} \vert U^\dagger
    $}
    \\
    &
    \scalebox{\termscale}{$
      \vert \mbox{-} \rangle
      \otimes
      \vert \mbox{-} \rangle
      \langle \phi \vert \psi \rangle
      \langle \mbox{-} \vert
    $}
    \ar[
      rr,
      phantom,
      "{ \longmapsto }"
    ]
    &&
    \scalebox{\termscale}{$
      \vert \mbox{-} \rangle
      \otimes
      U \vert \mbox{-} \rangle
      \langle \phi \vert \psi \rangle
      \langle \mbox{-} \vert U^\dagger
    $}
    \ar[
      ur,
      equals
    ]
    \\
    (\mbox{-})
    \otimes
    \HilbertSpace{H}_1
    \otimes
    \HilbertSpace{H}_1^\ast
    \ar[
      rrrrr,
      "{
        (\mbox{-})
        \,\otimes\,
        U \otimes {U^{\dagger}}^\ast
      }"{swap}
    ]
    &&&&&
    (\mbox{-})
    \otimes
    \HilbertSpace{H}_2
    \otimes
    \HilbertSpace{H}_2^\ast
  \end{tikzcd}
$$
  $$
    \begin{tikzcd}[
      row sep=7pt,
      column sep=20pt
    ]
      (\mbox{-})
      \otimes
      \HilbertSpace{H}_1
      \otimes
      \HilbertSpace{H}_1^\ast
      \ar[
        rrrrr,
        "{
          (\mbox{-})
          \,\otimes\,
          U \otimes U^{\dagger^\ast}
        }"
       ]
      \ar[
        ddddd,
        "{
          \counit
            { \HilbertSpace{H}_1^\ast\mathrm{Store} }
            { (\mbox{-}) }
        }"{swap}
      ]
      &[-30pt]
      &&
      &[-50pt]
      &[-40pt]
      (\mbox{-})
      \otimes
      \HilbertSpace{H}_2
      \otimes
      \HilbertSpace{H}_2^\ast
      \ar[
        ddddd,
        "{
          \counit
            { \HilbertSpace{H}_2^\ast\mathrm{Store} }
            { (\mbox{-}) }
        }"
      ]
      \\[-10pt]
      &
      \scalebox{\termscale}{$
        \vert\mbox{-}\rangle
        \otimes
        \vert \psi \rangle
        \langle \phi \vert
      $}
      \ar[
        ddd,
        phantom,
        "{\longmapsto}"{sloped}
      ]
      \ar[
        rrr,
        phantom,
        "{ \longmapsto }"
      ]
      &&&
      \scalebox{\termscale}{$
        \vert\mbox{-}\rangle
        \otimes
        U \vert \psi \rangle
        \langle \phi \vert U^\dagger
      $}
      \ar[
        dd,
        phantom,
        "{ \longmapsto }"{sloped}
      ]
      \\
      \\
      &&&&
      \scalebox{\termscale}{$
        \vert\mbox{-}\rangle
        \otimes
        \langle \phi \vert U^\dagger
        U \vert \psi \rangle
     $}
     \\
     &
     \scalebox{\termscale}{$
       \vert\mbox{-}\rangle
       \,
       \langle \phi \vert \psi \rangle
     $}
      \ar[
        rr,
        phantom,
        "{\longmapsto}"
      ]
      &&
      \scalebox{\termscale}{$
        \vert\mbox{-}\rangle
        \,
        \langle \phi \vert \psi \rangle
      $}
      \ar[
        ur,
        equals
      ]
      \\[-12pt]
      (\mbox{-})
      \ar[
        rrrrr,
        equals
      ]
      &&&&&
      (\mbox{-})
    \end{tikzcd}
$$
  $$
    \begin{tikzcd}[
      row sep=7pt,
      column sep=20pt
    ]
      (\mbox{-})
      \otimes
      \HilbertSpace{H}_1
      \otimes
      \HilbertSpace{H}_1^\ast
      \ar[
        rrrrr,
        "{
          (\mbox{-})
          \,\otimes\,
          U \otimes U^{\dagger^\ast}
        }"
       ]
      \ar[
        ddddd,
        "{
          \duplicate
            { \HilbertSpace{H}_1^\ast\mathrm{Store} }
            { (\mbox{-}) }
        }"{swap}
      ]
      &[-40pt]
      &&
      &[-110pt]
      &[-50pt]
      (\mbox{-})
      \otimes
      \HilbertSpace{H}_2
      \otimes
      \HilbertSpace{H}_2^\ast
      \ar[
        ddddd,
        "{
          \duplicate
            { \HilbertSpace{H}_2^\ast\mathrm{Store} }
            { (\mbox{-}) }
        }"
      ]
      \\[-10pt]
      &
      \scalebox{\termscale}{$
        \vert\mbox{-}\rangle
        \otimes
        \vert \psi \rangle
        \langle \phi \vert
      $}
      \ar[
        ddd,
        phantom,
        "{\longmapsto}"{sloped}
      ]
      \ar[
        rrr,
        phantom,
        "{ \longmapsto }"
      ]
      &&&
      \scalebox{\termscale}{$
        \vert\mbox{-}\rangle
        \otimes
        U \vert \psi \rangle
        \langle \phi \vert U^\dagger
      $}
      \ar[
        dd,
        phantom,
        "{ \longmapsto }"{sloped}
      ]
      \\
      \\
      &&&&
      \scalebox{\termscale}{$
        \vert\mbox{-}\rangle
        \otimes
        \underset{w}{\sum}
        U \vert \psi \rangle
        \langle w \vert
        \otimes
        \vert w \rangle
        \langle \phi \vert U^\dagger
      $}
     \\
     &
      \scalebox{\termscale}{$
        \vert\mbox{-}\rangle
        \otimes
        \underset{w}{\sum}
        \vert \psi \rangle
        \langle w \vert
        \otimes
        \vert w \rangle
        \langle \phi \vert
      $}
      \ar[
        rr,
        phantom,
        "{\longmapsto}"
      ]
      &&
      \scalebox{\termscale}{$
        \vert\mbox{-}\rangle
        \otimes
        \underset{w}{\sum}
        U \vert \psi \rangle
        \langle w \vert U^\dagger
        \otimes
        U \vert w \rangle
        \langle \phi \vert U^\dagger
      $}
      \ar[
        ur,
        equals,
        "{
          \scalebox{.6}{
            \color{gray}
            \eqref{DualConjugationOfAnInsertedIdentity}
          }
        }"{sloped}
      ]
      \\[-12pt]
      (\mbox{-})
      \ar[
        rrrrr,
        equals
      ]
      &&&&&
      (\mbox{-})
    \end{tikzcd}
$$
\vspace{-5mm}

\end{proof}

Here and in the following we make repeated use of the following elementary but important relations
for linear maps $E \,\isa\, \HilbertSpace{H}_1 \to \HilbertSpace{H}_2$:
\vspace{-2mm}
\begin{equation}
  \label{ConjugationOfAnInsertedIdentity}
  \begin{tikzcd}[row sep=-2pt, column sep=large]
    \HilbertSpace{H}_2
    \otimes
    \HilbertSpace{H}_2^\ast
    \ar[
      rr,
      "{ \sim }"
    ]
    &&
    \big(
    \HilbertSpace{H}_2
    \maplin
    \HilbertSpace{H}_2
    \big)
    \\
  \scalebox{\termscale}{$    \underset{w}{\sum}
    \,
    E
    \vert w \rangle \langle w \vert
    E^\dagger
    \;=\;
    E
    \Big(
    \underset{w}{\sum}
    \,
    \vert w \rangle \langle w \vert
    \Big)
    E^\dagger
    $}
    &\longmapsto&
  \scalebox{\termscale}{$    E
    \cdot
    \mathrm{id}_{\HilbertSpace{H}_1}
    \cdot
    E^\dagger
    \;=\;
    E
    \cdot
    E^\dagger
    $}
  \end{tikzcd}
\end{equation}
\begin{equation}
  \label{DualConjugationOfAnInsertedIdentity}
  \begin{tikzcd}[row sep=-2pt, column sep=large]
    \HilbertSpace{H}^\ast_2
    \otimes
    \HilbertSpace{H}_2
    \ar[
      rr,
      "{ \sim }"
    ]
    &&
    \big(
      \HilbertSpace{H}^\ast_2
      \maplin
      \HilbertSpace{H}^\ast_2
    \big)
    \\
 \scalebox{\termscale}{$     \underset{w}{\sum}
    \,
    \langle w \vert E^\dagger
    \otimes
    E \vert w \rangle
    \;=\;
    \underset{w}{\sum}
    \,
    \langle w \vert E^\dagger
    \otimes
    E \vert w \rangle
    $}
    &\longmapsto&
  \scalebox{\termscale}{$    \big(
      E \cdot E^\dagger
    \big)^\ast
    $}
  \end{tikzcd}
\end{equation}

The following Prop. \ref{QuantumStateEvolutionIsHeisenbergEvolution} invokes the covariant action  \eqref{CovariantFunctorOnModalesInducedByMonadTransformation} of monad transformations \eqref{MonadTransformation} on free modales, but restricted to the special case where the monad transformation is an isomorphism. In order to amplify the canonicity of this construction, the following Lemma \ref{EvolutionOfFreeModalesAlongIsomorphicTransformationOfMonads} highlights that in this case the transformation is equal to the inverse of the {\it contra}variant action \eqref{ExtensionOfModalesAlongMonadTransformations}
of monad morphisms of general modales (which is more commonly discussed in the monad-literature), restricted to free modales.

\begin{lemma}[\bf Evolution of free modales along isomorphic transformations of monads]
\label{EvolutionOfFreeModalesAlongIsomorphicTransformationOfMonads}
$\,$

\noindent {\bf (i)}  On \emph{isomorphic} monad transformation, $\mathrm{trans} \isa \mathcal{E} \xrightarrow{\sim} \mathcal{E}'$ \eqref{MonadTransformation},
the induced contravariant functor
$\mathrm{trans}^\ast$
\eqref{ExtensionOfModalesAlongMonadTransformations}
on general modales
is naturally isomorphic to the inverse $\circ \mathrm{trans}^{-1}$
of the induced covariant functor \eqref{CovariantFunctorOnModalesInducedByMonadTransformation}
on free
 modales \eqref{CategoriesOfModales},  via the  natural isomorphism whose components are just the
 components $\mathrm{trans}_{\scalebox{.7}{$(\mbox{-})$}}$ of
 the natural transformation $\mathrm{trans}$:
 \vspace{-1mm}
 \begin{equation}
   \label{NaturalityTransformationBetweenTransExtensionAndItsRestrictionToFreeModales}
   \begin{tikzcd}[
     row sep=20pt, column sep=large
   ]
     \mathcal{E}'
     \ar[
       from=rr,
       "{
         \mathrm{trans}
       }"{swap},
       "{
         \sim
       }"
     ]
     &&
     \mathcal{E}
     \\[-17pt]
     \Types_{\mathcal{E}'}
     \ar[
       d,
       hook
     ]
     \ar[
       rr,
       "{
         \mathrm{frtrans}^\ast
       }",
       "{
         \sim
       }"{swap}
     ]
     &&
     \Types_{\mathcal{E}}
     \ar[
       d,
       hook
     ]
     \ar[
       dll,
       Rightarrow,
       shorten=7pt,
       start anchor={[xshift=-15pt]},
       end anchor={[xshift=+15pt]},
       "{
         \mathrm{trans}_{
           \scalebox{.7}{$(\mbox{-})$}
         }
       }"{sloped, swap, yshift=-1pt}
     ]
     \\
     \Types^{\mathcal{E}'}
     \ar[
       rr,
       "{
         \mathrm{trans}^\ast
       }"{swap},
       "{
         \sim
       }"
     ]
     &&
     \Types^{\mathcal{E}}
   \end{tikzcd}
 \end{equation}

 \vspace{-1mm}
\noindent {\bf (ii)}  In that on Kleisli morphisms \eqref{KleisliEquivalence} this is given by postcomposition with the {\it inverse} transformation  $\mathrm{trans}^{-1}_{\scalebox{.7}{$(\mbox{-})$}}$ and as such
\vspace{-2mm}
\begin{equation}
  \label{EvolutionOfKleisliMorphisms}
  \mathrm{frtrans}^\ast
  \bind
    { \mathcal{E}' }
    {}
   \Big(
     D_1 \xrightarrow{ f' } \mathcal{E}'(D_2)
   \Big)
  \;\;
    =
  \;\;
  \bind
    { \mathcal{E} }
    {}
   \Big(
     D_1
       \xrightarrow{ f' }
     \mathcal{E}'(D_2)
       \xrightarrow{
         \mathrm{trans}^{-1}_{D_2}
       }
      \mathcal{E}(D_2)
   \Big).
\end{equation}
\end{lemma}
\begin{proof}
 First, notice the following diagram,
 which commutes by the defining properties of $\mathrm{trans}$ \eqref{MonadTransformationConditions}
 and the very definition of $\mathrm{trans}^\ast$ \eqref{ExtensionOfModalesAlongMonadTransformations}.
 \vspace{-3mm}
\begin{equation}\label{IsomorphicMonadTransformationActingOnFreeModales}
 \hspace{2cm}
 \mathllap{
  D \,\isa\, \Type
  \hspace{.8cm}
  \yields
  \hspace{.8cm}
  }
  \begin{tikzcd}[column sep=large]
    \mathcal{E}
    \mathcal{E}(D)
    \ar[
      rr,
      "{
        \mathcal{E}(\mathrm{trans}_D)
      }",
      "{
        \sim
      }"{swap}
    ]
    \ar[
      d,
      "{
        \join
          { \mathcal{E} }
          { D }
      }"
    ]
    &&
    \mathcal{E}
    \mathcal{E}'(D)
    \ar[
      rr,
      "{
        \mathrm{trans}_{\mathcal{E}'(D)}
      }"
    ]
    \ar[
      d,
      "{
        \rho
        \,\defneq\,
        \mathrm{trans}^\ast
        \rho'
      }"
    ]
    &&
    \mathcal{E}'
    \mathcal{E}'(D)
    \ar[
      d,
      "{
        \rho'
        \,\defneq\,
        \join
          { \mathcal{E}' }
          { D }
      }"
    ]
    \\
    \mathcal{E}(D)
    \ar[
      rr,
      "{
        \sim
      }",
      "{
        \mathrm{trans}_D
      }"{swap}
    ]
    &&
    \mathcal{E}'(D)
    \ar[
      rr,
      equals
    ]
    &&
    \mathcal{E}'(D)
    \\[-20pt]
    \mathclap{
      \scalebox{.7}{
        \color{darkblue}
        \bf
        free $\mathcal{E}$-modale
      }
    }
    \ar[
      rr,
      phantom,
      "{
        \scalebox{.7}{
          \color{darkgreen}
          \bf
          isomorphic to
        }
      }"{pos=.61, yshift=-1pt}
    ]
    &&
    \hspace{-10pt}
    \mathrlap{
      \scalebox{.7}{
        \color{darkblue}
        \bf
        transformation of
      }
    }
    &&
    \mathclap{
      \;\;\;
      \scalebox{.7}{
        \color{darkblue}
        \bf
        free $\mathcal{E}'$-modale
      }
    }
  \end{tikzcd}
\end{equation}

\vspace{-2mm}
\noindent But the left square now exhibits
$
  \mathrm{trans}_D
  \isa
  \mathcal{E}(D)
  \xrightarrow{\;\sim \;}
  \mathrm{trans}^\ast \mathcal{E}'(D)
$
as a homomorphism of modales \eqref{ModaleHomomorphism} from the free $\mathcal{E}$-modale on $D$ to the transformation
of the free $\mathcal{E}'$-modale on $D$; and this
homomorphism is an isomorphism by the assumption that $\mathrm{trans}$ is an isomorphism, as shown.
Therefore the claimed natural transformation in \eqref{NaturalityTransformationBetweenTransExtensionAndItsRestrictionToFreeModales}
is given in components as follows:
\vspace{-3mm}
\begin{equation}
  \label{NaturalIsomorphismBetweenInverseCovariantTransformAndContravariantTransform}
  \begin{tikzcd}[row sep=12pt]
    \Types_{\mathcal{E}'}
    \ar[
      rrr,
      bend left=15,
      "{
        \mathrm{frtrans}^\ast
      }"{description},
      "{\ }"{swap, name=s}
    ]
    \ar[
      rrr,
      bend right=15,
      "{
        \mathrm{trans}^\ast
      }"{description},
      "{\ }"{name=t}
    ]
    \ar[
      from=s,
      to=t,
      Rightarrow,
      "{
        \mathrm{trans}_{\scalebox{.7}{$(\mbox{-})$}}
      }"
    ]
    &&&
   \Types^{\mathcal{E}}
    \\
      \scalebox{\termscale}{$   \mathcal{E}'(D_1) $}
    \ar[
      ddd,
      "{
        \phi
      }"
    ]
    &\mapsto&
       \scalebox{\termscale}{$  \mathcal{E}(D_1) $}
    \ar[
      d,
      "{
        \mathrm{trans}_{D_1}
      }"
    ]
    \ar[
      rr,
      "{
        \mathrm{trans}_{D_1}
      }",
      "{
        \sim
      }"{swap}
    ]
    &&
     \scalebox{\termscale}{$    \mathcal{E}'(D_1) $}
    \ar[
      ddd,
      "{
        \phi
      }"
    ]
    \\
    &&
    \scalebox{\termscale}{$     \mathcal{E}'(D_1) $}
    \ar[
      d,
      "{
        \phi
      }"
    ]
    \\
    &&
      \scalebox{\termscale}{$   \mathcal{E}'(D_2) $}
    \ar[
      d,
      "{
        \mathrm{trans}_{D_2}^{-1}
      }"
    ]
    \\
      \scalebox{\termscale}{$   \mathcal{E}'(D_2) $}
    &\mapsto&
     \scalebox{\termscale}{$    \mathcal{E}(D_2) $}
    \ar[
      rr,
      "{
        \sim
      }",
      "{
        \mathrm{trans}_{D_2}
      }"{swap}
    ]
    &&
       \scalebox{\termscale}{$  \mathcal{E}'(D_2) $}
  \end{tikzcd}
\end{equation}
From this, we get the following commuting diagram, where the left square commutes by the transformation property \eqref{ConditionsOnMonadTransformer} while the right square commutes by \eqref{NaturalIsomorphismBetweenInverseCovariantTransformAndContravariantTransform}:
\vspace{-2mm}
$$
  \begin{tikzcd}[row sep=40pt,
   column sep=65pt
  ]
    D_1
    \ar[
      r,
      "{
        \unit
          { \mathcal{E} }
          { D_1 }
      }"{sloped}
    ]
    \ar[
      d,
      equals
    ]
    &
    \mathcal{E}(D_1)
    \ar[
      r,
      "{
        \mathrm{frtrans}^\ast
        \bind
          { \mathcal{E}' }
          { }
        f'
      }"
    ]
    \ar[
      d,
      "{
        \mathrm{trans}_{D_1}
      }"{description}
    ]
    &[15pt]
    \mathcal{E}(D_2)
    \ar[
      from=d,
      "{
        \mathrm{trans}^{-1}_{D_2}
      }"{description}
    ]
    \\
    D_1
    \ar[
      r,
      "{
        \unit
          { \mathcal{E}' }
          { D_1 }
      }"{swap}
    ]
    \ar[
      rr,
      rounded corners,
      to path={
           ([yshift=-00pt]\tikztostart.south)
        -- ([yshift=-11pt]\tikztostart.south)
        -- node[yshift=+6pt]{
             \scalebox{.7}{$
               f'
            $}
        }
            ([yshift=-10pt]\tikztotarget.south)
        -- ([yshift=-00pt]\tikztotarget.south)
      }
    ]
    &
    \mathcal{E'}(D_1)
    \ar[
      r,
      "{
        \bind
          { \mathcal{E}' }
          {}
        f'
      }"{swap}
    ]
    &
    \mathcal{E}'(D_2)
  \end{tikzcd}
$$
and the claim \eqref{EvolutionOfKleisliMorphisms} is the image under $\bind{\mathcal{E}}{}$ of this equality.
\end{proof}

As we apply (in Prop. \ref{QuantumStateEvolutionIsHeisenbergEvolution}) Lem. \ref{EvolutionOfFreeModalesAlongIsomorphicTransformationOfMonads} to QuantumStore-contextful maps, hence to Kleisli maps for a {\it co}monad, beware that the role of covariant and contravariant functors gets interchanged.

\begin{proposition}[\bf QuantumState evolution is Heisenberg evolution]
  \label{QuantumStateEvolutionIsHeisenbergEvolution}
  For $U \,\isa\, \HilbertSpace{H}_1 \to \HilbertSpace{H}_2$ a unitary linear map,
  the canonical evolution according to Lem. \ref{EvolutionOfFreeModalesAlongIsomorphicTransformationOfMonads}
  \begin{itemize}
  \item of quantum observables  regarded a QuantumState-contextful scalars $\mathcal{O}_A$ (via Ex. \ref{AlgebraOfQuantumObservablesAsQuantumStoreContextfulMaps})

  \item along the unitary quantum channel
  $\mathrm{chan}^U$
  regarded as a QuantumState transformation (via Prop. \ref{UnitaryQuantumChannelsAreQuantumStateTransformations})
  \end{itemize}
  is Heisenberg evolution \eqref{HeisenbergEvolution}

  \vspace{-.6cm}
  \begin{equation}
    \label{HeisenbergEvolutionAsModaleTransformation}
    \begin{tikzcd}[
      row sep=-2pt,
      column sep=10pt
    ]
      \HilbertSpace{H}_2
        \otimes
      \HilbertSpace{H}_2^\ast
      \ar[
        rr,
        "{
         \mathrm{chan}^{U^{-1}}
        }"
      ]
      \ar[
        rrrr,
        rounded corners,
        to path={
             ([yshift=+00pt]\tikztostart.north)
          -- ([yshift=+14pt]\tikztostart.north)
          -- node[yshift=6pt] {
             \scalebox{.8}{$
               \mathcal{O}_{ U \cdot A \cdot U^\dagger }
             $}
          }
             ([yshift=+15pt]\tikztotarget.north)
          -- ([yshift=+00pt]\tikztotarget.north)
        }
      ]
      &&
      \HilbertSpace{H}_1
        \otimes
      \HilbertSpace{H}_1^\ast
      \ar[
        rr,
        "{
          \mathcal{O}_A
        }"
      ]
      &&
      \TensorUnit
      \\
     \scalebox{\termscale}{$\rho $}
      &\longmapsto&
    \scalebox{\termscale}{$
      U^\dagger \cdot \rho \cdot U
    $}
    &\longmapsto&
    \scalebox{\termscale}{$
     \mathrm{tr}
      \big(
        (U^\dagger \cdot \rho \cdot U)
        \cdot A
      \big)
      $}
      \\
      && &=&
      \scalebox{\termscale}{$
       \mathrm{tr}
       \big(
         \rho \cdot
         ( U  \cdot A \cdot U^\dagger )
      \big)
      $}
      \\
      && &=&
      \scalebox{\termscale}{$
        \mathcal{O}_{ U \cdot A \cdot U^\dagger }
        ( \rho )
      $}
      \mathrlap{\,.}
    \end{tikzcd}
  \end{equation}
\end{proposition}

\smallskip

\noindent

\noindent
{\bf Quantum channels as QuantumState transformations.}
\begin{proposition}[\bf Uniform coupling channels are QuantumState transformations]
\label{UniformCouplingChannelsAreQuantumStateTransformations}
The quantum coupling channels to a uniform bath state \eqref{UniformCouplingQuantumChannel} of some system $\HilbertSpace{B}$
$$
  \begin{tikzcd}[
    row sep=0pt,
    column sep=10pt
  ]
    (\mbox{-})
    \otimes
    \HilbertSpace{H}
    \otimes
    \HilbertSpace{H}^\ast
    \ar[
      rrr,
      "{
        \mathrm{id}
        \otimes
        \unit
          {
            \HilbertSpace{B}\mathrm{State}
          }
          { \TensorUnit }
      }"
    ]
    &&&
    (\mbox{-})
    \otimes
    \HilbertSpace{H}
      \otimes
    \HilbertSpace{B}
    \otimes
    \HilbertSpace{B}^\ast
      \otimes
    \HilbertSpace{H}
    \ar[
      rr,
      "{ \sim }"
    ]
    &[-5pt]&[-10pt]
    (\mbox{-})
    \otimes
    (\HilbertSpace{H}
      \otimes
    \HilbertSpace{B})
    \otimes
    (\HilbertSpace{H}
      \otimes
    \HilbertSpace{B})^\ast
    \\
   \scalebox{.8}{$   \vert \mbox{-} \rangle
    \otimes
    \vert \psi \rangle
    \langle \psi' \vert
    $}
    &\longmapsto&&
 \scalebox{.8}{$     \vert \mbox{-} \rangle
    \otimes
    \vert \psi \rangle
    \Big(
      \underset{b}{\sum}
      \,
      \vert b \rangle
      \langle b \vert
    \Big)
    \langle \psi' \vert
    $}
    &=&
    \scalebox{.8}{$  \vert \mbox{-} \rangle
    \otimes
    \underset{b}{\sum}
    \vert \psi, b \rangle
    \langle b, \psi' \vert
    $}
  \end{tikzcd}
$$

\vspace{-2mm}
\noindent are monadic QuantumState transformations
\vspace{-2mm}
$$
  \mathllap{
    \mathrm{couple}^{\HilbertSpace{B}}
    \;\isa\;
  }
  \begin{tikzcd}
    \HilbertSpace{H}\mathrm{State}
    \ar[
      rr,
      "{
        \mathrm{mon}
      }"
    ]
    &&
    (\HilbertSpace{H}\otimes\HilbertSpace{B})\mathrm{State}
  \end{tikzcd}
$$

\vspace{-2mm}
\noindent
and as such the components of a pointed endofunctor \eqref{PointedEndofunctorOnCategoryOfMonads} on $\mathrm{Mnd}(\QuantumTypes)$.
\end{proposition}
\begin{proof}
Since the structure maps of the $(\HilbertSpace{H} \otimes \HilbertSpace{B})\mathrm{State}$-comonad are tensor products of structure maps of $\HilbertSpace{H}\mathrm{State}$ and $\HilbertSpace{B}\mathrm{State}$,
it is sufficient to show this for $\HilbertSpace{H} = \TensorUnit$, hence for the case that $\HilbertSpace{H}\mathrm{State} = \mathrm{Id}$.
But in this case $\mathrm{couple}^{\HilbertSpace{B}} \,=\, \unit{ \HilbertSpace{B}\mathrm{State} }{(\mbox{-})}$, which we know to be
a monadic transformation (in fact the initial one) according to \eqref{InitialMonadTransformation}.

Alternatively, it is immediate to explicitly check the required conditions. We have:
$$
  \begin{tikzcd}[row sep=15pt, column sep=large]
    (\mbox{-})
    \ar[
      rrrrr,
      equals
    ]
    \ar[
      ddddd,
      "{
        \unit
          { \HilbertSpace{H}\mathrm{State} }
        { (\mbox{-}) }
      }"{swap}
    ]
    &[-35pt]&&&[-60pt]&[-44pt]
    (\mbox{-})
    \ar[
      ddddd,
      "{
        \unit
          {( \HilbertSpace{H}\otimes \HilbertSpace{B})\mathrm{State} }
        { (\mbox{-}) }
      }"
    ]
    \\[-13pt]
    &
  \scalebox{0.8}{$  \vert \mbox{-} \rangle $}
    \ar[
      rrr,
      phantom,
      "{
        \longmapsto
      }"
    ]
    \ar[
      ddd,
      phantom,
      "{ \longmapsto }"{sloped}
    ]
    &&&
  \scalebox{0.8}{$    \vert \mbox{-} \rangle $}
    \ar[
      dd,
      phantom,
      "{ \longmapsto }"{sloped}
    ]
    \\
    \\
    &&&&
 \scalebox{0.8}{$     \vert \mbox{-} \rangle
    \otimes
    \underset{w,b}{\sum}
    \,
    \vert w, b \rangle
    \langle b, w \vert
    $}
    \\[-8pt]
    &
  \scalebox{0.8}{$    \vert\mbox{-}\rangle
    \otimes
    \underset{w}{\sum}
    \,
    \vert w \rangle
    \langle w \vert
    $}
    \ar[
      rr,
      phantom,
      "{ \longmapsto }"
    ]
    &&
   \scalebox{0.8}{$   \vert\mbox{-}\rangle
    \otimes
    \underset{w,b}{\sum}
    \,
    \vert w \rangle
    \otimes
    \vert b \rangle \langle b \vert
    \otimes
    \langle w \vert
    $}
    \ar[
      ur,
      equals
    ]
    \\[-20pt]
    (\mbox{-})
    \otimes
    \HilbertSpace{H}
    \otimes
    \HilbertSpace{H}^\ast
    \ar[
      rrrrr,
      "{
        \mathrm{couple}
          ^{\HilbertSpace{B}}
      }"{swap}
    ]
    &&&&&
    (\mbox{-})
    \otimes
    \HilbertSpace{H}
    \otimes
    \HilbertSpace{B}
    \otimes
    \HilbertSpace{B}^\ast
    \otimes
    \HilbertSpace{H}^\ast
  \end{tikzcd}
$$
$$
  \begin{tikzcd}[
   row sep=15pt, column sep=large
  ]
    (\mbox{-})
    \otimes
    \HilbertSpace{H}
    \otimes
    \HilbertSpace{H}^\ast
    \otimes
    \HilbertSpace{H}
    \otimes
    \HilbertSpace{H}^\ast
    \ar[
      ddddd,
      "{
        \join
          { \HilbertSpace{H}\mathrm{State} }
          { (\mbox{-}) }
      }"{swap}
    ]
    \ar[
      rrrrr,
      "{
        \mathrm{couple}
          ^{\HilbertSpace{B}}
          _{(\cdots)}
        \circ
        (
        \mathrm{couple}
          ^{\HilbertSpace{B}}
        \otimes
        \mathrm{id}
        )
      }"
    ]
    &[-55pt]&&&[-85pt]&[-93pt]
    (\mbox{-})
    \otimes
    \HilbertSpace{H}
    \otimes
    \HilbertSpace{B}
    \otimes
    \HilbertSpace{B}^\ast
    \otimes
    \HilbertSpace{H}^\ast
    \otimes
    \HilbertSpace{H}
    \otimes
    \HilbertSpace{B}
    \otimes
    \HilbertSpace{B}^\ast
    \otimes
    \HilbertSpace{H}^\ast
    \ar[
      ddddd,
      "{
        \join
          {
            (\HilbertSpace{H}\otimes\HilbertSpace{B})\mathrm{State}
          }
          { (\mbox{-}) }
      }"
    ]
    \\[-10pt]
    &
  \scalebox{0.8}{$      \vert \psi \rangle
      \langle \psi ' \vert
      \otimes
      \vert \phi \rangle
      \langle \phi' \vert
      $}
    \ar[
      rrr,
      phantom,
      "{ \longmapsto }"
    ]
    \ar[
      ddd,
      phantom,
      "{ \longmapsto }"{sloped}
    ]
    &&&
 \scalebox{0.8}{$     \underset{b,b'}{\sum}
      \vert \psi, b \rangle
      \langle b, \psi ' \vert
      \otimes
      \vert \phi, b' \rangle
      \langle b', \phi' \vert
      $}
    \ar[
      dd,
      phantom,
      "{ \longmapsto }"{sloped}
    ]
    \\
    \\
    &&&&
   \scalebox{0.8}{$   \langle \psi' \vert \phi \rangle
    \underset{b,b'}{\sum}
    \vert \psi, b \rangle
    \langle b \vert b' \rangle
    \langle b', \phi' \vert
    $}
    \\[-5pt]
    &
   \scalebox{0.8}{$   \vert \psi
    \langle \psi' \vert \phi \rangle
    \langle \phi' \vert
    $}
    \ar[
      rr,
      phantom,
      "{ \longmapsto }"
    ]
    &&
   \scalebox{0.8}{$   \langle \psi'  \vert \phi \rangle
    \underset{b}{\sum}
    \,
    \vert \psi, b \rangle
    \langle b, \phi' \vert
    $}
    \ar[
      ur,
      equals
    ]
    \\[-15pt]
    (\mbox{-})
    \otimes
    \HilbertSpace{H}
    \otimes
    \HilbertSpace{H}^\ast
    \ar[
      rrrrr,
      "{
        \mathrm{couple}^{\HilbertSpace{B}}
      }"{swap}
    ]
    &&&&&
    (\mbox{-})
    \otimes
    \HilbertSpace{H}
    \otimes
    \HilbertSpace{B}
    \otimes
    \HilbertSpace{B}^\ast
    \otimes
    \HilbertSpace{H}^\ast
    \mathrlap{\,.}
  \end{tikzcd}
$$
Alternatively, with Rem. \ref{QuantumStateEffectAsQuantumWriter} it is sufficient to observe that tensoring with an identity matrix $A \mapsto A \otimes I_{\HilbertSpace{B}}$ is an algebra homomorphism.

Finally, it is immediate that the naturality squares \eqref{PointedEndofunctorOnCategoryOfMonads} for a pointed endofunctor commute, by functoriality of the tensor product.
\end{proof}

Dually, we have:

\begin{proposition}[\bf Averaging quantum channels are QuantumStore transformations]
\label{AveragingQuantumChannelAsQuantumStateTransformation}
The {\it averaging quantum channel} \eqref{AveraginQuantumChannel}
\vspace{-2mm}
$$
  \begin{tikzcd}[
    row sep=0pt
  ]
    (\mbox{-})
    \otimes
    \big(
      \HilbertSpace{H}
      \otimes
      \HilbertSpace{B}
    \big)
    \otimes
    \big(
      \HilbertSpace{H}
      \otimes
      \HilbertSpace{B}
    \big)^\ast
    \ar[
      rr,
      "{
        \sim
      }"
    ]
    &[-20pt]&[-20pt]
    (\mbox{-})
    \otimes
      \HilbertSpace{H}
      \otimes
      \HilbertSpace{B}
      \otimes
      \HilbertSpace{B}^\ast
      \otimes
      \HilbertSpace{H}^\ast
    \ar[
      rr,
      "{
        \mathrm{id}
        \,\otimes\,
        \counit
          { \HilbertSpace{B}^\ast\mathrm{Store} }
          { \TensorUnit }
        \,\otimes\,
        \mathrm{id}
      }"{yshift=1pt}
    ]
    &&
    (\mbox{-})
     \,\otimes\,
    \big(
      \HilbertSpace{H}
      \otimes
      \HilbertSpace{H}^\ast
    \big)
    \\
    \scalebox{.8}{$  \vert \mbox{-} \rangle
    \otimes
    \vert \psi, \beta \rangle
    \langle \beta', \psi' \vert
    $}
    &=&
    \scalebox{.8}{$  \vert \mbox{-} \rangle
    \otimes
    \vert \psi \rangle
    \otimes
    \vert \beta \rangle
    \langle \beta' \vert
    \otimes
    \langle \psi' \vert
    $}
    &\longmapsto&
     \scalebox{.8}{$
    \vert \mbox{-} \rangle
    \otimes
    \vert \psi \rangle
    \langle \beta' \vert \beta \rangle
    \langle \psi' \vert
    $}
  \end{tikzcd}
$$
is a comonadic QuantumState-transformation
$$
  \mathrm{Tr}^{\HilbertSpace{B}}
  \,\isa\,
  \begin{tikzcd}
  (
    \HilbertSpace{H}
    \otimes
    \HilbertSpace{B}
  )\mathrm{State}
  \ar[
    rr,
    "{
      \mathrm{comon}
    }"
  ]
  &&
  \HilbertSpace{H}\mathrm{State}
  \end{tikzcd}
$$
and as such the component of a pointed endofunctor \eqref{PointedEndofunctorOnCategoryOfMonads} on $\mathrm{Mnd}(\QuantumTypes)$.
\end{proposition}
\begin{proof}
Since the structure maps of the $(\HilbertSpace{H} \otimes \HilbertSpace{B})\mathrm{State}$-comonad are tensor products of structure maps of $\HilbertSpace{H}\mathrm{State}$ and $\HilbertSpace{B}\mathrm{State}$,
it is sufficient to show this for $\HilbertSpace{H} = \TensorUnit$, hence for the case that $\HilbertSpace{H}\mathrm{State} = \mathrm{Id}$.
But in this case $\mathrm{Tr}^{\HilbertSpace{B}} \,=\, \counit{ \HilbertSpace{B}^\ast\mathrm{Store} }{ (\mbox{-}) }$, which we know to be a comonadic transformation according to \eqref{InitialMonadTransformation}.

Alternatively, it is immediate to explicitly check the required conditions. We have:
  $$
    \begin{tikzcd}[
      row sep=6pt,
      column sep=20pt
    ]
      (\mbox{-})
      \otimes
      \HilbertSpace{H}
      \otimes
      \HilbertSpace{B}
      \otimes
      \HilbertSpace{B}^\ast
      \otimes
      \HilbertSpace{H}^\ast
      \ar[
        rrrrr,
        "{
          \mathrm{Tr}^{\mathcal{B}}
           _{{ (\mbox{-}) }}
        }"
      ]
      \ar[
        ddddd,
        "{
          \duplicate
            {
              (
                \HilbertSpace{H}
                \otimes
                \HilbertSpace{B}
              )^\ast
              \mathrm{Store}
            }
            { (\mbox{-}) }
        }"{swap}
      ]
      &[-100pt]
      &&
      &[-60pt]
      &[-50pt]
      (\mbox{-})
      \otimes
      \HilbertSpace{H}
      \otimes
      \HilbertSpace{H}^\ast
      \ar[
        ddddd,
        "{
          \duplicate
            {
             \HilbertSpace{H}^\ast\mathrm{Store}
            }
            { (\mbox{-}) }
        }"
      ]
      \\[-10pt]
      &
      \scalebox{\termscale}{$
        \vert \psi, \beta \rangle
        \langle \beta', \psi' \vert
      $}
      \ar[
        ddd,
        phantom,
        "{\longmapsto}"{sloped}
      ]
      \ar[
        rrr,
        phantom,
        "{ \longmapsto }"
      ]
      &&&
       \scalebox{\termscale}{$
         \vert \psi \rangle
         \langle \beta' \vert \beta \rangle
         \langle \psi' \vert
       $}
      \ar[
        dd,
        phantom,
        "{ \longmapsto }"{sloped}
      ]
      \\
      \\
      &&&&
      \scalebox{\termscale}{$
        \mathllap{
        \langle \beta' \vert \beta \rangle
        \,
        \underset{w}{\sum}
        \,
        }
        \vert \psi \rangle
        \langle w \vert
        \otimes
        \vert w \rangle
        \langle \psi' \vert
      $}
      \\
      &
      \scalebox{\termscale}{$
        \underset{w,b}{\sum}
        \vert \psi, \beta \rangle
        \langle  b, w \vert
        \otimes
        \vert w, b \rangle
        \langle \beta', \psi' \vert
      $}
      \ar[
        rr,
        phantom,
        "{\longmapsto}"
      ]
      &&
      \scalebox{\termscale}{$
        \underset{w, b}{\sum}
        \vert \psi \rangle
        \langle b \vert \beta \rangle
        \langle w \vert
        \otimes
        \vert w \rangle
        \langle \beta' \vert b \rangle
        \langle \beta' \vert
      $}
      \ar[
        ur, equals
      ]
      \\[-5pt]
      (\mbox{-})
        \otimes
      \HilbertSpace{H}
        \otimes
      \HilbertSpace{B}
        \otimes
      \HilbertSpace{B}^\ast
        \otimes
      \HilbertSpace{H}^\ast
        \otimes
      \HilbertSpace{H}
        \otimes
      \HilbertSpace{B}
        \otimes
      \HilbertSpace{B}^\ast
        \otimes
      \HilbertSpace{H}^\ast
      \ar[
        rrrrr,
        "{
          \mathrm{Tr}^{\HilbertSpace{B}}
            _{ (\cdots) }
          \circ
          (
            \mathrm{Tr}^{\mathcal{B}}
               _{ (\mbox{-}) }
            \otimes
            \mathrm{id}
          )
        }"{swap}
      ]
      &&&&&
      (\mbox{-})
      \otimes
      \HilbertSpace{H}
      \otimes
      \HilbertSpace{H}^\ast
      \otimes
      \HilbertSpace{H}
      \otimes
      \HilbertSpace{H}^\ast
    \end{tikzcd}
  $$
  and
  $$
    \begin{tikzcd}[
      row sep=8pt,
      column sep=20pt
    ]
      (\mbox{-})
      \otimes
      \HilbertSpace{H}
        \otimes
      \HilbertSpace{B}
        \otimes
      \HilbertSpace{B}^\ast
        \otimes
      \HilbertSpace{H}^\ast
      \ar[
        rrrrr,
        "{
          \mathrm{Tr}^{\HilbertSpace{B}}
           _{ (\mbox{-}) }
        }"
      ]
      \ar[
        ddddd,
        "{
          \counit
            {
               (\HilbertSpace{H}\otimes\HilbertSpace{B})^\ast \mathrm{Store}
            }
            { (\mbox{-}) }
        }"{swap}
      ]
      &[-52pt]
      &&
      &[-40pt]
      &[-30pt]
      (\mbox{-})
      \otimes
      \HilbertSpace{H}
        \otimes
      \HilbertSpace{H}^\ast
      \ar[
        ddddd,
        "{
          \counit
            { \HilbertSpace{H}^\ast\mathrm{Store} }
            {(\mbox{-})}
        }"
      ]
      \\[-5pt]
      &
      \scalebox{\termscale}{$
        \vert \psi, \beta \rangle
        \langle \beta', \psi' \vert
      $}
      \ar[
        ddd,
        phantom,
        "{\longmapsto}"{sloped}
      ]
      \ar[
        rrr,
        phantom,
        "{ \longmapsto }"
      ]
      &&&
      \scalebox{\termscale}{$
        \vert \psi\rangle
        \langle \beta' \vert \beta \rangle
        \langle \psi' \vert
      $}
      \ar[
        dd,
        phantom,
        "{ \longmapsto }"{sloped}
      ]
      \\
      \\
      &&&&
      \scalebox{\termscale}{$
        \langle \beta' \vert \beta \rangle
        \langle \psi' \vert \psi \rangle
      $}
      \\
      &
      \scalebox{\termscale}{$
        \langle \beta', \psi'
        \vert
        \psi, \beta \rangle
      $}
      \ar[
        rr,
        phantom,
        "{\longmapsto}"
      ]
      &&
      \scalebox{\termscale}{$
        \langle \beta', \psi'
        \vert
        \psi, \beta \rangle
      $}
      \ar[
        ur, equals
      ]
      \\[-8pt]
      (\mbox{-})
      \ar[
        rrrrr,
        equals
      ]
      &&&&&
      (\mbox{-})
    \end{tikzcd}
  $$
Alternatively, with Rem. \ref{QuantumStateEffectAsQuantumWriter} it is sufficient to observe that partial tracing is a coalgebra homomorphism.

Finally, it is again immediate that the naturality squares \eqref{PointedEndofunctorOnCategoryOfMonads} for a pointed endofunctor commute,
by functoriality of the tensor product.
\end{proof}
\begin{remark}[Partial trace]
On the other hand, partial trace is not a monadic QuantumState transformation beyond the trivial case of $\mathrm{dim}(\HilbertSpace{B}) =  1$:
\vspace{-2mm}
  $$
    \begin{tikzcd}[
      row sep=6pt,
      column sep=14pt
    ]
      (\mbox{-})
      \ar[rrrrr, equals]
      \ar[
        ddddd,
        "{
          \unit
            { (\HilbertSpace{H} \otimes \HilbertSpace{B}) \mathrm{State} }
            { (\mbox{-}) }
        }"{swap}
      ]
      &[-30pt]
      &&
      &[-70pt]
      &[-40pt]
      (\mbox{-})
      \ar[
        ddddd,
        "{
          \unit
            { (\HilbertSpace{H} \otimes \HilbertSpace{B})\mathrm{State} }
            { (\mbox{-}) }
        }"
      ]
      \\[-10pt]
      &
       \scalebox{\termscale}{$    \vert\mbox{-}\rangle
       $}
      \ar[
        ddd,
        phantom,
        "{\longmapsto}"{sloped}
      ]
      \ar[
        rrr,
        phantom,
        "{ \longmapsto }"
      ]
      &&&
        \scalebox{\termscale}{$   \vert\mbox{-}\rangle
        $}
      \ar[
        dd,
        phantom,
        "{ \longmapsto }"{sloped}
      ]
      \\
      \\
      &&&&
      \scalebox{\termscale}{$     \vert\mbox{-}\rangle
        \otimes
      \underset{w}{\sum}
      \,
      \vert w \rangle
      \langle w \vert
      $}
      \\
      &
        \scalebox{\termscale}{$   \vert\mbox{-}\rangle
      \otimes
      \underset{w, b}{\sum}
      \,
      \vert w, b \rangle
      \langle b, w \vert
      $}
      \ar[
        rr,
        phantom,
        "{\longmapsto}"
      ]
      &&
       \scalebox{\termscale}{$    \vert\mbox{-}\rangle
        \otimes
      \mathrm{dim}(\HilbertSpace{B})
      \underset{w}{\sum}
      \,
      \vert w \rangle
      \langle w \vert
      $}
      \ar[
        ur,
        phantom,
        "{ \neq }"{sloped, pos=.8}
      ]
      \\[-12pt]
      (\mbox{-})
        \otimes
      \HilbertSpace{H}
        \otimes
      \HilbertSpace{B}
        \otimes
      \HilbertSpace{B}^\ast
        \otimes
      \HilbertSpace{H}^\ast
      \ar[
        rrrrr,
        "{
          (\mbox{-})
          \otimes
          \mathrm{Tr}_{\HilbertSpace{B}}
        }"{swap}
      ]
      &&&&&
      (\mbox{-})
        \otimes
      \HilbertSpace{H}
        \otimes
      \HilbertSpace{H}^\ast
    \end{tikzcd}
  $$
\end{remark}

\begin{corollary}[Quantum states as transformations]
  Every unistochastic quantum channel \eqref{UniformCouplingQuantumChannel} is a monadic QuantumState transformation
  (coupling and unitary evolution) followed by a comonadic QuantumState transformation (evolution and averaging).
\end{corollary}

\noindent
{\bf Interaction between QuantumState and QuantumEnvironment.} Recall from \cref{QuantumEpistemicLogicViaDependentLinearTypes} the
monadic indefiniteness modality (QuantumReader) $\indefinitely_W$ and the comonadic randomness modality (QuantumCoreader) $\randomly_W$.

\begin{remark}[\bf QuantumEnvironment monad]
  In its interaction with the QuantumState-monad, the epistemic modality $\indefinitely_W$/$\randomly_W$
  or $W$-{\it Reader} (co)monad is suggestively referred to under its alternative name $W$-{\it environment} (co)-monad, and as such we will denote it  ``$W\Environment$'' and understand it as a Frobenius monad. Hence all the following names refer to the same monadic structure on linear types (cf. Prop. \ref{QuantumCoEffectsViaFrobeniusAlgebra}):
  \vspace{-2mm}
  $$
    \begin{tikzcd}[
      row sep=2pt,
      column sep=5pt
    ]
      &&
      \quantized W \mathrm{Writer}
      \ar[
        d,
        phantom,
        "{ \simeq }"{sloped}
      ]
      \\[+4pt]
      \underset{W}{\indefinitely}
      &&
      W\Environment
      \ar[rr, phantom, "{\simeq}"]
      \ar[ll, phantom, "{\simeq}"]
      &&
      \underset{W}{\randomly}
      \\[-2pt]
      \mathrm{Monads}
      &&
      \mathrm{FrobMonad}
      \ar[rr]
      \ar[ll]
      &&
      \mathrm{Comonad}
    \end{tikzcd}
  $$
\end{remark}

\begin{proposition}[\bf
QuantumState and QuantumEnvironment distribute]
\label{QuantumStoreDistributesOverQuantumRreader} $\,$

\noindent   For $\HilbertSpace{H} \,\isa\, \DualizableQuantumType$ and $W \,\isa\, \FiniteType$
\begin{itemize}
\item[{\bf (i)}] the natural isomorphism
\vspace{-2mm}
\begin{equation} \label{DistributivityIsomorphismBetweenQuantumStateAndQuantumReader}
\hspace{-6mm}
  \begin{tikzcd}[
    row sep=-3pt,
    column sep=25pt
  ]
    \HilbertSpace{H}\State
    \Big(
    \underset{W}{\indefinitely}
    \,
    \HilbertSpace{K}
    \Big)
    \ar[
      r,
      phantom,
      "{ \defneq }"
    ]
    &
    \Big(
    \underset{W}{\osum}
    \,
    \HilbertSpace{K}
    \Big)
    \otimes
    \HilbertSpace{H}
    \otimes
    \HilbertSpace{H}^\ast
    \ar[
      rr,
      "{
        \distribute
          {
            \HilbertSpace{H}\State
            ,\,
            \indefinitely_W
          }
          { \HilbertSpace{K} }
      }",
      "{ \sim }"{swap}
    ]
    &&
    \underset{W}{\osum}
    \big(
    \HilbertSpace{K}
    \otimes
    \HilbertSpace{H}
    \otimes
    \HilbertSpace{H}^\ast
    \big)
    \ar[
      r,
      phantom,
      "{ \defneq }"
    ]
    &
    \underset{W}{\indefinitely}
    \big(
      \HilbertSpace{H}\State
      (\HilbertSpace{K})
    \big)
    \\
    &
    \hspace{-6pt}
    \scalebox{\termscale}{$
    \big(
      w,
      \,
      \vert \kappa \rangle
    \big)
    \otimes
    \vert \psi \rangle
    \langle \psi' \vert
    $}
    &\longleftrightarrow&
    \scalebox{\termscale}{$
    \big(
      w
      ,\,
      \vert \kappa \rangle
      \otimes
      \vert \psi \rangle
      \langle \psi' \vert
    \big)
    $}
    \\
    {\HilbertSpace{H}^\ast}\Store
    \Big(
    \underset{W}{\randomly}
    \,
    \HilbertSpace{K}
    \Big)
    \ar[
      r,
      phantom,
      "{ \defneq }"
    ]
    &
    \Big(
    \underset{W}{\osum}
    \,
    \HilbertSpace{K}
    \Big)
    \otimes
    \HilbertSpace{H}
    \otimes
    \HilbertSpace{H}^\ast
    \ar[
      from=rr,
      "{
        \distribute
          {
            \randomly_W
            ,\,
            {\HilbertSpace{H}^\ast}\Store
          }
          { \HilbertSpace{K} }
      }",
      "{ \sim }"{swap}
    ]
    &&
    \underset{W}{\osum}
    \big(
    \HilbertSpace{K}
    \otimes
    \HilbertSpace{H}
    \otimes
    \HilbertSpace{H}^\ast
    \big)
    \ar[
      r,
      phantom,
      "{ \defneq }"
    ]
    &
    \underset{W}{\randomly}
    \big(
      \HilbertSpace{H}^\ast\Store
      (\HilbertSpace{K})
    \big)
  \end{tikzcd}
\end{equation}
constitutes a distributivity transformation \eqref{DistributivityOfMonadOverMonad} for
\begin{itemize}
\item
the $\HilbertSpace{H}\State$ monad over the $W$-indefiniteness monad,
\item the $W$-randomness comonad distributing over the $\HilbertSpace{H}^\ast\Store$-comonad.
\end{itemize}
\item[{\bf (ii)}]
the same natural isomorphism, but understood as
\vspace{-2mm}
\begin{equation} \label{DistributivityIsomorphismBetweenQuantumStoreAndQuantumReader}
\hspace{-6mm}
  \begin{tikzcd}[
    row sep=-3pt,
    column sep=25pt
  ]
    {\HilbertSpace{H}^\ast}\Store
    \Big(
    \underset{W}{\indefinitely}
    \,
    \HilbertSpace{K}
    \Big)
    \ar[
      r,
      phantom,
      "{ \defneq }"
    ]
    &
    \Big(
    \underset{W}{\osum}
    \,
    \HilbertSpace{K}
    \Big)
    \otimes
    \HilbertSpace{H}
    \otimes
    \HilbertSpace{H}^\ast
    \ar[
      rr,
      "{
        \distribute
          {
            {\HilbertSpace{H}^\ast}\Store
            ,\,
            \indefinitely_W
          }
          { \HilbertSpace{K} }
      }",
      "{ \sim }"{swap}
    ]
    &&
    \underset{W}{\osum}
    \big(
    \HilbertSpace{K}
    \otimes
    \HilbertSpace{H}
    \otimes
    \HilbertSpace{H}^\ast
    \big)
    \ar[
      r,
      phantom,
      "{ \defneq }"
    ]
    &
    \underset{W}{\indefinitely}
    \big(
      \HilbertSpace{H}^\ast\Store
      (\HilbertSpace{K})
    \big)
    \\
    &
    \hspace{-6pt}
    \scalebox{\termscale}{$
    \big(
      w,
      \,
      \vert \kappa \rangle
    \big)
    \otimes
    \vert \psi \rangle
    \langle \psi' \vert
    $}
    &\longleftrightarrow&
    \scalebox{\termscale}{$
    \big(
      w
      ,\,
      \vert \kappa \rangle
      \otimes
      \vert \psi \rangle
      \langle \psi' \vert
    \big)
    $}
    \\
    {\HilbertSpace{H}}\State
    \Big(
    \underset{W}{\randomly}
    \,
    \HilbertSpace{K}
    \Big)
    \ar[
      r,
      phantom,
      "{ \defneq }"
    ]
    &
    \Big(
    \underset{W}{\osum}
    \,
    \HilbertSpace{K}
    \Big)
    \otimes
    \HilbertSpace{H}
    \otimes
    \HilbertSpace{H}^\ast
    \ar[
      from=rr,
      "{
        \distribute
          {
            \randomly_W
            ,\,
            {\HilbertSpace{H}}\State
          }
          { \HilbertSpace{K} }
      }",
      "{ \sim }"{swap}
    ]
    &&
    \underset{W}{\osum}
    \big(
    \HilbertSpace{K}
    \otimes
    \HilbertSpace{H}
    \otimes
    \HilbertSpace{H}^\ast
    \big)
    \ar[
      r,
      phantom,
      "{ \defneq }"
    ]
    &
    \underset{W}{\randomly}
    \big(
      \HilbertSpace{H}\State
      (\HilbertSpace{K})
    \big)
  \end{tikzcd}
\end{equation}

\vspace{-2mm}
\noindent
constitutes a distributivity transformation \eqref{DistributivityTransformation} for
\begin{itemize}
\item
the quantum $\HilbertSpace{H}^\ast\Store$ comonad over the $W$-indefiniteness monad,
\item the $W$-randomness comonad distributing over the $\HilbertSpace{H}\State$-monad.
\end{itemize}
\end{itemize}
\end{proposition}
\begin{proof}
  The required conditions
  \eqref{MonadOverMonadDistributivityAxioms} and \eqref{CoMonadDistributivityConditions} all hold rather immediately due to the ordinary distributivity
  of the tensor product (being a left adjoint) over the direct sum (being a coproduct, using here that $W$ is a finite type).

\ifdefined\monadology
\else
  For definiteness, we spell this out.
  For \eqref{DistributivityIsomorphismBetweenQuantumStateAndQuantumReader} we check \eqref{MonadOverMonadDistributivityAxioms} in one direction:
$$
\def\arraystretch{10}

$$

\newpage

and in the other direction
$$
\def\arraystretch{9}
%
$$

\newpage
For \eqref{DistributivityIsomorphismBetweenQuantumStoreAndQuantumReader} we check \eqref{CoMonadDistributivityConditions} in one direction
\vspace{-.4cm}
$$
\label{DistributivityConditionsOfQuantumStoreOverQuantumReader}
\hspace{-4cm}
\def\arraystretch{9}
%
\hspace{-4cm}
$$
and in the other direction:
$$
 \label{DistributivityConditionOfRandomnessOverQuantumState}
 \hspace{-15pt}
 \hspace{-4cm}
 \def\arraystretch{10}
%
\hspace{-4cm}
$$
\vspace{-1cm}

\fi
\end{proof}

\begin{remark}[Distributivity is purely structural]
  \label{DistributivityIsPurelyStructural}
  Since the distributivity laws in Prop. \ref{QuantumStoreDistributesOverQuantumRreader} are given just by the structure
  isomorphism of the underlying distributive monoidal category, we may and will leave it notationally implicit, writing
  $
    \underset{W}{\osum}
    \,
    \HilbertSpace{K}
    \otimes
    \HilbertSpace{H}
    \otimes
    \HilbertSpace{H}^\ast
  $
  as usual,
  without any parenthesis.
\end{remark}

In generalization of Prop. \ref{QuantumStoreComonad}, we have:

\medskip

\begin{proposition}[\bf Category of QuantumStore-context-dependent and Indefiniteness-effectful maps]
\label{QuantumStoreIndefinitenesKleisliCategory}
For $\HilbertSpace{H} \,\isa\, \DualizableQuantumType$ and $W \,\isa\, \FiniteClassicalType$,
the jointly
$\HilbertSpace{H}^\ast\Store$-contextful and $\indefinitely_W$-effectful morphisms
\eqref{ContextEffectfulProgram} are in bijection with $W$-indexed sets of linear operators
\vspace{-2mm}
\begin{equation}
  \label{WIndexedLinearOperatorsAsKleisliMaps}
  \begin{tikzcd}[
    row sep=-4pt
  ]
    \HilbertSpace{K}
    \otimes
    \HilbertSpace{H}
    \otimes
    \HilbertSpace{H}^\ast
    \ar[
      rr,
      "{
        \mathcal{O}_{A_\bullet}
      }"
    ]
    &&
    \underset{W}{\indefinitely}
    \,
    \HilbertSpace{K}'
    \\[-2pt]
    \scalebox{\termscale}{$
    \vert \kappa \rangle
    \otimes
    \vert \psi \rangle
    \langle \psi' \vert
    $}
    &\longmapsto&
    \scalebox{\termscale}{$
    \Big(
      w
      \,\mapsto\,
      \langle \psi', -
      \vert A_w
      \vert \kappa, \psi \rangle
    \Big)
    $}
    \\
    &=&
    \scalebox{\termscale}{$
    \Big(
      w
      \,\mapsto\,
      \underset{k'}{\sum}
      \,
      \langle \psi', k'
      \vert A_w
      \vert \kappa, \psi \rangle
      \,
      \vert k' \rangle
    \Big)
    $}
  \end{tikzcd}
  \hspace{.7cm}
  \longleftrightarrow
  \hspace{.7cm}
  \scalebox{\termscale}{$
  \Big(
  A_w
  \,\isa\,
  \HilbertSpace{K}
  \otimes
  \HilbertSpace{H}
  \to
  \HilbertSpace{K}'
  \otimes
  \HilbertSpace{H}
  \Big)_{w \isa W}
  $}
\end{equation}

\vspace{-2mm}
\noindent and their $\HilbertSpace{H}^\ast\Store/\indefinitely_W$-Kleisli composition \eqref{CoMonadicKleisliComposition}
under the distributivity transformation \eqref{QuantumStoreDistributesOverQuantumRreader} corresponds
to the $W$-component wise operator products:
$$
  \big(
  \bind
    { \indefinitely_W  }
    { \HilbertSpace{K}'' }
  \mathcal{O}_{B_\bullet}
  \big)
  \;\circ\;
  \distribute
    {
      \HilbertSpace{H}^\ast\Store,
      \indefinitely_W
    }
    { \HilbertSpace{K}; }
  \;\circ\;
  \big(
  \extend
    {
     \HilbertSpace{H}^\ast\Store
    }
    {
      \HilbertSpace{K}
    }
  \mathcal{O}_{A_\bullet}
  \big)
  \;\;\;
  =
  \;\;\;
  \mathcal{O}_{ (B \cdot A)_\bullet }
  \mathrlap{\,.}
$$
\end{proposition}
\begin{proof}
By the general formula
\eqref{CoMonadicKleisliComposition}
and with Rem. \ref{DistributivityIsPurelyStructural}:
$$
  \begin{tikzcd}[
    column sep=48pt
  ]
  &[-65pt]
  \HilbertSpace{K}
  \otimes
  \HilbertSpace{H}
  \otimes
  \HilbertSpace{H}^\ast
  \otimes
  \HilbertSpace{H}
  \otimes
  \HilbertSpace{H}^\ast
  \ar[
    rr,
    "{
      \mathcal{O}_{A_\bullet}
      \otimes
      \HilbertSpace{H}
      \otimes
      \HilbertSpace{H}^\ast
    }"
  ]
  &[-28pt]&[-28pt]
  \underset{W}{\osum}
  \,
  \HilbertSpace{K}'
  \otimes
  \HilbertSpace{H}
  \otimes
  \HilbertSpace{H}^\ast
  \ar[
    rr,
    "{
      \osum_W
      \,
      \mathcal{O}_{B_\bullet}
    }"
  ]
  \ar[
    drrr,
    "{
      \bind
        { \indefinitely_W }
        {  }
      \mathcal{O}_{B_\bullet}
    }"{sloped, swap}
  ]
  &[-35pt]&[-35pt]
  \underset{W}{\osum}
  \,
  \underset{W}{\osum}
  \,
  \HilbertSpace{K}''
  \ar[
    dr,
    shorten <=-8pt,
    "{
      \join
        { \indefinitely_W }
        { \HilbertSpace{K} }
    }"{sloped}
  ]
  &
  [-10pt]
  \\
  \HilbertSpace{K}
  \otimes
  \HilbertSpace{H}
  \otimes
  \HilbertSpace{H}^\ast
  \ar[
    ur,
    "{
      \duplicate
        { \HilbertSpace{H}^\ast\Store }
        { \HilbertSpace{K} }
    }"{sloped, pos=.3}
  ]
  \ar[
    urrr,
    "{
      \extend
        { \HilbertSpace{H}^\ast\Store }
        { \HilbertSpace{K} }
      \mathcal{O}_{A_\bullet}
    }"{sloped, swap}
  ]
  && && &&
  \underset{W}{\osum}
  \,
  \HilbertSpace{K}''
  \\[-20pt]
  &
  \scalebox{\termscale}{$
  \vert \kappa \rangle
  \otimes
  \underset{h}{\sum}
  \,
  \vert \psi \rangle
  \langle h \vert
  \otimes
  \vert h \rangle
  \langle \psi' \vert
  $}
  \ar[
    rr,
    phantom,
    "{ \longmapsto }"
  ]
  &&
  \scalebox{\termscale}{$
  \Big(
  w
  \mapsto
  \underset{h, k'}{\sum}
  \,
  \langle h, k'
  \vert A_w
  \vert \kappa, \psi \rangle
  \,
  \vert k' \rangle
  \otimes
  \vert h \rangle
  \langle \psi' \vert
  \Big)
  $}
  \ar[
    drr,
    phantom,
    "{ \longmapsto }"{sloped, pos=.3}
  ]
  \\
  \scalebox{\termscale}{$
  \vert \kappa \rangle
  \otimes
  \vert \psi \rangle
  \langle \psi' \vert
  $}
  \ar[
    ur,
    phantom,
    "{ \longmapsto }"{sloped}
  ]
  && && &
  \mathclap{
  \scalebox{\termscale}{$
  \Big(
  w
  \mapsto
  \underset{
    \langle \psi', -
    \vert B_w \cdot A_w \vert
    \kappa, \psi \rangle
  }{
  \underbrace{
  \underset{h, k'}{\sum}
  \,
  \langle \psi', -
  \vert B_w
  \vert k',  h \rangle
  \langle h, k'
  \vert A_w \vert
  \kappa, \psi \rangle
  }
  }
  \Big)
  $}
  }
  \end{tikzcd}
$$
\vspace{-1cm}

\end{proof}

\medskip

\noindent

\begin{example}[\bf State preparation with Probability weights]
\label{StatePreparationWithProbabilityWeights}
Given $W \,\isa\, \FiniteClassicalType$
we have the following basic examples of $W$-environment-contextful and $\quantized W$-effective maps:

\begin{itemize}

\item[{\bf (i)}]
The map $\mathrm{prep}$ which at environmental parameter $w \isa W$ produces (``prepares'') the corresponding pure
basis state $\vert h \rangle \langle h \vert$;
\item[{\bf (ii)}]
for $p \,\isa\, W \to \RealNumbers_{\geq 0}$ a (probability) measure, the map
$\mathrm{weigh}_{p_\bullet}$ which at environmental parameter $w \isa W$ produces the identity (density) matrix with coefficient $p_w$.
\end{itemize}
\vspace{1mm}
$$
  \begin{tikzcd}[
    sep=0pt
  ]
      \mathllap{
      \mathrm{prep}
      \;\isa\;\;
    }
    W\Environment(\TensorUnit)
    \ar[
      rr
    ]
    &&
    \quantized W \State(\TensorUnit)
    \\
    \underset{W}{\randomly}
    \,
    \ComplexNumbers
    \ar[
      rr
    ]
    &&
    \quantized W
    \otimes
    \quantized W^\ast
    \\
   \scalebox{.8}{$   (w,1) $}
    &\longmapsto&
   \scalebox{.8}{$   \vert h \rangle \langle h \vert
   $}
  \end{tikzcd}
  \hspace{2cm}
  \begin{tikzcd}[
    sep=0pt
  ]
      \mathllap{
      \mathrm{weigh}_{p_\bullet}
      \;\isa\;\;
    }
    W\Environment(\TensorUnit)
    \ar[
      rrrr
    ]
    &&&&
    \quantized W \State(\TensorUnit)
    \\
    \underset{W}{\randomly}
    \,
    \ComplexNumbers
    \ar[
      rr,
      "{
        p_\bullet
      }"
    ]
    &&
    \ComplexNumbers
    \ar[
      rr,
      "{
        \unit
          {
            \HilbertSpace{H}\State
          }
          {
            \TensorUnit
          }
      }"
    ]
    &&
    \quantized W
    \otimes
    \quantized W^\ast
    \\
  \scalebox{.8}{$    (w,1) $}
    &\longmapsto&
    p_w
    &\longmapsto&
 \scalebox{.8}{$     p_w
    \cdot
    \underset{h}{\sum}
    \,
    \vert h \rangle \langle h \vert
    $}
    \,.
  \end{tikzcd}
$$
Their two-sided Kleisli composition prepares the mixed state in which the pure state $\vert h \rangle$ appears with weight $p_w$:

\vspace{-.6cm}
$$
  \begin{tikzcd}[
   row sep=0pt,
   column sep=15pt
  ]
    \underset{W}{\randomly}
    \,
    \ComplexNumbers
    \ar[
      rr,
      "{
        \duplicate
          { \randomly_W }
          { \TensorUnit }
      }"
    ]
    \ar[
      rrrrrrrr,
      rounded corners,
      to path={
           ([yshift=+00pt]\tikztostart.north)
        -- ([yshift=+14pt]\tikztostart.north)
        -- node{
             \scalebox{.9}{\colorbox{white}{$
               \mathrm{weight}_{p_\bullet}
               \;\;\mbox{\tt >=>}\;\;
               \mathrm{prep}
             $}}
           }
           ([yshift=+14pt]\tikztotarget.north)
        -- ([yshift=+00pt]\tikztotarget.north)
      }
    ]
    &&
    \underset{W}{\randomly}
    \,
    \underset{W}{\randomly}
    \,
    \ComplexNumbers
    \ar[
      rr,
      "{
        \randomly_W
        \mathrm{weigh}_{p_\bullet}
      }"
    ]
    &&
    \underset{W}{\randomly}
    \,
    \HilbertSpace{H}
    \otimes
    \HilbertSpace{H}^\ast
    \ar[
      rr,
      "{
        \mathrm{prep}
        \otimes
        \HilbertSpace{H}
        \otimes
        \HilbertSpace{H}^\ast
      }"{yshift=2pt}
    ]
    &&
    \underset{W}{\randomly}
    \,
    \HilbertSpace{H}
    \otimes
    \HilbertSpace{H}^\ast
    \otimes
    \HilbertSpace{H}
    \otimes
    \HilbertSpace{H}^\ast
    \ar[
      rr,
      "{
        \join
          { \HilbertSpace{H}\State }
          { \TensorUnit }
      }"
    ]
    &&
    \underset{W}{\randomly}
    \,
    \HilbertSpace{H}
    \otimes
    \HilbertSpace{H}^\ast
    \\
   \scalebox{.8}{$   (w, 1)
   $}
    &\longmapsto&
   \scalebox{.8}{$   \big(w,(w,1)\big)
   $}
    &\longmapsto&
   \scalebox{.8}{$   \big(
      w,
      p_w
      \,
      I_{\HilbertSpace{H}}
    \big)
    $}
    &\longmapsto&
   \scalebox{.8}{$     p_w
      \,
      I_{\HilbertSpace{H}}
      \otimes
      \vert w \rangle \langle w \vert
      $}
    &\longmapsto&
  \scalebox{.8}{$    p_w
    \,
    \vert w \rangle \langle w \vert
    $}
  \end{tikzcd}
$$
$$
  \begin{tikzcd}[
   row sep=0pt,
   column sep=15pt
  ]
    \underset{W}{\randomly}
    \,
    \ComplexNumbers
    \ar[
      rr,
      "{
        \duplicate
          { \randomly_W }
          { \TensorUnit }
      }"
    ]
    \ar[
      rrrrrrrr,
      rounded corners,
      to path={
           ([yshift=+00pt]\tikztostart.north)
        -- ([yshift=+14pt]\tikztostart.north)
        -- node{
             \scalebox{.9}{\colorbox{white}{$
               \mathrm{prep}
               \;\;\mbox{\tt >=>}\;\;
               \mathrm{weigh}_{p_\bullet}
             $}}
           }
           ([yshift=+14pt]\tikztotarget.north)
        -- ([yshift=+00pt]\tikztotarget.north)
      }
    ]
    &&
    \underset{W}{\randomly}
    \,
    \underset{W}{\randomly}
    \,
    \ComplexNumbers
    \ar[
      rr,
      "{
        \randomly_W
        \,
        \mathrm{prep}
      }"
    ]
    &&
    \underset{W}{\randomly}
    \,
    \HilbertSpace{H}
    \otimes
    \HilbertSpace{H}^\ast
    \ar[
      rr,
      "{
        \mathrm{weigh}_{p_\bullet}
        \,
        \otimes
        \HilbertSpace{H}
        \otimes
        \HilbertSpace{H}^\ast
      }"{yshift=2pt}
    ]
    &&
    \underset{W}{\randomly}
    \,
    \HilbertSpace{H}
    \otimes
    \HilbertSpace{H}^\ast
    \otimes
    \HilbertSpace{H}
    \otimes
    \HilbertSpace{H}^\ast
    \ar[
      rr,
      "{
        \join
          { \HilbertSpace{H}\State }
          { \TensorUnit }
      }"
    ]
    &&
    \underset{W}{\randomly}
    \,
    \HilbertSpace{H}
    \otimes
    \HilbertSpace{H}^\ast
    \\
   \scalebox{.8}{$   (w, 1)
   $}
    &\longmapsto&
   \scalebox{.8}{$   \big(w,(w,1)\big)
   $}
    &\longmapsto&
 \scalebox{.8}{$     \big(
      w, \,
      \vert w \rangle \langle w \vert
    \big)
    $}
    &\longmapsto&
    \scalebox{.8}{$    p_w
      \,
      \vert w \rangle \langle w \vert
      \otimes
      I_{\quantized W}
      $}
    &\longmapsto&
   \scalebox{.8}{$   p_w
    \,
    \vert w \rangle \langle w \vert
    $}
    \,
  \end{tikzcd}
$$
\end{example}

\medskip

\begin{lemma}[\bf Distributive monad transformations act on context/effectful-maps]
  \label{DistributiveMonadTransformationsActOnContextEffecfulMaps}
  For $\mathcal{C}$ a comonad distributing
  \eqref{DistributivityTransformation}
  over a pair of monads $\mathcal{E}$, $\mathcal{E}'$
  \vspace{-2mm}
  $$
    \distribute
      { \mathcal{C}, \mathcal{E} }
      {  }
    \,\colon\,
    \mathcal{C}\circ \mathcal{E}
    \longrightarrow
    \mathcal{E} \circ \mathcal{C}
    \,,
    \hspace{1cm}
    \distribute
      { \mathcal{C}, \mathcal{E}' }
      {  }
    \,\colon\,
    \mathcal{C}\circ \mathcal{E}'
     \longrightarrow
    \mathcal{E}' \circ \mathcal{C}
  $$

  \vspace{-2mm}
\noindent  then a monad transformation \eqref{ComponentOfMonadTransformation}
  $$
    \transform
      { \mathcal{E}  \longrightarrow \mathcal{E}' }
      {  }
    \,\colon\,
    \mathcal{E} \to \mathcal{E}'
  $$
  which is compatible with the two distributive laws in that it makes the following diagram commute
\begin{equation}
  \label{MonadTransformationRespectingDistributivity}
  \begin{tikzcd}[row sep=small, column sep=huge]
    \mathcal{C}\big(
      \mathcal{E}(-)
    \big)
    \ar[
      dd,
      "{
        \distribute
          { \mathcal{C}, \mathcal{E} }
          { (-) }
      }"{swap}
    ]
    \ar[
      rr,
      "{
        \mathcal{C}\big(
          \transform
            { \mathcal{E} \to \mathcal{E}' }
            { (-) }
        \big)
      }"
    ]
    &&
    \mathcal{C}\big(
      \mathcal{E}'(-)
    \big)
    \ar[
      dd,
      "{
        \distribute
          { \mathcal{C}, \mathcal{E}' }
          { (-) }
      }"
    ]
    \\
    \\
    \mathcal{E}\big(
      \mathcal{C}(-)
    \big)
    \ar[
      rr,
      "{
        \transform
          { \mathcal{E} \to \mathcal{E}' }
          { \mathcal{C}(-) }
      }"{swap}
    ]
    &&
    \mathcal{E}'\big(
      \mathcal{C}(-)
    \big)
    \mathrlap{\,,}
  \end{tikzcd}
\end{equation}

\vspace{-2mm}
\noindent
respects the Kleisli composition of context-effectful maps \eqref{CoMonadicKleisliComposition} just as it does respect \eqref{MonadTransformationRespectsKleisliComposition} the plain Kleisli composition \eqref{KleisliComposition} of purely-effectful maps:
\vspace{-2mm}
\begin{equation}
\label{MonadTransformationRespectsTwoSidedKleisliComposition}
  \left.
  \def\arraystretch{1.2}
  \begin{array}{l}
    \mathrm{prog}_{12}
    \,\colon\,
    \mathcal{C}(D_1) \to
    \mathcal{E}(D_2)
    \\
    \mathrm{prog}_{23}
    \,\colon\,
    \mathcal{C}(D_2) \to
    \mathcal{E}(D_3)
  \end{array}
  \right\}
  \hspace{.7cm}
  \yields
  \hspace{.7cm}
  \def\arraystretch{2}
  \begin{array}{l}
  \big(
  \transform
    { \mathcal{E} \to \mathcal{E}' }
  { D_2 }
  \circ
  \mathrm{prog}_{12}
  \big)
  \;\;
  \mbox{\tt >=>}
  \;\;
  \big(
  \transform
    { \mathcal{E} \to \mathcal{E}' }
  { D_3 }
  \circ
  \mathrm{prog}_{23}
  \big)
  \\
  =
  \;\;\;\;
  \transform
    { \mathcal{E} \to \mathcal{E}' }
  { D_2 }
  \circ
  \big(
  \mathrm{prog}_{12}
  \;\;
    \mbox{\tt >=>}
  \;\;
  \mathrm{prog}_{23}
  \big)
  \mathrlap{\,.}
  \end{array}
\end{equation}
\end{lemma}
\begin{proof}
Consider the following diagram:
$$
  \begin{tikzcd}[
    column sep=38pt,
    row sep=45pt
  ]
    \mathcal{C}(D_1)
    \ar[
      r,
      "{
        \duplicate
          { \mathcal{C} }
          { D_1 }
      }"
    ]
    \ar[
      drr,
      end anchor={[xshift=-4pt, yshift=-9pt]},
      bend right=10,
      "{
        \extend
          { \mathcal{C} }
          {}
        \big(
        \transform
          {\mathcal{E} \to \mathcal{E}'}
          { D_2 }
        \circ
        \,
        \mathrm{prog}_{12}
        \big)
      }"{swap, sloped}
    ]
    \ar[
      rr,
      rounded corners,
      to path={
           ([yshift=+00pt]\tikztostart.north)
        -- ([yshift=+10pt]\tikztostart.north)
        -- node[yshift=7pt] {
              \scalebox{.7}{$
                \extend
                  { \mathcal{C} }
                  {  }
                \mathrm{prog}_{12}
              $}
           }
           ([yshift=+10pt]\tikztotarget.north)
        -- ([yshift=+00pt]\tikztotarget.north)
      }
    ]
    &
    \mathcal{C}\big(
      \mathcal{C}(D_1)
    \big)
    \ar[
      r,
      "{
        \mathcal{C}(\mathrm{prog}_{12})
      }"{yshift=2pt}
    ]
    \ar[
      dr,
      "{
        \mathcal{C}\big(
          \transform
            {\mathcal{E} \to \mathcal{E}'}
            {D_2}
          \circ
          \,
          \mathrm{prog}_{12}
        \big)
      }"{swap, sloped, pos=.42}
    ]
    &[15pt]
    \mathcal{C}\big(
      \mathcal{E}(D_2)
    \big)
    \ar[
      r,
      "{
        \distribute
          { \mathcal{C}, \mathcal{E} }
          { D_2 }
      }"{yshift=1.5pt}
    ]
    \ar[
      d,
      "{
        \mathcal{C}\big(
        \transform
          { \mathcal{E} \to \mathcal{E}' }
          { D_2 }
        \big)
      }"{description, pos=.45}
    ]
    &
    \mathcal{E}\big(
      \mathcal{C}(D_2)
    \big)
    \ar[
      r,
      "{
        \mathcal{E}(\mathrm{prog}_{23})
      }"{yshift=1.5pt}
    ]
    \ar[
      d,
      "{
        \transform
          { \mathcal{E} \to \mathcal{E}' }
          { \mathcal{C}(D_2) }
      }"{description, pos=.45}
    ]
    \ar[
      rrd,
      rounded corners,
      to path={
           ([yshift=+00pt]\tikztostart.north)
        -- ([yshift=+10pt]\tikztostart.north)
        -- node[yshift=7pt] {
              \scalebox{.7}{$
                \bind
                  { \mathcal{E} }
                  {  }
                \mathrm{prog}_{23}
              $}
           }
           ([yshift=+74pt]\tikztotarget.north)
        -- ([yshift=+00pt]\tikztotarget.north)
      }
    ]
    &[10pt]
    \mathcal{E}\big(
      \mathcal{E}(D_3)
    \big)
    \ar[
      d,
      "{
        \transform
          { \mathcal{E} \to \mathcal{E}' }
          { \mathcal{E}(D_3) }
      }"{description, pos=.45}
    ]
    \ar[
      dr,
      "{
        \join
          { \mathcal{E} }
          { D_3 }
      }"{sloped}
    ]
    \\
    &&
    \mathcal{C}\big(
      \mathcal{E}'(D_3)
    \big)
    \ar[
      r,
      "{
        \distribute
          { \mathcal{C}. \mathcal{E'} }
          { D_2 }
      }"{yshift=1.5pt}
    ]
    &
    \mathcal{E}'\big(
      \mathcal{C}(D_2)
    \big)
    \ar[
      r,
      "{
        \mathcal{E}'(\mathrm{prog}_{23})
      }"{yshift=1.5pt}
    ]
    \ar[
      dr,
      "{
        \mathcal{E}'\big(
          \transform
            { \mathcal{E} \to \mathcal{E}' }
            { D_2 }
          \circ\,
          \mathrm{prog}_{23}
        \big)
      }"{swap, sloped, pos=.37}
    ]
    \ar[
      ddrr,
      bend right=20,
      shift right=15pt,
      shorten=-10pt,
      end anchor={[yshift=10pt]},
      "{
        \bind
          { \mathcal{E}' }
          { }
        \big(
          \transform
            { \mathcal{E} \to \mathcal{E}' }
            { D_2 }
          \circ
          \,
          \mathrm{prog}_{23}
        \big)
      }"{sloped, swap}
    ]
    &
    \mathcal{E}'\big(
      \mathcal{E}(D_3)
    \big)
    \ar[
      d,
      "{
        \mathcal{E}'\big(
          \transform
            { \mathcal{E} \to \mathcal{E}' }
            { D_3 }
        \big)
      }"{description}
    ]
    &
    \mathcal{E}(D_3)
    \ar[
      dd,
      "{
        \transform
          { \mathcal{E} \to \mathcal{E}' }
          { D_3 }
      }"{description}
    ]
    \\
    &&&&
    \mathcal{E}'\big(
      \mathcal{E}'(D_3)
    \big)
    \ar[
      dr,
      "{
        \join
          { \mathcal{E}' }
          { D_3 }
      }"{swap, sloped}
    ]
    &
    \\
    &&&&&
    \mathcal{E}'(D_3)
    \mathrlap{\,.}
  \end{tikzcd}
$$
Here the middle square commutes by the distributivity assumption \eqref{MonadTransformationRespectingDistributivity}, the square to the right of
it due to naturality of the transformation $\transform{\mathcal{E} \to \mathcal{E}'}{}$ and the far right square due to its monad transformation property \eqref{MonadTransformationConditions}. Therefore the total diagram commutes. But its total top and right composite morphism is the right-hand side of \eqref{MonadTransformationRespectsTwoSidedKleisliComposition}, while its total bottom left (diagonal) composite morphism is the left-hand side of \eqref{MonadTransformationRespectsTwoSidedKleisliComposition}, thus proving their equality.
\end{proof}

It follows immediately that:
\begin{lemma}[{\bf QuantumState transformations compatible with distributivity over Quantum Reader}]
  \label{QuantumStateTransformationsCompatibleWithDistributivityOverQuantumReader}
  Every transformation \eqref{ComponentOfMonadTransformation}
  between quantum state monads \eqref{QuantumStateInducedFromLinearHomAdjunction}
  is compatible \eqref{MonadTransformationRespectingDistributivity} with the canonical distributivity \eqref{DistributivityIsomorphismBetweenQuantumStoreAndQuantumReader} over the QuantumReader monads.
\end{lemma}
\begin{proof}
  Use Lem. \ref{NaturalTransformationsBetweenTensoringFunctors}.
\end{proof}

As a corollary of Lem. \ref{DistributiveMonadTransformationsActOnContextEffecfulMaps} and Lem. \ref{QuantumStateTransformationsCompatibleWithDistributivityOverQuantumReader}:
\begin{proposition}[{\bf Preserving Quantum Kleisli composition}]
  Given a quantum channel which acts as a QuantumState transformation \emph{(such as unitary channels by Prop. \ref{UnitaryQuantumChannelsAreQuantumStateTransformations} and uniform coupling channels by Prop. \ref{UniformCouplingChannelsAreQuantumStateTransformations})}
  \vspace{-1mm}
  $$
    \mathrm{chan}
    \,\colon\,
    \HilbertSpace{H}\State
    \to
    \HilbertSpace{H}'\State
  $$
  \vspace{-1mm}

\noindent
then composition of this channel with maps that are Randomness-contextful and QuantumState-effectful preserves their Kleisli composition \eqref{CoMonadicKleisliComposition}, in that:
\vspace{-2mm}
\begin{equation}
\label{QuantumStateTransformationRespectsTwoSidedKleisliComposition}
  \left.
  \def\arraystretch{1.7}
  \begin{array}{l}
    \mathrm{prog}_{12}
    \,\colon\,
    \underset{W}{\randomly}
    \HilbertSpace{K}
    \to
    \HilbertSpace{K}'
    \otimes
    \HilbertSpace{H}
    \otimes
    \HilbertSpace{H}^\ast
    \\
    \mathrm{prog}_{23}
    \,\colon\,
    \underset{W}{\randomly}
    \HilbertSpace{K}'
    \to
    \HilbertSpace{K}''
    \otimes
    \HilbertSpace{H}
    \otimes
    \HilbertSpace{H}^\ast
  \end{array}
  \right\}
  \hspace{.7cm}
  \yields
  \hspace{.7cm}
  \def\arraystretch{2}
  \begin{array}{l}
  \big(
  \mathrm{chan}_{\HilbertSpace{K}'}
  \circ
  \mathrm{prog}_{12}
  \big)
  \;\;
  \mbox{\tt >=>}
  \;\;
  \big(
  \mathrm{chan}_{\HilbertSpace{K}''}
  \circ
  \mathrm{prog}_{23}
  \big)
  \\
  =
  \;\;\;\;
  \mathrm{chan}_{\HilbertSpace{K}''}
  \circ
  \big(
  \mathrm{prog}_{12}
  \;\;
    \mbox{\tt >=>}
  \;\;
  \mathrm{prog}_{23}
  \big)
  \mathrlap{\,.}
  \end{array}
\end{equation}
\end{proposition}

\medskip

\noindent
{\bf Indefinite QuantumStates.} We may now combine the indefiniteness-effects which model quantum measurement and classical control (\cref{ControlledQuantumGates}) with the QuantumState-effects that model mixed states:

\begin{definition}[\bf Category of Quantum State Effects]
  \label{CategoryOfQuantumStateEffects}
  We write
  $$
    \QuantumEffects
    \;\;
    \defneq
    \;\;
    \mathrm{Mnd}\big(
      \QuantumTypes
    \big)
  $$
  for the category of monads -- with monad transformations \eqref{ComponentOfMonadTransformation}
  between them -- on the category of quantum types. And we write
  \begin{equation}    \label{QuantumStateEffectsAmongAllQuantumEffects}
    \begin{tikzcd}
      \QuantumStateEffects
      \ar[
        rr,
        hook
      ]
      &&
      \QuantumEffects
    \end{tikzcd}
  \end{equation}
  for its full subcategory on the QuantumState monads $\HilbertSpace{H}\State$ for $\HilbertSpace{H} \,\isa\, \DualizableQuantumType$.
\end{definition}
\begin{lemma}[{\bf Natural transformations between tensoring functors}]
\label{NaturalTransformationsBetweenTensoringFunctors}
  For $\HilbertSpace{V}_1, \HilbertSpace{V}_2 \,\isa\, \DualizableQuantumTypes$, with
  $$
    (-)
    \otimes
    \HilbertSpace{V}_i
    \,\isa\,
    \QuantumTypes
    \to
    \QuantumTypes
  $$
  the functors of tensoring with these objects, then all natural transformations between them
  $$
    f_{(-)}
    \;\isa\;
    (-)\otimes
    \HilbertSpace{V}_1
    \to
    (-)\otimes
    \HilbertSpace{V}_1
  $$
  are
  given by tensoring with the linear map that is their value on the tensor unit:
  $$
    f_{
      \HilbertSpace{K}
    }
    \;\simeq\;
    \HilbertSpace{K}
      \otimes
    f_{
      \scalebox{.7}{$
      \TensorUnit
      $}
    }
    \,.
  $$
\end{lemma}
\begin{proof}
This follows by the $\QuantumTypes$-enriched Yoneda lemma after observing that the tensor functors $(-) \otimes \HilbertSpace{V}_i$ are representable
  $$
    (-)
    \otimes
    \HilbertSpace{V}_i
    \;\;\simeq\;\;
    (-)
    \otimes
    \big(\HilbertSpace{V}_i^\ast\big)^\ast
    \;\;\simeq\;\;
    \HilbertSpace{V}_i^\ast
    \linmap (-)
    \,.
  $$

  \vspace{-3mm}
  \end{proof}

\begin{lemma}[\bf QuantumState transformations are algebra homomorphisms]
\label{QuantumStateTransformationsAreMonoidHomomorphisms}
  QuantumState transformations are in natural bijection to monoid homomorphisms
  \vspace{-2mm}
  $$
    \begin{tikzcd}[
      sep = 0pt
    ]
      \mathrm{QuantumStateEffects}
      \ar[
        rr,
        hook
      ]
      &&
      \mathrm{Mon}\big(
        \QuantumTypes
      \big)
      \\
      \HilbertSpace{H}\State
      &\mapsto&
      \HilbertSpace{H}
        \otimes
      \HilbertSpace{H}^\ast
    \end{tikzcd}
  $$
\end{lemma}
\begin{proof}
  Via Rem. \ref{QuantumStateEffectAsQuantumWriter}, it is clear that natural transformations of tensor form
  $$
    \begin{tikzcd}[
      row sep=3pt
    ]
      \HilbertSpace{H}_1\State
      \ar[
        rr,
        "{
          (-)
          \otimes
          \phi
        }"
      ]
      &&
      \HilbertSpace{H}_2\State
      \\
      (-)
      \otimes
      \HilbertSpace{H}_1
      \otimes
      \HilbertSpace{H}_1^\ast
      \ar[
        rr,
        "{
          \mathrm{id} \otimes \phi
        }"
      ]
      &&
      (-)
      \otimes
      \HilbertSpace{H}_2
      \otimes
      \HilbertSpace{H}_2^\ast
    \end{tikzcd}
  $$
  are monad transformations if and only if $\phi$ is an algebra homomorphism. Therefore, it only remains to observe that all natural transformations
  are necessarily of this tensor form, which is the statement of Lem. \ref{NaturalTransformationsBetweenTensoringFunctors}.
\end{proof}

\newpage
As a corollary:
\begin{lemma}[{\bf QuantumState and linear maps}]
  The isomorphisms of QuantumState effects are given by conjugation with invertible linear maps. In particular,
  a natural transformation of the form
  \vspace{-2mm}
  $$
    \begin{tikzcd}[
      row sep=4pt, column sep=large
    ]
      \mathllap{
        \mathrm{chan}^H
        \;\isa\;\;
      }
      \HilbertSpace{H}_1\State
      \ar[
        rr
      ]
      &&
      \HilbertSpace{H}_2\State
      \\
      (-)\otimes
      \HilbertSpace{H}_1
      \otimes
      \HilbertSpace{H}_1^\ast
      \ar[
        rr,
        "{
          (-)
          \otimes
          (
          U \otimes {U^\dagger}^\ast
          )
        }"
      ]
      &&
      (-)\otimes
      \HilbertSpace{H}_2
      \otimes
      \HilbertSpace{H}_2^\ast
    \end{tikzcd}
  $$
  is a QuantumState-transformation if and only if $U \,\isa\, \HilbertSpace{H}_1 \to \HilbertSpace{H}_2$ is unitary.
\end{lemma}

\begin{definition}[\bf IndefiniteQuantumState-monad]
  \label{IndefiniteQuantumStateMonad}
  For $W \,\isa\, \FiniteClassicalType$ and $\HilbertSpace{H} \,\isa\, \DualizableQuantumType$, we have the
  composite monad \eqref{CompositeMonad} of the QuantumState- with the Indefiniteness-monad:
  $$
    \begin{tikzcd}[
      row sep=-3pt
    ]
      \mathllap{
        \underset{W}{\indefinitely}
          \circ
         \HilbertSpace{H}\State
        \;\isa\;\;
      }
      \QuantumTypes
      \ar[
        rr
      ]
      &&
      \QuantumTypes
      \\
   \scalebox{0.8}{$    \HilbertSpace{K}$}
      &\longmapsto&
    \scalebox{0.8}{$     \underset{W}{\osum}
      \,
      \HilbertSpace{K}
      \otimes
      \HilbertSpace{H}
      \otimes
      \HilbertSpace{H}^\ast
      $}
    \end{tikzcd}
  $$
\end{definition}
\begin{proposition}[\bf IndefiniteQuantumState-effectful transformations]
  \label{IndefiniteQuantumStateEffectfulTransformations}
  The monad transformations \eqref{ComponentOfMonadTransformation} from a QuantumState-monad (Def. \ref{QuantumStateCoMonads}) to an IndefiniteQuantumState-monad (Def. \ref{IndefiniteQuantumStateMonad})
  $$
    \begin{tikzcd}[
    ]
    \mathllap{
      f
      \;\isa\;\;
    }
    \HilbertSpace{H}_1\State
    \ar[
      r
    ]
    &
    \indefinitely_W \circ \HilbertSpace{H}_2\State
    \end{tikzcd}
  $$
    are in natural bijection to $W$-tuples of algebra homomorphisms.
\end{proposition}
\begin{proof}
  By Lem. \ref{NaturalTransformationsBetweenTensoringFunctors},
  the underlying natural transformation is given by tensoring
  $$
    f_{\HilbertSpace{K}}
    \;\simeq\;
    \HilbertSpace{K}
    \otimes
    f_{\scalebox{.7}{$\TensorUnit$}}
  $$
  with a linear map
  $$
    \begin{tikzcd}[sep=0pt]
    \mathllap{
    f_{
       \scalebox{.7}{$
      \TensorUnit
      $}
    }
    \;\isa\;\;
    }
    \HilbertSpace{H}_1
    \otimes
    \HilbertSpace{H}_1^\ast
    \ar[rr]
    &&
    \underset{W}{\osum}
    \,
    \HilbertSpace{H}_1
    \otimes
    \HilbertSpace{H}_1^\ast
    \\
    A &\mapsto&
    \underset{w}{\oplus}
    \,
    f_{\scalebox{.7}{$\TensorUnit$}}(A)_w
    \,.
    \end{tikzcd}
  $$
  In terms of this, the monad-transformation property of $f_{(-)}$
  $$
    \begin{tikzcd}
      \HilbertSpace{K}
      \ar[
        rr,
        equals
      ]
      \ar[
        dd,
        "{
          \unit
            { \HilbertSpace{H}\State }
            { \HilbertSpace{K} }
        }"{swap}
      ]
      &&
      \HilbertSpace{K}
      \ar[
        dd,
        "{
          \unit
            {
              \indefinitely_W
              \circ
              \HilbertSpace{H}\State
            }
            { \HilbertSpace{K} }
        }"
      ]
      \\
      \\
      \HilbertSpace{K}
      \otimes
      \HilbertSpace{H}_1
      \otimes
      \HilbertSpace{H}_1^\ast
      \ar[
        rr,
        "{
          f_{ \HilbertSpace{K} }
        }"{swap}
      ]
      &&
      \underset{W}{\indefinitely}
      \,
      \HilbertSpace{K}
      \otimes
      \HilbertSpace{H}_2
      \otimes
      \HilbertSpace{H}_2^\ast
    \end{tikzcd}
  $$
  $$
    \begin{tikzcd}[column sep=28pt]
      \HilbertSpace{K}
      \otimes
      \HilbertSpace{H}_1
      \otimes
      \HilbertSpace{H}^\ast_1
      \otimes
      \HilbertSpace{H}_1
      \otimes
      \HilbertSpace{H}^\ast_1
      \ar[
        rr,
        "{
          f_{
            \HilbertSpace{K}
            \otimes
            \HilbertSpace{H}_1
            \otimes
            \HilbertSpace{H}_1^\ast
          }
        }"
      ]
      \ar[
        dd,
        "{
          \join
            { \HilbertSpace{H}_1\State }
            { \HilbertSpace{K} }
        }"{swap}
      ]
      &&
      \underset{W}{\indefinitely}
      \,
      \HilbertSpace{K}
      \otimes
      \HilbertSpace{H}_1
      \otimes
      \HilbertSpace{H}^\ast_1
      \otimes
      \HilbertSpace{H}_2
      \otimes
      \HilbertSpace{H}^\ast_2
      \ar[
        rr,
        "{
          \underset{W}{\indefinitely}
          \,
          f_{
            \HilbertSpace{K}
          }
          \otimes
          \HilbertSpace{H}_2
          \otimes
          \HilbertSpace{H}^\ast_2
        }"
      ]
      &&
      \underset{W}{\indefinitely}
      \,
      \underset{W}{\indefinitely}
      \,
      \HilbertSpace{K}
      \otimes
      \HilbertSpace{H}_2
      \otimes
      \HilbertSpace{H}^\ast_2
      \otimes
      \HilbertSpace{H}_2
      \otimes
      \HilbertSpace{H}^\ast_2
      \ar[
        dd,
        "{
          \join
            {
              \indefinitely_W
              \circ
              \HilbertSpace{H}_2\State
            }
            { \HilbertSpace{K} }
        }"
      ]
      \\
      \\
      \HilbertSpace{K}
      \otimes
      \HilbertSpace{H}_1
      \otimes
      \HilbertSpace{H}_1^\ast
      \ar[
        rrrr,
        "{
          f_{
            \HilbertSpace{K}
          }
        }"{swap}
      ]
      &&&&
      \underset{W}{\indefinitely}
      \,
      \HilbertSpace{K}
      \otimes
      \HilbertSpace{H}_2
      \otimes
      \HilbertSpace{H}_2^\ast
    \end{tikzcd}
  $$
  translates to the condition for $W$-indexed monoid homomorphisms, as claimed:
  $$
    \begin{tikzcd}[row sep=12pt, column sep=large]
      \TensorUnit
      \ar[
        rrrr,
        equals
      ]
      \ar[
        dddd
      ]
      &[-30pt]
      &&
      &[-36pt]
      \TensorUnit
      \ar[
        dddd
      ]
      \\[-15pt]
      &
  \scalebox{0.8}{$     1
  $}
      \ar[
        rr,
        phantom,
        "{ \longmapsto }"
      ]
      \ar[
        dd,
        phantom,
        "{ \longmapsto }"{sloped}
      ]
      &&
  \scalebox{0.8}{$     1 $}
      \ar[
        dd,
        phantom,
        "{ \longmapsto }"{sloped}
      ]
      \\
      \\
      &
    \scalebox{0.8}{$   1_1 $}
      \ar[
        rr,
        phantom,
        "{ \longmapsto }"
      ]
      &&
   \scalebox{0.8}{$    \underset{w}{\oplus}
      1_2
      $}
      \\[-17pt]
      \HilbertSpace{H}_1
      \otimes
      \HilbertSpace{H}^\ast_1
      \ar[
        rrrr,
        "{
          f_{\scalebox{.7}{$\TensorUnit$}}
        }"{swap}
      ]
      &&&&
      \underset{W}{\indefinitely}
      \,
      \HilbertSpace{H}_2
      \otimes
      \HilbertSpace{H}^\ast_2
    \end{tikzcd}
  $$
  $$
    \begin{tikzcd}[
      column sep=15pt
    ]
      \HilbertSpace{H}_1
      \otimes
      \HilbertSpace{H}_1
      \otimes
      \HilbertSpace{H}_1
      \otimes
      \HilbertSpace{H}_1
      \ar[
        rrrr,
        "{
          f_{
            \scalebox{.7}{$
              \TensorUnit
            $}
          }
          \otimes
          f_{
            \scalebox{.7}{$
              \TensorUnit
            $}
          }
        }"
      ]
      \ar[
        dddd,
        "{
          (-)\cdot (-)
        }"
        {swap}
      ]
      &[-40pt]
      &&
      &[-55pt]
      \big(
      \underset{W}{\osum}
      \,
      \HilbertSpace{H}_2
      \otimes
      \HilbertSpace{H}_2
      \big)
      \otimes
      \big(
      \underset{W}{\osum}
      \,
      \HilbertSpace{H}_2
      \otimes
      \HilbertSpace{H}_2
      \big)
      \ar[
        dddd,
        "{
          \underset{w}{\oplus}
          \big(
          (-)_w \cdot (-)_w
          \big)
        }"
      ]
      \\[-20pt]
      &
  \scalebox{0.8}{$  A \otimes B $}
      \ar[
        rr,
        phantom,
        "{ \longmapsto }"
      ]
      \ar[
        dd,
        phantom,
        "{ \longmapsto }"{sloped}
      ]
      &&
     \scalebox{0.8}{$  \big(
      \underset{w}{\oplus}
      \,
      f_{\scalebox{.7}{$\TensorUnit$}}(
       A
      )_w
      \big)
      \otimes
      \big(
      \underset{w}{\oplus}
      \,
      f_{\scalebox{.7}{$\TensorUnit$}}(
       B
      )_w
      \big)
      $}
      \ar[
        dd,
        phantom,
        "{ \longmapsto }"{sloped}
      ]
      \\
      \\
      &
   \scalebox{0.8}{$    A \cdot B
   $}
      \ar[
        rr,
        phantom,
        "{ \longmapsto }"
      ]
      &&
   \scalebox{0.8}{$    \underset{w}{\oplus}
      \,
      f_{\scalebox{.7}{$\TensorUnit$}}(
       A \cdot B
      )_w
      $}
      \\[-20pt]
      \HilbertSpace{H}_1
      \otimes
      \HilbertSpace{H}_1^\ast
      \ar[
        rrrr,
        "{
          f_{
            \scalebox{.7}{$
              \TensorUnit
            $}
          }
        }"{swap}
      ]
      &&&&
      \underset{W}{\indefinitely}
      \,
      \HilbertSpace{H}_2
      \otimes
      \HilbertSpace{H}_2^\ast
    \end{tikzcd}
  $$

  \vspace{-5mm}
\end{proof}

\begin{proposition}[\bf Indefiniteness-effect on QuantumState-effects]
  \label{IndefinitenessEffectOnQuantumStateEffect}
  For $W \,\isa\, \FiniteClassicalType$
  and $\HilbertSpace{H} \,\isa\, \DualizableQuantumType$,
  the construction of the
  IndefiniteQuantumState-monad $\underset{W}{\indefinitely} \circ \HilbertSpace{H}\State$ (Def. \ref{IndefiniteQuantumStateMonad}) extends to a relative monad \eqref{RelativeBindOperation} on, in turn, the category of quantum state effects (Def. \ref{CategoryOfQuantumStateEffects}):
  \vspace{-2mm}
  $$
    \begin{tikzcd}[
      sep=2pt
    ]
      \mathllap{
        \indefinitely_W \,\circ\,
        (-)
        \;\isa\;\;
      }
      \QuantumStateEffects
      \ar[
        rr,
        "{}"
      ]
      &&
      \QuantumEffects
      \\
      \HilbertSpace{H}\State
      &\longmapsto&
      \underset{W}{\indefinitely}
      \circ
      \HilbertSpace{H}\State
    \end{tikzcd}
  $$

  \vspace{-2mm}
\noindent
relative to the full inclusion \eqref{QuantumStateEffectsAmongAllQuantumEffects}.
\end{proposition}
\begin{proof}
  The return-operation is
  \vspace{-2mm}
  \begin{equation}
    \label{ReturnForIndefinitenessMonadOnQuantumStateEffects}
    \begin{tikzcd}[
      row sep=0pt, column sep=large
    ]
        \unit
          {
            \indefinitely_W
            \circ
          }
          {
            \HilbertSpace{H}\State
          }
      \ar[
        r,
        phantom,
        "{ \isa }"
      ]
      &
      \HilbertSpace{H}\State
      \ar[
        rr
      ]
      &&
      \underset{W}{\indefinitely}
      \circ
      \HilbertSpace{H}\State
      \\
      \big(
        \unit
          {
            \indefinitely_W
            \circ
          }
          {
            \HilbertSpace{H}\State
          }
        \big)_{\HilbertSpace{K}}
      \ar[
        r,
        phantom,
        "{ \isa }"
      ]
      &
      \HilbertSpace{K}
      \otimes
      \HilbertSpace{H}
      \otimes
      \HilbertSpace{H}^\ast
      \ar[
        rr,
        "{
          \unit
            { \indefinitely_W }
            {
              \HilbertSpace{K}
              \otimes
              \HilbertSpace{H}
              \otimes
              \HilbertSpace{H}^\ast
            }
        }"{yshift=2pt}
      ]
      &&
      \underset{W}{\osum}
      \,
      \HilbertSpace{K}
      \otimes
      \HilbertSpace{H}
      \otimes
      \HilbertSpace{H}^\ast
      \\
      &
   \scalebox{0.8}{$    \vert \kappa \rangle
      \otimes
      \vert \psi \rangle
      \langle \psi' \vert
      $}
      &\longmapsto&
    \scalebox{0.8}{$   \underset{w}{\sum}
      \big(
      w,\,
      \vert \kappa \rangle
      \otimes
      \vert \psi \rangle
      \langle \psi' \vert
      \big)
      $}
    \end{tikzcd}
  \end{equation}

  \vspace{-2mm}
\noindent   and the bind-operation takes a monad transformation
  $f \,\isa\,
  \HilbertSpace{H}_1\State \to \underset{W}{\indefinitely} \circ \HilbertSpace{H}_2\State$
  to
  $
    \join
     { \indefinitely_W }
     {}
    \circ
    \underset{W}{\indefinitely}
    \,
    f
    \,.
  $
That this satisfies the axioms of a relative monad follows immediately from the monad structure on $\indefinitely_W \,\isa\, \QuantumTypes \to \QuantumTypes$.
But for this to be well-defined as a monad on monads, we do in addition need to check that the return- and bind-operations now are actually morphisms in $\QuantumEffects$:

That \eqref{ReturnForIndefinitenessMonadOnQuantumStateEffects} is a monad transformation follows by the definition \eqref{StructureMapsOfCompositeMonad}
of the composite monad alone, which immediately shows that these diagrams commute:
\vspace{-2mm}
$$
  \begin{tikzcd}[row sep=small, column sep=large]
    \HilbertSpace{K}
    \ar[
      rr,
      equals
    ]
    \ar[
      dd,
      "{
        \unit
          {
            \HilbertSpace{H}\State
          }
          {
            \HilbertSpace{K}
          }
      }"{swap}
    ]
    &&
    \HilbertSpace{K}
    \ar[
      dd,
      "{
        \unit
          {
            \indefinitely_W
            \circ
            \HilbertSpace{H}\State
          }
          {
            \HilbertSpace{K}
          }
      }"
    ]
    \\
    \\
    \HilbertSpace{K}
    \otimes
    \HilbertSpace{H}
    \otimes
    \HilbertSpace{H}^\ast
    \ar[
      rr,
      "{
        \unit
          { \indefinitely_W }
          {
            \HilbertSpace{K}
            \otimes
            \HilbertSpace{H}
            \otimes
            \HilbertSpace{H}^\ast
          }
      }"
    ]
    &&
    \underset{W}{\indefinitely}
    \,
    \HilbertSpace{K}
    \otimes
    \HilbertSpace{H}
    \otimes
    \HilbertSpace{H}^\ast
  \end{tikzcd}
$$
$$
  \begin{tikzcd}[column sep=29pt]
    \HilbertSpace{K}
    \otimes
    \HilbertSpace{H}
    \otimes
    \HilbertSpace{H}^\ast
    \otimes
    \HilbertSpace{H}
    \otimes
    \HilbertSpace{H}^\ast
    \ar[
      rr,
      "{
        \unit
          { \indefinitely_W }
          {
            \HilbertSpace{K}
            \otimes
            \HilbertSpace{H}
            \otimes
            \HilbertSpace{H}^\ast
            \otimes
            \HilbertSpace{H}
            \otimes
            \HilbertSpace{H}^\ast
          }
      }"{yshift=1pt}
    ]
    \ar[
      dd,
      "{
        \join
          {
            \HilbertSpace{H}\State
          }
          {
            \HilbertSpace{K}
          }
      }"{swap}
    ]
    &&
    \underset{W}{\indefinitely}
    \big(
    \HilbertSpace{K}
    \otimes
    \HilbertSpace{H}
    \otimes
    \HilbertSpace{H}^\ast
    \otimes
    \HilbertSpace{H}
    \otimes
    \HilbertSpace{H}^\ast
    \big)
    \ar[
      rr,
      "{
      \scalebox{0.8}{$   \underset{W}{\indefinitely}
        \,
        \unit
          { \indefinitely_W }
          {
            \HilbertSpace{K}
            \otimes
            \HilbertSpace{H}
            \otimes
            \HilbertSpace{H}^\ast
          }
            \otimes
            \HilbertSpace{H}
            \otimes
            \HilbertSpace{H}^\ast
            $}
      }"{yshift=2pt, pos=.5}
    ]
    \ar[
      ddrr,
      shorten <= -6pt,
      "{
        \underset{W}{\indefinitely}
        \,
        \join
          {
            \HilbertSpace{H}\State
          }
          {
            \HilbertSpace{K}
          }
      }"{description, sloped}
    ]
    &&
    \underset{W}{\indefinitely}
    \Big(
    \big(
    \underset{W}{\indefinitely}
    \,
    \HilbertSpace{K}
    \otimes
    \HilbertSpace{H}
    \otimes
    \HilbertSpace{H}^\ast
    \big)
    \otimes
    \HilbertSpace{H}
    \otimes
    \HilbertSpace{H}^\ast
    \Big)
    \ar[
      dd,
      "{
        \join
          {
            \indefinitely_W
            \circ
            \HilbertSpace{H}\State
          }
          {
            \HilbertSpace{K}
          }
      }"
    ]
    \\
    \\
    \HilbertSpace{K}
    \otimes
    \HilbertSpace{H}
    \otimes
    \HilbertSpace{H}^\ast
    \ar[
      rrrr,
      "{
        \unit
         { \indefinitely_W }
         {
           \HilbertSpace{K}
           \otimes
           \HilbertSpace{H}
           \otimes
           \HilbertSpace{H}^\ast
         }
      }"
    ]
    &&&&
    \underset{W}{\indefinitely}
    \,
    \HilbertSpace{K}
    \otimes
    \HilbertSpace{H}
    \otimes
    \HilbertSpace{H}^\ast
    \otimes
    \HilbertSpace{H}
    \otimes
    \HilbertSpace{H}^\ast
  \end{tikzcd}
$$
That the effect-binding of $f$ is still a monad transformation follows from the fact that $f$ itself is assumed to be
a monad transformation and using Prop. \ref{IndefiniteQuantumStateEffectfulTransformations}:
$$
  \begin{tikzcd}
    \HilbertSpace{K}
    \ar[
      rrrr,
      equals
    ]
    \ar[
      dddd,
      "{
        \unit
          {  \HilbertSpace{H}_1\State}
          { \HilbertSpace{K} }
      }"{swap}
    ]
    &[-40pt]&&&[-40pt]
    \HilbertSpace{K}
    \ar[
      dddd,
      "{
        \unit
          {
            \indefinitely_W
            \circ
            \HilbertSpace{H}_1\State
          }
          { \HilbertSpace{K} }
      }"
    ]
    \\[-15pt]
    &
 \scalebox{0.8}{$    \vert \kappa \rangle
 $}
    \ar[
      rr,
      phantom,
      "{ \longmapsto }"
    ]
    \ar[
      dd,
      phantom,
      "{ \longmapsto }"{sloped}
    ]
    &&
   \scalebox{0.8}{$  \vert \kappa \rangle
   $}
    \ar[
      dd,
      phantom,
      "{ \longmapsto }"{sloped}
    ]
    \\
    \\
    &
  \scalebox{0.8}{$   \vert \kappa \rangle
    \otimes
    1_1
    $}
    \ar[
      rr,
      phantom,
      "{ \longmapsto }"
    ]
    &&
  \scalebox{0.8}{$   \vert \kappa \rangle
    \otimes
    \underset{w}{\oplus}
        1_2
    $}
    \\[-20pt]
    \HilbertSpace{K}
    \otimes
    \HilbertSpace{H}_1
    \otimes
    \HilbertSpace{H}_1^\ast
    \ar[
      rrrr,
      "{
        f_{\HilbertSpace{K}}
      }"{swap}
    ]
    &&&&
    \underset{W}{\indefinitely}
    \,
    \HilbertSpace{K}
    \otimes
    \HilbertSpace{H}_2
    \otimes
    \HilbertSpace{H}_2^\ast
  \end{tikzcd}
$$
$$
  \begin{tikzcd}[
    column sep=15pt,
    row sep=15pt
  ]
    \underset{W}{\indefinitely}
    \,
    \underset{W}{\indefinitely}
    \,
    \HilbertSpace{K}
    \otimes
    \HilbertSpace{H}_1
    \otimes
    \HilbertSpace{H}_1^\ast
    \otimes
    \HilbertSpace{H}_1
    \otimes
    \HilbertSpace{H}_1^\ast
    \ar[
      dd,
      "{
        \underset{W}{\indefinitely}
        \,
        f_{
          \scalebox{.7}{$
          \indefinitely_W
          \,
          \HilbertSpace{K}
          \otimes
          \HilbertSpace{H}_1
          \otimes
          \HilbertSpace{H}_1^\ast
          $}
        }
      }"{description}
    ]
    \ar[
      rrrr,
      "{
        \join
          {
            \indefinitely_W
            \circ
            \HilbertSpace{H}_1\State
          }
          {
            \HilbertSpace{K}
          }
      }"
    ]
    &[-20pt]
    &&
    &[-35pt]
    \underset{W}{\indefinitely}
    \,
    \HilbertSpace{K}
    \otimes
    \HilbertSpace{H}_1
    \otimes
    \HilbertSpace{H}_1^\ast
    \ar[
      dddd,
      "{
        \underset{W}{\indefinitely}
        \,
        f_{\HilbertSpace{K}}
      }"{description}
    ]
    \\
    &
  \scalebox{0.8}{$   \vert \kappa \rangle
    \otimes
      \big(
        w
        ,\,
        A
      \big)
      \otimes
      \big(
        w'
        ,\,
        A'
      \big)
      $}
    \ar[
      ddd,
      phantom,
      "{ \longmapsto }"{sloped}
    ]
    \ar[
      rr,
      phantom,
      "{ \longmapsto }"
    ]
    &&
 \scalebox{0.8}{$    \vert \kappa \rangle
    \otimes
      \big(
        w
        ,\,
        \delta_w^{w'}
        A \cdot A'
      \big)
      $}
    \ar[
      dddddd,
      phantom,
      "{ \longmapsto  }"{sloped}
    ]
    \\
    \underset{W}{\indefinitely}
    \,
    \underset{W}{\indefinitely}
    \,
    \underset{W}{\indefinitely}
    \,
    \HilbertSpace{K}
    \otimes
    \HilbertSpace{H}_1
    \otimes
    \HilbertSpace{H}_1^\ast
    \otimes
    \HilbertSpace{H}_2
    \otimes
    \HilbertSpace{H}_2^\ast
    \ar[
      dd,
      "{
        \join
          { \indefinitely_W }
          {
            \indefinitely_W
            \,
            \HilbertSpace{K}
            \otimes
            \HilbertSpace{H}
            \otimes
            \HilbertSpace{H}^\ast
            \otimes
            \HilbertSpace{H}
            \otimes
            \HilbertSpace{H}^\ast
          }
      }"{description}
    ]
    \\
    \\
    \underset{W}{\indefinitely}
    \,
    \underset{W}{\indefinitely}
    \,
    \HilbertSpace{K}
    \otimes
    \HilbertSpace{H}_1
    \otimes
    \HilbertSpace{H}_1^\ast
    \otimes
    \HilbertSpace{H}_2
    \otimes
    \HilbertSpace{H}_2^\ast
    \ar[
      dd,
      "{
        \underset{W}{\indefinitely}
        \,
        \underset{W}{\indefinitely}
        f_{
          \HilbertSpace{K}
        }
        \,
        \otimes
        \HilbertSpace{H}_2
        \otimes
        \HilbertSpace{H}_2
      }"{description}
    ]
    &
\scalebox{0.8}{$     \vert \kappa \rangle
    \otimes
      \big(
        w
        ,\,
        A
      \big)
      \otimes
      \big(
        w'
        ,\,
        f_{
          \scalebox{.7}{$\TensorUnit$}
        }
        (A')_{w'}
      \big)
      $}
    \ar[
      ddd,
      phantom,
      "{ \longmapsto }"{sloped}
    ]
    &&&
    \underset{W}{\indefinitely}
    \,
    \underset{W}{\indefinitely}
    \,
    \HilbertSpace{K}
    \otimes
    \HilbertSpace{H}_2
    \otimes
    \HilbertSpace{H}^\ast_2
    \ar[
      dddd,
      "{
        \join
          {
            \indefinitely_W
          }
          {
            \HilbertSpace{K}
            \otimes
            \HilbertSpace{H}_2
            \otimes
            \HilbertSpace{H}_2^\ast
          }
      }"{description}
    ]
    \\
    \\
    \underset{W}{\indefinitely}
    \,
    \underset{W}{\indefinitely}
    \,
    \underset{W}{\indefinitely}
    \,
    \HilbertSpace{K}
    \otimes
    \HilbertSpace{H}_2
    \otimes
    \HilbertSpace{H}_2^\ast
    \otimes
    \HilbertSpace{H}_2
    \otimes
    \HilbertSpace{H}_2^\ast
    \ar[
      dd,
      "{
        \indefinitely_W
        \join
          { \indefinitely_W }
          {
            \HilbertSpace{K}
            \otimes
            \HilbertSpace{H}_2
            \otimes
            \HilbertSpace{H}_2^\ast
          }
        \otimes
        \HilbertSpace{H}_2
        \otimes
        \HilbertSpace{H}_2^\ast
      }"{description}
    ]
    \\
    &
\scalebox{0.8}{$     \vert \kappa \rangle
    \otimes
      \big(
        w
        ,\,
        f_{
          \scalebox{.7}{$\TensorUnit$}
        }
        (A)_w
      \big)
      \otimes
      \big(
        w'
        ,\,
        f_{
          \scalebox{.7}{$\TensorUnit$}
        }
        (A')_{w'}
      \big)
      $}
    \ar[
      rr,
      phantom,
      "{ \longmapsto }"
    ]
    &&
   \scalebox{0.8}{$  \vert \kappa \rangle
    \otimes
      \big(
        w
        ,\,
        \delta_w^{w'}
        f_{
          \scalebox{.7}{$\TensorUnit$}
        }
        (A \cdot A')_{w}
      \big)
      $}
    \\
    \underset{W}{\indefinitely}
    \,
    \underset{W}{\indefinitely}
    \,
    \HilbertSpace{K}
    \otimes
    \HilbertSpace{H}_2
    \otimes
    \HilbertSpace{H}_2^\ast
    \otimes
    \HilbertSpace{H}_2
    \otimes
    \HilbertSpace{H}_2^\ast
      \ar[
      rrrr,
      "{
        \join
          {
            \indefinitely_W
            \circ
            \HilbertSpace{H}_1\State
          }
          {
            \HilbertSpace{K}
          }
      }"{swap}
    ]
    &&&&
    \underset{W}{\indefinitely}
    \,
    \HilbertSpace{K}
    \otimes
    \HilbertSpace{H}_2
    \otimes
    \HilbertSpace{H}_2^\ast
\end{tikzcd}
$$

\end{proof}

\smallskip

\begin{proposition}[\bf Enhancing parameterized quantum circuits to parameterized quantum channels]
\label{LiftingParamaterizedQuantumCircuitsToParameterizedQuantumChannels}
  The $W$-componentwise unitary $\indefiniteness_W$-effectful maps of $\QuantumTypes$ lift
  via \eqref{TensorPairingOfIndefinitenessEffectfulMaps}
  to  $\indefiniteness_W \circ$-effectful maps on $\mathrm{QuEffect}$
  $$
    \begin{tikzcd}
      \QuantumTypes
        ^{\mathrm{untr}}
        _{\indefinitely_W}
      \ar[
        rr
      ]
      &&
      \QuantumEffects
        _{\indefinitely_W \circ}
      &[-10pt]
      \\[-3pt]
      \HilbertSpace{H}_1
      \ar[
        rr,
        phantom,
        "{ \longmapsto }"
      ]
      \ar[
        dd,
        "{
          U_\bullet
        }"
      ]
      &&
      \HilbertSpace{H}_1\State
      \ar[
        dd,
        "{
          \mathrm{chan}^{U_\bullet}
        }"
      ]
      &
      (-) \otimes
      \HilbertSpace{H}_1
      \otimes
      \HilbertSpace{H}^\ast_1
      \ar[
        d,
        "{
          (-)
          \otimes
          U_\bullet
          \otimes
          {U_\bullet^\dagger}^\ast
        }"
      ]
      \\
      && &
      (-)
      \otimes
      \underset{W}{\osum}
      \,
      \HilbertSpace{H}_1
      \otimes
      \underset{W}{\osum}
      \,
      \HilbertSpace{H}^\ast_1
      \ar[
        d,
        "{
          (-)
          \otimes
          \pair
            { \indefinitely_W }
            {
              \HilbertSpace{H}_2
              ,
              \HilbertSpace{H}_2^\ast
            }
        }"
      ]
      \\
      \underset{W}{\indefinitely}
      \,
      \HilbertSpace{H}_2
      &&
      \underset{W}{\indefinitely}
      \circ
      \HilbertSpace{H}_2\State
      &
      \underset{W}{\osum}
      \,
      (-) \otimes
      \HilbertSpace{H}_2
      \otimes
      \HilbertSpace{H}^\ast_2
    \end{tikzcd}
  $$
\end{proposition}
\begin{proof}
  It remains to see that the paired Kleisli maps \eqref{TensorPairingOfIndefinitenessEffectfulMaps} are indeed QuantumState monad-transformations. This is ensured by the unitarity assumption, as in Prop. \ref{UnitaryQuantumChannelsAreQuantumStateTransformations}.
\end{proof}

\ifdefined\monadology

\section{Quantum Language}
\label{Pseudocode}

With all quantum effects identified --
\ifdefined\monadology
in  \cref{QuantumEffects}
\else
in  \cite{QuantumTypeSemantics} and \cref{QuantumEffects}
\fi
-- as (co)monads definable through the Motivic Yoga (Def. \ref{MotivicYoga}), we may follow established language paradigms for monadic effects (Lit. \ref{LiteratureProgrammingSyntaxForMonadicEffects})
to obtain a natural
quantum language -- to be called
{\QS} \footnote{We call this language ``\QS'', both as shorthand for ``Quantum Systems Language'' as well as alluding to the remarkable
fact that (the semantics of) its universe of quantum data types goes far beyond the usual (Hilbert-) vector spaces to include
``higher homotopy'' linear types (``spectra''): Over the ground ``field'' $\mathbb{F}_1$,  the quantization modality $\quantized$
takes the spherical homotopy types $S^n$ to the ``sphere spectrum'' traditionally denoted ``$Q S^n$''.} --
that should be be embeddable as a domain-specific language (Lit. \ref{LiteratureComputationalEffectsAndModalities}) into any dependent linear type theory which verifies the Motivic Yoga, notably into  {\LHoTT} (Lit. \ref{LiteratureLHoTT}).

\medskip

\medskip

\medskip

\cref{QSLanguageDesign}: Pseudocode Design

\cref{ExampleCode}: Example Pseudocode

\else
\fi

\subsection{Pseudocode Design}
\label{QSLanguageDesign}

In the spirit of traditional \mbox{\tt do}-notation for monadic computational effects (Lit. \ref{LiteratureProgrammingSyntaxForMonadicEffects})
our ambition is to find (sugaring to) an accurate but neatly intelligible formal language for the monadic quantum effects which is close to a natural description of the coded processes.
For that purpose, we employ syntactic sugar both for effect binding and for pure effects \eqref{BindingAndReturning}:

\begin{itemize}
  \item[{\bf (i)}] {\bf Syntactic sugar for effect-binding.}

  For effect-binding we use traditional {\mbox{\tt do}}-notation but in the more verbose form of {\mbox{\tt for}...\mbox{\tt do}}-blocks
  \eqref{ForDoNotationForEffectBinding},
 \item[{\bf (ii)}]
   {\bf Syntactic sugar for pure effects.}

  We furthermore sugar the {\mbox{\tt return}}-operation of each effect such as to notationally indicate the nature of the pure datum that
  is being returned \eqref{PseudocodeForEpistemicModalities}.
\end{itemize}

\medskip

First we discuss the declaration of plain linear maps (quantum gates). Recall our convention \eqref{LinearTypeDeclaration} to write an ``open colon'' ``$\isalin$'' for typing judgements in the context of the linear tensor unit, which we will use throughout.

\smallskip

\noindent
{\bf Declaration of linear maps out of the tensor unit.}
To start with, in declaring linear maps out of the linear tensor unit it should, by linearity, be sufficient to declare the value on the unit element
\begin{equation}
  \label{LinearMapOutOfTensorUnit}
  \declarelin
    { \phi }
    { \TensorUnit \linmap \HilbertSpace{H}}
    {
      1
        \,\mapsto\,
      \phi(1)
      \mathrlap{\,.}
    }
\end{equation}
Self-evident as this may seem, this is ultimately a consistency demand on the ambient linear type theory, which must provide the corresponding elimination rule for the tensor unit. In {\LHoTT} this is the case: \cite[p. 55]{Riley22} speaks of the {\it $\mathbb{S}$-elimination}- or  {\it $\mathbb{S}$-induction}-rule (where the notation ``$\mathbb{S}$'' alludes to the sphere spectrum, which is the tensor unit in the expected model of {\LHoTT} in parameteroized plain spectra, aka $\mathbb{S}$-modules.)

\smallskip

\noindent
{\bf Declaration of linear maps out of a linear span.}
Recall that the quantization modality $\quantized$ (Def. \ref{QuantizationModality}) is just the quantumly-modality $\quantumly$ restricted to classical types along the operation $\TensorUnit \times (-)$
\[
  \quantized
  \;\defneq\;
  \quantumly\big(
    (-)
    \times
    \TensorUnit
  \big)
  \,.
\]
Regarded as a restriction of $\quantumly$, it binds not just $\quantized$-effects but generally $\quantumly$-effects, cf. \eqref{QuantizationBind}. Now, $\quantumly$ is idempotent \eqref{SubcategoriesOfBundleTypes}, meaning that for every linear type is a free $\quantumly$-modale: $\HilbertSpace{H} \,=\, \quantumly \HilbertSpace{H}$.

In conclusion this means that do-notation applies to to declare linear maps (quantum gates) of the form $G \,\isalin\, \quantized W_1 \multimap \HilbertSpace{H}$, whose
domain is equipped with a linear basis $W$ with corresponding basis vectors are denoted $\vert w \rangle \,\isalin\, \quantized W$
\eqref{BindReturnForQuantizationModality}, while the codomain may be any linear type.

In natural language, we would describe such a map by declaring what it does {\it for} a given
basis vector $\vert w \rangle$  -- namely sending it to $\vert G_w\rangle \,:=\, G \vert w\rangle$ -- and we want this natural description
to essentially already be our syntax, as follows:
\vspace{-3mm}
\begin{equation}
  \label{LinearMapInForDoNotation}
  \declarelinLeftAligned
    { G }
    { \quantized W \multimap \HilbertSpace{H} }
    {
      \fordo
        { \vert w \rangle  }
        {
          G \vert w \rangle
          \mathrlap{\,.}
        }
    }
\end{equation}
Indeed, this is the traditional \mbox{\tt do}-notation
(Lit. \ref{LiteratureProgrammingSyntaxForMonadicEffects})
in \mbox{\tt for}...\mbox{\tt do}-form \eqref{ForDoNotationForEffectBinding},
applied to the quantum modality, except for a further sugaring of the plain ``$w$'' to its pure-effect incarnation ``$\vert w \rangle$''.
This notation naturally reflects that $\quantized W$ is freely {\it generated}
\begin{itemize}
\item[{\bf (i)}] in the sense of generating sets of vector spaces:
by the vectors $\vert w \rangle$
\item [{\bf (ii)}] in the sense of free $\quantumly$-modales: by the elements $(w,1) \,\isalin\, W \times \TensorUnit$,
\end{itemize}
and the operation which relates these two incarnations of the generators is $\return{\quantized}{W}$ \eqref{BindReturnForQuantizationModality}, namely:
\vspace{-2mm}
\begin{equation}
  \label{KetAsQReturn}
  \vert w \rangle
  \;\;\;\defneq\;\;\;
  \return{\quantized}{W}(w)
  \;\;\defneq\;\;
  \return{\quantumly}{
    W \!\times\! \TensorUnit
  }\big(
    (w,1)
  \big)
  \;\;\isalin\;\;
  \quantized W
  \,.
\end{equation}
Therefore the natural do-notation for the $\quantumly$-bind operation on a linear map
$$
  G\vert-\rangle
  \;\isalin\;
  W \times \TensorUnit
  \linmap
  \HilbertSpace{H}
  \,,
$$
-- which according to \eqref{LinearMapOutOfTensorUnit} is specified by its value on the elements $(w,1)$ whose natural name in $\quantized W$ is $\vert w \rangle$ -- is the above \eqref{LinearMapInForDoNotation}.

\noindent
{\bf Declaration of linear maps out of a tensor product.}
In the same vein, for declaring a linear map out of a tensor product, one would naturally want the following syntax, defining its value {\it for} each decomposable tensor:
\begin{equation}
  \label{DoNotationForMapsOutOfTensorProduct}
  \declarelinLeftAligned
    { G }
    {
      \quantized W_1
        \otimes \quantized W_2
      \multimap
      \HilbertSpace{H}
    }
    {
      \fordo
        {
          \vert w_1 \rangle
            \otimes
          \vert w_2 \rangle
        }
        {
          G
          \big(
          \vert w_1 \rangle
            \otimes
          \vert w_2 \rangle
          \big)
        }
    }
\end{equation}
Now understanding
$$
  \quantized W_1
  \otimes
  \quantized W_2
  \;\;=\;\;
  \quantumly
  \big(
  \quantized W_1
  \otimes
  \quantized W_2
  \big)
$$
again as a restriction of the quantum modality -- to the entanglement relative monad \eqref{TensorProductBind} -- we may indeed take this as the coresponding do-notation subject only to the further convention that, as before, we refer to the argument via its pure effect incarnation:
$$
  \vert w_1 \rangle
    \otimes
  \vert w_2 \rangle
  \;\;\defneq\;\;
  \return
    {
      \quantized(\mbox{-})
      \otimes
      \quantized(\mbox{-})
    }
    { (W_1, W_2) }
    (w_1, w_2)
  \,.
$$

For example, with \eqref{LinearMapInForDoNotation} and \eqref{DoNotationForMapsOutOfTensorProduct} the operations which witness the strong $\otimes$-monoidal property of $\quantized$ may thus be coded as follows:
\vspace{-2mm}
\begin{equation}
  \declarelinLeftAligned
  {\mu}
  {
    \quantized W_1 \otimes \quantized W_2
    \maplin
    \quantized( W_1 \times W_2 )
  }
  {
    \fordo
      {
        \vert w_1 \rangle
          \otimes
        \vert w_2 \rangle
      }
      {
        \vert w_1, w_2 \rangle
      }
  }
  \hspace{1cm}
  \declarelinLeftAligned
  {\mu^{-1}}
   {
    \quantized( W_1 \times W_2 )
    \maplin
    \quantized W_1
      \otimes
    \quantized W_2
   }
   {
    \fordo
      { \vert w_1, w_2 \rangle }
      {
        \vert w_1 \rangle
        \otimes
        \vert w_2 \rangle
      }
   }
\end{equation}

\vspace{-2mm}
\noindent
and the tensor product on maps
out of linear spans
is given by
\vspace{-2mm}
$$
  \def\arraystretch{1.4}

  \;\;\;\;\;\;
  \yields
  \;\;\;\;\;\;
  \declarelin
    { G \otimes G' }
    {
      \quantized W
        \otimes
      \quantized W'
      \maplin
      \HilbertSpace{H}
        \otimes
      \HilbertSpace{H}'
    }
    {
     \fordo
       {
         \vert w \rangle
           \otimes
          \vert w' \rangle
       }
       {
         G\vert w \rangle
           \otimes
          G'\vert w' \rangle
          \mathrlap{\,,}
       }
    }
$$

\vspace{-2mm}
\noindent
and the ``pipe''-notation ``$\pipe$`` for the sequential composition of maps may be declared as follows:
\vspace{-2mm}
$$
 \def\arraystretch{1.5}
 %
   \hspace{.7cm}
   \yields
   \hspace{.7cm}
   \declarelinLeftAligned
     {
       \Phi_{12}
       \pipe
       \Phi_{23}
     }
     {
       \quantized W_1
         \maplin
        \HilbertSpace{H}_3
     }
     {
       \fordo
         {
           \vert w \rangle
         }
         {
           \Phi_{23}\big(
             \Phi_{12}\vert w \rangle
           \big)
           \mathrlap{\,.}
         }
     }
$$

In summary, to obtain neat pseudo-code we adopt and adapt traditional \mbox{\tt do}-notation as follows:

\begin{equation}
\label{ComparisonOfMonadicDeclarationsOfLinearMaps}
\adjustbox{}{
\def\arraystretch{1.6}
%
}
\end{equation}

\medskip

In  linguistic generalization of this situation we therefore proceed to identify similarly suggestive verbalization of the structure maps of the other monadic effects from \cref{QuantumEpistemicLogicViaDependentLinearTypes}

\medskip

\begin{center}
\def\arraystretch{1.5}
%
$
}
\end{center}

This syntax is to closely reflect the fact that

-- {\it for} an input of the form $\return{\mathcal{E}}{D}(d) \isa \mathcal{E}D$,

-- which may appear as a
{\it pure effect} {\it generator in}  the input data

-- the operation $\bind{\mathcal{E}}{\mathrm{prog}}$ {\it does} produce the output $\mathrm{prog}(d)$,

which prescription completely defines it, by linearity.

\medskip

\begin{equation}
  \label{PseudocodeForEpistemicModalities}
  \def\arraystretch{2}
  \arraycolsep=3pt
  %

\noindent
\begin{remark}[Towards natural language]
$\,$

\begin{itemize}
\item[{\bf (i)}]
The above sugared \mbox{\tt for}...\mbox{\tt do}-notation for classically-controlled quantum gates again neatly
expresses the actual physical process in almost natural language: In general, the input state of a $W$-controlled
quantum gate is itself a $W$-dependent quantum state $\vert \psi_w \rangle$, whence the epistemic declaration of
$G_\bullet$ is of the form
$$
  w \isa W
  \;\;\;
    \yields
  \;\;\;
  G_w
  \;\defneq\;
  \big(
  \vert \psi_w \rangle
  \;\mapsto\;
  G_w \vert \psi_w \rangle
  \big)
  \,,
$$
but for describing the action of $G_w$ on a generic state it does not matter whether this state carries a $w$-index,
and this is what the \mbox{\tt for}...\mbox{\tt do}-notation reflects: It is sufficient to define $G_w$ assuming that
we are {\it definitely} presented with the state $\vert \psi \rangle$ (no matter the value of $w$), hence sufficient
to define it {\it for} states of the form $\definitely\, \vert \psi \rangle$.

\item[{\bf (ii)}]
With the components of the classically-controlled quantum gate themselves being coherent quantum gates, the latter
may in turn be declared on basis states as before, which gives the following further nested declaration of a
classically-controlled quantum gate, reducing to its component output states
$\big( {\color{purple}G_w \vert b \rangle} \,\isa\, \HilbertSpace{K}\big)_{(w,b) \isa W \times B}$:
\vspace{-2mm}
$$
  \adjustbox{
  }{
    \begin{minipage}{6cm}
      \small
      Declaration of a measurement-controlled
      quantum gate in terms of its component values
      on each basis state $\vert b \rangle$ for each
      measurement result $w$.
    \end{minipage}
  }
  \hspace{1cm}
  \declarelinLeftAligned
    { G_\bullet }
    {
      \indefinitely_W \quantized B
      \maplin
      \indefinitely_W \HilbertSpace{K}
    }
    {
      \fordo
      {
        \definitely
        \,
        \vert \psi \rangle
      }
      {
        \ifmeasuredthen
          { w }
          {
            \fordo
              {
                \vert b \rangle
                \among
                \vert \psi \rangle
              }
              {
                 {\color{purple}
                  G_w \vert b \rangle
                 }
              }
          }
      }
    }
$$

\item[{\bf (iii)}] The \mbox{\tt return}-sugaring in the \mbox{\tt for}...\mbox{\tt do}-blocks is just that: The semantics of all notations in \eqref{ComparisonOfMonadicDeclarationsOfLinearMaps} are exactly identical.
In particular, a declaration ``\mbox{\tt for} $\vert b \rangle$'' has access to the actual variable $b \isa B$. For instance we can declare linear maps that duplicate the given basis states (needed in \cref{QuantumBitFlipCode} below, for purposes of constructing ``logical qbits'') as follows
\vspace{-2mm}
$$
  \declarelin
    { \mathrm{encode} }
    {
      \quantized W
      \maplin
      \quantized (W \times W \times W)
    }
    {
      \fordo
        {
          \vert w \rangle
        }
        {
          \vert w,\,w,\,w \rangle
        }
    }
$$

\end{itemize}

\end{remark}

\subsection{Example Pseudocode}
\label{ExampleCode}

\subsubsection{Standard $\mathrm{QBit}$-gates}

For reference we show a few basic quantum gates declared in {\QS}-pseudocode, all of which examples of the general scheme \eqref{LinearMapInForDoNotation},
according to which a general linear map on $\QBit$ is coded by:
$$
  \declarelinLeftAligned
    {\Phi}
    { \QBit \multimap \QBit }
    {
     \fordo
       { \vert b \rangle  }
       { \Phi\vert b \rangle }
    }
$$

\medskip

The quantum $\mathrm{NOT}$-gate:

\begin{equation}
\label{QSCodeForNOTGate}
\declarelinLeftAligned
  {\mathrm{X}}
  {\QBit \multimap \QBit}
  {
    \fordo
      {
        \vert b \rangle
      }
      {
        \vert 1 - b \rangle
      }
  }
\end{equation}

\medskip

\vspace{1cm}

The $\mathrm{CNOT}$-gate \eqref{TraditionalCNOTGate}

\begin{equation}
\label{QSCodeForCNOTGate}
\declarelinLeftAligned
  { \mathrm{CNOT} }
  {
    \quantized(\Bit \times \Bit)
      \multimap
    \quantized(\Bit \times \Bit)
  }
  {
    \fordo
    {
      \vert b_1 ,\, b_2 \rangle
    }
    {
      \vert
        b_1
        ,\;
        b_1 \,\mbox{\tt xor}\, b_2
      \rangle
    }
  }
\end{equation}

\medskip

The Hadamard gate:\footnote{
  The irrtraditional factor $1/\sqrt{2}$ in the Hadamard gate -- whose implementation in a formal language like {\LHoTT}, while  certainly possible, opens a can of worms (cf. \cite[pp. 71]{TQP})  --
  has the purpose of making the map be unitary with respect to the canonical Hermitian inner product structure on $\QBit$\ifdefined\monadology\else(cf. \cref{QuantumProbability})\fi. But since we are not imposing the Hermitian structure in the $\QBit$ data type, for the time being, the factor could as well be omitted for ease of full formalization of the pseudo-code, at the small cost of picking up some irrelevant factors of 2 in subsequent expressions. For example, the
  quantum teleportation protocol \cref{QuantumTeleportationProtocol} without these prefactors in $\Hadamard$
  will not strictly reproduce the input state $\vert \psi \rangle$, but return it multiplied by 2 -- which is physically still the same state, of course, up to normalization.
}

\begin{equation}
\label{QSCodeForHadamardGate}
\declarelinLeftAligned
  {\Hadamard}
  {\QBit \multimap \QBit}
  {
    \fordo
      {
        \vert b \rangle
      }
      {
        \tfrac{1}{\sqrt{2}}
        \Big(
        \vert 0 \rangle
        +
        (-1)^b
        \vert 1 \rangle
        \Big)
      }
  }
\end{equation}

\medskip

The Bell state:

\begin{equation}
  \label{TheBellState}
  \declarelinLeftAligned
    { \mathrm{BellState} }
    { \QBit \otimes \QBit }
    {
      \tfrac{1}{\sqrt{2}}
      \big(
        \vert 0 \rangle \otimes \vert 0 \rangle
        +
        \vert 1 \rangle \otimes \vert 1 \rangle
      \big)
    }
\end{equation}

In typical discussion of $\QBit$-circuits, the initial $\QBit$-states are all assumed to be $\vert 0 \rangle$, and the Bell state \eqref{TheBellState} is prepared by sending $\vert 0 \rangle \otimes \vert 0 \rangle$  through the quantum circuit $(\Hadamard \otimes \mathrm{id}) \pipe \CNOT$ (cf. the first step in the circuit shown on page \pageref{ProblemOfCertifyingClassicalControl}). With the identification types available in {\LHoTT} it is possible to construct a formal certificate that this indeed yields the intended state:
$$
      \mbox{
        \tt
        verify\_Bell\_preparation
      }
      \;\;\;\;
      \isa
      \;\;\;\;
      \mathrm{BellState}
        \;\;\;=\;\;\;
      \vert 0 \rangle
      \otimes
      \vert 0 \rangle
      \pipe
      (\Hadamard \otimes \mathrm{id})
      \pipe
      \CNOT
$$

\newpage

\subsubsection{Quantum Teleportation Protocol}
\label{QuantumTeleportationProtocol}

In combined exposition
of {\QS}-pseudocode
and
of the quantum teleportation protocol (as shown in the circuit diagram in page \pageref{ProblemOfCertifyingClassicalControl}, originally due to \cite{BennetEtAl93}, see \cite[\S 1.3.7]{NielsenChuang10}\cite[\S 3.3]{BEZ20})
we narrate the logic of quantum teleportation by perpetually switching between natural and {\QS}-language:

\medskip

The punchline of quantum teleportation is to send a quantum state $\vert \psi \rangle$ (typically: a qbit) into a process ``Alice''
$$
  \vert \psi \rangle \pipe \mathrm{Alice}(\cdot)
$$
which itself only records classical measurement results (concretely: a pair of bits):
$$
  \mathrm{Alice}(\cdot)
  \;\isalin\;
  \overset{
    \mathclap{
      \raisebox{5pt}{
        \scalebox{.7}{
          \color{darkblue}
          \bf
          \def\arraystretch{.9}
          \begin{tabular}{c}
            quantum
            \\
            input
          \end{tabular}
        }
      }
    }
  }{
    \QBit
  }
    \maplin
  \overset{
    \mathclap{
      \raisebox{5pt}{
        \scalebox{.7}{
          \color{darkblue}
          \bf
          \def\arraystretch{.9}
          \begin{tabular}{c}
            classical
            \\
            output
          \end{tabular}
        }
      }
    }
  }{
  \underset{\Bit^2}{\indefinitely}
  \TensorUnit
  }
$$
and yet such that the transmission of this purely {\it classical} information $\indefinitely_{\Bit^2}$ to a further process ``Bob``:
$$
  \mathrm{Alice}(\cdot)
  \pipe
  \mathrm{Bob}(\cdot)
$$
allows the latter to re-construct a quantum state
$$
  \mathrm{Bob}(\cdot)
  \;\isalin\;
  \underset{\Bit^2}{\indefinitely}
  \TensorUnit
  \maplin
  \underset{
    \Bit^2
  }{\indefinitely}
  \QBit
$$
which is {\it definitely} equal to the initial state (ie. independently of Alice's intermediate measurement intermedi results):
$$
  \mathrm{verify}
  \;\;\;\isa\;\;\;
  \vert \psi \rangle
  \pipe
  \mathrm{Alice}(\cdot)
  \pipe
  \mathrm{Bib}(\cdot)
  \;\;\;\;\;\;
    \overset{{\color{gray}?}}{=}
  \;\;\;\;\;\;
  \mbox{\tt definitely}_{{}_{\Bit^2}}
  \;
  \vert \psi \rangle
  \,.
$$
For this to really work we need to fill in one missing ingredient indicated by ``$(\cdot)$'', namely the two processes need to ``share an entanglement source'' up front, in that they need to share the two ``halfs'' of a Bell state pair of maximally entangle qbits \eqref{TheBellState}, like this:
$$
  \mathrm{verify}
  \;\;\;\isa\;\;\;
  \fordo
    {
      \vert {\color{blue}\mathrm{bell}_A} \rangle
      \otimes
      \vert {\color{orange}\mathrm{bell}_B} \rangle
      \among
      \BellState
    }
    {
  \vert \psi \rangle
  \pipe
  \mathrm{Alice}({\color{blue}\mathrm{bell}_A})
  \pipe
  \mathrm{Bob}({\color{orange}\mathrm{bell}_B})
  }
  \;\;\;\;\;\;
    \overset{}{=}
  \;\;\;\;\;\;
  \mbox{\tt definitely}_{{}_{\Bit^2}}
  \;\;
  \vert \psi \rangle
  \,.
$$
Thus, the global structure of the quantum teleportation protocol is given by the following code:
\displaycode
{
  \declarelinLeftAligned
    { \mathrm{teleport} }
    {
      \QBit
      \multimap
      \underset{
        \Bit^2
      }{\indefinitely}
      \QBit
    }
    {
    {
      \fordo
        { \vert b \rangle }
      {
      \fordo
        {
          \vert
            {\color{blue}\mathrm{bell}_1}
          \rangle
            \otimes
          \vert
            {\color{orange}\mathrm{bell}_2}
          \rangle
          \among
          \mathrm{BellState}
        }
        {
          \vert b \rangle
          \pipe
          \mathrm{Alice}
          \scalebox{1.3}{$($}
            \vert
              {\color{blue}\mathrm{bell}_1}
            \rangle
          \scalebox{1.3}{$)$}
          \pipe
          \mathrm{Bob}
          \scalebox{1.3}{$($}
            \vert
              {\color{orange}\mathrm{bell}_2}
            \rangle
          \scalebox{1.3}{$)$}
        }
    }
    }
    }
}
and it remains to declare the sub-processes $\mathrm{Alice}$ and $\mathrm{Bob}$.

The procedure of Alice's protocol is to
\begin{itemize}[leftmargin=2cm]
\item[(1.)] entangle the input state with the Bell state

\item[(2.)] feed the result through a suitable quantum gate and then

\item[(3.)] measure in the $\Bit^2$-basis and return the measurement result
\end{itemize}
like this:
\displaycode{
  \declarelinLeftAligned
    { \mathrm{Alice} }
    {
      \QBit
      \maplin
      \big(
        \QBit
        \maplin
        \underset{\Bit^2}{\indefinitely}
        \mathbbm{1}
      \big)
    }
    {
  \fordo
    {
      \vert \mathrm{bell}_1 \rangle
    }
    {
      \fordo
        {
          \vert b \rangle
        }
        {
          \big(
            \vert b \rangle
              \otimes
            \vert \mathrm{bell}_1 \rangle
          \big)
          \pipe
          \mathrm{CNOT}
          \pipe
          (\Hadamard \otimes \mathrm{id})
          \pipe
          \collapse
       }
    }
    }
}
The crux is that with the classical information received from Alice,
Bob can apply quantum gates to his part of the Bell-state {\it conditioned on} this classical information, like this:
\displaycode{
  \declarelinLeftAligned
    { \mathrm{Bob} }
    {
      \QBit
      \maplin
      \big(
        \underset{\Bit^2}{\indefinitely}
        \mathbbm{1}
        \maplin
        \underset{\Bit^2}{\indefinitely}
        \QBit
      \big)
    }
    {
  \fordo
    { \vert \mathrm{bell}_2 \rangle }
    {
      \ifmeasuredthen
        { (b_1, b_2) }
        {
            \vert \mathrm{bell}_2 \rangle
            \pipe
            \mathrm{X}^{b_1}
            \pipe
            \mathrm{Z}^{b_2}
        }
    }
    }
}
The categorical semantics of this code, when in turn expressed in string diagram notation, gives the usual circuit-diagram for the quantum teleportation protocol as shown on page \pageref{ProblemOfCertifyingClassicalControl}. But now the correct encoding of the protocol becomes formally {\it verifiable}:

\medskip
If these procedures $\mathrm{Alice}$ and $\mathrm{Bob}$ are correctly coded, then the quantum state which $\mathrm{Bob}$ re-constructs from his Bell-state is definitely equal to the one that Alice originally received (independent of the random measurement results that Alice obtained), and we will be able to certify this property at compile-time by constructing a term of the following identification-type:
\displaycode{
{\color{purple}\mathrm{verify}}
  \;\;\;\;\;
  :
  \;\;\;\;\;
\underset{
  \scalebox{.6}{$
  \vert \psi \rangle
  \isalin
  \QBit
  $}
}{\prod}
\Big(
\mathrm{teleport}\; \vert \psi \rangle
\;=\;
\definitely \; \vert \psi \rangle
\Big)
}

\newpage

\subsubsection{Quantum Bit Flip Code}
\label{QuantumBitFlipCode}

{\bf Bit flip error correction as {\QS}-pseudocode,} is another simple but instructive example (cf. \cite[\S 10.1.1]{NielsenChuang10}):

\vspace{2mm}
\hspace{-.9cm}
\adjustbox{scale=.92}{

$
\\
\hline
\end{tabular}
}
\vspace{.1cm}

\begin{remark}
\label{VerificationOfBitFlipCode}
The last line asserts a term of identification type which {\it formally certifies} that any single bit flip on a logically encoded qbit is {\it always} corrected by the code (i.e.: no matter the measurement outcome). The construction of such certificates in {\LHoTT} (not shown here, but straightforward in the present case) provides the desired formal verification of classically controlled quantum algorithms and protocols.
\end{remark}


\end{document}